\documentclass[11pt]{article}

\usepackage[english]{babel}
\usepackage[a4paper,top=2.5cm,bottom=2.5cm,left=2.5cm,right=2.5cm,marginparwidth=1.75cm]{geometry}

\usepackage{amsmath,amsfonts,amssymb}
\usepackage{bbm}
\numberwithin{equation}{section}

\usepackage{comment}

\usepackage[svgnames,x11names,table]{xcolor}
\usepackage{hyperref}
\hypersetup{
colorlinks=true,final=true,
        linkcolor=DodgerBlue4,
        citecolor=DodgerBlue4,
        filecolor=DodgerBlue4,
        urlcolor=DodgerBlue4,
}

\usepackage{graphicx}

\usepackage{enumitem}

\usepackage{textpos}
\usepackage{array}
\newcommand{\vpos}{-1.4}
\newcommand{\hpos}{11}

\newcommand{\av}[1]{\ensuremath{\left\langle #1 \right\rangle}}
\newcommand{\qv}{\mathbf{q}}
\newcommand{\kv}{\mathbf{k}}
\DeclareMathOperator{\Tr}{Tr}

\usepackage{tikz}
\usetikzlibrary{calc}
\usetikzlibrary{decorations.pathmorphing}
\usetikzlibrary{decorations.pathreplacing}
\usetikzlibrary{decorations.markings}

\tikzset{->-/.style={decoration={
  markings,
  mark=at position #1 with {\arrow{>}}},postaction={decorate}}}
\tikzset{-<-/.style={decoration={
  markings,
  mark=at position #1 with {\arrow{<}}},postaction={decorate}}}

\usepackage[square,numbers,sort&compress,comma]{natbib}

\begin{document}

\thispagestyle{empty}

\begin{textblock}{4}(-2.4,-1.55)
    \textblockcolour{blue}
    \includegraphics[height=850pt]{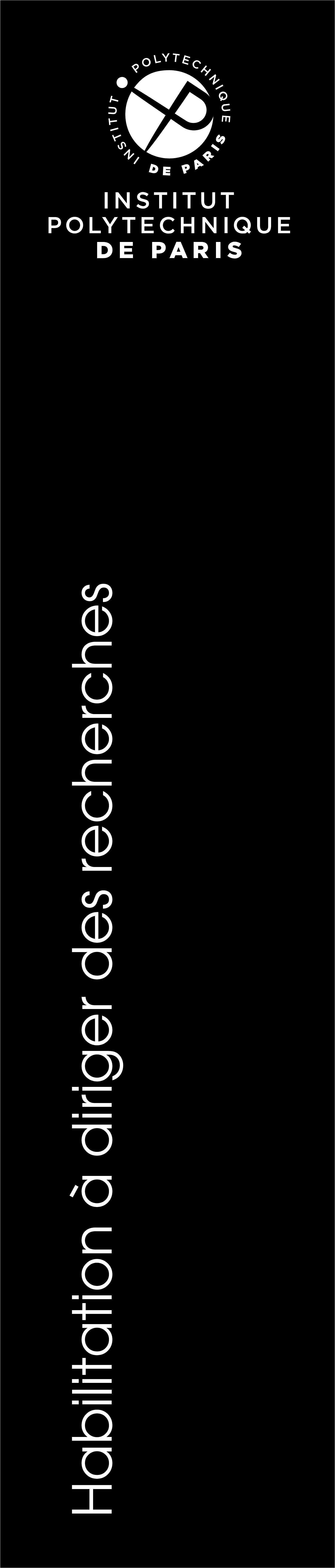}
    \vspace{300mm}
\end{textblock}

\begin{textblock}{1}(\hpos,\vpos)
    \textblockcolour{white}
    \includegraphics[scale=1]{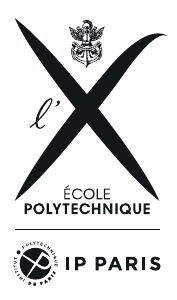}
\end{textblock}

\begin{textblock}{9}(4,3)
    \textblockcolour{white}

    \begin{center}
        \huge{Diagrammatics in the Dual Space,\\ {\it or}\\ There and Back Again}
    \end{center}

    \bigskip
    \bigskip

    \begin{flushright}
        \normalsize{Habilitation \`a diriger des recherches \\de l'Institut Polytechnique de Paris}

        \smallskip

        \small{Sp\'ecialit\'e : Physique}

        \bigskip

        \footnotesize{Habilitation soutenue \`a Palaiseau, le 25 Novembre 2025, par}

        \bigskip

        \Large{Evgeny Stepanov}
    \end{flushright}

    \bigskip

    \begin{flushleft}
        \small Composition du jury :
    \end{flushleft}

    \small

    \newcolumntype{L}[1]{>{\raggedright\let\newline\\\arraybackslash\hspace{0pt}}m{#1}}
    \newcolumntype{R}[1]{>{\raggedleft\let\newline\\\arraybackslash\hspace{0pt}}rm{#1}}

    \begin{flushleft}
    \begin{tabular}{@{} L{9.65cm} @{ } R{4.2cm}}
    Luca Perfetti \\ Professeur, LSI, \'Ecole Polytechnique, France & Pr\'esident \\
    Massimo Capone \\ Professeur, SISSA, Trieste, Italy & Rapporteur \\
    Luca de' Medici \\ Professeur, ESPCI Paris, France & Rapporteur \\
    Marco Schir\`o \\ Directeur de recherche CNRS, Coll\`ege de France, Paris, France & \hspace{-0.15cm}Rapporteur \\
    Roser Valent\'i \\ Professeure, Goethe Universit\"at Frankfurt, Allemagne  & Examinatrice
    \end{tabular}
    \end{flushleft}

\end{textblock}

\pagenumbering{roman}
\clearpage
~
\clearpage
\section*{Abstract}

Accurately describing many-body effects in multi-orbital systems remains a major challenge in theoretical condensed matter physics. 
Beyond the collective charge, spin, and superconducting fluctuations already present in effective single-orbital models, realistic materials also host orbital degrees of freedom, which greatly enrich the range of physical phenomena they display. 
In strongly interacting electronic systems, these fluctuating degrees of freedom become entangled, making it difficult to predict which collective effects dominate the system's behavior. 
At present, there is a significant methodological gap between the numerical tools used in \emph{ab initio} computational materials science and those developed to study strong electronic correlations. 
The former can treat realistic, large-scale systems but typically neglect many-body effects, while the latter focus on simplified models with only a few degrees of freedom, as only such models can be solved accurately in the presence of strong interactions.

The purpose of this thesis is to bridge these two approaches and establish a systematic theoretical framework for realistic correlated electronic materials. 
This involves a full-cycle methodology that begins with constructing \emph{ab initio} interacting models from density-functional theory (DFT), solving them using dynamical mean-field theory (DMFT) to capture local correlations, and extending beyond DMFT to incorporate non-local collective electronic fluctuations. 
To this end, we introduce the ``dual'' approach to strong correlations, which includes the dual fermion (DF), dual boson (DB), and dual triply irreducible local expansion (\mbox{D-TRILEX}) methods. 
The central idea of the dual theories is to shift the reference point of the conventional Feynman diagrammatic expansion from a non-interacting electronic system to an interacting but exactly solvable one, such as the effective local impurity problem of DMFT. 
Integrating out this reference system recasts the expansion in an effective dual space, where all diagrammatic building blocks are renormalized by the corresponding impurity quantities. This procedure transforms a non-perturbative expansion in the original variables into a perturbative expansion in terms of dual fermionic and bosonic fields, exact in both the weak- and strong-coupling limits.

In this thesis, we collect and systematize the major developments of dual techniques achieved to date. 
We begin with the general derivation of the dual approach and discuss its advantages compared to conventional perturbative diagrammatic methods. 
We then introduce the DF and DB approximations, which restrict the electronic interaction to the two-particle level and focus on classes of Feynman diagrams believed to be essential for describing the leading particle-hole and particle-particle scattering processes in correlated electronic systems. 
These approximations are benchmarked against exact solutions where available, and their most prominent applications are presented.
Next, we introduce a partially bosonized approximation of the renormalized electronic interaction (the four-point vertex function), which represents the latter as electronic scattering mediated by a single bosonic fluctuation. 
This approximation drastically simplifies the diagrammatic expansion and forms the basis of the \mbox{D-TRILEX} approach. 
Benchmarks demonstrate that this approximation preserves numerical accuracy while substantially reducing computational cost, thereby enabling efficient implementation in multi-band frameworks. 
The capabilities of \mbox{D-TRILEX} are illustrated on both model systems and realistic materials.
We further show that adopting an instantaneous approximation for the fermion-boson coupling reduces the diagrammatic structure of \mbox{D-TRILEX} to a $GW$-like form. 
Unlike standard $GW$, however, this framework allows the simultaneous treatment of multiple fluctuating channels (charge, spin, etc.), enabling a real time formulation called dual $GW$ (\emph{D-GW}) theory. 
This method holds particular promise for addressing correlated systems under time-dependent perturbations, especially in transport calculations.
Finally, we demonstrate that the dual approach allows mapping the original interacting electronic problem onto an effective bosonic model describing charge and spin degrees of freedom. 
This not only makes it possible to compute all relevant exchange interactions between charge and spin densities, but also to derive the equations of motion that correctly describe the system's dynamics - including spin precession through the Landau-Lifshitz-Gilbert equation and Higgs-like fluctuations of magnetic moments and charge densities.

\clearpage
\markboth{}{}
\tableofcontents
\clearpage

\pagenumbering{arabic}

\newpage
\section{Introduction}

Materials with strong Coulomb correlations display signatures of a collective electronic behavior that manifests itself in fascinating electronic, magnetic, optical, and transport properties. 
These materials, typically characterized by partially filled $d$ or $f$ atomic shells, include transition metal oxides, heavy fermions, low-dimensional systems, and notable high-temperature superconducting compounds such as cuprates, ruthenates, iron-based superconductors, and nickelates. 
In the correlated metal regime the electron dynamics in these systems is characterized by enhanced effective masses of electrons and a strong renormalization of the electronic dispersion. 
Such effects have been observed in many correlated materials by low-T specific heat, optical conductivity, angle-resolved photoemission spectroscopy (ARPES), and quantum oscillations measurements~\cite{PhysRevLett.112.177001, deMedici2017}. 
These features are a direct consequence of many-body renormalization effects due to strong electronic correlations. The local Coulomb repulsion between electrons in these materials favors a single electron occupancy of an orbital and drives the system towards a Mott-insulating state~\cite{Mott, RevModPhys.70.1039}. 
Hund's exchange coupling (intra-atomic exchange) lowers the cost in repulsive Coulomb energy when placing two electrons in different orbitals with parallel spin instead of with antiparallel spin. 
In materials with non-degenerate orbitals the Hund's coupling plays a role of a band decoupler, which means that correlation effects in the system become strongly dependent on the individual filling and the electronic structure of each band~\cite{PhysRevB.83.205112, Hunds_metals1}. 
Under certain circumstances the combined effect of the Coulomb interaction and the Hund's coupling may result in an orbital-selective Mott phase, which is characterized by the coexistence of localized electrons in some orbitals with itinerant electrons in other orbitals~\cite{PhysRevLett.101.126401, Kou_2009, de2009genesis, Hackl_2009}. 

The theoretical description of correlated materials requires to go beyond single-particle approaches (such as the Kohn-Sham construction of density functional theory (DFT)) that implicitly approximate the interaction between electrons by an effective potential~\cite{RevModPhys.87.897}. 
Moreover, simple perturbative (weak-coupling) methods, such as $GW$~\cite{GW1, GW2, GW3} or fluctuation-exchange (FLEX)~\cite{PhysRevLett.62.961, Bickers89} approaches, are also not suitable for this task, because the essential role of local correlations in correlated materials can be captured with sufficient accuracy only via elaborate non-perturbative methods. 
Indeed, even a rather advanced two-particle self-consistent (TPSC) approach~\cite{Vilk94, Vilk97}, that is based on a non-perturbative treatment of the local Coulomb interaction, is unable to describe the most prominent effect of local electronic correlations, i.e., the Mott transition, and is of limited accuracy already in a weak-coupling regime if magnetic fluctuations in the system are strong~\cite{PhysRevX.11.011058}.
 
\begin{figure}[t!]
\centering
\includegraphics[width=0.95\linewidth]{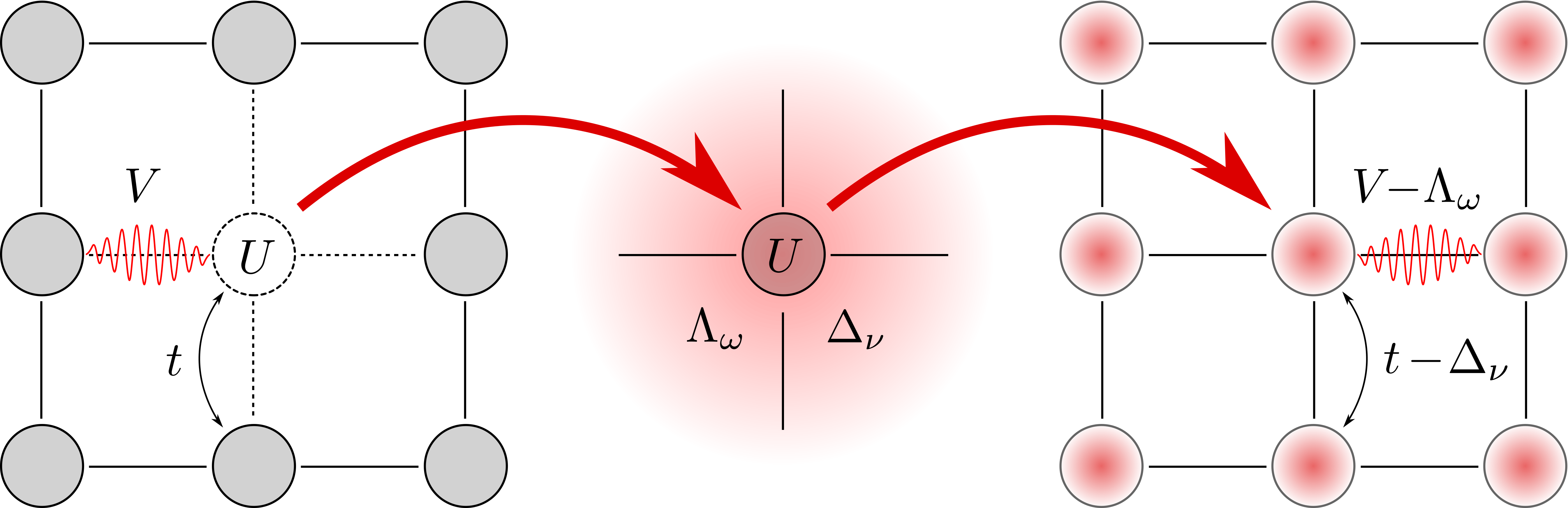}
\caption{EDMFT scheme of mapping the initial
lattice model (left) onto an effective impurity
problem (center) and, consequently, onto a lattice
of identical non-interacting impurities with the
modified hopping and non-local interaction
parameters (right).
\label{fig:EDMFT}}
\end{figure}

Complex computational methods based on many-body theory have started their rapid development since the emergence of high-performance computational systems. 
One of the important steps in this development was the appearance of non-perturbative techniques, such as the dynamical mean field theory (DMFT)~\cite{RevModPhys.68.13}. Nowadays, the DMFT and its generalizations to the extended dynamical mean-field theory (EDMFT)~\cite{PhysRevB.52.10295, PhysRevLett.77.3391, PhysRevB.61.5184, PhysRevLett.84.3678, PhysRevB.63.115110} and the non-equilibrium DMFT~\cite{RevModPhys.86.779} have become the most popular state-of-the-art approaches to correlated materials that provide an approximate solution to the correlated electronic problem. 
The concept of (E)DMFT can be illustrated in the context of the (extended) Hubbard model, which is a minimal model that accounts for the interplay between the kinetic energy (hopping processes of electrons between lattice sites with the amplitude $t$) and the (on-site $U$ and non-local $V$) Coulomb interaction of electrons, as sketched in the left part of Fig.~\ref{fig:EDMFT}. 
The key idea of (E)DMFT is to isolate one site of a lattice and embed it into an effective bath described by the frequency-dependent local fermionic $\Delta_\nu$ and bosonic $\Lambda_\omega$ hybridization functions that mimic the influence of electrons from other lattice sites on the considered site. 
Thus, the isolated lattice site is mapped onto an effective local impurity problem shown in the middle part of Fig.~\ref{fig:EDMFT}. 
Note, that the impurity problem is an interacting many-body problem, as it accounts for the local Coulomb interaction $U$ explicitly. 
Since all lattice sites are equivalent, (E)DMFT effectively replaces the initial lattice problem by a set of identical impurities that, however, do not interact with each other (right part of Fig.~\ref{fig:EDMFT}). 
The advantage of this approach is that the local impurity problem is solved numerically exactly, which provides a means to account for the local electronic correlations in a non-perturbative manner. 
This allowed (E)DMFT to uncover key features of correlated materials associated with local correlation effects, including the formation of Hubbard bands~\cite{Hubbard63, Hubbard64}, the Mott insulator transition~\cite{Mott, RevModPhys.70.1039}, and Hund's metal behavior~\cite{Hunds_metals2}.

Despite these successes, DMFT has clear limitations when non-local correlations are non-negligible, preventing this theory from becoming a universal method with truly predictive power. Indeed, correlated materials are often characterized by strong collective electronic fluctuations, such as plasmons, magnons, Cooper pairs, etc., that can be observed, for instance, in scattering experiments~\cite{JPSJ.81.011007, RevModPhys.83.705}. 
These fluctuations are intrinsically non-local and often long-ranged, and can strongly affect various electronic, magnetic, optical, and transport properties of the system. 
For this reason, an accurate theoretical description of correlated materials can be performed only via an advanced approach that allows for a simultaneous consideration of the feedback of the leading collective electronic fluctuations onto single-particle quantities and {\it vice versa}. 
Indeed, describing collective (usually two-particle) excitations requires the knowledge of the correct electronic spectral function (a single-particle quantity), which, in turn, is strongly renormalized by these collective fluctuations. Strong electronic correlations may also lead to the formation of various ordered states of matter, such as magnetic~\cite{RevModPhys.87.855, de2008magnetic, yin2011kinetic}, superconducting~\cite{RevModPhys.75.657, doi:10.1126/science.1200182, doi:10.1143/JPSJ.81.011009, 10.1063/1.4752092, mackenzie2017even, HOSONO2018278, pickett2021dawn, GU2022100202, Nomura_2022}, nematic~\cite{PhysRevB.91.155106, baek2015orbital, li2017nematic, PhysRevLett.124.157001, doi:10.1126/science.1134796, doi:10.1073/pnas.1921713117}, orbital ordered~\cite{Hotta_2006, fernandes2014drives, yi2017role}, and charge ordered~\cite{doi:10.1021/ja984015x, PhysRevLett.82.3871, PhysRevLett.88.126402, 10.1002/adma.202100593} phases, that can possibly be exploited for applications. 
To explore the ordered phase, one has to account for a spontaneous symmetry breaking mediated by electronic correlations, which is a highly-nontrivial task. 
Besides, nowadays experimental techniques allow one to selectively excite different collective electronic fluctuations, which offers an outstanding possibility to explore unique transient states of matter, such as light-induced magnetism~\cite{kimel2005ultrafast, RevModPhys.82.2731, koopmans2010explaining, satoh2012directional, schellekens2014ultrafast, stupakiewicz2017ultrafast} and superconductivity~\cite{Fausti189, mankowsky2014nonlinear, hu2014optically, doi:10.1073/pnas.1908368116}. Describing these effects requires to solve a quantum many-body problem in the presence of an external perturbation at the ultrafast timescale, which represents one of the most prominent challenges in modern science. 
Furthermore, teh proximity to ordered phases can strongly influence the transport properties of the system. 
For instance, in some cases the unusual resistivity (or conductivity) behavior can be related to the superconducting mechanism~\cite{doi:10.1126/science.abh4273, jin2011link, yuan2022scaling, doi:10.1126/sciadv.aav6753}. 
Accurately addressing the unconventional transport properties requires accounting for spatial correlation effects and complex multi-electron scattering processes known as ``vertex corrections,'' which can drastically alter transport and optical properties of a system. 
Notably, the local vertex corrections included in DMFT are irrelevant for conductivity calculations due to symmetry constraints. 
In contrast, non-local vertices, which are crucial, cannot be captured by the local DMFT framework.
All these aspects represent an outstanding challenge even for state-of-the-art methods that, in the case of multi-orbital systems, are often either perturbative, or based on local approximations to electronic correlations.

There have been many attempts to go beyond DMFT in order to incorporate spatial correlations in the theory. 
Let us briefly highlight the most representative methods that strongly contributed to the development of this field. 
The most straightforward way of accounting for spatial correlations has been proposed by various cluster extension of DMFT~\cite{PhysRevB.58.R7475, PhysRevB.62.R9283, PhysRevLett.87.186401, RevModPhys.77.1027, doi:10.1063/1.2199446, RevModPhys.78.865}, which essentially replace the single-site impurity problem of DMFT by a collection (a cluster) of lattice sites. 
The advantage of these methods lies in their exact treatment of short-range correlation effects within the cluster. 
This allows the method to predict the renormalization of the critical interaction value for the metal-insulator transition predicted by DMFT~\cite{PhysRevLett.101.186403} and to capture the effect of vertex corrections in conductivity~\cite{PhysRevLett.123.036601}. 
At the same time, correlations beyond the cluster are discarded, which does not allow one to accurately describe long-range collective excitations, such as plasmons and magnons. 
An obvious disadvantage of cluster theories is that solving the cluster problem numerically is significantly more time-consuming than solving the single-site impurity problem in DMFT. 
These solutions are usually based either on the quantum Monte Carlo approach, that often suffers from a fermionic sign problem in multi-orbital and/or large cluster calculations~\cite{PhysRevB.80.155132}, or on the exact diagonalization method, which complexity scales exponentially with the number of orbitals/sites~\cite{RevModPhys.66.763, PhysRevB.92.245135}. 
For this reason, in the multi-orbital case, cluster extensions of DMFT can account only for narrow-range correlation effects~\cite{PhysRevLett.101.256404, PhysRevB.89.195146, PhysRevB.91.235107}. 
Cluster extensions of DMFT for non-equilibrium problems have also been developed~\cite{eckstein2016ultra, PhysRevB.94.245114, PhysRevB.101.085127, PhysRevB.102.235169}, but they are limited to 4-site clusters of single-orbital sites and have not yet been applied to multi-orbital systems.

While cluster methods are designed for incorporating short-range correlations, long-range collective electronic fluctuations are commonly treated by diagrammatic extensions of (E)DMFT~\cite{RevModPhys.90.025003, Lyakhova_review}. 
Some of these methods have been developed to a very advanced level and can provide a very accurate solution of effective single-band model systems~\cite{PhysRevX.11.011058}. 
However, despite a remarkable progress in this direction, describing realistic multi-orbital systems by means of the most accurate diagrammatic techniques remains prohibitively expensive computationally~\cite{PhysRevLett.107.137007, PhysRevLett.113.266403, PhysRevB.95.115107, Boehnke_2018, acharya2019evening, PhysRevB.100.125120}. 
On the other hand, using less sophisticated diagrammatic approaches carries the risk of missing important physical effects. 

A combined $GW$ many-body perturbation theory with EDMFT ($GW$+EDMFT)~\cite{PhysRevLett.90.086402} is one of the simplest diagrammatic approaches that is actively used for solving realistic multi-orbital~\cite{PhysRevLett.113.266403, acharya2019evening, PhysRevB.95.041112, Tomczak_2012, PhysRevB.88.165119, PhysRevB.88.235110, Tomczak14, ryee2020nonlocal, PhysRevLett.109.237010, PhysRevX.8.021038}, and time-dependent~\cite{PhysRevLett.118.246402, PhysRevB.100.041111, PhysRevB.100.235117} correlated problems. 
This method takes the best of two popular state-of-the-art methods: the itinerant physics of long-range collective electronic fluctuations is captured via the non-local $GW$ self-energy~\cite{GW1, GW2, GW3} written in terms of the electronic Green's function ($G$) and the renormalized interaction ($W$), and local correlations are treated exactly by solving the EDMFT impurity problem. 
A consistent description of single- and two-particle observables in this theory is achieved by introducing an analog of the Almbladh functional~\cite{doi:10.1142/S0217979299000436} that relates the self-energy and the polarization operator. 
However, among various collective electronic fluctuations $GW$+EDMFT accounts for only charge excitations, and thus misses important effects related to spatial magnetic fluctuations.  
The first attempt to include long-range magnetic fluctuations and the exact local Hedin three-point vertex correction~\cite{GW1} in a $GW$+DMFT-like theory has been performed by the triply irreducible local expansion (TRILEX)~\cite{PhysRevB.92.115109, PhysRevB.93.235124}. 
Magnetic fluctuations are often the main source of instability in the system. 
While in standard situations charge fluctuations usually screen the interaction between electrons, magnetic fluctuations may strongly impact the electronic self-energy by enhancing the renormalized interaction $W$. 
Considering vertex corrections in diagrams for the self-energy and the polarization operator is also important for an accurate description of magnetic~\cite{PhysRevB.95.041112, Aryasetiawan08}, optical~\cite{doi:10.1143/JPSJ.75.013703, PhysRevB.80.161105} and transport~\cite{PhysRevLett.123.036601, PhysRevB.84.085128} properties. 
Incorporating these two ingredients allowed TRILEX to improve the $GW$+EDMFT result for the electronic spectral function. 
However, using TRILEX requires to resolve a double-counting issue (a famous Fierz ambiguity problem~\cite{PhysRevB.65.245118, Borejsza_2003, PhysRevD.68.025020, PhysRevB.70.125111, Bartosch_2009}) when charge and spin fluctuations are considered simultaneously. 
Unfortunately, so far this problem can be mitigated by introducing a more complex cluster extension of the theory~\cite{PhysRevLett.119.166401}. 

Much more advanced diagrammatic theories, such as the dynamical vertex approximation (D$\Gamma$A)~\cite{PhysRevB.75.045118, PhysRevB.80.075104, PhysRevB.95.115107, doi:10.7566/JPSJ.87.041004, PhysRevB.103.035120}, the dual fermion (DF)~\cite{PhysRevB.77.033101}, the dual boson (DB)~\cite{Rubtsov20121320, PhysRevB.90.235135, PhysRevB.93.045107}, and the dual TRILEX (\mbox{D-TRILEX})~\cite{PhysRevB.100.205115, PhysRevB.103.245123, 10.21468/SciPostPhys.13.2.036} methods allow for an accurate equal-footing description of the leading collective electronic fluctuations without facing the Fierz ambiguity. 
These theories have a complex diagrammatic structure that is based on the exact local three- and/or four-point vertex functions. 
These vertices are obtained numerically from the DMFT impurity problem and represents the generalized screened local electronic interaction. 
As a matter of fact, the calculation of vertex functions is the most time-consuming part of any diagrammatic extension of DMFT. 
In addition, using the four-point vertex in multi-orbital calculations simulations requires to solve the Bethe-Salpeter equation in the frequency-orbital space, which is extremely expensive numerically. 
For this reason, among these four theories, only the D$\Gamma$A~\cite{PhysRevB.95.115107} and \mbox{D-TRILEX}~\cite{10.21468/SciPostPhys.13.2.036} are currently available in a multi-orbital realization. 
Multi-orbital D$\Gamma$A calculations are very demanding and so far have only been performed in the weakly correlated regime, where the results are similar to those of DMFT~\cite{doi:10.7566/JPSJ.87.041004, PhysRevB.103.035120}. 
It is also important to mention a recent multi-orbital TPSC extension of DMFT~\cite{PhysRevB.107.235101}, which incorporates static (constant) four-point vertices for spatial electronic fluctuations via TPSC, while the local correlations are handled by the DMFT self-energy. 
This scheme improves the single-particle quantities compared to the original TPSC approach. 
However, the two-particle quantities, particularly the susceptibilities, do not show significant improvement, especially in the strongly correlated regime.

This brief overview of state-of-the-art methods highlights that an accurate theoretical description of the many-body effects in realistic multi-orbital framework remains a central challenge in condensed matter physics.
The condensed matter community urgently needs a comprehensive computational approach that strikes a proper balance between computational efficiency and the effectiveness of the theory in describing and predicting various properties of correlated electronic materials.
In this work, we focus on dual diagrammatic techniques and demonstrate that these methods provide a powerful and broadly applicable theoretical framework for identifying, quantitatively describing, and exploring the tunability of correlated materials’ properties driven by collective electronic behavior, both in and out-of-equilibrium.

\subsection{Structure and Scope}

\vspace{0.3cm}
\begin{center}
\begin{minipage}{0.79\textwidth}
\begin{itemize}
\item[{\it Gandalf:}] {\it I am looking for someone to share in an adventure that I am arranging, and it's very difficult to find anyone.}
\item[{\it Bilbo:}] {\it I should think so -- in these parts! We are plain quiet folk and have no use for adventures. Nasty disturbing uncomfortable things! Make you late for dinner! I can't think what anybody sees in them \ldots}
\item[{\it Gandalf:}] {\it You'll have a tale or two to tell when you come back.}
\item[{\it Bilbo:}] {\it You can promise that I'll come back?}
\item[{\it Gandalf:}] {\it No. And if you do, you will not be the same.}
\end{itemize}
\end{minipage}
\end{center}

This thesis embarks on an adventure into the intricate world of dual diagrammatic techniques. 
The journey takes the reader through a series of \emph{nasty, disturbing, uncomfortable transformations} of the initial interacting electronic problem into a new, dual space. 
These steps are mathematically demanding and conceptually subtle, yet they open the door to a consistent treatment of local correlations and spatial collective electronic fluctuations that is not straightforward in the original formulation. 
The journey continues with a back-transformation from the dual space to the physical variables describing fluctuations of the charge and spin densities. 
The resulting bosonic problem is no longer the same as the initial electronic system and provides an efficient framework for describing the full dynamics of charge and spin degrees of freedom.

The thesis is organized as follows:\\
Section~\ref{sec:Action_Dual}: We provide a detailed derivation of the dual action starting from the initial interaction electronic problem.
This section guides the reader through the transformations into the dual space, highlighting the conceptual and technical challenges, the choice of the reference system, and the benefits of working with dual variables.\\
Section~\ref{sec:DF_DB}: We introduce the dual fermion (DF) and dual boson (DB) approaches, derived from the dual action, and discuss exact (via the diagrammatic Monte Carlo) and approximate diagrammatic expansions within this framework. Representative applications and benchmarks against exact solutions are presented.\\
Section~\ref{sec:PBDT}: A partially bosonized approximation for the exact four-point vertex function of the reference problem is introduced. This allows one to eliminate the fermion-fermion interaction from the dual action and obtain a much simpler dual problem, expressed in terms of fermionic and bosonic fields and their mutual coupling. \\
Section~\ref{sec:DTRILEX}: We formulate the simplest diagrammatic approximation for the partially bosonized dual action, called the dual triply irreducible local expansion (\mbox{D-TRILEX}). This approximation is significantly simpler than its parent dual boson theory, enabling efficient multi-band calculations. Benchmarks show that it preserves accuracy while substantially reducing computational cost. 
Applications to both model systems and realistic materials demonstrate the capabilities of the method.\\
Section~\ref{sec:DGW}: We discuss the real-time implementation of the dual $GW$ method (D-GW), derived from the D-TRILEX theory using an instantaneous approximation for the fermion-boson coupling. This approach provides a promising framework for studying non-equilibrium dynamics and transport phenomena.\\
Section~\ref{sec:Exchange}: We perform the back-transformation of the dual variables to the physical bosonic degrees of freedom describing fluctuations of charge and spin densities. The resulting effective bosonic model incorporates local correlations and allows the determination of all relevant exchange interactions between spin and charge densities, effectively mapping the original interacting electronic problem onto Heisenberg- or Ising-like models, respectively. 
This framework also enables the derivation of the fundamental equations of motion governing the system's dynamics, including spin precession described by the Landau-Lifshitz-Gilbert equation and Higgs-like fluctuations of magnetic moments and charge densities.\\
Section~\ref{sec:Conclusions}: We summarize the main achievements, highlight the broader implications of dual diagrammatic methods, and outline future directions.

\newpage
\section{Action formalism based on an interacting reference system}
\label{sec:Action_Dual}

In this Section, we present a general approach for constructing a diagrammatic expansion for a quantum many-body fermionic lattice problem, based on an arbitrary interacting reference system, within the action formalism.
The key motivation is that, if the reference system is chosen wisely and solved accurately, it already captures the essential non-perturbative correlation effects of the original problem.
Such a reference system therefore serves as an excellent starting point for a perturbative diagrammatic expansion aimed at incorporating correlation effects beyond those included in the reference problem.
It is natural to expect that all building blocks of the modified expansion are ``aware'' of the correlations contained in the reference system and are renormalized accordingly.
In particular, the bare Green’s function in this framework should already be dressed by the self-energy of the reference problem, while the bare interaction should correspond to the electron-electron interaction of the original problem screened by all many-body effects of the reference system, yielding an effective vertex function.
A central requirement of this construction is to avoid double counting of correlations treated within and beyond the reference problem.
As we explicitly demonstrate below, such a scheme can be consistently derived within the so-called ``dual'' approach.
  
\subsection{Lattice action}  
\label{sec:Lattice_action}

We start with a general action of a multi-band extended Hubbard model:
\begin{align}
{\cal S} =& - \sum_{\substack{k,\{l\},\\\{\sigma\}}} c^{*}_{k \sigma l} \left[(i\nu+\mu)\delta^{\phantom{*}}_{\sigma\sigma'}\delta^{\phantom{*}}_{ll'} - \varepsilon^{\sigma\sigma'}_{\kv, ll'}\right] c^{\phantom{*}}_{k \sigma' l'} 
+ \frac12 \sum_{\substack{q, \{l\}, \\ \{k\}, \{\sigma\}}} U^{pp}_{l_1 l_2 l_3 l_4} c^{*}_{k \sigma l_1} c^{*}_{q-k, \sigma' l_2} c^{\phantom{*}}_{q-k',\sigma' l_4} c^{\phantom{*}}_{k' \sigma l_3}
\notag\\
&+ \frac12 \sum_{\substack{q,\{l\}, \\ \varsigma=d,m}} V^{\varsigma}_{q,\, l_1 l_2,\, l_3 l_4} \, \rho^{\varsigma}_{-q,\, l_1 l_2} \, \rho^{\varsigma}_{q,\, l_4 l_3}
+ \sum_{\substack{q,\{l\}, \\ \vartheta = s, t}} V^{\vartheta}_{q,\, l_1 l_2,\, l_3 l_4}\, \rho^{*\,\vartheta}_{q,\, l_1 l_2} \, \rho^{\vartheta}_{q,\, l_3 l_4}.
\label{eq:actionlatt}
\end{align}
In this equation, $c^{(*)}_{k\sigma{}l}$ is the Grassmann variable that describes the annihilation (creation) of an electron with momentum $\kv$, fermionic Matsubara frequency $\nu$, and spin projection ${\sigma \in \left\{\uparrow, \downarrow \right\}}$.
The label $l$ numerates the orbital and the site within the unit cell.
To simplify notations, we use a combined index ${k\in\{\kv,\nu\}}$.
Summations over momenta and frequencies are defined as:
\begin{align}
\sum_{k} = \frac{1}{\beta} \sum_{\nu} \, \frac{1}{N_{k}}\sum_{{\bf k}}
\end{align}
where ${\beta=T^{-1}}$ is the inverse temperature and $N_{k}$ is the number of ${\bf k}$-points in the discretized Brillouin zone (BZ).
The single-particle part of the lattice action (first term in Eq.~\eqref{eq:actionlatt}) contains the chemical potential $\mu$ and the single-particle Hamiltonian term ${\varepsilon^{\sigma\sigma'}_{\kv,ll'}}$ that has the following structure in the spin space:
${\varepsilon^{\sigma\sigma'}_{\kv,ll'} = \varepsilon_{\kv,ll'}\delta_{\sigma\sigma'} + i\,\vec{\gamma}_{\kv,ll'}\cdot\vec{\sigma}_{\sigma\sigma'}}$.
The diagonal part in the spin space $\varepsilon_{\kv,ll'}$ of this matrix contains the momentum- and orbital-space representation of the hopping amplitudes between different lattice sites, and may also account for the effect of the crystal field splitting (CFS) and of the external electric field.
The non-diagonal contribution in spin space $\vec{\gamma}_{\kv,ll'}$ describes the effect of the external magnetic field and the spin-orbit coupling (SOC), that is usually expressed in the Rashba form~\cite{Rashba}. The latter corresponds to a Fourier transform of the effective spin-dependent hopping amplitudes~\cite{PhysRevB.52.10239}. $\vec{\sigma}=\{\sigma^{x}, \sigma^{y}, \sigma^{z}\}$ is a vector of Pauli matrices.

The on-site Coulomb potential is written in the conventional (particle-particle) form:
\begin{align}
U^{pp}_{l_1 l_2 l_3 l_4} = \int dr dr' \psi^{*}_{l_1}(r) \psi^{*}_{l_2}(r') V(r-r') \psi^{\phantom{*}}_{l_3}(r) \psi^{\phantom{*}}_{l_4} (r'),
\end{align}
where ${V(r-r')}$ is the screened Coulomb interaction and $\psi_{l}(r)$ are localized on-site basis functions. 
The local interaction can also be rewritten in the particle-hole representation using the following relation:
\begin{align}
&\sum_{\omega, \{\nu\}}\sum_{\{l\}, \{\sigma\}} U^{pp}_{l_1 l_2 l_3 l_4} c^{*}_{\nu \sigma l_1} c^{*}_{\omega-\nu, \sigma' l_2} c^{\phantom{*}}_{\omega-\nu',\sigma' l_4} c^{\phantom{*}}_{\nu' \sigma l_3} = \notag\\
&\sum_{\omega, \{\nu\}}\sum_{\{l\}, \{\sigma\}} U^{ph}_{l_1 l_2 l_3 l_4} c^{*}_{\nu \sigma l_1} c^{\phantom{*}}_{\nu+\omega, \sigma l_2} c^{*}_{\nu'+\omega, \sigma' l_4} c^{\phantom{*}}_{\nu'\sigma' l_3}
- \sum_{\nu, \{l\}, \sigma} U^{pp}_{l_1 l_2 l_3 l_4} c^{*}_{\nu \sigma l_1} c^{\phantom{*}}_{\nu \sigma l_4} \delta^{\phantom{*}}_{l_2l_3},
\label{eq:Upp_to_Uph}
\end{align}
where the particle-hole representation for the interaction satisfies: ${U^{ph}_{l_1 l_2 l_3 l_4} = U^{pp}_{l_1 l_4 l_2 l_3}}$.

The remaining part of the interaction ${V^{r}_{q}}$ in Eq.~\eqref{eq:actionlatt} is written in the channel representation ${r\in\{\varsigma,\vartheta\}}$, where ${\varsigma\in\{d,m\}}$ denotes charge ($d$) and magnetic (${m\in\{x,y,z\}}$) channels, and $\vartheta\in\{s,t\}$ depicts singlet ($s$) and triplet ($t$) channels.
This interaction can have an arbitrary momentum ${\bf q}$ and bosonic Matsubara frequency $\omega$ dependence as depicted by a combined index ${q\in\{{\bf q},\omega\}}$.
Usually, ${V^{r}_{q}}$ corresponds to the non-local interaction.
However, it may also contain the frequency-dependent part of the local interaction that is not included in the $U_{l_1l_2l_3l_4}$ term.
Composite fermionic variables $\rho^{r}_{q,\,l_1l_2}$ for the considered bosonic channels describe fluctuations of corresponding densities around their average values ${\rho^{r}_{q,\,l_1l_2} = n^{r}_{q,\,l_1l_2} - \langle{n^{r}_{q,\,l_1l_2}\rangle}}$. 
The orbital-dependent charge and magnetic densities can be introduced as follows:
\begin{align} 
n^{d}_{q,\, l_1 l_2} = \sum_{k,\sigma} c^{*}_{k+q, \sigma l_1} c^{\phantom{*}}_{k \sigma l_2}\,, ~~~~~~
\vec{n}^{\,m}_{q,\, l_1 l_2} = \sum_{k,\{\sigma\}} c^{*}_{k+q, \sigma l_1} \vec\sigma_{\sigma\sigma'} c^{\phantom{*}}_{k \sigma' l_2}\,.
\label{eq:nph_channels}
\end{align}
Densities for the particle-particle channel $n^{(*)\,\vartheta}_{q,\,l_1l_2}$ are defined as:
\begin{alignat}{2}
n^{s}_{q,\, l_1 l_2} &= \frac12\sum_{k}
\left( c^{\phantom{*}}_{q-k, \downarrow l_2} c^{\phantom{*}}_{k \uparrow l_1} - c^{\phantom{*}}_{q-k, \uparrow l_2} c^{\phantom{*}}_{k \downarrow l_1} \right)\,, ~~~~~~
&&n^{*\,s}_{q,\, l_1 l_2} = \frac12\sum_{k} \left( c^{*}_{k \uparrow l_1} c^{*}_{q-k, \downarrow l_2} - c^{*}_{k \downarrow l_1} c^{*}_{q-k, \uparrow l_2} \right)\,, \notag\\
n^{t0}_{q,\, l_1 l_2} &= \frac12\sum_{k}
\left( c^{\phantom{*}}_{q-k, \downarrow l_2} c^{\phantom{*}}_{k \uparrow l_1} + c^{\phantom{*}}_{q-k, \uparrow l_2} c^{\phantom{*}}_{k \downarrow l_1} \right)\,,~~~~~~
&&n^{*\,t0}_{q,\, l_1 l_2} = \frac12\sum_{k} \left( c^{*}_{k \uparrow l_1} c^{*}_{q-k, \downarrow l_2} + c^{*}_{k \downarrow l_1} c^{*}_{q-k, \uparrow l_2} \right)\,, \notag\\
n^{t+}_{q,\, l_1 l_2} &= \frac{1}{\sqrt{2}}\sum_{k} c_{q-k, \uparrow l_2} c_{k \uparrow l_1}\,,~~~~~~
&&n^{*\,t+}_{q,\, l_1 l_2} = \frac{1}{\sqrt{2}}\sum_{k} c^{*}_{k \uparrow l_1} c^{*}_{q-k, \uparrow l_2}\,, \notag\\ 
n^{t-}_{q,\, l_1 l_2} &= \frac{1}{\sqrt{2}}\sum_{k} c_{q-k, \downarrow l_2} c_{k \downarrow l_1}\,,~~~~~~
&&n^{*\,t-}_{q,\, l_1 l_2} = \frac{1}{\sqrt{2}}\sum_{k} c^{*}_{k \downarrow l_1} c^{*}_{q-k, \downarrow l_2}\,.
\label{eq:npp_app}
\end{alignat}

\subsection{Reference system}
\label{sec:reference_system}

The choice of reference system for a diagrammatic expansion depends on the specific form of the lattice problem and on computational constraints.
For a single-site reference system, the most natural choice is the impurity problem of (extended) dynamical mean-field theory (E)DMFT~\cite{RevModPhys.68.13, PhysRevB.52.10295, PhysRevLett.77.3391, PhysRevB.61.5184, PhysRevLett.84.3678, PhysRevB.63.115110}.
DMFT maps the original lattice model onto an auxiliary local impurity problem via a self-consistent, frequency-dependent hybridization function $\Delta_{\nu}$, which accounts for the renormalization effect of the surrounding electrons on the impurity.
This approach captures emergent strongly correlated phenomena such as the Mott metal-insulator transition~\cite{Mott, RevModPhys.70.1039} and Hund's metal behavior~\cite{Hunds_metals1, Hunds_metals2}, which lie beyond weak-coupling descriptions.
In EDMFT, a local frequency-dependent bosonic hybridization function $Y_{\omega}$ is additionally introduced to describe the screening of local correlations by the non-local part of Coulomb interaction.
If the lattice unit cell contains multiple atoms, the reference system can be formulated as a multi-impurity problem, with one isolated impurity for each atom in the unit cell (see, e.g., Refs.~\cite{PhysRevLett.94.026404, PhysRevB.97.115150, vandelli2024doping}).
Finally, a cluster reference problem~\cite{PhysRevB.58.R7475, PhysRevB.62.R9283, PhysRevLett.87.186401, RevModPhys.77.1027, doi:10.1063/1.2199446, RevModPhys.78.865, PhysRevLett.101.186403} makes it possible to capture important non-perturbative short-range correlation effects, such as singlet formation and local magnetic moment physics, which are believed to be crucial for accurately describing strong-coupling antiferromagnetism, pseudogap formation, and high-temperature superconductivity~\cite{PhysRevB.62.R9283, PhysRevB.76.104509, PhysRevLett.100.046402, PhysRevB.94.125133, PhysRevB.101.045119, Danilov2022, dong2022, PhysRevX.8.021048, Cuprates}.

Let us introduce the reference system in the form of an EDMFT-like impurity problem:
\begin{align}
{\cal S}_{\rm imp} = & - \hspace{-0.1cm} \sum_{\substack{\nu,\{l\}, \\ \sigma\sigma'}} c^{*}_{\nu \sigma{}l} \left[(i\nu+\mu)\delta^{\phantom{*}}_{\sigma\sigma'}\delta^{\phantom{*}}_{ll'} - \Delta^{\sigma\sigma'}_{\nu, ll'}\right] c^{\phantom{*}}_{\nu\sigma'{}l'} 
+ \frac12 \hspace{-0.15cm} \sum_{\substack{q, \{k\}, \\ \{l\}, \{\sigma\}}} \hspace{-0.1cm} U^{pp}_{l_1 l_2 l_3 l_4} c^{*}_{k, \sigma, l_1} c^{*}_{q-k, \sigma', l_2} c^{\phantom{*}}_{q-k',\sigma', l_4} c^{\phantom{*}}_{k', \sigma, l_3}
\notag\\
&+ \frac12 \sum_{\omega,\{l\},\varsigma} Y^{\varsigma}_{\omega,\, l_1 l_2,\, l_3 l_4} \, \rho^{\varsigma}_{-\omega,\, l_1 l_2} \, \rho^{\varsigma}_{\omega,\, l_4 l_3}
+ \sum_{\omega,\{l\},\vartheta} Y^{\vartheta}_{\omega,\, l_1 l_2,\, l_3 l_4}\, \rho^{*\,\vartheta}_{\omega,\, l_1 l_2} \, \rho^{\vartheta}_{\omega,\, l_3 l_4}.
\label{eq:actionimp_app}
\end{align}
It can be isolated from the initial lattice action by adding and subtracting the fermionic $\Delta^{\sigma\sigma'}_{\nu, ll'}$ and bosonic $Y^{r}_{\omega,\, l_1 l_2,\, l_3 l_4}$ hybridization functions from Eq.~\eqref{eq:actionlatt} and retaining only the momentum-independent parts of the lattice action.
In Eq.~\eqref{eq:actionimp_app} the hybridizations are expressed in their most general form, depending on frequency $\nu$ ($\omega$), band $l$, spin $\sigma$, and channel $r\in\{\varsigma,\vartheta\}$.
They are usually determined by a self-consistency condition, and their explicit form depends on both the choice of the reference system and the specifics of the original problem.
The impurity problem is assumed to be identical for every unit cell of the lattice.
Therefore, the remaining part of the lattice action, ${{\cal S}_{\rm rem} = {\cal S} - \sum_{i}{\cal S}^{(i)}_{\rm imp}}$, where the sum runs over the number of unit cells, reads:
\begin{align}
{\cal S}_{\rm rem} =& \sum_{k,\{l\}}\sum_{\sigma\sigma'} c^{*}_{k \sigma l} \tilde{\varepsilon}^{\sigma\sigma'}_{k, ll'} c^{\phantom{*}}_{k \sigma' l'} 
+ \frac12 \sum_{q,\{l\},\varsigma} \tilde{V}^{\varsigma}_{q,\, l_1 l_2,\, l_3 l_4} \, \rho^{\varsigma}_{-q,\, l_1 l_2} \, \rho^{\varsigma}_{q,\, l_4 l_3}
+ \sum_{q,\{l\},\vartheta} \tilde{V}^{\vartheta}_{q,\, l_1 l_2,\, l_3 l_4}\, \rho^{*\,\vartheta}_{q,\, l_1 l_2} \, \rho^{\vartheta}_{q,\, l_3 l_4},
\label{eq:actionrem_app}
\end{align}
where we have introduced ${\tilde{\varepsilon}^{\sigma\sigma'}_{k, ll'} = \varepsilon^{\sigma\sigma'}_{\kv, ll'} - \Delta^{\sigma\sigma'}_{\nu, ll'}}$ and ${\tilde{V}^{r}_{q,\, l_1 l_2,\, l_3 l_4} = V^{r}_{q,\, l_1 l_2,\, l_3 l_4} - Y^{r}_{\omega,\, l_1 l_2,\, l_3 l_4}}$.

\subsection{Transformation to the dual space}
\label{sec:Dual_transformation}

To account exactly for the influence of the correlations captured by the reference problem on the lattice quantities, the reference problem should be integrated out.
We note that the reference system and the remaining part of the action are expressed in terms of the same variables $c^{(*)}$, which makes this integration non-trivial.
Therefore, the reference system can be integrated out only after performing a change of variables in the remaining part of the action.
This can be achieved by applying Hubbard-Stratonovich transformations to the partition function of ${{\cal S}_{\rm rem}}$, ${{\cal Z}_{\rm rem} = \int D[c^{*},c]\,e^{-{\cal S}_{\rm rem}}}$, thereby rewriting it in terms of new fermionic $f^{(*)}$ and bosonic $\varphi^{r}$ variables of an effective ``dual'' space:
\begin{align}
&\exp\left\{-\sum_{k,\{l\}}\sum_{\sigma\sigma'} c^{*}_{k \sigma l} \, \tilde{\varepsilon}^{\sigma\sigma'}_{k, ll'} \, c^{\phantom{*}}_{k \sigma' l'} \right\} = \notag\\
&{\cal D}_{f} \int D[f^{*},f] \exp\left\{\sum_{k,\{l\}}\sum_{\sigma\sigma'} 
\left( f^{*}_{k \sigma{}l} \left[\tilde{\varepsilon}_{k}^{-1}\right]^{\sigma\sigma'}_{ll'} f^{\phantom{*}}_{k\sigma'{}l'} 
- f^{*}_{k\sigma{}l} c^{\phantom{*}}_{k\sigma{}l}
- c^{*}_{k\sigma{}l} f^{\phantom{*}}_{k\sigma{}l}
\right)\right\}\,,
\label{eq:HSf}\\
&\exp\left\{ -\frac12\sum_{q,\{l\},\varsigma} \rho^{\varsigma}_{-q,\, l_1 l_2} \tilde{V}^{\varsigma}_{q,\,l_1 l_2,\, l_3 l_4} \, \rho^{\varsigma}_{q,\, l_4 l_3}\right\} = \notag\\
&{\cal D}_{\varphi}\int D[\varphi^{\varsigma}] \exp\left\{ \sum_{q,\{l\},\varsigma} \left( \frac12\varphi^{\varsigma}_{-q,\, l_1 l_2} \left[\left(\tilde{V}^{\varsigma}_{q}\right)^{-1}\right]_{l_1 l_2,\, l_3 l_4} \varphi^{\varsigma}_{q,\, l_4 l_3} - 
\varphi^{\varsigma}_{-q,\, l_1 l_2} \rho^{\varsigma}_{q,\, l_2 l_1}
\right)\right\}\,,
\label{eq:HSb}
\end{align}
\begin{align}
&\exp\left\{ -\sum_{q,\{l\},\vartheta} \rho^{*\,\vartheta}_{q,\, l_1 l_2} \tilde{V}^{\vartheta}_{q,\,l_1 l_2,\, l_3 l_4} \, \rho^{\vartheta}_{q,\, l_3 l_4} \right\} = \notag\\ 
&{\cal D}_{\varphi}\int D[\varphi^{*\,\vartheta}, \varphi^{\vartheta}] \exp\left\{ \sum_{q,\{l\},\vartheta} \left( \varphi^{*\,\vartheta}_{q,\, l_1 l_2} \left[\left(\tilde{V}^{\vartheta}_{q}\right)^{-1}\right]_{l_1 l_2,\, l_3 l_4} \varphi^{\vartheta}_{q,\, l_3 l_4} - 
\varphi^{*\,\vartheta}_{q,\, l_1 l_2} \rho^{\vartheta}_{q,\, l_1 l_2} - \rho^{*\,\vartheta}_{q,\, l_1 l_2} \varphi^{\vartheta}_{q,\, l_1 l_2}
\right)\right\}\,.
\label{eq:HSbs}
\end{align}
The terms ${\cal D}_{f} = -{\rm det}\left[\tilde{\varepsilon}_{k}\right]$ and ${\cal D}^{-1}_{\varphi} = -\sqrt{{\rm det}\left[\tilde{V}_{q}\right]}$ can be neglected, because they do not affect the calculation of expectation values.
We note that the dual fermionic $f^{(*)}$ and bosonic $\phi^{r}$ fields have no direct physical meaning.
They are introduced solely to facilitate integrating out the reference system.
After these transformations the lattice action takes the following form:
\begin{align} 
{\cal S}'
= &- \sum_{k,\{l\}}\sum_{\sigma\sigma'} f^{*}_{k \sigma{}l} \left[\tilde{\varepsilon}_{k}^{-1}\right]^{\sigma\sigma'}_{ll'} f^{\phantom{*}}_{k\sigma'{}l'}
+ \sum_{k,\sigma,l} 
\Big\{\left(f^{*}_{k\sigma{l}} + \eta^{*}_{k\sigma{l}} \right) c^{\phantom{*}}_{k\sigma{}l}
+ c^{*}_{k\sigma{}l} \left(f^{\phantom{*}}_{k\sigma{}l} + \eta^{\phantom{*}}_{k\sigma{}l} \right) \Big\} + {\cal S}_{\rm imp} 
\notag\\
&- \hspace{-0.1cm}\sum_{q,\{l\},\vartheta} \hspace{-0.05cm} \left\{ \varphi^{*\,\vartheta}_{q,\, l_1 l_2} \left[\left(\tilde{V}^{\vartheta}_{q}\right)^{-1}\right]_{l_1 l_2,\, l_3 l_4} \varphi^{\vartheta}_{q,\, l_3 l_4} 
- \left( \varphi^{*\,\vartheta}_{q,\, l_1 l_2} + j^{*\,\vartheta}_{q,\, l_1 l_2} \right) \rho^{\vartheta}_{q,\, l_1 l_2} - 
\rho^{*\,\vartheta}_{q,\, l_1 l_2} \left(\varphi^{\vartheta}_{q,\, l_1 l_2} + j^{\vartheta}_{q,\, l_1 l_2} \right) \right\} \notag\\
&- \hspace{-0.1cm}\sum_{q,\{l\},\varsigma} \left\{ \frac12\varphi^{\varsigma}_{-q,\, l_1 l_2} \left[\left(\tilde{V}^{\varsigma}_{q}\right)^{-1}\right]_{l_1 l_2,\, l_3 l_4} \varphi^{\varsigma'}_{q,\, l_4 l_3} 
- \left(\varphi^{\varsigma}_{-q,\, l_1 l_2} + j^{\varsigma}_{-q,\, l_1 l_2} \right) \rho^{\varsigma}_{q,\, l_2 l_1} \right\}\,,
\label{eq:Sprime_action}
\end{align}
where we additionally introduced source fields $\eta^{(*)}$ and $j^{(*)}$ for fermionic $c^{(*)}$ and composite $\rho^{(*)}$ variables, respectively.
Below, these fields will be used to derive the connection between the dual and lattice quantities.
Now, we shift fermionic ${f^{(*)} \to \hat{f}^{(*)} = f^{(*)} - \eta^{(*)}}$ and bosonic ${\varphi^{(*)} \to \hat{\varphi}^{(*)} = \varphi^{(*)} - j^{(*)}}$ variables to decouple the source fields from the original Grassmann variables $c^{(*)}$.
After that the lattice action can be decomposed into the three contributions:
\begin{align}
{\cal S}'[c^{(*)}, f^{(*)}, \varphi] = {\cal S}_{\rm imp}[c^{(*)}] + {\cal S}_{\rm mix}[c^{(*)}, f^{(*)}, \varphi] + {\cal S}_{\rm dual}[f^{(*)}, \varphi], 
\end{align}
where the dual part contains only new fermionic and bosonic variables:
\begin{align}
{\cal S}_{\rm dual}[f^{(*)}, \varphi] = &- \sum_{k,\{l\}}\sum_{\sigma\sigma'} \hat{f}^{*}_{k \sigma{}l} \left[\tilde{\varepsilon}_{k}^{-1}\right]^{\sigma\sigma'}_{ll'} \hat{f}^{\phantom{*}}_{k\sigma'{}l'} - \frac12\sum_{q,\{l\},\varsigma} \hat{\varphi}^{\varsigma}_{-q,\, l_1 l_2} \left[\left(\tilde{V}^{\varsigma}_{q}\right)^{-1}\right]_{l_1 l_2,\, l_3 l_4} \hat{\varphi}^{\varsigma}_{q,\, l_4 l_3} \notag\\
&- \sum_{q,\{l\},\vartheta} \hat{\varphi}^{*\,\vartheta}_{q,\, l_1 l_2} \left[\left(\tilde{V}^{\vartheta}_{q}\right)^{-1}\right]_{l_1 l_2,\, l_3 l_4} \hat{\varphi}^{\vartheta}_{q,\, l_3 l_4}\,,
\end{align}
and the mixed part describes the coupling between the original degrees of freedom and the dual variables:
\begin{align}
{\cal S}_{\rm mix}[c^{(*)}, f^{(*)}, \varphi] = \sum_{k,\sigma,l} 
\left(f^{*}_{k\sigma{l}} c^{\phantom{*}}_{k\sigma{}l}
+ c^{*}_{k\sigma{}l} f^{\phantom{*}}_{k\sigma{}l} \right)
+ \sum_{q,\{l\},r}\left( \varphi^{\varsigma}_{-q,\, l_1 l_2} \rho^{\varsigma}_{q,\, l_2 l_1} + \varphi^{*\,\vartheta}_{q,\, l_1 l_2} \rho^{\vartheta}_{q,\, l_1 l_2} + 
\rho^{*\,\vartheta}_{q,\, l_1 l_2} \varphi^{\vartheta}_{q,\, l_1 l_2} \right).
\end{align}
At this step, we can integrate out the reference problem ${\cal S}_{\rm imp}$ by explicitly taking the path integral over the original fermionic $c^{(*)}$ variables in the partition function of the lattice problem:
\begin{align}
{\cal Z} = \int D[f^{*},f,\varphi] \, e^{-{\cal S}_{\rm dual}[f^{(*)}, \varphi]} \int D[c^{*},c] \, e^{-{\cal S}_{\rm imp}[c^{(*)}]} \, e^{-{\cal S}_{\rm mix}[c^{(*)}, f^{(*)}, \varphi]}\,.
\label{eq:Z_lattice}
\end{align}
Evaluating the path integral over $c^{(*)}$ means taking the average over the impurity degrees of freedom:
\begin{align}
\frac{1}{{\cal Z}_{\rm imp}} \int D[c^{*},c] \, e^{-{\cal S}_{\rm imp}[c^{(*)}]} \, e^{-{\cal S}_{\rm mix}[c^{(*)}, f^{(*)}, \varphi]} = \left\langle e^{-{\cal S}_{\rm mix}[c^{(*)}, f^{(*)}, \varphi]} \right\rangle_{\rm imp} 
= e^{-\Phi[f^{(*)}, \varphi]}\,,
\end{align}
where ${{\cal Z}_{\rm imp} = \int D[c^{*},c]\,e^{-{\cal S}_{\rm imp}}}$ is the partition function of the reference impurity problem~\eqref{eq:actionimp_app}.
In this expression $\Phi[f^{(*)}, \varphi]$ is the generating functional, where the dual variables $f^{(*)}$ and $\varphi$ play a role of the source fields for the corresponding original degrees of freedom $c^{(*)}$ and $\rho$.
One can formally expand the generating functional to all orders in the dual variables and get the following result:
\begin{align}
\Phi[f^{(*)}, \varphi] &= - \left\langle e^{-{\cal S}_{\rm mix}[c^{(*)}, f^{(*)}, \varphi]} \right\rangle_{\rm imp} \notag\\
&= \sum_{k,\{l\},\{\sigma\}} f^{*}_{k\sigma l_1} g^{\sigma\sigma'}_{\nu,l_1l_2} f^{\phantom{*}}_{k\sigma'l_2} 
+ \frac12 \sum_{q,\{l\},\{\varsigma\}} \varphi^{\varsigma}_{-q,\,l_1l_2} \chi^{\varsigma\varsigma'}_{\omega,\,l_1l_2,\,l_3l_4} \varphi^{\varsigma'}_{q,\,l_4l_3} \notag\\
&+ \sum_{q,\{l\},\{\vartheta\}} \varphi^{*\,\vartheta}_{q,\,l_1l_2} \chi^{\vartheta\vartheta'}_{\omega,\,l_1l_2,\,l_3l_4} \varphi^{\vartheta'}_{q,\,l_3l_4} 
+ \tilde{\cal F}[f^{(*)},\varphi].
\label{eq:Generating_functional}
\end{align}
Note, that expanding the logarithm in Eq.~\eqref{eq:Generating_functional} corresponds to a cumulant expansion, yielding the \emph{connected} correlation functions that are coupled to the dual variables.
Thus, the lowest-order correlation functions are given by the Green's function $g$ and susceptibility $\chi$ of the impurity problem, that are defined as:
\begin{align}
g^{\sigma\sigma'}_{\nu,\,ll'} &= - \langle c^{\phantom{*}}_{\nu\sigma{}l} c^{*}_{\nu\sigma'l'} \rangle\,, \\
\chi^{\varsigma\varsigma'}_{\omega,\,l_1 l_2,\, l_3 l_4} &= - \langle \rho^{\varsigma}_{\omega,\, l_2 l_1} \rho^{\varsigma'}_{-\omega,\, l_3 l_4} \rangle\,, \\
\chi^{\vartheta\vartheta'}_{\omega,\,l_1 l_2,\, l_3 l_4} &= - \langle \rho^{\vartheta}_{\omega,\, l_1 l_2} \rho^{*\,\vartheta'}_{\omega,\, l_3 l_4} \rangle\,.
\end{align}
The higher-order contributions that correspond to all possible fermion-fermion, fermion-boson and boson-boson connected correlation functions~\cite{PhysRevB.77.033101, Rubtsov20121320, PhysRevB.90.235135, PhysRevB.93.045107}, are contained in the interaction part of the action $\tilde{\cal F}[f,\varphi]$.
For convenience, let us rescale the fermionic $f^{(*)}$ and bosonic $\varphi^{(*)}$ fields in the partition function~\eqref{eq:Z_lattice} by the parameters ${B^{-1}_{\nu}}$ and ${\alpha^{-1}_{\omega}}$, respectively.
These parameters will be determined below.
After that, the lowest-order contributions to $\tilde{\cal F}[f,\varphi]$, which correspond to the two-particle connected correlations functions, take the following form:
\begin{align}
&\tilde{\cal F}[f,\varphi]
\simeq \sum_{q,\{k\}}\sum_{\{\nu\}, \{l\}}\sum_{\{\sigma\}, \varsigma/\vartheta} \times \notag\\
&\times\Bigg\{ \frac14 \left[\Gamma_{\nu\nu'\omega}\right]^{\sigma_1\sigma_2\sigma_3\sigma_4}_{l_1 l_2 l_3 l_4} f^{*}_{k\sigma_1l_1} f^{\phantom{*}}_{k+q, \sigma_2 l_2} f^{*}_{k'+q, \sigma_4 l_4} f^{\phantom{*}}_{k'\sigma_3 l_3} 
+ \Lambda^{\sigma\sigma'\varsigma}_{\nu\omega,\, l_1,\, l_2,\, l_3 l_4} \, f^{*}_{k\sigma{}l_1} f^{\phantom{*}}_{k+q,\sigma', l_2} \varphi^{\varsigma}_{q,\, l_4 l_3} \notag\\
&\hspace{0.6cm} + \frac12\left(\Lambda^{\sigma\sigma'\vartheta}_{\nu\omega,\,l_1,\, l_2,\, l_3 l_4} \, f^{*}_{k\sigma{}l_1} f^{*}_{q-k,\sigma', l_2} \varphi^{\vartheta}_{q,\, l_3 l_4} 
+ \Lambda^{*\,\sigma\sigma'\vartheta}_{\nu\omega,\,l_1,\, l_2,\, l_3 l_4} \, \varphi^{*\,\vartheta}_{q,\, l_3 l_4} f^{\phantom{*}}_{q-k,\sigma', l_2} f^{\phantom{*}}_{k\sigma{}l_1}  \right)\Bigg\}\,.
\label{eq:lowestint}
\end{align}
Here, $\Gamma$ is the exact four-point (fermion-fermion) vertex function of the reference problem, which is defined as:
\begin{align}
\left[\Gamma_{\nu\nu'\omega}\right]^{\sigma_1\sigma_2\sigma_3\sigma_4}_{l_1 l_2 l_3 l_4} 
= &\sum_{\{l'\},\{\sigma'\}} \av{c^{\phantom{*}}_{\nu\sigma'_1l'_1} c^{*}_{\nu+\omega,\sigma'_2l'_2} c^{*}_{\nu'\sigma'_3l'_3} c^{\phantom{*}}_{\nu'+\omega,\sigma'_4l'_4}}_{\rm connected} \times \notag\\
&\times \left[B^{-1}_{\nu}\right]^{\sigma'_1\sigma_1}_{l'_1 l_1} \left[B^{-1}_{\nu+\omega}\right]^{\sigma'_2\sigma_2}_{l'_2 l_2} \left[B^{-1}_{\nu'}\right]^{\sigma'_3\sigma_3}_{l'_3 l_3} \left[B^{-1}_{\nu'+\omega}\right]^{\sigma'_4\sigma_4}_{l'_4 l_4}\,.
\label{eq:GammaPH}
\end{align}
The exact three-point (fermion-boson) vertex functions $\Lambda^{\sigma\sigma'r}_{\nu\omega}$ have the following form:
\begin{align}
\Lambda^{\sigma_1\sigma_2\varsigma}_{\nu\omega,\,l_1,\, l_2,\, l_3 l_4} &=
\sum_{\{\sigma'\},\{l'\},\varsigma'} \av{c^{\phantom{*}}_{\nu\sigma'_1l'_1} c^{*}_{\nu+\omega,\sigma'_2l'_2}\,\rho^{\varsigma'}_{-\omega,\, l'_3 l'_4}} \left[B^{-1}_{\nu}\right]^{\sigma'_1\sigma_1}_{l'_1l_1} \left[B^{-1}_{\nu+\omega}\right]^{\sigma'_2\sigma_2}_{l'_2l_2} \left[\alpha^{-1}_{\omega}\right]^{\varsigma'\varsigma}_{l'_3 l'_4 l_3 l_4} \,,
\label{eq:Vertex_PH_app}\\
\Lambda^{\sigma_1\sigma_2\vartheta}_{\nu\omega,\,l_1,\, l_2,\, l_3 l_4} &= 
\sum_{\{\sigma'\},\{l'\},\vartheta'} \av{c^{\phantom{*}}_{\nu\sigma'_1l'_1} c^{\phantom{*}}_{\omega-\nu,\sigma'_2l'_2}\,\rho^{*\,\vartheta'}_{\omega,\, l'_3 l'_4}} \left[B^{-1}_{\nu}\right]^{\sigma'_1\sigma_1}_{l'_1l_1} \left[B^{-1}_{\omega-\nu}\right]^{\sigma'_2\sigma_2}_{l'_2l_2} \left[\alpha^{-1}_{\omega}\right]^{\vartheta'\vartheta}_{l'_3 l'_4 l_3 l_4}\,, \label{eq:Vertex_PP_app}\\
\Lambda^{*\,\sigma_1\sigma_2\vartheta}_{\nu\omega,\,l_1,\, l_2,\, l_3 l_4} &= 
\sum_{\{\sigma'\},\{l'\},\vartheta'} 
\av{\rho^{\vartheta'}_{\omega,\,l'_3 l'_4} \, c^{*}_{\omega-\nu,\sigma_2'l'_2} c^{*}_{\nu\sigma'_1l'_1}} \left[B^{-1}_{\nu}\right]^{\sigma'_1\sigma_1}_{l'_1l_1} \left[B^{-1}_{\omega-\nu}\right]^{\sigma'_2\sigma_2}_{l'_2l_2} \left[\alpha^{-1}_{\omega}\right]^{\vartheta'\vartheta}_{l'_3 l'_4 l_3 l_4}\,.
\label{eq:Vertex_PPP_app}
\end{align}
Therefore, after integrating out the reference system ${\cal S}_{\rm imp}$ and neglecting the source fields, the lattice action transforms into the dual action expressed solely in terms of dual variables:
\begin{align} 
{\cal \tilde{S}}
= &- \sum_{k,ll'}\sum_{\sigma\sigma'} f^{*}_{k\sigma{}l} [\tilde{\cal G}^{-1}_{k}]^{\sigma\sigma'}_{ll'} f^{\phantom{*}}_{k\sigma'{}l'} 
+ \tilde{\cal F}[f,\varphi] \notag\\
&- \frac12\sum_{q,\{l\},\{\varsigma\}}
\varphi^{\varsigma}_{-q,\, l_1 l_2} \left[\tilde{\cal X}_{q}^{-1}\right]^{\varsigma\varsigma'}_{l_1 l_2;\, l_3 l_4} \varphi^{\varsigma'}_{q,\, l_4 l_3} 
- \sum_{q,\{l\},\{\vartheta\}}
\varphi^{*\,\vartheta}_{q,\, l_1 l_2} \left[\tilde{\cal X}_{q}^{-1}\right]^{\vartheta\vartheta'}_{l_1 l_2;\, l_3 l_4} \varphi^{\vartheta'}_{q,\, l_3 l_4}\,,
\label{eq:DB_action_app}
\end{align}
where the bare dual fermion and boson propagators explicitly read:
\begin{align}
\tilde{\cal G}^{\,\sigma_1\sigma_2}_{k,\,l_1 l_2} &= \sum_{\{l'\},\{\sigma'\}}B^{\sigma_1\sigma'_1}_{\nu,\,l_1 l'_1} \left[ \left(\left(\varepsilon_{\bf k} - \Delta_{\nu}\right)^{-1} - g_{\nu}\right)^{-1} \right]^{\sigma'_1\sigma'_2}_{l'_1l'_2} B^{\sigma'_2\sigma_2}_{\nu,\,l'_2 l_2}\,,
\label{eq:bare_dual_G}\\
\tilde{\cal X}^{\,r_1r_2}_{q,\,l_1 l_2;\, l_3 l_4} &= \sum_{\{l'\},\{r'\}}\alpha^{r_1r'_1}_{\omega,\,l_1 l_2;\,l'_1 l'_2} \left[ \left(\big(\tilde{V}_{q}\big)^{-1} - \chi_{\omega} \right)^{-1}\right]^{r'_1r'_2}_{l'_1l'_2;\,l'_3l'_4} \alpha^{r'_2r_2}_{\omega,\,l'_3 l'_4;\,l_3l_4}\,.
\label{eq:bare_dual_X}
\end{align}

\subsection{Choice for the scaling parameters}
\label{sec:Scaling_parameters}

The parameters $B^{\sigma\sigma'}_{\nu,ll'}$ and $\alpha^{rr'}_{\omega,\,l_1l_2,\,l'_1l'_2}$ enter the theory as scaling factors of the dual fermionic $f^{(*)}$ and bosonic $\varphi^{r(*)}$ fields, introduced via the exact Hubbard-Stratonovich transformations~\eqref{eq:HSf}--\eqref{eq:HSb}. 
Consequently, the choice of these scaling parameters does not affect the lattice quantities expressed in terms of the original fermionic variables $c^{(*)}$. 
At the same time, specific choices of $B^{\sigma\sigma'}_{\nu,ll'}$ and $\alpha^{rr'}_{\omega,\,l_1l_2,\,l'_1l'_2}$ make calculations in the dual space more convenient.

If the scaling parameter for the fermionic variables is set to ${B^{\sigma\sigma'}_{\nu,ll'} = g^{\sigma\sigma'}_{\nu,ll'}}$, the fermion-fermion vertex functions, e.g., the four-point vertex~\eqref{eq:GammaPH}, take the conventional form of the vertex function, where the division by $B$ corresponds to amputation of a fermionic leg $g$ in the correlation function.
Furthermore, the bare dual fermion propagator~\eqref{eq:bare_dual_G} becomes the difference between the DMFT and impurity Green's functions: ${\tilde{\cal G} = G^{\rm DMFT} - g}$.
Here, the DMFT Green's function, which is the bare Green's function dressed in the impurity self-energy $\Sigma^{\rm imp}_{\nu}$, can be found from the usual expression:
\begin{align}
\left[\left(G^{\rm DMFT}_{k}\right)^{-1}\right]^{\sigma\sigma'}_{ll'} = \delta_{ll'}\delta_{\sigma\sigma'}(i\nu + \mu) - [\varepsilon_{\bf k}]^{\sigma\sigma'}_{ll'} - \left[\Sigma^{\rm imp}_{\nu}\right]^{\sigma\sigma'}_{ll'}. 
\end{align}
The self-energy of the impurity problem is related to the impurity Green's function as:
\begin{align}
\left[g^{-1}_{\nu}\right]^{\sigma\sigma'}_{ll'} = \delta_{ll'}\delta_{\sigma\sigma'}(i\nu + \mu) - [\Delta_{\nu}]^{\sigma\sigma'}_{ll'} - \left[\Sigma^{\rm imp}_{\nu}\right]^{\sigma\sigma'}_{ll'}\,.
\label{eq:Dyson_g_imp}
\end{align}

On the other hand, dividing the four-point correlation function by the impurity Green's functions in Eq.~\eqref{eq:GammaPH} makes the result very noisy at high frequencies.
In this respect, choosing ${B^{\sigma\sigma'}_{\nu,ll'} = \delta_{\sigma\sigma'} \delta_{ll'}}$ allows one to avoid working with noisy objects and simplifies calculation by removing unnecessary divisions in the implementation.

There also exist several convenient choices for the bosonic scaling parameter.
In the equations~\eqref{eq:Vertex_PH_app}, \eqref{eq:Vertex_PP_app}, and \eqref{eq:Vertex_PPP_app} for the three-point vertex $\Lambda$, the division by  $\alpha$ corresponds to amputation of a bosonic leg from the fermion-boson correlation function. 
Therefore, if ${\alpha^{rr'}_{\omega,\,l_1l_2;\,l_3l_4} = \chi^{rr'}_{\omega,\,l_1l_2;\,l_3l_4}}$, the vertex function corresponds to a renormalized coupling of fermions to the bosonic susceptibility $\chi$. 
Consequently, the bare dual boson propagator~\eqref{eq:bare_dual_X} takes the form of the susceptibility and is given by the difference between the EDMFT and impurity susceptibilities: ${\tilde{\cal X} = X^{\rm EDMFT} - \chi}$.
The EDMFT susceptibility corresponds to the renormalization of the impurity polarization operator $\Pi^{\rm imp}_{\omega}$ by the electron-electron interaction and can be found as follows:
\begin{align}
\left[\left(X^{\rm EDMFT}_{q}\right)^{-1}\right]^{rr'}_{l_1l_2;\,l_3l_4} = \left[\left(\Pi^{\rm imp}_{\omega}\right)^{-1}\right]^{rr'}_{l_1l_2;\,l_3l_4} - \delta_{rr'}\big[U^{r}_{\bf q}\big]_{l_1l_2;\,l_3l_4}\,.
\end{align}
In this expression, ${U^{r}_{\bf q} = U^{r} + V^{r}_{\bf q}}$ is the total bare interaction and $U^{r}$ is the local part of the bare interaction in a given channel $r$. 
The polarization operator of the impurity problem is related to the impurity susceptibility as:
\begin{align}
\left[\chi_{\omega}^{-1}\right]^{rr'}_{l_1l_2;\,l_3l_4} = \left[\left(\Pi^{\rm imp}_{\omega}\right)^{-1}\right]^{rr'}_{l_1l_2;\,l_3l_4} - \delta_{rr'}\big[{\cal U}^{r}_{\omega}\big]_{l_1l_2;\,l_3l_4}\,,
\label{eq:Dyson_chi_imp}
\end{align}
where ${{\cal U}^{r}_{\omega,\,l_1 l_2,\, l_3 l _4} = U^{r}_{l_1 l_2,\, l_3 l _4} + Y^{r}_{\omega,\,l_1 l_2,\, l_3 l _4}}$ is the bare interaction of the impurity problem.
The explicit form of the bare interaction in the channel representation, $U^{r}$, will be discussed below. 

Alternatively, one can follow the approach of $GW$ theory~\cite{GW1, GW2, GW3}, where the bosonic propagator corresponds to a renormalized interaction $W$ instead of the susceptibility $X$.
The $GW$-like form for the dual diagrammatic expansion can be achieved by the following choice of the bosonic scaling factor:
\begin{align}
\alpha^{rr'}_{\omega,\,l_1l_2,\,l'_1l'_2} = \delta_{l_1,l'_1} \delta_{l_2,l'_2} \delta_{rr'} + \sum_{l_3, l_4} {\cal U}^{r}_{\omega,\,l_1 l_2,\, l_3 l _4} \, \chi^{rr'}_{\omega,\,l_3 l_4,\, l'_1 l'_2}\,.
\label{eq:alpha_app}
\end{align}
In this case, the three-point vertex~\eqref{eq:Vertex_PH_app}--\eqref{eq:Vertex_PPP_app} corresponds to the renormalized coupling of fermions to the interaction~\cite{ayral:tel-01247625, PhysRevB.94.205110, PhysRevB.100.205115}, and the bare dual bosonic propagator takes the form of the dressed interaction, which corresponds to the difference between the EDMFT-like and impurity renormalized interactions: ${\tilde{\cal X} = W^{\rm EDMFT} - w}$.
The EDMFT-like renormalized interaction can be defined as the bare electron-electron interaction dressed by the polarization operator of the impurity problem and can be found as follows:
\begin{align}
\left[\left(W^{\rm EDMFT}_{q}\right)^{-1}\right]^{rr'}_{l_1l_2;\,l_3l_4} = \delta_{rr'}\left[\left(U^{r}_{\bf q}\right)^{-1}\right]_{l_1l_2;\,l_3l_4} - \left[\Pi^{\rm imp}_{\omega}\right]^{rr'}_{l_1l_2;\,l_3l_4}\,.
\label{eq:W_EDMFT}
\end{align}
In turn, the renormalized interaction of the impurity problem is defined as:
\begin{align}
\left[w^{-1}_{\omega}\right]^{rr'}_{l_1l_2;\,l_3l_4} = \delta_{rr'}\left[\big({\cal U}^{r}_{\omega}\big)^{-1}\right]_{l_1l_2;\,l_3l_4} - \left[\Pi^{\rm imp}_{\omega}\right]^{rr'}_{l_1l_2;\,l_3l_4}\,.
\label{eq:w_imp}
\end{align}

\begin{figure}[t!]
\centering
\includegraphics[width=0.7\linewidth]{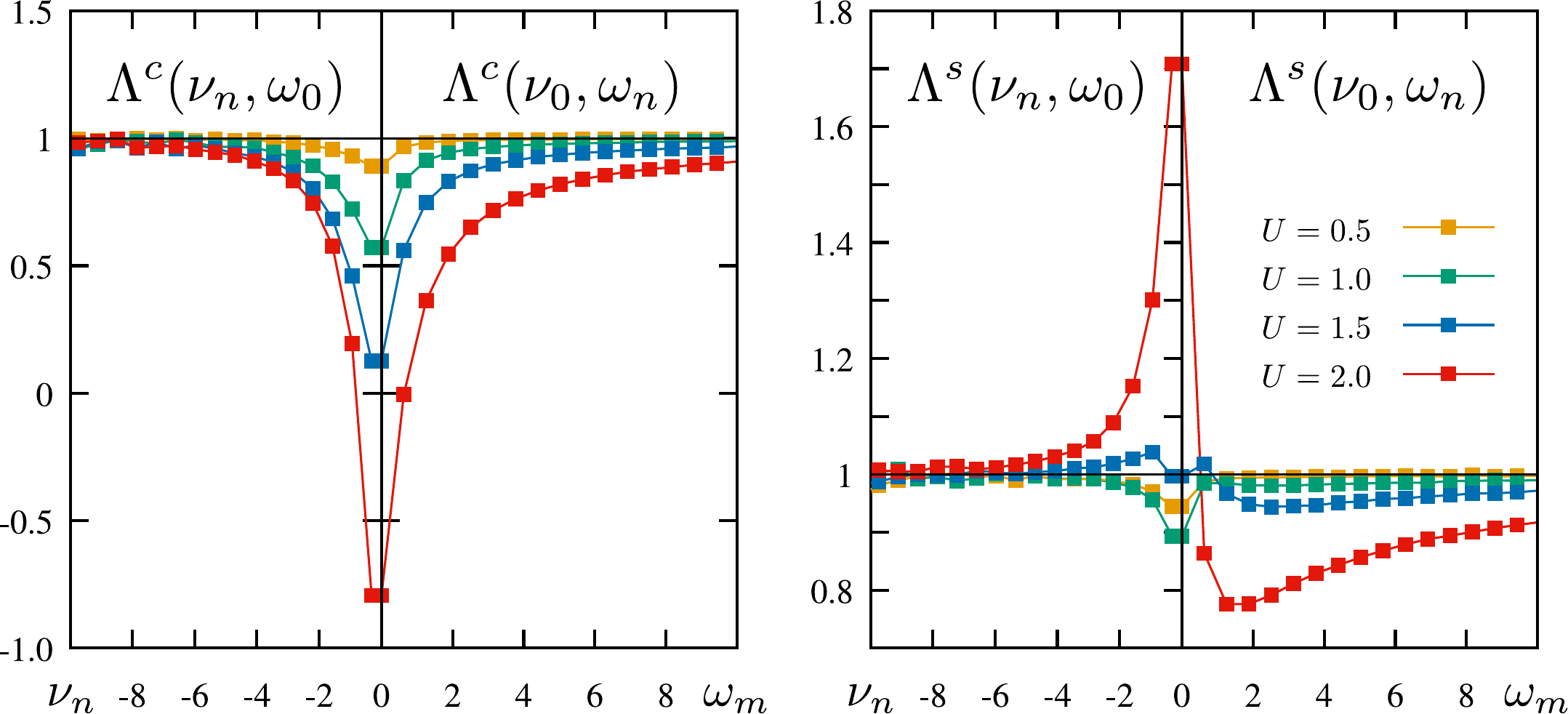}
\caption{The fermion-boson vertex function $\Lambda_{\nu\omega}$ in the charge (left panel) and spin (right panel) channels as a function of fermionic $\nu_n$ and bosonic $\omega_{m}$ frequencies. The result is obtained for the half-filled Hubbard model on a square lattice with the nearest-neighbor hopping ${t=0.25}$ for different values of the local Coulomb interaction $U$. The Figure is taken from Ref.~\cite{PhysRevB.100.205115}.}
\label{fig:Lambda}
\end{figure}

Finally, one can also benefit from the fact that by choosing ${B=g}$ and $\alpha$ in the form of Eq.~\eqref{eq:alpha_app}, the three-point vertex function acquires a useful high-frequency asymptotics:
\begin{align}
\Lambda^{\sigma_1\sigma_2r}_{\nu\omega,\,l_1,l_2,l_3l_4} \to \delta_{l_1l_3} \delta_{l_2l_4}, ~\text{if}~ \nu\to\infty ~\text{or}~ \omega\to\infty\,.
\end{align}
This asymptotic behavior of the three-point vertex in the charge (left panel) and spin (right panel) channels is illustrated in Figure.~\ref{fig:Lambda}. 

\subsection{Advantages of the dual space expansion}
\label{sec:Advantages_dual}

Working with the dual action~\eqref{eq:DB_action_app} offers several important advantages.
First, since the reference problem is integrated out through the transformation of the lattice problem to the dual space, the dual diagrammatic expansion naturally accounts only for correlations beyond those already included in the reference problem. 
This avoids any double counting of correlation effects.
Furthermore, all building blocks of the dual diagrammatic expansion, namely the bare propagators and vertex functions, are already renormalized by the correlations of the reference problem.
The correlations within the reference problem are usually treated exactly using numerical solvers based on continuous-time quantum Monte Carlo techniques~\cite{PhysRevB.72.035122, PhysRevLett.97.076405}. 
This enables an accurate non-perturbative description of local, and in the case of a cluster reference system, short-range electronic correlations.
Correlation effects beyond the reference problem are then incorporated diagrammatically (typically approximately), but in a way that allows leading collective electronic fluctuations in different channels (charge, spin, orbital, etc.) to be treated simultaneously and on equal footing, without restrictions on their spatial range.
As we demonstrate below, the diagrammatic calculations are performed self-consistently at both the single- and two-particle levels, a procedure referred to as ``inner self-consistency'' in the computational scheme described in Section~\ref{sec:DB_scheme}.

Another advantage is that the dual theory is perturbative in both the weak- and strong-coupling limits.
The building blocks for the diagrammatic expansion in the dual space are the bare fermionic $\tilde{\cal G}_{k}$ and bosonic $\tilde{\cal X}^{\varsigma}_{q}$ propagators, along with the exact impurity four-point $\Gamma^{\varsigma}_{\nu\nu'\omega}$ and three-point $\Lambda^{\varsigma}_{\nu\omega}$ vertex functions, which correspond to the renormalized fermion-fermion and fermion-boson interactions of the reference problem.
In the weak-coupling limit, the bare electron-electron interactions $U$ and $V_{\bf q}$ are small, leading to an even smaller four-point vertex $\Gamma^{\varsigma}_{\nu\nu'\omega}$, since it represents an interaction screened by the local correlations of the reference problem. 
The bosonic propagator $\tilde{\cal X}^{\varsigma}_{q}$ is also small, being proportional to $V_{\bf q}$~\eqref{eq:bare_dual_X}.
Thus, at small interaction strengths, the dual diagrammatic expansion reduces to a conventional weak-coupling perturbation theory with ${U/t\ll1}$.
In the strong-coupling limit, where electrons are localized, the bare fermionic propagator $\tilde{\cal G}_{k}$, which corresponds to the difference between the (E)DMFT and impurity Green's functions, becomes small, and the diagrammatic expansion in the dual space reduces to a strong-coupling expansion with ${t/U\ll1}$ (see also the discussion in Section~\ref{sec:strong_coupling}).
In this way, the transformation of the lattice problem to the dual space enables a unified framework that efficiently combines both weak- and strong-coupling diagrammatic expansions.

In the following, we discuss several applications of dual theories that make use of different choices for the scaling factors, depending on the context of the problem.

\newpage
\section{Dual fermion/boson approach}
\label{sec:DF_DB}

The dual fermion (DF)~\cite{PhysRevB.77.033101} and dual boson (DB)~\cite{Rubtsov20121320, PhysRevB.90.235135, PhysRevB.93.045107} approaches provide an approximate solution to the dual problem~\eqref{eq:DB_action_app} by truncating the interaction term $\tilde{\cal F}$ at the two-particle level given by Eq.~\eqref{eq:lowestint}.
Various comparisons of the DF and DB results against the exact benchmark solutions of the (extended) Hubbard model suggest that the contribution of the higher-order vertex functions to the self-energy is indeed negligibly small in a broad range of model parameters~\cite{PhysRevB.96.035152, PhysRevB.103.245123}. 
We note, that if the non-local interaction $V^{r}_{\bf q}$ in a certain bosonic channel $r$ is absent in the theory and the self-consistency on the bosonic quantities does not generate the bosonic hybridization function $Y^{r}_{\omega}$, the bare bosonic propagator is identically equal to zero, ${\tilde{\cal X}^{r}_{q}=0}$.
For this reason, the bosonic fields corresponding to a superconducting channel $\vartheta$ are absent in the DB theory. 
Due to a rather complex diagrammatic structure of the method, the actual dual fermion/boson calculations have been performed on the basis of a paramagnetic reference system and were limited to a single-orbital case.
The only exception is the multi-orbital implementation of the second-order dual fermion approximation that was introduced in Ref.~\cite{van_Loon_2021}.
Taking all these into account leads to the DB action:
\begin{align} 
{\cal \tilde{S}}_{\rm DB}
= &- \sum_{k,\sigma} f^{*}_{k\sigma} \tilde{\cal G}^{-1}_{k} f^{\phantom{*}}_{k\sigma} 
- \frac12\sum_{q,\varsigma}
\varphi^{\varsigma}_{-q} \tilde{\cal X}_{q,\varsigma}^{-1} \varphi^{\varsigma}_{q}
+ \tilde{\cal F}[f,\varphi]
\notag\\
&- \sum_{k,\sigma} \left\{ \eta^{*}_{k\sigma} \tilde{\varepsilon}^{-1}_{k} f^{\phantom{*}}_{k\sigma}
- \eta^{*}_{k\sigma} \tilde{\varepsilon}^{-1}_{k} B^{-1}_{\nu} f_{k\sigma} - f^{*}_{k\sigma} B^{-1}_{\nu} \tilde{\varepsilon}^{-1}_{k} \eta_{k\sigma} \right\} \notag\\
&- \sum_{q,\varsigma} \left\{ \frac12 j^{\varsigma}_{-q} \big(\tilde{V}^{\varsigma}_{q}\big)^{-1} j^{\varsigma}_{q} - \varphi^{\varsigma}_{-q} (\alpha^{\varsigma}_{\omega})^{-1}\big(\tilde{V}^{\varsigma}_{q}\big)^{-1} j^{\varsigma}_{q}  \right\} ,
\label{eq:DB_action}
\end{align}
where we neglected the orbital indices and assumed paramagnetism.
In the latter case, the bosonic propagator becomes diagonal in the channel space, ${\tilde{X}^{\varsigma\varsigma'}_{q}=\delta_{\varsigma,\varsigma'}\tilde{X}^{\varsigma}_{q}}$, and the interaction term simplifies to:
\begin{align}
{\cal F}[f,b]
&=\sum_{k,q}\sum_{\varsigma,\sigma,\sigma'}
\Lambda^{\varsigma}_{\nu\omega} \, f^{*}_{k\sigma}\sigma^{\varsigma}_{\sigma\sigma'} f^{\phantom{*}}_{k+q,\sigma'} \, \varphi^{\varsigma}_{q} + \frac18 \sum_{k,k',q}\sum_{\varsigma,\{\sigma\}} \Gamma^{\,\varsigma}_{\nu\nu'\omega} f^{*}_{k\sigma}\sigma^{\varsigma}_{\sigma\sigma'}f^{\phantom{*}}_{k+q,\sigma'}\,f^{*}_{k'+q,\sigma''} \sigma^{\varsigma}_{\sigma''\sigma'''} f^{\phantom{*}}_{k'\sigma'''}\,,
\label{eq:Wfull}
\end{align}
where the vertex functions in the charge and spin channels are defined as follows:
\begin{align}
\Gamma^{d/m}_{\nu\nu'\omega} &= \Gamma_{\nu\nu'\omega}^{\uparrow\uparrow\uparrow\uparrow} \pm \Gamma_{\nu\nu'\omega}^{\uparrow\uparrow\downarrow\downarrow}\,, \\
\Lambda^{\varsigma}_{\nu\omega} &= \Lambda^{\uparrow\uparrow\varsigma}_{\nu\omega}\,.
\label{eq:Lambda_para}
\end{align}
The dual fermion action, which corresponds to the case of ${V^{\varsigma}_{\bf q}=0}$, can be obtained from Eq.~\eqref{eq:DB_action} by neglecting the bosonic degrees of freedom. 

\subsection{Physical quantities}

In the last two lines of the DB action~\eqref{eq:DB_action} we have explicitly restored the source terms. 
They are needed to derive the connection between the effective dual and physical lattice correlation functions, because the original Grassman variables $c^{(*)}$ have been already integrated out from the dual action.

By definition, the fermionic Green's function can be found as:
\begin{align}
G_{k\sigma} = - \langle c^{\phantom{*}}_{k\sigma} c^{*}_{k\sigma} \rangle = -\frac{1}{{\cal Z}} \frac{\partial^{2}{\cal Z}}{\partial\eta^{*}_{k\sigma}\partial\eta^{\phantom{*}}_{k\sigma}}\,,
\label{eq:Gsource}
\end{align}
where ${{\cal Z} = \int D[c^{*},c]\,e^{-{\cal S}}}$ is the partition function of the initial lattice problem~\eqref{eq:actionlatt}.
Taking into account that the source fields enter only the DB action~\eqref{eq:DB_action} yields the following relation for the lattice Green's function:
\begin{align}
G_{k} &= - \tilde{\varepsilon}^{-1}_{k} 
-\tilde{\varepsilon}^{-1}_{k} B^{-1}_{\nu} \langle f^{\phantom{*}}_{k\sigma} f^{*}_{k\sigma}\rangle^{\phantom{q}}_{\tilde{\cal S}_{\rm DB}} B^{-1}_{\nu} \tilde{\varepsilon}^{-1}_{k} 
= -\tilde{\varepsilon}^{-1}_{k} + 
\tilde{\varepsilon}^{-1}_{k} B^{-1}_{\nu} \tilde{G}_{k} B^{-1}_{\nu} \tilde{\varepsilon}^{-1}_{k}\,,
\label{eq:GF_relation}
\end{align}
where the average is taken with respect to the DB action~\eqref{eq:DB_action} and corresponds to the dressed dual Green's function ${\tilde{G}_{k} = - \langle f^{\phantom{*}}_{k\sigma} f^{*}_{k\sigma}\rangle_{\tilde{\cal S}_{\rm DB}}}$. 
The obtained relation for the lattice Green's function can be rewritten through the dual self-energy $\tilde{\Sigma}_{k}$ as:
\begin{align}
G^{-1}_{k} = \left[g_{\nu} + B_{\nu}\tilde\Sigma_{k}B_{\nu}\right]^{-1} - \tilde\varepsilon_{k}\,.
\label{eq:GtoSigma}
\end{align}
The dual self-energy is related to the dressed dual Green's function through the Dyson equation: 
\begin{align}
\tilde{G}^{-1}_{k} = \tilde{\cal G}^{-1}_{k} - \tilde{\Sigma}_{k}\,.
\label{eq:Gt_Dyson}
\end{align}
The lattice charge and spin susceptibilities can be obtained in a similar way:
\begin{align}
X^{\varsigma}_{q} &= - \langle \rho^{\varsigma}_{q} \, \rho^{\varsigma}_{-q} \rangle =  -\frac{1}{{\cal Z}} \frac{\partial^2{\cal Z}}{\partial j^{\,\varsigma}_{-q} \partial j^{\,\varsigma}_{q}}\,.
\label{eq:Xsource}
\end{align}
Note, that the physical susceptibilities are usually defined with the opposite sign to the correlation function introduced above.
This leads to the following expressions:
\begin{align}
X^{\varsigma}_{q} &= -\big(\tilde{V}^{\varsigma}_{q}\big)^{-1} + \big(\tilde{V}^{\varsigma}_{q}\big)^{-1} (\alpha^{\varsigma}_{\omega})^{-1}\tilde{X}^{\varsigma}_{q}(\alpha^{\varsigma}_{\omega})^{-1}\big(\tilde{V}^{\varsigma}_{q}\big)^{-1}\,, 
\label{eq:X_relation}\\
\big(X^{\varsigma}_{q}\big)^{-1} &= \left[\chi^{\varsigma}_{\omega} + \alpha^{\varsigma}_{\omega}\tilde{\Pi}^{\varsigma}_{q}\alpha^{\varsigma}_{\omega} \right]^{-1} - \tilde{V}^{\varsigma}_{q}\,.
\label{eq:XtoPi}
\end{align}
The dual polarization operator $\tilde{\Pi}^{\varsigma}_{q}$ is related to the dressed dual susceptibility ${\tilde{X}^{\varsigma}_{q} = - \langle \varphi^{\varsigma}_{q} \, \varphi^{\varsigma}_{-q} \rangle_{\tilde{\cal S}_{\rm DB}}}$ through the Dyson equation:
\begin{align}
\big(\tilde{X}^{\varsigma}_{q}\big)^{-1} = \big(\tilde{\cal X}^{\varsigma}_{q}\big)^{-1} - \tilde{\Pi}^{\varsigma}_{q}\,.
\label{eq:Xt_Dyson}
\end{align}

\subsubsection{Role of the ``dual denominator''}
\label{sec:dual_demoninator}

Let us discuss the derived relations between the dual and lattice quantities in more detail. 
In particular, it is insightful to analyze the relation between the lattice and dual self-energies, as well as between the lattice and dual polarization operators:
\begin{align}
\Sigma_{k} &= \Sigma^{\rm imp}_{\nu} + \frac{g^{-1}_{\nu}B_{\nu}\tilde{\Sigma}_{k}B_{\nu}g^{-1}_{\nu}}{1 + g_{\nu} \left[g^{-1}_{\nu}B_{\nu}\tilde{\Sigma}_{k}B_{\nu}g^{-1}_{\nu}\right]}\,,
\label{eq:Sigma_relation}\\
\Pi^{\varsigma}_{q} &= \Pi^{\varsigma\,\rm imp}_{\omega} + \frac{(1-\Pi^{\varsigma\,\rm imp}_{\omega}{\cal U}^{\varsigma}_{\omega}) \alpha^{\varsigma}_{\omega} \tilde{\Pi}^{\varsigma}_{q} \alpha^{\varsigma}_{\omega} (1-\Pi^{\varsigma\,\rm imp}_{\omega}{\cal U}^{\varsigma}_{\omega})}{1 + w^{\varsigma}_{\omega} \left[(1-\Pi^{\varsigma\,\rm imp}_{\omega}{\cal U}^{\varsigma}_{\omega}) \alpha^{\varsigma}_{\omega} \tilde{\Pi}^{\varsigma}_{q} \alpha^{\varsigma}_{\omega} (1-\Pi^{\varsigma\,\rm imp}_{\omega}{\cal U}^{\varsigma}_{\omega})\right]}\,.
\label{eq:Pi_relation}
\end{align}
The expression~\eqref{eq:Sigma_relation} can be obtained upon substituting the Dyson equations for the impurity~\eqref{eq:Dyson_g_imp} and lattice 
\begin{align}
G^{-1}_{k} = i\nu + \mu - \varepsilon_{\bf k} - \Sigma_{k}
\end{align}
Green's functions into Eq.~\eqref{eq:GtoSigma}.
The expression~\eqref{eq:Pi_relation} can be obtained in a similar way upon substituting the Dyson equations for the impurity~\eqref{eq:Dyson_chi_imp} and lattice
\begin{align}
\big(X^{\varsigma}_{q}\big)^{-1} = \big(\Pi^{\varsigma}_{q}\big)^{-1} - U^{\varsigma}_{\bf q}
\end{align}
susceptibilities into Eq.~\eqref{eq:X_relation}.

The relations~\eqref{eq:Sigma_relation} and~\eqref{eq:Pi_relation} immediately suggest the most convenient for interpretation choice of the scaling factors: ${B_{\nu}=g_{\nu}}$ and $\alpha^{\varsigma}_{\omega} = (1-\Pi^{\varsigma\,\rm imp}_{\omega}{\cal U}^{\varsigma}_{\omega})^{-1}$, which is equivalent to Eq.~\eqref{eq:alpha_app}.
As we discussed above, these choices give the most physically intuitive form of the bare dual fermion and boson propagators, which respectively correspond to the difference between the (E)DMFT and impurity Green's functions ${\tilde{\cal G}_{k} = G^{\rm DMFT}_{k} - g_{\nu}}$ and renormalized interactions ${\tilde{\cal X}^{\varsigma}_{q} = W^{\varsigma\,\rm EDMFT}_{q} - w^{\varsigma}_{\omega}}$. 
Upon substituting the scaling factors, one gets the following expressions:
\begin{align}
\Sigma_{k} &=\Sigma^{\rm imp}_{\nu} + \frac{\tilde{\Sigma}_{k}}{1 + g_{\nu} \tilde{\Sigma}_{k}}\,,
\label{eq:Sigma_latt_denom}\\
\Pi^{\varsigma}_{q} &= \Pi^{\varsigma\,\rm imp}_{\omega} + \frac{\tilde{\Pi}^{\varsigma}_{q} }{1 + w^{\varsigma}_{\omega} \tilde{\Pi}^{\varsigma}_{q}}\,.
\label{eq:Pi_latt_denom}
\end{align}
Importantly, one finds that the total lattice self-energy~\eqref{eq:Sigma_latt_denom} is not given by a sum of the self-energy of the reference problem $\Sigma^{\rm imp}_{\nu}$ and the dual self-energy $\tilde{\Sigma}_{k}$, which accounts for the correlation effects beyond the reference problem. 
Instead, the correction to the reference self-energy is given by $\tilde{\Sigma}_{k}$ divided by a, so called, ``dual denominator'' $(1+g_{\nu}\tilde{\Sigma}_{k})$.
A similar relation holds true for the polarization operator~\eqref{eq:Pi_latt_denom}.

We would like to emphasize that the derived relations between the dual and lattice spaces are exact and do not depend on the approximation that are used to obtain the dual quantities.
Furthermore, the use of the source fields allows one to establish direct relations~\eqref{eq:Gsource} and~\eqref{eq:Xsource} only between the correlation functions in the dual and lattice spaces, i.e. between the Green's functions~\eqref{eq:GF_relation} or the susceptibilities~\eqref{eq:X_relation}. 
The relations between the self-energies~\eqref{eq:Sigma_latt_denom} and the polarization operators~\eqref{eq:Pi_latt_denom}, which are not the correlation functions, are obtained from the corresponding relations for the correlation functions via Dyson equations.
These observations suggest that the appearance of dual denominators in Eqs.~\eqref{eq:Sigma_latt_denom} and~\eqref{eq:Pi_latt_denom} has nothing to do with the specific form of the dual self-energy $\tilde{\Sigma}$ and polarization operator $\tilde{\Pi}$ and is a general property of the theory, which emphasizes that correlations within and beyond the reference problem are intertwined in a complex way.
As a consequence, the total lattice self-energy (or polarization operator) cannot be simply obtained as a sum of corresponding contributions $\Sigma^{\rm imp}_{\nu}$ and $\tilde{\Sigma}_{k}$ (or $\Pi^{\varsigma\,\rm imp}_{\nu}$ and $\tilde{\Pi}^{\varsigma}_{k}$), contrary to a common belief of diagrammatic theories formulated on the basis of the DMFT impurity reference system.

The role of the dual denominators is not very straightforward to see if one derives the dual theory from path integral formalism.
However, it is clearly visible if one uses topological analysis of Feynman diagrams, which has been performed in Ref.~\cite{BRENER2020168310} for the self-energy.
This analysis shows that if one considers all possible skeleton diagrams for the self-energy $\Sigma_{k}$ in the conventional
weak-coupling diagrammatic technique for the original system and isolates parts of the diagram that consist of only local impurity Green's functions $g_{\nu}$ and interaction $U$, one immediately arrives at the expression~\eqref{eq:Sigma_latt_denom}.
This happens, because the diagrammatic expansion contains not only two separate parts that can be expressed via only impurity $g_{\nu}$ or only bare dual ${\tilde{\cal G}_{k}}$ Green's function, but there also exist the contributions that contain the complex mixture of these two fermionic lines.
It is important to note, that the diagrammatic expansion in dual theories is defined by the action~\eqref{eq:DB_action_app}, which is derived without approximation from the original lattice problem~\eqref{eq:actionlatt}. 
This diagrammatic expansion is formulated in terms of the exact impurity vertex functions, contained in the interaction term ${{\cal F}[f,\varphi]}$, that are connected by the bare fermionic ${\tilde{\cal G}_{k} = G^{\rm DMFT}_{k}-g_{\nu}}$ and bosonic ${\tilde{\cal X}^{\varsigma}_{q} = W^{\varsigma\,\rm EDMFT}_{q} - w^{\varsigma}_{\omega}}$ propagators. 
Therefore, this form of the diagrammatic expansion does not allow one to connect two impurity vertices by the fermionic $g_{\nu}$ or bosonic $w^{\varsigma}_{\omega}$ lines, which are not present in the dual theory.
This has been pointed out already in the Ref.~\cite{DUPUIS2001617} in the context of a strong-coupling diagrammatic expansion. 
If such a connection would occur, the resulting object itself would be the impurity vertex function.
Since all the vertex functions are already present in ${{\cal F}[f,\varphi]}$, this would lead to a double counting problem in the exact diagrammatic expansion.
Since the dual diagrammatic expansion does not contain this problem by construction, the only place where these term can occur is the Dyson equation for the Green's function.
Therefore, the role of the dual denominator in Eq.~\eqref{eq:Sigma_latt_denom} is to remove the contributions from the lattice Green's function, where the two impurity vertex functions are connected by the impurity Green's function $g_{\nu}$. Note, that the latter appears in the Dyson equation for the lattice Green's function as the local part of the DMFT Green's function $G^{\rm DMFT}_{k}$. 
This can be clearly seen by rewriting the lattice Green's function through the T-matrix:
\begin{align}
G_{k} = \frac{{\cal G}_{k}}{1 - \Sigma_{k}{\cal G}_{k}} 
= \frac{G^{\rm DMFT}_{k}}{1 - \overline{\Sigma}_{k}G^{\rm DMFT}_{k}}
= G^{\rm DMFT}_{k} + G^{\rm DMFT}_{k} T_{k} G^{\rm DMFT}_{k}\,,
\label{eq:G_T_matrix}
\end{align}
where ${\cal G}_{k} = (i\nu + \mu - \varepsilon_{\bf k})^{-1}$ is the bare lattice Green's function and ${\overline{\Sigma}_{k} = \Sigma_{k} - \Sigma^{\rm imp}_{\nu} = \frac{\tilde{\Sigma}_{k}}{1 + g_{\nu}\tilde{\Sigma}_{k}}}$ is the correction to the impurity self-energy.
The introduced T-matrix can be rewritten in the following form:
\begin{align}
T_{k} = \frac{\overline{\Sigma}_{k}}{1 - \overline{\Sigma}_{k}G^{\rm DMFT}_{k}} = \frac{\tilde{\Sigma}_{k}}{1 - \tilde{\Sigma}_{k}(G^{\rm DMFT}_{k} - g_{\nu})} = \frac{\tilde{\Sigma}_{k}}{1 - \tilde{\Sigma}_{k}\tilde{\cal G}_{k}}\,,
\end{align}
which means that the T-matrices for the lattice Green's function~\eqref{eq:G_T_matrix} and the dual Green's function:
\begin{align}
\tilde{G}_{k} =  \tilde{\cal G}_{k} + \tilde{\cal G}_{k} T_{k} \tilde{\cal G}_{k}
\label{eq:G_T_matrix_dual}
\end{align}
coincide identically. 
Therefore, since the diagrammatic expansion for the dual $T$ matrix does not contain the terms, where the impurity vertices are connected by the impurity propagators, these contributions are also not present in the lattice Green's function. This is due to them being excluded from the lattice T-matrix with the help of the dual denominator. 
A similar conclusion applies to the bosonic quantities.

\begin{figure}[t!]
\centering
\includegraphics[width=1\linewidth]{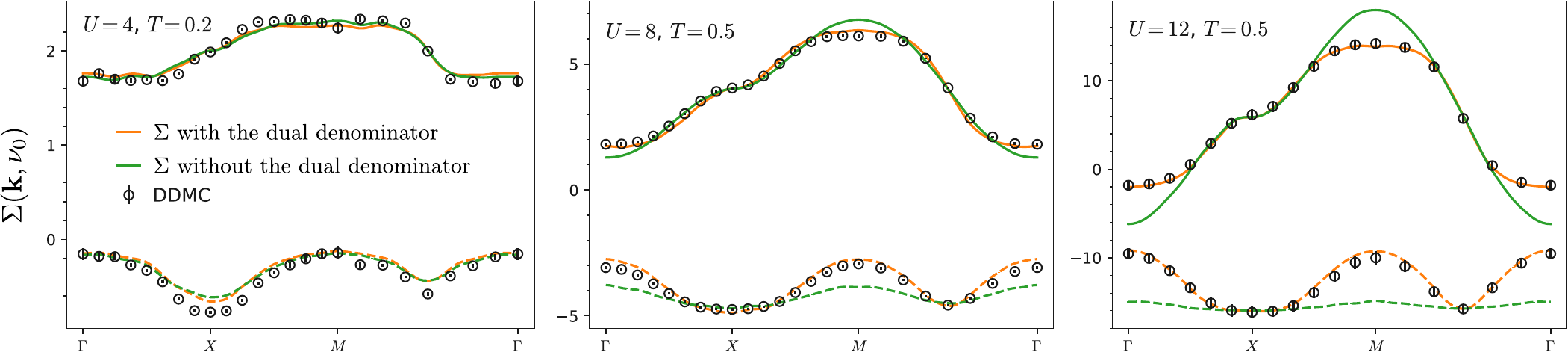}
\caption{The lattice self-energy~\eqref{eq:Sigma_latt_denom} calculated at the lowest Matsubara frequency $\nu_0$ along the high-symmetry path in the BZ. The results are obtained using DiagMC@DF with (orange lines) and without (green lines) the dual denominator and are compared to the DDMC benchmarks (black dots). 
Calculations are performed for a half-filled Hubbard model on a square lattice with the nearest-neighbor hopping ${t=1}$ for three values of the local interaction $U$ and temperature $T$.
Stochastic errors are displayed as vertical bars on all data, but often indiscernible because they are smaller than the line width.
The Figure is adopted from Ref.~\cite{PhysRevB.100.205115}.}
\label{fig:DiagMC_Sigma}
\end{figure}

The importance of the dual denominator can also be confirmed numerically by solving the DB action~\eqref{eq:DB_action} exactly using the diagrammatic Monte Carlo technique (see Section~\ref{sec:DiagMC}) and comparing the result for the lattice self-energy~\eqref{eq:Sigma_latt_denom} with and without the dual denominator against the exact diagrammatic determinant Monte Carlo (DDMC) benchmarks.
The result of this comparison, performed in Ref.~\cite{PhysRevB.100.205115}, is shown in Fig.~\ref{fig:DiagMC_Sigma} for three values of the local Coulomb interaction $U$.
At weak coupling (${U = 4}$), the two alternative self-energies, with (orange lines) and without (green lines) the dual denominators, can hardly be distinguished, because the dual denominator is very close to 1. 
At strong coupling, the self-energy with the dual denominator is in perfect agreement with the DDMC benchmarks (black dots), confirming that the contribution of vertex functions beyond the two-particle level, which are included in the DB scheme, is small in the considered parameter regimes.
In contrast, the self-energy obtained without the dual denominator shows significantly worse agreement with the exact result. 
These findings confirm that the dual denominator should be included in DF/DB calculations by default.

\subsection{Computational workflow}
\label{sec:DB_scheme}

\begin{figure}[b!]
\centering
\begin{tikzpicture}
[
box/.style = {rectangle,draw=black,thick,inner sep=5pt,minimum size=0mm,align=center,text width=1.8cm},
abovebox/.style = {above}
]
 \coordinate (hybridization) at (0,0) ;
 \coordinate (impurity) at ($(hybridization)+(4,0)$) ;
 \coordinate (dualpt) at ($(impurity)+(2.0,-4.0)$) ;
 \coordinate (dualsigma) at ($(dualpt)+(-4,-3.5)$) ;
 \coordinate (lattice) at ($(hybridization)+(-1.0,-3.5)$) ;
 
 \node (hybridization_box) at (hybridization) [box] {$\Delta_{\nu}$, $Y^{\varsigma}_\omega$} ;
 \node (impurity_box) at (impurity) [box] {$g_{\nu}$, $\chi^{\varsigma}_\omega$\\[5pt]$\Gamma^{\varsigma}_{\nu\nu'\omega}$, $\Lambda^{\varsigma}_{\nu\omega}$} ;
 \node (dualpt_box) at (dualpt) [box] {$\tilde{\mathcal{G}}_{k}$, $\tilde{\mathcal{X}}_{q}$\\[5pt]$\Gamma^{\varsigma}_{\nu\nu'\omega}$, $\Lambda^{\varsigma}_{\nu\omega}$} ; 
 \node (dualsigma_box) at (dualsigma) [box] {$\tilde{G}_{k}$ ~~~ $\tilde{\Sigma}_{k}$
 $\tilde{X}^{\varsigma}_{q}$ ~~~ $\tilde{\Pi}^{\varsigma}_{q}$} ; 
 \node (lattice_box) at ($(dualpt_box.north)+(-8,0)$) [box,anchor=north] {$G_{k}$,\! $X^{\varsigma}_{q}$} ; 

 \node (hybridization_text) at (hybridization_box.north) [abovebox] {} ;
 \node (impurity_text) at (impurity_box.north) [abovebox] {} ;
 \node (dualpt_text) at (dualpt_box.north) [abovebox] {} ; 
 \node (dualsigma_text) at ($(dualsigma_box.south)$) [below] {} ; 
 \node (lattice_text) at (lattice_box.north) [abovebox] {} ; 
 
 \node[above] at ($(hybridization)!0.5!(impurity)$) {IS};
 
 \coordinate (helper1) at ($(lattice_box.north east)!0.5!(lattice_box.south east)$) ;
 \coordinate (helper) at (dualpt_box.west |- helper1) ;
 
 \draw[->,very thick] ($(impurity_box.west)!.9!(hybridization_box.east)$) to ($(impurity_box.west)!.1!(hybridization_box.east)$) ;
 \draw[->,very thick] ($(impurity_box.south)!0.1!(dualpt_text.north)$) to node[midway,sloped,above]{~(\ref{eq:bare_dual_G},\ref{eq:bare_dual_X})} (dualpt_text);
 \draw[->,very thick] ($(dualsigma_box.east)!0.9!(dualpt_box.south)$) to node[midway,sloped,below]{dual diagrammatics} ($(dualsigma_box.east)!0.1!(dualpt_box.south)$) ;
 \draw[->,very thick] ($(lattice_box.south)!0.9!(dualsigma_box.west)$) to node[midway,sloped,below]{(\ref{eq:GtoSigma},\ref{eq:XtoPi})} ($(lattice_box.south)!0.1!(dualsigma_box.west)$) ;
 \draw[->,very thick] (lattice_text) to node[midway,sloped,above]{self-consistency~~} ($(hybridization_box.south)!0.1!(lattice_text.north)$) ;
 \draw[dashed,->,very thick] (helper) to node [auto,swap]{EDMFT} ($(lattice_box.north east)!0.5!(lattice_box.south east)$) ;
 
 \draw[densely dotted,thin] (dualpt_box.west) to (dualpt_box.east) ;
 \draw[densely dotted,thin] (impurity_box.west) to (impurity_box.east) ;
 \draw[densely dotted,thin] (dualsigma_box.north) to (dualsigma_box.south) ;

 \draw[->,very thick] ($(dualsigma_box.north west)!0.25!(dualsigma_box.north east)$) to [bend left=120, looseness=2] node [auto,align=center]{Dyson Eqs.\\ (\ref{eq:Gt_Dyson},\ref{eq:Xt_Dyson})} ($(dualsigma_box.north east)!0.25!(dualsigma_box.north west)$) ;
 
 \end{tikzpicture}
 \caption{Summary of the computational workflow. The loop is started with an initial guess for the hybridization function and retarded interaction. Local observables are calculated in the impurity solver (IS) step. Working with bare dual propagators and neglecting diagrammatic corrections is equivalent to EDMFT (dashed arrow). Corrections beyond EDMFT are taken into account by evaluating diagrams for the dual self-energies, which involve dual propagators $\tilde{\cal G}$ and $\tilde{\cal X}$, and the vertex functions $\Lambda$ and $\Gamma$. The diagrams are renormalized self-consistently using Dyson equations. From the dual Green's functions one obtains the physical lattice Green's functions and, in turn, an update for the hybridization and retarded interaction. The Figure was adopted from Ref.~\cite{PhysRevB.90.235135}.}
 \label{fig:summary}
\end{figure}
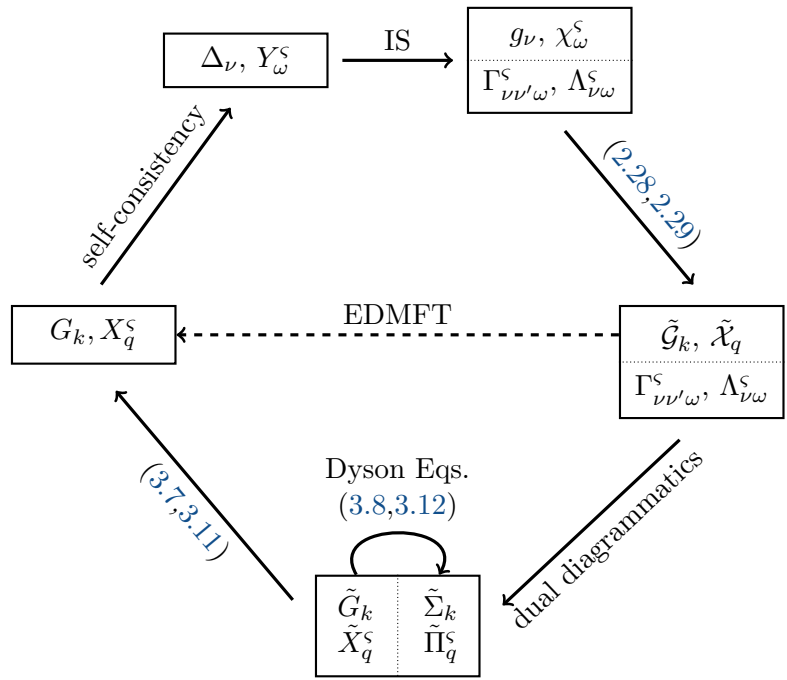

The general computational workflow of the DB theory defined by the action~\eqref{eq:DB_action}, independent of the approximation for the dual self-energy $\tilde{\Sigma}_{k}$ and polarization operator $\tilde{\Pi}^{\varsigma}_{q}$, is depicted in Fig.~\ref{fig:summary} and can be summarized as follows:

\begin{enumerate}
 \item[(1)] Generate an initial guess for the hybridization functions $\Delta_{\nu}$ and $Y^{\varsigma}_{\omega}$. 
 \item[(2)] \label{enum:imp} Solve the impurity problem based on $\Delta_{\nu}$ and $Y^{\varsigma}_{\omega}$, and compute $g_{\nu}$, $\chi^{\varsigma}_\omega$, $\Lambda^{\varsigma}_{\nu\omega}$, and $\Gamma^{\varsigma}_{\nu\nu'\omega}$.
 \item[(3)] Calculate the bare dual propagators $\tilde{\mathcal{G}}_{k}$ and $\tilde{\mathcal{X}}^{\varsigma}_{q}$ according to Eqs.~\eqref{eq:bare_dual_G} and~\eqref{eq:bare_dual_X}.
 \item[(4)] \label{enum:dual} Evaluate diagrams for $\tilde{\Sigma}_{k}$ and $\tilde{\Pi}^{\varsigma}_{q}$ using dual perturbation theory.
 \item[(5)] \label{enum:innersc} Compute renormalized dual propagators $\tilde{G}_{k}$ and $\tilde{X}^{\varsigma}_{q}$ using Dyson equations~\eqref{eq:Gt_Dyson} and~\eqref{eq:Xt_Dyson}. 
 \item[(6)] Go back to step (4) and loop until convergence (inner self-consistency).
 \item[(7)] Once the inner loop is converged, calculate the physical propagators $G_{k}$ and $X^{\varsigma}_{q}$ according to Eqs.~\eqref{eq:GtoSigma} and~\eqref{eq:XtoPi}.
 \item[(8)] Update the hybridization functions $\Delta_{\nu}$ and $Y^{\varsigma}_{\omega}$. 
 \item[(9)] Go back to step (2) and repeat until convergence is reached (outer self-consistency).
\end{enumerate}
\noindent
If steps (4-6) are skipped, the calculation proceeds with the bare dual propagators, yielding the EDMFT solution. 
This situation is indicated by the dashed arrow in Fig.~\ref{fig:summary}, and in this case the vertices do not need to be computed.

\subsubsection{Self-consistency condition}
\label{sec:self-consistency}

The computational workflow of the dual theories allows for several variations. 
First, one may notice that applying the outer self-consistency, which modifies the reference problem through the update of the hybridization functions, drastically increases the costs of numerical calculations, because one has to recalculate the two-particle quantities of the reference problem (the susceptibilities and vertex functions) at every iteration of the outer loop.
Therefore, to make the computations more feasible, in many cases the DF/DB calculations are done without the outer self-consistency starting from the hybridization functions of the converged (E)DMFT solution at the step (1).   
The DMFT self-consistency condition on the fermionic hybridization function $\Delta_{\nu}$ equates the local part of the lattice Green's function to the impurity Green's function, which is equivalent to requiring that the local part of the bare dual Green's function is zero:
\begin{align}
\sum_{\bf k} G^{\rm DMFT}_{{\bf k}\nu} = g^{\phantom{*}}_{\nu} ~~~ \equiv ~~~ \sum_{\bf k} \tilde{\cal G}^{\phantom{*}}_{{\bf k}\nu} = 0\,.
\label{eq:DMFT_condition}
\end{align}
The EDMFT self-consistency condition on the bosonic hybridization $Y^{\varsigma}_{\omega}$ equates the local part of the lattice susceptibility (or of the renormalized interaction) to the impurity susceptibility (or to the renormalized interaction), which is equivalent to requiring that the local part of the bare dual bosonic propagator vanishes:
\begin{align}
\sum_{\bf q} X^{\varsigma\,\rm EDMFT}_{{\bf q}\omega} = \chi^{\varsigma}_{\omega} ~~~ \equiv ~~~ \sum_{\bf q} W^{\varsigma\,\rm EDMFT}_{{\bf q}\omega} = w^{\varsigma}_{\omega} ~~~ \equiv ~~~ \sum_{\bf q} \tilde{\cal X}^{\varsigma}_{{\bf q}\omega}=0\,.
\label{eq:EDMFT_condition}
\end{align}
After the converged (E)DMFT solution is found, vertex functions and susceptibilities of the impurity problem are evaluated only once, and the computational workflow truly corresponds to a diagrammatic expansion around the (E)DMFT impurity problem.

On the other hand, introducing the outer self-consistency has a clear advantage of providing a better starting point for the diagrammatic expansion and also, depending on the condition, can impose consistency between the local quantities calculated within the lattice and impurity problems. 
The dual theories introduce the non-local diagrammatic corrections to the self-energy and polarization operator.
This makes the self-consistency conditions~\eqref{eq:DMFT_condition} and~\eqref{eq:EDMFT_condition} for the dressed lattice and dual propagators nonequivalent: 
\begin{gather}
\sum_{\bf k} G_{{\bf k}\nu} = g^{\phantom{*}}_{\nu} ~~~ \not\equiv ~~~ \sum_{\bf k} \tilde{G}^{\phantom{*}}_{{\bf k}\nu} = 0\,, \\
\sum_{\bf q} X^{\varsigma}_{{\bf q}\omega} = \chi^{\varsigma}_{\omega} ~~~ \not\equiv ~~~ \sum_{\bf q} W^{\varsigma}_{{\bf q}\omega} = w^{\varsigma}_{\omega} ~~~ \not\equiv ~~~ \sum_{\bf q} \tilde{X}^{\varsigma}_{{\bf q}\omega}=0\,.
\end{gather}
At the moment, there is no consensus on wether the ``lattice'' or ``dual'' self-consistency performs better, especially since the majority of actual calculations within the dual approaches is performed without the outer self-consistency, which greatly reduces computational costs.
Moreover, each of these conditions has its own advantages.

For example, the lattice conditions on the Green's function, ${\sum_{\bf k} G_{{\bf k}\nu} = g^{\phantom{*}}_{\nu}}$, and the susceptibility, ${\sum_{\bf q} X^{\varsigma}_{{\bf q}\omega} = \chi^{\varsigma}_{\omega}}$, equates the local correlation functions calculated within the lattice and impurity problems.
In particular, this makes the calculation of the local electronic density and potential energy/double occupancy consistent between these two problems.
Using the lattice self-consistency condition for the susceptibility thus enforces the Mermin-Wagner theorem, which forbids phase transitions associated with continuous symmetry breaking in finite systems or in low-dimensions ${d<3}$ at finite temperatures~\cite{Mermin66}.
According to Vilk and Tremblay~\cite{refId0}, a sufficient criterion for satisfying the Mermin-Wagner theorem is that the double
occupancy ${D = \langle n_{\uparrow}n_{\downarrow}\rangle}$ obtained from taking the local, equaltime part of the susceptibility stays within the physical
range ${[0, n^2/2]}$. 
The lattice self-consistency on the susceptibility ensures that the double occupancy of the lattice is equal to that of a reference impurity model, which is solved exactly. 
The latter stays within the physical bounds, so that the former does as well and the method satisfies the Mermin-Wagner theorem.
Additionally, considering the lattice condition on the susceptibility allows one to remove some contributions of the high-order vertex functions from the dual self-energy based on the ``superline'' argument (see Ref.~\cite{PhysRevB.93.045107} for details).

In turn, the dual conditions ${\sum_{\bf k} \tilde{G}^{\phantom{*}}_{{\bf k}\nu} = 0}$ and ${\sum_{\bf q} \tilde{X}^{\varsigma}_{{\bf q}\omega}=0}$ simplify the dual diagrammatics by removing the diagrams that have local fermionic and bosonic lines in their structure, respectively.
Furthermore, one can prove that using the self-consistency condition on the dual Green's function leads to a causal solution. 
Indeed, applying this self-consistency condition to Eq.~\eqref{eq:GF_relation} gives:
\begin{align}
\Delta_{\nu} = \sum_{\bf k} \tilde{\varepsilon}_{{\bf k}\nu} G_{{\bf k}\nu} \tilde{\varepsilon}_{{\bf k}\nu}
= \sum_{\bf k} \frac{(\varepsilon_{\bf k} - \Delta_{\nu})^2}{i\nu + \mu - \Sigma_{{\bf k}\nu}}\,,
\end{align}
which, according to Ref.~\cite{PhysRevB.105.245115}, is causal.

Note, that the presented above computational workflow enables construction of the diagrammatic expansion in the dual space based on an arbitrary choice of the hybridization functions.
Indeed, the latter are added and subtracted from Eq.~\eqref{eq:actionlatt} in order to isolate the reference system~\eqref{eq:actionimp_app} from the initial lattice action.
In particular, this means that one is allowed to perform DB calculations without considering the bosonic hybridization, ${Y^{\varsigma}_{\omega}=0}$, even if the non-local interaction $V^{\varsigma}_{\bf q}$ is present in the system. 
In this case, the effects related to $V^{\varsigma}_{\bf q}$ will be accounted for diagrammatically through the bosonic propagator $\tilde{X}^{\varsigma}_{q}$.
Since the non-local interaction is typically much smaller than the local Coulomb potential $U$, neglecting the bosonic hybridization usually does not lead to any substantial loss of accuracy, except for very large values of $V^{\varsigma}_{\bf q}$ (see, e.g., Ref.~\cite{10.21468/SciPostPhys.6.1.009}), while significantly reduces the computational cost of solving the reference problem.
Additionally, considering the frequency-dependent bosonic hybridizations in the spin channel, $Y^{m}_{\omega}$ may lead to the breaking of local conservation laws in the reference system~\cite{PhysRevB.96.075155}. 
Furthermore, the use of the bosonic hybridization function is hindered by the lack of efficient impurity solvers capable of handling bosonic hybridizations, especially in the multi-orbital case.

\begin{figure}[t!]
\centering
\includegraphics[width=1\linewidth]{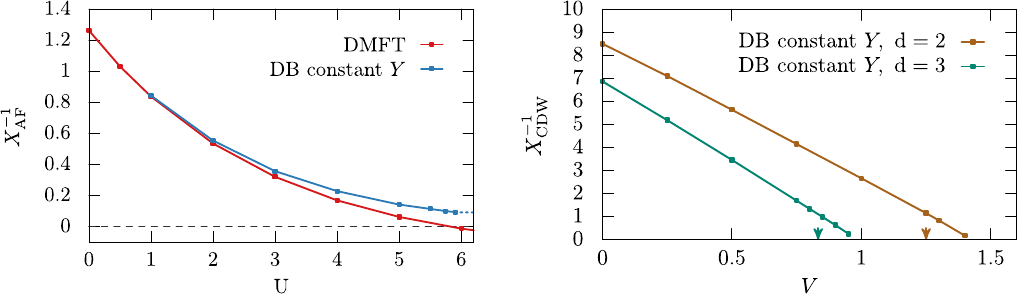}
\caption{Left panel: The inverse of the static [${\omega=0}$] antiferromagnetic [${{\bf q} = (\pi,\pi)}$] susceptibility ${X_{\rm AF} = X^{z}_{{\bf q},\omega}}$ of the square lattice Hubbard model with the nearest-neighbor hopping ${t=1}$ calculated at the inverse temperature ${\beta = 3}$ as a function of the local Coulomb interaction $U$. 
The N\'eel transition to an antiferromagnetically ordered state occurs when this inverse
susceptibility is equal to zero (dashed black line).
The DMFT is unstable towards antiferromagnetism after ${U = 6}$. In DB, the self-consistency condition~\eqref{eq:MW_condition} pushes the inverse susceptibility away from zero.
Right panel: Comparison of the charge density wave (CDW) transition in the extended Hubbard model on a square (${d=2}$) and cubic (${d=3}$) lattices with the nearest-neighbor hopping ${t=1}$. 
The inverse of the static [${\omega=0}$] charge susceptibility ${X_{\rm CDW} = X^{\rm d}_{{\bf Q},\omega}}$ at the CDW wave vector ${Q=(\pi,\pi)}$ (${d=2}$) and ${Q=(\pi,\pi,\pi)}$ (${d=3}$) is calculated using the DB approach with instantaneous interaction for ${U = 5}$ as a function of a nearest-neighbor Coulomb interaction $V$. The square lattice simulations correspond to ${\beta = 3}$, the cubic lattice -- to ${\beta = 2.5}$. 
The arrows indicate the mean-field estimates ${V = U/4}$ and ${V = U/6}$ for the CDW transition on a square and cubic lattices, respectively.
The Figure is adopted from Ref.~\cite{PhysRevB.100.165128}.
\label{fig:MW}}
\end{figure}

However, one can still consider the instantaneous (frequency-independent) bosonic hybridization functions in the computational scheme.
As has been shown in Ref.~\cite{PhysRevB.100.165128}, accounting for such hybridizations in the charge and spin $z$ channels renormalizes the local Coulomb interaction as ${U\to{}U+Y^{d} - Y^{z}}$, which does not complicate numerical solution of the impurity problem.
The bosonic hybridization functions can be determined from the following condition:
\begin{align}
\sum_{{\bf q},\omega}X^{d/z}_{{\bf q},\omega} = \sum_{\omega}\chi^{d/z}_{\omega}\,,
\label{eq:MW_condition}
\end{align}
which enforces the Mermin-Wagner theorem, as discussed above.
In particular, this approach prevents the N\'eel transition associated with the formation of the antiferromagnetic order in a two-dimensional system, as demonstrated in the left panel in Figure~\ref{fig:MW}. 
At the same time, it does not affect the phase transition to a charge ordered state, which is allowed by the Mermin-Wagner theorem as it is associated with the spontaneous breaking of a discrete symmetry (see right panel of Figure~\ref{fig:MW}).

\begin{figure}[b!]
\centering
\includegraphics[width=0.5\linewidth]{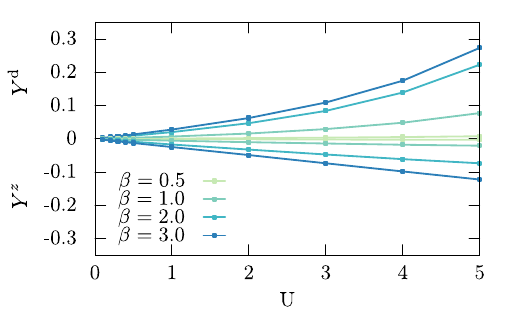}~~~
\includegraphics[width=0.5\linewidth]{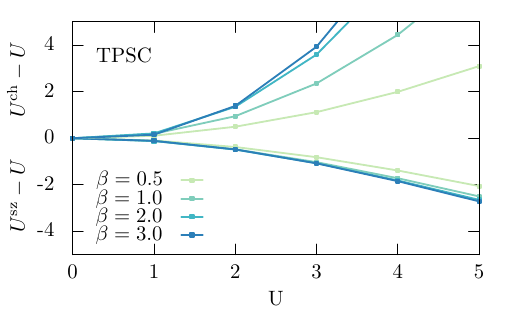}
\caption{Left panel: Instanteneous bosonic hybridizations $Y^{d/z}$ as a function of temperature and interaction strength. Bottom panel: Effective interactions in the two-particle
self-consistent method. Calculations are performed for the Hubbard model on a square lattice with the nearest-neighbor hopping ${t=1}$. The Figure is adopted from Ref.~\cite{PhysRevB.100.165128}.
\label{fig:Ueff_DB}}
\end{figure}

Interestingly, the static charge and spin hybridizations obtained within the proposed scheme have opposite sign, but similar amplitude ${Y^{d}\simeq-Y^{z}}$ that is proportional to $U$ at small interaction strengths (left panel of Fig.~\ref{fig:Ueff_DB}).
Additionally, the renormalization of the local Coulomb interaction in the static DB scheme is substantially smaller than that in the two-particle self-consistent approach (TPSC)~\cite{PhysRevB.55.3870, refId0} (right panel of Fig.~\ref{fig:Ueff_DB}), where the bare interaction in the charge and spin channels is also adjusted by imposing the sum rules for the corresponding susceptibilities that enforce the Pauli exclusion principle and Mermin-Wagner theorem.

\subsection{Perturbative expansion}
\label{sec:strong_coupling}

In addition to the fact that the dual expansion is perturbative in both weak- and strong-coupling limits (see Section~\ref{sec:Advantages_dual}), one can show~\cite{hafermann2010numerical} that this expansion coincides with the strong-coupling expansion in terms of physical fermions~\cite{PhysRevB.72.045111} in the limit of ${\tilde{\varepsilon}_{k}\to0}$. 
In particular, if the fermionic hybridization is zero, ${\Delta_{\nu}=0}$, the ${\tilde{\varepsilon}_{k}}$ reduces to the electronic dispersion ${\tilde{\varepsilon}_{k}\to\varepsilon_{\bf k}}$, and the expansion becomes the strong-coupling expansion around the atomic limit~\cite{DUPUIS2001617}.
Here, we will introduce a bit different proof to the one presented in Ref.~\cite{hafermann2010numerical}.
The electronic Green's function $G_{k}$ can be obtained as~\eqref{eq:Gsource}: 
\begin{align}
G_{k\sigma} = - \frac{1}{{\cal Z}} \int D[c^{*},c] \, c^{\phantom{*}}_{k\sigma} c^{*}_{k\sigma} \, e^{-{\cal S}}\,,
\end{align}
where ${\cal Z}$ is the partition function corresponding to the lattice action~\eqref{eq:actionlatt}.
Taking the action in the form ${{\cal S} = \sum_{i}{{\cal S}^{(i)}_{\rm imp}} + {\cal S}_{\rm rem}}$ defined in Section~\ref{sec:reference_system} allows one to directly expand the Green's function to the first order in $\tilde{\varepsilon}$ as:
\begin{align}
G_{k\sigma} &\simeq - \frac{1}{{\cal Z}[c^{*},c]} \int D[c^{*},c] \, c^{\phantom{*}}_{k\sigma} c^{*}_{k\sigma} \, e^{-{\cal S}_{\rm imp}} + \frac{1}{{\cal Z}[c^{*},c]} \int D[c^{*},c] \sum_{k',\sigma'} c^{\phantom{*}}_{k\sigma} c^{*}_{k\sigma}  c^{*}_{k'\sigma'} c^{\phantom{*}}_{k'\sigma'} \, \tilde{\varepsilon}_{k'\sigma'} \, e^{-{\cal S}_{\rm imp}} \notag\\
&\hspace{0.4cm}- \frac{1}{{\cal Z}[c^{*},c]} \int D[c^{*},c] \, c^{\phantom{*}}_{k\sigma} c^{*}_{k\sigma}\, e^{-{\cal S}_{\rm imp}} \times \frac{1}{{\cal Z}[c^{*},c]} \int D[c^{*},c] \sum_{k',\sigma'} c^{*}_{k'\sigma'}c^{\phantom{*}}_{k'\sigma'} \, \tilde{\varepsilon}_{k'\sigma'} \, e^{-{\cal S}_{\rm imp}} \notag\\
&= g_{\nu\sigma} + \sum_{k',\sigma'} \tilde{\varepsilon}_{k'\sigma'} \big\{- \langle c^{\phantom{*}}_{\nu\sigma} c^{*}_{\nu'\sigma'} c^{*}_{\nu\sigma} c^{\phantom{*}}_{\nu'\sigma'} \rangle_{\rm connected} - g_{\nu\sigma}g_{\nu'\sigma'} +  g_{\nu\sigma}g_{\nu\sigma} \delta_{\nu,\nu'} \delta_{\sigma,\sigma'} + g_{\nu\sigma}g_{\nu'\sigma'} \big\} \notag\\
&= g_{\nu\sigma} + g_{\nu\sigma} \tilde{\varepsilon}_{k\sigma} g_{\nu\sigma} - \sum_{q,\sigma'} g_{\nu\sigma} \left[\Gamma^{\sigma\sigma'\sigma\sigma'}_{\nu\nu\omega} g_{\nu+\omega,\sigma'}\tilde{\varepsilon}_{k+q,\sigma'}g_{\nu+\omega,\sigma'}\right] g_{\nu\sigma}\,.
\label{eq:Strong_coupling_exact}
\end{align}
Note, that in the last line of this expression the four-point vertex $\Gamma$~\eqref{eq:GammaPH} is defined with the scaling parameter ${B_{\nu\sigma}=g_{\nu\sigma}}$ and that we have changed the index $k'$ to ${k+q}$. 
The obtained expression is identical to the strong-coupling expansion in terms of the physical fermionic fields $c^{(*)}$ derived in Ref.~\cite{PhysRevB.72.045111}.

One can notice, that the term in square brackets in the last line of Eq.~\eqref{eq:Strong_coupling_exact} reminds of the lowest-order self-energy that can be introduced for the dual action~\eqref{eq:DB_action}:
\begin{align}
\tilde{\Sigma}^{(1)}_{\nu\sigma} = -\sum_{q,\sigma'}\tilde{\cal G}^{\phantom{f}}_{k+q,\sigma'}\Gamma^{\sigma\sigma'\sigma\sigma'}_{\nu \nu \omega}\,,
\end{align}
with the approximate bare dual Green's function~\eqref{eq:bare_dual_G} expanded to the first order in $\tilde{\varepsilon}_{k\sigma}$:
\begin{align}
\tilde{\cal G}_{k\sigma} \simeq g_{\nu\sigma}\tilde{\varepsilon}_{k\sigma}g_{\nu\sigma}\,.
\end{align}
Indeed, Eq.~\eqref{eq:Strong_coupling_exact} can be obtained from the dual diagrammatic expansion by varying the corresponding expression for the lattice Green's function~\eqref{eq:GtoSigma} to the first order in $\tilde{\varepsilon}$:
\begin{align}
G_{k\sigma} \simeq g_{\nu\sigma} + g_{\nu\sigma} \tilde{\varepsilon}_{k\sigma} g_{\nu\sigma} +  g_{\nu\sigma} \left[\sum_{k',\sigma'}\frac{\partial\tilde{\Sigma}_{k\sigma}}{\partial\tilde{\varepsilon}_{k'\sigma'} }\Bigg|_{\tilde{\varepsilon}=0}\tilde{\varepsilon}_{k'\sigma'} \right]g_{\nu\sigma}.
\end{align}
Taking into account that the variation of the dual self-energy $\tilde{\Sigma}_{k\sigma}$ with respect to $\tilde{\varepsilon}_{k'\sigma'}$ affects only the dual Green's functions, and that ${\tilde{G}_{k\sigma}=0}$ if ${\tilde{\varepsilon}_{k\sigma}=0}$, reduces the dual self-energy to the lowest order and leads to the following relation:
\begin{align}
\frac{\partial\tilde{\Sigma}_{k\sigma}}{\partial\tilde{\varepsilon}_{k'\sigma'} }\Bigg|_{\tilde{\varepsilon}=0} = \frac{\partial\tilde{\Sigma}^{(1)}_{\nu\sigma}}{\partial\tilde{\varepsilon}_{k'\sigma'} }\Bigg|_{\tilde{\varepsilon}=0} = - \sum_{q,\sigma''}  \frac{\partial\tilde{\cal G}_{k+q,\sigma''}}{\partial\tilde{\varepsilon}_{k'\sigma'} }\Bigg|_{\tilde{\varepsilon}=0} \Gamma^{\sigma\sigma''\sigma\sigma''}_{\nu \nu \omega}.
\end{align}
Upon using Eq.~\eqref{eq:bare_dual_G} with the  scaling parameter ${B_{\nu\sigma}=g_{\nu\sigma}}$, this gives the final expression for the lattice Green's function:
\begin{align}
G_{k\sigma} \simeq g_{\nu\sigma} + g_{\nu\sigma} \tilde{\varepsilon}_{k\sigma} g_{\nu\sigma} - \sum_{q,\sigma'} g_{\nu\sigma} \left[\Gamma^{\sigma\sigma'\sigma\sigma'}_{\nu\nu\omega} \left(g_{\nu+\omega,\sigma'}\tilde{\varepsilon}_{k+q,\sigma'}g_{\nu+\omega,\sigma'}\right)\right] g_{\nu\sigma}
\end{align}
that identically coincides with Eq.~\eqref{eq:Strong_coupling_exact}.

\subsection{Diagrammatic Monte Carlo calculations}
\label{sec:DiagMC}

The truncation of the interaction term ${\tilde{\cal F}[f,b]}$ at the two-particle level~\eqref{eq:lowestint} together with a perturbative form of the diagrammatic expansion in the dual space enables the exact numerical solution of the dual problem~\eqref{eq:DB_action} by summing up all Feynman diagrams by means of the diagrammatic Monte Carlo (DiagMC) technique~\cite{PhysRevLett.81.2514}. 

The bare DiagMC method has been successfully applied to the Hubbard model at weak and moderate Coulomb interactions~\cite{Kozik_2010}. 
This method relies on a weak-coupling expansion in terms of the local Coulomb interaction $U$ and constructs all Feynman diagrams up to some finite but high order in $U$. The algorithm enables sampling all possible diagrams without any restriction to specific topologies.
Efficient algorithms that express all connected diagrams of the perturbative expansion up to a given order by means of determinants~\cite{PhysRevLett.119.045701} have been developed for various observables and correlation functions ~\cite{PhysRevB.100.121102, PhysRevB.97.085117, sigmaDet, PhysRevLett.124.117602}, significantly reducing the computational cost of the calculation.
Approaches based on a small-coupling expansion work very well in the regime of small to moderate couplings, but start to fail when $U$ is of the order of half of the bandwidth~\cite{PhysRevX.5.041041, PhysRevB.91.235114}. 
These failure is related to the finite convergence radius of the diagrammatic series and can be improved using resummation techniques~\cite{PhysRevB.100.121102}.

To allow for a non-perturbative treatment of strong correlation effects, a diagrammatic Monte Carlo scheme based on dual fermion (DiagMC@DF) approach was proposed in Refs.~\cite{PhysRevB.94.035102, PhysRevB.96.035152}. The advantage of this method in comparison with diagrammatic expansions in terms of the bare Coulomb interaction $U$ is that the reference impurity problem for the dual diagrammatic expansion already accounts for the main effects of local correlations that strongly screen the bare interaction $U$. The dual expansion is thus performed in terms of the renormalized local interaction vertex function, which appears to be naturally more convenient at moderate and strongly interacting regime.
Additionally, the diagrams are sampled in continuously in the momentum space without the discretization of the Brillouin zone. Hence, the result of the calculation is not influenced by any finite-size effects.
In Ref.~\cite{PhysRevB.102.195109} this approach was generalized to the Dual Boson Diagrammatic Monte Carlo (DiagMC@DB) method by incorporating the non-local Coulomb interaction in the original DiagMC@DF approach.

\subsubsection{Self-energy calculations}

\begin{figure}[t!]
\centering
\includegraphics[scale=0.8]{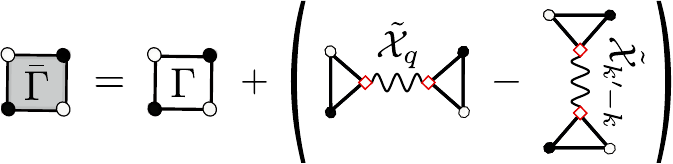}
\caption{\label{fig:vertex} Schematic diagrammatic interpretation of Eq.~\ref{DBvertex}. The full antisymmetrized fermion-fermion interaction $\overline{\Gamma}$ (gray box) is a combination of the impurity vertex ${\Gamma}$ (white box) and processes involving a boson exchange (wiggly line). The full vertex acquires a momentum dependence due to the presence of the bosonic lines. White and black dots represent incoming and outgoing particles, respectively. Triangles represent $\Lambda_{\nu\omega}$ vertices. The exact dependence on the channel indices and prefactors is shown in Eq.~\ref{DBvertex}. The Figure is adopted from Ref.~\cite{PhysRevB.102.195109}.}
\end{figure}

To implement the DiagMC@DB method, one can start with the DB action~\eqref{eq:DB_action} and integrate out the bosonic degrees of freedom.
This integration can be done analytically, because the DB action is quadratic in the bosonic fields, which results in a modified dual fermion action~\cite{PhysRevB.102.195109}:
\begin{align}
&\tilde{\mathcal{S}} = -\sum\limits_{k,\sigma} f^{*}_{k\sigma} \tilde{\mathcal{G}}^{-1}_{k} f^{\phantom{s}}_{k\sigma} + {\frac{1}{8}} \sum_{k,q,\varsigma,\{\sigma\}}\overline{\Gamma}^{\varsigma}_{kk'q} f^*_{k\sigma} \sigma_{\sigma\sigma'}^\varsigma f^{\phantom{*}}_{k+q,\sigma'} f^*_{k'+q,\sigma''} \sigma_{\sigma''\sigma'''}^\varsigma f^{\phantom{*}}_{k'\sigma'''}\,.
\label{finalaction}
\end{align}
In this expression, we introduced a new momentum dependent fermion-fermion vertex $\overline{\Gamma}^{\varsigma}_{kk'q}$ that combines the vertex function of the local impurity problem $\Gamma^{\varsigma}_{\nu\nu'\omega}$ and the non-local interaction between fermions ${\tilde{M}_{\nu,\nu',\omega}^{\varsigma, \qv} = \Lambda^{\varsigma}_{\nu,\omega} \tilde{\cal X}^{\varsigma}_{q}\Lambda^{\varsigma}_{\nu'+\omega, -\omega}}$ mediated by the dual bosonic fluctuation $\tilde{\cal X}^{\varsigma}_{q}$:
\begin{align}
\overline{\Gamma}^{{\rm d}}_{kk'q} &= \Gamma^{\rm d}_{\nu\nu'\omega} + 2 \tilde{M}_{\nu,\nu',\omega}^{{\rm d},\qv} - \tilde{M}_{\nu,\nu+\omega,\nu'-\nu}^{{\rm d}, \kv'-\kv} - 3\tilde{M}_{\nu,\nu+\omega,\nu'-\nu}^{{\rm m}, \kv'-\kv}\,, \notag\\
\overline{\Gamma}^{{\rm m}}_{kk'q} &= \Gamma^{{\rm m}}_{\nu\nu'\omega} + 2\tilde{M}_{\nu,\nu',\omega}^{{\rm m},\qv} + \tilde{M}_{\nu,\nu+\omega,\nu'-\nu}^{{\rm m}, \kv'-\kv} - \tilde{M}_{\nu,\nu+\omega,\nu'-\nu}^{{\rm d}, \kv'-\kv}\,.
\label{DBvertex}
\end{align}
The structure of the new fermion-fermion vertex function~\eqref{DBvertex} is schematically shown in Fig.~\ref{fig:vertex}.

\begin{figure}[b!]
\centering
\includegraphics[width=0.6\linewidth]{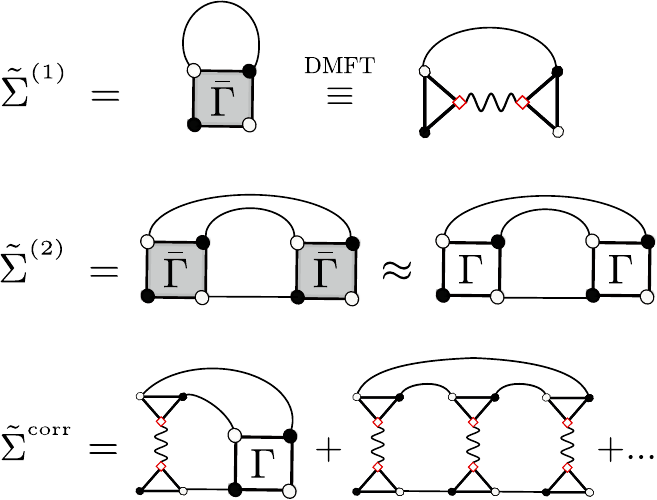}
\caption{\label{fig:diagrams} Most important self-energy diagrams. Top row shows the only nonzero contribution to the first order diagram, taking into account that we can not connect two slots of the same local vertex with a propagator line due to DMFT self-consistency condition. The middle row shows the second-order diagrams $\tilde{\Sigma}^{(2)}$. If $V$ is small compared to $U/4$, it can be approximated by the second-order dual fermion diagram on the right hand side. The last term $\tilde{\Sigma}^{ {\rm corr}}$ shows few diagrams that enter $\tilde{\Sigma}$ in our calculations, but are not included in the ladder DB.}
\end{figure}

Integrating out of bosonic fields from the DB action~\eqref{eq:DB_action} is very important for the calculation of diagrams, because it allows to eliminate the bosonic degrees of freedom from the theory analytically and to avoid their sampling in diagrammatic Monte Carlo. 
The DF action does not contain the bosonic degrees of freedom by construction, and the interaction vertex $\overline{\Gamma}^{\varsigma}_{kk'q}$ reduces to the one of the impurity problem $\Gamma^{\varsigma}_{\nu\nu'\omega}$.
The derived action~\eqref{finalaction} allows one to construct all the Feynman diagrams for the dual self-energy $\tilde{\Sigma}_{k}$ up to any finite order in the interaction vertex and to sum over them using Markov chain Monte Carlo (for details see Refs.~\cite{PhysRevB.94.035102, PhysRevB.96.035152, PhysRevB.102.195109}). 
The leading diagrammatic contributions to $\tilde{\Sigma}_{k}$ are shown in Fig.~\ref{fig:diagrams}.
The single-particle observables of the lattice problem are further obtained using the exact relations~\eqref{eq:GtoSigma} and~\eqref{eq:Sigma_relation}.

\begin{figure}[t!]
\centering
\includegraphics[width=0.75\linewidth]{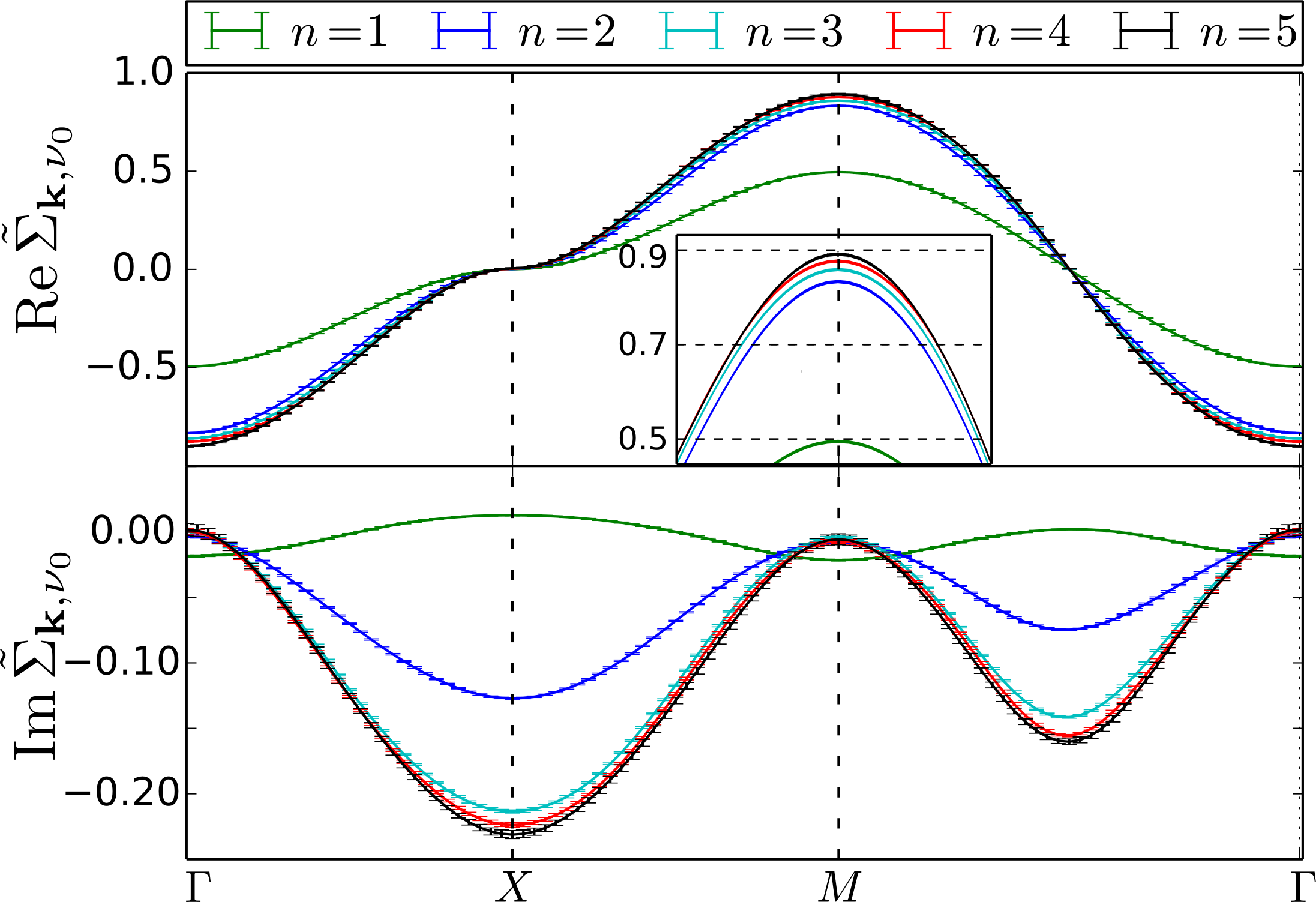}   
\caption{\label{dualsigma} Convergence of the real (top panel) and imaginary (bottom panel) parts of the dual self-energy $\tilde{\Sigma}_{\kv,\nu}$ obtained for the zeroth Matsubara frequency $\nu_0$. The result is plotted along the high-symmetry path in momentum space $\kv$ as a function of the expansion order $n$. The parameters are $U=5$, $V=1.25$, and $\beta=2$ in the units of the hopping amplitude. The inset shows the convergence of the real part around the $M=\{\pi, \pi\}$ point. }
\end{figure}

Importantly, thanks to its perturbative nature, the DiagMC@DB approach converges at low expansion orders.
Fig.~\ref{dualsigma} shows an output of the calculation performed for the extended Hubbard model on a square lattice with the nearest neighbor hopping ${t=1}$, the local Coulomb interaction ${U=5}$, and the nearest-neighbor Coulomb interaction ${V=1.25}$. 
The result is obtained at the inverse temperature ${\beta=2}$ at the lowers Matsubara frequency $\nu_0$ along the high-symmetry path in the Brillouin zone that goes through the ${\Gamma=(0,0)}$, ${\text{X}=(\pi,0)}$, and ${\text{M}=(\pi,\pi)}$ points. 
We find, that the dual self-energy $\tilde{\Sigma}_{k}$ obtained at the 5th order of expansion differs from the 4th order by ${\sim1\%}$. 
Practically, this means that away from collective electronic instabilities the DiagMC@DB calculations can be considered converged already at order 4.
Additionally, we observe that the main contribution to the real part of the self-energy comes from two kind of diagrams that are shown in Fig.~\ref{fig:diagrams}. 
The first is the single boson diagram $\tilde{\Sigma}^{(1)}$ that contains only one factor $\tilde{M}^{\varsigma,\,\qv}_{\nu,\nu'\omega}$, which is the only non-zero contribution at the first order of the diagrammatic expansion in terms of the vertex function.
This can be already seen in Fig.~\ref{dualsigma}, where the first order contribution accounts for around 50\% of the real part of the dual self-energy. 
The second important contribution is the second order dual fermion diagram $\tilde{\Sigma}^{(2)}$, that contains two fermion-fermion vertices connected to each other.
At values of $V$ far away from the CDW instability, other contribution to ${{\rm Re}\,\tilde{\Sigma}}$ are rather small compared to these ones. 
On the other hand, the imaginary part of the self-energy is much more sensitive to higher order corrections. 
In Fig.~\ref{dualsigma} we can see that the second order is way off compared to the third order, accounting for only around 50\% of the contributions to ${{\rm Im}\,\tilde{\Sigma}}$.
Interestingly, the third order already accounts for most of the contributions. 
We deduce, that the inclusions of third-order diagrams in our expansion that contain multiple fermion-fermion scattering and bosonic exchanges are important for the imaginary part of the self-energy. 
These diagrams contribute to around 40\% of ${{\rm Im}\,\tilde{\Sigma}}$ at high symmetry points for ${U=5}$, and their impact on dual quantities becomes even more important at larger $U$. 
Orders larger than the third typically amount to a correction of less than 10\% of ${{\rm Im}\,\tilde{\Sigma}}$ at high symmetry points.

\begin{figure}[t!]
\centering
\includegraphics[width=1\linewidth]{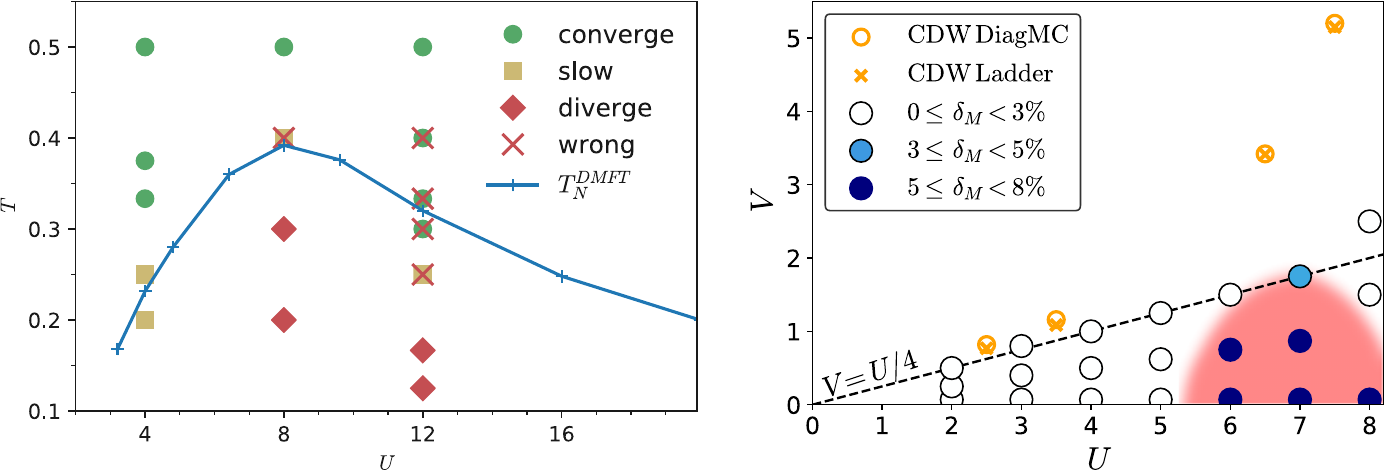}   
\caption{Left panel: The $T$-$U$ phase diagram of the half-filled Hubbard model on a square lattice with the nearest-neighbor hopping ${t=1}$ obtained through the analysis of the DiagMC@DF diagrammatic expansion.
Red diamonds mark parameters where the DF series appears to diverge. 
Green circles (yellow squares) indicate rapid (slow) series convergence, red crosses convergence to incorrect results. For comparison, the blue curve shows DMFT N\'eel
temperatures from Ref.~\cite{PhysRevB.83.085102}. The Figure is adopted from Ref.~\cite{PhysRevB.96.035152}.
Right panel: The $V$-$U$ phase diagram of the extended Hubbard model on a square lattice with the nearest-neighbor hopping ${t=1}$ obtained using the DiagMC@DB method and the ladder dual boson (LDB) approximation (see Section~\ref{sec:LDB}). 
Results for ${U=4}$ are obtained at ${\beta=4}$. 
For ${U > 4}$ the calculations are performed at ${\beta=2}$. 
The mismatch parameter $\delta_{M}$ between the LDB and DiagMC@DB results is depicted by color.
Points correspond to physical parameters for which calculations are
performed. The red area highlights the region where the mismatch
parameter is larger. Transition points between the normal and charge density-wave (CDW)
phases obtained in DiagMC@DB and LDB calculations are
depicted by an orange circle and cross, respectively. 
The dashed black line ${V = U/4}$ represents a mean-field estimation of the CDW phase boundary. The Figure is taken from Ref.~\cite{PhysRevB.102.195109}.
\label{fig:DiagMC_Phase}}
\end{figure}

\subsubsection{Detection of phase transitions}
\label{sec:DiagMC_phase}

Besides accurate calculation of the single-particle quantities, such as the self-energies shown, e.g., in Figs.~\ref{fig:DiagMC_Sigma} and~\ref{dualsigma}, the DiagMC scheme in the dual space enables detection of phase transitions to the spin~\cite{PhysRevB.96.035152} and charge~\cite{PhysRevB.102.195109} ordered states by analyzing the divergence of the diagrammatic expansion. 
The resulting phase diagrams can be found in Fig.~\ref{fig:DiagMC_Phase}.

The left panel in Fig.~\ref{fig:DiagMC_Phase} shows the analysis of the divergence of the DiagMC@DF expansion for the square lattice Hubbard model in the vicinity of the N\'eel transition to the antiferromagnetically ordered state.
For all considered interactions, convergence of the DF expansion with diagram order becomes slower at low temperatures and eventually the series starts to diverge.
The crossover from fast to slow convergence and further to divergence is visible in the
succession of green circles, yellow squares, and red diamonds.
Red crosses indicate cases where the series converges towards a result that differs significantly from the exact solution. 
In all cases, the breakdown regime agrees well with the temperature range where the system develops significant magnetic correlations, as confirmed by comparison with the N\'eel transition temperatures predicted by DMFT~\cite{PhysRevB.83.085102} (blue curve).

The right panel in Fig.~\ref{fig:DiagMC_Phase} shows the $V$-$U$ phase diagram of the extended Hubbard model on the square lattice obtained using the DiagMC@DB approach~\cite{PhysRevB.102.195109}.
The transition to the charge density-wave (CDW) state, obtained through the analysis of the DiagMC@DB expansion and within the ladder DB (LDB) approximation (see Section~\ref{sec:LDB}), is respectively depicted by the orange circles and crosses~\cite{PhysRevB.102.195109}. 
The CDW state on a square lattice is characterized by a checkerboard configuration of local electronic densities in the real space with alternating empty sites and doubly occupied sites. 
The phase transition to this state occurs when the nearest-neighbor interaction $V$ is large enough to overcome the effect of the on-site Coulomb repulsion $U$ that favors a single-electron occupation of lattice sites. 
A perturbative expansion at small values of $U$ predicts the onset of the CDW phase to be located at $V\simeq{}U/8+\text{const}$~\cite{PhysRevB.99.115112}. 
A mean-field estimate based on RPA or $GW$ theories gives the transition point at $V\simeq{}U/4$ \cite{PhysRevB.95.245130}. 
This behavior is reproduced at moderate interaction strength by the dynamical cluster approximation (DCA) calculations~\cite{PhysRevB.99.245146, PhysRevB.97.115117}. 
Finally, for large values of $U$ and large temperatures the position of the onset of the CDW phase appears to shift towards the value $V\simeq{}U$ that can be found, for example, using the Peierls-Feynman-Bogoliubov variational principle~\cite{PhysRevLett.111.036601}. 
The DiagMC@DB and LDB results shown in the right panel of Fig.~\ref{fig:DiagMC_Phase} and presented in Refs.~\cite{PhysRevB.90.235135, PhysRevB.94.205110} reproduce all these distinct trends in the different regimes.

\subsection{Ladder approximation}
\label{sec:LDB}

A standard DiagMC computation in the dual space for a single-band extended Hubbard model requires around 12 hours with a hundred parallel runs in order to obtain a converged result at the 5th order. 
For a converged result at the 6th order, the required computational time increases to more than 24 hours in order to reach a reasonable accuracy.
Although this is still a reasonable time for a single calculation, exploring phase diagrams with the DiagMC approach for DF/DB becomes inefficient.
Additionally, a drastic increase in the computational time is expected if the method is applied in a realistic multi-orbital framework.
In this regard, the ladder approximation for the dual self-energy and polarization operator is a good compromise.
This approximation accounts for the particle-hole excitations in the charge and spin channels, which are usually the leading collective electronic fluctuations in the case of extended Hubbard model with repulsive interactions.
Additionally, this approximation provides rather accurate results for the single- and two-particle quantities in a broad range of model parameters.

\subsubsection{Diagrammatic expressions}
\label{sec:LDB_diagrammatics}

\begin{figure}[t!]
\centering
\includegraphics[width=0.85\linewidth]{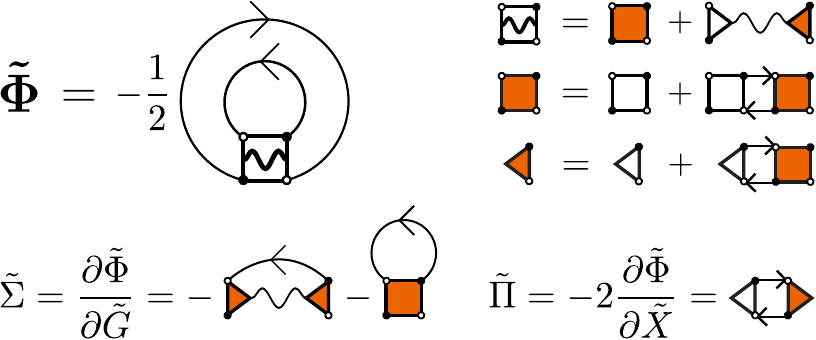}
\caption{The diagrammatic expression for the dual functional $\tilde{\Phi}$, self-energy $\tilde{\Sigma}$, and polarization operator $\tilde{\Pi}$ in the ladder approximation.
The three- and four-point vertex functions of the reference problem are denoted by the white triangle and square, respectively. The vertices renormalized by the particle-hole excitations in the ladder approximation are shown in orange. The solid lines with arrows correspond to the dressed dual Green's functions $\tilde{G}$. The wiggly lines are the dressed bosonic propagators $\tilde{X}$. The Figure is adopted from Ref.~\cite{PhysRevB.93.045107}. 
\label{fig:DB_functional}}
\end{figure}

Within the ladder DF (LDF) and ladder DB (LDB) approximations, the consistency between the single- and two-particle quantities is ensured by the generating functional $\tilde{\Phi}$ in the dual space~\cite{PhysRevB.93.045107} shown in Fig.~\ref{fig:DB_functional}.
The dual self-energy and polarization operator can be found as usual, by cutting the respective line in the functional:
\begin{align}
\tilde{\Sigma}_{k\sigma} = \frac{\partial\tilde{\Phi}}{\partial\tilde{G}_{k\sigma}}\,;~~~~
\tilde{\Pi}_{q}^{\varsigma} = - 2 \frac{\partial\tilde{\Phi}}{\partial\tilde{X}^{\varsigma}_{q}}\,.
\end{align}
In these expressions, we have restored the dependence of single-particle quantities on the spin index $\sigma$ in order to make the diagrammatic expressions more clear.
The self-energy of LDB consists of the LDF contribution, that contains only the four-point vertices and Green's functions, and the mixed contribution, which additionally includes the bosonic propagators and three-point vertex functions: 
\begin{align}
\tilde{\Sigma}^{\rm LDB}_{k\sigma} =
\tilde{\Sigma}^{\rm LDF}_{k\sigma} +
\tilde{\Sigma}^{\rm mix}_{k\sigma}\,.
\label{eq:Sigma_LDB}
\end{align}
The LDF self-energy:
\begin{align}
\tilde{\Sigma}^{\rm LDF}_{k\sigma} =
\tilde{\Sigma}^{\rm ladd}_{k\sigma} -\tilde{\Sigma}^{(2)}_{k\sigma}
\label{eq:Sigma_LDF}
\end{align}
is given by the two-particle ladder diagram:
\begin{align}
\tilde{\Sigma}^{\rm ladd}_{k\sigma} = -\sum_{q,\sigma'}\tilde{G}^{\phantom{f}}_{k+q,\sigma'}{\rm P}^{\sigma\sigma'\sigma\sigma'}_{\nu \nu q} = -\frac12 \sum_{q,\varsigma} \tilde{G}^{\phantom{f}}_{k+q,\sigma}{\rm P}^{\varsigma}_{\nu \nu q}\,.
\label{eq:Sigma_ladd}
\end{align}
As follows from the Schwinger-Dyson equation for the dual self-energy, the ladder diagram accounts for twice the contribution of the second-order self-energy~\cite{hafermann2010numerical}:
\begin{align}
\Sigma_{k\sigma}^{(2)} &= -\frac{1}{2}\sum_{q, k',\sigma^{\{\prime\}}} \Gamma^{\sigma \sigma' \sigma'' \sigma'''}_{\nu\nu'\omega} \tilde{G}_{k',\sigma''} \tilde{G}_{k'+q, \sigma'''} \tilde{G}_{k+q,\sigma'} \Gamma^{\sigma'' \sigma''' \sigma\sigma'}_{\nu'\nu\omega},
\label{Sigma_2nd}
\end{align}
which has to be excluded from the expression~\eqref{eq:Sigma_LDF} in order to avoid the double-counting.
The mixed diagram, that additionally appears in the LDB theory due to the presence of the bosonic propagator $\tilde{X}^{\varsigma}_{q}$, is following:
\begin{align}
\tilde{\Sigma}^{\rm mix}_{k\sigma} = -\sum_{q,\varsigma} L^{\varsigma}_{\nu q} \tilde{G}_{q+k,\sigma} \tilde{X}^{\varsigma}_{q} L^{\varsigma}_{\nu+q,-q}\,.
\label{eq:Sigma_mix}
\end{align}
The dual polarization operator in the ladder approximation reads:
\begin{align}
\tilde{\Pi}^{\varsigma}_{q}
= \sum_{k,\sigma} \Lambda^{\varsigma}_{\nu+\omega,-\omega} \tilde{G}_{k\sigma} \tilde{G}_{q+k,\sigma} L^{\varsigma}_{\nu q}\,.
\label{eq:DualPi}
\end{align}
The diagrammatic expressions for ${\tilde{\Sigma}}$ and ${\tilde{\Pi}}$ in the ladder approximation are shown in Fig.~\ref{fig:DB_functional}.
There, the orange triangles and squares correspond respectively to the three-point ($L^{\varsigma}_{\nu q}$) and four-point (${\rm P}^{\varsigma}_{\nu\nu'q}$) vertex functions that are renormalized by the particle-hole ladders:
\begin{align}
L^{\varsigma}_{\nu q} &= \Lambda^{\varsigma}_{\nu \omega} + \sum_{k_1} {\rm P}^{\varsigma}_{\nu \nu_1 q} \tilde{G}^{\phantom{\varsigma}}_{k_1\sigma} \tilde{G}^{\phantom{\varsigma}}_{k_1+q, \sigma} \Lambda^{\varsigma}_{\nu_1 \omega}\,,
\label{eq:Lvertex}\\
{\rm P}^{\varsigma}_{\nu\nu'q} &= 
\Gamma^{\varsigma}_{\nu\nu'\omega} + 
\sum_{k_1} {\rm P}^{\varsigma}_{\nu \nu_1 q} \tilde{G}^{\phantom{\varsigma}}_{k_1\sigma} \tilde{G}^{\phantom{\varsigma}}_{k_1+q, \sigma}
\Gamma^{\varsigma}_{\nu_1 \nu' \omega}\,.
\label{eq:Pvertex}
\end{align}

\subsubsection{Benchmarking against the DiagMC results}

Benchmarking the ladder dual fermion/boson approach against the exact DiagMC@DF/DB solution of the dual problem~\eqref{eq:DB_action} performed in Refs.~\cite{PhysRevB.94.035102, PhysRevB.96.035152, PhysRevB.102.195109} confirms that the particle-hole ladder diagrams give the leading contribution to the self-energy in a broad range of temperatures, interacting strengths and doping levels, and accounting for these contributions is already sufficient to obtain accurate results for the single- and two-particle correlation functions, such as the electronic Green's function and the charge and spin susceptibilities.

\begin{figure}[t!]
\centering   
\includegraphics[width=0.9\linewidth]{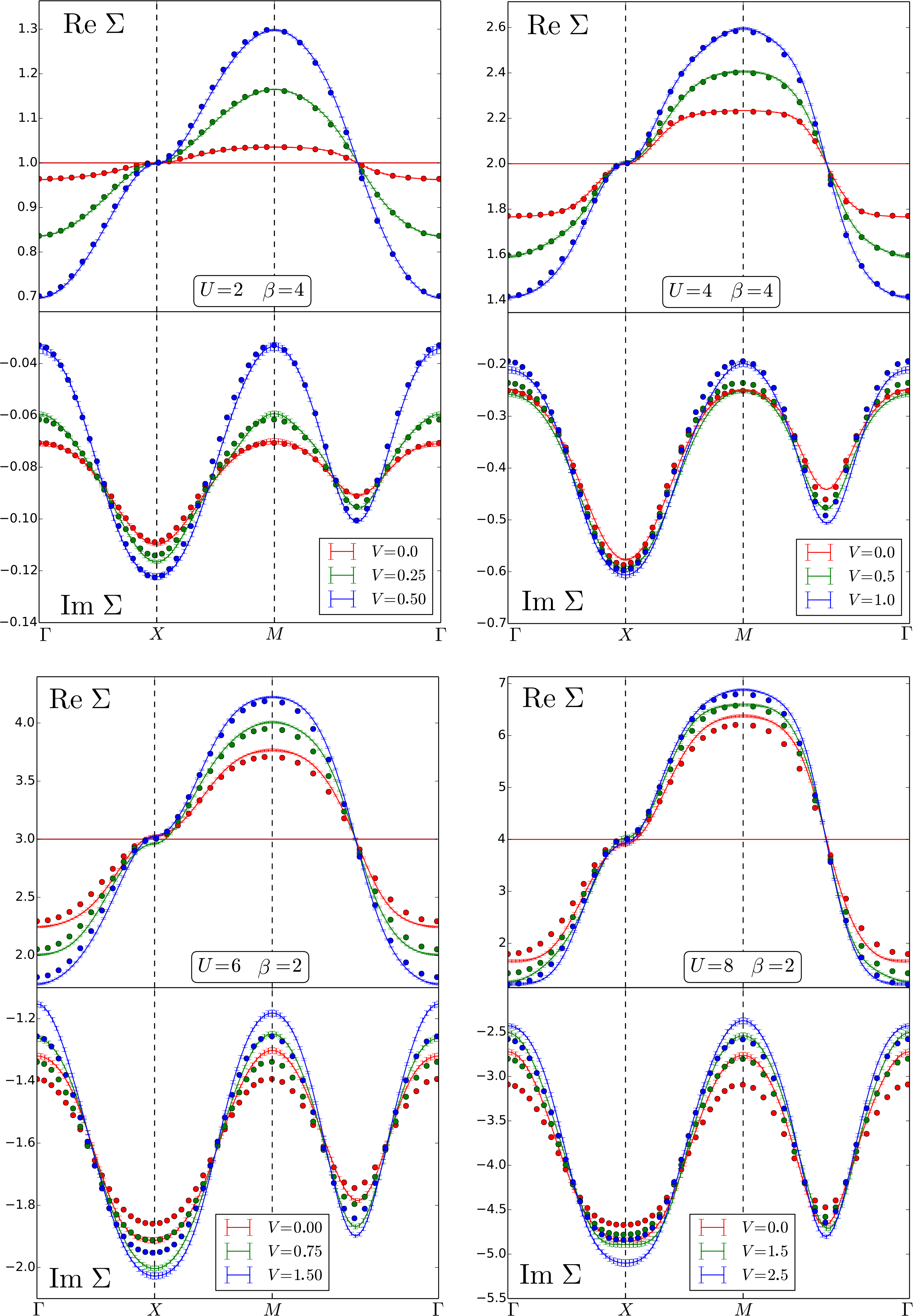}
\caption{\label{fig:comparison2468} 
Comparison between DiagMC@DB (solid lines with error bars) and LDB (dots) results for the real (top panel) and imaginary (bottom panel) parts of the lattice self-energy $\Sigma_{{\bf k}, \nu_0}$ calculated at the lowest Matsubara frequency $\nu_0$ along the high-symmetry path in the BZ. The result is obtained for a half-filled extended Hubbard model on a square lattice with the nearest-neighbor hopping ${t=1}$ at ${U=2}$, ${\beta=4}$ (top left), ${U=2}$, ${\beta=4}$ (top right), ${U=2}$, ${\beta=4}$ (bottom left), and ${U=2}$, ${\beta=4}$ (bottom right), and different values of the non-local Coulomb interaction $V$ specified in the legend. The Figure is taken from Ref.~\cite{PhysRevB.102.195109}.}
\vspace{-1cm}
\end{figure}

Figure~\ref{fig:comparison2468} shows a comparison of the DiagMC@DB and LDB calculations for the extended Hubbard model on a square lattice with the nearest-neighbor hoipping ${t=1}$. 
The results are obtained for different values of the local $U$ and nearest-neighbor $V$ Coulomb interactions, from a quarter of the bandwidth ($U=2$) up to the bandwidth ($U=8$). 
We find that the agreement between these two methods is substantially exact up to a half of the bandwidth for all considered values of the non-local interaction $V$. 
In fact, in this regime the LDB result for the lattice self-energy lies inside the error bars of the DiagMC@DB calculation. 
For larger values of the on-site Coulomb interaction exceeding the half of the bandwidth, the difference between the two theories is more noticeable, especially for a small strength of the non-local interaction $V$.
In order to quantify the difference between these two methods, we look at the $M=(\pi,\pi)$ point in the momentum space and calculate the following quantity:
\begin{align}
\delta_{M} ={\rm Re}\left[ \frac{\overline{\Sigma}_{M,\nu_0}^{\rm{DiagMC}}-\overline{\Sigma}_{M,\nu_0}^{\rm{ladd.}\phantom{g}}}{\overline{\Sigma}_{M,\nu_0}^{\rm{DiagMC.}\phantom{g}}}\right],
\label{eq:Delta_LDB}
\end{align}
where $\overline{\Sigma}_{M,\nu_0}$ is the difference between the self-energy in the specified method and the DMFT impurity self-energy $\Sigma_{\nu_0}^{\rm imp}$, and $\nu_0$ is the lowest positive Matsubara frequency.
We measure differences with respect to DMFT self-energy, because the latter is constant in momentum space and quite large. 
If we want to resolve relatively small differences in momentum space, we have to exclude its contribution.
Additionally, we choose the $M$ point, because it shows the largest difference between the two curves in the Brillouin zone. In this way, we are sure that the $\delta_M$ parameter contains information only about the maximum mismatch coming from the dual corrections. The reason for taking the real part of this quantity comes from the fact that the imaginary part of the dual self-energy ${\rm Im} \tilde{\Sigma}$ shows a systematic shift already at ${V=0}$, i.e. at the dual fermion level (see Ref.~\cite{PhysRevB.96.035152}).
Here, we aim to assess the behavior of the self-energy as a function of the non-local $V$ rather than to investigate this aspect.

The result for the mismatch parameter $\delta_{M}$ is summarised in a tentative phase diagram shown in the right panel of Fig.~\ref{fig:DiagMC_Phase}. 
We can conclude that the difference between DiagMC@DB and LDB approaches is negligible at small $U$ below the half of the bandwidth and further increases with the local interaction. 
This behavior can be explained considering that for small local Coulomb interaction $U$ the regime is still perturbative in the dual boson theory, so we expect all the methods to give quantitatively similar results. 
This finding is also in agreement with the result of DiagMC@DF calculations~\cite{PhysRevB.96.035152} obtained for the zero non-local Coulomb interaction. 
On the other hand, we observe that the mismatch is more severe at $V=0$ and decreases as $V$ increases. 
Indeed, when the non-local Coulomb interaction is large, charge fluctuations in the particle-hole channel are expected to give the main contribution to physical observables such as self-energy and susceptibility~\cite{PhysRevB.99.115124}, because the system lies close to the charge density wave (CDW) phase. 
Ladder DB approach accounts for this kind of fluctuations by construction, and for this reason the mismatch $\delta_M$ decreases. 
From the right panel of Fig.~\ref{fig:DiagMC_Phase}, we find that the values of $U$ at which the largest mismatch occurs (red area) lie in the region where the N\'eel temperature is highest~\cite{PhysRevB.90.235132} and the spin fluctuations are characterized by the largest correlation length~\cite{PhysRevB.91.125109, PhysRevLett.124.017003, PhysRevLett.124.117602, PhysRevResearch.4.043201}.
This result suggests, that in this regime magnetic fluctuations become strongly anharmonic~\cite{PhysRevB.102.224423, PhysRevB.103.245123}, and their accurate description requires considering the particle-hole scattering on the transverse momentum-dependent fluctuations that are not accounted for by the ladder approximation but are included in DiagMC@DB (see, e.g. $\tilde{\Sigma}^{\rm corr}$ in Fig.~\ref{dualsigma}). 
Nevertheless, largest mismatch between the DiagMC@DB and LDB results does not exceed $\delta_{\rm M}=8\%$ in the region of the parameter space where the DiagMC@DB series converges, which offers a further validation of the accuracy of the LDB technique over a very wide range of interaction strengths. 
Another consideration that emerges from this analysis as a function of $U$ and $V$ is that up to a half the bandwidth a momentum dependence of the real part of the self-energy at $V=U/4$ is dominated by the non-local interaction $V$.
It plays a very important role even for $U=6$, where we would expect the local interaction to give the most important contribution.

\subsubsection{Electronic spectral function}

\begin{figure}[t!]
\centering   
\includegraphics[width=0.7\linewidth]{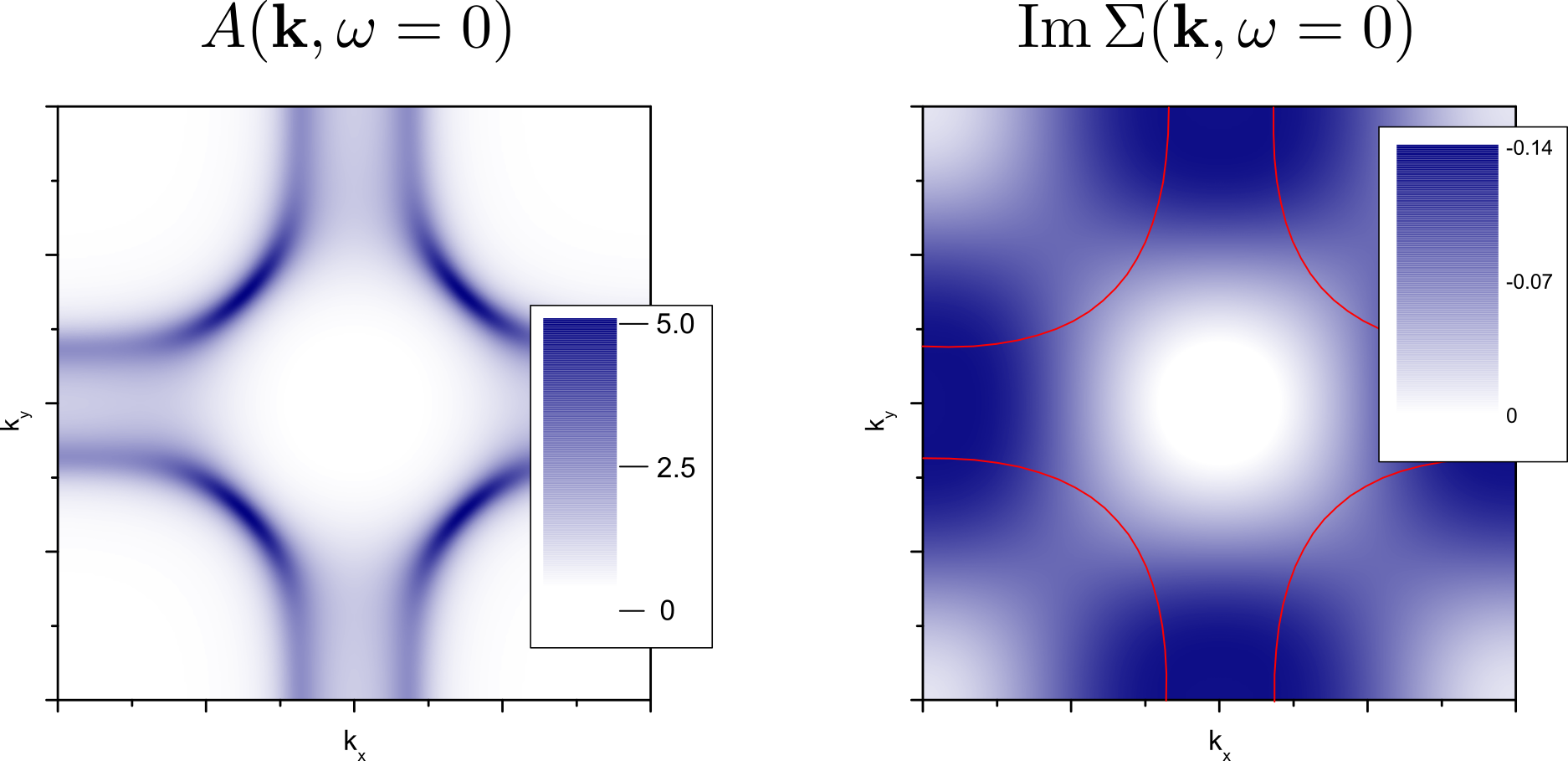}
\caption{\label{fig:A_ImS} 
The electronic spectral function ${A({\bf k},\omega=0)}$ (left panel) and the imaginary part of the electronic self-energy ${{\rm Im}\,\Sigma({\bf k},\omega=0)}$ (left panel) obtained at Fermi level (${\omega=0}$).
The calculation is performed for the square lattice Hubbard model for the set of parameters corresponding to the pseudogap regime of high-temperature superconducting cuprates: the nearest-neighbor hopping ${t
= 0.25}$, the next-nearest-neighbor hopping ${t'=-0.075}$, the local Coulomb interaction ${U=4}$, the inverse temperature ${\beta=80}$, and the 14\% of hole doping. The red line in the right panel depicts the non-interacting Fermi surface.
The Figure is adopted from Ref.~\cite{PhysRevB.79.045133}.}
\end{figure}

The advantage of the ladder DF/DB method over DMFT is that it incorporates the effect of spatial collective electronic fluctuations in the momentum-dependent self-energy. 
The leading collective electronic instability of the Hubbard model in a broad range of interaction strength is associated with strong magnetic fluctuations. 
These fluctuations can strongly modify the electronic spectral function.
In particular, at low temperatures these fluctuations may lead to a momentum-selective  destruction of the Fermi surface leading to formation of Fermi arcs in the momentum resolved electronic spectral function that are observed, e.g., in cuprates in the pseudogap regime.
These effects can be captured only via a momentum-dependent self-energy and are reproduced in the DF approach though the incorporation of the particle-hole ladder diagrams, describing collective spin fluctuations, in the dual self-energy.

\begin{figure}[t!]
\centering   
\includegraphics[width=0.65\linewidth]{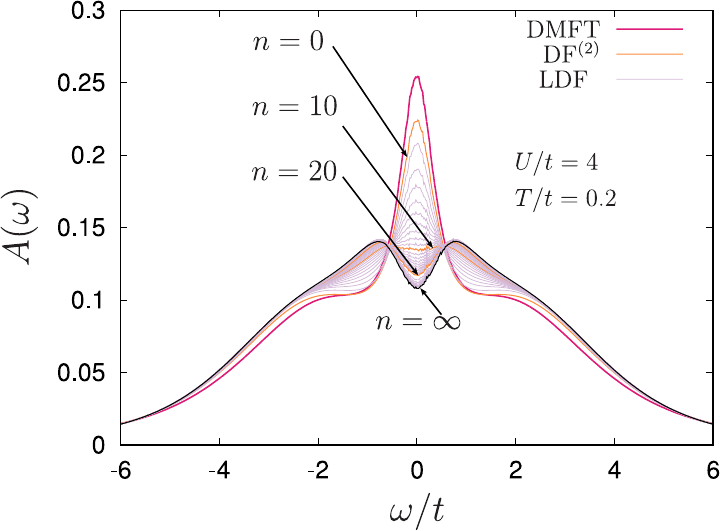}
\caption{\label{fig:Infinite_PG} 
The local electronic spectral function ${A_\omega}$ calculated using DMFT and LDF for the half-filled square lattice Hubbard model with nearest-neighbor hopping $t$. 
For a given $n$, the dual self-energy includes ladder
diagrams up to order ${n + 2}$ in the four-point vertex $\Gamma$. 
The case ${n = 0}$ corresponds to the second-order approximation DF$^{(2)}$.
The Figure is taken from Ref.~\cite{PhysRevB.97.085125}.}
\end{figure}

Fig.~\ref{fig:A_ImS} shows the DF calculation performed in Ref.~\cite{PhysRevB.79.045133} for the square lattice Hubbard model for the set of parameters corresponding to the pseudogap regime of high-temperature superconducting cuprates. 
One finds that in the presence of strong magnetic fluctuations only the parts of Fermi surface near the nodal direction remain well defined at low temperature. 
In the antinodal direction, the spectral function at the Fermi level is vanishingly small in accordance with the experimental results.
The reduction of the spectral weight in the antinodal region of the FS is directly connected to the large, compared to the nodal point, value of the imaginary self-energy shown in the right panel.
Interestingly, the momentum-selective  destruction of the Fermi surface is captured already by the second-order self-energy~\eqref{Sigma_2nd}, which is the lowest-order ${\bf k}$-dependent self-energy in the DF diagrammatic expansion.
However, we note that in some cases, an accurate description of the pseudogap in the electronic spectral function requires considering the infinite-ladder contribution to the self-energy~\cite{PhysRevB.97.085125}.
As illustrated in Fig.~\ref{fig:Infinite_PG}, in the regime of strong magnetic fluctuations the paramagnetic DMFT solution exhibits a quasiparticle peak in the local spectral fucntion $A_{\omega}$.
Considering the non-local self-energy within the DF approach reduces the spectral weight. 
The reduction is small in second-order DF (labeled DF$^{(2)}$), but increases with order of the ladder diagrams. 
Remarkably, diagrams at all orders contribute to the pseudogap. 
As has been found in Ref.~\cite{PhysRevB.97.085125}, the low-order diagrams mediate short-range correlations, while long-range correlations require high diagram orders. 

\subsubsection{Phase transitions to the ordered states}

The ladder DF/DB method is efficient in detecting phase transitions to the spin and charge ordered states.  
According to Landau phenomenology~\cite{LL_V}, a transition from a normal to an ordered state characterized by an order parameter $\rho$ occurs due to a spontaneous symmetry breaking. 
The latter results in the change of the free energy ${\cal S}[\rho]$ from a parabolic-like form with a minimum at ${\rho=0}$ to a double-well potential characterized by minima at ${\rho\neq0}$.
This change in the free energy can be seen in the sign change of the second variation of the free energy $\partial^{2}_{\rho}{\cal S}[\rho]\Big|_{\rho=0}$ with respect to the corresponding order parameter $\rho^{\varsigma}_{Q,\omega=0}$ (see e.g. Ref.~\cite{PhysRevB.102.224423}). 
This second variation corresponds to the inverse of the static susceptibility $X^{\varsigma}_{Q,\omega=0} = -\langle \rho^{\phantom{*}}_{Q,\omega=0} \rho^{\phantom{*}}_{-Q,\omega=0} \rangle$, which becomes zero at the transition point:
\begin{align}
-\frac{\partial^2{\cal S}[\rho]}{\partial\rho^{\varsigma}_{Q,\omega=0}\partial\rho^{\varsigma}_{-Q,\omega=0}} = \left[X^{\varsigma}_{Q,\omega=0}\right]^{-1}.
\label{eq:AFM_criterion}
\end{align}

The infinite-order ladder DF/DB expansion is the minimal approximation that is able to capture the transition to the ordered state through the divergence of susceptibility~\eqref{eq:X_relation}.
In DF, this divergence occurs when the leading eigenvalue of the Bethe-Salpeter equation (BSE) for the four-point vertex~\eqref{eq:Pvertex} renormalized by the particle-hole ladder reaches unity~\cite{PhysRevB.77.195105}.
The wave vector $Q$ at which divergence occurs defines to the ordering vector.
Eq.~\eqref{eq:X_relation} derived for the DB approach, and the corresponding one for the DF approach~\cite{PhysRevB.90.235132, hafermann2010numerical}, prove that the lattice and dual susceptibilities diverge at the same time.
This fact allows one to efficiently detect collective electronic instabilities already in the dual space by the divergence of the corresponding diagrammatic expansion without the need to calculate physical observables for this purpose.

\begin{figure}[t!]
\centering   
\includegraphics[width=0.65\linewidth]{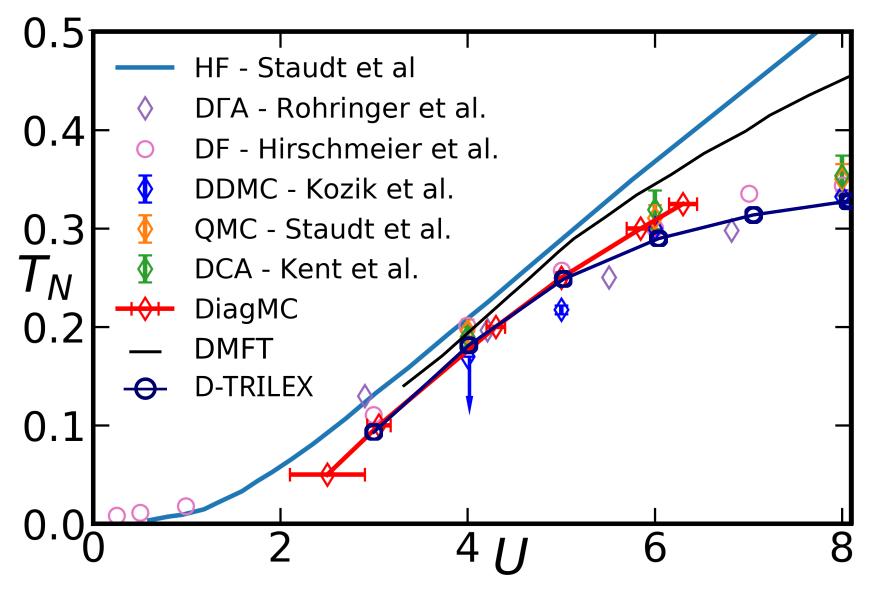}
\caption{\label{fig:Neel_3D} 
Comparison of the N\'eel temperature of the half-filled Hubbard model on a cubic lattice with the nearest-neighbor hopping ${t=1}$. The results are obtained using Hartree-Fock (HF)~\cite{staudt2000phase}, dynamical vertex approximation (D$\Gamma$A)~\cite{PhysRevLett.107.256402}, LDF~\cite{PhysRevB.92.144409}, diagrammatic determinant Monte Carlo (DDMC)~\cite{PhysRevB.87.205102}, quantum Monte Carlo (QMC)~\cite{staudt2000phase}, dynamical cluster approximation (DCA)~\cite{PhysRevB.72.060411}, diagrammatic Monte Carlo (DiagMC)~\cite{PhysRevLett.129.107202}, DMFT~\cite{PhysRevB.92.144409}, and \mbox{D-TRILEX}~\cite{vandelli2022quantum}.
The Figure is taken from Ref.~\cite{vandelli2022quantum}.}
\end{figure}

The LDF approach provides a very accurate result for the N\'eel temperature of the Hubbard model on a cubic lattice~\cite{PhysRevB.92.144409}.
Fig.~\ref{fig:Neel_3D} shows that the phase boundary between the normal and antiferromagnetically (AFM) ordered state predicted by LDF (violet circles) is the most accurate among the approximated methods considered, and coincides within errorbars with the exact DiagMC solution (red curve) obtained in Ref.~\cite{PhysRevLett.129.107202}. 

\begin{figure}[b!]
\centering   
\includegraphics[width=1\linewidth]{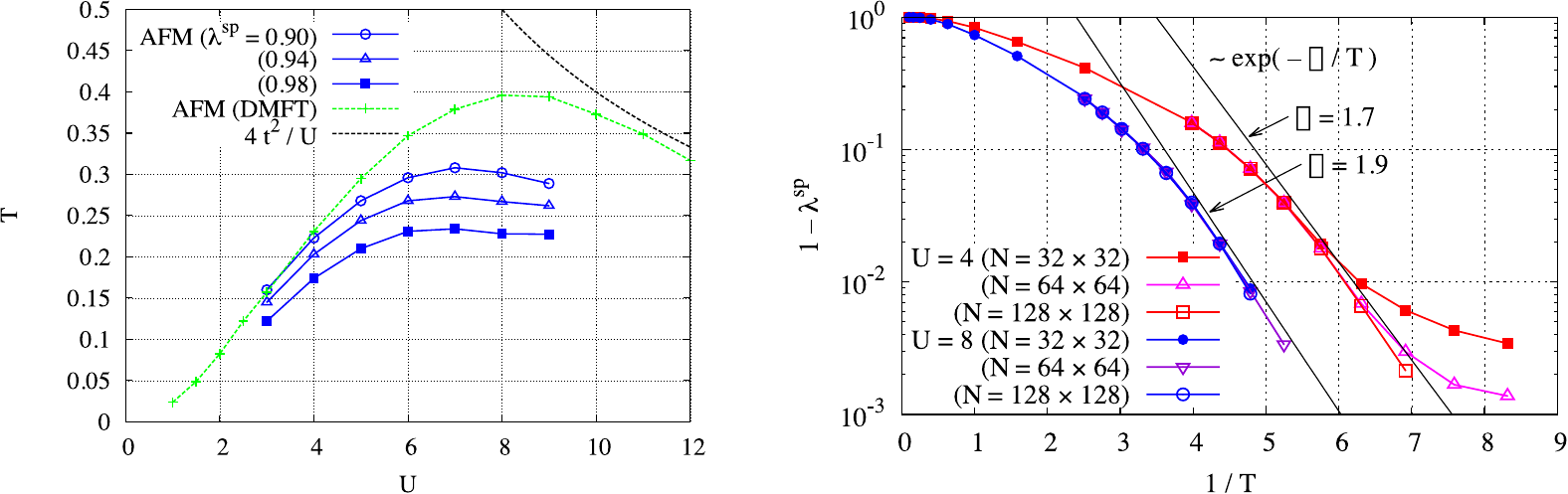}
\caption{\label{fig:Neel_2D} 
Left panel: $T$-$U$ magnetic phase diagram for the half-filled Hubbard model on a square lattice with the nearest-neighbor hopping ${t=1}$. The green curve corresponds to the N\'eel temperature predicted by DMFT. 
The blue curves depict temperatures corresponding to the leading eigenvalues $\lambda^{\rm sp}$ of the BSE of the LDF approach in the spin channel approaching unity.
Right panel:  A scaling plot ${1-\lambda^{\rm sp}}$ as a function of the inverse temperature $1/T$. 
Results for different system sizes, ${N = 32\times32}$, ${64\times64}$, and ${128\times128}$, are shown for comparison. 
The solid lines indicate the scaling ${1-\lambda^{\rm sp} \propto\exp(-\Delta/T)}$.
The Figure is taken from Ref.~\cite{PhysRevB.90.235132}.}
\end{figure}

In two dimensions, the N\'eel transition at finite temperatures is forbidden by the Mermin-Wagner theorem. 
Nevertheless, approximate methods frequently predict a finite N\'eel temperature for two dimensional systems, which should be seen as a crossover associated with the formation of a short-range AFM ordering, which is difficult to distinguish from the true long-range order. 
In Ref.~\cite{PhysRevB.90.235132}, the authors demonstrated that the LDF approach with the outer self-consistency on the dual Green's function ${\sum_{\bf k}\tilde{G}_{{\bf k}\nu}=0}$ (see Section~\ref{sec:self-consistency}) respects the Mermin-Wagner theorem.
The left panel of Fig.~\ref{fig:Neel_2D} shows that DMFT predicts the finite N\'eel temperature $T^{\rm DMFT}_{N}$ for the half-filled Hubbard model on a square lattice. 
The temperatures, at which the leading eigenvalues $\lambda^{\rm sp}$ of the BSE of the LDF approach in the spin channel are close to unity are shown in blue.
It turns out that $\lambda^{\rm sp}$ approaches 1 with decreasing T in the ladder approximation.
However, one can show that in the regime ${T<T^{\rm DMFT}_{N}}$, where magnetic fluctuations are strong,  the AFM susceptibility diverges exponentially toward ${T = 0}$: ${X^{\rm sp}_{Q,\omega=0}\sim e^{\beta\Delta}}$. 
It follows that $\lambda^{\rm sp}$ approaches 1 according to ${1-\lambda^{\rm sp} \propto\exp(-\Delta/T)}$. 
In order to check this behavior, in the right panel of Fig.~\ref{fig:Neel_2D} we plot ${1-\lambda^{\rm sp}}$ as a function of $1/T$.
It turns out that the data for different system sizes deviate from each other at low temperatures such that ${1-\lambda^{\rm sp}\lesssim10^{-2}}$. 
It indicates that the slow decays for ${N = 32\times32}$ and ${64\times64}$ observed at ${1/T\gtrsim7}$ are artifacts
due to a finite-size effect. 
Apart from the finite-size effect, the results agree with the expected scaling ${1-\lambda^{\rm sp} \propto\exp(-\beta\Delta)}$ indicated by the solid lines. 
We thus conclude that the LDF approximation correctly reproduces the N\'eel temperature of ${T_{N} = 0}$ required from the Mermin-Wagner theorem.
However, in some calculations capturing the exponential divergence of the spin susceptibility can be prohibitively difficult, as the numerical calculations close to the phase transition become unstable. 
Note also, that the LDB approach provides another way to enforce the Mermin-Wagner theorem by imposing an outer self-consistency on the lattice susceptibility, as discussed in Section~\ref{sec:self-consistency}.

\begin{figure}[b!]
\centering
\includegraphics[width=0.65\linewidth]{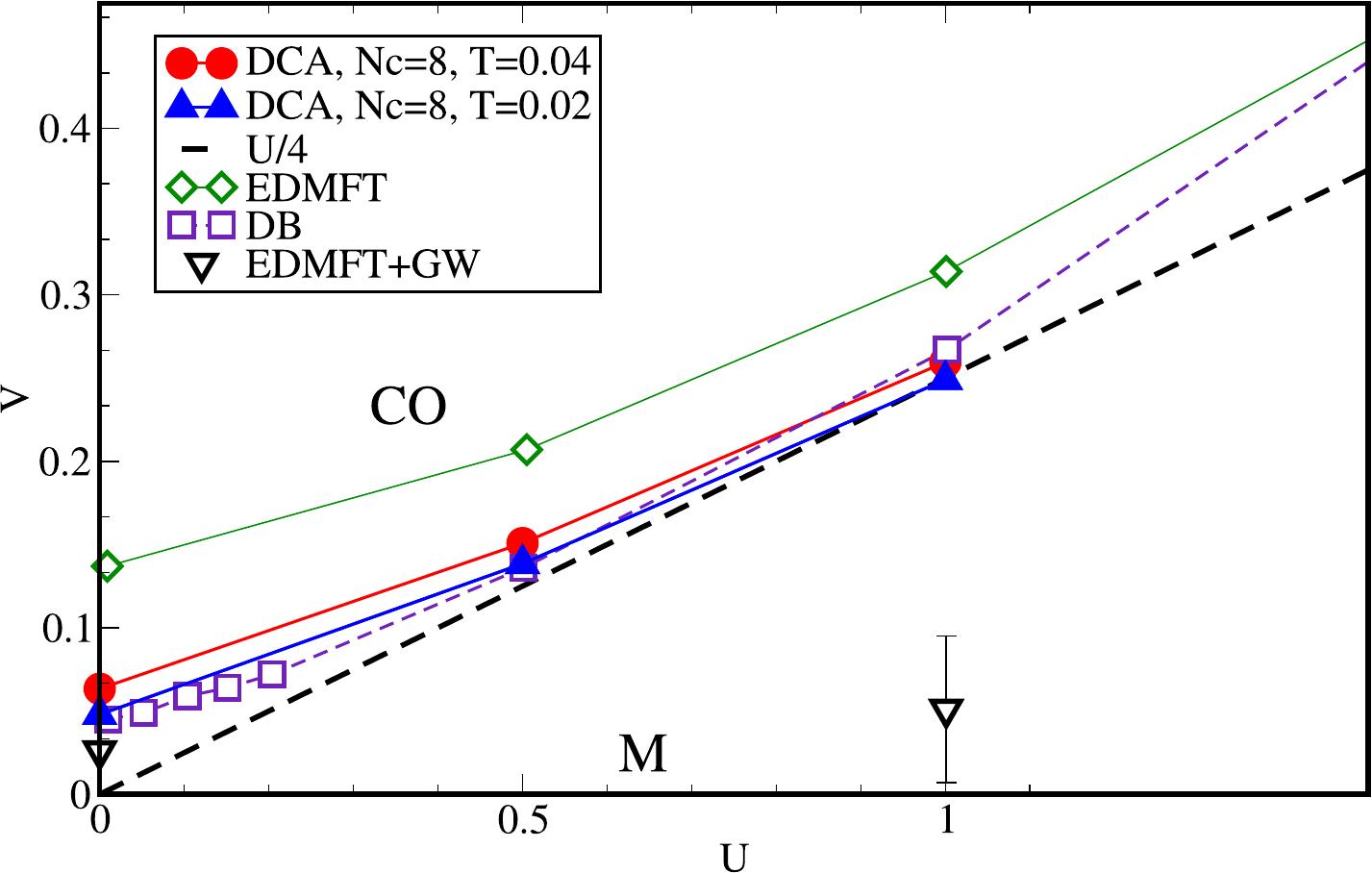}   
\caption{\label{fig:CDW_DB_DCA_GW} 
$V$-$U$ phase diagram of the half-filled extended Hubbard
model on a square lattice. 
${N_{c} = 8}$ DCA data~\cite{PhysRevB.95.115149} are shown for
${T/4t = 0.04}$ (red circles) and ${T/4t = 0.02}$ (blue triangles). 
The DB result for ${T/4t = 0.02}$~\cite{PhysRevB.90.235135} is depicted by magenta squares. 
The EDMFT (green squares) and EDMFT+$GW$ (black triangles) data are shown for ${T/4t = 0.01}$~\cite{PhysRevB.87.125149}. 
The dashed line corresponds to the mean-field estimate ${V=U/4}$ for the CDW phase boundary. 
Energies are shown in units of $t$ with ${4t = 1}$.
The Figure is taken from Ref.~\cite{PhysRevB.95.115149}.}
\end{figure}

The phase transition to the charge density-wave state is not forbidden by the Mermin-Wagner theorem, as this transition is associated with breaking the discrete lattice symmetry. 
The formation of the CDW state can be captured in a similar way, by the divergence of the charge susceptibility~\cite{PhysRevB.90.235135, PhysRevB.94.205110, PhysRevB.102.195109}.  
As discussed in Section~\ref{sec:DiagMC_phase}, one possible source of the CDW order is the non-local Coulomb interaction that disfavors the single occupation of the lattice sites.  
Fig.~\ref{fig:CDW_DB_DCA_GW} shows the $V$-$U$ phase diagram of the extended Hubbard model on a square lattice with the nearest-neighbor hopping $t$ and Coulomb interaction $V$.
These results show that the LDB approach predicts the CDW phase boundary in a very good agreement with another approximate but rather accurate DCA method~\cite{PhysRevB.95.115149}.
Comparison with the DiagMC@DB results~\cite{PhysRevB.102.195109} presented in Fig.~\ref{fig:DiagMC_Phase} additionally confirm the high accuracy of the LDB approach in predicting the phase transition to the charge ordered state.
The comparison between the CDW phase transitions in two and three dimensions obtained within the LDB scheme with instanteneous interaction~\cite{PhysRevB.100.165128} can be found in the right panel of Fig.~\ref{fig:MW}.

\subsubsection{Plasmons and magnons}

The total charge ${N = \sum_{i} n_i}$ and spin ${M = \sum_{i} m^{z}_i}$ densities are conserved quantities in the extended Hubbard model.
This can be seen from the fact that these quantities commute with the model Hamiltonian: ${[N,H]=0}$, ${[M,H]=0}$.
This leads to the following relation for the susceptibility:
\begin{align}
X^{\varsigma}_{{\bf q}=0,\omega} = 0 ~~~ \text{if} ~~~ \omega\neq0\,,
\label{eq:X_conservation}
\end{align}
which means that the system develops a Goldstone mode with vanishing energy in the limit of small momenta. 
It turns out, that finding an approximation that fulfills the relation~\eqref{eq:X_conservation} is a challenging task. 
In particular, as most of the approximate diagrammatic techniques, the general form of the LDB approach violates the charge and spin conservation law, 
because the momentum-dependent self-energy included in the theory is incompatible with the local four-point vertex function considered in the BSE for the susceptibility~\cite{Rubtsov20121320, PhysRevB.90.235105, PhysRevB.93.045107, PhysRevB.96.075155}. 
However, there exists one version of the ladder approximation that fulfills the charge ans spin conservation law.
The sufficient conditions for the charge and spin conservation are~\cite{PhysRevB.93.045107, PhysRevB.96.075155}:
\begin{itemize}
\item[1.] The reference impurity problem shares the same Ward identities as the original lattice problem and is solved in such a way, e.g. numerically exactly, such that these Ward identities are fulfilled. 
\item[2.] The lattice self-energy is momentum-independent and coincides with the impurity self-energy: ${\Sigma_{\nu} = \Sigma^{\rm imp}_{\nu}}$.
\item[3.] The fermionic hybridization function $\Delta_{\nu}$ is chosen according to the dual self-consistency condition ${\sum_{\bf k}\tilde{G}_{{\bf k}\nu}=0}$.
\item[4.] The dual polarization is calculated in the ladder approximation~\eqref{eq:DualPi}.
\end{itemize}
Importantly, no explicit condition is applied to the bosonic hybridization function, except that it should not break the local Ward identities of the reference impurity problem~\cite{PhysRevB.96.075155}.
We also note, that the LDF susceptibility (${V^{\varsigma}_{\bf q}=0}$, ${Y^{\varsigma}_{\omega}=0}$), calculated within a single DF iteration of the inner self-consistent loop (see Section~\ref{sec:self-consistency}) starting from the fermionic hybridization of the converged DMFT solution, coincides with the DMFT susceptibility with local vertex corrections~\cite{RevModPhys.68.13} and fulfills the charge and spin conservation laws according to presented above conditions.

\begin{figure}[t!]
\centering
\includegraphics[width=1\linewidth]{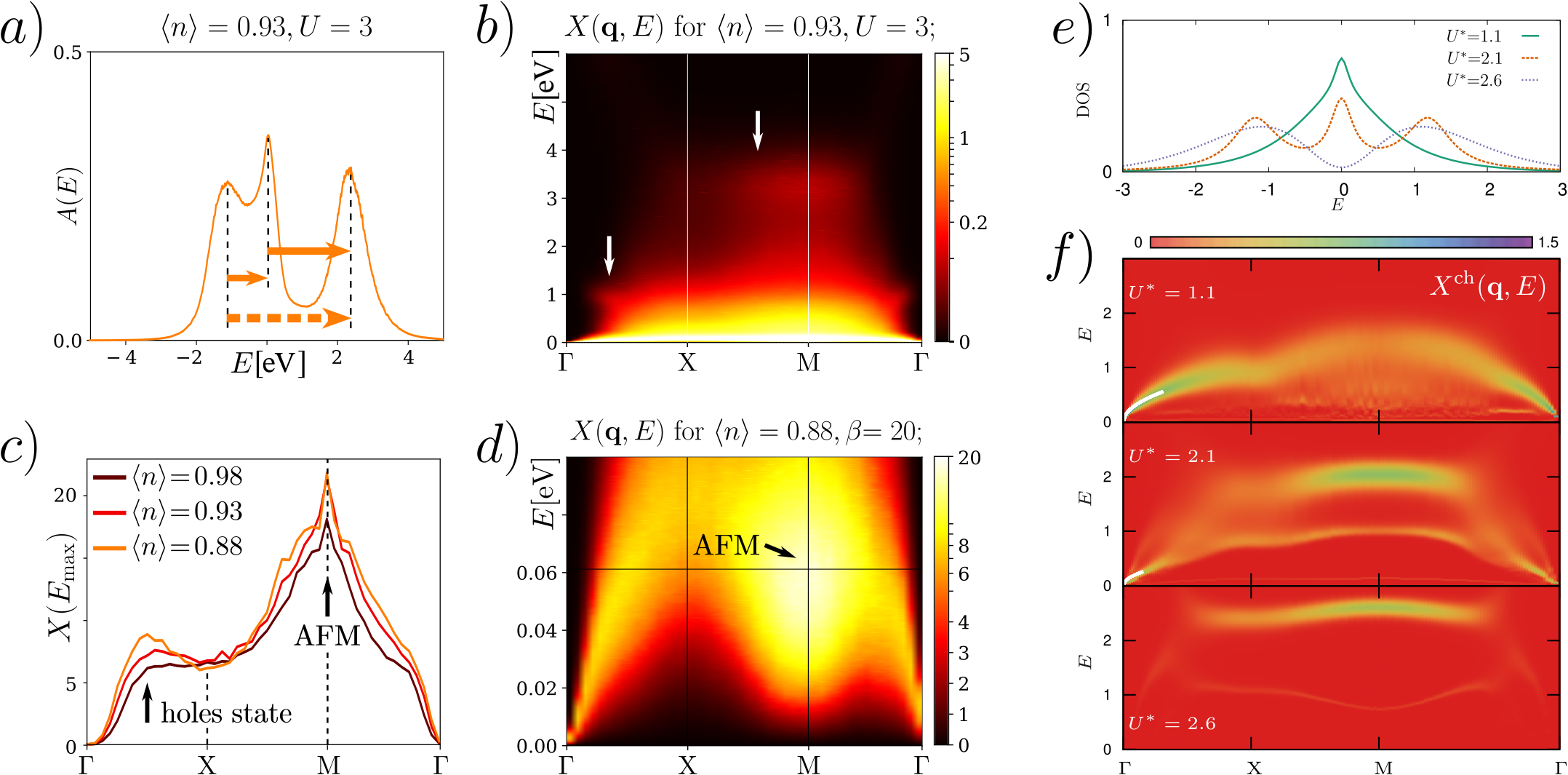}   
\caption{\label{fig:plasmons_magnons} 
Panels $a)$-$d)$ show the LDB calculation for the doped extended Hubbard model corresponding to the high-temperature superconducting cuprates with the nearest-neighbor hopping ${t=0.3}$, the nearest neighbor hopping ${t'=-0.15t}$, the local Coulomb interaction ${U=3}$, and the nearest-neighbor Coulomb interaction ${V=0.5}$. 
Panel $a)$ shows the local electronic density of states ${A(E)}$. The arrows illustrate resonant particle-hole excitations between the quasi-particle peak at Fermi energy (${E=0}$) and the lower and upper Hubbard bands.
Panel $b)$ shows the momentum-resolved spectral function of collective spin excitation in the full energy range. The white arrows indicate the resonant excitations illustrated in panel $a)$ by the solid orange arrows.
Panel $d)$ shows the lower part of the energy spectrum plotted in panel $b)$, corresponding to magnon excitations.
Panel $c)$ shows the cut of the spin excitation spectrum at the energy ${E=60\,\text{meV}}$. The higher peak at the ${M=(\pi,\pi)}$ point displays to the AFM mode. The lower peak corresponds to spin fluctuations of holes.
The Figure is adopted from Ref.~\cite{Stepanov18-2}.
Panels $e)$ and $f)$ show the LDB calculations for the half-filled Hubbard model on the square lattice with ${t=0.25}$ and the local $U$ and long-range $V_{\bf q}=V_0/|{\bf q}|$ Coulomb interaction, where ${U^{*}=U+\sum_{\bf q}V_{\bf q}}$.
Panel $e)$ shows the local electronic spectral function calculated for different values of $U^{*}$.
Panel $f)$ displays the momentum-resolved spectral functions of collective charge excitations corresponding to different $U^{*}$.
The Figure is adopted from Ref.~\cite{PhysRevLett.113.246407}.}
\end{figure}

In Fig.~\ref{fig:plasmons_magnons} we show the momentum-resolved spectral functions of collective spin~\cite{Stepanov18-2} (panels $b$-$d$) and charge~\cite{PhysRevLett.113.246407} (panel $f$) excitations. The results are obtained within a conserving LDB scheme discussed above.
The spectral functions are obtained using the analytical continuation of the corresponding susceptibility $X^{\rm ch/sp}_{{\bf q},\omega}$ from the bosonic Matsubara frequency ${\omega}$ to real energy $E$.
The calculation for the spin susceptibility is performed for the doped extended Hubbard model on a square lattice. 
The parameters of the model are taken to be relevant for high-temperature superconducting cuprates, specifically for the La$_2$CuO$_4$ material: the nearest-neighbor hopping ${t=0.3}$, the nearest neighbor hopping ${t'=-0.15t}$, the local Coulomb interaction ${U=3}$, and the nearest-neighbor Coulomb interaction ${V=0.5}$~\cite{ANDERSEN19951573, PhysRevB.53.8751, PhysRevB.92.245113}.
At half-filling, the considered system lies in the Mott insulating state. 
Upon small hole doping, the local electronic spectral function displays a typical three-peak structure with the two Hubbard bands below and above the Fermi level, and the central quasiparticle peak, that appears at the Fermi energy ${E=0}$.
Panel $b)$ shows the momentum-resolved spectral function of spin excitations in the full energy range. The lower part of the energy spectrum is shown in panel $d)$ and corresponds to magnon excitations.
Apart of a dispersive magnon band, the calculated spectral function reveals dispersiveless high-energy bands of low intensity, indicated by white arrows, corresponding to the resonant particle-hole excitations between the peaks in the electronic spectral function depicted by the solid orange arrows in panel $a)$.
Remarkably, the intensity of these excitations vanish at ${{\bf q}\to0}$, and only the magnon band displays the finite intensity in the limit ${{\bf q}\to0}$ and ${E\to0}$ in agreement with the spin conservation law.
Panel $c)$ displays the finite energy (${E=60\,\text{meV}}$) cut of the spin excitation spectrum.
It shows that apart of the main AFM fluctuation, upon doping the system develops another mode, corresponding to collective spin excitations of excessive charge carriers (holes). 

The calculation of charge fluctuations was performed using the conserved LDB scheme for the extended Hubbard model on a square lattice with ${t=0.25}$~\cite{PhysRevLett.113.246407}. 
The Coulomb interaction was considered in the long-range form ${U + V_0/|{\bf q}|}$ with ${V_0 = 2}$ and the local interaction ${U^{*} = U + \sum_{\bf q}V_0/|{\bf q}|}$.
The long-range form of the Coulomb interaction is crucial for describing plasmonic branch of collective charge excitations.
Panel $e)$ shows the local electronic density of states (DOS) for three different values of $U^{*}$.
For weak interaction ${U^{*}=1.1}$, the DOS exhibits a single quasiparticle peak at the Fermi level (green curve).
As the interaction is increased to ${U^{*}=2.1}$, the peak is renormalized as the spectral weight is partially moved to Hubbard bands that are formed at energies ${E\simeq\pm{}U}$ (orange curve). 
At ${U^{*}=2.6}$, the system lies in the Mott insulating regime, where the quasiparticle peak is destroyed and the spectral function shows only   
the Hubbard bands (gray curve).
Panel $f)$ shows the momentum-resolved spectral function for the charge excitations calculated for these three values of interaction.
At ${U^{*}=1.1}$ the spectral function can be understood within an itinerant
electron picture. 
In the proximity of the $\Gamma=(0,0)$ point, the spectrum exhibits a dispersive single plasmon branch whose energy vanishes in the long-wavelength limit. 
At ${U^{*}=2.1}$ the dispersion is split into two branches except for small wave vectors.
These features appear simultaneously with the Hubbard bands in the density of states shown in panel $e)$.
The lower branch originates mainly from particle-hole excitations for which the electron is excited from the Hubbard band to the quasiparticle-peak
(or vice versa), whereas the upper branch stems from excitations between the Hubbard bands.
At ${U^{*}=2.1}$ the two-particle excitation corresponds to a creation of a doublon and a holon, which costs an energy $U^{*}$. Such an excitation is expected to be highly localized, which corresponds to a weakly dispersing branch at an energy ${E\simeq{}U^{*}}$. At this interaction strength the low-energy plasmon mode has disappeared together with the quasiparticle peak.
We note that all spectral functions of charge excitations shown in panel $f)$ fulfill the charge conservation law.
An extended discussion of charge and spin excitations in two and three dimensions within the LDB approach can be found in Ref.~\cite{PhysRevB.90.235105}. 

\subsection{Application to superconductivity}
\label{sec:SC_DB}

The leading collective electronic fluctuations in most regimes of the extended Hubbard model correspond to spin and charge excitations, which are sufficiently accurately captured within the ladder DF/DB approximation.
For this reason, the LDB self-energy~\eqref{eq:Sigma_LDB} accounts for the three-point~\eqref{eq:Lvertex} and four-point~\eqref{eq:Pvertex} vertices that are screened only in the particle-hole ($\varsigma$) channel.
The particle-particle (pp) fluctuations in the ladder form can be incorporated in the theory straightforwardly~\cite{hafermann2010numerical}.
However, in the ladder approximation based on the local four-point vertex $\Gamma$~\eqref{eq:Vertex_PH_app} these fluctuations correspond to the $s$-wave pairing and are typically weak (see, e.g., Ref.~\cite{PhysRevB.103.245123}).
Considering the particle-particle scattering on the momentum-dependent vertex function drastically complicates numerical calculations, because it requires solving the BSE in the momentum and frequency space.
A particular example is the parquet approximation~\cite{Bickers91} that can be formulated in the dual space~\cite{PhysRevB.101.075109, PhysRevB.101.165101, PhysRevB.102.195131, PhysRevB.102.235133} in order to account for various competing instabilities.

A more convenient way to address the superconducting instability within the dual approach was proposed in Refs.~\cite{hafermann2009dual, PhysRevB.90.235132}.
The idea relies on calculating the four-point vertex ${\rm P}$~\eqref{eq:Pvertex} renormalized by the particle-hole fluctuations in the ladder approximation from the converged DF calculation and using it as a two-particle irreducible (bare) vertex in the BSE in the particle-particle channel (note, that here we use different notations for the four-point vertex than in Refs.~\cite{hafermann2009dual, PhysRevB.90.235132}): 
\begin{align}
\Gamma^{\rm pp}_{kk'} = {\rm P}^{\uparrow\uparrow\downarrow\downarrow}_{\nu,-\nu',k'-k} - {\rm P}^{\uparrow\downarrow\uparrow\downarrow}_{\nu,\nu',-k-k'}  
- \Gamma^{\uparrow\uparrow\downarrow\downarrow}_{\nu,-\nu',\nu'-\nu}\,.
\end{align}
The dimension of the matrices for the vertex function is too large to solve the BSE, similar to Eq.~\eqref{eq:Pvertex}, for the vertex function screened in the particle-particle channel.
Instead, one can solve the corresponding eigenvalue problem:
\begin{align}
-\sum_{k'} \Gamma^{\rm pp}_{kk'} \tilde{G}^{\phantom{\varsigma}}_{k'\uparrow} \tilde{G}^{\phantom{\varsigma}}_{-k', \downarrow} \phi_{k'} = \lambda^{\rm SC} \phi_{k}
\label{eq:dualBSE}
\end{align}
to determine the transition temperature and to extract the dominant pairing fluctuations.

\begin{figure}[t!]
\centering
\includegraphics[width=0.7\linewidth]{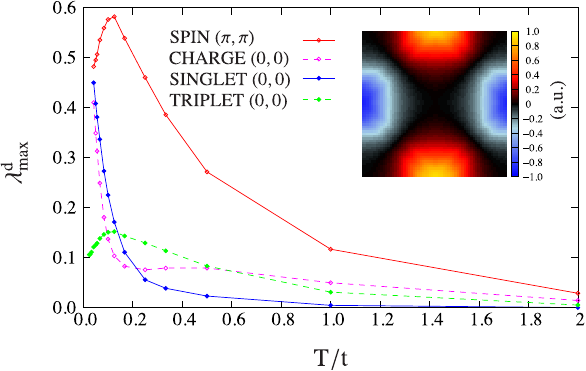}   
\caption{\label{fig:SCDF} 
Leading dual eigenvalue in various channels calculated for the Hubbard model on a square lattice at ${U/t=4}$ and 15\% of hole doping as a function of temperature. The inset shows the momentum dependence of the dominant eigenfunction in the singlet pairing channel corresponding to the $d$-wave symmetry (${T/t = 0.042}$). The Figure is taken from Ref.~\cite{hafermann2009dual}.}
\end{figure}

Fig.~\ref{fig:SCDF} shows the leading dual eigenvalues of the BSE in various channels as a function of temperature calculated using the DF scheme for the Hubbard model on a square lattice with the nearest-neighbor hopping $t$ at 15\% hole doping~\cite{hafermann2009dual}. 
An eigenvalue of ${\lambda=1}$ would indicate a transition to an ordered state. 
One finds that the dominant contribution to superconductivity corresponds to the singlet pairing channel. 
As shown in the inset, the momentum dependence of the corresponding eigenfunction exhibits $d$-wave symmetry, as expected. 

\subsubsection{Nambu space representation}
\label{sec:DF_Nambu}

Calculating the leading eigenvalue in the particle-particle channel makes it possible to detect the superconducting phase transition by approaching it from the normal state.
To study the ordered state directly, one must explicitly introduce the possibility of spontaneous symmetry breaking into the formalism.
In the case of superconducting state, one can introduce the superconducting field $h^{\rm sc}_{\bf k}$ in the lattice problem, which enters the remaining part of the lattice action~\eqref{eq:actionrem_app} as:
\begin{align}
{\cal S}_{\rm rem} =& \sum_{k,\sigma} c^{*}_{k \sigma} \tilde{\varepsilon}^{\phantom{*}}_{k} c^{\phantom{*}}_{k \sigma} 
+ \sum_{k} h^{\rm sc}_{\bf k} \left( n^{*\, \rm sc}_{k} + n^{\rm sc}_{k} \right)
+ \frac12 \sum_{q,\varsigma} \tilde{V}^{\varsigma}_{q} \, \rho^{\varsigma}_{-q} \, \rho^{\varsigma}_{q}\,,
\label{eq:actionrem_sc}
\end{align}
where we have introduced the densities in the superconducting channel, similarly to Eq.~\eqref{eq:npp_app}:
\begin{align}
n^{\rm sc}_{k} = 
\frac12
\sum_{\sigma}
c^{\phantom{*}}_{-k, \overline{\sigma}} \, \sigma^{\,z}_{\sigma\sigma} c^{\phantom{*}}_{k\sigma}\,, 
~~~
n^{*\,\rm sc}_{k} = 
\frac12
\sum_{\sigma}
c^{*}_{k\sigma} \, \sigma^{\,z}_{\sigma\sigma} c^{*}_{-k, \overline{\sigma}}\,.
\label{eq:n_sc}
\end{align}
Following Ref.~\cite{Cuprates}, we limit ourselves to a single-band case and omit the interaction in the particle-particle channel from the problem. 
Note, that in the single-orbital case the superconducting densities~\eqref{eq:n_sc} have a singlet form~\eqref{eq:npp_app}.

It is convenient to work in the Nambu representation for fermionic degrees of freedom by introducing spinors 
$\Psi^{*}_{k} = \left( c^{*}_{k,\uparrow}; c^{\phantom{*}}_{-k,\downarrow}; c^{*}_{k,\downarrow}; c^{\phantom{*}}_{-k,\uparrow} \right)$.
In this representation the action~\eqref{eq:actionrem_sc} has the block diagonal matrix form, so one can focus only on the subspace 
$\Psi^{*}_{k} = \left( c^{*}_{k,\uparrow}; c^{\phantom{*}}_{-k,\downarrow} \right)$ and $\Psi^{\phantom{*}}_{k} = \left( c^{\phantom{*}}_{k,\uparrow}; c^{*}_{-k,\downarrow} \right)^{T}$.
After performing the usual transformation to the dual space (see Section~\ref{sec:Dual_transformation}), one gets the following form for the dual boson action~\eqref{eq:DB_action}:
\begin{align}
\tilde{\cal S}_{\rm DB} = &-\Tr\sum_{k} \varphi^{*}_{k} \tilde{\cal G}_{k}^{-1}\varphi^{\phantom{*}}_{k} 
- \frac12\sum_{q,\varsigma}
\varphi^{\varsigma}_{-q} \tilde{\cal X}_{q,\varsigma}^{-1} \varphi^{\varsigma}_{q}
+ \tilde{\cal F}[f,\varphi]\,,
\label{eq:SC_DB_action}
\end{align}
where $\varphi^{*}_{k} = \left( f^{*}_{k,\uparrow}; f^{\phantom{*}}_{-k,\downarrow} \right)$ and $\varphi^{\phantom{*}}_{k} = \left( f^{\phantom{*}}_{k,\uparrow}; f^{*}_{-k,\downarrow} \right)^{T}$, and the trace is taken over the ${2\times2}$ matrices in the Nambu subspace.
The interaction term $\tilde{\cal F}[f,\varphi]$ remains unchanged and has the form defined by Eq.~\eqref{eq:Wfull}.

The bare dual Green's function~\eqref{eq:bare_dual_G} (with the choice of ${B^{\sigma\sigma'}_{\nu} = \delta_{\sigma,\sigma'}}$) transforms to:
\begin{align}
\tilde{\cal G}_{k} = 
\begin{bmatrix}
\tilde{\cal G}^{\uparrow\uparrow}_{k} & \tilde{\cal G}^{\uparrow\downarrow}_{k} \\
\tilde{\cal G}^{\downarrow\uparrow}_{k} & - \tilde{\cal G}^{\downarrow\downarrow}_{-k}
\end{bmatrix} =
\left[
\begin{bmatrix}
\tilde\varepsilon_{k} & h^{\rm sc}_{\bf k} \\
h^{\rm sc\,*}_{\bf k} & -\tilde\varepsilon_{-k}
\end{bmatrix}^{-1}
-
\begin{bmatrix}
g^{\uparrow\uparrow}_{\nu} & 0 \\
0 & -g^{\downarrow\downarrow}_{-\nu}
\end{bmatrix}
\right]^{-1}.
\label{eq:Gnambu}
\end{align}
Note, that due to SU(2) symmetry and lattice inversion symmetry ${\tilde{\cal G}^{\uparrow\downarrow}_{k} = \tilde{\cal G}^{\downarrow\uparrow}_{k}}$, but we will use a more general relation, namely: ${\tilde{\cal G}^{\uparrow\downarrow}_{k} = \left[\tilde{\cal G}^{\downarrow\uparrow}_{k}\right]^{*}}$.
The normal and anomalous components of the Green's function are defined as:
\begin{align}
\tilde{\cal G}^{\uparrow\uparrow}_{k} &= - \langle f^{\phantom{*}}_{k\uparrow} f^{*}_{k\uparrow} \rangle\,, ~~~~~~~~~~~
\tilde{\cal G}^{\uparrow\downarrow}_{k} = - \langle f^{\phantom{*}}_{k\uparrow} f^{\phantom{*}}_{-k\downarrow} \rangle\,, \notag\\
\tilde{\cal G}^{\downarrow\uparrow}_{k} &= - \langle f^{*}_{-k\downarrow} f^{*}_{k\uparrow} \rangle\,, ~~~~~~
-\tilde{\cal G}^{\downarrow\downarrow}_{-k} = - \langle f^{*}_{-k\downarrow} f^{\phantom{*}}_{-k\downarrow} \rangle\,.
\end{align}
The Nambu-Gor'kov lattice Green's function can be found using the relation~\eqref{eq:GtoSigma} written in the matrix form:
\begin{align}
\hat{G}_{k} = 
\begin{pmatrix}
G_{k} & F_{k} \\
F^{*}_{k} & -G_{-k}
\end{pmatrix}
=
\left[
\begin{bmatrix}
g^{\uparrow\uparrow}_{\nu} + \tilde\Sigma^{\uparrow\uparrow}_{k} & \tilde\Sigma^{\uparrow\downarrow}_{k} \\
\tilde\Sigma^{\downarrow\uparrow}_{k} & -g^{\downarrow\downarrow}_{-\nu} - \tilde\Sigma^{\downarrow\downarrow}_{-k}
\end{bmatrix}^{-1}
-
\begin{bmatrix}
\tilde\varepsilon_{\bf k} & h^{\rm sc}_{\bf k} \\
h^{\rm sc}_{\bf k} & -\tilde\varepsilon_{\bf -k}
\end{bmatrix}
\right]^{-1},
\label{eq:Nambu_GF}
\end{align}
where
\begin{align}
\tilde{\Sigma}_{k} = 
\begin{bmatrix}
\tilde\Sigma^{\uparrow\uparrow}_{k} & \tilde\Sigma^{\uparrow\downarrow}_{k} \\
\tilde\Sigma^{\downarrow\uparrow}_{k} & -\tilde\Sigma^{\downarrow\downarrow}_{-k}
\end{bmatrix}.
\label{eq:Sigma_Nambu}
\end{align}

In Ref.~\cite{Cuprates} this method was applied to a single-orbital Hubbard model (${V=0}$) relevant for high-temperature cuprate superconductors: 
the nearest-neighbor hopping ${t = 1}$ (which defines the unit of energy) the next-nearest-neighbor (NNN) hopping ${t' = -0.3}$, and the local Coulomb interaction ${U = 5.6}$. 
The choice of the interaction strength is based on a highly degenerate point in energy spectrum observed in small cluster calculations, which favors superconductivity~\cite{Harland16, Harland20, Danilov2022}. 
The same value of $U$, related to the optimal nodal-antinodal dichotomy near the Lifshitz transition, was reported in Ref.~\cite{Wei_point}.
The calculations were performed using the Dual Fermion lattice DQMC scheme for a ${16\times16}$ square lattice with periodic boundary conditions.
Importantly, the reference problem for this calculation had the same lattice size, but was considered half-filled and particle-hole symmetric (${t'=0}$), whereas the lattice problem was studied at ${\simeq15\%}$ hole doping with the adjusted chemical potential $\mu$ and a non-zero value of ${t' = -0.3}$. 
In this case, the dual diagrammatic expansion becomes perturbative, because the difference between the lattice and reference problems ${\tilde\varepsilon_{\bf k} = t'_{\bf k}-\mu}$, where $t'$ is a Fourier transform of the NNN hopping, is indeed small compared to the bandwidth.
A detailed discussion of the Dual Fermion lattice DQMC method can be found in Section~\ref{sec:DFQMC}.
In this perturbative case one can truncate the dual expansion at the lowest order diagram for the self-energy, as discussed in Section~\ref{sec:strong_coupling}.
The matrix components of the dual self-energy~\eqref{eq:Sigma_Nambu} in the Nambu space read:
\begin{align}
\tilde\Sigma^{\uparrow\uparrow}_{k} &=  \sum_{k'} \left[ \tilde{\cal G}^{\uparrow\uparrow}_{k'} \Gamma^{\uparrow\uparrow\uparrow\uparrow}_{k,k',0} - 
(-\tilde{\cal G}^{\downarrow\downarrow}_{-k'}) \Gamma^{\uparrow\uparrow\downarrow\downarrow}_{k,-k',0} \right],
\label{eq:Sigma_norm}\\
-\tilde\Sigma^{\downarrow\downarrow}_{-k} &= \sum_{k'} \left[
(-\tilde{\cal G}^{\downarrow\downarrow}_{-k'}) \Gamma^{\downarrow\downarrow\downarrow\downarrow}_{-k,-k',0} - \tilde{\cal G}^{\uparrow\uparrow}_{k'}\Gamma^{\downarrow\downarrow\uparrow\uparrow}_{-k,k',0} \right],
\\
\tilde\Sigma^{\uparrow\downarrow}_{k} &= -\sum_{k'}\tilde{\cal G}^{\uparrow\downarrow}_{k'} \Gamma^{\uparrow\downarrow\uparrow\downarrow}_{k,k',-k-k'}\,,
\\
\tilde\Sigma^{\downarrow\uparrow}_{k} &= -\sum_{k'}\tilde{\cal G}^{\downarrow\uparrow}_{k'} \Gamma^{\downarrow\uparrow\downarrow\uparrow}_{k,k',-k-k'}\,.
\label{eq:Sigma_anom}
\end{align}
Note, that in the case of a lattice reference problem, the four-point vertex function depends on both Matsubara frequencies and momenta, as indicated by the combined index ${k\in\{\nu,{\bf k}\}}$.

\begin{figure}[t!]
\centering
\includegraphics[width=0.85\textwidth]{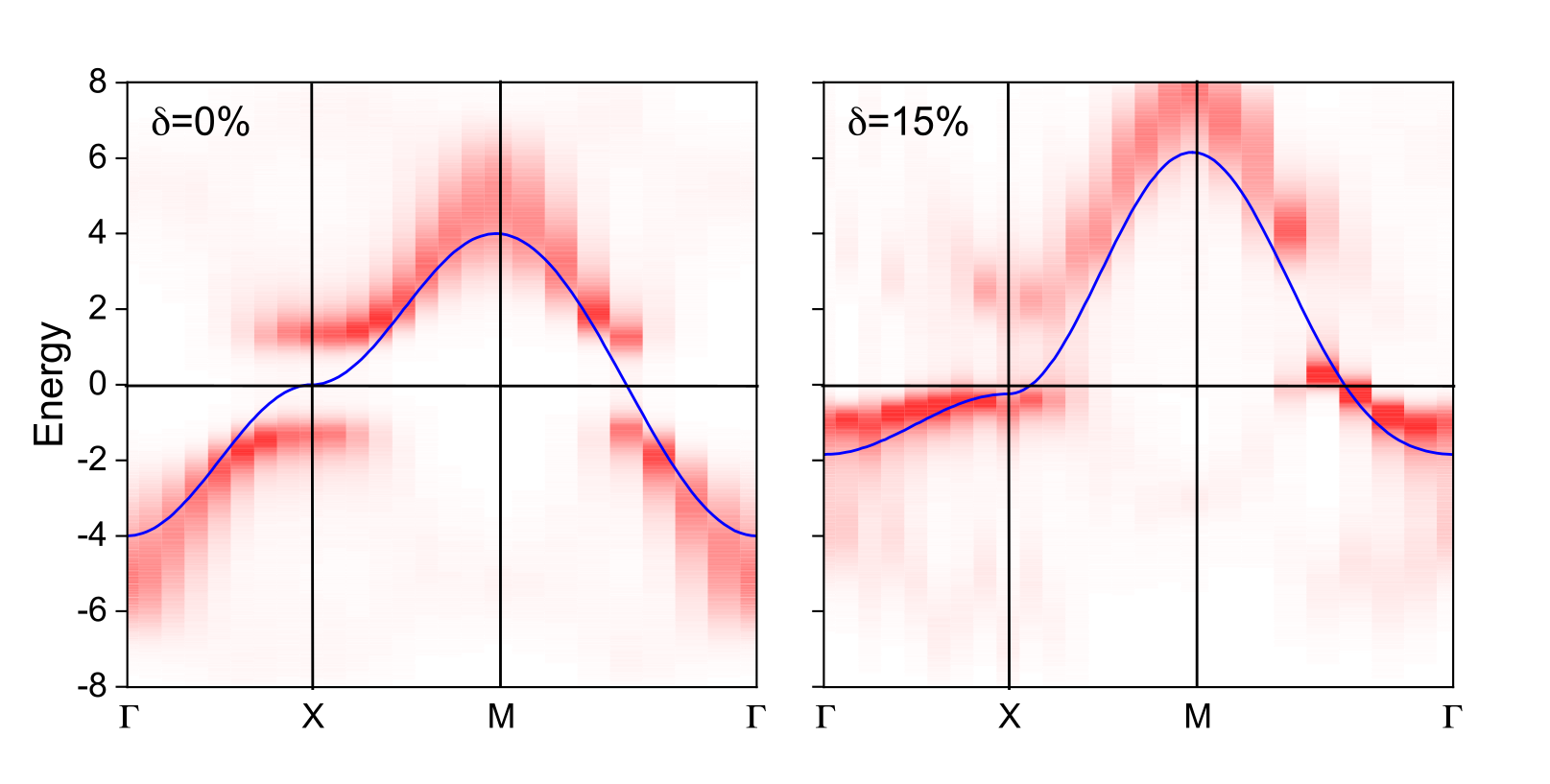}
\caption{The electronic spectral function of the $t$\,-\,$t'$ Hubbard model calculated for ${U=5.6}$ at ${\beta=5}$
relative to the chemical potential together with the non-interacting dispersions (blue line). The results are obtained for the half-filled particle-hole symmetric case (${t'=0}$, left panel) and for ${\delta=15\%}$ hole doping with ${t'=-0.3}$ (right panel). 
The Figure is takem from Ref.~\cite{Cuprates}.}
\label{fig:Ak}
\end{figure}

Calculations in Ref.~\cite{Cuprates} were performed in the presence of a small superconducting field:
\begin{align}
h^{\rm sc}_{\bf k} = -2h \left( \cos k_x - \cos k_y \right).
\end{align}
In Fig.~\ref{fig:Ak} we show the transformation of the electronic spectral function from the half-filled reference case (left panel) to the hole doped system with ${\mu=-1.45}$ and ${t'=-0.3}$ (right panel).
The results are calculated at the inverse temperature ${\beta=1/T=5}$ using stochastic analytical continuation from Matsubara space to real energy~\cite{SOM2}. 
For the considered model parameters the reference systems lies in the Mott insulating phase.
The corresponding spectral function (left panel) reveals the formation of broad Hubbard bands around the energy ${E=\pm6}$, and shadow antiferromagnetic bands at ${E\simeq-4}$ in the vicinity of the ${\text{M}=(\pi,\pi)}$ point. 
Upon ${\delta=15\%}$ hole doping, the spectral function changes dramatically (right panel).
One can clearly see a strong effect of $t'$ on the van Hove singularity that results in the formation of a narrow, almost flat band in the $\Gamma$--X direction and the appearance of a pseudogap near the antinodal ${\text{X}=(\pi,0)}$ point, which signals the quasi-localized behavior of electrons related to the formation of local magnetic moments.
On the other hand, the spectrum remains metallic near the nodal point ($\Gamma$--M)/2. 
Flattening of the bands and enhancement of van Hove singularities near the Fermi energy due to correlation effects were earlier found and studied by weak-coupling renormalization group~\cite{IKK2002} and strong-coupling DF approach~\cite{Yudin2014}. 

\begin{figure}[t!]
\centering
\includegraphics[width=0.9\textwidth]{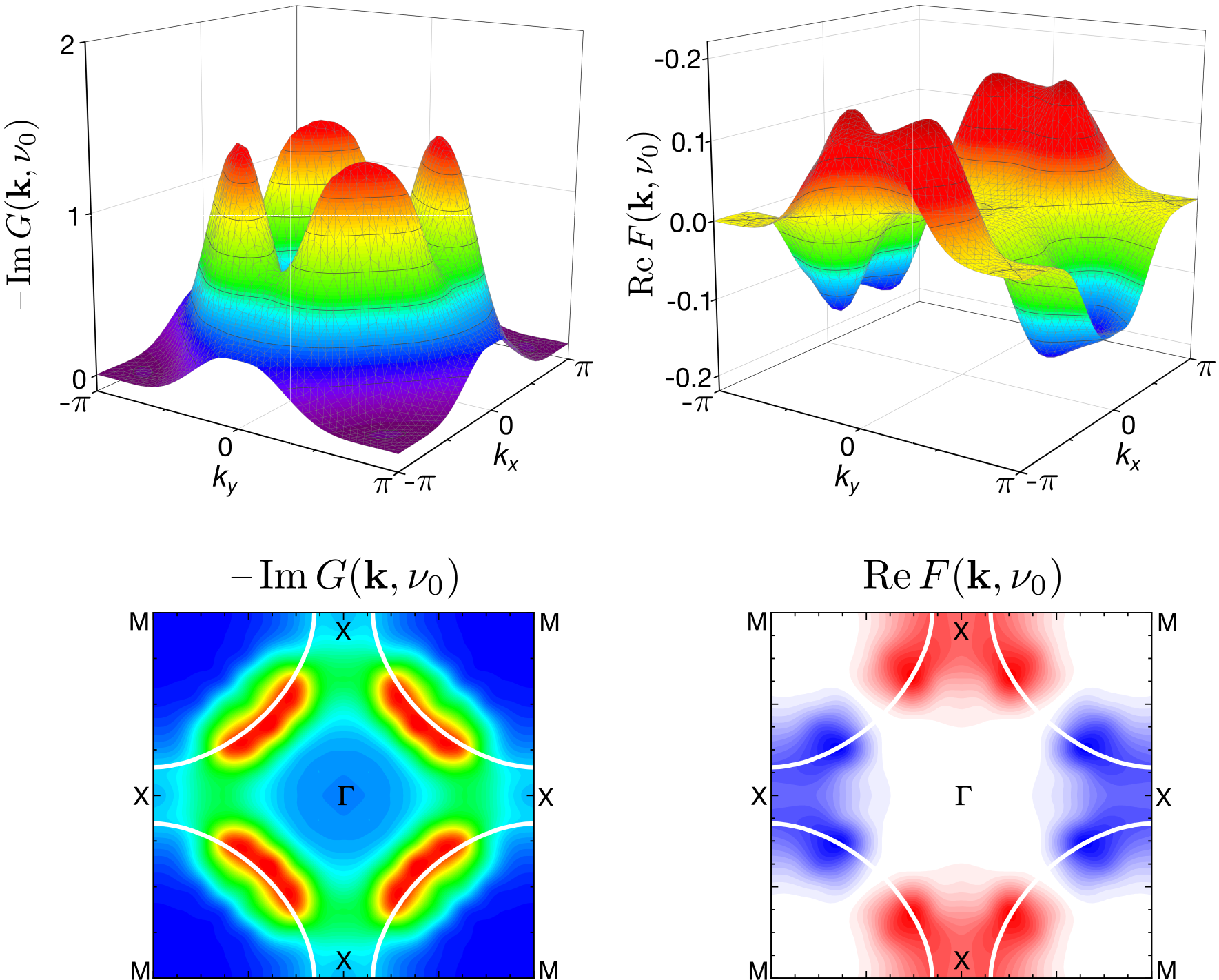}
\caption{The imaginary part of the normal Green function $G$ (left panels) and the real part of the anomalous Green function $F$ (right panels) calculated at the lowest Matsubara frequency ${\nu_0=\pi/\beta}$ for 12\% hole doping and ${\beta=8}$.
The Figure is adopted from Ref.~\cite{Cuprates}.}
\label{fig:Gk}
\end{figure}

In Fig.~\ref{fig:Gk} we plot the imaginary part of the normal Green's function $G({\bf k})$ (left panels), as well as the real part of the anomalous Green's function $F({\bf k})$ (right panels).
The results are obtained at the lowest Matsubara frequency ${\nu_0=\pi{}T}$ for the first BZ in the presence of a small external superconducting $d$-wave field with the amplitude ${h=0.05}$. 
These calculations clearly capture the formation of a large pseudogap in the electronic spectral function ${-\frac{1}{\pi}\text{Im}\,G({\bf k})}$ at the antinodal X point, which exists at already relatively high temperatures (${\beta=5}$ corresponds to ${T\simeq700\,\text{K}}$ for the realistic hopping amplitude ${t=0.3}$\,eV).
Additionally, we find that the anomalous Green's function $F({\bf k})$ is relatively large and has a very unusual shape that features a two-gap structure.
Indeed, ${\text{Re}\,F({\bf k})}$ reveals a suppressed spectral weight at the X points, related to the pseudogap formation in the spectral function, which shifts its extrema in the direction toward the nodal point. 
We attribute such a strong deviation of the anomalous Green's function from a usual $(\cos{k_x}-\cos{k_y})$ form of an applied external $d_{x^2-y^2}$ field to a fingerprint of a strongly-correlated superconductivity.
The emergence of two distinct gaps, found in our calculations, is confirmed by recent ARPES experiments~\cite{Fujimori_ARPES, ZX_2Gaps, Kondo_2Gap_2009, Hashimoto_2Gaps, Kordyuk_3G, Campuzano2G, Hashimoto2014, He_ZX_2G}.

\subsection{Dual Fermion lattice DQMC method}
\label{sec:DFQMC}

As anticipated in Section~\ref{sec:reference_system}, the dual diagrammatic expansion can be formulated on the basis of a generic interacting reference problem~\cite{BRENER2020168310}. 
Until this moment, we limited ourselves to a single-site DMFT impurity problem as a reference.
In the case of lattice geometries with several atoms in the unit cell, such as the honeycomb lattice, the dual formalism can be extended to a multi-impurity reference system.
In this multi-impurity formulation, different atoms within the unit cell are treated as distinct DMFT impurity problems. 
The correlations captured in the reference problem remain local but may differ between impurities, while correlations between the impurities are neglected.
Furthermore, the dual diagrammatic expansion can be formulated based on a cluster reference problem.
This approach enables to consistently combine exact treatment of short-range correlation effects within the considered cluster with the diagrammatic description of long-range collective electronic fluctuations.
The multi-impurity DF calculations applied to the Hubbard model on a honeycomb lattice have been performed in Ref.~\cite{PhysRevB.97.115150}.
Various cluster extension of the DF method can be found in Refs.~\cite{hafermann2008, PhysRevB.84.155106, PhysRevB.97.125114, BRENER2020168310}.
We will not go into the details of these methods here, but rather turn to an example of a reference system that differs most from the single-site reference impurity problem.

\begin{figure}[t!]
\centering
\includegraphics[width=0.7\textwidth]{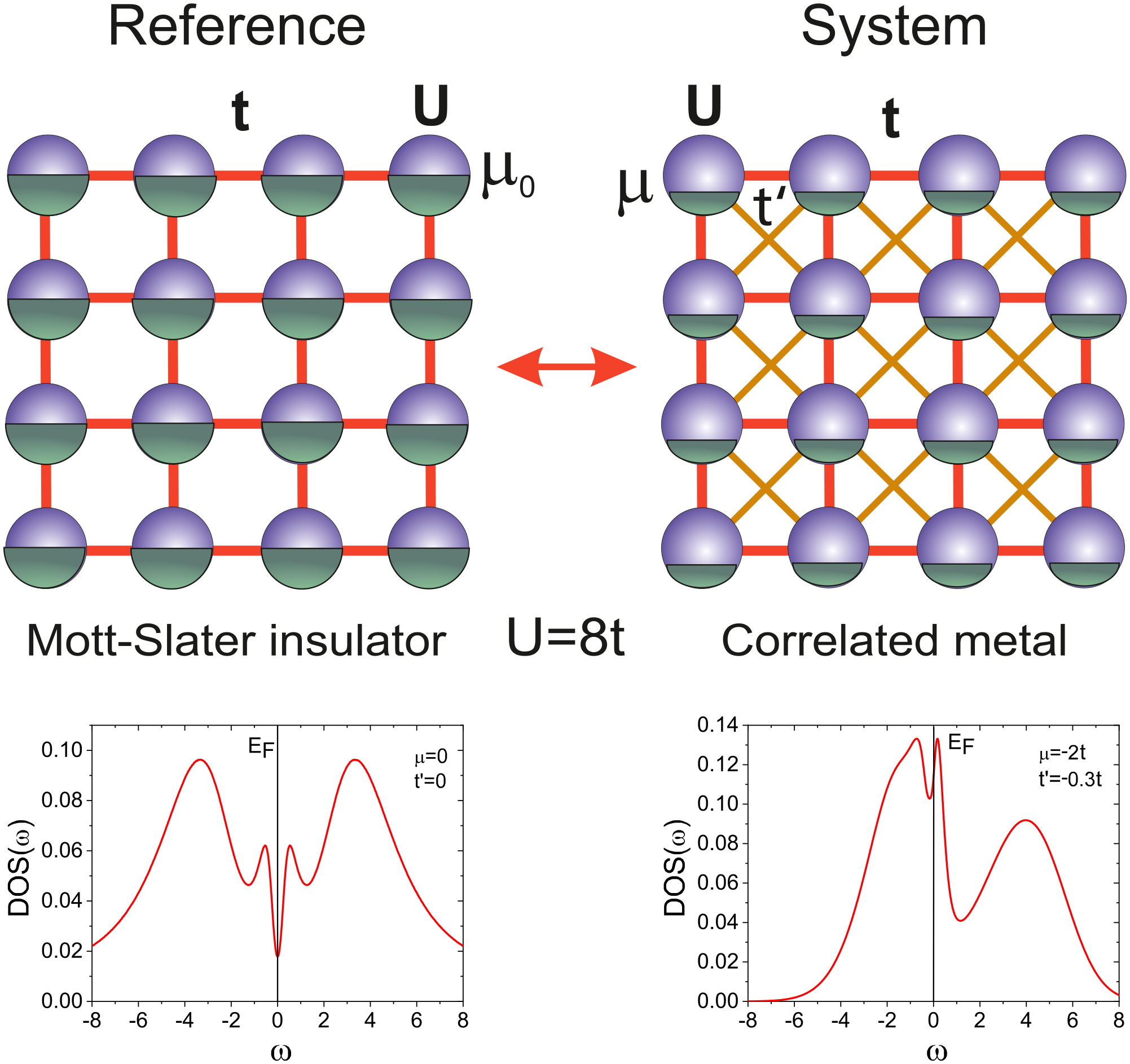}
\caption{Schematic representation of a half-filled particle-hole symmetric reference system with ${\mu=0}$ and ${t'=0}$ (top left) for the doped square lattice with ${\mu=-2t}$ and ${t'=-0.3t}$ (top right). 
The corresponding local density of states (DOS) calculated using the DF lattice DQMC scheme scheme calculated for ${U=8t}$ at the inverse temperature ${\beta=10/t}$ are shown in bottom panels.
The Figure is taken from Ref.~\cite{DFQMC}.}
\label{fig:DFQMC_system}
\end{figure}

In Ref.~\cite{DFQMC}, the authors proposed a novel strong-coupling perturbation scheme based on the lattice determinantal Quantum Monte Carlo method, dubbed ``Dual Fermion lattice DQMC approach''.
In this approach, the reference system is considered of the same lattice size as the original problem, e.g. a Hubbard model on a ${N\times{}N}$ square lattice with periodic boundary conditions.
The difference is that, while the hopping matrix and filling of the original problem can be chosen arbitrarily, the reference system is assumed to be half-filled and particle-hole symmetric, so that only the nearest-neighbor hopping $t$ enters the reference problem.
In particular, in Ref.~\cite{DFQMC} the hopping amplitudes were taken in the following form:
\begin{align}
t^{(\alpha)}_{ij}=
\begin{cases}
t, & \text{for $(i, j) \in$ NN}\\
\alpha t', & \text{for $(i, j) \in$ NNN}\\
\alpha \mu, & \text{for $i=j$} \\
0, & \text{otherwise}
\end{cases}
\label{tij}
\end{align}
where $t$ is the nearest-neighbor (NN) and $t'$ is the next-nearest-neighbor (NNN) hopping amplitudes on a square lattice, and the $\alpha$ parameter was introduced to distinguish hopping amplitudes of the reference (${\alpha=0}$) and original (${\alpha=1}$) system. 
The advantage of choosing such a reference system is that it is free from the fermionic sign problem and can therefore be solved numerically exactly for relatively large number of lattice sites over a broad temperature range, using the lattice DQMC approach~\cite{PhysRevB.44.10502}.
An example of the reference lattice and the original lattice systems is shown in Fig.~\ref{fig:DFQMC_system}.
Importantly, the lattice reference system already incorporates a significant portion of spatial correlations, in particular, non-perturbative effects related to formation of a singlet state. 
Therefore, the local density of states of the reference problem, shown in the bottom left panel of Fig.~\ref{fig:DFQMC_system}, captures not only the existence of high-energy Hubbard bands but also the presence of lower-energy antiferromagnetic Slater peaks. 
Such a ``four-peak'' structure is the characteristic feature of the half-filled Hubbard model for large interaction strength~\cite{Rost_QMC}. It is important to note that the Slater peaks cannot be captured by the paramagnetic single-site reference problem in the DMFT-like approximation. 

Another advantage is that the difference between the original lattice and the reference lattice problems, 
${\tilde{\varepsilon} = t^{(1)} - t^{(0)}}$, which corresponds to the difference between the hopping matrices including the chemical potentials, for reasonable values of $t'$ and $\mu$ is small compared to the bandwidth ${W=8t}$ and can be used as a perturbative parameter.
In this case, according to discussions presented in Section~\ref{sec:strong_coupling}, the DF diagrammatic expansion can be truncated at the first order, resulting in the following form of the dual self-energy, similar to Eq.~\eqref{eq:Sigma_norm}:
\begin{align}
\tilde\Sigma_{k} &=  \sum_{k'} \tilde{\cal G}_{k'} \left[\Gamma^{\uparrow\uparrow\uparrow\uparrow}_{k,k',0} + \Gamma^{\uparrow\uparrow\downarrow\downarrow}_{k,k',0} \right].
\label{eq:Sigma_QMC}
\end{align}
Calculating and storing the four-point vertex $\Gamma_{k,k',0}$ in the case of large lattices is computationally inefficient. 
Instead, the DQMC scheme is used to stochastically evaluate the self-energy~\eqref{eq:Sigma_QMC} directly.
Within the DQMC scheme, which employs auxiliary Ising fields ``${s}$,'' the Wick theorem can be applied to compute the four-point correlation functions:
\begin{align}
 \langle{c_{1}^{\phantom{*}}c_{2}^{*}c_{3}^{\phantom{*}} c_{4}^{*}} \rangle _s =
 \langle{c_{1}^{\phantom{*}}c_{2}^{*}} \rangle _s \, 
 \langle{c_{3}^{\phantom{*}}c_{4}^{*}} \rangle_s - \langle{c_{1}^{\phantom{*}}c_{4}^{*}}\rangle_s \, 
 \langle{c_{3}^{\phantom{*}}c_{2}^{*}} \rangle _s\,.
\label{gamma4qmc}
\end{align}
The four-point vertex function $\Gamma$ corresponds to the connected part of the four-point correlation function.
In order to subtract the disconnected part of the correlation function from Eq.~\eqref{gamma4qmc}, we subtract the exact Green's function of the reference system, ${g_{12} = -\langle c^{\phantom{*}}_{1} c_{2}^{\ast} \rangle}$, obtained from the separate DQMC run, as follows:
\begin{align}
\overline{g}^s_{12}=g^{s}_{12}-g^{\phantom{s}}_{12}.
\label{gdfqmc}
\end{align}
We also utilize Fourier space for the efficient evaluation of Eq.~\eqref{eq:Sigma_QMC}.
Within the DQMC framework, the lattice auxiliary Green's function is not translationally invariant, i.e. $g^s_{12} = -\langle c_{1}^{\phantom{*}} c_{2}^{\ast} \rangle_s$. 
To calculate $\overline{g}^s_{kk'}$, we employ a double fast Fourier transform to momentum ${\bf k}$ and Matsubara frequency $\nu$ space, with ${k \in \{{\bf k}, \nu}\}$.
It is important to note that the dual Green's function, $\tilde{\cal G}_{k}$, is translationally invariant. Additionally, after performing the Monte Carlo summation over the auxiliary spins ${s}$, the self-energy also becomes translationally invariant. The final expression for the dual self-energy in momentum space reads:
\begin{align}
\tilde\Sigma_{k} = -\sum_{k'}\tilde{G}_{k'} \sum_{\rm QMC}\left(  \overline{g}^{\uparrow\,s}_{k,k}\overline{g}^{\uparrow\,s}_{k',k'} - \overline{g}^{\uparrow\,s}_{k,k'}\overline{g}^{\uparrow\,s}_{k',k} + \overline{g}^{\uparrow\,s}_{k,k} \overline{g}^{\downarrow\,s}_{k',k'}\right).
\label{eq:Sigma_DQMC}
\end{align}
Similar expressions can also be derived for the normal and anomalous components~\eqref{eq:Sigma_norm}-\eqref{eq:Sigma_anom} of the dual self-energy in the Nambu space presented in Section~\ref{sec:DF_Nambu}.

The result for the local DOS obtained within the DF lattice DQMC scheme is shown in the bottom right panel of Fig.~\ref{fig:DFQMC_system}.
Taking into account the first-order DF self-energy transforms the Mott insulating reference system into a correlated metal with a typical three-peak structure that features a quasi-particle peak at Fermi energy.
Moreover, for the model parameters relevant for hole-doped cuprate superconductors, the presented scheme allows one to capture formation of the pseudogap in the energy spectrum and nodal-antinodal dichotomy (that is, well-defined quasiparticles in the nodal part of the Fermi surface and strong quasiparticle damping for the antinodal part), as discussed in Section~\ref{sec:DF_Nambu}.
The DF lattice DQMC scheme is able to capture these effects, because the first-order DF self-energy relies on the lattice four-point vertex, which is obtained numerically exactly and contains the information about the collective spin and charge fluctuations of the lattice.
Remarkably, the calculation for the DOS of the hole-doped lattice presented in the bottom panel of Fig.~\ref{fig:DFQMC_system} are performed at the temperature ${T = 0.1t}$, which is challenging to access using straightforward lattice QMC calculations.

\newpage
\section{Partially bosonized dual theory}
\label{sec:PBDT}

As we have demonstrated in Section~\ref{sec:LDB}, the ladder dual fermion/boson approach enables an accurate solution of the single-band (extended) Hubbard model across a broad range of model parameters. 
This approach relies on solving the Bethe-Salpeter equation (BSE) with a frequency-dependent vertex function. 
However, solving this BSE in a multi-orbital framework becomes numerically expensive, making practical calculations for realistic materials highly inefficient. 
The four-point vertex function represents the bare local Coulomb interaction screened by the many-body effects of the reference problem. It is an essential ingredient of the theory that cannot be simply neglected~\cite{PhysRevB.94.205110}. 
In this Section, we introduce a way to eliminate the four-point vertex directly from the dual action~\eqref{eq:DB_action}, leading to a simplified, partially bosonized dual theory that involves only the three-point (fermion-boson) vertex function in the interaction term $\tilde{\cal F}$.

\subsection{Partially bosonized approximation for the four-point vertex function}
\label{sec:PB_approx}

\begin{figure}[t!]
\centering
\includegraphics[width=0.9\textwidth]{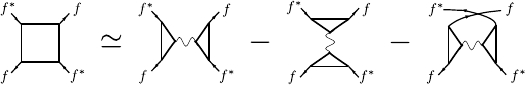} 
\caption{Diagrammatic representation of the partially bosonized approximation of the four-fermion vertex corresponding to Eq.~\eqref{eq:Gamma_approx_app}. Coefficients and channel indices are omitted for clarity. The pictorial representation explains the distinction between the horizontal contributions $\left[M^{\varsigma\varsigma'}_{\nu\nu'\omega}\right]^{\sigma_1\sigma_2\sigma_3\sigma_4}_{l_1 l_2 l_3 l_4}$, the vertical contributions $\left[M^{\varsigma\varsigma'}_{\nu,\nu+\omega,\nu'-\nu}\right]^{\sigma_1\sigma_3\sigma_2\sigma_4}_{l_1 l_3 l_2 l_4}$ and the particle-particle contributions $\left[M^{\vartheta\vartheta'}_{\nu,\nu',\omega+\nu+\nu'}\right]^{\sigma_1\sigma_4\sigma_3\sigma_2}_{l_1 l_4 l_3 l_2}$. Fermionic legs are attached to the vertices for clarity.}
\label{fig:partial_bos}
\end{figure}

First, let us introduce the partially bosonized approximation for the exact four-point vertex function of the reference problem.
In this approximation, the renormalized electronic interaction, represented by the four-point vertex, is expressed in terms of electronic scattering via a single bosonic fluctuation (bosonic propagator), as shown in Fig.~\ref{fig:partial_bos}.
We note that the partially bosonized form of the vertex is similar to that of the vertex function~\eqref{DBvertex} (Fig.~\ref{fig:vertex}) of the modified DF action~\eqref{finalaction} in the DiagMC@DB approach (see Section~\ref{sec:DiagMC}).
As argued in Ref.~\cite{PhysRevB.100.205115}, this is the only form of the vertex that can be generated by transforming the DB action~\eqref{eq:DB_action}.

The four-point vertex function corresponds to the antisimmetrized form of the electron-electron interaction.
In order to obtain the bare four-point vertex we first antisymmetrize the bare local Coulomb interaction~\eqref{eq:Upp_to_Uph} of the initial lattice problem~\eqref{eq:actionlatt} as (quadratic terms in Grassmann $c^{(*)}$ variables are neglected for simplicity):
\begin{align}
&\frac12 \sum_{\omega, \{\nu\}}\sum_{\{l\}, \{\sigma\}} U^{ph}_{l_1 l_2 l_3 l_4} c^{*}_{\nu \sigma l_1} c^{\phantom{*}}_{\nu+\omega, \sigma l_2} c^{*}_{\nu'+\omega, \sigma' l_4} c^{\phantom{*}}_{\nu'\sigma' l_3} = \notag\\ 
&\frac18 \sum_{\omega, \{\nu\}}\sum_{\{l\}, \{\sigma\}} 
\Gamma^{0\,d}_{l_1 l_2 l_3 l_4} \left(c^{*}_{\nu \sigma l_1} c^{\phantom{*}}_{\nu+\omega, \sigma l_2}\right) \left(c^{*}_{\nu'+\omega, \sigma', l_4} c^{\phantom{*}}_{\nu', \sigma', l_3}\right) + \notag\\
&\frac18 \sum_{\omega, \{\nu\}}\sum_{\{l\}, \{\sigma\}} \Gamma^{0\,m}_{l_1 l_2 l_3 l_4} \left(c^{*}_{\nu \sigma_1 l_1} \vec{\sigma}_{\sigma_1\sigma_2} c^{\phantom{*}}_{\nu+\omega, \sigma_2 l_2}\right) \left(c^{*}_{\nu'+\omega, \sigma_4 l_4} \vec{\sigma}_{\sigma_4\sigma_3} c^{\phantom{*}}_{\nu' \sigma_3 l_3}\right)\,,
\label{eq:G0ph}
\end{align}
where the bare four-point vertex functions in the particle-hole density ($d$) and magnetic ($m$) channels read:
\begin{align}
\Gamma^{0\,d}_{l_1 l_2 l_3 l_4} &= 2U^{ph}_{l_1 l_2 l_3 l_4} - U^{ph}_{l_1 l_3 l_2 l_4}
= 2U^{pp}_{l_1 l_4 l_2 l_3} - U^{pp}_{l_1 l_4 l_3 l_2}\,, \label{eq:Uph_d}\\
\Gamma^{0\,m}_{l_1 l_2 l_3 l_4} &= - U^{ph}_{l_1 l_3 l_2 l_4} = - U^{pp}_{l_1 l_4 l_3 l_2}\,. \label{eq:Uph_m}
\end{align}
Alternatively, the interaction can also be antisymmetrized in the particle-particle channel:
\begin{align}
&\frac12 \sum_{\omega, \{\nu\}}\sum_{\{l\}, \{\sigma\}} U^{ph}_{l_1 l_2 l_3 l_4} c^{*}_{\nu \sigma l_1} c^{\phantom{*}}_{\nu+\omega, \sigma l_2} c^{*}_{\nu'+\omega, \sigma' l_4} c^{\phantom{*}}_{\nu'\sigma' l_3} = \notag\\
&\frac14 \sum_{\omega,\{\nu\},\{l\}} 
\Gamma^{0\,s}_{l_1 l_2 l_3 l_4} \left(c^{*}_{\nu \uparrow l_1} 
c^{*}_{\omega-\nu \downarrow l_2} - c^{*}_{\nu \downarrow l_1} 
c^{*}_{\omega-\nu, \uparrow l_2}\right) 
\left(c^{\phantom{*}}_{\omega-\nu', \downarrow l_4}
c^{\phantom{*}}_{\nu' \uparrow l_3} - 
c^{\phantom{*}}_{\omega-\nu', \uparrow l_4} c^{\phantom{*}}_{\nu' \downarrow l_3}\right) + \notag\\
&\frac14 \sum_{\omega,\{\nu\},\{l\}} \Gamma^{0\,t}_{l_1 l_2 l_3 l_4} \left(c^{*}_{\nu \uparrow l_1} 
c^{*}_{\omega-\nu, \downarrow l_2} + c^{*}_{\nu \downarrow l_1}
c^{*}_{\omega-\nu, \uparrow l_2}\right) 
\left(c^{\phantom{*}}_{\omega-\nu', \downarrow l_4}
c^{\phantom{*}}_{\nu' \uparrow l_3} + 
c^{\phantom{*}}_{\omega-\nu', \uparrow l_4} c^{\phantom{*}}_{\nu' \downarrow l_3}\right) + \notag\\
&\frac12 \sum_{\omega,\{\nu\},\{l\}} \Gamma^{0\,t}_{l_1 l_2 l_3 l_4} \Bigg\{ \left(c^{*}_{\nu \uparrow l_1} 
c^{*}_{\omega-\nu, \uparrow l_2}\right) 
\left(c^{\phantom{*}}_{\omega-\nu', \uparrow l_4}
c^{\phantom{*}}_{\nu' \uparrow l_3}\right) + \left(c^{*}_{\nu \downarrow l_1} 
c^{*}_{\omega-\nu, \downarrow l_2}\right)
\left(c^{\phantom{*}}_{\omega-\nu', \downarrow l_4} c^{\phantom{*}}_{\nu' \downarrow l_3}\right) \Bigg\},
\label{eq:G0pp}
\end{align}
which gives the corresponding bare vertex functions in the singlet ($s$) and triplet ($t$) channels:
\begin{align}
\label{eq:UppS}
\Gamma^{0\,s}_{l_1 l_2 l_3 l_4} &= \frac12\left(U^{ph}_{l_1 l_3 l_4 l_2} + U^{ph}_{l_1 l_4 l_3 l_2} \right) = \frac12\left(U^{pp}_{l_1 l_2 l_3 l_4} + U^{pp}_{l_1 l_2 l_4 l_3} \right), \\
\Gamma^{0\,t}_{l_1 l_3 l_4 l_2} &= \frac12\left(U^{ph}_{l_1 l_3 l_4 l_2} - U^{ph}_{l_1 l_4 l_3 l_2} \right) = \frac12\left(U^{pp}_{l_1 l_2 l_3 l_4} - U^{pp}_{l_1 l_2 l_4 l_3}\right).\label{eq:UppT}
\end{align}
Note that the obtained expressions~\eqref{eq:Uph_d},~\eqref{eq:Uph_m},~\eqref{eq:UppS}, and~\eqref{eq:UppT} coincide with the standard definition for the bare vertex functions in the fluctuation exchange (FLEX) approach (see e.g. Ref.~\cite{PhysRevB.57.6884}).

In order to express the vertex function in terms of the bosonic fluctuations, one must first introduce the bare interaction for the relevant bosonic channels.
To do so, one can decouple the bare Coulomb interaction $U^{ph}_{l_1 l_2 l_3 l_4}$ into different channels ($r$) and rewrite the bare interaction of the reference system~\eqref{eq:actionimp_app} in the channel representation as:
\begin{align}
&\frac12 \sum_{\omega, \{\nu\}}\sum_{\{l\}, \{\sigma\}} U^{ph}_{l_1 l_2 l_3 l_4} c^{*}_{\nu \sigma l_1} c^{\phantom{*}}_{\nu+\omega, \sigma l_2} c^{*}_{\nu'+\omega, \sigma' l_4} c^{\phantom{*}}_{\nu'\sigma' l_3} +
\notag\\
&\frac12 \sum_{\omega,\{l\},\varsigma} Y^{\varsigma}_{\omega,\, l_1 l_2,\, l_3 l_4} \, \rho^{\varsigma}_{-\omega,\, l_1 l_2} \, \rho^{\varsigma}_{\omega,\, l_4 l_3}
+ \sum_{\omega,\{l\},\vartheta} Y^{\vartheta}_{\omega,\, l_1 l_2,\, l_3 l_4}\, \rho^{*\,\vartheta}_{\omega,\, l_1 l_2} \, \rho^{\vartheta}_{\omega,\, l_3 l_4} = \notag\\
&\frac12\sum_{\omega,\{l\},\varsigma} \tilde{U}^{\varsigma}_{\omega,\,l_1 l_2,\, l_3 l_4} \rho^{\varsigma}_{-\omega,\, l_1 l_2} \, \rho^{\varsigma}_{\omega,\, l_4 l_3}
+ \sum_{q,\{l\},\vartheta} \tilde{U}^{\vartheta}_{\omega,\,l_1 l_2,\, l_3 l_4} \rho^{*\,\vartheta}_{q,\, l_1 l_2} \, \rho^{\vartheta}_{q,\, l_3 l_4}\,,
\label{eq:Uchannel}
\end{align}
where ${\tilde{U}^{r}_{\omega,\,l_1l_2,\,l_3l_4} = U^{r}_{l_1l_2,\,l_3l_4} + Y^{r}_{\omega,\,l_1l_2,\,l_3l_4}}$ and $U^{r}_{l_1l_2,\,l_3l_4}$ is the bare Coulomb interaction in the channel representation.
Importantly, the decoupling of the local Coulomb interaction is an exact mathematical procedure.
However, there is some freedom in choosing the bare interaction in different channels. 
For instance, in the single-band case, the decoupling reads~\cite{PhysRevLett.119.166401}:
\begin{align}
U n_{\uparrow} n_{\downarrow} = \frac12 \sum_{\varsigma} U^{\varsigma} \rho^{\varsigma} \rho^{\varsigma},
\end{align}   
where the bare interaction in the charge and spin channels is given by:
\begin{align}
U^{d} = (3a-1)U ~~~ \text{and} ~~~ U^{m} = (a-2/3)U\,,
\end{align}
with the parameter $a$ being arbitrary.
If the decoupled interaction is treated approximately, for example in a $GW$-like fashion, the results become highly sensitive to the specific choice of decoupling (see, e.g., Ref.~\cite{PhysRevLett.119.166401}), leading to the well-known Fierz ambiguity problem~\cite{PhysRevB.65.245118, Borejsza_2003, PhysRevD.68.025020, PhysRevB.70.125111, Bartosch_2009}.

Importantly, in \mbox{D-TRILEX} the form of the bare interaction in different channels in not obtained by decoupling the local Coulomb interaction.
Instead, it is derived from the most accurate partially bosonized approximation of the exact four-point vertex of the impurity problem. 
This vertex corresponds to the two-particle correlation function and is therefore free from the Fierz ambiguity problem.
This is already evident from the form of the bare vertex function. 
Following Refs.~\cite{PhysRevB.100.205115, PhysRevB.103.245123}, upon antisymmetrizing the bare interaction of the impurity problem~\eqref{eq:Uchannel}, one obtains the following bare vertex functions, which take a form similar to Eqs.~\eqref{eq:Uph_d},~\eqref{eq:Uph_m}:
\begin{align}
\left[\Gamma^{0\,d}_{\nu\nu'\omega}\right]_{l_1 l_2 l_3 l_4}
&= 2\tilde{U}^{d}_{\omega,\,l_1 l_2, l_3 l_4} - \tilde{U}^{d}_{\nu'-\nu,\,l_1 l_3, l_2 l_4} - 3\tilde{U}^{m}_{\nu'-\nu,\,l_1 l_3, l_2 l_4} + \tilde{U}^{s}_{\omega+\nu+\nu',\,l_1 l_4, l_3 l_2} - 3\tilde{U}^{t}_{\omega+\nu+\nu',\,l_1 l_4, l_3 l_2}  \notag\\
&= 2U^{ph}_{l_1 l_2 l_3 l_4} - U^{ph}_{l_1 l_3 l_2 l_4} + o\left(Y\right)\,,
\label{eq:BareGammaCharge}\\
\left[\Gamma^{0\,m}_{\nu\nu'\omega}\right]_{l_1 l_2 l_3 l_4} 
&= 2\tilde{U}^{m}_{\omega,\,l_1 l_2, l_3 l_4} + \tilde{U}^{m}_{\nu'-\nu,\,l_1 l_3, l_2 l_4} - \tilde{U}^{d}_{\nu'-\nu,\,l_1 l_3, l_2 l_4} - \tilde{U}^{s}_{\omega+\nu+\nu',\,l_1 l_4, l_3 l_2} - \tilde{U}^{t}_{\omega+\nu+\nu',\,l_1 l_4, l_3 l_2} \notag\\ 
&= - U^{ph}_{l_1 l_3 l_2 l_4} + o\left(Y\right)\,.
\label{eq:BareGammaSpin}
\end{align}
The second lines in Eqs.~\eqref{eq:BareGammaCharge} and~\eqref{eq:BareGammaSpin} are obtained using the antisymmetrized forms of the bare interactions~\eqref{eq:Uph_d} and~\eqref{eq:Uph_m}, respectively, and demonstrate that the bare vertex is already independent of the particular choice of decoupling procedure.

In order to derive the most accurate partially bosonized approximation for the four-point vertex, we follow Ref.~\cite{PhysRevB.100.205115} and associate the static parts $U^{ph}$ of the vertex functions~\eqref{eq:BareGammaCharge} and~\eqref{eq:BareGammaSpin} with the longitudinal contributions $\tilde{U}^{\varsigma}_{\omega}$.
This allows one to exclude the bare Coulomb interaction $U^{ph}$ from the transverse contributions $\tilde{U}_{\nu'-\nu}$. 
As discussed in Ref.~\cite{PhysRevB.100.205115}, these transverse terms generate irreducible in the bosonic fluctuation contributions to the exact (renormalized) four-point vertex $\Gamma$ of the reference problem that cannot be reproduced by the partially bosonized approximation, which is reducible with respect to the bosonic propagator (interaction).  
After doing that, we immediately get the following result for the bare local interaction in the charge and spin channel:
\begin{align}
U^{d}_{l_1 l_2,\, l_3 l_4} &= \frac12 \left( 2U^{ph}_{l_1 l_2 l_3 l_4} - U^{ph}_{l_1 l_3 l_2 l_4} \right) = \frac12 \left(2U^{pp}_{l_1 l_4 l_2 l_3} - U^{pp}_{l_1 l_4 l_3 l_2}\right), 
\label{eq:Ud_app}\\
U^{m}_{l_1 l_2,\, l_3 l_4} &= - \frac12 U^{ph}_{l_1 l_3 l_2 l_4} = -\frac12 U^{pp}_{l_1 l_4 l_3 l_2} \,.
\label{eq:Um_app}
\end{align}
The same procedure can be performed for the particle-particle channel. 
Since the variables $\rho^{(*)\,\vartheta}_{\omega}$ are already defined in the antisymmetrized form~\eqref{eq:npp_app}, the bare interaction in the particle-particle channel simply coincides with the bare vertex defined in Eqs.~\eqref{eq:UppS} and~\eqref{eq:UppT}:
\begin{align}
U^{s}_{l_1 l_2,\, l_3 l_4} &= \frac12\left(U^{ph}_{l_1 l_3 l_4 l_2} + U^{ph}_{l_1 l_4 l_3 l_2} \right) = \frac12\left(U^{pp}_{l_1 l_2 l_3 l_4} + U^{pp}_{l_1 l_2 l_4 l_3} \right), \label{eq:Us_app}\\ 
U^{t}_{l_1 l_2,\, l_3 l_4} &= \frac12\left(U^{ph}_{l_1 l_3 l_4 l_2} - U^{ph}_{l_1 l_4 l_3 l_2} \right) = \frac12\left(U^{pp}_{l_1 l_2 l_3 l_4} - U^{pp}_{l_1 l_2 l_4 l_3} \right). \label{eq:Ut_app}
\end{align}

Following the derivation presented in Refs.~\cite{PhysRevB.100.205115, PhysRevB.103.245123} and inspired by Refs.~\cite{PhysRevLett.121.037204, PhysRevB.99.115124}, the partially bosonized approximation for the exact four-point vertex function $\Gamma$ of the reference problem can be obtained by renormalizing the bare vertices~\eqref{eq:BareGammaCharge} and~\eqref{eq:BareGammaSpin} by the two-particle excitation in the particle-hole and particle-particle channels and retaining the contributions that are reducible in the interaction.
In the multi-band case this gives the following form for the fermion-fermion vertex~\cite{10.21468/SciPostPhys.13.2.036} illustrated diagrammatically in Fig.~\ref{fig:partial_bos}:
\begin{align}
\left[\Gamma_{\nu\nu'\omega}\right]^{\sigma_1\sigma_2\sigma_3\sigma_4}_{l_1 l_2 l_3 l_4} = \sum_{\{r\}} \left\{\left[M^{\varsigma\varsigma'}_{\nu\nu'\omega}\right]^{\sigma_1\sigma_2\sigma_3\sigma_4}_{l_1 l_2 l_3 l_4}
- \left[M^{\varsigma\varsigma'}_{\nu,\nu+\omega,\nu'-\nu}\right]^{\sigma_1\sigma_3\sigma_2\sigma_4}_{l_1 l_3 l_2 l_4} - \left[M^{\vartheta\vartheta'}_{\nu,\nu',\omega+\nu+\nu'}\right]^{\sigma_1\sigma_4\sigma_3\sigma_2}_{l_1 l_4 l_3 l_2} \right\},
\label{eq:Gamma_approx_app}
\end{align}
where the partially bosonized collective electronic fluctuations in different channels are:
\begin{align}
\left[M^{\varsigma\varsigma'}_{\nu\nu'\omega}\right]^{\sigma_1\sigma_2\sigma_3\sigma_4}_{l_1 l_2 l_3 l_4} &= \sum_{\{l'\}} \Lambda^{\sigma_1\sigma_2\varsigma}_{\nu\omega,\,l_1,\, l_2,\, l'_1 l'_2} \, \bar{w}^{\varsigma\varsigma'}_{\omega,\,l'_1 l'_2,\, l'_3 l'_4} \, \Lambda^{\sigma_4\sigma_3\varsigma'}_{\nu'+\omega,-\omega,\,l_4,\, l_3,\, l'_4 l'_3} \,,
\label{eq:ExactGscreenedvertPH_app}\\
\left[M^{\vartheta\vartheta'}_{\nu\nu'\omega}\right]^{\sigma_1\sigma_2\sigma_3\sigma_4}_{l_1 l_2 l_3 l_4} &= \sum_{\{l'\}} \Lambda^{\sigma_1\sigma_2\vartheta}_{\nu\omega,\,l_1,\, l_2,\, l'_1 l'_2} \bar{w}^{\vartheta\vartheta'}_{\omega,\,l'_1 l'_2,\, l'_3 l'_4} \Lambda^{*\,\sigma_3\sigma_4\vartheta'}_{\nu'\omega,\,l_3,\, l_4,\, l'_3 l'_4}\,.
\label{eq:ExactGscreenedvertPP_app}
\end{align}
The bosonic fluctuation that connects the three-point vertices in Eqs.~\eqref{eq:ExactGscreenedvertPH_app} and~\eqref{eq:ExactGscreenedvertPP_app} is:
\begin{align}
\bar{w}^{rr'}_{\omega,\,l_1 l_2,\, l_3 l_4} = w^{rr'}_{\omega,\,l_1 l_2,\, l_3 l_4} - \bar{u}^{r}_{l_1 l_2 l_3 l_4} \delta^{\phantom{*}}_{rr'}\,.
\label{eq:w_bar-u_app}
\end{align}
It corresponds to the renormalized interaction $w^{rr'}$ of the reference system~\eqref{eq:w_imp} that can also be obtained from the corresponding susceptibility as:
\begin{align}
w^{rr'}_{\omega,\,l_1l_2,\,l_3l_4} = \tilde{U}^{r}_{l_1l_2,\,l_3l_4}\delta^{\phantom{*}}_{rr'} + \sum_{\{l'\}} \tilde{U}^{r}_{l_1l_2,\,l'_1l'_2} \, \chi^{rr'}_{\omega,\,l'_1l'_2,\,l'_3l'_4} \, \tilde{U}^{r'}_{l'_3l'_4,\,l_3l_4}\,.
\end{align}
As has been discussed in Refs.~\cite{PhysRevB.100.205115, PhysRevB.103.245123}, the choice~\eqref{eq:Ud_app}--\eqref{eq:Ut_app} for the bare interaction $U^{r}$ in different $r$ channels provides the best possible partially bosonized approximation for the four-point vertex function~\eqref{eq:Gamma_approx_app}.
However, it leads to the double-counting of the bare interaction of the reference problem. 
This double-counting is removed by the $\bar{u}^{r}$ term that enters Eq.~\eqref{eq:w_bar-u_app}.
The expression for this term can be obtained in the same way as in Refs.~\cite{PhysRevB.100.205115, PhysRevB.103.245123}, and for the multi-band case explicitly reads:
\begin{align}
{\bar{u}^{\varsigma}_{l_1l_2,\,l_3l_4} = \frac12 U^{\varsigma}_{l_1l_2,\,l_3l_4}}\,,~~~~~~\bar{u}^{\vartheta}_{l_1l_2,\,l_3l_4} = U^{\vartheta}_{l_1l_2,\,l_3l_4}\,.
\label{eq:U_correction}
\end{align}

\subsection{Effect of the approximation on the vertex}

\begin{figure}[ht!]
\centering
\includegraphics[width=1\textwidth, trim={0 1cm 0 1.5cm},clip]{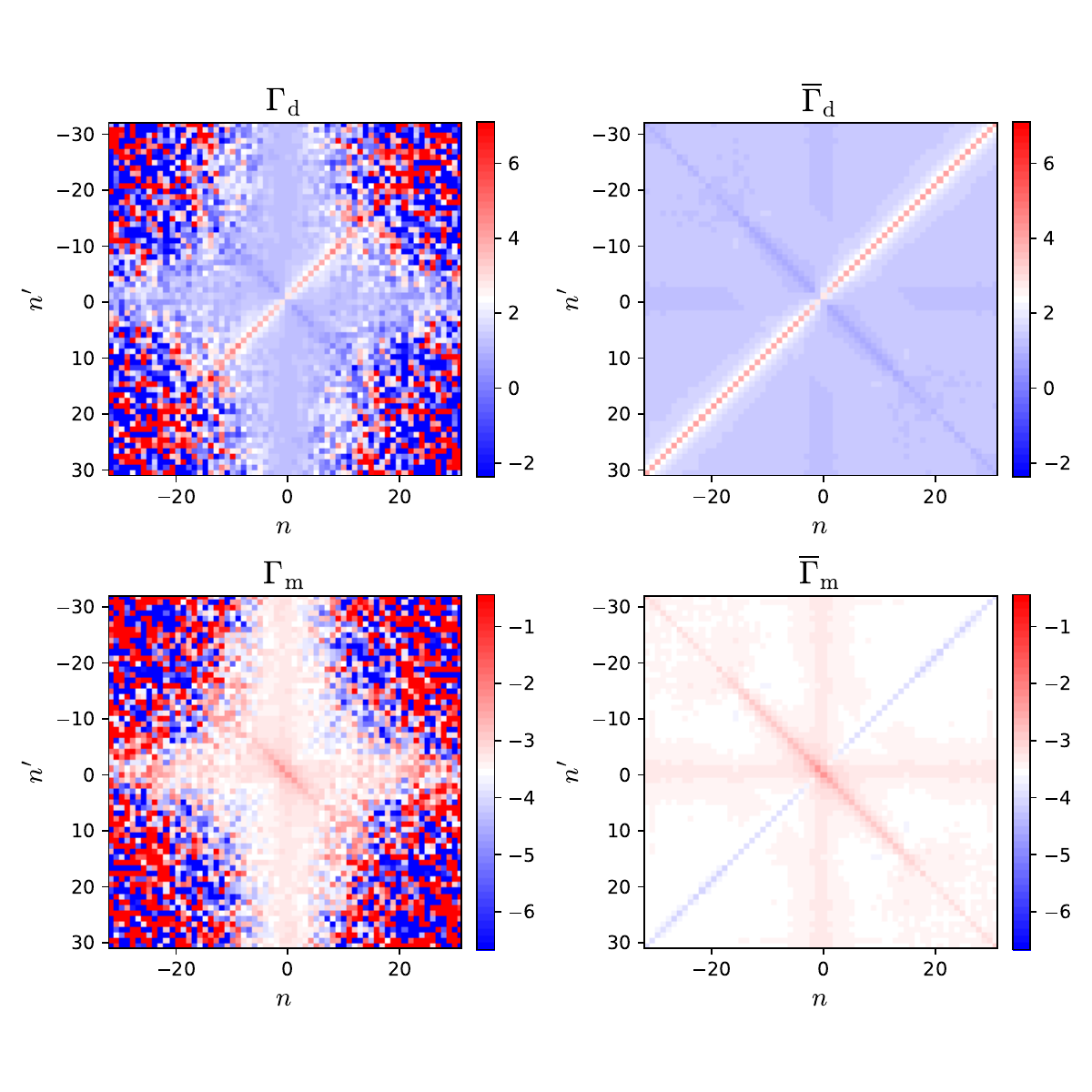} 
\includegraphics[width=1\textwidth]{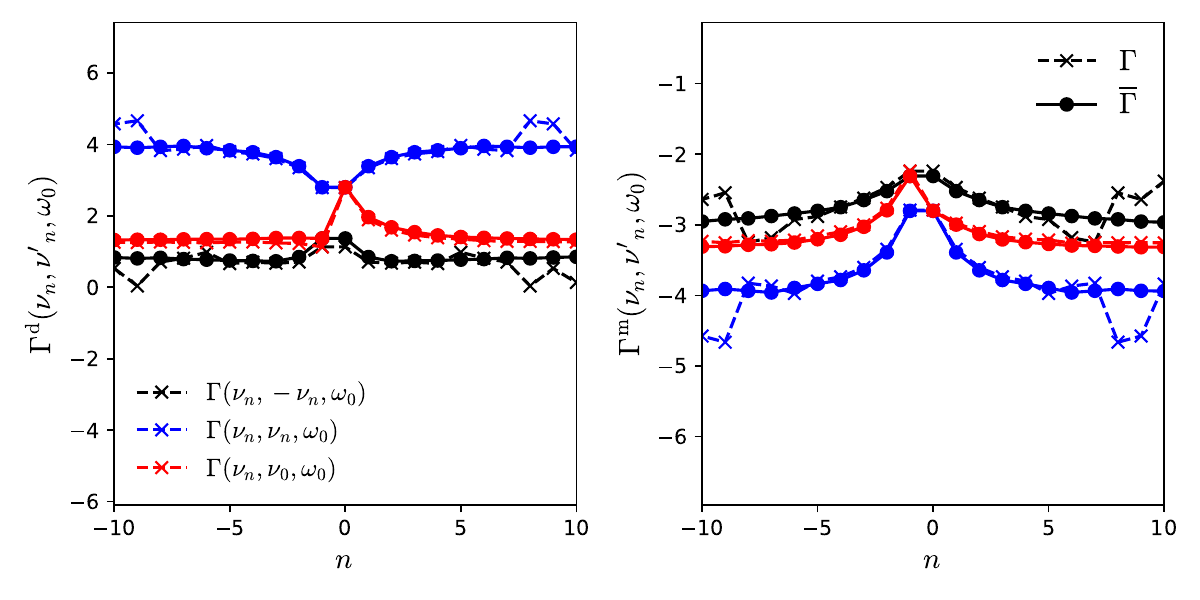} 
\caption{Comparison between the exact vertex $\Gamma_{\nu\nu'\omega}$ and its partially bosonized approximation $\overline{\Gamma}_{\nu\nu'\omega}$ for the impurity obtained solving self-consistently the DMFT problem for the Hubbard model on the square lattice at half-filling. The parameters are $U=2$ and $\beta=4$, which means weakly interacting regime. The bosonic frequency is $\omega=0$. The Figure is taken from Ref.~\cite{vandelli2022quantum}.
\label{fig:approx_vertex_u2}}
\end{figure}

\begin{figure}[ht!]
\centering
\includegraphics[width=1\textwidth, trim={0 1cm 0 1.5cm},clip]{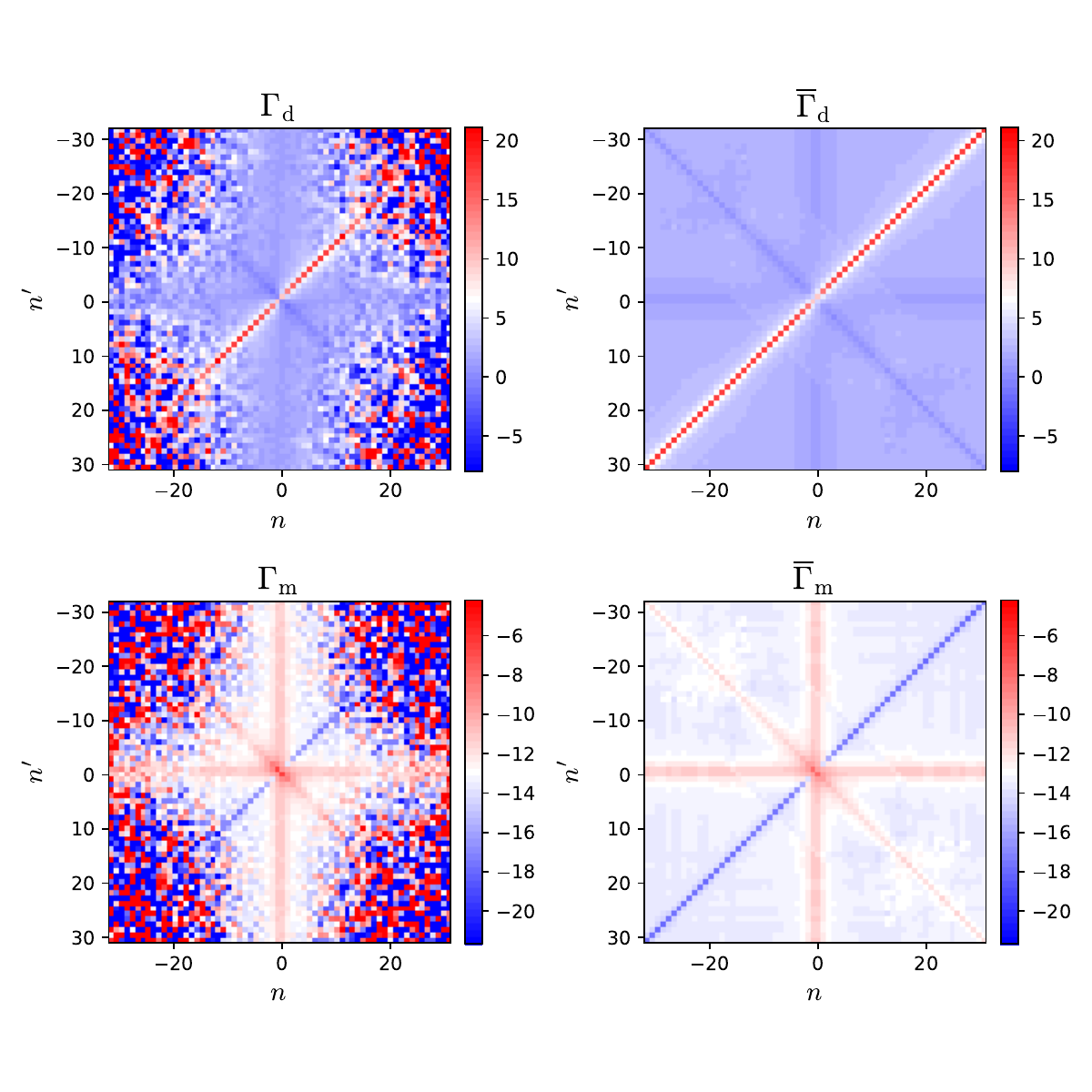} 
\includegraphics[width=1\textwidth]{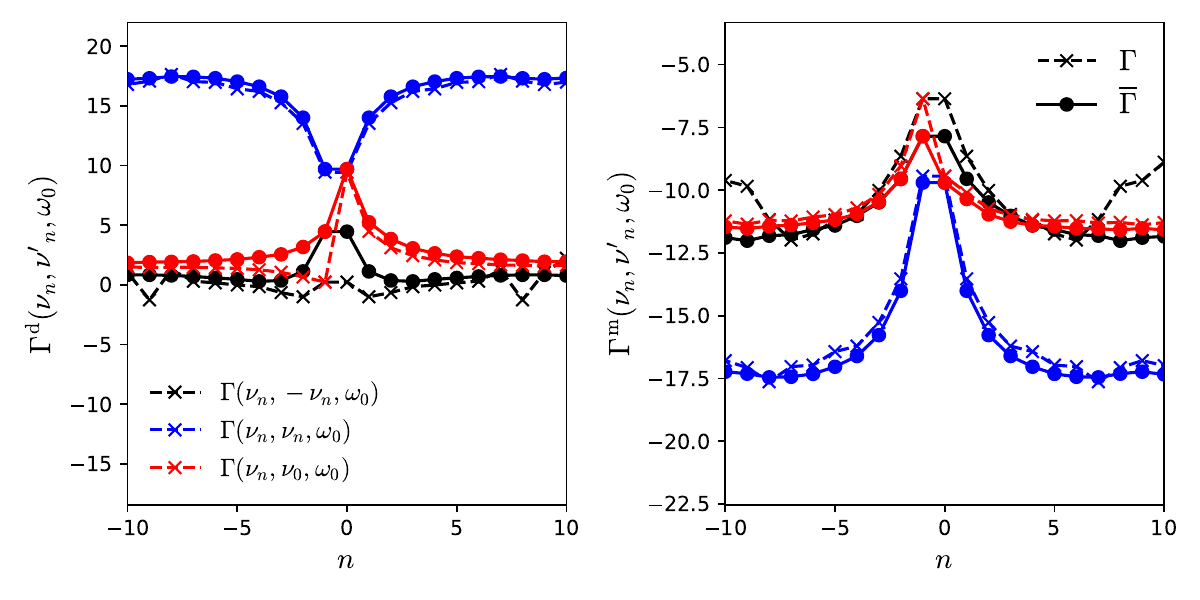} 
\caption{Comparison between the exact vertex $\Gamma_{\nu\nu'\omega}$ and its partially bosonized approximation $\overline{\Gamma}_{\nu\nu'\omega}$ for the impurity obtained solving self-consistently the DMFT problem for the Hubbard model on the square lattice at half-filling. In this plot, we consider the intermediate coupling regime with parameters $U=4$ and $\beta=4$. The Figure is taken from Ref.~\cite{vandelli2022quantum}.
\label{fig:approx_vertex_u4}}
\end{figure}

\begin{figure}[ht!]
\centering
\includegraphics[width=1\textwidth, trim={0 1cm 0 1.5cm},clip]{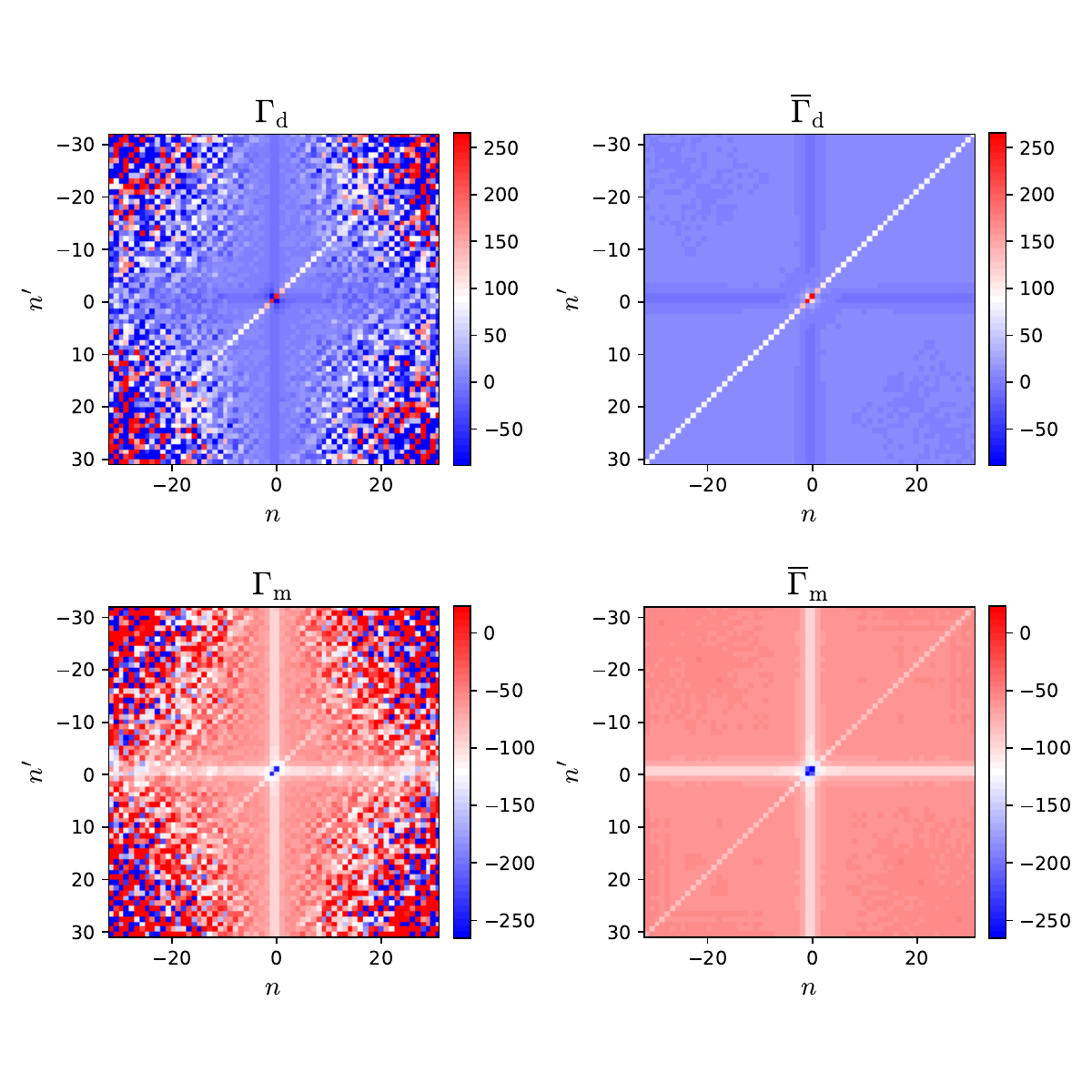} 
\includegraphics[width=1\textwidth]{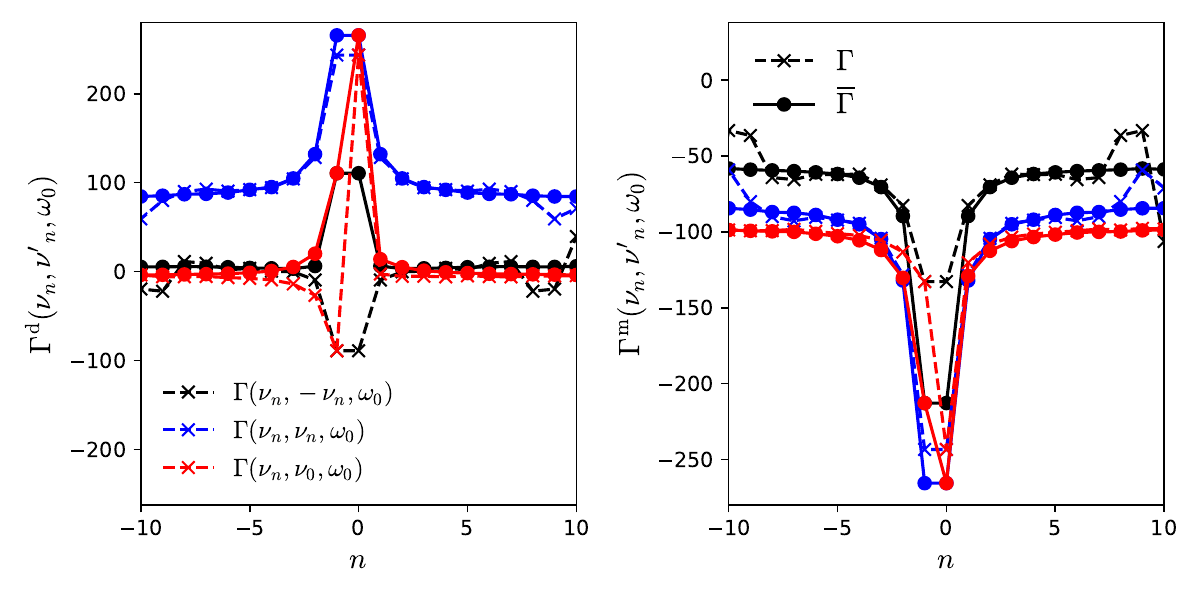} 
\caption{Same vertices as in Fig~\ref{fig:approx_vertex_u2} for a stronger coupling regime with parameters $U=8$ and $\beta=2$. Here, the interaction strength equals the bandwidth of the lattice system of the converged DMFT. The Figure is taken from Ref.~\cite{vandelli2022quantum}.
\label{fig:approx_vertex_u8}}
\end{figure}

\begin{figure}[ht!]
\centering
\includegraphics[width=1\textwidth, trim={0 1cm 0 1.5cm},clip]{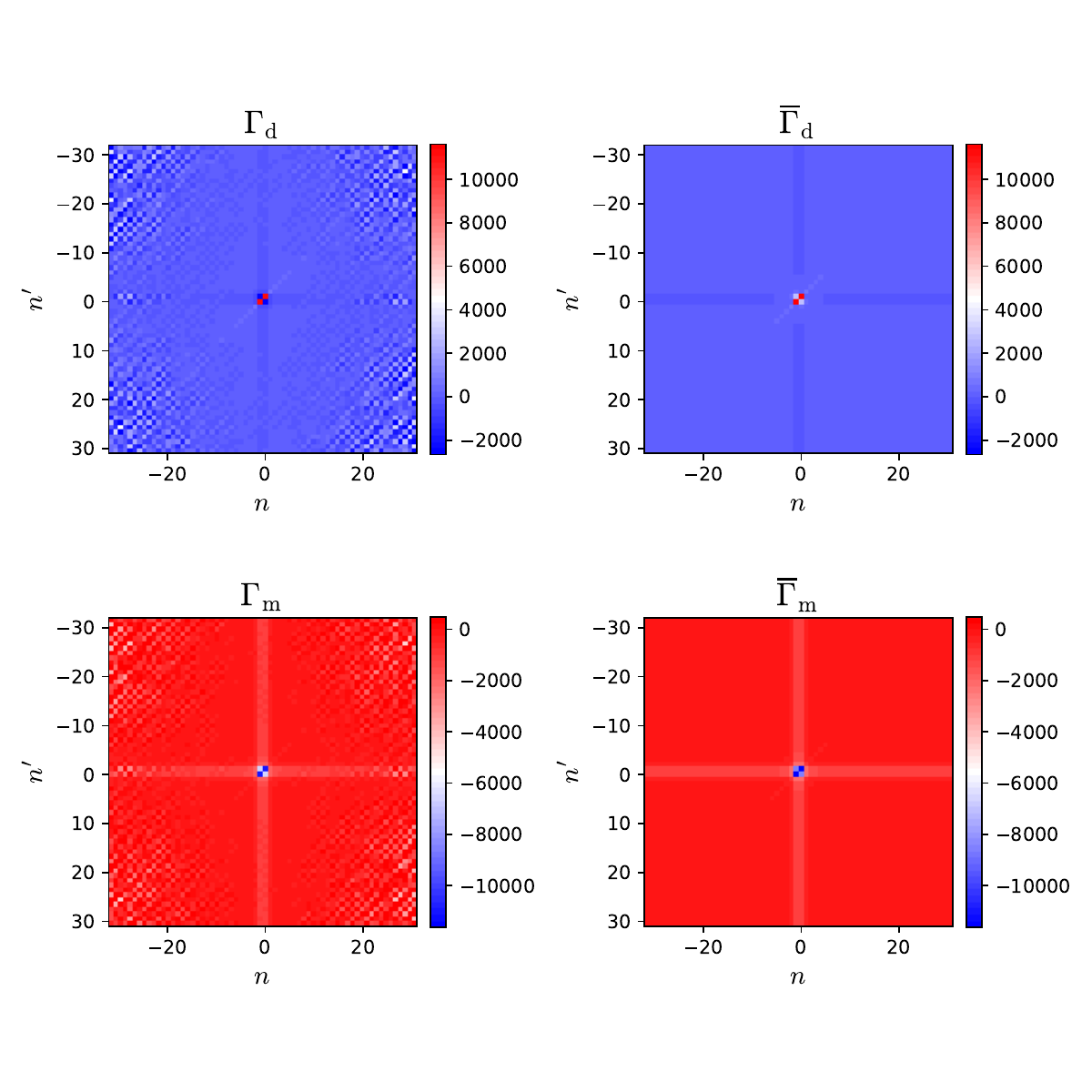} 
\includegraphics[width=1\textwidth]{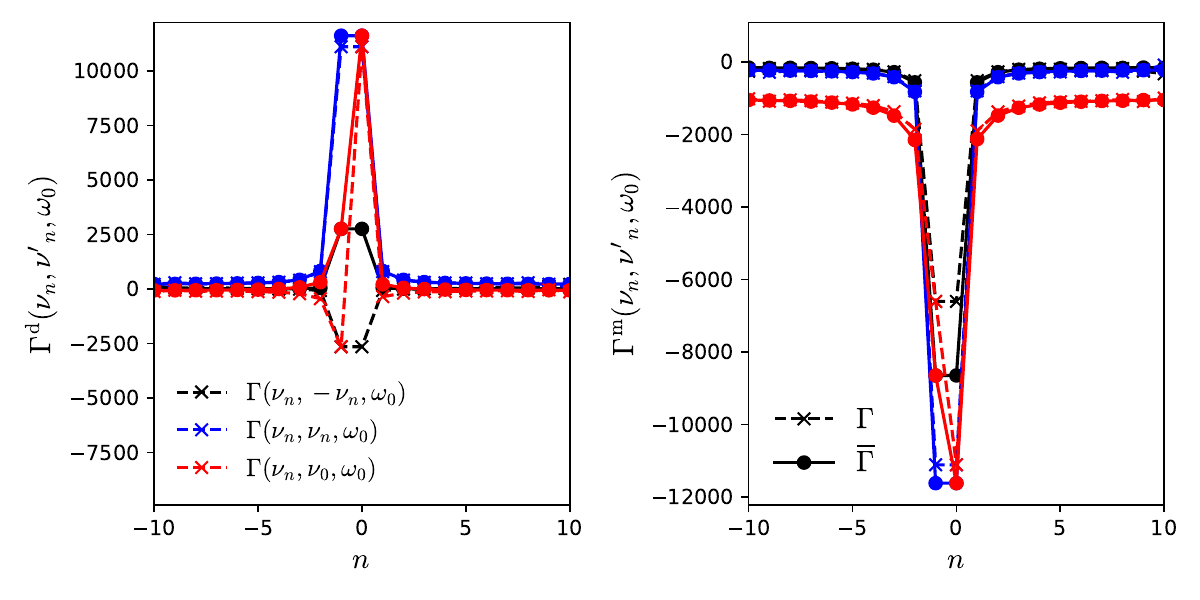} 
\caption{Same vertices as in Fig~\ref{fig:approx_vertex_u2} for the strong coupling regime with parameters $U=12$ and $\beta=2$. The interaction strength is considerably larger than the bandwidth of the lattice system of the converged DMFT. The Figure is taken from Ref.~\cite{vandelli2022quantum}.
\label{fig:approx_vertex_u12}}
\end{figure}

To check whether the introduced partially bosonized approximation reproduces well the exact four-point vertex function, in Figs.~\ref{fig:approx_vertex_u2}--\ref{fig:approx_vertex_u12} (taken from Ref.~\cite{vandelli2022quantum}) we compare the exact vertex $\Gamma_{\nu\nu', \omega=0}$ with its approximated version $\overline{\Gamma}_{\nu\nu', \omega=0}$. 
In particular, in Ref.~\cite{vandelli2022quantum} the impurity problem was taken from the converged DMFT calculation for the half-filled Hubbard model on a square lattice with the nearest-neighbor hopping ${t=1}$ in a broad region of the local interaction $U$.
In all the figures, in the upper panels we show the vertex in the density channel for the exact calculation $\Gamma^d$ (left) and in the partially bosonized approximation $\overline{\Gamma}^d$ (right). In the middle panels, the results for ${\Gamma}^m$ (left) and $\overline{\Gamma}^m$ (right) are displayed for the magnetic channel $m$. 
Finally, the bottom row shows the density (left) and magnetic (right) vertices in the exact case (crosses with dashed lines) and partially bosonized case (dots with full lines) along some symmetry path in the ($\nu$, $\nu'$)-plane. 

In all the figures, there is a very evident noise at large frequencies in the exact vertex function. This is related to the fact that we compute the impurity single-particle Green's function and impurity four-point correlation function using continuous-time QMC. 
In particular, to obtain these plots, the author of in Ref.~\cite{vandelli2022quantum} used the CT-HYB method in its implementation provided in the ALPS software package~\cite{Bauer_2011, HAFERMANN20131280}. 
The four-point vertex was obtained from the four-point correlation function using the Eq.~\eqref{eq:GammaPH} with the conventional choice of the vertex function, putting $B_\nu = g_\nu$, hence the noisy data at large frequencies. 
However, dual calculations based on this vertex functions are still accurate, since the dual Green's function is $\tilde{G}_{\kv\nu} = {G}^{\rm DMFT}_{\kv\nu} - g_\nu$, so that it decays as $\sim \nu^{-2}$ as a function of the Matsubara frequency.

The figures show, that the partially bosonized approximation reproduces the frequency-dependence of the exact vertex rather accurately in all considered regimes.
The results obtained with the approximated $\overline{\Gamma}$ at the two lower values of the interaction ${U=2}$ and ${U=4}$ are in a good agreement with the exact result $\Gamma$ both in the density ($d$) and in the magnetic ($m$) channels. 
At large values of the interaction, $U=8$ and $U=12$, the approximation reproduces the structure of the vertex rather well, but shows a noticeable difference in the value at the lowest Matsubara frequencies along the ${\nu'_n=-\nu_n}$ and ${\nu'_n=\nu_o}$ cuts.
Based on these observations, one can expect that replacing the exact vertex with its partially bosonized approximation may noticeably affect the calculation of observables in the strong-coupling regime (${U \gtrsim 8}$). 
On the other hand, as we demonstrate below, using the approximate vertex drastically reduces the computational cost and enables the application of dual techniques in the multi-orbital framework. 
This is a significant achievement that far outweighs the loss in accuracy when using the approximate vertex, considering the substantial challenges posed by multi-orbital calculations.

\clearpage

\subsection{Partially bosonized dual action}
\label{sec:PBDAction}

The four-point vertex function can be eliminated from the DB action~\eqref{eq:DB_action_app} by generating a counter term corresponding to the partially bosonized approximation~\eqref{eq:Gamma_approx_app}. 
To do so, we add and subtract the following terms: 
\begin{align}
\frac12\sum_{q,\{l\}}\sum_{\varsigma\varsigma'} \varphi^{\varsigma}_{-q,\,l_1l_2} \left[\bar{w}^{-1}_{\omega}\right]^{\varsigma\varsigma'}_{l_1l_2,\, l_3l_4} \varphi^{\varsigma'}_{q,\, l_4l_3} +
\sum_{q,\{l\}}\sum_{\vartheta\vartheta'} \varphi^{*\,\vartheta}_{q,\,l_1l_2} \left[\bar{w}^{-1}_{\omega}\right]^{\vartheta\vartheta'}_{l_1l_2,\, l_3l_4} \varphi^{\vartheta'}_{q,\, l_3l_4}
\label{eq:W_bar_terms_app}
\end{align}
from the dual action~\eqref{eq:DB_action_app}. 
The dual boson action becomes:
\begin{align}
&{\cal \tilde{S}}
= - \Tr_{\sigma,l}\sum_{k} \left\{ \hat{f}^{*}_{k}  \tilde{\varepsilon}^{-1}_{k} \hat{f}^{\phantom{*}}_{k}
- f^{*}_{k} g^{\phantom{*}}_{\nu} f^{\phantom{*}}_{k} \right\} + \tilde{\cal F}[f,\varphi] 
+ \frac12 \Tr_{\varsigma,l} \sum_{q} \varphi^{\phantom{*}}_{-q} \bar{w}^{-1}_{\omega} \varphi^{\phantom{*}}_{q}
+ \Tr_{\vartheta,l} \sum_{q} \varphi^{*}_{q} \bar{w}^{-1}_{\omega} \varphi^{\phantom{*}}_{q}
\notag\\
& - \Tr_{\vartheta,l} \sum_{q} \Bigg\{ \varphi^{*}_{q} \alpha^{-1}_{\omega} \left( \tilde{V}^{-1}_{q} - \chi^{\phantom{*}}_{\omega} + \alpha^{\phantom{*}}_{\omega} \bar{w}^{-1}_{\omega} \alpha^{\phantom{*}}_{\omega} \right) \alpha^{-1}_{\omega} \varphi^{\phantom{*}}_{q}
+ j^{*}_{q} \tilde{V}^{-1}_{q} j^{\phantom{*}}_{q} 
- \varphi^{*}_{q} \alpha^{-1}_{\omega}\tilde{V}^{-1}_{q} j^{\phantom{*}}_{q} - j^{*}_{q} \tilde{V}^{-1}_{q} \alpha^{-1}_{\omega} \varphi^{\phantom{*}}_{q} \Bigg\} \notag\\
& - \Tr_{\varsigma,l} \sum_{q} \Bigg\{ \frac12 \varphi^{\phantom{*}}_{-q} \alpha^{-1}_{\omega} \left( \tilde{V}^{-1}_{q} - \chi^{\phantom{*}}_{\omega} + \alpha^{\phantom{*}}_{\omega} \bar{w}^{-1}_{\omega} \alpha^{\phantom{*}}_{\omega} \right) \alpha^{-1}_{\omega} \varphi^{\phantom{*}}_{q}
+ \frac12 j^{\phantom{*}}_{-q} \tilde{V}^{-1}_{q} j^{\phantom{*}}_{q}
- \varphi^{\phantom{*}}_{-q} \alpha^{-1}_{\omega}\tilde{V}^{-1}_{q} j^{\phantom{*}}_{q} \Bigg\},
\label{eq:dual_sources_app}
\end{align}
where we explicitly restored the terms that contain the bosonic source fields $j^{(*)}$.
Note, that as in the DB approach, the interaction term $\tilde{\cal F}[f,\varphi]$ is truncated at the two-particle level~\eqref{eq:lowestint}.
To shorten the expression, we omit band, spin, and channel indices that can be easily restored using the fact that all multiplications in Eq.~\eqref{eq:dual_sources_app} are performed in the matrix form.
Now, we perform the following Hubbard-Stratonovich transformations:
\begin{align}
&\exp\left\{\frac12\Tr_{\varsigma,l}\sum_{q} \varphi^{\phantom{*}}_{-q} \alpha^{-1}_{\omega} \left[ \tilde{V}^{-1}_{q} - \chi^{\phantom{*}}_{\omega} + \alpha^{\phantom{*}}_{\omega} \bar{w}^{-1}_{\omega} \alpha^{\phantom{*}}_{\omega} \right] \alpha^{-1}_{\omega} \varphi^{\phantom{*}}_{q} \right\} = {\cal D}_{b}
\int D[b^{\varsigma}] \, \times \notag\\
&\times\exp\left\{-\Tr_{\varsigma,l}\sum_{q} \left( \frac12 b^{\phantom{*}}_{-q} \bar{w}^{-1}_{\omega}
\alpha^{\phantom{*}}_{\omega} \left[ \tilde{V}^{-1}_{q} - \chi^{\phantom{*}}_{\omega} + \alpha^{\phantom{*}}_{\omega} \bar{w}^{-1}_{\omega} \alpha^{\phantom{*}}_{\omega} \right]^{-1} \alpha^{\phantom{*}}_{\omega}
\bar{w}^{-1}_{\omega} b^{\phantom{*}}_{q} 
- \varphi^{\phantom{*}}_{-q} \bar{w}^{-1}_{\omega} b^{\phantom{*}}_{q} \right)\right\}, \\
&\exp\left\{\Tr_{\vartheta,l}\sum_{q} \varphi^{*}_{q} \alpha^{-1}_{\omega} \left[ \tilde{V}^{-1}_{q} - \chi^{\phantom{*}}_{\omega} + \alpha^{\phantom{*}}_{\omega} \bar{w}^{-1}_{\omega} \alpha^{\phantom{*}}_{\omega} \right] \alpha^{-1}_{\omega} \varphi^{\phantom{*}}_{q} \right\} = {\cal D}_{b} \int D[b^{\vartheta}] \, \times \notag\\
&\times\exp\left\{-\Tr_{\vartheta,l}\sum_{q} \left( b^{*}_{q} \bar{w}^{-1}_{\omega}
\alpha^{\phantom{*}}_{\omega} \left[ \tilde{V}^{-1}_{q} - \chi^{\phantom{*}}_{\omega} + \alpha^{\phantom{*}}_{\omega} \bar{w}^{-1}_{\omega} \alpha^{\phantom{*}}_{\omega} \right]^{-1} \alpha^{\phantom{*}}_{\omega}
\bar{w}^{-1}_{\omega} b^{\phantom{*}}_{q} 
- \varphi^{*}_{q} \bar{w}^{-1}_{\omega} b^{\phantom{*}}_{q} - b^{*}_{q} \bar{w}^{-1}_{\omega} \varphi^{\phantom{*}}_{q} \right)\right\},
\label{eq:HSb_DT}
\end{align}
where the terms ${\cal D}^{-1}_{b} = \sqrt{{\rm det}\left[\bar{w}\alpha^{-1} \left[\tilde{V}^{-1} - \chi + \alpha \bar{w}^{-1} \alpha\right]\alpha^{-1} \bar{w}\right]}$ can again be neglected, because they also do not affect the calculation of correlation functions.
The dual action becomes:
\begin{align}
&\tilde{\cal S}'
= - \Tr_{\sigma,l}\sum_{k} \left\{ \hat{f}^{*}_{k}  \tilde{\varepsilon}^{-1}_{k} \hat{f}^{\phantom{*}}_{k}
- f^{*}_{k} g^{\phantom{*}}_{\nu} f^{\phantom{*}}_{k} \right\} + \tilde{\cal F}[f,\varphi] \notag\\
&+ \Tr_{\vartheta,l}\sum_{q} b^{*}_{q} \bar{w}^{-1}_{\omega}
\alpha^{\phantom{*}}_{\omega} \left[ \tilde{V}^{-1}_{q} - \chi^{\phantom{*}}_{\omega} + \alpha^{\phantom{*}}_{\omega} \bar{w}^{-1}_{\omega} \alpha^{\phantom{*}}_{\omega} \right]^{-1} \alpha^{\phantom{*}}_{\omega}
\bar{w}^{-1}_{\omega} b^{\phantom{*}}_{q} 
- \Tr_{\vartheta,l} \sum_{q} j^{*}_{q} \tilde{V}^{-1}_{q} j^{\phantom{*}}_{q} \notag\\
&+ \frac12\Tr_{\varsigma,l}\sum_{q} b^{\phantom{*}}_{-q} \bar{w}^{-1}_{\omega}
\alpha^{\phantom{*}}_{\omega} \left[ \tilde{V}^{-1}_{q} - \chi^{\phantom{*}}_{\omega} + \alpha^{\phantom{*}}_{\omega} \bar{w}^{-1}_{\omega} \alpha^{\phantom{*}}_{\omega} \right]^{-1} \alpha^{\phantom{*}}_{\omega}
\bar{w}^{-1}_{\omega} b^{\phantom{*}}_{q}
- \frac12\Tr_{\varsigma,l} \sum_{q} j^{\phantom{*}}_{-q} \tilde{V}^{-1}_{q} j^{\phantom{*}}_{q} \notag\\
& + \Tr_{\vartheta,l} \sum_{q} \varphi^{*}_{q} \bar{w}^{-1}_{\omega} \varphi^{\phantom{*}}_{q}
- \Tr_{\vartheta,l}\sum_{q} \Bigg\{ \varphi^{*}_{q} \bar{w}^{-1}_{\omega} \left(b^{\phantom{*}}_{q} - \bar{w}^{\phantom{*}}_{\omega} \alpha^{-1}_{\omega}\tilde{V}^{-1}_{q} j^{\phantom{*}}_{q} \right) + \left(b^{*}_{q} - j^{*}_{q} \tilde{V}^{-1}_{q}\alpha^{-1}_{\omega}  \bar{w}^{\phantom{*}}_{\omega} \right) \bar{w}^{-1}_{\omega} \varphi^{\phantom{*}}_{q} \Bigg\} \notag\\
&+ \frac12 \Tr_{\varsigma,l} \sum_{q} \varphi^{\phantom{*}}_{-q} \bar{w}^{-1}_{\omega} \varphi^{\phantom{*}}_{q}
- \Tr_{\varsigma,l}\sum_{q} \varphi^{\phantom{*}}_{-q} \bar{w}^{-1}_{\omega} \left( b^{\phantom{*}}_{q} - \bar{w}^{\phantom{*}}_{\omega} \alpha^{-1}_{\omega}\tilde{V}^{-1}_{q} j^{\phantom{*}}_{q}\right).
\label{eq:prelastS_app}
\end{align}
We shift bosonic variables as ${b^{(*)} \to \hat{b}^{(*)} = b^{(*)} + \bar{w} \alpha^{-1}\tilde{V}^{-1} j^{(*)}}$ to decouple the sources $j^{(*)}$ from the dual bosonic fields $\varphi^{(*)}$. 
After that the fields $\varphi^{(*)}$ can be integrated out as:
\begin{align}
&\int D[\varphi^{\varsigma}]\,\exp\left\{-\frac12 \Tr_{\varsigma,l} \sum_{q} \varphi^{\phantom{*}}_{-q} \bar{w}^{-1}_{\omega} \varphi^{\phantom{*}}_{q} 
+ \Tr_{\varsigma,l}\sum_{q} \left(b^{\phantom{*}}_{-q} \bar{w}^{-1}_{\omega} 
- \sum_{k,\sigma\sigma'}f^{*}_{k\sigma} f^{\phantom{*}}_{k+q,\sigma'}\Lambda^{\sigma\sigma'}_{\nu\omega} \right) \varphi^{\phantom{*}}_{q} \right\} = \notag\\
&{\cal Z}_{\varphi} \, \exp\Bigg\{ \frac12 \Tr_{\varsigma,l} \sum_{q} b^{\phantom{*}}_{-q} \bar{w}^{-1}_{\omega} b^{\phantom{*}}_{q} 
- \Tr_{\varsigma,l}\sum_{q,k}\sum_{\sigma\sigma'} \Lambda^{\sigma\sigma'}_{\nu\omega} \, f^{*}_{k\sigma} f^{\phantom{*}}_{k+q,\sigma'} b^{\phantom{*}}_{q} \notag \\
&\hspace{1.4cm}+ \frac12 \Tr_{\varsigma,l}\sum_{q,\{\sigma\}}\sum_{k,k'} \Lambda^{\sigma_1\sigma_2}_{\nu\omega} \bar{w}^{\phantom{*}}_{\omega} \Lambda^{\sigma_4\sigma_3}_{\nu'+\omega,-\omega} f^{*}_{k\sigma_1} f^{\phantom{*}}_{k+q,\sigma_2} f^{*}_{k'+q,\sigma_4} f^{\phantom{*}}_{k'\sigma_3} \Bigg\},
\label{eq:integrationphiPH}\\
&\int D[\varphi^{*\,\vartheta}, \varphi^{\vartheta}]\,\exp\Bigg\{-\Tr_{\vartheta,l} \sum_{q} \varphi^{*}_{q} \bar{w}^{-1}_{\omega} \varphi^{\phantom{*}}_{q} 
+ \Tr_{\vartheta,l}\sum_{q} \left(b^{*}_{q} \bar{w}^{-1}_{\omega} \varphi^{\phantom{*}}_{q} + \varphi^{*}_{q} \bar{w}^{-1}_{\omega} b^{\phantom{*}}_{q} \right) \notag\\
&\hspace{2.5cm}- \frac12\Tr_{\vartheta,l}\sum_{q,k}\sum_{\sigma\sigma'} \left(f^{*}_{k\sigma} f^{*}_{q-k,\sigma'}\Lambda^{\sigma\sigma'}_{\nu\omega} \varphi^{\phantom{*}}_{q} + \varphi^{*}_{q} \Lambda^{*\,\sigma\sigma'}_{\nu\omega} f^{\phantom{*}}_{q-k,\sigma'} f^{\phantom{*}}_{k\sigma}\right) \Bigg\} = \notag\\
&{\cal Z}_{\varphi} \, \exp\Bigg\{ \Tr_{\vartheta,l} \sum_{q} b^{*}_{q} \bar{w}^{-1}_{\omega} b^{\phantom{*}}_{q} 
- \frac12\Tr_{\vartheta,l}\sum_{q,k}\sum_{\sigma\sigma'} \left(f^{*}_{k\sigma} f^{*}_{q-k,\sigma'}\Lambda^{\sigma\sigma'}_{\nu\omega} b^{\phantom{*}}_{q} + b^{*}_{q} \Lambda^{*\,\sigma\sigma'}_{\nu\omega} f^{\phantom{*}}_{q-k,\sigma'} f^{\phantom{*}}_{k\sigma}\right) \notag \\
&\hspace{1.4cm}+ \frac14\Tr_{\vartheta,l}\sum_{q,\{\sigma\}}\sum_{k,k'} \Lambda^{\sigma_1\sigma_2}_{\nu\omega} \bar{w}^{\phantom{*}}_{\omega} \Lambda^{\sigma_3\sigma_4}_{\nu'\omega} f^{*}_{k\sigma_1} f^{*}_{q-k,\sigma_2} f^{\phantom{*}}_{q-k',\sigma_4} f^{\phantom{*}}_{k'\sigma_3} \Bigg\},
\label{eq:integrationphiPP}
\end{align}
where ${\cal Z}_{\varphi}$ is a partition function of the Gaussian part of the bosonic action. 
Remarkably, the quartic terms in the last line of Eqs.~\eqref{eq:integrationphiPH} and~\eqref{eq:integrationphiPP} identically coincide with the partially bosonized approximation for the four-point vertex function~\eqref{eq:Gamma_approx_app}.
Importantly, these terms appear with the opposite sign to the exact four-point vertex $\Gamma$ and therefore approximately cancel the fermion-fermion ($\Gamma$) part of the interaction~\eqref{eq:lowestint} from the dual action.
As the result, the problem reduces to a partially bosonized dual action:
\begin{align}
&\tilde{\cal S}_{fb}
= - \Tr_{\sigma,l}\sum_{k} \left\{ \hat{f}^{*}_{k}  \tilde{\varepsilon}^{-1}_{k} \hat{f}^{\phantom{*}}_{k}
- f^{*}_{k} g^{\phantom{*}}_{\nu} f^{\phantom{*}}_{k} \right\} \notag\\
&+ \frac12\Tr_{\varsigma,l}\sum_{q} \hat{b}^{\phantom{*}}_{-q} \bar{w}^{-1}_{\omega}
\alpha^{\phantom{*}}_{\omega} \left[ \tilde{V}^{-1}_{q} - \chi^{\phantom{*}}_{\omega} + \alpha^{\phantom{*}}_{\omega} \bar{w}^{-1}_{\omega} \alpha^{\phantom{*}}_{\omega} \right]^{-1} \alpha^{\phantom{*}}_{\omega}
\bar{w}^{-1}_{\omega} \hat{b}^{\phantom{*}}_{q}
- \frac12\Tr_{\varsigma,l} \sum_{q} j^{\phantom{*}}_{-q} \tilde{V}^{-1}_{q} j^{\phantom{*}}_{q} \notag\\
&-\frac12 \Tr_{\varsigma,l} \sum_{q} b^{\phantom{*}}_{-q} \bar{w}^{-1}_{\omega} b^{\phantom{*}}_{q} 
+ \Tr_{\varsigma,l}\sum_{q,k}\sum_{\sigma\sigma'} \Lambda^{\sigma\sigma'}_{\nu\omega} \, f^{*}_{k\sigma} f^{\phantom{*}}_{k+q,\sigma'} b^{\phantom{*}}_{q} \notag \\
&+ \Tr_{\vartheta,l}\sum_{q} \hat{b}^{*}_{q} \bar{w}^{-1}_{\omega}
\alpha^{\phantom{*}}_{\omega} \left[ \tilde{V}^{-1}_{q} - \chi^{\phantom{*}}_{\omega} + \alpha^{\phantom{*}}_{\omega} \bar{w}^{-1}_{\omega} \alpha^{\phantom{*}}_{\omega} \right]^{-1} \alpha^{\phantom{*}}_{\omega}
\bar{w}^{-1}_{\omega} \hat{b}^{\phantom{*}}_{q} 
- \Tr_{\vartheta,l} \sum_{q} j^{*}_{q} \tilde{V}^{-1}_{q} j^{\phantom{*}}_{q} \notag\\
&-\Tr_{\vartheta,l} \sum_{q} b^{*}_{q} \bar{w}^{-1}_{\omega} b^{\phantom{*}}_{q} 
+ \frac12\Tr_{\vartheta,l}\sum_{q,k}\sum_{\sigma\sigma'} \left(f^{*}_{k\sigma} f^{*}_{q-k,\sigma'}\Lambda^{\sigma\sigma'}_{\nu\omega} b^{\phantom{*}}_{q} + b^{*}_{q} \Lambda^{*\,\sigma\sigma'}_{\nu\omega} f^{\phantom{*}}_{q-k,\sigma'} f^{\phantom{*}}_{k\sigma}\right)
\label{eq:fbaction_app}
\end{align}
that, upon neglecting fermionic $\eta^{(*)}$ and bosonic $j^{(*)}$ sources, takes the simple form:
\begin{align}
{\cal S}_{fb} = &- \sum_{k,\{l\}}\sum_{\sigma\sigma'} f^{*}_{k\sigma{}l} \left[\tilde{\cal G}^{-1}_{k}\right]^{\sigma\sigma'}_{ll'} f^{\phantom{*}}_{k\sigma'l'} 
- \frac12\sum_{q, \{l\}}\sum_{\varsigma\varsigma'}  b^{\varsigma}_{-q,\, l_1 l_2}
\left[\tilde{\cal W}^{-1}_{q}\right]^{\varsigma\varsigma'}_{l_1 l_2,\, l_3 l_4} b^{\varsigma'}_{q,\, l_4 l_3} \notag\\
&- \sum_{q, \{l\}}\sum_{\vartheta\vartheta'}  b^{*\,\vartheta}_{q,\, l_1 l_2}
\left[\tilde{\cal W}^{-1}_{q}\right]^{\vartheta\vartheta'}_{l_1 l_2,\, l_3 l_4} b^{\vartheta'}_{q,\, l_3 l_4} + \mathcal{F}\,[f, b].
\label{eq:fbaction}
\end{align}
The bare fermionic Green's function for this action remains the same as in the DB approach~\eqref{eq:bare_dual_G}.
The bare dual bosonic propagator has the following form:
\begin{align}
\tilde{\cal W}^{r_1r_2}_{q,\,l_1l_2,\,l_3l_4} = \sum_{\{r'\},\{l'\}}\alpha^{r_1r'_1}_{\omega,\,l_1l_2,\,l'_1l'_2} \left[ \left( \tilde{V}^{-1}_{q} - \chi_{\omega} \right)^{-1} \right]^{r'_1r'_2}_{l'_1l'_2,\,l'_3l'_4} \alpha^{r'_2r_2}_{\omega,\,l'_3l'_4,\,l_3l_4} + \bar{w}^{r_1r_2}_{\omega,\,l_1l_2,\,l_3l_4}\,.
\end{align}
Substituting the explicit expression~\eqref{eq:w_bar-u_app} for $\bar{w}^{rr'}_{\omega}$ and using the bosonic scaling factor $\alpha$ in the form of Eq.~\eqref{eq:alpha_app} leads for the final form of the bare bosonic propagator:
\begin{align}
\tilde{\cal W}^{rr'}_{q,\, l_1 l_2,\, l_3 l_4} &= W^{{\rm EDMFT}\,rr'}_{q,\, l_1 l_2,\, l_3 l_4} - \bar{u}^{r}_{l_1 l_2,\, l_3 l_4}\delta_{rr'}\,.
\label{eq:bareWgeneral}
\end{align}
The $\bar{u}^{r}_{l_1 l_2,\, l_3 l_4}$ terms, that prevent the double counting of the bare interaction between different channels, are defined in Eq.~\eqref{eq:U_correction}.
The renormalized interaction of EDMFT, $W^{{\rm EDMFT}}_{q}$, is defined in Eq.~\eqref{eq:W_EDMFT}.
The bare interaction in the channel representation, $U^{r}$, is defined in Eqs.~\eqref{eq:Ud_app}--\eqref{eq:Ut_app}.
Therefore, the bosonic fields $b^{(*)}$ in the partially bosonized theory~\eqref{eq:fbaction} are more physical than the dual boson variables $\varphi^{(*)}$ in the DB action~\eqref{eq:DB_action}, because the propagator of the $b^{(*)}$ fields corresponds to the physical renormalized interaction. 

Since the four-point vertex function was eliminated from the dual theory, the interacting term of the partially bosonized dual action~\eqref{eq:fbaction} simplifies to:
\begin{align}
\mathcal{F}[f, b] = &\sum_{q,\{k\}}\sum_{\{\nu\}, \{\sigma\}}\sum_{\{l\}, \varsigma/\vartheta}
\Bigg\{ \Lambda^{\sigma\sigma'\varsigma}_{\nu\omega,\, l_1,\, l_2,\, l_3 l_4} \, f^{*}_{k\sigma{}l_1} f^{\phantom{*}}_{k+q,\sigma', l_2} b^{\varsigma}_{q,\, l_4 l_3} \notag \\ 
&+ \frac12\left(\Lambda^{\sigma\sigma'\vartheta}_{\nu\omega,\,l_1,\, l_2,\, l_3 l_4} \, f^{*}_{k\sigma{}l_1} f^{*}_{q-k,\sigma', l_2} b^{\vartheta}_{q,\, l_3 l_4} 
+ \Lambda^{*\,\sigma\sigma'\vartheta}_{\nu\omega,\,l_1,\, l_2,\, l_3 l_4} \, b^{*\,\vartheta}_{q,\, l_3 l_4} f^{\phantom{*}}_{q-k,\sigma', l_2} f^{\phantom{*}}_{k\sigma{}l_1}  
\right) \Bigg\}
\label{eq:fbinteraction}
\end{align}
and contains only the three-point vertex function $\Lambda^{\hspace{-0.05cm}(*)}_{\nu\omega}$ of the reference system.
The three-point vertex remains the same as in the DB theory, as defined in Eqs.~\eqref{eq:Vertex_PH_app}--\eqref{eq:Vertex_PP_app}.

\subsection{Impact of partially bosonized collective fluctuations on electronic degrees of freedom}
\label{sec:Impact}

The action~\eqref{eq:fbaction} of the partially bosonized dual theory (PBDT) contains bosonic fluctuations in the charge, spin, and pairing channel. 
To evaluate the importance of different fluctuations, we investigate their impact to the electronic self-energy.
For simplicity, we restrict ourselves to a paramagnetic case as in Ref.~\cite{PhysRevB.103.245123}.
We note, that the action~\eqref{eq:fbaction} is quadratic in bosonic fields and thus can be solved numerically exactly using the DiagMC approach (DiagMC@PBDT) similar to the one applied to the DB action~\eqref{eq:DB_action} (see Section~\ref{sec:DiagMC}).
To do so, we integrate out the bosonic fields, which leads to a modified dual fermion action~\eqref{finalaction} with the four-point vertex that is similar to the one defined by Eq.~\eqref{DBvertex}:
\begin{align}
\overline{\Gamma}^{\rm d}_{kk'q} &\simeq 2M^{\rm d}_{\nu\nu'q} - M^{\rm d}_{\nu,\nu+\omega,k'-k} - 3M^{\rm m}_{\nu,\nu+\omega,k'-k} + M^{\rm s}_{\nu,\nu',q+k+k'}\,, \notag\\
\overline{\Gamma}^{\rm m}_{kk'q} &\simeq 2M^{\rm m}_{\nu\nu'q} + M^{\rm m}_{\nu,\nu+\omega,k'-k} - M^{\rm d}_{\nu,\nu+\omega,k'-k} - M^{\rm s}_{\nu,\nu',q+k+k'}\,, \notag\\
\overline{\Gamma}^{\rm s}_{kk'q} &\simeq M^{\rm s}_{\nu\nu'q} + \frac12 \left(M^{\rm d}_{\nu,\nu',q-k-k'} + M^{\rm d}_{\nu,\omega-\nu',k'-k}\right) - \frac32 \left(M^{\rm m}_{\nu,\nu',q-k-k'} + M^{\rm m}_{\nu,\omega-\nu',k'-k}\right),
\label{eq:PBDT_vertex}
\end{align}
where ${M_{\nu,\nu',\omega}^{\vartheta} = \Lambda^{\varsigma}_{\nu,\omega} \tilde{\cal W}^{\varsigma}_{q}\Lambda^{\varsigma}_{\nu'+\omega, -\omega}}$~\cite{PhysRevB.103.245123}.

To consistently investigate the effect of different contributions to the four-point vertex of the reference problem on the electronic self-energy, we consider a half-filled single-band Hubbard model on a square lattice with the neareast-neighbor hopping amplitude ${t=1}$ and different values of the on-site Coulomb potential $U$. 
All non-local interactions are set to zero $V^{\vartheta}_{\qv}=0$.
In this case, the bare bosonic propagator~\eqref{eq:bareWgeneral} coinsides with $\bar{w}^{r}_{\omega}$ and the DB action~\eqref{eq:DB_action} reduced to the DF problem.
The calculations are performed on the basis of the converged DMFT solution of the lattice problem~\eqref{eq:actionlatt}.
The corresponding single-site impurity problem of DMFT~\eqref{eq:actionimp_app} is solved numerically exactly using the open source CT-HYB solver~\cite{HAFERMANN20131280, PhysRevB.89.235128} based on ALPS libraries~\cite{Bauer_2011}.

\begin{figure}[t!]
\centering
\includegraphics[width=0.95\linewidth]{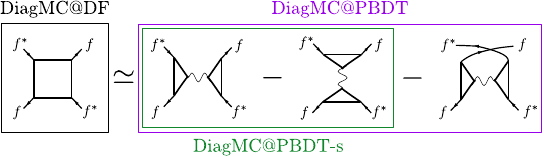}
\caption{\label{fig:DiagMC_vertices} Summary of the approximations for the four-point vertex function of the reference problem used in different DiagMC calculations.}
\end{figure}

We consider different levels of approximation for the exact four-point vertex function $\Gamma$ of the reference problem.
The contribution to the self-energy that stems from the irreducible part of $\Gamma$, which is not accounted for by the partially bosonized approximation~\eqref{eq:Gamma_approx_app}, can be identified by comparing the exact DiagMC@DF solution of the dual problem~\eqref{eq:DB_action} with the result of the DiagMC@PBDT calculation with approxmiate vertex.
The next level of approximation that allows to observe the effect of collective fluctuations in the singlet channel can be achieved by performing DiagMC@PBDT-s calculations using the partially bosonized vertex~\eqref{eq:PBDT_vertex} without singlet contributions $M^{\rm s}$.
There approximations are summarized in Fig.~\ref{fig:DiagMC_vertices}.

\begin{figure}[t!]
\centering
\includegraphics[width=1\linewidth]{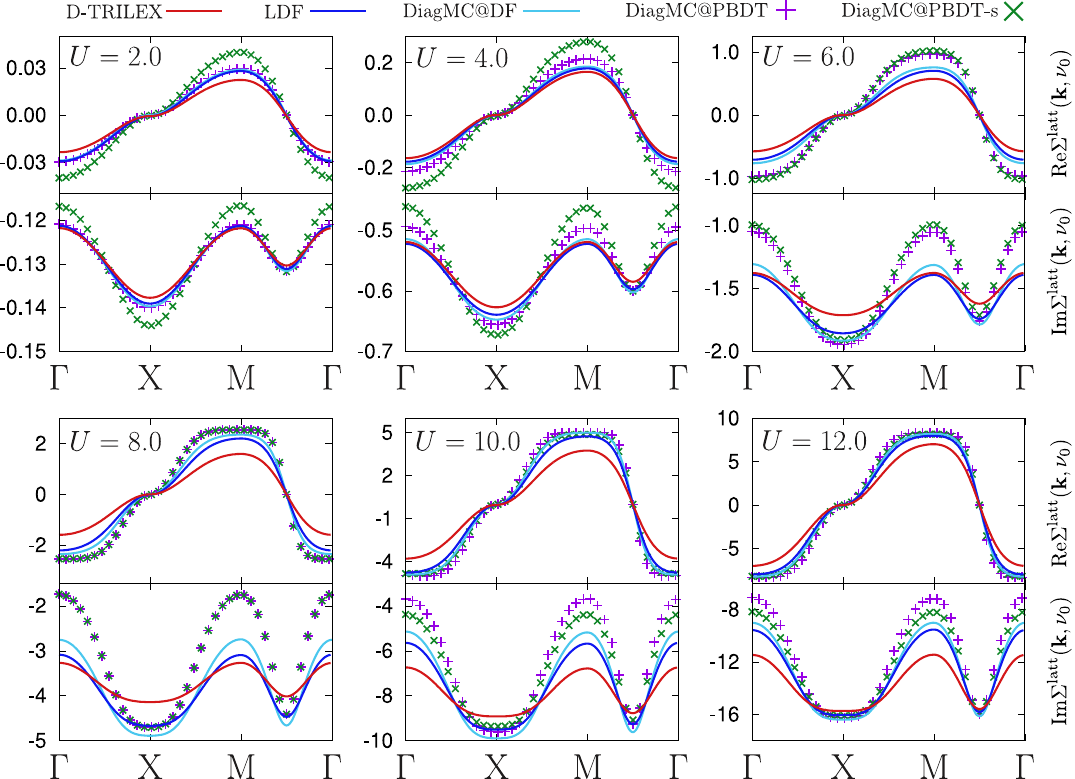}
\caption{\label{fig:compare_sigma} The lattice self-energy $\Sigma^{\rm latt}_{{\bf k},\nu_0}$ calculated at the lowest Matsubara frequency $\nu_0=\pi/\beta$ for ${\beta=2}$ along the high-symmetry path in ${\bf k}$-space. Upper and lower part of each panel corresponds to real and imaginary part of the self energy, respectively. Results are obtained using $\text{D-TRILEX}$ (red line), LDF (dark blue line), $\text{DiagMC@DF}$ (light blue line), DiagMC@PBDB (purple crosses), and DiagMC@PBDB-s (green crosses). The Figure is taken from Ref.~\cite{PhysRevB.103.245123}.}
\vspace{-1cm}
\end{figure}
\begin{figure}[b!]
\centering
\includegraphics[width=0.55\linewidth]{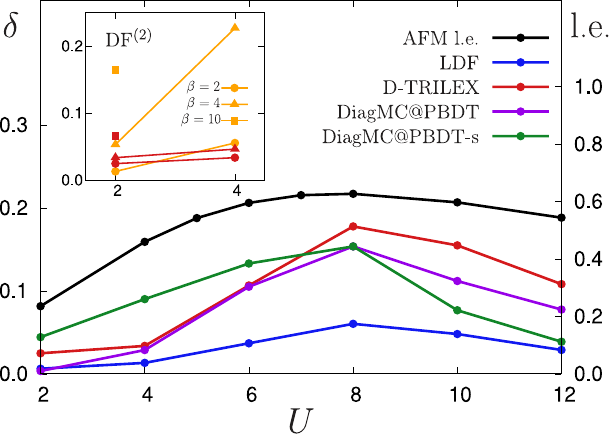}
\caption{\label{fig:Ev_and_deviation} The normalized deviation $\delta$ calculated for the LDF (blue), $\text{D-TRILEX}$ (red), $\text{DiagMC@PBDT}$ (purple), and $\text{DiagMC@PBDT-s}$ (green) methods with respect to the reference Diag@DF result. The black line shows the leading eigenvalue (l.e.) of AFM fluctuations. The vertical left axis shows the scale for the normalized deviation, while the vertical right axis displays l.e. The inset compares $\delta$ obtained for $\text{D-TRILEX}$ and second-order DF$^{(2)}$ for $\beta=2$ (circles), $\beta=4$ (triangles), and $\beta=10$ (squares). The Figure is taken from Ref.~\cite{PhysRevB.103.245123}.}
\end{figure}

Fig.~\ref{fig:compare_sigma} shows the lattice self-energy $\Sigma^{\rm latt}_{{\bf k},\nu_0}$ calculated at the lowest Matsubara frequency ${\nu_0=\pi/\beta}$ along the high symmetry path that connects $\Gamma=(0,0)$, ${\rm X}=(0, \pi)$, and ${\rm M}=(\pi, \pi)$ points in momentum space using the above-mentioned approaches.
The result is obtained at the inverse temperature ${\beta=2}$ for different values of the local Coulomb interaction $U$.
Let us first consider the effect of the irreducible part of the four-point vertex comparing the self-energy of $\text{DiagMC@DF}$ (light blue line) and $\text{DiagMC@PBDT}$ (purple crosses) approaches.
We find that at ${U=2}$ both methods produce identical results, which means that in a weakly-correlated regime the irreducible contributions to the vertex do not affect the self-energy.
Upon increasing the local interaction, the discrepancy between these two methods also increases and is noticeable the most in the strongly-correlated regime at ${U=8}$, which is equal to the bandwidth.
After that, at very large interactions ${U=10}$ and ${U=12}$ the real part of the $\text{DiagMC@PBDT}$ self-energy again nearly coincides with the one of the $\text{DiagMC@DF}$ approach.
The agreement in the imaginary part of the self-energy also improves, but the discrepancy between these two methods remains noticeable.

To quantify the difference of the given self-energy from the reference $\text{DiagMC@DF}$ result we calculate the following normalized deviation: 
\begin{align}
\delta = \sum_{\bf k}\left|\frac{\Sigma^{\rm ref}_{{\bf k}, \nu_0} - \Sigma^{\phantom{f}}_{{\bf k}, \nu_0}}{\Sigma^{\rm ref}_{{\bf k}, \nu_0}}\right|.
\label{eq:norm_diff}
\end{align}
A similar quantity but for only one ${\bf k}$-point was introduced in Eq.~\eqref{eq:Delta_LDB} in order to benchmark the ladder DB approximation.
The corresponding result for all considered approaches is presented in Fig.~\ref{fig:Ev_and_deviation}.
We find that the normalized deviation of the $\text{DiagMC@PBDT}$ method reaches its maximum value ${\delta = 15\%}$ at ${U=8}$.  
As has been pointed out in Ref.~\cite{PhysRevB.100.205115}, the irreducible part can be excluded from the four-point vertex by the specific form~\eqref{eq:Ud_app}--\eqref{eq:Ut_app} for the bare interaction $U^{r}$ only in the ladder approximation.
In the strongly-correlated regime non-ladder diagrams become important~\cite{PhysRevB.94.035102, PhysRevB.96.035152, PhysRevB.102.195109}, which is also confirmed by the increase of the normalized deviation of the LDF approach (blue line in Fig.~\ref{fig:Ev_and_deviation}).
Consequently, the contribution of the irreducible part of the vertex to the electronic self-energy also becomes noticeable.  
We would like to emphasize that by the strength of electronic correlations we mean not only the strength of the interaction, but also the proximity of the system to an instability.
The latter can be estimated by the leading eigenvalue (l.e.) of the Bethe-Salpeter equation of the LDF theory~\cite{PhysRevLett.102.206401, PhysRevB.90.235132} (black line in Fig.~\ref{fig:Ev_and_deviation}), which in our case indicates the strength of antiferromagnetic (AFM) fluctuations.

In the next step we investigate the effect of an additional exclusion of all singlet contributions from the partially bosonized four-point vertex~\eqref{eq:PBDT_vertex}.
At small (${U=2}$) and moderate (${U=4}$) interactions this immediately leads to a large discrepancy between $\text{DiagMC@PBDT-s}$ (green crosses) and $\text{DiagMC@PBDT}$ (purple crosses) results for the self-energy presented in Fig.~\ref{fig:compare_sigma}.
In addition, from Fig.~\ref{fig:Ev_and_deviation} we find that for these values of the interaction the $\text{DiagMC@PBDT-s}$ strongly differs from the reference result, while the $\text{DiagMC@PBDT}$ performs reasonably well.
Increasing the interaction strength to ${U=6}$, makes the discrepancy between $\text{DiagMC@PBDT}$ and $\text{DiagMC@PBDT-s}$ results rapidly decrease, and at ${U=8}$ both methods produce identical results.
Remarkably, for $U=10$ and $U=12$ the $\text{DiagMC@PBDT-s}$ method shows the best agreement with the $\text{DiagMC@DF}$ result among all considered DiagMC-based approximations.
This result suggests that in the regime of very large interactions contributions to the self-energy that stem from the irreducible and singlet parts of the renormalized four-point vertex, which are not considered in the $\text{DiagMC@PBDT-s}$ theory, nearly cancel each other.

At a first glance, the observation that singlet fluctuations are found to be more important in weakly- and moderately-correlated regimes is in a contradiction with the statement that particle-particle fluctuations are believed to be negligibly small at standard fillings~\cite{Pao94}.
To explain this result, let us first note that at ${U \leq{} 4}$ the LDF method is in a very good agreement with the $\text{DiagMC@DF}$ theory.
Therefore, in the weakly- and moderately-correlated regime ladder diagrams provide the most important contribution to the self-energy.
This fact allows for a direct comparison of the self-energies produced by ladder DF and $\text{D-TRILEX}$ methods with the result of DiagMC@ methods that account for all diagrammatic contributions.
Note however, that all DiagMC-based schemes tend to overestimate the reference result, while LDB approache underestimates it. 
Therefore, the normalized deviation presented in Fig.~\ref{fig:Ev_and_deviation} should be compared cautiously.
Fig.~\ref{fig:compare_sigma} shows that LDF and $\text{DiagMC@PBDT}$ self-energies obtained at ${U=2}$ and ${U=4}$ are very close to the reference result.
Compared to the $\text{DiagMC@PBDT}$, the LDF approach neglects all transverse particle-hole and particle-particle modes beyond the ladder approximation.
Keeping in mind that for these interaction strengths the exclusion of only singlet fluctuations leads to a large overestimation of the self-energy, we can conclude that transverse particle-hole and particle-particle fluctuations partially screen each other. 
This means that the exclusion of both types of vertical insertions in diagrams, as it is done in the ladder theories, turns out to be a good approximation in the weakly- and moderately-correlated regime.
On the other hand, excluding only one channel leaves the other channel unscreened, which results in a large contribution to the self-energy. 

Remarkably, the normalized deviation for all considered approximations shown in Fig.~\ref{fig:Ev_and_deviation} resembles the behavior of l.e. in the magnetic channel (black line).
For instance, the LDF methods show the largest discrepancy with the $\text{DiagMC@DF}$ result exactly in the region where the l.e. is maximal.
As has been pointed out in Ref.~\cite{PhysRevB.102.224423}, approaching an instability leads to collective fluctuations becoming strongly anharmonic, which cannot be captured by simple diagrammatic theories.
Consequently, in this regime transverse momentum-dependent fluctuations are expected to be important.

To conclude, we have found that irreducible contributions that are not accounted for by the partially bosonized vertex function can be excluded from the theory in a broad range of physical parameters. 
Indeed, they can be completely eliminated in the ladder approximation by a special choice~\eqref{eq:Ud_app}--\eqref{eq:Ut_app} of the bare local interaction in different channels.
In a weakly-interacting regime, the remaining non-ladder contributions have only a minor effect on the electronic self-energy, and at large interactions these contributions are nearly cancelled by transverse singlet fluctuations.
In turn, these transverse singlet modes partially cancel transverse particle-hole fluctuations in weakly- and moderately-interacting regimes.
Finally, longitudinal singlet bosonic modes have been found to be negligibly small in all considered cases.
All these results confirm that in a broad regime of physical parameters the leading contribution to the self-energy is given by the longitudinal particle-hole bosonic modes.
This important statement allows for a drastic simplification of the diagrammatic expansion, which implies a huge reduction of computational efforts.

\newpage
\section{\mbox{D-TRILEX} approach}
\label{sec:DTRILEX}

In this Section, we introduce the dual triply irreducible local expansion (\mbox{D-TRILEX}) approach~\cite{PhysRevB.100.205115, PhysRevB.103.245123, 10.21468/SciPostPhys.13.2.036}.
This diagrammatic approximation was inspired by the TRILEX~\cite{PhysRevB.92.115109, PhysRevB.93.235124} method, which allows for a simultaneous treatment of charge and spin fluctuations in a much simpler way than it is done in the D$\Gamma$A~\cite{PhysRevB.75.045118, PhysRevB.80.075104, PhysRevB.95.115107, doi:10.7566/JPSJ.87.041004, PhysRevB.103.035120}, DF~\cite{PhysRevB.77.033101}, and DB~\cite{Rubtsov20121320, PhysRevB.90.235135, PhysRevB.93.045107} theories.
\mbox{D-TRILEX} considers the simplest set of diagrams that can be constructed from the partially bosonized dual action~\eqref{eq:fbaction}, in a very similar fashion to the $GW$ formalism~\cite{GW1, GW2, GW3}.
The method is formulated in the dual space and therefore shares all advantages of the dual theories --
the theory is perturbative in both weak- (itinerant) and strong-coupling (localized) regimes, and avoids double-counting of correlations effects by integrating our the reference problem.
The diagrammatic structure of \mbox{D-TRILEX} is similar to that of $GW$+EDMFT~\cite{PhysRevLett.90.086402} and TRILEX, preserving consistency between single- and two-particle correlations.
However, in contrast to these methods, \mbox{D-TRILEX} enables a simultaneous and unambiguous treatment of the leading collective electronic fluctuations in different (charge, spin, orbital, etc.) channels without suffering from the Fierz ambiguity problem~\cite{PhysRevB.65.245118, Borejsza_2003, PhysRevD.68.025020, PhysRevB.70.125111, Bartosch_2009}.
This advantage arises because, in \mbox{D-TRILEX}, the bare interactions in the different bosonic channels are determined not by decoupling the local Coulomb interaction, but by using the most accurate partially bosonized approximation of the four-point vertex (see Section~\ref{sec:PB_approx}).
In addition, unlike more complex D$\Gamma$A, DF, and DB methods that use the four-point vertex corrections, the diagrammatic expansion in \mbox{D-TRILEX} approach is formulated in terms of the exact local three-point vertex functions only. 
The same vertex is also used in the TRILEX method, but there it is introduced only at one side of the $GW$-like diagrams for the self-energy and polarization operator. 
Instead, having symmetric vertex corrections at both sides of the diagrams allows \mbox{D-TRILEX} to preserve a correct orbital structure of the single- and two-particle quantities. 
Furthermore, by eliminating the more complex four-point vertices from the theory via a path-integral transformation of bosonic variables (see Section~\ref{sec:PBDT}), the cost of the numerical calculations is drastically diminished.\\

\mbox{D-TRILEX} has several decisive advantages over other DMFT extensions:
\begin{itemize}
\setlength\itemsep{0pt}
\item[$\bullet$] The self-energy and polarization operator in \mbox{D-TRILEX} depend on a single momentum and frequency. 
Calculating these quantities does not require solving the Bethe-Salpeter equation in momentum and frequency space. 
This is a key advantage that enables efficient calculations within multi-orbital~\cite{PhysRevLett.127.207205, PhysRevResearch.5.L022016, PhysRevLett.129.096404, PhysRevLett.132.226501, stepanov2023charge, Ruthenates} and time-dependent~\cite{vglv-2rmv, Conductivity} frameworks.
\item[$\bullet$] The method has been successfully benchmarked against much more advanced and computationally demanding approaches, demonstrating its great performance for a broad range of model parameters, from small-scale models to multi-orbital lattice systems~\cite{PhysRevB.103.245123, 10.21468/SciPostPhys.13.2.036}.
\item[$\bullet$] \mbox{D-TRILEX} is formulated with a well-defined action derived using the Feynman path integral formalism. 
This provides a clear and systematic framework for improving the theory by introducing additional diagrams that capture the physical processes required to describe specific phenomena. 
The \mbox{D-TRILEX} diagrammatic expansion is not restricted to the DMFT impurity problem and can be formulated for an arbitrary reference system, such as a cluster problem, which enables studying symmetry-broken phases~\cite{fossati2025cluster}. 
\item[$\bullet$] The method has already been applied to a broad range of correlated electronic problems, from model systems~\cite{PhysRevLett.127.207205, PhysRevLett.129.096404, PhysRevLett.132.226501, PhysRevLett.132.236504} to realistic materials~\cite{stepanov2021coexisting, vandelli2024doping, stepanov2023charge, j6bj-gz7j, Ruthenates}.
\end{itemize}

\mbox{D-TRILEX} has significant potential to become a game-changing theoretical framework with broad applicability to a wide range of realistic correlated problems. 
Below, we present a detailed derivation of the method and discuss some benchmark calculations and the most representative applications of the method to various model systems and realistic materials calculations.

\subsection{Diagrammatic structure}
\label{sec:DT_diagrams}

Following the conclusion of Section~\ref{sec:Impact}, we neglect the particle-particle fluctuations from the partially-bosonized dual action~\eqref{eq:fbaction}.
To be consistent with the applications of the method discussed in this section, we restrict ourselves to the paramagnetic regime and do not consider the spin-orbit coupling. 
A general (spin-dependent) form of the \mbox{D-TRILEX} equations can be found in Ref.~\cite{10.21468/SciPostPhys.13.2.036}.
In the paramagnetic case all single-particle quantities are diagonal in the spin space and do not depend on the spin projection.
For instance, the bare dual Green's function becomes ${\tilde{\cal G}^{\,\sigma\sigma'}_{k,\,ll'} = \tilde{\cal G}_{k,\,ll'}\,\delta_{\sigma\sigma'}}$.
Consequently, the two-particle quantities are diagonal in the channel indices, as, e.g., holds true for the bare bosonic propagator ${\tilde{\cal W}^{\varsigma\varsigma'}_{q} = \tilde{\cal W}^{\varsigma}_{q}\,\delta_{\varsigma\varsigma'}}$.
One can also introduce spin-independent three-point vertex functions for the charge and spin channel as in Eq.~\eqref{eq:Lambda_para}: ${\Lambda^{\varsigma}_{\nu\omega} = \Lambda^{\uparrow\uparrow\varsigma}_{\nu\omega}}$~ (see e.g. Refs.~\cite{PhysRevB.102.195109, PhysRevB.103.245123}). 
This results in the following form of the partially bosonized dual action:
\begin{align}
{\cal S}_{fb} = &- \sum_{k,\sigma,\{l\}} f^{*}_{k\sigma{}l} \left[\tilde{\cal G}^{-1}_{k}\right]_{ll'} f^{\phantom{*}}_{k\sigma l'} 
- \frac12\sum_{q,\varsigma, \{l\}}  b^{\varsigma}_{-q,\, l_1 l_2}
\left[\tilde{\cal W}^{-1}_{q,\varsigma}\right]_{l_1 l_2;\, l_3 l_4} b^{\varsigma}_{q,\, l_4 l_3} \notag\\
&+\sum_{q,\{k\}}\sum_{\sigma,\sigma'}\sum_{\{l\}, \varsigma}
\Lambda^{\varsigma}_{\nu\omega,\, l_1,\, l_2,\, l_3 l_4} \, f^{*}_{k\sigma{}l_1} \sigma^{\varsigma}_{\sigma\sigma'} f^{\phantom{*}}_{k+q,\sigma', l_2} b^{\varsigma}_{q,\, l_4 l_3}\,.
\label{eq:DTRILEX_action}
\end{align}
We stick to the specific choice of the fermionic ${B_{\nu,ll'} = g_{\nu,ll'}}$ and bosonic~\eqref{eq:alpha_app} scaling parameters, which results in the following form of the bare dual propagators:
\begin{align}
\tilde{\cal G}^{\phantom{D}}_{k,\,l l'} &= G^{\rm DMFT}_{k,\,l l'} - g^{\phantom{D}}_{\nu,\,ll'},
\label{eq:bare_dual_G_DT}\\
\tilde{\cal W}^{\varsigma}_{q,\, l_1 l_2,\, l_3 l_4} &= W^{{\rm EDMFT}\,\varsigma}_{q,\, l_1 l_2,\, l_3 l_4} - \bar{u}^{\varsigma}_{l_1 l_2,\, l_3 l_4}\,.
\label{eq:bare_dual_W_DT}
\end{align}
The dressed dual Green's function $\tilde{G}_{k}$ and renormalized interaction $\tilde{W}^{\varsigma}_{q}$ can be found via the usual Dyson equations: 
\begin{align}
\left[\tilde{G}^{-1}_{k}\right]_{ll'} &= \left[\tilde{\cal G}^{-1}_{k}\right]_{ll'} -  \tilde{\Sigma}_{k,ll'}\,,
\label{eq:fermionc_dyson}\\
\left[\left(\tilde{W}^{\varsigma}_{q}\right)^{-1}\right]_{l_1l_2,\,l_3l_4} &= \left[\left(\tilde{\cal W}^{\varsigma}_{ q}\right)^{-1}\right]_{l_1 l_2,\, l_3 l_4} - \tilde{\Pi}^{\varsigma}_{q,\,l_1 l_2,\, l_3 l_4}\,.
\label{eq:bosonic_dyson}
\end{align}
In \mbox{D-TRILEX}, the self-energy $\tilde{\Sigma}$ and the polarization operator $\tilde{\Pi}$ are obtained self-consistently from the functional that corresponds to the partially-bosonized dual action~\eqref{eq:DTRILEX_action}:
\begin{align}
\Phi[\tilde{G},\tilde{W},\Lambda] = &-\frac12 \sum_{q,k}\sum_{\{l\},\sigma,\varsigma} 
\left[\Lambda^{\varsigma}_{\nu\omega}\right]_{l_1,l_2,\, l_3 l_4} \left[\tilde{G}^{\sigma}_{k+q}\right]_{l_2l_8} \left[\tilde{G}^{\sigma}_{k}\right]_{l_7l_1} \left[\tilde{W}^{\varsigma}_{q}\right]_{l_3 l_4,\, l_5 l_6} \left[\Lambda^{\varsigma}_{\nu+\omega,-\omega}\right]_{l_8,l_7,\, l_6 l_5} \notag\\
&+\frac12 \sum_{k,k'}\sum_{\{l\},\varsigma}\sum_{\sigma,\sigma'} 
\left[\Lambda^{\varsigma}_{\nu,\omega=0}\right]_{l_1,l_2,\, l_3 l_4} \left[\tilde{G}^{\sigma}_{k}\right]_{l_2l_1} \left[\tilde{G}^{\sigma'}_{k'}\right]_{l_7l_8} \left[\tilde{\cal W}^{\varsigma}_{q=0}\right]_{l_3 l_4,\, l_5 l_6} \left[\Lambda^{\varsigma'}_{\nu',\omega=0}\right]_{l_8,l_7,\, l_6 l_5}.
\end{align}
The self-energy and polarization operator can be found by varying the functional as:
\begin{align}
\tilde{\Sigma}_{k, ll'}^{\sigma} = \frac{\partial\Phi[\tilde{G},\tilde{W},\Lambda]}{\partial\tilde{G}^{\sigma}_{k,l'l}}\,\Bigg|_{\tilde{W},\Lambda}\,,~~~~~~
\tilde{\Pi}_{q,\, l_1l_2,\,l_3l_4}^{\varsigma} = -\,2\, \frac{\partial\Phi[\tilde{G},\tilde{W},\Lambda]}{\partial\tilde{W}^{\varsigma}_{q,\,l_3l_4,\,l_1l_2}}\,\Bigg|_{\tilde{G},\Lambda}\,.
\end{align}
The dual self-energy $\tilde{\Sigma}$ in the \mbox{D-TRILEX} approximation consists of the tadpole (TP) and $GW$-like diagrams $\tilde{\Sigma}_{k}=\tilde{\Sigma}_{k}^{\rm TP} + \tilde{\Sigma}_{k}^{GW}$.
The explicit expressions for these contributions are:
\begin{align}
\left[\tilde{\Sigma}^{\rm TP}_{\nu}\right]_{l_1 l_7} 
&= 2\sum_{k',\{l\}} 
\left[\Lambda^{d}_{\nu,\omega=0}\right]_{l_1, l_7,\, l_3 l_4}
\left[\tilde{\cal W}^{d}_{q=0}\right]_{l_3 l_4,\,l_5 l_6} 
\left[\Lambda^{d}_{\nu',\omega=0}\right]_{l_8, l_2,\, l_6 l_5} \left[\tilde{G}^{\phantom{d}}_{k'}\right]_{l_2 l_8}\,,
\label{eq:dual_sigma_td} \\
\left[\tilde{\Sigma}^{GW}_{k}\right]_{l_1 l_7} 
&= -\sum_{q,\{l\},\varsigma} \left[\Lambda^{\hspace{-0.05cm}\varsigma}_{\nu\omega}\right]_{l_1, l_2,\, l_3 l_4} \left[\tilde{G}^{\phantom{\varsigma}}_{k+q}\right]_{l_2 l_8} \left[\tilde{W}^{\varsigma}_{q}\right]_{l_3 l_4,\,l_5 l_6} \left[\Lambda^{\hspace{-0.05cm}\varsigma}_{\nu+\omega,-\omega}\right]_{l_8, l_7,\, l_6 l_5}\,.
\label{eq:dual_sigma_ex}
\end{align}
The polarization operator $\tilde{\Pi}$ of the \mbox{D-TRILEX} approach is the following:
\begin{align}
\left[\tilde{\Pi}^{\varsigma}_{q}\right]_{l_1 l_2,\, l_7 l_8}  
&= \,2\sum_{k,\{l\}} \left[\Lambda^{\hspace{-0.05cm}\varsigma}_{\nu+\omega,-\omega}\right]_{l_4, l_3,\, l_2 l_1} \left[\tilde{G}^{\phantom{\varsigma}}_{k}\right]_{l_3 l_5} \left[\tilde{G}^{\phantom{\varsigma}}_{k+q}\right]_{l_6 l_4} \left[\Lambda^{\hspace{-0.05cm}\varsigma}_{\nu\omega}\right]_{l_5, l_6,\, l_7 l_8}.
\label{eq:dual_pol}
\end{align}
Note that the \mbox{D-TRILEX} diagrams~\eqref{eq:dual_sigma_td}--\eqref{eq:dual_pol} represent the leading contribution to the self-energy and the polarization operator of the partially bosonized dual action~\eqref{eq:DTRILEX_action} in both the weak and the strong coupling limits. This is independent of the dimensionality of the problem. 
Indeed, at weak coupling the \mbox{D-TRILEX} diagrammatic expansion is perturbative in terms of the renormalized interaction~\eqref{eq:bare_dual_W_DT}. 
On the other hand, in the strong coupling limit, as in the DB approach, the small parameter of the diagrammatic expansion is the bare dual Green's function~\eqref{eq:bare_dual_G_DT}, which is purely non-local (see Sections~\ref{sec:Advantages_dual} and~\ref{sec:strong_coupling} for discussion).

\begin{figure}[t!]
\centering
\includegraphics[width=0.78\linewidth]{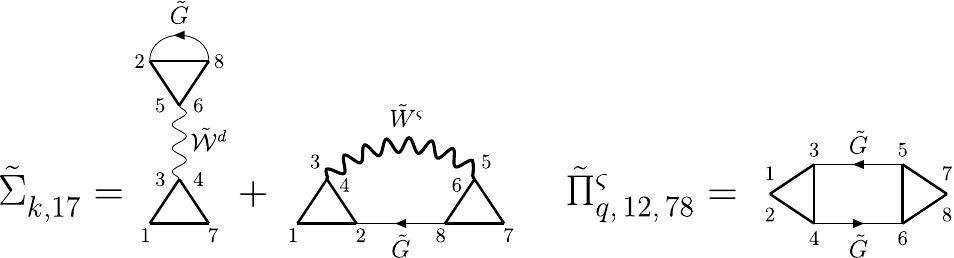} 
\caption{Diagrammatic representation for the dual self-energy $\tilde{\Sigma}$ (left) and polarization operator $\tilde{\Pi}$ (right). 
Wavy lines represent dual bosonic propagators and straight lines depict dual Green's function, as explicitly indicated in the Figure. 
Triangles represent three-point vertex functions $\Lambda^{\varsigma}_{\nu\omega}$. 
Numbers correspond to band indices. The Figure is taken from Ref.~\cite{10.21468/SciPostPhys.13.2.036}.
\label{fig:DT_diagrams}}
\end{figure}

The diagrammatic expressions for $\tilde{\Sigma}$ and $\tilde{\Pi}$ are shown in Fig.~\ref{fig:DT_diagrams}.
These expressions illustrate that in the \mbox{D-TRILEX} approach the single- and two-particle quantities are treated self-consistently, which allows one to account for the effect of collective electronic fluctuations onto the electronic spectral function and {\it vice versa}~\cite{PhysRevB.103.245123, stepanov2021coexisting, PhysRevResearch.5.L022016, PhysRevLett.129.096404}.

\begin{figure}[b!]
\centering
\includegraphics[width=0.85\linewidth]{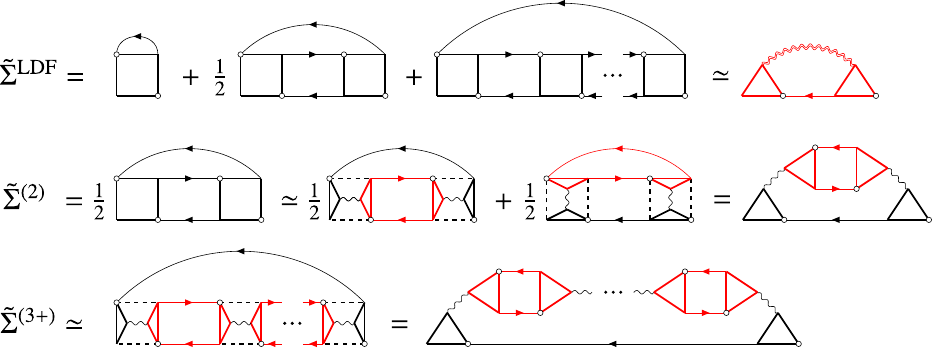}
\caption{\label{fig:DT_DB_relation} Top row shows the ladder dual fermion self-energy $\tilde{\Sigma}^{\rm LDF}$, which consists of the first-order Hartree-like term, second-order diagram $\tilde{\Sigma}^{(2)}$, and the rest of the ladder $\tilde{\Sigma}^{(3+)}$. Keeping only longitudinal modes of the partially bosonized representation for the four-point vertex $\Gamma$ (black squares), the LDF self-energy reduces to the self-energy of the $\text{D-TRILEX}$ approach (red diagram at the end of the top row). For $\tilde{\Sigma}^{(2)}$ and $\tilde{\Sigma}^{(3+)}$ the result of this approximation is explicitly shown in the middle and bottom rows, respectively. Red parts of these diagrams that consist of two triangles (three-point vertices) connected by two solid lines (dual Green's functions) is the polarization operator of the $\text{D-TRILEX}$ theory. Wiggly line corresponds to the renormalized interaction. The Figure is taken from Ref.~\cite{PhysRevB.103.245123}.}
\end{figure}

\subsubsection{Relation to the ladder dual boson/fermion approach}

The $GW$-like diagrammatic structure~\eqref{eq:dual_sigma_td}--\eqref{eq:dual_pol} of the $\text{D-TRILEX}$ approach can also be related to its parental DB theory. 
For simplicity, Fig.~\ref{fig:DT_DB_relation} shows the sketch of this relation in the absence of the non-local interaction (${V^{\varsigma}_{\qv}=0}$).
In this case, the DB theory~\eqref{eq:DB_action} identically coincides with the DF approach, and the dual self-energy in the ladder approximation takes the form of $\tilde{\Sigma}^{\rm LDF}$~\eqref{eq:Sigma_LDF} displayed in the first line of Fig.~\ref{fig:DT_DB_relation}.
Note, that, the second-order contribution $\tilde{\Sigma}^{(2)}$ has a ``1/2'' prefactor that does not appear for the rest of the ladder self-energy $\tilde{\Sigma}^{(3+)}$ (see Section~\ref{sec:LDB_diagrammatics}).
As two subsequent lines in Fig.~\ref{fig:DT_DB_relation} show, if one uses the partially bosonized representation~\eqref{eq:Gamma_approx_app} for every vertex function that enters $\tilde{\Sigma}^{\rm LDF}$ and keeps only longitudinal contributions in this approximation, the dual self-energy immediately reduces to the $\text{D-TRILEX}$ form~\eqref{eq:dual_sigma_ex}.
By longitudinal contributions we refer to the $M^{\varsigma}_{\nu\nu'\omega}$ terms, where the bosonic propagator $\bar{w}^{\varsigma}_{\omega}$ carries the main bosonic frequency $\omega$. 
The explicit analytical derivation of the relation between $\text{D-TRILEX}$ and ladder DB self-energies for the general case when the non-local interaction is not neglected can be found in Ref.~\cite{PhysRevB.103.245123}.
This result demonstrates the important advantage of the $\text{D-TRILEX}$ theory over its parental DB method, which drastically reduces costs of numerical calculations.
Thus, although the $\text{D-TRILEX}$ approach accounts for the main longitudinal part of the full two-particle ladder fluctuation, the calculation of the self-energy~\eqref{eq:dual_sigma_ex} and polarization operators~\eqref{eq:dual_pol} does not require the inversion of the Bethe-Salpeter equation in the momentum-frequency space.
In the ladder DF/DB theory the inversion of the Bethe-Salpeter equation in the frequency space cannot be avoided due to a three-frequency dependence of the local vertex function $\Gamma_{\nu\nu'\omega}$.

\subsection{Relation between physical and dual quantities}
\label{sec:Lattice_dual_DT}

As in the DB approach, the \mbox{D-TRILEX} diagrammatic expansion is performed in the dual space that describes electronic correlations beyond the ones of the reference system.
This formulation of the theory allows one to avoid double-counting of correlation effect that are already taken into account by the reference problem. 
The Green's function and susceptibility of the initial lattice problem~\eqref{eq:actionlatt} can be obtained from the  quantities of the partially bosonized dual action~\eqref{eq:fbaction_app} using the exact relations~\eqref{eq:Gsource} and~\eqref{eq:Xsource}.
Since the dual fermionic fields remain the same in the DB and \mbox{D-TRILEX} methods, the lattice Green's function and self-energy in the \mbox{D-TRILEX} approach are given by the same relations~\eqref{eq:GtoSigma} and~\eqref{eq:Sigma_relation} as in the DB theory (${B_{\nu,ll'}=g_{\nu,ll'}}$):
\begin{align}
\left[G^{-1}_{k}\right]_{ll'} &= \left[\left(g_{\nu} + g_{\nu}\cdot\tilde{\Sigma}_{k}\cdot{}g_{\nu}\right)^{-1}\right]_{ll'} - \tilde{\varepsilon}_{k,ll'}\,,
\label{eq:Glatt_DT}\\
\Sigma^{\phantom{ip}}_{k, ll'} &= \Sigma^{\rm imp}_{\nu,ll'} + \sum_{l_1} \tilde{\Sigma}^{\phantom{ip}}_{k,l l_1}\left[\left(\mathbbm{1} + g_\nu \cdot \tilde{\Sigma}_k \right)^{-1} \right]_{l_1 l'}.
\label{eq:Sigmalatt_DT}
\end{align}
The bosonic fields in the DB and \mbox{D-TRILEX} approach are different, which results in a bit different expression for the susceptibility that was explicitly derived in Ref.~\cite{10.21468/SciPostPhys.13.2.036}:
\begin{align}
\left[\left(X^{\varsigma}_{q}\right)^{-1}\right]_{l_1 l_2,\, l_3 l_4} &= \left[\left(\Pi^{\varsigma}_{q}\right)^{-1}\right]_{l_1 l_2,\, l_3 l_4} - \left[U^{\varsigma}_{\bf q}\right]_{l_1 l_2,\, l_3 l_4}
\label{eq:Xlatt_DT}
\end{align}
that involves the bare interaction ${U^{\varsigma}_{\bf q} = U^{\varsigma} + V^{\varsigma}_{\bf q}}$ and the polarization operator of the lattice problem:
\begin{align}
\Pi^{\varsigma}_{q,\,l_1 l_2,\, l_3 l_4} = \Pi^{{\rm imp}\,\varsigma}_{\omega,\,l_1 l_2,\, l_3 l_4} + \sum_{l',l''} \tilde{\Pi}^{\varsigma}_{q,\,l_1 l_2,\, l' l''} \left[\left(\mathbbm{1} + \bar{u}^{\varsigma} \cdot \tilde{\Pi}^{\varsigma}_{q}  \right)^{-1}\right]_{l'l'',\,l_3l_4}.
\label{eq:Pilatt_DT}
\end{align}
Importantly, as shown in Ref.~\cite{10.21468/SciPostPhys.13.2.036}, the divergence in the lattice susceptibility $X^{\varsigma}_{q}$~\eqref{eq:Xlatt_DT} occurs at the same time as in the dual renormalized interaction $\tilde{W}^{\varsigma}_{q}$~\eqref{eq:bosonic_dyson}.
This allows the \mbox{D-TRILEX} approach to capture the formation of collective electronic instabilities directly in the dual space, as in the DB approach.
We stress that the obtained relations between the dual and the lattice quantities are valid for any form of the dual self-energy $\tilde{\Sigma}$ and polarization operator $\tilde{\Pi}$ of the partially bosonized dual action~\eqref{eq:DTRILEX_action}. 
In particular, the role of the ``dual denominator'' in the expression for the lattice self-energy~\eqref{eq:Sigmalatt_DT} is discussed in detail in Section~\ref{sec:dual_demoninator}. 

\begin{figure}[t!]
\centering
\includegraphics[width=1\linewidth]{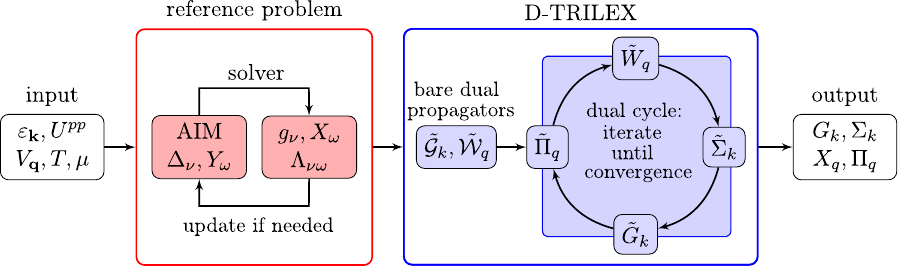} 
\caption{Computational workflow of the {\mbox D-TRILEX} method. The input consists in the parameters of the electronic lattice problem (Init.1-2 in the main text). The red box indicates the solution of the reference impurity problem (Init.3), that in some case has to be updated until self-consistency is reached (for instance in DMFT, EDMFT and cluster DMFT). The blue box contains the operations performed in the dual space, i.e. the calculation of the bare dual propagators (Init.4) and the self-consistency cycle on the dual quantities (St.2). The output consists in the Green's function, self-energy, susceptibility and polarization operator of the lattice problem, obtained by applying the exact relations between lattice and dual quantities (St.3). The Figure is taken from Ref.~\cite{10.21468/SciPostPhys.13.2.036}.
\label{fig:DT_scheme}}
\end{figure}

\subsection{Computational workflow}
\label{sec:WF_DT}

The computational workflow of the \mbox{D-TRILEX} approach is divided into several parts, as shown in Fig~\ref{fig:DT_scheme}. 
The first step (St.1) consists in solving the reference system, e.g. the DMFT impurity problem.
This produces the inputs necessary for the initialization of the diagrammatic part of the calculation. 
Hence, the inner steps are denoted with (Init.). 
The second step (St.2) takes care of the self-consistent dressing of the dual Green's function $\tilde{G}$ and the renormalized interaction $\tilde{W}$. 
The inner steps (I.) of the self-consistent
diagrammatic iteration are highlighted below.
After the dressed dual quantities are determined, the single- and two-particle quantities for the initial (lattice) problem are evaluated at the third step (St.3).\\

The computational workflow has the following form: 
\begin{enumerate}[label=({St.}{{\arabic*}})]
\item Input initialization:
    \begin{enumerate}[label=({Init.}{{\arabic*}})]
    \item Specify the single-particle term $\varepsilon_{\kv, ll'}$, the interactions $U^{pp/ph}_{l_1l_2l_3l_4}$ and {$V^{\varsigma}_{q,\,l_1l_2,\,l_3l_4}$}, and the temperature $T$ that enter the initial action~\eqref{eq:actionlatt}.
    \item Define the chemical potential $\mu$ and the hybridization function $\Delta_{\nu, ll'}$. 
    \item Solve the reference system and get the corresponding Green's function $g_{\nu, ll'}$, the susceptibility $\chi^{\varsigma}_{\omega,\,l_1l_2,\,l_3l_4}$, and the vertex function $\Lambda^{\varsigma}_{\nu\omega,\,l_1,l_2,\,l_3l_4}$.
    \item  Compute the bare fermionic $\tilde{\cal G}_{k,ll'}$ and bosonic $\tilde{\cal W}^{\varsigma}_{q,\,l_1l_2,\,l_3l_4}$ propagators of the effective partially bosonized dual action~\eqref{eq:DTRILEX_action} according to Eqs.~\eqref{eq:bare_dual_G_DT} and~\eqref{eq:bare_dual_W_DT}, respectively.
    \end{enumerate}
\item Self-consistent calculation of \mbox{D-TRILEX} diagrams:
    \begin{enumerate}[label=(I.{{\arabic*}})]
    \item Compute the dual polarization operator $\tilde{\Pi}$ using Eq.~\eqref{eq:dual_pol}.
    \item Compute the dual renormalized interaction $\tilde{W}$ using Eq.~\eqref{eq:bosonic_dyson}.
    \item Compute the diagrams $\tilde{\Sigma}^{\rm TP}$~\eqref{eq:dual_sigma_td} and $\tilde{\Sigma}^{GW}$~\eqref{eq:dual_sigma_ex} for the dual self-energy.
    \item Compute the dressed dual Green's function $\tilde{G}$ using Eq.~\eqref{eq:fermionc_dyson}.
    \item If the desired accuracy $\delta$ for the self-consistent condition is reached, go to (St.3). Otherwise go back to (I.1).
    \end{enumerate}
\item Evaluation of lattice quantities:
    \begin{itemize}
    \item[] Compute the dressed Green's function $G_{k,ll'}$~\eqref{eq:Glatt_DT}, the self-energy $\Sigma_{k,ll'}$~\eqref{eq:Sigmalatt_DT}, the susceptibility $X^{\varsigma}_{q,\,l_1l_2,\,l_3l_4}$~\eqref{eq:Xlatt_DT}, and the polarization operator $\Pi^{\varsigma}_{q,\,l_1l_2,\,l_3l_4}$~\eqref{eq:Pilatt_DT} for the lattice problem~\eqref{eq:actionlatt}. 
    Determine the orbital-resolved average density ${\langle n_{l} \rangle}$.
    If one aims at the specific density $\langle n \rangle$, it is possible to update the chemical potential $\mu$ and go back to the beginning of the outer loop. In that case, go to (St.1 of Init.2) and fix the new $\mu$ and update the hybridization function $\Delta_{\nu, ll'}$ if needed.
    \end{itemize}
\end{enumerate}

\noindent
A detailed description of the current numerical implementation of the \mbox{D-TRILEX} method can be found in Refs.~\cite{10.21468/SciPostPhys.13.2.036, vandelli2022quantum}.

\subsubsection{Details of the numerical calculation}
The complexity of the diagrammatic part of the D-TRILEX calculation is estimated as 
\begin{align}
\mathcal{O}\left(N_\nu N_{\omega}\right) \, \mathcal{O}\left( N_{\rm imp}^2 \, \times \, \sum_{i=1}^{N_{\rm imp}}N^8_{l_i}\right) \, \mathcal{O}\left(N_k \log N_k\right)
\label{eq:ccomplexity}
\end{align}
where ${N_{\nu \, (\omega)}}$ is the number of fermionic (bosonic) Matsubara frequencies, $N_{\rm imp}$ is the number of impurities in the reference system, $N_{l_i}$ is the number of orbitals for the $i$-th impurity and $N_k$ is the total number of ${\bf k}$-points.
In this context, $N_{\rm imp}$ is the number of independent impurities in the unit cell of the reference problem. 
Note that the case of ${N_{\rm imp} > 1}$ corresponds to a collection of impurities, as explained in Ref.~\cite{PhysRevB.97.115150}, and not to a cluster of $N_{\rm imp}$ sites.
If the impurities are all identical, then the reference system reduces to a single site impurity problem.
If some of them are different, it is sufficient to solve an impurity problem only for the non-equivalent ones.
In the multi-impurity case, fluctuations between the impurities are taken into account diagrammatically in the framework of \mbox{D-TRILEX} approach. 
On the other hand, a cluster reference system corresponds to a multi-orbital problem with ${N_{\rm imp} = 1}$.
In this case, $N_{l}$ is the total number of orbitals and sites of the considered cluster. 
The separation between orbitals and sites that we introduce is useful to reduce the computational complexity when addressing problems with several atoms in the unit cells.

The scaling as a function of ${\bf k}$-points is determined from the fact that we utilise the fast-Fourier transform (FFT) algorithm for computing convolutions in momentum space.
This shows that the multi-impurity calculation has a quadratic scaling with respect to the number of impurities. 
In our current implementation, 
the local Coulomb matrix is considered as a non-sparse matrix within each site subspace, hence the scaling to the 8th power in the number of orbitals. 
However, before running the actual calculations, we introduced a check to assess which components of the vertices are zero. These components are automatically skipped in order to avoid unnecessary calculations and to automatically take advantage of a possible sparsity of the Coulomb matrix, effectively reducing the complexity~\eqref{eq:ccomplexity} in most cases. 
The summation over frequencies and band indices can be efficiently parallelized both in a shared-memory framework (as done in the current implementation) and in an message-passing interface (MPI) framework. 

To measure the accuracy at the $n$-th iteration of the self-consistent cycle, we use the relative Frobenius norm of the Green's function ${F = ||\tilde{G}_{n} - \tilde{G}_{n-1}||/||\tilde{G}_{n-1}||}$ as a metric, where $||...||$ is the square root of the squared sum over all the components of the array. If $F$ is smaller than some predefined accuracy value $\delta$, the self-consistent cycle stops. 
The cycle stops also if a specified maximum number of iterations is reached. 
The stability of the bosonic Dyson equation~\eqref{eq:bosonic_dyson} can be problematic in regimes of parameters, where one or more of the eigenvalues of the quantity ${\tilde{\Pi} \cdot \tilde{\cal W}}$ become equal or larger than 1. 
In particular, this happens when the system is close to a phase transition or if the correlation length in some channel of instability exceeds a critical value.
This issue appears in similar forms in other diagrammatic extensions of DMFT (see, e.g., Ref.~\cite{Otsuki14}). 
In one- and two-dimensional systems, where Mermin-Wagner theorem forbids the breaking of continuous symmetries~\cite{Mermin66}, the issue can be mitigated by imposing that the eigenvalues $\lambda_i$ of the $\tilde{\Pi} \cdot \tilde{\cal W}$ matrix in the orbital space for a physical meaningful solution are always smaller than 1.
In our implementation, we check whether any eigenvalue ${\lambda_i \geq 1}$ (${i \in \{k, \, \varsigma\}}$). If this happens, the eigenvalue can be rescaled as described in Ref.~\cite{Otsuki14} in order to improve convergence. 

Several strategies can be used to improve the stability of the self-consistent procedure in the general case. 
The first strategy is implemented when updating the \mbox{D-TRILEX} self-energy. 
The updated dual self-energy at the $n$-th iteration is computed as ${\tilde{\Sigma}_{n} = (1-\xi)\tilde{\Sigma}_{n-1} +  \xi\tilde{\Sigma}}$ for ${\xi \in (0, 1)}$, where $\tilde{\Sigma}_{n-1}$ is the value of the dual self-energy computed at the previous (${n-1}$) iteration, and $\tilde{\Sigma}$ is computed using the propagators $\tilde{G}_{n-1}$ and $\tilde{W}_{n-1}$ obtained at the previous iteration.
This procedure was shown to improve stability in $GW$-like theories~\cite{PhysRevB.92.115125}. A similar mixing scheme can be applied to the dual polarization.
To the same aim, we also introduce multiplicative factors for the dual self-energy and the dual polarization at the first iteration. 
Of course, no rescaling is expected to work in the presence of the symmetry breaking due to a true phase transition. 
The latter case should be addressed using a suitable cluster or multi-impurity reference problem.

It is worth mentioning that the efficiency of the whole scheme is strongly affected by the computational cost of the impurity solver. 
In our tests, the time needed to solve the impurity problem and to obtain the required correlation functions of the reference system using continuous time quantum Monte Carlo solvers~\cite{PhysRevB.72.035122, PhysRevLett.97.076405, PhysRevLett.104.146401, RevModPhys.83.349} always exceeds the computational cost for the diagrammatic part of the calculation, even by several orders of magnitude. 
For example, a single iteration of the self-consistent diagrammatic cycle for a two-orbital case takes only few minutes.

\subsection{Benchmarking of the method}

\subsubsection{Single-band Hubbard model}

\paragraph{Self-energy of the half-filled Hubbard model on a square lattice:}

The comparison of the ladder dual fermion (LDF) approach against the exact DiagMC@DF solution of the dual action~\eqref{eq:DB_action} for the case of a half-filled Hubbard model on a square lattice with the nearest-neighbor hopping ${t=1}$ was done in Ref.~\cite{PhysRevB.103.245123} and was discussed in Section~\ref{sec:Impact}. 
In the same work, we also performed \mbox{D-TRILEX} calculations.
The corresponding results can be found in Figs.~\ref{fig:compare_sigma} and~\ref{fig:Ev_and_deviation}.
The best agreement between the \mbox{D-TRILEX} (red line) and the reference $\text{DiagMC@DF}$ (light blue line) results for the imaginary part of the self-energy occurs at ${U=2}$.
At small and moderate values of $U$, the \mbox{D-TRILEX} self-energy seems to be pinned to the LDF result (dark blue line) at $\Gamma$ and M points. 
Therefore, the difference between these two methods is mostly visible around local minima located at antinodal ${\text{AN}=(0, \pi)}$ and nodal ${\text{N}=(\pi/2, \pi/2)}$ points.
This difference increases with the interaction, and the observed trend persists up to $U=6$. 
At larger interactions, when the value of the self-energy at local minima becomes similar, the $\text{D-TRILEX}$ result shifts downwards, and at ${U=12}$ becomes pinned to the LDF result at N and AN points.

The discrepancy between the $\text{D-TRILEX}$ and the reference results for the real part of the self-energy also increases with the interaction up to $U=8$, and after that decreases again for very large interaction strengths. 
However, here the best agreement with the exact result is achieved at $U=4$ (see red line in Fig.~\ref{fig:compare_sigma}). 
It can be explained by the fact, that in the perturbative regime of small interactions ($U=2$) and high temperatures ($\beta=2$) the second-order dual self-energy $\tilde{\Sigma}^{(2)}$ gives the main contribution to the nonlocal part of the total self-energy~\cite{PhysRevB.91.235114, PhysRevB.94.035102, PhysRevB.96.035152, PhysRevB.102.195109}.  
The $\text{D-TRILEX}$ theory is not based on a perturbation expansion, because it takes into account only a particular ($GW$-like) subset of diagrams. 
For this reason, this simple theory does not fully reproduce the second-order self-energy $\tilde{\Sigma}^{(2)}$, which leads to a slight underestimation of the result as discussed in Ref.~\cite{PhysRevB.103.245123}. 
On the contrary, the $\text{D-TRILEX}$ approach correctly accounts for the screening of the interaction that is represented by longitudinal part of the infinite two-particle ladder in all bosonic channels.
At lower temperatures and/or larger interactions, when the system enters the correlated regime, these types of diagrams become more important than the second-order self-energy.
To illustrate this point, we also obtained the normalized deviation for the $\text{D-TRILEX}$ approach for $\beta=4$ (for $U=2$ and $U=4$) and $\beta=10$ (for $U=2$), and compared it with $\delta$ calculated for the second-order DF (DF$^{(2)}$) approximation that considers only $\tilde{\Sigma}^{(2)}$ contribution to the dual self-energy. 
The corresponding result is shown in the inset of Fig.~\ref{fig:Ev_and_deviation}. 
As expected, the accuracy of the DF$^{(2)}$ approximation rapidly decreases with the temperature and becomes $\delta=16.5\%$ (for $\beta=10$ and $U=2$) and $\delta=22.5\%$ (for $\beta=4$ and $U=4$) in the regime, which is yet above the DMFT N\'eel point $\beta_{N}\simeq12.5$ for $U=2$ and $\beta_{N}\simeq4.3$ for $U=4$.
At the same time, the $\text{D-TRILEX}$ theory remains in a reasonable agreement with the reference result.

We note, that the largest discrepancy between the $\text{D-TRILEX}$ and reference $\text{DiagMC@DF}$ result $\delta = 18\%$ corresponds to the most correlated regime ($U=8$). 
At small and moderate interactions the normalized difference does not exceed 3.5\% ($U=4$). 
At the same time, the maximal difference from the parental LDF is only around 10\% ($U=8$), which can be considered as relatively good result for such a simple theory. 
We also looked at the contribution of the longitudinal particle-particle fluctuations to the $\text{D-TRILEX}$ self-energy and, as expected, found it to be negligibly small. 
We observed that the part of the self-energy that stems from the singlet bosonic mode makes only $3\%$ of the $\text{D-TRILEX}$ self-energy at $U=2$, and does not exceed $1\%$ for other interaction strength. 
Taking into account all above discussions, this result confirms that all particle-particle fluctuations can indeed be safely excluded from the simple $\text{D-TRILEX}$ theory in the repulsive $U$ regime, which however does not hold for every diagrammatic approach.

Our results confirm that in a broad regime of physical parameters the leading contribution to the self-energy is given by the longitudinal particle-hole bosonic modes.
This important statement allows for a drastic simplification of the diagrammatic expansion, which implies a huge reduction of computational efforts.
Consequently, the $\text{D-TRILEX}$ theory, which appears as the result of this simplification, looks as a very promising and powerful tool for solving a broad class of interacting electronic problems.
At the same time, the theory should not be oversimplified. 
Thus, we have shown that considering only second-order dual self-energy does not provide good result even at weak and moderate interactions when the system enters the correlated regime lowering the temperature.

\paragraph{N\'eel transition in the half-filled Hubbard model on a cubic lattice:}

In Ref.~\cite{vandelli2022quantum}, the \mbox{D-TRILEX} approach was benchmarked against the exact DiagMC solution for estimating the N\'eel temperature of the single-band Hubbard model on a cubic lattice with nearest-neighbor hopping ${t=1}$. 
In three dimensions, the N\'eel transition to the antiferromagnetically (AFM) ordered state is not forbidden by the Mermin-Wagner theorem.

\begin{figure}[t!]
\centering
\includegraphics[width=0.49\linewidth]{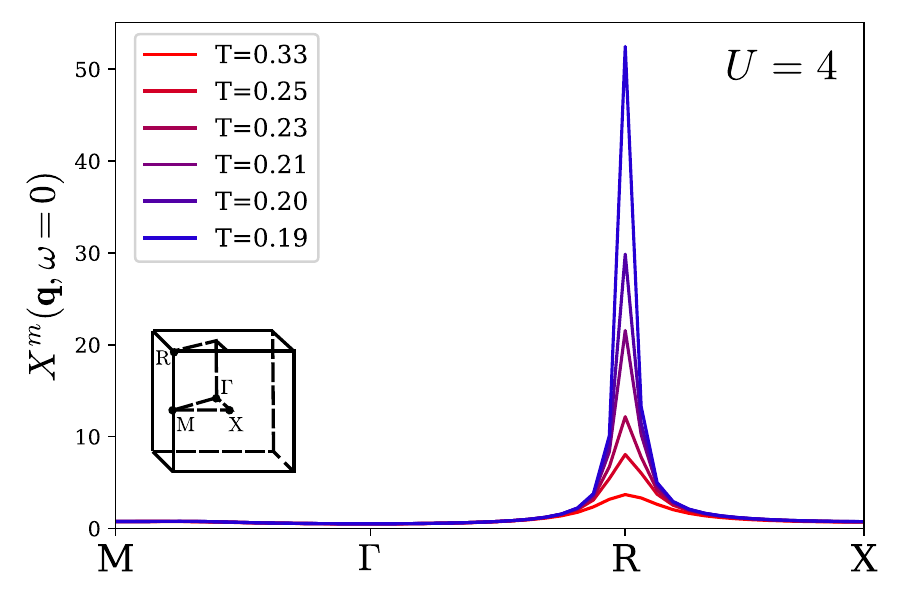}~~
 \includegraphics[width=0.49\linewidth]{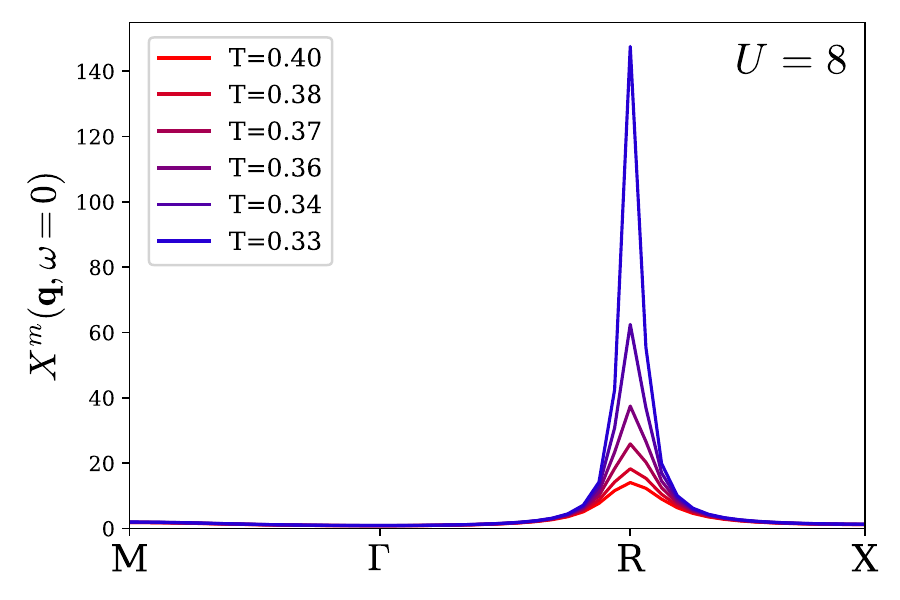}
\caption{The plot shows the static magnetic susceptibility $X^m(\qv, \omega=0)$ along the high-symmetry path in the BZ of the cubic lattice (sketched in the upper panel). The upper panel is computed at the value of the Hubbard interaction $U=4$, the lower panel at $U=8$. Different colors correspond to different temperatures. The peak at $\qv = $R is clearly visible. The Figure is taken from Ref.~\cite{vandelli2022quantum}.
\label{fig:X_vs_T_cubic}}
\end{figure}

To confirm that \mbox{D-TRILEX} accurately describes the AFM phase transition on a cubic lattice, we performed calculations for several values of the Hubbard interaction $U$. 
In our calculations, we use a ${32 \times 32 \times 32}$ mesh in the Brillouin Zone, 48 fermionic frequencies and 32 bosonic frequencies. 
The AFM order on a cubic lattice corresponds to a wave vector $\qv_{\rm AFM} = {\rm R} = (\pi, \pi, \pi)$ and can be revealed by the respective peak appearing in the momentum-resolved static spin susceptibility $X^{m}_{\qv, \omega=0}$, as shown in  Fig.~\ref{fig:X_vs_T_cubic}.
As discussed in Section~\ref{sec:Lattice_dual_DT}, the divergence of this susceptibility occurs simultaneously with the divergence of the dual renormalized interaction $\tilde{W}^{m}_{\qv, \omega=0}$ and indicates the phase transition. 

A comparison of N\'eel temperatures ($T_N$) obtained by various methods is shown in Fig.~\ref{fig:Neel_3D}.
We find, that $T_N$ predicted by \mbox{D-TRILEX} is in very good agreement with the exact methods (QMC, DDMC and DiagMC) within the stochastic error bars of these methods as well as with the LDF method. 
This good agreement is related to the fact that the AFM transition is well-described by longitudinal spin fluctuations which are accounted for in \mbox{D-TRILEX}. 
We observe, that \mbox{D-TRILEX} consistently improves the DMFT results even in the most correlated regime. 
For instance, in Ref.~\cite{PhysRevB.104.235128} it is shown that $T_N$ predicted by DMFT at ${U=8}$ is around ${T_N \approx 0.45}$~\cite{PhysRevB.92.144409}, while \mbox{D-TRILEX} predicts a value much closer to the more advanced methods. 
  
\begin{figure}[b!]
\centering
\begin{tabular}{ccc}
\hspace{1.2cm}$\delta = 0\%$ & \hspace{1.2cm}$\delta = 5\%$ \\
\includegraphics[width=0.4\linewidth]{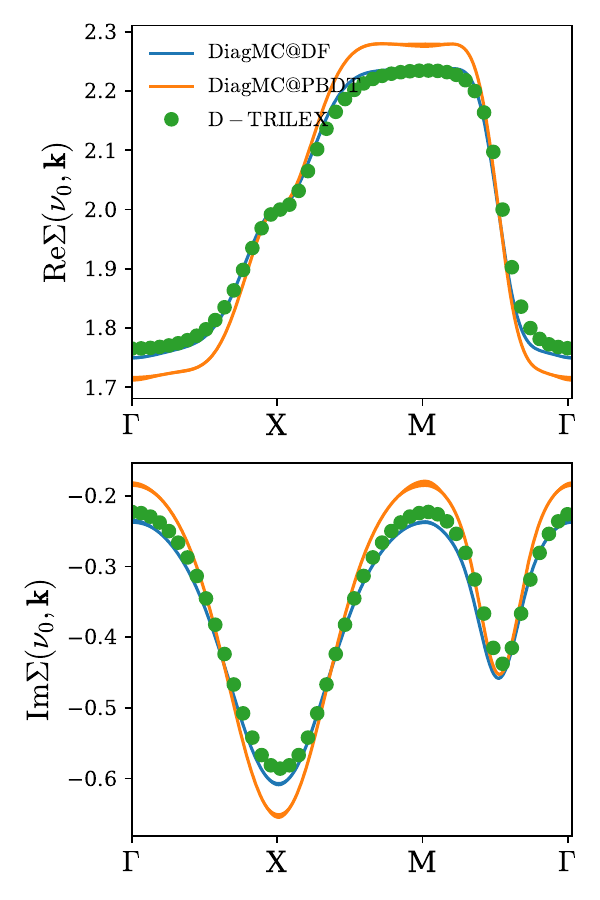}  & 
\includegraphics[width=0.4\linewidth]{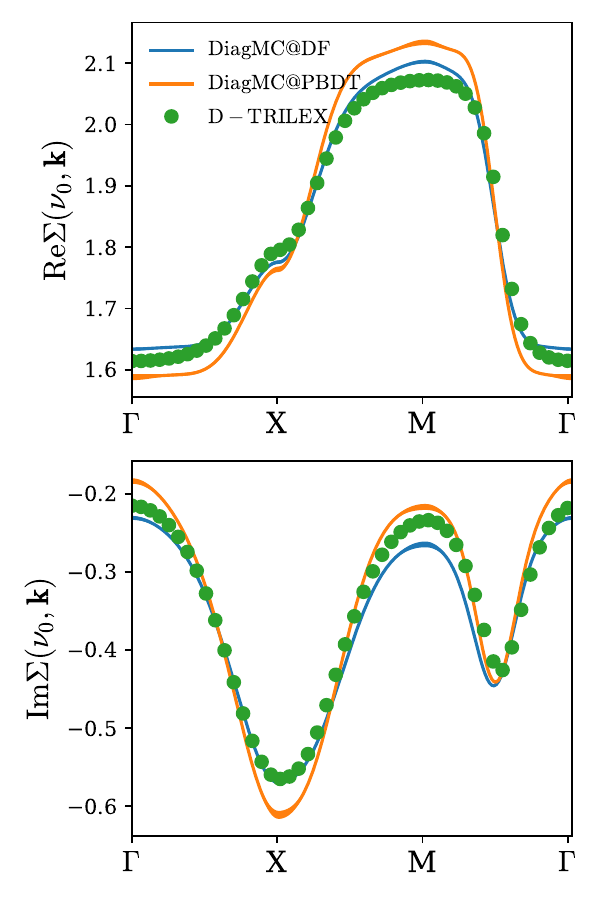}
\end{tabular}
\caption{The lattice self-energy $\Sigma_{\nu_0 \kv}$ of the doped Hubbard model obtained for the lowest Matsubara frequency ${\nu_0=\pi/\beta}$ along the high-symmetry path in the Brillouin zone. The \mbox{D-TRILEX} calculation (green dots) is compared with the DiagMC@DF (blue line) and DiagMC@PBDT (orange line) results at half-filling and the small value of hole-doping $\delta=5\%$. The model parameters are ${t=1}$, ${t'=0}$, ${U=4}$ and $\beta=4$. The Figure is taken from Ref.~\cite{vandelli2022quantum}.
\label{fig:small_doping}
}
\end{figure}

\paragraph{Hole-doped Hubbard model on a square lattice:}

\begin{figure}[t!]
\centering
\begin{tabular}{ccc}
\hspace{1.2cm}$\delta = 10\%$ & \hspace{1.2cm}$\delta = 15\%$ \\
\includegraphics[width=0.4\linewidth]{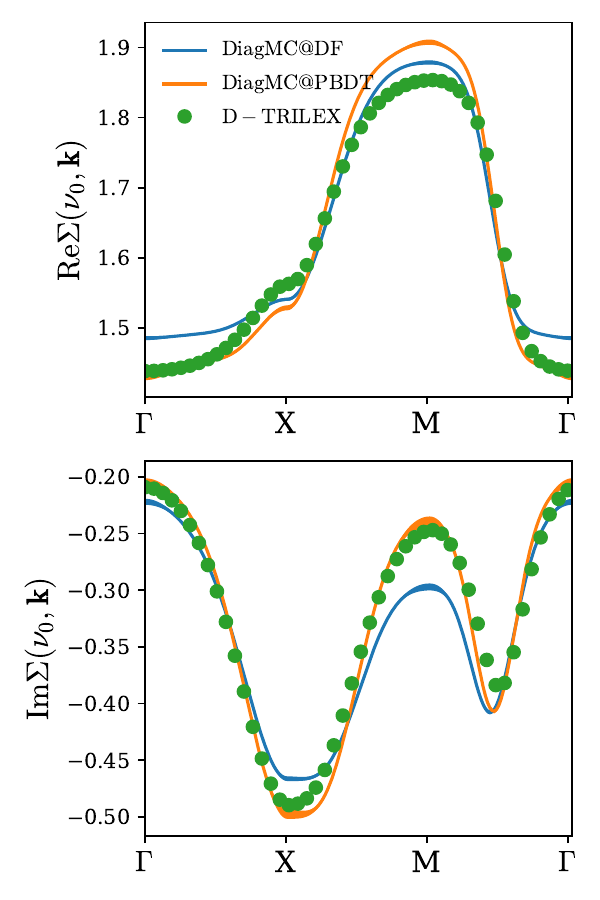}  &
\includegraphics[width=0.4\linewidth]{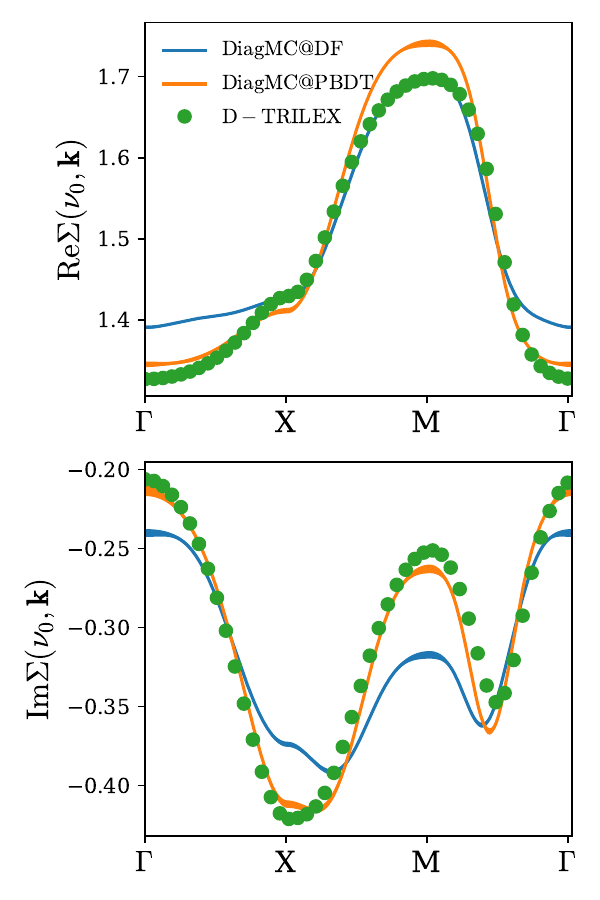} 
\end{tabular}
\caption{The lattice self-energy $\Sigma_{\nu_0 \kv}$ of the doped Hubbard model obtained for the lowest Matsubara frequency ${\nu_0=\pi/\beta}$ along the high-symmetry path in the Brillouin zone. The \mbox{D-TRILEX} calculation (green dots) is compared with the DiagMC@DF (blue line) and DiagMC@PBDT (orange line) results at intermediate values of hole-doping ${\delta=10\%}$ and ${\delta=15\%}$. The model parameters are ${t=1}$, ${t'=0}$, ${U=4}$ and ${\beta=4}$. The Figure is taken from Ref.~\cite{vandelli2022quantum}.
\label{fig:inter_doping}}
\end{figure}

\begin{figure}[t!]
\centering
\[\delta=20\%\]
\includegraphics[width=0.8\linewidth]{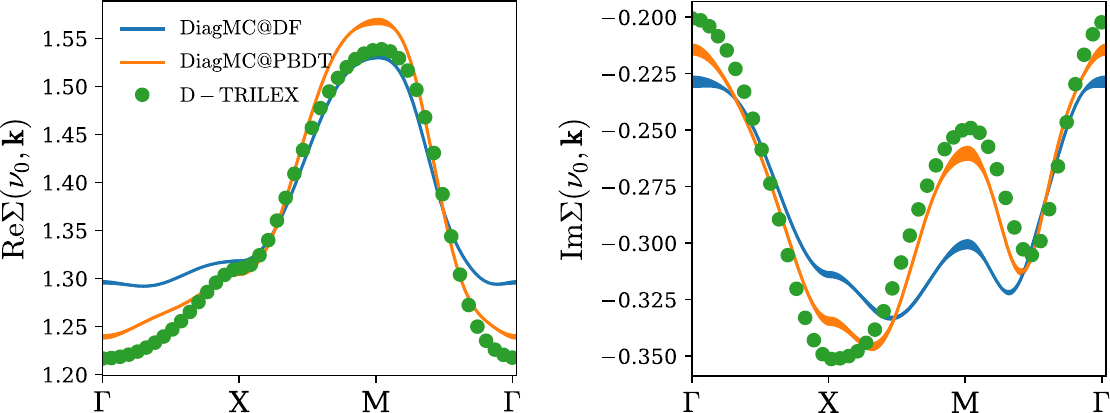}
\caption{The lattice self-energy $\Sigma_{\nu_0 \kv}$ of the doped Hubbard model obtained for the lowest Matsubara frequency ${\nu_0=\pi/\beta}$ along the high-symmetry path in the Brillouin zone. The \mbox{D-TRILEX} calculation (green dots) is compared with the DiagMC@DF (blue line) and DiagMC@PBDT (orange line) results at the large value of hole-doping ${\delta=20\%}$. The model parameters are ${t=1}$, ${t'=0}$, ${U=4}$, and ${\beta=4}$. The Figure is taken from Ref.~\cite{vandelli2022quantum}.
\label{fig:large_doping}}
\end{figure}

The doped two-dimensional Hubbard model on a square lattice with the nearest-neighbor $t$ and next-nearest-neighbor $t'$ hopping amplitudes is widely known as a prototype model for high-temperature superconducting cuprate compounds. 
We start the discussion of this benchmark by assessing the range of validity of \mbox{D-TRILEX} as a function of doping only and, to that aim, we initially we set ${t'=0}$. 
We study the discrepancy between the \mbox{D-TRILEX} method and the reference \mbox{DiagMC@DF} result. 
Additionally, we perform the \mbox{DiagMC@PBDT} calculations (see Section~\ref{sec:Impact}) to obtain an exact solution of the partially bosonized dual action~\eqref{eq:fbaction}, in order to clarify the source of the mismatch with DiagMC@DF. 
In Figs.~\ref{fig:small_doping}--\ref{fig:large_doping}, we show the lattice self-energy calculated in Ref.~\cite{vandelli2022quantum} for the set of model parameters ${t=1}$, ${t'=0}$, ${U=4}$, and ${\beta=4}$, and different levels of hole-doping $\delta$. 
The green dots represent the \mbox{D-TRILEX} results, while the blue and orange lines represent \mbox{DiagMC@DF} and \mbox{DiagMC@PBDT}, respectively. 
The results were obtained summing all the diagrams up to 6th order for both DiagMC schemes.

Fig.~\ref{fig:small_doping} shows that in the regime close to half-filling \mbox{D-TRILEX} accurately describes the self-energy of the system. 
In this regime \mbox{D-TRILEX} even outperforms the \mbox{DiagMC@PBDT} approach. 
In the small-doping regime, the minimum of the imaginary part of the self-energy lies at the X point and a secondary minimum appears at the nodal point N, lying in the middle between M and $\Gamma$.
This picture changes for larger values of doping. 
Indeed, for doping between $\delta=10\%$ and $\delta=15\%$, the exact \mbox{DiagMC@DF} solution displays a change in the position of the imaginary self-energy minimum, which is shifted in an intermediate incommensurate position between X- and M-points. 
The \mbox{DiagMC@PBDT} result mimics this behavior even if the effect is smaller, while the \mbox{D-TRILEX} result fails to capture this effect and instead shows a minimum at the X-point. Apart from the small shift in the minimum around the X-point, the \mbox{D-TRILEX} and \mbox{DiagMC@PBDT} results appear to be in quantitative good agreement up to ${\delta=15\%}$. 
The largest discrepancy between the \mbox{D-TRILEX} and \mbox{DiagMC@DF} results for the ${{\rm Im}\,\Sigma}$ appears at the M-point and increases with increasing the doping level.

These results show that the \mbox{D-TRILEX} method appears to be in good agreement with DiagMC@DF up to some value of doping, around ${10-15\%}$. 
In this regime, the longitudinal spin fluctuations provide the largest contribution to the self-energy. 
This is signaled by a large value of the leading AFM eigenvalue of the dual Dyson equation, which goes from ${\lambda=0.81}$ at half-filling to ${\lambda =0.72}$ at ${\delta=10\%}$. 
The AFM eigenvalue drops significantly above this value of doping and becomes ${\lambda =0.62}$ at ${\delta=15\%}$ and ${\lambda =0.48}$ at ${\delta=20\%}$, signaling that the spin fluctuations are strongly suppressed by doping. 
The discrepancy between the \mbox{DiagMC@DF} and \mbox{DiagMC@PBDT} results suggests that the non-local transverse fluctuations are responsible for the shift of the minimum away from the X point, but they are not sufficient to accurately describe the behavior of the imaginary self-energy at the M-point. Therefore, the mismatch at the M point is related to irreducible contributions to the fermion-fermion vertex that are not accounted for by the partially bosonized approximation for the four-point vertex function~\eqref{eq:Gamma_approx_app}.

Our results suggest that the \mbox{D-TRILEX} method provides reliable results in the vicinity of the pseudogap regime where spin fluctuations are large. 
This regime is the most difficult regime to perform calculations with weak coupling methods like FLEX~\cite{PhysRevLett.62.961, Bickers89}. 
The opening of a pseudogap and the dichotomy between the N and AN points in this model has been studied recently in Ref.~\cite{PhysRevB.96.041105} in the framework of the exact DiagMC method. 
In that work, the authors considered the following set of model parameters ${t=1}$, ${t'=-0.3}$, ${U=5.6}$, ${\beta=5}$, and $4\%$ hole-doping that leads to a largest onset temperature for the pseudogap.
In Ref.~\cite{PhysRevB.103.245123}, we addressed this physically interesting regime for a comparable hole doping level of $3.4\%$ within the \mbox{D-TRILEX} and \mbox{DiagMC@DF} approaches. 
For this calculation, we could not converge a \mbox{D-TRILEX} calculation based on the DMFT impurity problem and introduced an additional outer loop to update self-consistently the parameters of the impurity problem. 
To stress this difference with the other calculations, we introduce the label \mbox{scD-TRILEX} for this specific calculation. 
The obtained self-energies are compared with the exact DiagMC result that was provided by the authors of Ref.~\cite{PhysRevB.96.041105}.
For the sake of consistency, the \mbox{DiagMC@DF} expansion was performed based on the impurity problem determined within the \mbox{scD-TRILEX} approach. 

\begin{figure}[t!]
\centering
\includegraphics[width=0.7\textwidth]{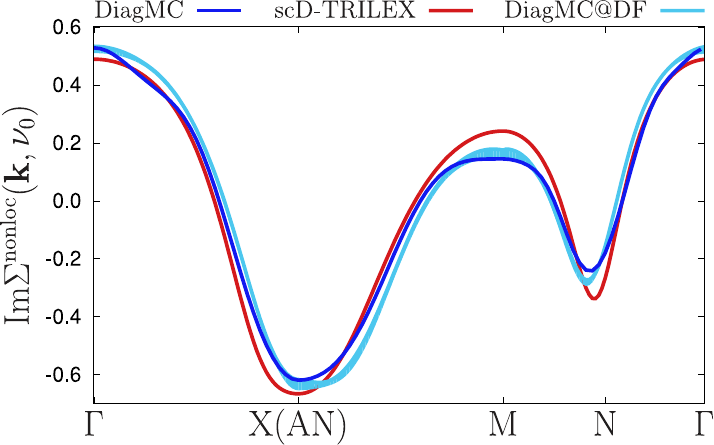}
\caption{\label{fig:U5.6_gmx} The imaginary part of the non-local self-energy obtained for the lowest Matsubara frequency ${\nu_0=\pi/\beta}$ along the high symmetry path in the Brillouin zone. Calculations are performed for ${t=1}$, ${t'=-0.3}$, ${U=5.6}$, and ${\beta=5}$ using \mbox{scD-TRILEX} (red line) and \mbox{DiagMC@DF} (light blue line) methods for $3.4\%$ hole-doping. 
The \mbox{DiagMC} result (dark blue line) for $4\%$ doping is provided by the authors of Ref.~\cite{PhysRevB.96.041105}. The Figure is taken from Ref.~\cite{PhysRevB.103.245123}.}
\end{figure}

In Fig.~\ref{fig:U5.6_gmx} we compare the imaginary part of the non-local self-energy $\Sigma^{\rm nonloc}_{{\bf k},\nu_0}$ calculated for the lowest Matsubara frequency along the high-symmetry path in momentum-space for all three approaches.
To obtain this quantity we subtract the local part $\Sigma^{\rm loc}_{{\bf k},\nu_0}$ from the lattice self energy $\Sigma^{\rm latt}_{{\bf k},\nu_0}$, where ${\Sigma^{\rm loc}_{{\bf k},\nu_0} = \sum_{\bf k}\Sigma^{\rm latt}_{{\bf k},\nu_0}}$.
Due to the lack of reference DiagMC data, the sum over the Brillouin zone in this expression is approximated by the sum over the high-symmetry path in momentum space.
We find that the nonlocal part of the \mbox{DiagMC@DF} self-energy is in a very good agreement with the reference DiagMC result. 
The \mbox{scD-TRILEX} approach also performs remarkably good in this physically nontrivial regime, especially given that the considered value of the local Coulomb interaction ${U=5.6}$ exceeds the half of the bandwidth.  
This good agreement in ${{\rm Im}\,\Sigma^{\rm nonloc}_{{\bf k},\nu_0}}$ indicates that the simple ladder-like \mbox{scD-TRILEX} method accurately captures the N/AN dichotomy in the formation of a pseudogap in this regime~\cite{PhysRevB.96.041105}.
This fact additionally confirms our finding that going away from the Slater regime allows to use less sophisticated methods to capture the effect of collective fluctuations. 

At the same time we find that the \mbox{DiagMC@DF} and the \mbox{scD-TRILEX} methods do not provide a good value for the local part of the lattice self-energy.
Indeed, ${{\rm Im}\,\Sigma^{\rm loc}_{{\bf k},\nu_0}}$ of \mbox{DiagMC@DF} calculated for the lowest Matsubara frequency is equal to $-0.77$.
The corresponding value for the \mbox{scD-TRILEX} approach is $-0.80$, while the exact DiagMC result reads $-1.04$.   
This discrepancy can again be explained by the fact that DMFT impurity problem does not provide a good starting point for a diagrammatic expansion already for moderate interactions.
To address this issue, we exploited the dual self-consistency condition to update the fermionic hybridization as an attempt for the improvement of the reference system.
However, the result obtained in this Section clearly demonstrates the need for an even better starting point, which should be able to provide more accurate local quantities to reproduce the exact result.

\subsubsection{Single-band extended Hubbard model at half filling}

Let us now investigate the performance of \mbox{D-TRILEX} in the case of a half-filled single-orbital extended Hubbard model on a square lattice with the nearest-neighbor hopping ${t=1}$.
To benchmark our results, we compare the lattice self-energy $\Sigma$ calculated using \mbox{D-TRILEX} with the exact solution of the dual action~\eqref{eq:DB_action} provided by the dual boson diagrammatic Monte Carlo (\mbox{DiagMC@DB}) method (see Section~\ref{sec:DiagMC}).
In Ref.~\cite{10.21468/SciPostPhys.13.2.036}, the calculations were performed for different strengths of the local interaction ${U=2}$, ${U=4}$, and ${U=6}$. 
For each value of $U$ we considered three different values of the nearest-neighbor Coulomb interaction $V^{d}$: ${V^{d}=0}$, ${V^{d}=U/8}$, and ${V^{d}=U/4}$.
To simplify notations, in the following the superscript ``$d$'' for the non-local interaction is omitted.
For ${U=2}$ and ${U=4}$ the temperature is set to ${T=0.25}$, for ${U=6}$ to ${T=0.50}$. 

\begin{figure}[t!]
\centering
\includegraphics[width=1\linewidth]{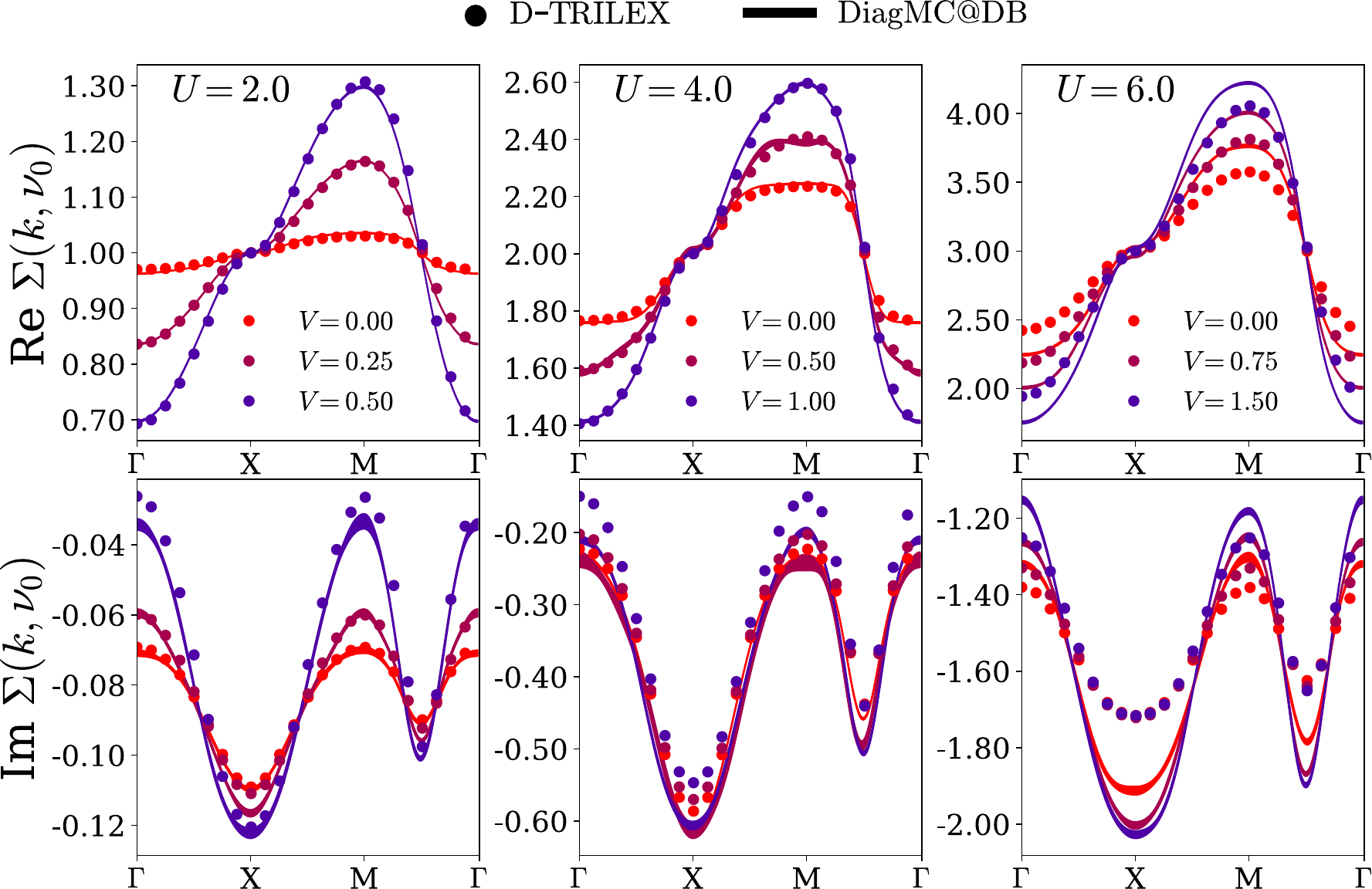}
\caption{Real (top row) and imaginary (bottom row) parts of the lattice self-energy for the half-filled single-band extended Hubbard model on a square lattice. The result is obtained for ${U=2.0}$ (left column), ${U=4.0}$ (middle column), and ${U=6.0}$ (right column) for three different values of the nearest-neighbor interaction ${V=0.0}$ (light red), ${V=U/8}$ (dark red), and ${V=U/4}$ (purple). The \mbox{D-TRILEX} result is depicted by dots. The \mbox{DiagMC@DB} data is taken from Ref.~\cite{PhysRevB.102.195109} and is represented by solid lines with the width that corresponds to the estimated stochastic error. The Figure is taken from Ref.~\cite{10.21468/SciPostPhys.13.2.036}.
\label{fig:U-V_Sigma}}
\end{figure}

The obtained results for the lattice self-energy are shown in Fig.~\ref{fig:U-V_Sigma}.
We find that for the smallest value of the Hubbard interaction ${U=2}$ the agreement between the two methods is almost perfect.
A slight difference appears only in the imaginary part of the self-energy in the vicinity of the ${X=(\pi, 0)}$ point for ${V=0.25}$ and near the ${\Gamma = (0, 0)}$ point for ${V=0.5}$. 
When the interaction reaches the value of the half-bandwidth ${U=4}$, the real part of the \mbox{D-TRILEX} self-energy remains very close to the \mbox{DiagMC@DB} result for all values of $V$ considered here. 
On the other hand, we observe a constant shift in the imaginary part of the self-energy that increases with the strength of the non-local interaction $V$.
A constant but smaller shift was also observed between DF and \mbox{DiagMC@DF} results~\cite{PhysRevB.96.035152,PhysRevB.103.245123}, hence it does not seem to be a feature of only the \mbox{D-TRILEX} method. Finally, at ${U=6.0}$ we observe that \mbox{D-TRILEX} does not agree with \mbox{DiagMC@DB} as accurately as for smaller values of the interaction.
In the real part, the difference between the two methods is not very large and appears to be independent on the value of $V$.
On the contrary, the imaginary part of the self-energy displays a rather large mismatch already at ${V=0}$, and the agreement seems to become worse as $V$ increases. 
This result comes as no surprise and agrees with the findings of Refs.~\cite{PhysRevB.96.035152, PhysRevB.102.195109, PhysRevB.103.245123} that the ladder-like dual approximations become less accurate in the regime of strong magnetic fluctuations. 
The reason is that magnetic fluctuations become strongly non-linear close to a magnetic instability (see, e.g., Ref.~\cite{PhysRevB.102.224423}).
This non-linear behavior originates from the mutual interplay between different bosonic modes as well as from an anharmonic fluctuation of the single mode itself.
The description of these effects requires to consider much more complex diagrammatic structures that account for vertical (transverse) insertions of momentum- and frequency-dependent bosonic fluctuations, which are present in the DiagMC@DB approach but are not considered in ladder-like dual approximations including the \mbox{D-TRILEX} approach.
However, despite the quantitative disagreement, at ${U=6.0}$ \mbox{D-TRILEX} qualitatively captures the correct momentum dependence of the self-energy, which is completely missing in DMFT. 

\subsubsection{Two-orbital Hubbard-Kanamori dimer}
\label{sec:dimer}

After investigating the performance of the \mbox{D-TRILEX} approach in the single-band case we turn to a multi-orbital two-site model, also known as dimer, considered in Ref.~\cite{vandelli2022quantum}. 
Due to a small size of this system, the exact solution for the dimer problem for small number of orbitals can be achieved by ED.
This makes the dimer an ideal platform to benchmark various approximate methods.
To test our multi-orbital \mbox{D-TRILEX} implementation, we consider a particular case of a Hubbard-Kanamori dimer, where each of the two identical sites has two degenerate orbitals. 
The single-particle part of the corresponding Hamiltonian reads:
\begin{align}
H_0 = - t \sum_{l,\sigma}\sum_{j\neq{}j'} c^{\dagger}_{j\sigma{}l} c^{\phantom{\dagger}}_{j'\sigma{}l}\,.
\label{eq:H0_dimer}
\end{align}
The single-particle Hamiltonian~\eqref{eq:H0_dimer} can be diagonalized in the site-space.
After that, the dimer problem can be effectively considered as a periodic system with the dispersion:
\begin{align}
\varepsilon_{\kv, ll'} = - 2 t \cos({\bf k}) \, \delta_{ll'}\,,
\label{eq:kanamori_dimer}
\end{align}
defined for ${N_k = 2}$ points in momentum space that correspond to ${{\bf k}=0}$ (symmetric solution) and ${{\bf k}=\pi}$ (anti-symmetric solution).
Based on this consideration, we can apply our multi-band \mbox{D-TRILEX} method, that is designed for solving periodic lattice models, to this benchmark system.
The interacting part is considered in the Kanamori form~\cite{10.1143/PTP.30.275, Hunds_metals1} as: 
\begin{align}
H_{U} = U \sum_{l} n_{l\uparrow} n_{l\downarrow} + \sum_{l \neq l'} \left\{ U'n_{l\uparrow} n_{l'\downarrow} + 
\frac12(U'-J) \sum_{\sigma} n_{l\sigma} n_{l'\sigma}
- J c^{\dagger}_{l\uparrow} c^{\phantom{\dagger}}_{l\downarrow} c^{\dagger}_{l'\downarrow} c^{\phantom{\dagger}}_{l'\uparrow}
+ J c^{\dagger}_{l\uparrow} c^{\dagger}_{l\downarrow} c^{\phantom{\dagger}}_{l'\downarrow} c^{\phantom{\dagger}}_{l'\uparrow} \right\}.
\label{eq:S_Kanamori}
\end{align}
Here, $c^{(\dagger)}_{l\sigma}$ is the annihilation (creation) operator for an electron at the orbital $l$ with the spin projection $\sigma\in\{\uparrow,\downarrow\}$.
${n_{l\sigma} = c^{\dagger}_{l\sigma} c^{\phantom{\dagger}}_{l\sigma}}$ is the local spin-dependent density.
In the notation of Eq.~\eqref{eq:actionlatt}, the non-zero components of the Kanamori interaction are
\begin{alignat}{2}
    U^{pp}_{llll\phantom{''}} &= U \quad &&\text{intraorbital density-density} \notag\\
    U^{pp}_{ll'll'} &= U' \quad &&\text{interorbital density-density}\notag\\
    U^{pp}_{ll'l'l} &= J \quad &&\text{pair hopping} \notag\\
    U^{pp}_{lll'l'} &= J \quad &&\text{spin flip}
    \label{eq:U_kanamori_int}
\end{alignat}
We also fix ${U' = U - 2J}$ to ensure rotational invariance~\cite{Hunds_metals1}.
In the single-orbital case, the on-site Coulomb interaction reduces to a Hubbard form given by the first term in Eq.~\eqref{eq:U_kanamori_int}.

We chose the single-site two-orbital impurity problem of DMFT as the reference system for the \mbox{D-TRILEX} calculation.
Since interorbital hoping processes are not taken into account, different orbitals do not hybridize.
For the case of degenerate orbitals considered here it implies that the Green's function is diagonal in the orbital space and has identical components for both orbitals (${G_{ll'}=G\delta_{ll'}}$). 
To compare the \mbox{D-TRILEX} result with the exact solution for the dimer problem we perform ED calculations using the pomerol package~\cite{krivenko_igor_2021_5739623}.
The total number of degrees of freedom for the two-orbital Hubbard-Kanamori dimer for the ED calculation is ${2 N_{ l} N_{\rm imp} = 8}$ and the total number of states is ${N_{\rm tot} = 2^{8} = 256}$.
This makes the ED calculation numerically inexpensive.

\begin{figure}[t!]
\centering
\includegraphics[width=1\linewidth]{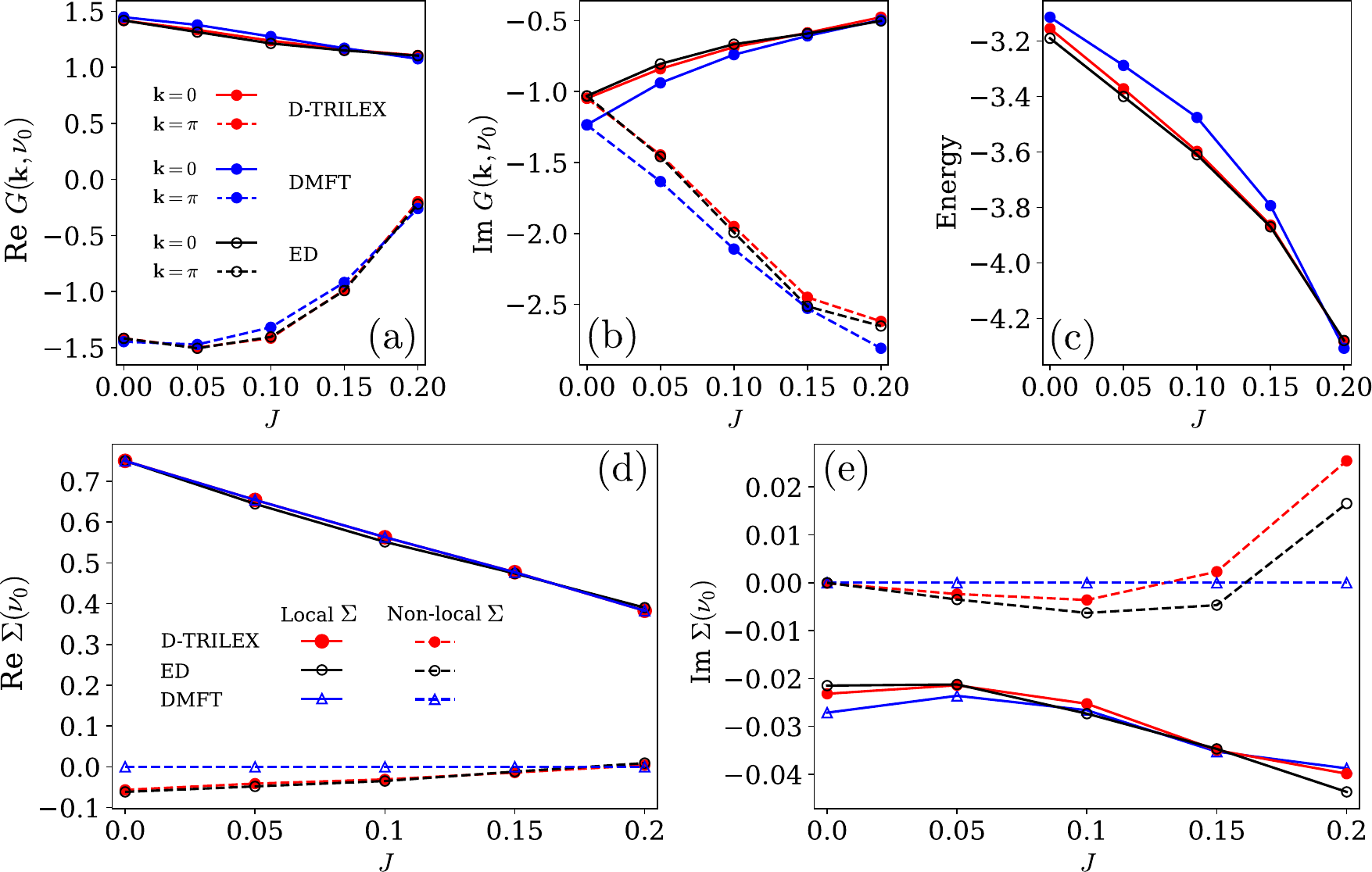} 
\caption{Panels (a) and (b) respectively show the real and the imaginary parts of the Green's function for the Hubbard-Kanamori dimer calculated for the frequency ${\nu_0 = \pi/\beta}$ at momenta ${{\bf k}=0}$ (solid line) and ${{\bf k}=\pi}$ (dashed line). 
The panel (c) shows the average energy of the system. 
Panels (d) and (e) respectively show the real and the imaginary parts of the self-energy $\Sigma$ for the frequency ${\nu_0 = \pi/\beta}$. The local component is denoted by a thick line, while the non-local component is represented by a dashed line. 
Non-local components are identically zero for DMFT, but are displayed for consistency.
Results obtained using \mbox{D-TRILEX} (red), DMFT (blue), and ED (black) methods for different values of the Hund's coupling $J$.
Model parameters for these calculations are ${t=0.2}$, ${U=0.5}$, ${\beta=10}$, and ${\mu=0.75}$, and are equal across all panels. The Figure is taken from Ref.~\cite{vandelli2022quantum}.
\label{fig:gf_vs_J}
}
\end{figure}

First, we focus on the effect of the Hund's exchange coupling $J$. 
To this aim we perform calculations for different values of $J$ fixing other model parameters to
${t=0.2}$, ${U=0.5}$, ${\beta=10}$, and ${\mu=0.75}$.
A very similar set of model parameters for a single-orbital dimer problem was recently used in Ref.~\cite{van_Loon_2021} to benchmark another diagrammatic extension of DMFT.
In Fig.\ref{fig:gf_vs_J}, we show the real (${{\rm Re}\,G}$, left panel) and imaginary (${{\rm Im}\,G}$, middle panel) parts of the Green's function produced by \mbox{D-TRILEX} (red), DMFT (blue), and ED (black) methods. 
We find that the \mbox{D-TRILEX} result for the ${{\rm Re}\,G}$ lies on top of the exact solution in the whole range of values for the Hund's coupling considered here.
DMFT is also rather accurate in calculating the real part of the Green's function, but the discrepancy between the DMFT and ED results is noticeable. 
The \mbox{D-TRILEX} solution for ${{\rm Im}\,G}$ is very close to the one provided by ED, while the DMFT result becomes substantially different from the exact solution, especially for small values of $J$.
A very good agreement between \mbox{D-TRILEX} and ED methods is also confirmed by analyzing the result for the average energy $\left\langle E \right\rangle$ (right panel in Fig.~\ref{fig:gf_vs_J}). 
The average energy for ED is obtained as 
   $\left\langle E \right\rangle_{\rm ED} = \sum_i (E_i-\mu) e^{-\beta (E_i-\mu)} $ where the index $i$ runs over the eigenstates of the system.
The average energy in \mbox{D-TRILEX} and DMFT is calculated as explained in Ref.~\cite{10.21468/SciPostPhys.13.2.036}. 
We show that the mismatch in \mbox{D-TRILEX} and ED results for the energy is ${1.1 \%}$ (${\delta E=0.034}$) at ${J=0}$ and decreases as $J$ increases. 
The largest difference between DMFT and ED results is found at $J=0.1$ and amounts to ${3.7 \%}$ (${\delta E=0.134}$), which is approximately four times larger than the one of the \mbox{D-TRILEX} approach. 
Nevertheless, we observe that in this case DMFT is surprisingly close to the exact result. 
The reason is that for the considered set of model parameters the system lies very far away from half-filling, hence the non-local fluctuations between the two sites of the dimer are suppressed.
This fact can be confirmed by looking at the self-energy $\Sigma$ shown in panels (d) and (e) of Fig.~\ref{fig:gf_vs_J}. 
The local contribution to the self-energy ${2\Sigma^{\rm local} = \Sigma({\bf k}=0) + \Sigma({\bf k}=\pi)}$ is dominant and is in a very good agreement among all three methods.
The non-local part ${2\Sigma^{\rm non-local} = \Sigma({\bf k}=0) - \Sigma({\bf k}=\pi)}$, which is completely missing in DMFT, is relatively small and is also well reproduced by \mbox{D-TRILEX} approach.

\begin{figure}[t!]
\centering
\includegraphics[width=1\linewidth]{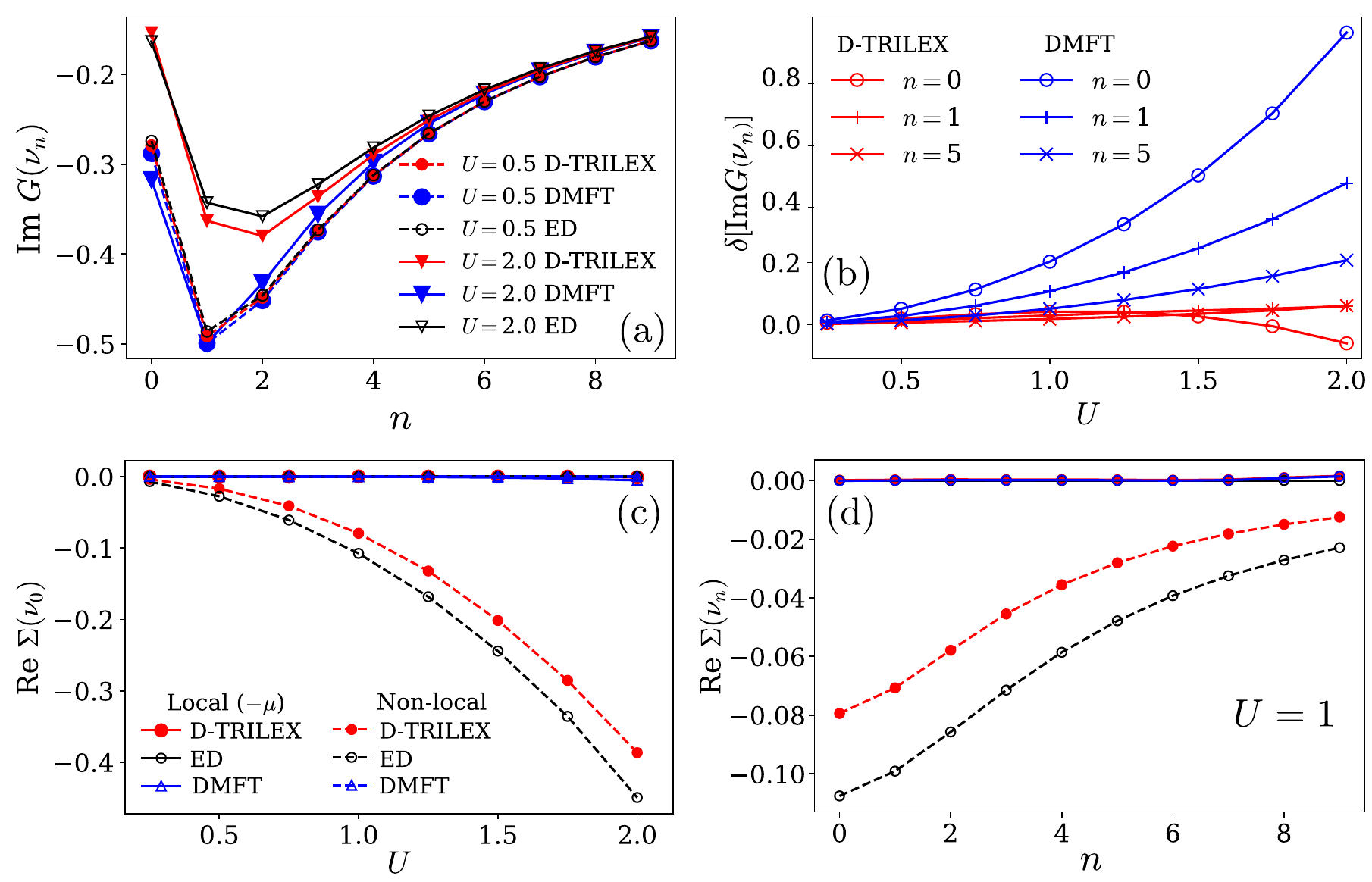} 
\caption{(a) Imaginary part of the local Green's function ${{\rm Im}\,G(\nu_{n})}$
calculated as a function of the Mastubara frequency index $n$ for two values of the interaction ${U=0.5}$ (dashed lines) and ${U=2.0}$ (solid lines). The result is obtained at half-filling for ${t=0.5}$, ${\beta=10}$, and ${J=U/4}$ using the \mbox{D-TRILEX} (red), the DMFT (blue), and the ED (black) methods. 
(b) Normalized difference ${\delta\left[{\rm Im} G(\nu_{n})\right]}$~\eqref{eq:norm_diff_dimer} with respect to the ED solution calculated for \mbox{D-TRILEX} (red), the DMFT (blue) methods as a function of $U$. The result is obtained for three difference Matsubara frequencies with indices ${n=0}$ (empty circles), ${n=1}$ (pluses), and ${n=5}$ (crosses). 
(c) Real part of the self-energy $\Sigma$ as a function of the interaction $U$ at the first Matsubara frequency $\nu_0$. 
(d) Real part of the self-energy $\Sigma(\nu_n)$ as a function of the Matsubara index obtained at ${U=1}$. In panels (c) and (d), the local and non-local parts are shown and the chemical potential $\mu$ is subtracted from the local part. 
\label{fig:gf_vs_UJ}
}
\end{figure}

At half-filling, DMFT ceases to be a good approximation. 
To illustrate that \mbox{D-TRILEX} is able to improve and even to cure a wrong behavior of the DMFT result, we perform calculations for ${t=0.5}$ and ${\beta=10}$ for different values of the Hubbard interaction $U$ for a fixed ratio ${U/J=4}$.
The chemical potential is set to ${\mu = (3 U - 5 J)/2}$ in order to ensure half-filling~\cite{PhysRevB.83.205112}.
Panel (a) of Fig.~\ref{fig:gf_vs_UJ} shows the imaginary part of the local Green's function as a function of the Matsubara frequency.
The result is obtained in a weak ($U=0.5$, dots) and strong coupling ($U=2.0$, triangles) regimes of the interaction.
At ${U=0.5}$, the \mbox{D-TRILEX} result coincides with the exact solution in the whole frequency range.
The DMFT result is also very accurate and only slightly deviates from the ED solution at lowest frequencies.
This situation changes completely at ${U=2.0}$, where the exact ${{\rm Im}\,G}$ provided by ED is strongly reduced at low frequencies.
Remarkably, DMFT does not capture this change and predicts approximately the same result for the ${{\rm Im}\,G}$ for both values of the interaction.
Instead, the \mbox{D-TRILEX} solution lies very close to exact result and reproduces the correct behavior of the ${{\rm Im}\,G}$. 
To confirm this fact, we compute the normalized difference from the ED result for \mbox{D-TRILEX} and DMFT as:
\begin{align}
\delta\left[{\rm Im}\,G(\nu_{n})\right] = {\rm Im}\left[ G(\nu_{n}) - G_{\rm ED}(\nu_{n})\right]/{\rm Im}\,G_{\rm ED}(\nu_{n})\,.
\label{eq:norm_diff_dimer}
\end{align}
The corresponding result obtained for three different frequencies as a function of $U$ is shown in the panel (b) of Fig.~\ref{fig:gf_vs_UJ}. 
We find that the normalized difference for DMFT is relatively large and drastically increases upon increasing the interaction strength.
At ${U=2.0}$, the ${{\rm Im}\,G({\nu})}$ calculated at the zeroth and the first Matsubara frequency using DMFT is respectively almost two and $1.5$ times larger than the exact result.
On the contrary, the ${{\rm Im}\,G({\nu})}$ of \mbox{D-TRILEX} lies very close to the ED result.
Indeed, the normalized difference for \mbox{D-TRILEX} calculated for the first and the fifth frequency does not exceed $2\%$.  
The difference for \mbox{D-TRILEX} calculated for the zeroth frequency becomes larger than $2\%$ at ${U>1.5}$ and reaches the maximum value of $7.6\%$ at ${U=2.0}$. 
We find that the DMFT result strongly deviates from the considered benchmark at moderate and large values of $U$.
By looking at the real part of the self-energy ${\rm Re}\Sigma$ (panels (c) and (d) in Fig.~\ref{fig:gf_vs_UJ}), we can immediately understand the origin of the large mismatch between ED and DMFT. As a matter of fact, the real part of the self-energy at moderate to large $U$ is dominated by the non-local contributions  (dashed lines), which are completely missing in DMFT, while local contributions (solid lines) are approximately zero. {\mbox D-TRILEX} does not exactly reproduce all the contributions to the non-local self-energy, as they correspond to roughly 25\% of the value of self-energy at ${U=1}$. However, it follows the same trend as the ED result and this ensures the correct behavior of the Green's function as $U$ is increased. We do not show the imaginary part of the self-energy, since it is at least an order of magnitude smaller than the real part in the whole range of parameters considered here.

\begin{figure}[t!]
\centering
\includegraphics[width=1\linewidth]{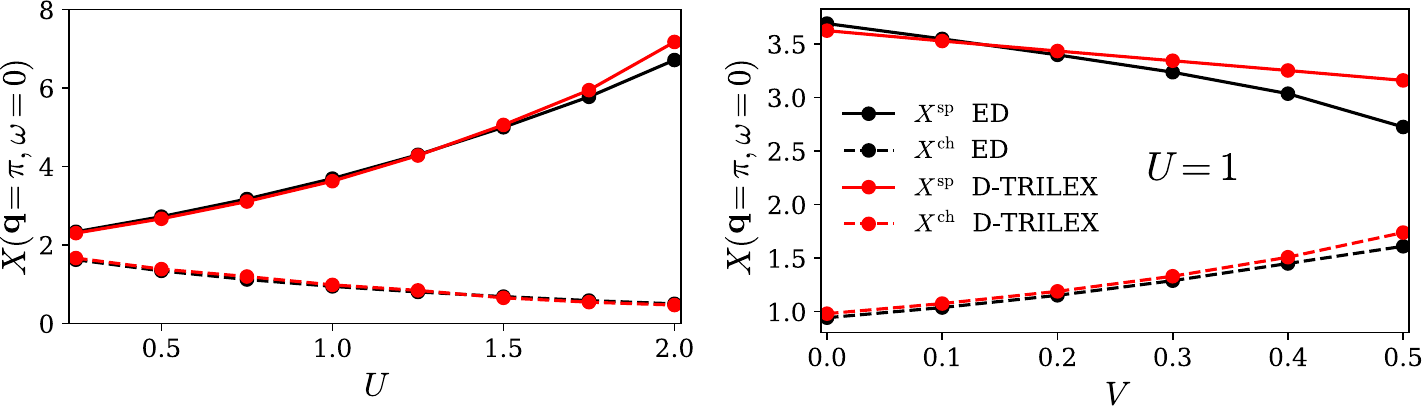} 
\caption{Spin (full lines) and charge (dashed lines) static (${\omega=0}$) susceptibility $X$ obtained at the ${{\bf q}=\pi}$ point for the half-filled system with model parameters ${t=0.5}$, ${J=U/4}$, and ${\beta=10}$. 
The left panel shows the result as a function of the local interaction strengths $U$ in the absence of the non-local interaction ($V=0$). 
The right panel illustrates the susceptibility as a function of $V$ calculated for the fixed value of the local interaction ${U=1}$.
\label{fig:X_vs_UJ}
}
\end{figure}

In addition to single-particle quantities we calculate the charge and spin susceptibilities defined as ${X^{\rm ch/sp} = - \sum_{ll'} X^{d/m}_{lll'l'}}$.
Fig.~\ref{fig:X_vs_UJ} shows the corresponding results for the static susceptibilities ${X^{\rm ch/sp}({\bf q},\omega=0)}$ obtained at the ${{\bf q}=\pi}$ point. 
The susceptibilities at the ${{\bf q}=0}$ point are very small in the whole range of considered parameters and are not shown here. 
In the left panel of Fig.~\ref{fig:X_vs_UJ}, we illustrate the results for the half-filled Hubbard-Kanamori dimer considered above. 
We find that the susceptibilities of the \mbox{D-TRILEX} approach are in a very good agreement with the exact ED solution in the whole range of local interaction strength ${0.25 \leq U \leq 2}$. 
In the right panel, we demonstrate the dependence of the static charge and spin susceptibilities on the value of the non-local interaction $V^{d}$ between electronic densities on neighboring sites $\langle i, j \rangle$~\eqref{eq:actionlatt}.
More explicitly, we consider the non-local interaction in the form:
\begin{align}
\frac{V^{d}}{2}\sum_{ll',i\neq j} \rho^{d}_{i, ll} \rho^{d}_{j, l'l'}\,.
\end{align}
To simplify notations, in the following the superscript ``$d$'' for the non-local interaction is omitted. 
We find that in the presence of the non-local interaction the susceptibilities obtained using ED and \mbox{D-TRILEX} methods are nearly identical up to ${V=0.3}$. 
Above that threshold, the \mbox{D-TRILEX} susceptibility starts to deviate from the exact ED result. 
At ${V>0.3}$ the \mbox{D-TRILEX} spin susceptibility continues to decrease almost linearly with increasing the value of $V$, while the exact result shows a stronger non-linear damping. 
This trend continues also above $V=0.5$, where the difference between the D-TRILEX result and the exact result continues to increase.
This behavior can be explained by the fact that the strong non-local interaction favors either full or zero occupancy of a lattice site.
This charge density wave instability strongly suppresses magnetic fluctuations. 
In this regime the \mbox{D-TRILEX} calculations break down, because they are performed on the basis of the DMFT impurity problem, which does not incorporate any effect of the non-local interaction.
The inclusion of the bosonic hybridization function in the spirit of EDMFT could improve the result, as in this case some contributions of the non-local interaction would be taken into account in the impurity problem via the bosonic hybridization.

These findings show that \mbox{D-TRILEX} improves the DMFT results in all considered regimes. 
Additionally, \mbox{D-TRILEX} reproduces the trends observed in ED calculations in all the cases, even when DMFT fails. 
This fact suggests that for the considered system the difference between DMFT and ED mostly stems from non-local correlations that have the form accounted for in the \mbox{D-TRILEX} diagrams.
These results are particularly remarkable taking into account that DMFT approximation is not very accurate in low dimensions, hence the DMFT impurity problem is probably not an optimal reference system for a diagrammatic expansion in this case.

\subsection{Investigation of model systems}

In this section we briefly highlight applications of the \mbox{D-TRILEX} approach to various model systems. 

\subsubsection{Orbital isotropy of magnetic fluctuations in correlated electron materials
induced by Hund's exchange coupling}

Perovskite materials with partially filled $t_{2g}$ orbitals reveal a high degree of anisotropy and may serve as an attractive playground for studying the interplay between orbital and spin degrees of freedom.
Prominent examples are LaTiO$_3$, SrRuO$_3$ and, to a lesser degree, Sr$_2$RuO$_4$. 
In Ref.~\cite{PhysRevLett.127.207205}, we address the spatial symmetry and orbital structure of magnetic fluctuations in perovskite materials in the framework of the \mbox{D-TRILEX} approach.
We find that non-local spin fluctuations enhanced by large Hund's exchange coupling strongly reduce the orbital anisotropy of the perovskite structure. 
As a consequence, magnetic fluctuations become isotropic in orbital space, as we show both at half-filling, as well as for the case of $4$ electrons per lattice site.
These results illustrate the important role that the local Hund's coupling plays not only for the local spin physics, but also for the symmetry and orbital structure of spatial magnetic excitations.

In Ref.~\cite{PhysRevLett.127.207205} we considered a realistic $t_{2g}$ tight-binding model for the perovskite materials described by the three-orbital Hamiltonian:
\begin{align}
{\cal H} = - \sum_{ij,l,\sigma} t^{ll}_{ij} c^{\dagger}_{il\sigma} c^{\phantom{\dagger}}_{jl\sigma} + \frac12 \sum_{i,ll'} \left( U^{\rm ch}_{ll'} n^{\phantom{\dagger}}_{il} n^{\phantom{\dagger}}_{il'} +  U^{\rm sp}_{ll'} m^{\phantom{\dagger}}_{il} m^{\phantom{\dagger}}_{il'} \right),
\label{eq:Hamiltonian}
\end{align}
where operator $c^{(\dagger)}_{il\sigma}$ annihilates (creates) an electron with spin projection ${\sigma=\{\uparrow, \downarrow\}}$ on site $i$ and orbital ${l=\{1, 2, 3\}}$. 
The anisotropy of this model originates from hopping parameters $t^{ll}_{ij}$ that are diagonal in the orbital space and have the following structure in momentum ({\bf k}) space~\cite{Pavarini_2005}:
\begin{align}
t_{ll}({\bf k}) = \epsilon + 2t_\pi ({\cal C}_\alpha + {\cal C}_\beta)+2t_\delta {\cal C}_\gamma +4t_\sigma {\cal C}_\alpha {\cal C}_\beta\,,
\label{eq:dispersion}
\end{align}
where $\epsilon$ is the center of bands and ${{\cal C}_{\alpha} = \cos k_{\alpha}}$. 
For simplicity, we introduce three non-equivalent $\alpha$, $\beta$, $\gamma$ indices, where the first two are defined by the orbital label ${l=\{\alpha\beta\}}$ with ${1 = yz}$, ${2 = zx}$, and ${3 = xy}$. 
The last index $\gamma$ takes the remaining value among ${\{x, y, z\}}$. 
In this model, orbital degrees of freedom are tied to a spatial motion of the electrons, because the latter can hop only within the same orbital in a strictly defined direction, which is different for every considered orbital. The
$t_{\pi, \delta, \sigma}$ matrix elements describe the main hopping processes that provide the $t_{2g}$ symmetry. 
We choose ${t_\pi=1}$, which defines the energy scale of the system, and a realistic value for ${t_\delta=0.12}$ for the SrVO$_3$ perovskite~\cite{Pavarini_2005}. 
We note that $t_\sigma$ plays the role of $t^\prime$ in a two-dimensional model for cuprates and shifts the van-Hove singularity (vHS) away from the Fermi level. 
The presence of the vHS at the Fermi energy results in a peak in the density of states, which enhances correlation effects in the system.
Thus, for the half-filled case (${\langle N_i \rangle = 3}$ electrons per site) we preserve the particle-hole symmetry for $t_{2g}$ bands and set $t_\sigma=0$. 
For the case of ${\langle N_i \rangle = 4}$ we choose the positive value for ${t_\sigma=0.35}$~\cite{Pavarini_2005}, which ensures that the vHS again appears at the Fermi level.
The interaction is parametrized in the Kanamori form~\cite{10.1143/PTP.30.275} with intraorbital $U$ and interorbital $U'$ Coulomb interactions, and the Hund's coupling $J$. 
This parametrization is rotationally invariant provided ${U'=U-2J}$. Given that the matrix of hopping amplitudes is diagonal in orbital space, we consider only the density-density part of the Kanamori interaction:  
\begin{align}
2U^{\rm ch} &= 
\begin{pmatrix}
U & U^{*} & U^{*} \\
U^{*} & U & U^{*} \\
U^{*} & U^{*} & U
\end{pmatrix}, ~~~~
2U^{\rm sp} = 
\begin{pmatrix}
-U & -J & -J \\
-J & -U & -J \\
-J & -J & -U
\end{pmatrix},
\end{align}
where ${U^{*} = 2U'-J}$. 

\begin{figure}[t!]
\centering
\includegraphics[width=0.7\linewidth]{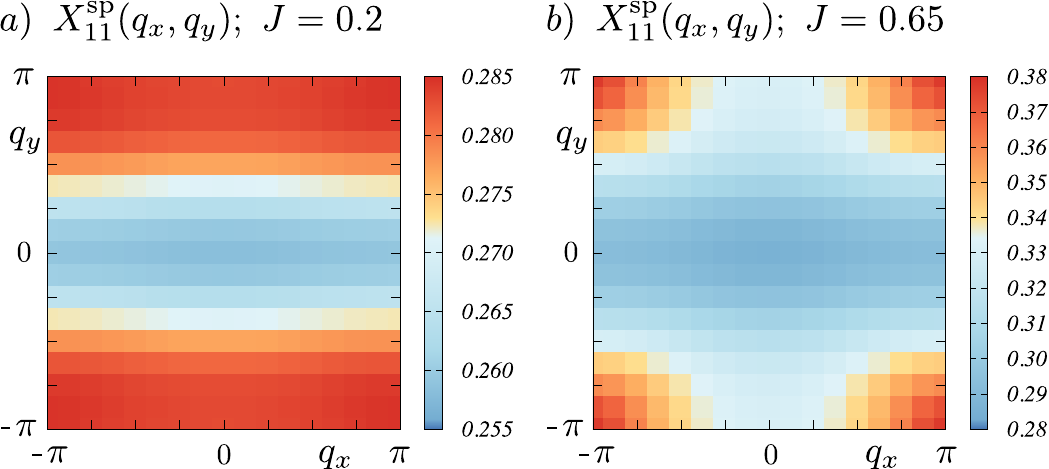}
\caption{\label{fig:W_hf} The absolute value of the diagonal $yz$ orbital component of the spin susceptibility $X^{\rm sp}_{11}(q_{x}, q_{y}; q_{z}=0, \omega=0)$ obtained for the half-filled $t_{2g}$ model for ${U=4}$. 
Color bars show the strength of $X^{\rm sp}$.
(a) In the case of small Hund's coupling $J=0.2$, the diagonal component of the susceptibility is highly anisotropic and is almost dispersionless along $q_{x}$ direction. (b) Increasing the Hund's coupling to $J=0.65$, intraorbital spin fluctuations become isotropic with a pronounced antiferromagnetic behavior depicted by the largest value of $X^{\rm sp}_{11}$ at corners of the Brillouin zone. The Figure is taken from Ref.~\cite{PhysRevLett.127.207205}.
}
\end{figure}

Remarkably, we find that the strength and orbital structure of spatial magnetic fluctuations are controlled by the value of the local Hund's coupling $J$.
To illustrate this point, let us first consider the interacting three-orbital model~\eqref{eq:Hamiltonian} at half-filling with ${U=4}$ and temperature ${T=1/2}$. 
For the specified parameters, the leading eigenvalue (l.e.) $\lambda$ of the Dyson equation for the  susceptibility $X_{ll'}$ indicates that strongest collective excitations in the system correspond to a magnetic instability channel.
We observe that for a relatively small ${J=0.2}$, the l.e. of magnetic fluctuations is not very large (${\lambda=0.78}$). 
In this case, the diagonal (intraorbital) part of the spin susceptibility $X^{\rm sp}_{ll}$, presented in Fig.~\ref{fig:W_hf}\,$a$ for the $yz$ orbital, is highly anisotropic in momentum space and is almost dispersionless along $q_{x}$ direction. 
This spatial structure of the spin susceptibility originates from the orbital symmetry of $t_{2g}$ hopping processes~\eqref{eq:dispersion}. 
This result indicates that for small Hund's coupling, orbital degrees of freedom are anisotropic.

\begin{figure}[b!]
\centering
\includegraphics[width=0.7\linewidth]{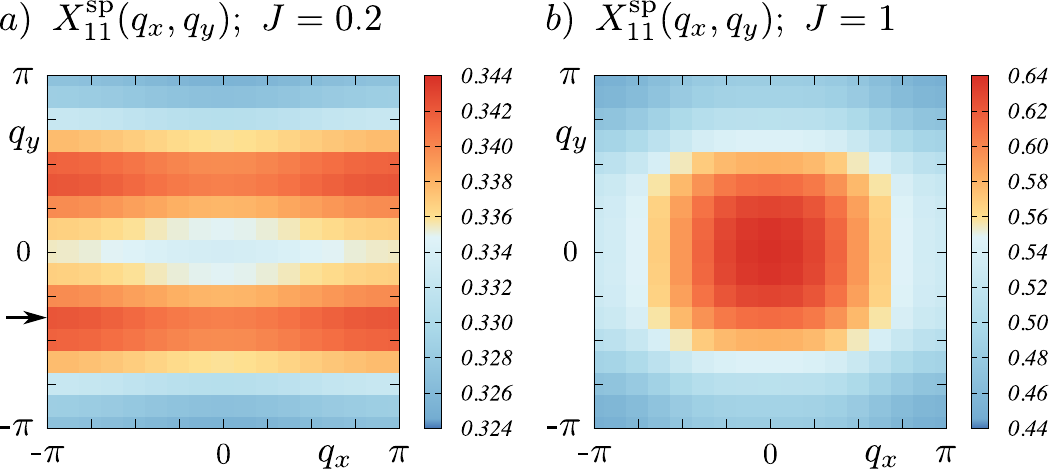}
\caption{\label{fig:W_dop} The absolute value of the diagonal $yz$ orbital component of the spin susceptibility ${X^{\rm sp}_{11}(q_{x}, q_{y}; q_{z}=0, \omega=0)}$ obtained for $U=5$ for the doped $t_{2g}$ model.
(a) For small $J=0.2$, the diagonal susceptibility $X^{\rm sp}_{11}$ is highly anisotropic, and spin fluctuations correspond to an incommensurate spiral state associated with the momentum depicted by the black arrow. (b) For large $J=1$, intraorbital spin fluctuations are isotropic and ferromagnetic is shown by the symmetric bright spot at the center of the Brillouin zone. The Figure is taken from Ref.~\cite{PhysRevLett.127.207205}.
}
\end{figure}

Increasing the value of the Hund's coupling to ${J=0.65}$, the magnetic l.e. approaches unity (${\lambda=0.99}$), which indicates that spin fluctuations are strongly enhanced. 
This can also be concluded from Fig.~\ref{fig:W_hf} comparing the amplitude of the susceptibility for two considered cases of $J$.
Moreover, at this large value of the Hund's coupling interorbital components of $X^{\rm sp}$ become comparable to intraorbital ones (see Ref.~\cite{PhysRevLett.127.207205}). 
This is the first signature of the isotropic orbital behavior of magnetic fluctuations. 
A proximity of the l.e. to unity indicates that all orders of an effective perturbation expansion given by the Dyson equation contribute almost equally to the total $X^{\rm sp}$.
This leads to a more thorough mixing of orbital and spatial degrees of freedom in the susceptibility.
As shows Fig.~\ref{fig:W_hf}\,$b$, this results in a highly isotropic form of spin fluctuations with a clearly distinguishable antiferromagnetic (AFM) behavior. 
This means, that orbital degrees of freedom are no more tied to a specific spatial direction defined by hopping parameters~\eqref{eq:dispersion} of the considered model. As a consequence, collective fluctuations in the magnetic channel become orbitally isotropic.

Interestingly, similar effects can also be observed away from half-filling, where strong magnetic fluctuations are related to a completely different type of magnetic instability. 
Let us repeat the calculations for the same $t_{2g}$ model~\eqref{eq:Hamiltonian} for the case of ${\av{N_{i}}=4}$ and $U=5$. 
The diagonal component of the susceptibility is presented in Fig.~\ref{fig:W_dop}. 
For a small value of the Hund's coupling $J=0.2$ (${\lambda=0.71}$) the susceptibility is again nearly diagonal in the orbital space and is highly anisotropic. 
We also find that for the case of $\av{N_{i}}=4$ electrons per lattice site the spatial structure of spin fluctuations is considerably different from the half-filled case and corresponds to an incommensurate spiral state with momentum indicated in Fig.~\ref{fig:W_dop}\,$a$ by the black arrow. 
Nevertheless, the $q_{x}$ direction still remains almost dispersionless. 
Increasing the value of the Hund's coupling to $J=1$, the magnetic l.e. again approaches unity (${\lambda=0.91}$).
Straightforwardly, magnetic fluctuations become isotropic and exhibit a pronounced peak at the center of the Brillouin zone (see Fig.~\ref{fig:W_dop}\,$b$), which is associated with strong ferromagnetic (FM) fluctuations. 
This finding is reminiscent of the case of itinerant FM fluctuations in the $d^{4}$ compounds SrRuO$_3$ in its high-temperature paramagnetic phase~\cite{RevModPhys.84.253, PhysRevB.74.094104}. 

As we demonstrate in Ref.~\cite{PhysRevLett.127.207205}, the three-point vertex corrections considered in \mbox{D-TRILEX} are the driving force for the orbitally isotropic state of the spin fluctuations.
In their absence, at half filling the l.e. of magnetic fluctuations approaches unity (${\lambda=0.99}$) already for a relatively small value of the Hund's coupling ${J=0.4}$. 
However, the diagonal susceptibility $X^{\rm sp}_{11}$ remains anisotropic in momentum space.
In the case of ${\av{N_{i}}=4}$ the effect of the vertex corrections is even more substantial. 
At large Hund's coupling ${J=1}$ the l.e. of magnetic fluctuations stays nearly the same for both considered approaches.
At the same time, if vertex corrections are neglected the corresponding magnetic mode remains incommensurate as in the regime of small Hund's coupling.
These result show that the $\text{D-TRILEX}$ approach indeed provides a minimal consistent multi-orbital diagrammatic extension of DMFT that correctly describes collective electronic fluctuations.

To conclude, we have studied collective spin fluctuations in a realistic strongly interacting highly anisotropic three-orbital model. 
We have found that the Hund's coupling enhances collective electronic effects in the spin channel. 
Strong magnetic fluctuations efficiently mix orbital and spatial degrees of freedom leading to orbitally isotropic behavior of the system. 
Remarkably, this effect emerges independently of the antiferromagnetic or ferromagnetic nature of spin fluctuations.
The vertex corrections considered in the \mbox{D-TRILEX} approach are found to be essential for describing the isotropic nature of the spin fluctuations described above: in their absence, this physics is not even qualitatively captured.

\subsubsection{Extended regime of metastable metallic and insulating phases in a two-orbital electronic system}

There are two main mechanisms responsible for the formation of an insulating phase in electronic materials: a gap at the Fermi energy in the non-interacting band structure and the many-body localization induced by strong electronic interactions, as for instance the Mott scenario.
The interplay between these different mechanisms can strongly affect the degree of electronic correlations and therefore the phase diagram of the material.
In Ref.~\cite{PhysRevResearch.5.L022016}, we investigated the effect of non-local correlations on the Mott transition in a two-orbital model with the crystal field splitting and the density-density approximation for the  interaction.
This model is the simplest multi-orbital system, where the influence of the orbital splitting on the Mott transition was studied in details using DMFT~\cite{PhysRevB.78.045115}.
We challenge this solution of the problem by using the \mbox{D-TRILEX} approach in order to account for the effect of the non-local collective electronic fluctuations on the spectral function in a self-consistent manner.
We find that, despite the apparent simplicity, the considered model displays a non-trivial behavior around the Mott transition.
In particular, considering the non-local correlations beyond DMFT reveals a broad coexistence region of meta-stable metallic and Mott insulating phases that extends from approximately the bandwidth to more than twice the bandwidth in the value of the interaction.
Our results illustrate that non-local correlations can have crucial consequences on the electronic properties in the strongly correlated regime of the simplest multi-orbital systems.

\begin{figure}[t!]
\centering
\includegraphics[width=0.7\linewidth]{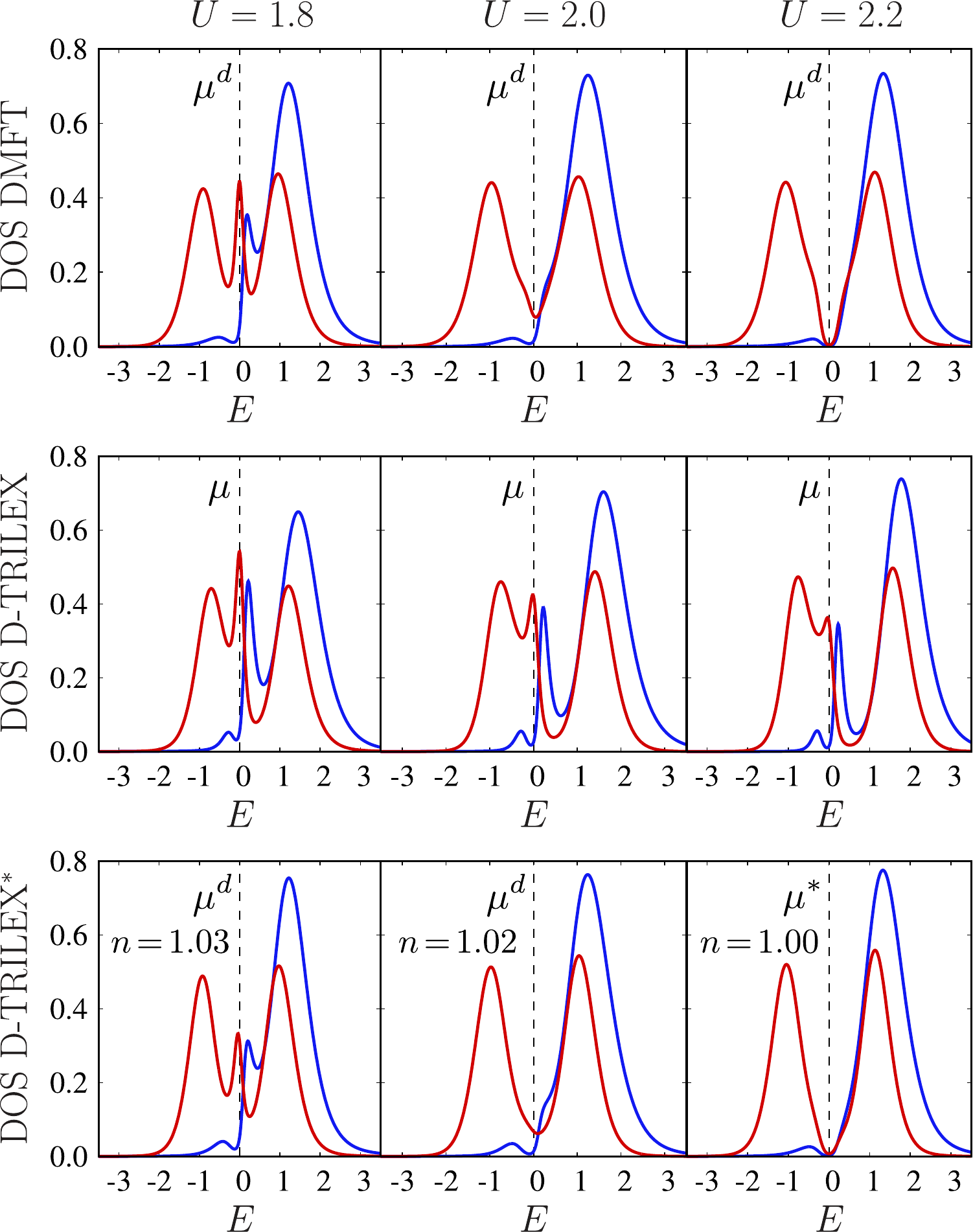}
\caption{DOS for the upper (${l=1}$, blue line) and lower (${l=2}$, red line) orbitals calculated for different interactions ${U=1.8}$ (left column), ${U=2.0}$ (middle column), and ${U=2.2}$ (right column).
Top row: DMFT solution at quarter-filling that corresponds to the chemical potential $\mu^{d}$.
Middle row: quarter-filled metallic \mbox{D-TRILEX} solution for the chemical potential $\mu$.
Bottom row: a further \mbox{D-TRILEX}$^{*}$ calculation based on the DMFT solution. 
Calculations for ${U=1.8}$ and ${U=2.0}$ are performed for $\mu^{d}$. The resulting ${\langle n \rangle > 1}$ is specified in panels.
At ${U=2.2}$ the quarter-filled D-TRILEX$^{*}$ solution appears at ${\mu^{*}\simeq\mu^{d}}$ and corresponds to the Mott insulating state. The Figure is taken from Ref.~\cite{PhysRevResearch.5.L022016}.
\label{fig:DOS_cfs}}
\end{figure}

In Ref.~\cite{PhysRevResearch.5.L022016} we considered a two-orbital model on a cubic lattice:
\begin{align}
H =
\sum_{jj',l,\sigma} c^{\dagger}_{\hspace{-0.05cm}jl\sigma}\left(t^{l}_{\hspace{-0.05cm}jj'} + \Delta^{\phantom{l}}_{l}\,\delta^{\phantom{l}}_{\hspace{-0.05cm}jj'}\right) c^{\phantom{\dagger}}_{\hspace{-0.05cm}j'l\sigma}
+ \frac{U}{2}\hspace{-0.05cm}\sum_{j,ll'} n_{\hspace{-0.05cm}jl}\hspace{0.05cm} n_{\hspace{-0.05cm}jl'}\,.
\end{align}
The nearest-neighbor hopping is set to ${t^{l}_{\langle jj'\rangle}=1/6}$ for each of the two orbitals ${l\in\{1,2\}}$.
Hereinafter, the energy is expressed in units of the half-bandwidth of the cubic dispersion ${6t=1}$.
The interaction $U$ between electronic densities ${n_{\hspace{-0.05cm}jl} = \sum_{\sigma}c^{\dagger}_{\hspace{-0.05cm}jl\sigma}c^{\phantom{\dagger}}_{\hspace{-0.05cm}jl\sigma}}$ describes both the intra- and interorbital Coulomb repulsion.
Calculations are performed at quarter-filling, which corresponds to the average density of ${\langle n \rangle = 1}$ electron per two orbitals.
In order to induce an orbital polarization ${\delta n = (\langle n_{2}\rangle - \langle n_{1}\rangle})/\langle n \rangle$, we take a relatively large value for the crystal field splitting 
${\Delta = 2\Delta_1=-2\Delta_2} = 0.3$. 
This case was studied in details in Ref.~\cite{PhysRevB.78.045115} using DMFT. 
It was demonstrated, that local electronic correlations enlarge the orbital splitting, resulting in a high degree of orbital polarization.
Consequently, the single electron mostly populates the lower orbital (${l=2}$) that undergoes the Mott transition at a critical value of the electronic interaction. 
We set the inverse temperature to ${T^{-1}=15}$, which ensures that the system is located outside the AFM phase but close to its boundary to observe strong magnetic fluctuations.

\begin{figure}[t!]
\centering
\includegraphics[width=0.7\linewidth]{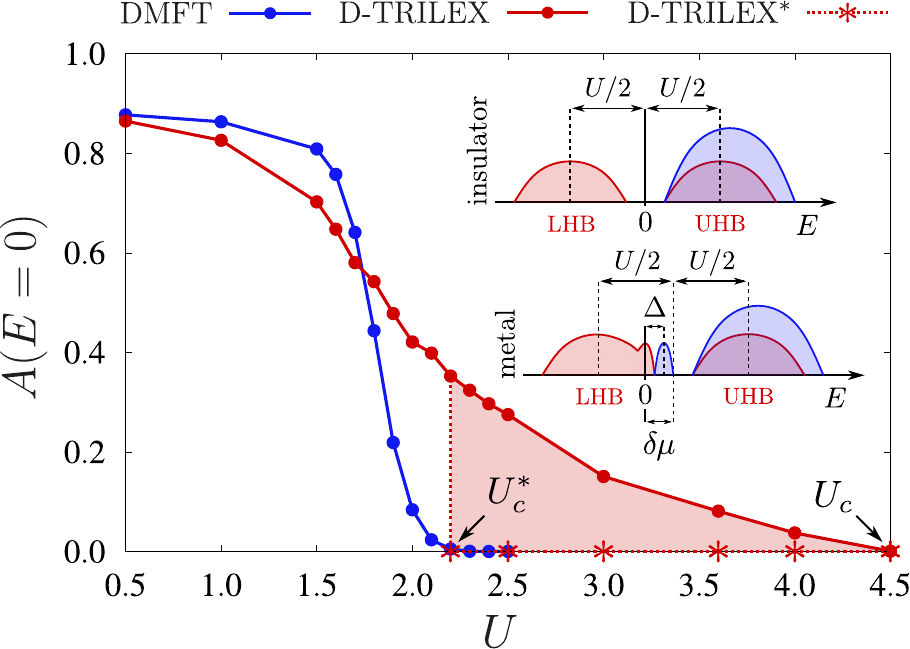}
\caption{Electronic density at Fermi energy ${A(E=0)}$ for the lower orbital (${l=2}$) as a function of the interaction $U$. The result is obtained from DMFT (blue dots), metallic \mbox{D-TRILEX} (red dots), and insulating \mbox{D-TRILEX}$^{*}$ (red asterisks) solutions. The red shaded area highlights the simultaneous existence of the metallic and the Mott insulating solutions.
The inset sketches the difference in the DOS between the insulating (top) and metallic (bottom) \mbox{D-TRILEX} solutions.
In the insulating case, the Fermi energy lies between the LHB and UHB that are split approximately by $U$. 
In the metallic case, the difference in the chemical potential ${\delta\mu=\mu^{*}-\mu}$ brings the upper part of the LHB to the Fermi energy, which results in the formation of the quasi-particle peak at ${E=0}$.
The splitting between the quasi-particle peaks coincides with the value of the crystal field splitting ${\simeq\Delta}$. The Figure is taken from Ref.~\cite{PhysRevResearch.5.L022016}.
\label{fig:Phase_cfs}}
\end{figure}

To illustrate the effect of non-local correlations on the Mott transition, we compare the DOS predicted by DMFT and \mbox{D-TRILEX} methods.
The result of these calculations is shown in Fig.~\ref{fig:DOS_cfs} for three different values of the interaction ${U=1.8}$, ${U=2.0}$, and ${U=2.2}$.
First, let us focus on the quarter-filled calculations presented in the two upper rows of this figure. 
We find that the results of the DMFT and \mbox{D-TRILEX} methods are different already at ${U=1.8}$. 
In both cases, the DOS is metallic.
The lower orbital (${l=2}$, red line) displays a three-peak structure consisting of the quasi-particle peak at Fermi energy ${E=0}$ and two side peaks that correspond to lower and upper Hubbard bands (LHB and UHB).
The upper orbital (${l=1}$, blue line) also exhibits the quasi-particle peak in the DOS that appears close to the Fermi energy at ${E \simeq \Delta}$.
However, the three-peak structure predicted by DMFT possesses a high degree of electron-hole symmetry.
Instead, the DOS of obtained for the same orbital (${l=1}$) using the \mbox{D-TRILEX} approach resembles the DOS of a hole-doped Mott insulator with the quasi-particle peak being shifted closer to the LHB~\cite{RevModPhys.68.13}.
The quasi-particle peaks in the DOS of DMFT vanish simultaneously between ${U=1.8}$ and ${U=2.0}$, which signals the tendency towards a Mott insulating state in a multi-orbital system at finite temperature.
A further increase of the interaction decreases the electronic density at Fermi energy ${A(E=0)}$.
The latter reaches zero at ${U^{\ast}_{c}\simeq2.2}$ (blue line in Fig.~\ref{fig:Phase_cfs}), and the DMFT solution enters the Mott insulating phase.
On the contrary, the \mbox{D-TRILEX} solution remains metallic for the discussed values of the interaction (middle row in Fig.~\ref{fig:DOS_cfs}).
Thus, even at ${U^{\ast}_{c}}$ it reveals pronounced quasi-particle peaks in the DOS for both orbitals.
Fig.~\ref{fig:Phase_cfs} shows that ${A(E=0)}$ in the metallic \mbox{D-TRILEX} solution also decreases upon increasing the interaction.
However, this solution turns into a Mott insulator only at a very strong critical interaction ${U_{c}\simeq4.5}$, which is larger than twice the bandwidth. 

To explain the observed effect, we note that quarter-filling in DMFT and \mbox{D-TRILEX} corresponds to different values of the chemical potential.
We point out that \mbox{D-TRILEX} calculations are based on the DMFT solution of the local impurity problem that plays a role of the reference system.
We find that the quarter-filled metallic \mbox{D-TRILEX} solution originates from the metallic impurity problem that has smaller average density (${\langle n \rangle < 1}$).
For this reason, the reference impurity problem remains metallic even at $U^{*}_{c}$.
This fact suggests that for a given value of the chemical potential the effect of non-local collective electronic fluctuations in the metallic regime consists in moving the spectral weight from above to below the Fermi energy, which brings the filling of the system to ${\langle n \rangle=1}$.

To confirm this statement, we perform \mbox{D-TRILEX} calculations for the chemical potential $\mu^{d}$ of the quarter-filled DMFT solution.
The corresponding result is shown in the bottom row of Fig.~\ref{fig:DOS_cfs} and is referred to as the \mbox{D-TRILEX}$^{*}$ calculation in order not to confuse it with the metallic solution.
We observe that the obtained DOS is again practically identical to the one of DMFT (bottom {\it vs.} top row in Fig.~\ref{fig:DOS_cfs}).
However, the \mbox{D-TRILEX}$^{\ast}$ calculations performed in the regime ${1.0\lesssim U < 2.2}$, where DMFT solution is metallic, correspond to ${\langle n \rangle>1}$.
Moreover, no quarter-filled \mbox{D-TRILEX}$^{\ast}$ solution is found near ${\mu^{d}}$ in this regime of interactions.
This fact supports our previous finding that in the metallic regime non-local correlations increase the average density of the considered system.

This physical picture changes when the DMFT solution becomes Mott insulating.
We find that the corresponding \mbox{D-TRILEX}$^{*}$ solution undergoes the Mott transition at the same critical interaction ${U^{\ast}_{c}}$ as in DMFT.
Moreover, at ${U\geq{}U^{\ast}_{c}}$ the average density for the \mbox{D-TRILEX}$^{*}$ solution becomes ${\langle n \rangle = 1}$ for ${\mu^{*}\simeq\mu^{d}}$.
We find, that the insulating DMFT and \mbox{D-TRILEX}$^{\ast}$ solutions are almost fully polarized, which results in electron-hole symmetric DOS for the lower orbital (top and bottom right panels of Fig.~\ref{fig:DOS_cfs}).
Consequently, the upper orbital becomes nearly unoccupied and thus cannot strongly interact with the lower one.
Therefore, no transfer of the spectral weight between the orbitals by means of the non-local fluctuations occurs in the insulating regime.

At ${U\geq{}U^{\ast}_{c}}$ the \mbox{D-TRILEX}$^{*}$ solution remains quarter-filled and Mott insulating, which is confirmed by the zero electronic density at Fermi energy (red asterisks in Fig.~\ref{fig:Phase_cfs}).
Therefore, both, the DMFT and the \mbox{D-TRILEX} methods predict the Mott transition for the considered system at the same value of the critical interaction $U^{\ast}_{c}$. 
However, including non-local collective electronic fluctuations beyond DMFT allows one to additionally capture the metallic solution that coexists with the Mott insulating one up to the second critical interaction $U_{c}$. 
For ${U>U_{c}}$ any value of the chemical potential inside the Mott gap gives the same average density, and the two solutions corresponding to $\mu$ and $\mu^{*}$ can be considered equivalent.

To conclude, we investigated the effect of non-local collective electronic fluctuations on the Mott transition in a two-orbital quarter-filled model with density-density interaction by comparing the results of the \mbox{D-TRILEX} and DMFT methods.
At the considered temperature, the DMFT solution of the problem remains metallic below the critical interaction ${U^{\ast}_{c}=2.2}$, and at this value of the interaction undergoes the Mott transition.
We find that the inclusion of non-local correlations by means of the \mbox{D-TRILEX} approach stabilizes the metallic phase up to the very large critical interaction ${U_{c}=4.5}$.
The \mbox{D-TRILEX} method also captures the appearance of Mott insulating phase at ${U^{\ast}_{c}}$ as a second meta-stable solution.
This leads to a remarkably broad coexistence region between the metallic and the Mott insulating phases that exist at the same filling, but with different values of the chemical potential between the $U^{\ast}_{c}$ and the $U_{c}$ critical interactions.
Our results show, that for a simple two-orbital model, DMFT cannot correctly interpolate between the moderately- and strongly-interacting regimes, in analogy with the single-orbital case.
This fact brings further evidence that non-local correlations may lead to non-trivial effects due to the presence of additional channels for collective electronic fluctuations also in multi-orbital systems.

\subsubsection{Eliminating orbital selectivity from the metal-insulator transition by strong magnetic fluctuations}

Orbital-selective phenomena that can be realized in electronic materials attract enormous interest. 
A prominent example is an orbital-selective pairing mechanism that is considered to be of crucial importance for the formation of a superconducting state in ruthenates and iron-based superconductors. 
An orbital-selective Mott phase, where itinerant and localized electrons live in different orbitals of the same material, is the other example of a state that can be potentially observed in experiments. 
Since its theoretical prediction, the orbital-selective Mott transition (OSMT) has been intensively studied by the state-of-the-art theoretical methods that are based on local approximations to electronic correlations, namely DMFT and the slave-spin approach. 
Nevertheless, the existence of the OSMT in realistic materials is still heavily debated and, to my knowledge, this state has not been realized experimentally yet. 

As a matter of fact, even the state-of-the-art methods sometimes fail to correctly describe the actual electronic behavior in the system. Especially, it happens when the material exhibits strong non-local collective electronic fluctuations that are discarded in DMFT and slave-spin methods. On the other hand, until very recently a consistent consideration of these fluctuations in a multi-orbital case was technically impossible. 
For these reasons, all present theoretical predictions for the OSMT are based solely on the local approximations and are thus questionable.

In Ref.~\cite{PhysRevLett.129.096404}, we have shown that consistently taking into account the non-local correlation effects completely changes the physical picture in the model, where the OSMT was predicted in the framework of the local theories. 
We have found that upon lowering temperature the considered system undergoes the N\'eel transition to an ordered antiferromagnetic phase before it experiences the OSMT. 
Importantly, the former occurs simultaneously for all orbitals, which eliminates the orbital selectivity from the metal-insulator transition.

A minimal model that allows one to address the OSMT is a half-filled Hubbard model with two orbitals that have different bandwidths.
The Hamiltonian for this model reads:
\begin{align}
H = \sum_{jj',ll',\sigma} t^{ll'}_{jj'} c^{\dagger}_{jl\sigma} c^{\phantom{\dagger}}_{j'l'\sigma} + \frac12 \hspace{-0.05cm} \sum_{j,\{l\},\sigma\sigma'} U_{l_1l_2l_3l_4} c^{\dagger}_{j l_1 \sigma} c^{\phantom{\dagger}}_{j l_2 \sigma} c^{\dagger}_{j l_4 \sigma'} c^{\phantom{\dagger}}_{j l_3 \sigma'}\,.
\label{eq:Hubbard}
\end{align}
For the convenience of describing magnetic fluctuations the interacting part of the Hamiltonian is written in the particle-hole representation. 
The local interaction matrix $U_{l_1l_2l_3l_4}$ is parametrized in the Kanamori form with the intraorbital ${U=U_{llll}}$ and the interorbital ${U'=U_{lll'l'}}$ Coulomb interactions, and the Hund's coupling ${J=U_{ll'll'}=U_{ll'l'l}}$.
Operators $c^{(\dagger)}_{jl\sigma}$ describe annihilation (creation) of an electron on the site $j$, at the orbital $l\in\{1, 2\}$, and with the spin projection $\sigma\in\{\uparrow, \downarrow \}$.

In the absence of the interorbital hopping $t^{12}$, the OSMT has been predicted in a large space of model parameters using DMFT and slave spin methods.
However, the recent DMFT study~\cite{PhysRevLett.129.096403} predicted that including the interorbital hopping between the metallic and the insulating orbitals destroys the OSMT and results in a metallic ground state of the system.
In order to address this problem beyond the local DMFT picture, we additionally consider the effect of spatial collective electronic fluctuations using the \mbox{D-TRILEX} approach.

\begin{figure}[t!]
\centering
\includegraphics[width=0.7\linewidth]{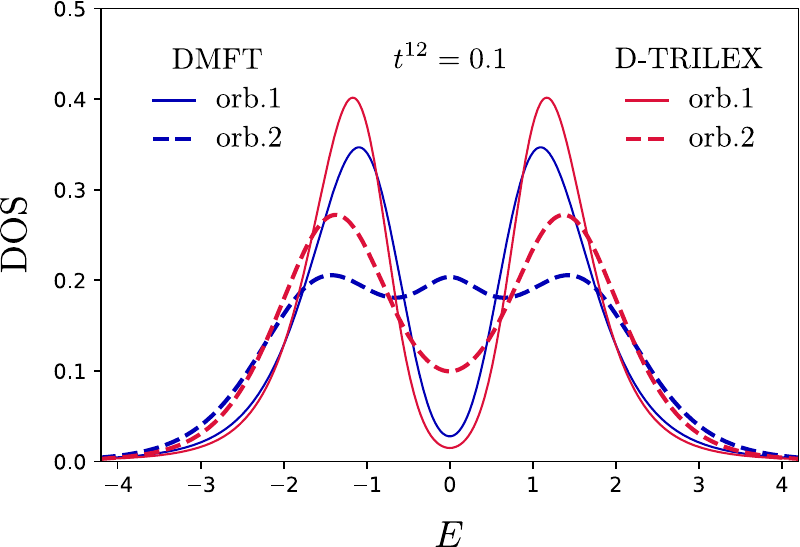}
\caption{The orbitally-resolved local electronic DOS calculated using DMFT (blue lines) and \mbox{D-TRILEX} (red lines) methods. Solid lines correspond to the first (narrow) orbital. Dashed lines depict the DOS of the second (wide) orbital. The results are obtained for the case of ${t^{12}=0.1}$ at the inverse temperature ${T^{-1}=8.2}$ close to the N\'eel transition point. The Figure is taken from Ref.~\cite{PhysRevLett.129.096404}.
\label{fig:DOS}}
\end{figure}

The signature of the metal-insulator transition in both orbitals can be identified in the electronic spectral function through the formation of a pseudogap.
Fig.~\ref{fig:DOS} shows the orbital-resolved local electronic DOS obtained for the case of ${t^{12}=0.1}$ at the inverse temperature ${T^{-1}=8.2}$.
This is the closest point to the N\'eel transition where we could converge numerical calculations.
DMFT (blue lines) predicts different behavior for the two orbitals.
The narrow orbital depicted by a solid line is almost in the Mott-insulating regime, but has a finite electronic density at Fermi energy according to the result of Ref.~\cite{PhysRevLett.129.096403}.
The wide orbital demonstrates a correlated metallic behavior (dashed line). 
It has a relatively large density of electrons at zero energy and two side peaks that correspond to the Hubbard sub-bands.

Taking into account the non-local collective electronic effects via the \mbox{D-TRILEX} approach strongly suppresses the electronic density at Fermi energy (red lines).
The most remarkable change in the DOS occurs for the wide orbital, where magnetic fluctuations turn the quasiparticle peak at the Fermi energy (${E=0}$) into a pseudogap.
The density of electrons at the Fermi energy is also diminished for the narrow orbital, but this change is not that striking, because in DMFT this orbital already exhibits a pseudogap.

Our results suggest that the OSMT should rather occur in systems, where the non-local collective electronic instabilities are suppressed.
For instance, one can explore doping the system, which usually reduces the strength of magnetic fluctuations.
In this context one should avoid having van Hove singularities at Fermi energy in the electronic spectrum, because they enhance collective electronic instabilities.  
On the other hand, the dynamical orbital-selective metal-insulator transitions that are driven by collective electronic fluctuations can probably be found in systems, where the orbital degrees of freedom in these two-particle fluctuations are decoupled.
However, as we illustrated in this work, this separation of the orbital degrees of freedom cannot be provided by the Hund's exchange coupling that acts as a band decoupler for the local electronic correlations.
These findings narrow the circle of possible systems, where the OSMT could be realized experimentally. In addition, the obtained results motivate further investigation of the role of the non-local electronic correlations in physical properties of the multi-orbital systems.

\subsubsection{Can orbital-selective N\'eel transitions survive strong nonlocal electronic correlations?}

Understanding the microscopic mechanism of the orbital-selective metal-insulator transitions (OSMIT) represents a great interest for the community, which is additionally motivated by recent observation of signatures of the orbital-selective states in nickelates, iron-based superconductors, and iron chalcogenides.
To date, theoretical studies of OSMIT are mainly limited to local approximations to electronic correlations provided by, e.g., DMFT. This limitation allows one to describe only the transition to the paramagnetic orbital-selective Mott phase. The orbital-selective Mott phase is however a rather fragile state that is unstable against inter-orbital hopping processes and can be destroyed by strong magnetic fluctuations. These facts question the existence of this state in realistic conditions. At least in iron-based superconductors, signatures of orbital-selective states are generally observed in the presence of magnetic fluctuations.

Can one realize the OSMIT in the presence of strong magnetic fluctuations? What would be the mechanism of an OSMIT to a magnetically ordered state? These are questions that we have addressed in this work. 
This requires considering long-range magnetic fluctuations, which is extremely challenging in the multi-orbital context. Indeed, the only currently existing theoretical studies on ``magnetic'' OSMITs are based on cluster DMFT calculations that cannot account for long-range electronic correlations. 
Remarkably, in our work we have shown that long-range magnetic fluctuations destroy the magnetic OSMIT predicted on the basis of DMFT calculations. 
We propose a novel mechanism of the formation of the orbital selective magnetic state. 
We argue, that this state can be realized only upon decoupling magnetic fluctuations of different orbitals. 
We demonstrate that this decoupling occurs in the absence of Hund's exchange coupling. 
Importantly, the proposed orbital-selective N\'eel state can be realized for arbitrary strength of the electronic interaction, but is formed differently in the weak- and strong-coupling regimes. 

In Ref.~\cite{PhysRevLett.132.226501}, we again consider the half-filled two-orbital Hubbard model on a cubic lattice described by the Hamiltonian~\eqref{eq:Hubbard}. We restrict ourselves to nearest-neighbor hoppings, and choose the half-bandwidth of the narrow
band as our unit of energy, i.e. $t^1_{jj^{\prime}}=1/6$. The second band is double as wide with 
$t^2_{jj^{\prime}}=1/3$.

\begin{figure}[t!]
\centering
\includegraphics[width=0.7\linewidth]{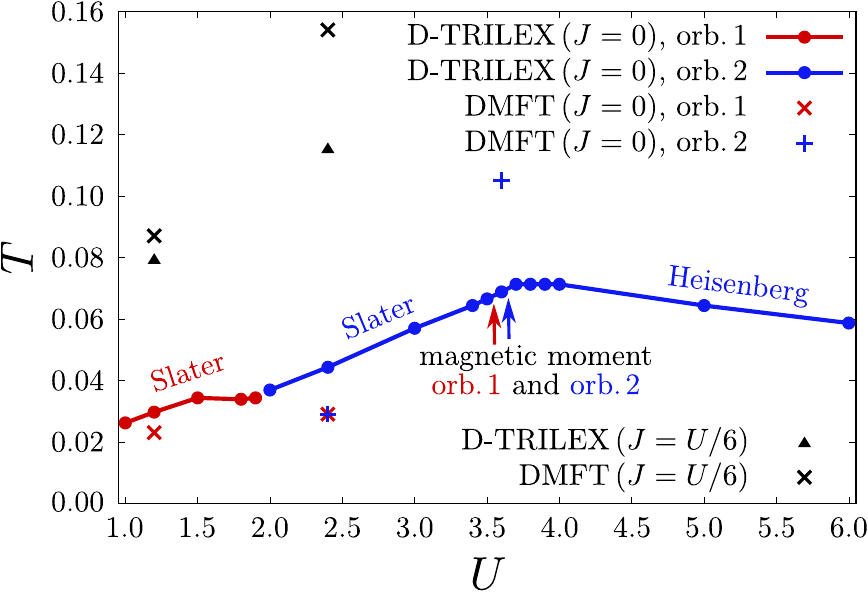}
\caption{N\'eel temperature for the two-orbital half-filled Hubbard model on a cubic lattice with different bandwidths of the two orbitals. Results are obtained using \mbox{D-TRILEX} and DMFT for different values of the Hund's exchange coupling $J$. At finite $J$ the N\'eel transition occurs simultaneously for both orbitals (``$\blacktriangle$'' and ``$\boldsymbol{\times}$'' markers). The OSNT scenario is realized in the absence of $J$: at ${U<1.95}$ upon decreasing the temperature the N\'eel transition occurs first for the narrow orbital (orb.\,1, red color), and at ${U>1.95}$ for the wide orbital (orb.\,2, blue color). Critical interactions at which the local magnetic moment is formed above the N\'eel transition are indicated by arrows. The Figure is taken from Ref.~\cite{PhysRevLett.132.226501}.
\label{fig:OSNT}}
\end{figure}

We have first performed \mbox{D-TRILEX} calculations in the absence of Hund's coupling (${J=0}$). 
We calculate the orbital-selective N\'eel temperatures ($T_N$) by identifying the divergences of the orbital components of the static (${\omega=0}$) spin susceptibility ${X^{sp}_{ll'}({\bf q},\omega)}$ obtained at the AFM wave vector ${{\bf Q}=\{\pi,\pi,\pi\}}$.
The results are summarised in Figure~\ref{fig:OSNT}. 

In the weak-coupling regime (${U<1.95}$) upon lowering the temperature the ${l=1}$ (narrow orbital) component of the AFM susceptibility diverges first, while the ${l=2}$ (wide orbital) component remains finite at the transition point.
This behaviour indicates the orbital-selective N\'eel transition (OSNT) to a phase, where electrons in the narrow orbital order antiferromagnetically, while the wide orbital stays itinerant.
At ${U>1.95}$ we observe the opposite situation: the transition to the ordered AFM state occurs first for the wide orbital, while the narrow orbital remains itinerant.
Remarkably, the system exhibits an OSNT for any value of the interaction, except for ${U\simeq1.95}$ where ${T_{N_1}=T_{N_2}}$.

The choice of vanishing Hund's coupling made above is essential for these results. Indeed, in a multi-orbital system magnetic fluctuations of different orbitals are coupled due to the presence of Hund's exchange coupling $J$.
This coupling is realized through inter-orbital three-point (Hedin) vertex corrections that are present in the self-energy and the polarization operator for any finite value of $J$.
These vertices connect the renormalized interaction in the spin channel to the electronic Green's function and thus are responsible for mixing different orbital contributions to the spin susceptibility.
Strong spatial magnetic fluctuations enhance this mixing, which leads to a simultaneous N\'eel transition for different orbitals, as has been described above in Ref.~\cite{PhysRevLett.129.096404}.
Therefore, realizing the OSNT necessarily requires magnetic fluctuations of different orbitals to decouple.
This happens in the absence of Hund's coupling, since in this case the inter-orbital components of the vertex function in the spin channel are identically zero.
In realistic materials, the Hund's coupling can be suppressed, e.g., through the Jahn-Teller effect of phonons, which, as has been demonstrated for fullerides, can result even in a sign change of $J$.

To confirm the fact that the presence of a non-zero value of the Hund's coupling destroys the OSNT, we perform \mbox{D-TRILEX} calculations for ${J=U/6}$.
As expected from the results of Ref.~\cite{PhysRevLett.129.096404}, in this case the OSNT transforms into an ordinary N\'eel transition that occurs simultaneously for both orbitals.
The corresponding N\'eel temperatures are shown in Fig.~\ref{fig:OSNT} for ${U=1.2}$ and ${U=2.4}$ by ``$\blacktriangle$''.
These results confirm, that the OSNT scenario is realized only in the absence of Hund's coupling, where the two orbitals have different N\'eel temperatures.

To conclude, in this work we have established strategies for realizing orbital-selective N\'eel-ordered magnetic states.
We have demonstrated that in the absence of Hund's exchange coupling $J$ the two-orbital Hubbard model with different bandwidths can indeed undergo an OSNT, at any interaction strength (except one specific value of the interaction).
Remarkably, the OSNT occurs differently in the weak and strong coupling regimes of interaction.
In the weak coupling regime, the N\'eel transition occurs first for the narrow orbital, while the wide orbital remains itinerant.
In the strong coupling regime, the localized behavior occurs first in the wide orbital, while the narrow orbital stays itinerant.
Most intriguingly, in the presence of Hund's coupling the OSNT is destroyed altogether: Hund's exchange effectively couples orbital degrees of freedom, and is thus detrimental
to orbital-selective behavior. 

The ubiquity of Hund's exchange in real materials may provide a natural explanation for OSNTs likely being rather the exception than the rule. 
Nevertheless, a systematic search for OSNT in real materials might prove worthwhile in view of potential applications in spintronics devices, e.g. for memory applications, spin valves or spin-charge converters. Our work strongly suggests a materials screening among materials with as low as possible Hund's exchange coupling.
A trivial corollary of this argument is obtained by replacing orbital indices by site indices: thanks to the intrinsically weak direct intersite exchange interaction, site-selective magnetic orderings in materials with several correlated sites per unit cell should be found more easily than orbital-selective OSNTs.

\subsubsection{Local and nonlocal electronic correlations at the metal-insulator transition in the Hubbard model in two dimensions}

The Hubbard model is the prototypical model for correlated materials, and progress on its theoretical understanding will have immediate impact on important materials problems such as the puzzle around the celebrated copper oxides.
In this work, we present a comprehensive study of the spectral properties and phase diagram of the single-band Hubbard model at half-filling, focusing on the challenging intermediate coupling regime where accurate theoretical calculations are most difficult. Analyzing the behavior of the momentum-resolved spectral function, we disentangle the contribution of antiferromagnetic (AFM) fluctuations and local electronic correlations to the formation of a depletion of spectral weight at the Fermi energy, connecting the weak- and strong-coupling limits. These two limits are separated by a crossover regime that starts when the local magnetic moment is formed and ends at the Mott transition. In this regime, spatial fluctuations are still notable, but the presence of a local magnetic moment (LMM) indicates the emergence of strong local electronic correlations in the system. We identify the weak-coupling region that precedes the crossover with a Slater regime. There, the metal-insulator transition is solely governed by AFM fluctuations, which results in a momentum-selective formation of the pseudogap with a pronounced nodal-antinodal (N-AN) dichotomy increasing upon an increasing interaction. In the intermediate-coupling (crossover) regime, increasing interaction enhances the value of the LMM and thus reduces the N-AN dichotomy that eventually disappears upon entering the Mott phase. Finally, the Mott phase can be associated with the Heisenberg regime of local magnetic moments. Indeed, we have found that the Mott transition occurs when the LMM reaches a critical value. The simultaneous disappearance of the quasiparticle peak along the Fermi surface at the Mott transition additionally illustrates the governing role of local electronic correlations for the physics of the system. As expected, in the Heisenberg regime, the N\'eel transition is driven by a decrease in the temperature, which separates the Mott phase into paramagnetic and AFM states.

In Ref.~\cite{PhysRevLett.132.236504} we studied the half-filled single-orbital Hubbard model:
\begin{align}
\hat{H} = t\sum_{\langle jj' \rangle, \sigma} c_{j\sigma}^{\dagger} c^{\phantom{\dagger}}_{j'\sigma} + U\sum_{i} n_{j\uparrow} n_{j\downarrow}
\end{align}
on a square lattice with dispersion ${\epsilon_{\bf k}=2t(\cos k_x+\cos k_y)}$ and the on-site Coulomb interaction $U$.
We use the half bandwidth as our unit of energy, i.e. ${t=0.25}$. 
Here, $c^{(\dagger)}_{\sigma}$ is the annihilation (creation) operator for an electron with spin ${\sigma\in\{\uparrow, \downarrow\}}$, and ${n_{j\sigma} = c^{\dagger}_{j\sigma}c^{\phantom{\dagger}}_{j\sigma}}$ is the spin-resolved density operator.

\begin{figure}[t!]
\centering
\includegraphics[width=0.25\linewidth]{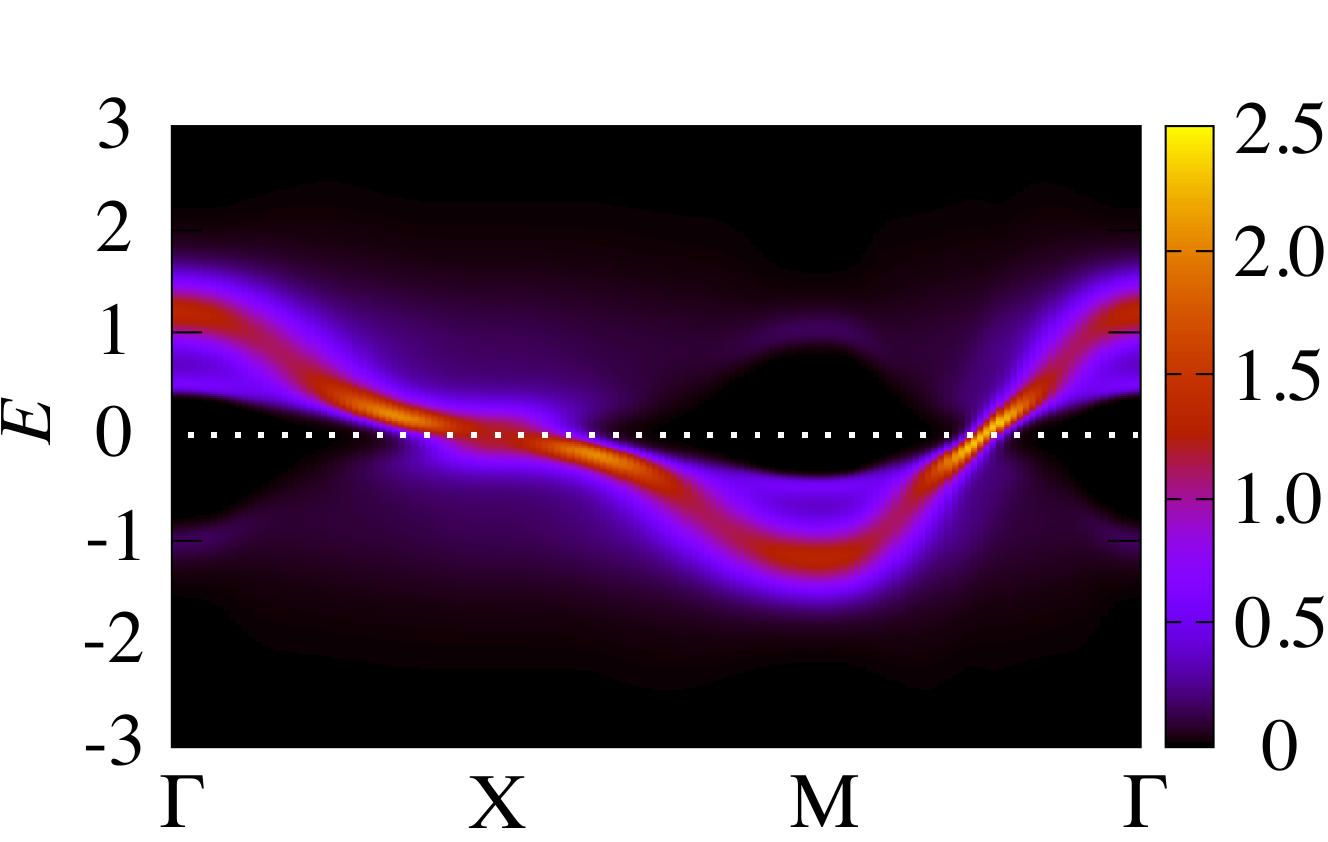}\,
\includegraphics[width=0.25\linewidth]{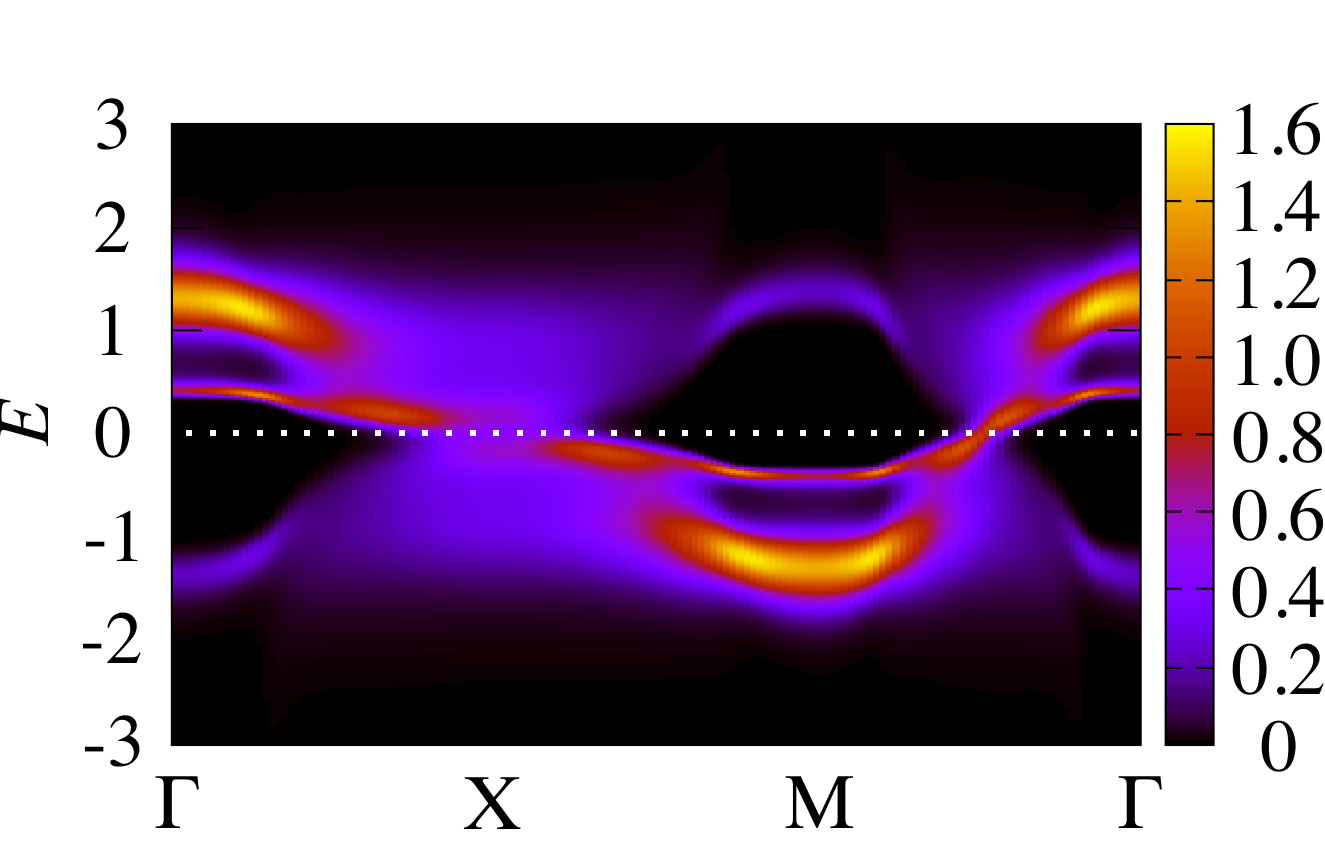}\,
\includegraphics[width=0.25\linewidth]{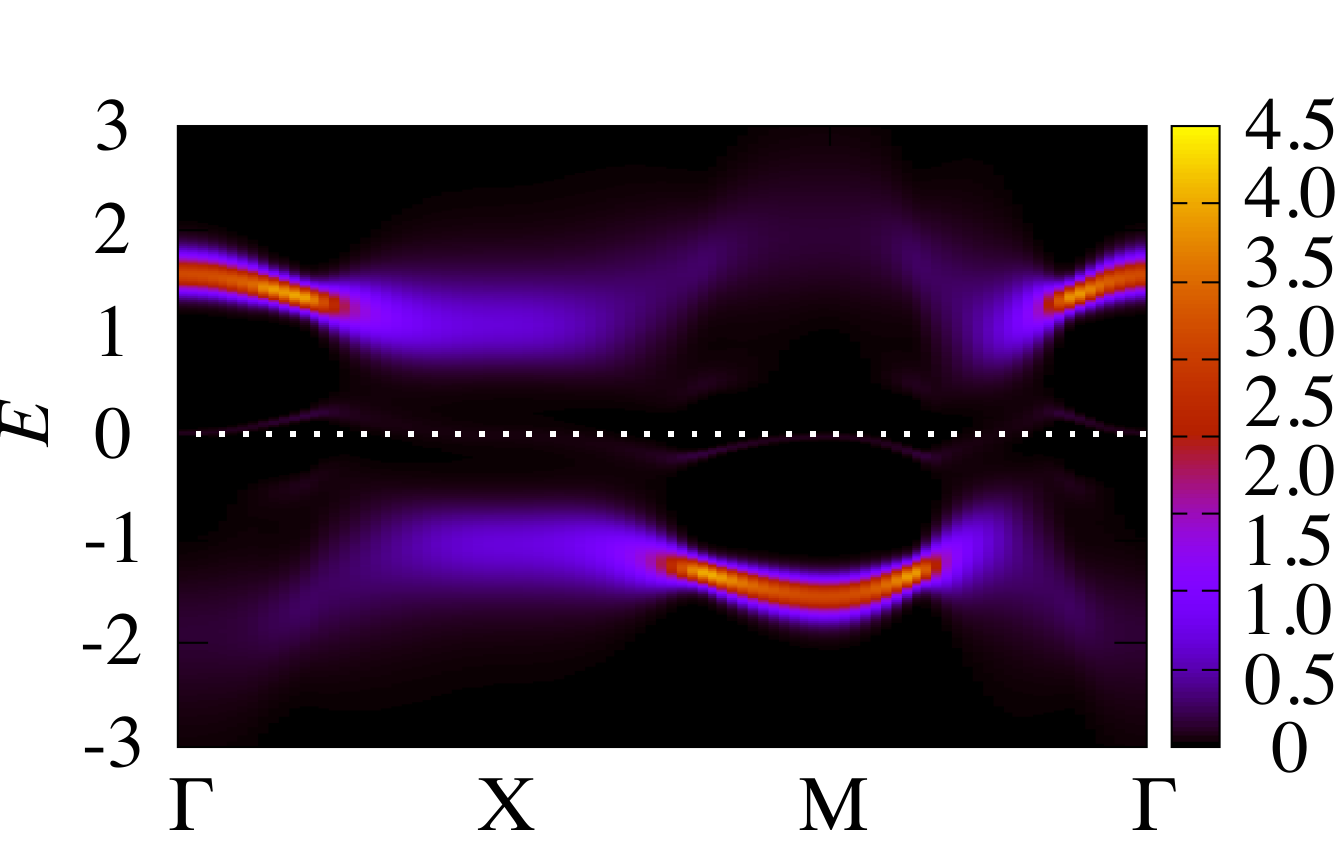}\,
\includegraphics[width=0.19\linewidth]{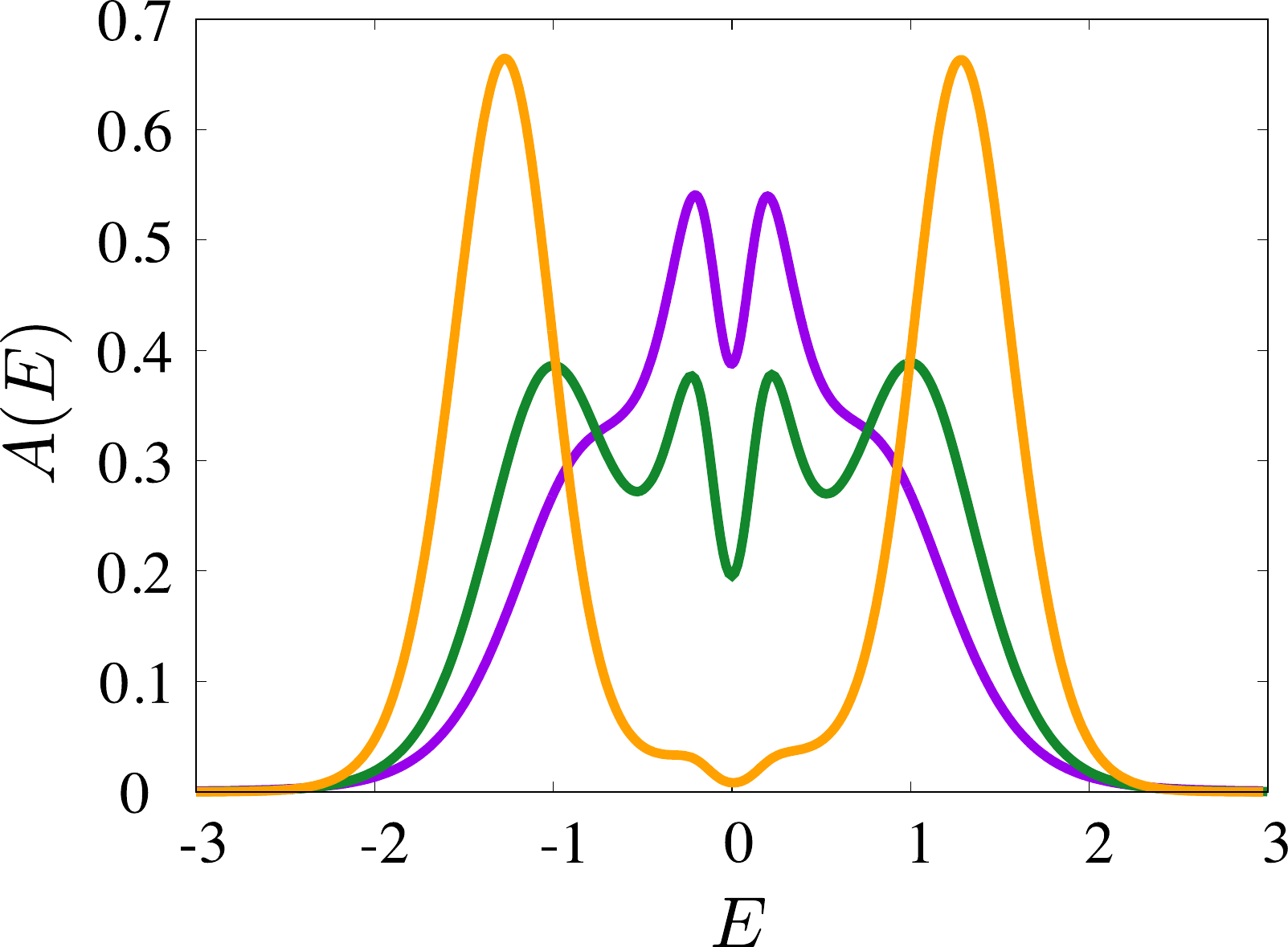}
\caption{Momentum-resolved spectral function of the Hubbard model at half filling calculated
at ${T=0.06}$ within \mbox{D-TRILEX} for
$U=1.2$ (left), $U=1.6$ (middle) and $U=2.4$ (right),
along the $k$-path $\Gamma-X-M-\Gamma$ with $\Gamma=(0,0)$, $X=(\pi, 0)$, and $M=(\pi, \pi)$.
The far right column displays the corresponding local spectral functions obtained for ${U=1.2}$ (purple), ${U=1.6}$ (green) and ${U=2.4}$ (orange). The Figure is taken from Ref.~\cite{PhysRevLett.132.236504}.
\label{fig:spectr_dtrilex}}
\end{figure}

Fig.~\ref{fig:spectr_dtrilex} displays the {\bf k}-resolved spectral function along a high symmetry path in the Brillouin zone (BZ) and its local counterpart.
The results are obtained using \mbox{D-TRILEX} in the low-temperature (${\beta=16}$) regime for three values of the interaction.
With increasing interaction strength, from left to right, the system evolves from a weakly correlated metal, with a spectral function resulting from the non-interacting one by mere temperature broadening, to a Mott insulator, where the Mott gap results from a splitting of the spectral features into Hubbard bands, separated by the Hubbard interaction ${\sim{}U}$.

We find that the spectral functions close to $E_F$ reveal characteristic {\bf k}-dependent features. 
Already at ${U=1.2}$ (bottom left panel) we observe a depletion of the spectral weight around the anti-nodal (${{\rm AN}={\rm X}=(\pi,0)}$) point, while the nodal (N) (${{\rm N}=(\pi/2,\pi/2)}$) point
remains unaffected. 
In the local spectral function, the reduction of the spectral weight at the AN point results in a minimum at the Fermi level. 
When increasing the interaction strength to ${U=1.6}$ the quasi-particle (QP) peak at $E_{F}$ completely 
disappears at the AN point, but yet remains at the N point.
This is a well-known effect of AFM fluctuations in the formation of an insulating state, which implies 
that the gap in the spectral function first opens at the AN point (middle panel at the bottom row), 
then propagates along the FS, and finally opens at the N point.

Inspection of the local spectral functions for the two lowest interaction values in the light
of the k-resolved spectra, calls for an important caveat: obviously, from the depletion of the
local spectral weight at the Fermi level one cannot establish insulating behavior of the system.
Indeed, in the presence of nonlocal correlations, the local spectral function is no longer
the appropriate quantity to look at, since a metallic regime with a momentum-selectively gapped
Fermi surface may not be distinguished from a thermally broadened insulator.

\begin{figure}[t!]
\centering
\includegraphics[width=0.68\linewidth]{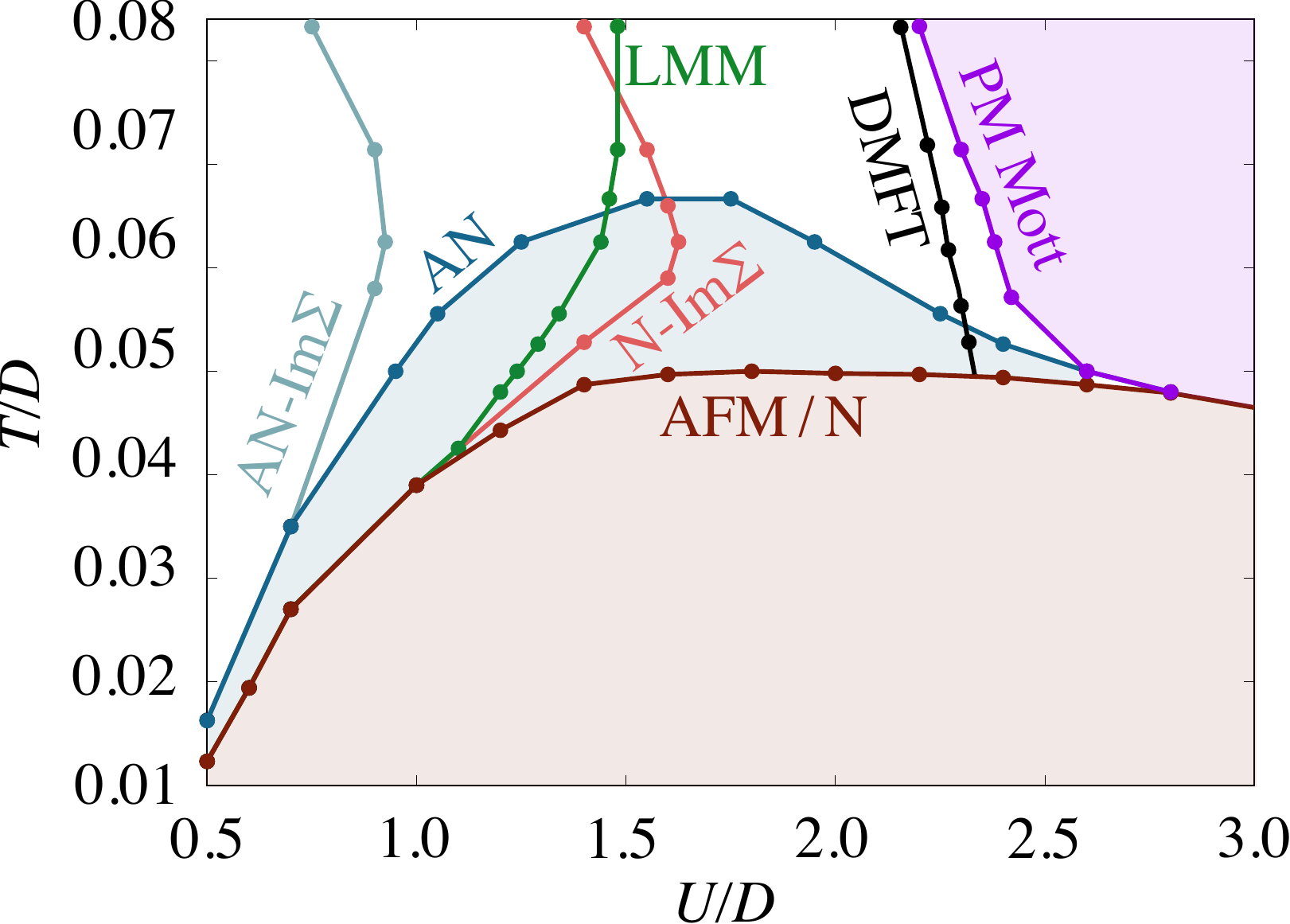}
\caption{Phase diagram of the 2D Hubbard model within \mbox{D-TRILEX}. 
The different lines have been determined as follows:
Divergence of the spin susceptibility at ${{\bf q}=(\pi,\pi)}$ (dark red line). Sign change of the slope of the self-energy at the AN point (light blue curve). Gap opening in the spectral function at the AN point (dark blue curve).
Sign change of the slope of the self-energy at the N point (light red curve). Formation of non-zero local
magnetic moment (green curve).
The purple line corresponds to the simultaneous disappearance of the quasi-particle peak at $E_{F}$ for both N and AN points, identified with the paramagnetic Mott transition. The black curve is the equivalent result from single-site DMFT calculations. The Figure is taken from Ref.~\cite{PhysRevLett.132.236504}. 
\label{fig:phase_diag_weakcoupling}}
\end{figure}

The observed momentum-selective disappearance of the quasiparticle peak at $E_{F}$ along the FS suggests to revisit the phase diagram and study the spectral function at the AN and N points separately.
In Fig.~\ref{fig:phase_diag_weakcoupling}, we show the resulting phase diagram.
The dark red line corresponds to the N\'eel transition to the AFM (quasi-)ordered state at 
lower $T$.  
Concomitantly, in the spectral function, at the N point, the QP peak at $E_F$ is lost,
and a gap opens.
The dark blue line depicts the formation of a minimum at $E_{F}$ in the spectral function at the AN point, which, as seen above, occurs at larger temperatures than at the N point. 
Interestingly, the AN curve (dark blue line) has a dome-like shape mimicking the form of the N\'eel phase boundary.
At ${U\simeq2.8}$ the AN and N curves coincide, and the disappearance of the quasi-particle peak at $E_{F}$ is not momentum-selective any more in the regime of large interactions ${U\gtrsim2.8}$.
Remarkably, the form of the dark blue curve suggests that at certain values of $T$ by increasing 
the interaction strength $U$ one can move the system from the regime where part of the FS is gapped 
(blue area in Fig.~\ref{fig:phase_diag_weakcoupling}) to a metallic regime (white area between the dark 
blue and purple curves), where QP are restored along the entire FS.

Increasing the interaction even more results in a simultaneous disappearance of the QP peak at $E_{F}$ and in the formation of the gap at both AN and N points, which is depicted in Fig.~\ref{fig:phase_diag_weakcoupling} by the purple line.
In the low-temperature regime the purple curve goes toward larger values of $U$ upon decreasing temperature, until it eventually merges with the AN (dark blue) and N (dark red) lines.
Remarkably, this simultaneous opening of the gap along the FS is non-momentum-selective and is thus governed by local electronic correlations.
For this reason, we identify the purple curve with the Mott transition. 
Thus, in the regime of weak-to-intermediate interactions AFM fluctuations highly affect the behavior of the system, introducing momentum-selectivity between the N and AN points, which results in the momentum-selective disappearance of the quasi-particle peak along the FS.
This behavior is consistent with a Slater-like scenario for the MIT.
On the contrary, at strong interactions local electronic correlations dominate.

How does the system interpolate between the two regimes of dominant spatial and local electronic correlations?
The first signature of the crossover between the two regimes is the decrease of the N-AN dichotomy at ${U\gtrsim1.5}$, which manifests itself in a decrease of the difference between the critical temperatures for the disappearance of the quasi-particle peak at $E_{F}$ at the AN and N points (dark blue and red lines in Fig.~\ref{fig:phase_diag_weakcoupling}).
This means that the local electronic correlations start to take over the spatial ones well before the system undergoes the Mott transition.
This behavior can be related to the formation of the local magnetic moment (LMM), which appearance is depicted in Fig.~\ref{fig:phase_diag_weakcoupling} by the green line. 
In the weak-coupling regime the green curve lies on top of the N\'eel phase boundary (dark red curve), which is consistent with a Slater mechanism of the N\'eel transition in this regime.
Remarkably, the LMM curve starts deviating from the N\'eel phase boundary at ${U=1.0}$, which means that at ${U>1.0}$ the MIT transition is no longer determined solely by magnetic fluctuations of itinerant electrons accounted for by the self-energy, as the LMM also starts contributing to the formation of the insulating state.
The formation of the LMM also reduces the momentum-selectivity in the disappearance of the quasiparticle peak at the $E_{F}$. 
Indeed, we find that the LMM curve crosses the AN line at ${U\simeq1.4}$ right before the N-AN dichotomy starts being suppressed. 
This effect can be explained by the fact that the LMM and spatial collective electronic fluctuations are formed by the same electrons. Upon increasing the interaction more electrons are involved in the formation of the LMM and thus less electrons remain for the fluctuations. Eventually, when reaching the critical value, the LMM completely screens the fluctuations and the system enters the Mott insulator regime. 
The formation of the LMM therefore indicates the beginning of the crossover regime separating the weak-coupling Slater part of the phase diagram from the strong-coupling Heisenberg one.
The end of the crossover regime occurs upon reaching the critical value of the LMM effectively when the system undergoes the Mott transition.

To conclude, we have studied the $T$-$U$ phase diagram of the half-filled single-orbital Hubbard model on a square lattice.
Analyzing the behavior of the momentum-resolved spectral function, we disentangle the contribution of AFM fluctuations and local electronic correlations to the formation of a depletion of spectral weight at the Fermi energy, connecting the weak (Slater) and strong coupling (Heisenberg) limits.
These two limits are separated by a crossover regime that starts when the local magnetic moment is formed and ends at the Mott transition.
Our work bridges between the weak and strong coupling pictures of the 2D Hubbard model on the square lattice at half filling, by disentangling the interplay among the different underlying mechanisms.
Importantly, we have demonstrated that the local spectral function is not a good quantity for determining the metal-insulator phase transition.
Instead, this problem has to be addressed by analyzing the momentum-resolved spectral function.

\subsection{Description of materials' properties}

In this Section, we provide a summary of the \mbox{D-TRILEX} calculations describing the properties of realistic materials.

\subsubsection{Coexisting charge density wave and ferromagnetic instabilities in monolayer InSe}

Recently fabricated InSe monolayers exhibit remarkable characteristics that indicate the potential of this material to host a number of many-body phenomena.
The electronic structure of ultrathin InSe features flat regions in the valence band dispersion leading to prominent van Hove singularities (vHS) in the hole density of states (DOS). 
If the vHS appears at the Fermi energy, it may result in numerous competing channels of instabilities such as magnetic, charge, or superconducting order with a very non-trivial interplay between them.
In Ref.~\cite{stepanov2021coexisting}, we addressed the problem of collective electronic effects in monolayer InSe.
For this purpose, we derived a realistic model that considers both, long-range Coulomb interactions and the electron-phonon coupling.
The introduced interacting electronic problem was further solved using the \mbox{D-TRILEX} approach.
In the regime of hole-doping, we find that charge ordering represents the main instability.
It is formed in a broad range of doping levels and corresponds to a commensurate CDW, which indicates that this instability is rather driven by strong electronic Coulomb correlations than by an electron-phonon mechanism as discussed previously.
Inside the CDW phase, we detect another collective effect that drives the system towards a ferromagnetic (FM) ordering.
This instability is formed in close proximity to the vHS in the DOS.
Finally, we observe that the electron-phonon coupling tends to suppress the FM ordering, enlarging the CDW phase.
However, the presence of the electron-phonon coupling does not qualitatively affect the observed effects. 

In the monolayer limit DFT calculations predict InSe to be a semiconductor with an indirect energy gap of ${\sim2\,{\rm eV}}$.
The electronic dispersion shows a single well-separated valence band, which has the shape of a Mexican hat.
In Ref.~\cite{stepanov2021coexisting}, we constructed a tractable tight-binding model that accurately reproduces this highest valence band. 
The corresponding single-band model Hamiltonian on an effective triangular lattice reads 
\begin{gather}
H = \sum_{ij,\sigma} t^{\phantom{\dagger}}_{ij} c^{\dagger}_{i\sigma} c^{\phantom{\dagger}}_{j\sigma} + U\sum_{i}n^{\phantom{\dagger}}_{i\uparrow} n^{\phantom{\dagger}}_{i\downarrow} + \frac12 \sum_{i\neq{}j,\sigma\sigma'} V^{\phantom{\dagger}}_{ij}\, n^{\phantom{\dagger}}_{i\sigma} n^{\phantom{\dagger}}_{j\sigma'} +\omega_{\rm ph}\sum_{i}b^{\dagger}_{i}b^{\phantom{\dagger}}_{i} + g\sum_{i,\sigma}n^{\phantom{\dagger}}_{i\sigma}\left(b^{\phantom{\dagger}}_{i}+b^{\dagger}_{i}\right),
\label{eq:H_latt}
\end{gather}
where $c^{(\dagger)}_{i\sigma}$ operator annihilates (creates) an electron on the site $i$ with the spin projection ${\sigma=\{\uparrow,\downarrow\}}$. 
The {\it ab initio} electronic dispersion is reproduced by five neighboring hopping amplitudes $t_{ij}$ (see Ref.~\cite{stepanov2021coexisting}).
The on-site screened Coulomb repulsion ${U=1.78\,{\rm eV}}$ greatly exceeds the bandwidth ${\approx1\,{\rm eV}}$, which usually indicates strong magnetic fluctuations in the system.
As expected for a 2D material, the non-local Coulomb interaction $V_{ij}$ in monolayer InSe is weakly-screened and long-ranged.
Moreover, in monolayer InSe the interaction ${V_{01}=1.04\,{\rm eV}}$ between nearest-neighbor electronic densities ${n_{i\sigma} = c^{\dagger}_{i\sigma} c^{\phantom{\dagger}}_{i\sigma}}$ is larger than the half of the on-site Coulomb repulsion $U$.
This suggests that the considered system may have a tendency to form a CDW phase due to the competition between local and non-local Coulomb interactions.
The effective electron-phonon coupling is dominated by a rather sharp resonance at low phonon frequency, which we can approximate with a local phonon model with the phonon energy ${\omega_{\rm ph}=8.5\,{\rm meV}}$ and its coupling ${g=34.7\,{\rm meV}}$ to the topmost valence band (see Ref.~\cite{stepanov2021coexisting}).
The strong local coupling of electrons to phonons renders the CDW formation even more favorable.
Indeed, upon integrating out bosonic operators $b^{(\dagger)}$ that correspond to phonon degrees of freedom one gets an effective local frequency-dependent attractive interaction
${U^{\rm ph}_{\omega} = 2g^2\frac{\omega_{\rm ph}}{\omega^2_{\rm ph}-\omega^2}}$ that reduces the repulsive on-site Coulomb potential as ${U \to U - U^{\rm ph}_{\omega}}$ and thus enhances the effect of the non-local Coulomb interaction $V_{ij}$.

One of the most remarkable features of the InSe monolayer is the presence of a Mexican-hat-like valence band in the electronic dispersion.
The top of this band exhibits flat regions that lead to a sharp vHS in the DOS.
However, under normal conditions this valence band is fully filled, making the material an indirect semiconductor.
To enhance correlation effects in the system we consider realistic hole dopings with the Fermi level close to the vHS.
First, we solve the many-body problem without considering the electron-phonon coupling in order to investigate purely Coulomb correlation effects.
For detecting main instabilities in the system we perform single-shot \mbox{D-TRILEX} calculations for the charge and spin susceptibilities $X^{\varsigma}_{{\bf q}\omega}$.
In this case, divergences of charge and spin susceptibilities do not affect each other through the self-energy, which allows one to detect instabilities inside broken-symmetry phases.

\begin{figure}[t!]
\centering
\includegraphics[width=0.7\linewidth]{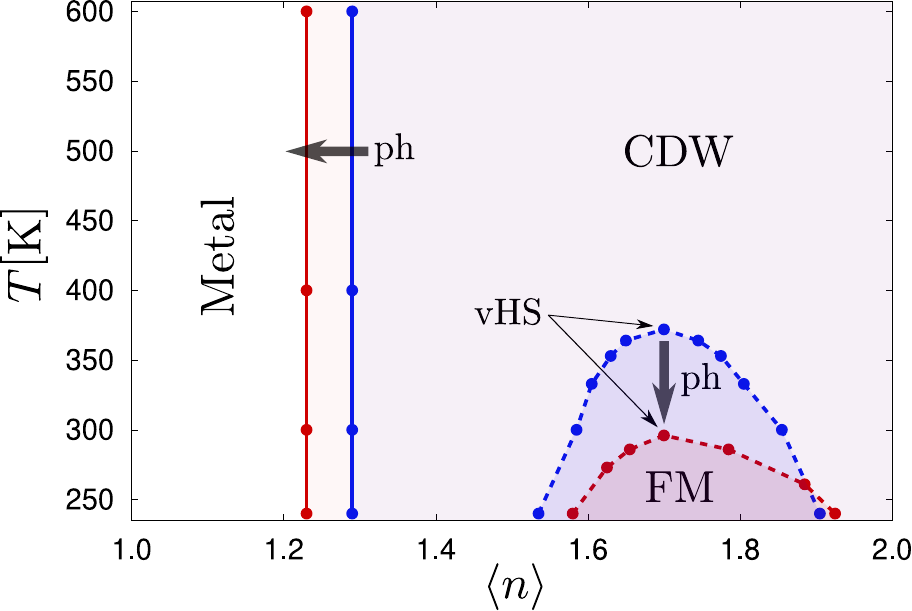}
\caption{Phase diagram for the monolayer InSe as a function of temperature and doping. 
Solid vertical lines correspond to the CDW phase boundaries, dashed lines depict the FM instabilities. 
Results are obtained in the presence (red line) and in the absence (blue line) of the electron-phonon coupling. 
The top of the FM dome corresponds to the filling ${\langle n \rangle = 1.70}$ at which the vHS appears at the Fermi energy. Black arrows with the label ``ph'' illustrate the effect of phonon degrees of freedom that tend to suppress the FM instability and favor the CDW ordering. The Figure is taken from Ref.~\cite{stepanov2021coexisting}. 
\label{fig:phase_InSe}}
\end{figure}

Figure~\ref{fig:phase_InSe} shows the obtained phase diagram for the InSe monolayer, where ${\langle n \rangle}$ is the filling of the considered valence band (${\langle n \rangle}=2$ in the fully-filled band that corresponds to the undoped case). 
Phase boundaries indicate points in the temperature $(T)$ vs. doping space, where corresponding susceptibilities diverge.
We find that the charge susceptibility diverges in a broad range of hole dopings ${\langle n \rangle\geq1.29}$, and the corresponding phase boundary is independent of temperature.
The corresponding Bragg peaks in the charge susceptibility appear at the K points of the Brillouin zone (BZ), which indicates the formation of a commensurate CDW ordering (see Ref.~\cite{stepanov2021coexisting}).
In turn, the spin susceptibility remains finite at the CDW transition point and diverges only inside the CDW phase.
The corresponding instability has a dome shape as depicted in Figure~\ref{fig:phase_InSe} by a blue dashed line. 
Remarkably, we find that the top of the dome corresponds to the filling ${\langle n \rangle = 1.70}$ at which the vHS appears exactly at the Fermi level. 
The momentum resolved spin susceptibility obtained close to the top of the dome (${T=375\,{\rm K}}$, ${\langle n \rangle = 1.70}$) reveals a sharp Bragg peak at the $\Gamma$ point of the BZ, which indicates the tendency towards FM ordering (see Ref.~\cite{stepanov2021coexisting}).

In order to investigate the effect of phonons on the observed instabilities we repeat the same calculation in the presence of the electron-phonon coupling.
As Figure~\ref{fig:phase_InSe} shows, in this case the CDW phase boundary is shifted to smaller values of the filling ${\langle n \rangle = 1.23}$. 
At the same time, the FM instability is pushed down to lower temperatures, but the top of the FM dome remains at the vHS filling ${\langle n \rangle = 1.70}$ as in the absence of phonons.
This result is consistent with the fact that the electron-phonon coupling effectively reduces the on-site Coulomb potential, which consequently decreases the critical temperature for the magnetic instability. 
The observed shift of the CDW phase boundary can also be explained by the same argument.
Indeed, the local Coulomb repulsion favors single occupation of lattice sites.
On the contrary, the non-local Coulomb interaction promotes the CDW ordering, which upon reducing the local Coulomb interaction becomes energetically preferable.

To conclude, we have systematically studied many-body effects in the hole-doped InSe monolayer.
We have found that this material displays coexisting instabilities that are mainly driven by non-local Coulomb correlations.
The commensurate CDW ordering represents the main instability in the system and is revealed in a broad range of doping levels and temperatures.
We also observed the tendency to a FM ordering that manifests itself only inside the CDW phase and is related to a vHS in the electronic spectrum.
The inclusion of the electron-phonon coupling results in a shift of the CDW and FM ordering phases on the phase diagram, which can be explained by an effective reduction of the local Coulomb interaction.
However, the qualitative physical picture does not change in the presence of phonons.
Our results suggest that monolayer InSe can serve as an attractive playground for investigation of coexisting many-body correlation effects and, in particular, of 2D magnetism, although in a bulk phase this material is non-magnetic.

\subsubsection{Doping-dependent charge- and spin-density wave orderings in a monolayer of Pb adatoms on Si(111)}
\label{sec:SiPb}

In Ref.~\cite{vandelli2024doping}, we have used modern computational methods to study many-body effects in a monolayer of lead (Pb) adatoms periodically arranged on a silicon (Si(111)) substrate. Such systems of adatoms artificially are a promising platform for quantum simulations. Indeed, the possibility of varying the geometry and the chemical composition of the adatoms allows for a high degree of tunability of materials properties. An accurate description of the collective electronic behavior arising in these systems is a highly non-trivial task. In particular, the adatom systems are characterized by strong and long-ranged Coulomb interaction and by a relatively large value of the spin-orbit coupling in the case of heavy adsorbants (Sn, Pb, etc.).

Si(111):Pb is one of the most poorly-understood compounds in this class because of many entangled effects that determine its physical properties. In particular, experiments on this system reveal a coexistence region between a chiral magnetic structure and a charge density wave ordering at low temperatures. The underlying mechanism leading to this phase is highly debated.

We have demonstrated that Si(111):Pb hosts a multitude of magnetic and charge ordered phases, that can also coexist with each other. We have found, that these phases are formed as a result of the dynamical symmetry breaking associated with strong electronic correlations, depending on the doping level and temperature. Our calculations show that the geometry of the spin and charge orderings is strongly affected by the doping. In addition, we demonstrate that the large spin-orbit coupling is responsible for the formation of chiral spin textures with potential topological structure.

According to density functional theory calculations, the Si(111):Pb system with 1/3 coverage in the high-temperature ${\sqrt{3} \times \sqrt{3}}$ phase exhibits a narrow half-filled band at the Fermi level, well separated from the rest of the bands.
This band has a $p_{z}$ character, and the corresponding Wannier orbitals are centered at the Pb adatom sites.
We thus employed the single-band interacting electronic model derived from the first principle DFT calculations.
The introduced model was solved using the \mbox{D-TRILEX} method. 
The instabilities related to collective electronic fluctuations in the charge ($c$) and spin ($s$) channels were detected via the momentum-dependent static susceptibility ${X^{c/s}({\bf q}, \omega=0)}$ obtained at zero frequency ${\omega}$. 
The divergence of the susceptibility at momenta ${\bf q= Q}$ indicates a transition to a symmetry-broken ordered state associated with Bragg peaks at ${\bf Q}$.
Transitions without symmetry-breaking, such as the metal to Mott insulator phase transition, were detected by inspecting the spectral function.

\begin{figure}[t!]
\centering
\includegraphics[width=0.7\linewidth]{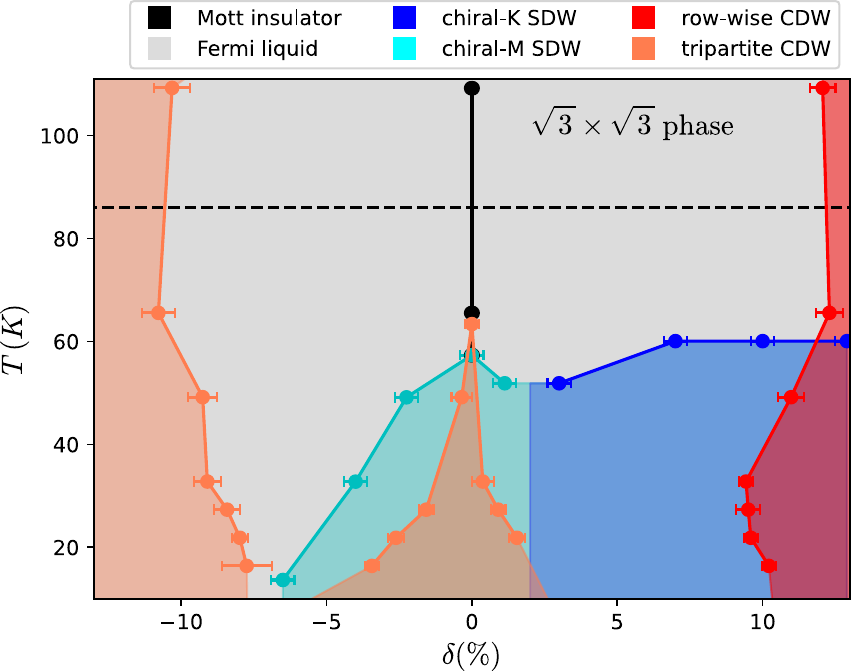} 
\caption{Phase diagram for Si(111):Pb in the ${\sqrt{3}\times\sqrt{3}}$ structure. Different phases as a function of doping $\delta$ and temperature $T$ are highlighted by colors (color code in the legend). Calculations have been performed by fixing the temperature and conducting a scan over doping levels on a finite grid, which defines the error bars. Positive (negative) values of $\delta$ correspond to electron (hole) doping. The horizontal dashed black line depicts the temperature ${T=86~\text{K}}$ at which the material exhibits a structural phase transition. The vertical line that divides the magnetic phases below the transition points is only meant as a guide to the eye, since we are not able to distinguish between the two different phases in symmetry-broken regime. The Figure is taken from Ref.~\cite{vandelli2024doping}.
\label{fig:Si}}
\end{figure}

The phase diagram for the Si(111):Pb material in the ${\sqrt{3}\times \sqrt{3}}$ structure is shown in Fig.~\ref{fig:Si} as a function of doping level $\delta$ and temperature $T$. In the considered system, the value of the local Coulomb interaction is approximately $3$ times larger than the electronic bandwidth. 
As a consequence, at high temperature the half-filled system lies deep in the Mott insulating phase (black line at ${\delta=0\%}$). 
A small amount of hole- or electron-doping causes a phase transition to a Fermi liquid regime (gray area). 
For this reason, the electronic behavior in doped Si(111):Pb is a characteristic manifestation of the physics of a doped Mott insulator.

Upon solving the many-body problem we have identified several different spin density wave (SDW) and CDW orderings at different values of doping, as illustrated in Fig.~\ref{fig:Si}. 
Since these phases are realized for a non-integer filling of electrons, they are likely metallic. 
Specifically, around half-filling we observe a CDW ordering (orange area around ${\delta=0\%}$) characterised by the divergence of the static charge structure factor at the ${{\bf Q}={\rm K}}$ point of the Brillouin zone (BZ).
This ordering is analogous to the 120$^\circ$-N\'eel phase of the Heisenberg model on a triangular lattice with three inequivalent sites in the unit cell.
For this reason, hereinafter we call this type of ordering a ``tripartite CDW''.
Additionally, we identify two other CDW phase transitions at dopings around ${\delta=\pm10\%}$.
These instabilities appear to be weakly temperature-dependent and approximately symmetric with respect to half-filling. 
At hole doping, the CDW ordering vector remains ${{\bf Q}={\rm K}}$ (orange area), as in the half-filled case.
However, in the electron-doped regime the divergence of the static charge structure factor occurs at the ${{\bf Q}={\rm M}}$ point of the BZ, which can be associated with a ``row-wise CDW'' ordering (red area). 

In addition to the CDW instabilities, we also observe magnetic structures with different ordering vectors depending on the doping level. 
Around half-filling, we observe a SDW characterized by Bragg peaks in the static spin structure factor that lie at an incommensurate point ${{\bf Q} \simeq \frac{2}{3} \,{\rm M}}$ of the BZ. 
At ${\delta\gtrsim2\%}$ of electron-doping the SDW ordering vector changes, and the peaks shift to another incommensurate position ${{\bf Q} \simeq \frac{3}{4} \,{\rm K}}$.
The appearance of the Bragg peaks at incommensurate points of the BZ signals the formation of a chiral magnetic order  that can be viewed as a superposition of spin spirals. 
According to the position of the Bragg peaks, we call these magnetic structures ``\mbox{chiral-M}'' (cyan area) and ``\mbox{chiral-K}'' (blue area) SDW, respectively. 

Remarkably, the obtained chiral SDW structures partially coexist with the CDW orderings.
In the considered Si(111):Pb material such coexistence was recently observed by means of STM measurements~\cite{PhysRevLett.120.196402}, but an estimate of the doping level in the system was not provided, presumably due to difficulties in the determination of the effective doping. 
Remarkably, we find that the \mbox{chiral-M} SDW structure coexists only with the tripartite CDW ordering, which appears around half-filling.
Instead, the row-wise CDW ordering coexists only with the \mbox{chiral-K} SDW at a relatively large electron doping.
This observation suggests a simple way for a qualitative estimation of the doping level in the experimentally measured material, which is difficult to probe directly. 
We have made a very crude estimation of the doping level by calculating the area of the Fermi surface that can be deduced from the STM map shown in Ref.~\cite{PhysRevLett.120.196402}.
The obtained result is compatible with up to $\simeq11\%$ electron-doping, which coincides with the region of coexisting chiral-K SDW and row-wise CDW orderings. This result appears to be consistent with use of an electron-doped substrate~\cite{PhysRevLett.120.196402}.

To conclude, we performed many-body calculations for a system of Pb adatoms on a Si(111) substrate, including the SOC and long-range Coulomb interactions. 
By investigating spatial collective electronic fluctuations in both, charge and spin channels, we observe a rich variety of different symmetry-broken charge- and spin-density wave phases in the low temperature regime by varying the doping level. 
We find that the strong SOC in this material results in the formation of \mbox{chiral-M} and \mbox{chiral-K} SDW phases, a signature of which have recently been observed in STM measurements~\cite{PhysRevLett.120.196402}. 
Tuning the doping level allows one to switch between the two chiral SDW phases and thus realize different kinds of spin structures with potential topological structure in one material. 

We also find that two different CDW orderings can appear in Si(111):Pb, and that their geometry is strongly affected by the doping level. 
In order to realize these theoretically predicted phases in experiment, it is necessary to use a probe sensitive to collective excitations, as well as to be able to give an accurate estimation of the doping level. 
Since the precise occupation of the isolated band is experimentally challenging to access, we propose an alternative way to identify the doping level. 
Using a probe sensitive to the underlying magnetic structure, such as spin-polarized STM, could prove a valid alternative to the measurements of the doping, since the magnetic textures appearing at different doping levels exhibit different geometry and also coexist with different types of CDW ordering. 

These results suggest the possibility of a controllable switching between different ordered phases that can be stabilized in Si(111):Pb at different doping levels, and confirm a high tunability that can possibly be exploited for applications.

\subsubsection{Charge density wave ordering in NdNiO$_2$: effects of multiorbital nonlocal correlations}

Layered nickel-oxide compounds have garnered significant attention since the discovery of superconductivity in this class of materials. In addition to superconductivity, these systems exhibit signatures of other non-trivial many-body effects, such as charge density wave and magnetic orderings. Specifically, these two instabilities are considered possible mediators for superconducting pairing, although the microscopic mechanism of the formation of the superconducting state in these materials is still undetermined.
Despite numerous efforts, a comprehensive understanding of the electronic properties of nickelates remains elusive. In particular, because the physics of nickelates is governed by a complex interplay between strong local Coulomb correlations, spatial collective electronic fluctuations, orbital degrees of freedom, and a non-trivial band structure. Until now, even the most advanced numerical calculations could not account for all these important effects simultaneously. This has led to a significant ongoing debate about whether the properties of these systems can be described within a single-band framework, or if it unavoidably requires taking several correlated orbitals into account.

Our work~\cite{stepanov2023charge} addresses these problems by presenting accurate many-body calculations for the layered NdNiO$_2$ compound. We, for the first time, self-consistently incorporate the effect of local correlations, nonlocal charge and magnetic fluctuations, and electron-phonon coupling within the framework of an ab-initio three-orbital model. This approach surpasses any numerical calculation ever performed for this class of materials. 
Our results demonstrate that both spatial electronic correlations and orbital degrees of freedom are important. Our findings reveal that nonlocal electronic correlations in NdNiO$_2$ result in a substantial momentum- and orbital-dependent renormalization of the electronic spectral function, which cannot be captured on the basis of local theories such as DMFT. In particular, we observe that all bands considered in our ab-initio model strongly hybridize with each other in a non-trivial and momentum-dependent way, preventing a single-band description of the system. Moreover, the nonlocal electronic correlations pin the electron pocket of the Ni-$d_{z^2}$ band to the Fermi energy even upon hole doping. These two effects thereby challenge the proposed mechanism of superconductivity based on a single-band model description. 

In Ref.~\cite{stepanov2023charge}, we examined the possibility of a CDW instability in the infinite-layer nickelate NdNiO$_2$, at stoichiometry as well as upon hole doping, by a model Hamiltonian study based on its realistic low-energy electronic structure. 
Specifically, we use a minimal 3-orbital ${\{d_{z^2}, d_{x^2-y^2}, \text{SD}\}}$ model that accounts for an almost occupied Ni-$d_{z^2}$ orbital, a half-filled Ni-$d_{x^2-y^2}$ orbital, and a nearly empty self-doping (SD) band.
This model has been derived from DFT calculations and was supplemented with local interactions.
The problem was solved using the \mbox{D-TRILEX} approach that enables accounting for the combined effect of strong local electronic correlations, nonlocal collective electronic fluctuations, and phonon degrees of freedom. 

\begin{figure}[t!]
\includegraphics[width=1\linewidth]{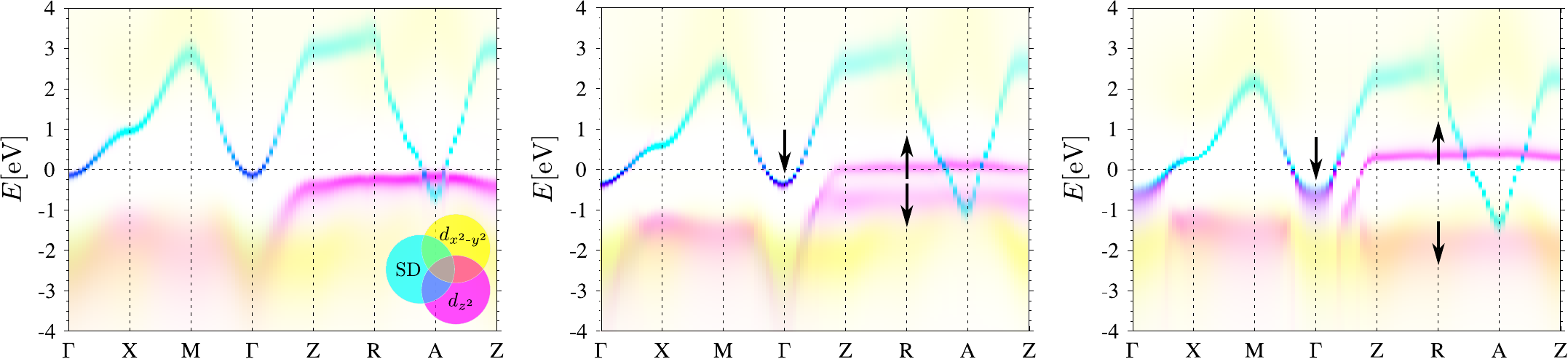}
\caption{Momentum-resolved electronic spectral functions. Results are obtained for the $d_{z^2}$, $d_{x^2-y^2}$, and SD bands (see the inset in the right panel for the color code) for ${\mu_{xy}}$ (left panel), ${\mu^{\rm DC}=11.2}$ (middle panel), and ${\mu_z}$ (right panel) along the high-symmetry path in the BZ.
The low-intensity part of the spectrum at high energies plotted in yellow corresponds to Hubbard bands of the half-filled $d_{x^2-y^2}$ orbital.
The high-intensity part of the spectrum around the Fermi energy plotted in magenta corresponds to the $d_{z^2}$ orbital that displays a flat band feature at ${k_{z}=\pi}$. 
The black arrows depict the shift of the $\Gamma$ point and of the flat part of the $d_{z^2}$ band upon decreasing $\mu^{\rm DC}$. The Figure is taken from Ref.~\cite{stepanov2023charge}. 
\label{fig:Ni}}
\end{figure}

To identify the source of the CDW instability let us look at the momentum-resolved electronic spectral function shown in Fig.~\ref{fig:Ni} for the three different values of the chemical potential $\mu^{\rm DC}$ (double-counting correction), that was introduced for the $d_{z^2}$ and $d_{x^2-y^2}$ to reproduce for the correct position of the correlated bands. 
The dispersiveless low-intensity ``yellow'' weight at small and large energies corresponds to the Hubbard $d_{x^2-y^2}$ bands.
The dispersive high-intensity ``magenta'' weight near the Fermi energy is the metallic $d_{z^2}$ band that displays a flat feature at ${k_{z}=\pi}$. 
We point out, that for all considered values of $\mu^{\rm DC}$ the $d_{z^2}$ band reveals a momentum-dependent hybridization with the other bands.
The hybridization of the $d_{z^2}$ and SD bands around the $\Gamma$ point results in the formation of the electron pocket.

We find that tuning $\mu^{\rm DC}$ leads to a strong momentum-dependent renormalization of the electronic spectral function, and, in particular, of the $d_{z^2}$ band.
This effect cannot be captured on the basis of local theories and requires an advanced combination of the band structure theory with the momentum-dependent many-body approach.
The most striking change concerns the flat part of the $d_{z^2}$ band.
At the largest DC correction $\mu_{xy}$ this part lies below $E_{F}$ (left panel in Fig.~\ref{fig:Ni}).
Upon decreasing ${\mu^{\rm DC}}$ the flat band splits into two parts that move in the opposite directions in energy as depicted by the black arrows in Fig.~\ref{fig:Ni}. 
One part moves toward the Fermi energy and at ${\mu^{\rm DC}=11.2}$ appears at $E_{F}$ (middle panel).
The other part moves toward lower energies and at $\mu_{z}$ hybridizes with the lower Hubbard $d_{x^2-y^2}$ band (right panel). 
At $\mu_{z}$ the system approaches the CDW transition point.
Remarkably, at this value of the chemical potential the upper part of the flat band moves above the Fermi energy and becomes unoccupied.
From this fact one can conclude that the CDW instability in NdNiO$_2$ is not related to the flat band feature.

Another drastic change in the spectral function occurs in the vicinity of the $\Gamma$ point, where the hybridized $d_{z^2}$ and the SD bands form an electron pocket. 
We note that this pocket contains a relatively large spectral weight.
Upon reducing $\mu^{\rm DC}$ the $\Gamma$ point moves to lower energies causing an enhancement of the electron pocket in order to compensate the shift of the upper part of the flat band in the opposite direction.
These observations suggest that the CDW instability originates from the nesting of the Fermi surface (FS) related to the electron pocket of the $d_{z^2}$ band, as the SD band is uncorrelated.

In our calculations, the CDW transition is found to occur for the chemical potential ${\mu_z}$.
However, at this value of ${\mu^{\rm DC}}$ the occupation of the SD band is already approximately two times larger and the occupation of the $d_{z^2}$ band is substantially lower than the ones obtained within the DMFT framework.
In addition, the local spectral function calculated for ${\mu_{xy}}$ (left panel in Fig.~\ref{fig:Ni}) much better reproduces the experimentally observed photoemission spectrum than the one obtained for ${\mu_z}$ (right panel in Fig.~\ref{fig:Ni}). 
Therefore, the choice of ${\mu_z}$ does not correspond to the pristine NdNiO$_2$, but is rather related to the doped case.
On the other hand, in the case of ${\mu_{xy}}$, which reproduces the DMFT occupation of bands, the charge fluctuations are absent in the system, as discussed above.
Therefore, one can conclude that the CDW ordering cannot be found in NdNiO$_2$ at stoichiometry, which is in line with the results of recent experiments, but can be achieved upon hole doping the system. 

To conclude, in this work we investigated the effect of collective electronic fluctuations in the infinite-layer NdNiO$_2$.
We have found that the strong CDW fluctuations are related to electronic correlations within the $d_{z^2}$ orbital.
Upon doping, these fluctuations drive the material toward the CDW ordered phase when considering only possible particle-hole instabilities. The mechanism of this transition relies on the strong momentum-dependent renormalization of the electronic spectral function.

We argue that this complex momentum-dependent renormalization of the electronic spectral function, which is associated with a rather large redistribution of the electronic density between the orbitals, is unlikely to happen in the stoichiometric case.
This means that the formation of the CDW ordering cannot be found in the pristine NdNiO$_2$, which is also confirmed by the most recent experiments.
Instead, we demonstrate that the same renormalization can be obtained upon hole doping leading to the CDW phase transition at a critical density ${n_{\rm total}\simeq2.79}$.
This allows us to speculate that the dip in the superconducting dome observed in the hole-doped NdNiO$_2$ at the density ${n_{\rm total}\simeq2.80}$ may arise due to a competition of the superconductivity with strong CDW fluctuations.

Our work not only improves the description of the electronic properties of layered nickelates but also communicates a more crucial message regarding the multi-orbital and momentum-dependent origin of electronic correlations in these materials. These aspects should be considered essential when addressing the problem of superconductivity and other collective electronic instabilities in these systems.

\subsubsection{On the nature of momentum- and orbital-dependent magnetic fluctuations in Sr$_2$RuO$_4$}

Sr$_2$RuO$_4$ is a paradigmatic correlated material that becomes an unconventional superconductor at low temperatures. Although the mechanism for its superconductivity is still not fully understood, it is believed to arise from strong magnetic fluctuations. However, previous theoretical descriptions of these fluctuations have been inadequate, and significant discrepancies with experimental observations have been reported.

Inelastic neutron scattering (INS) measurements reveal exceptionally strong spin excitations in this material that correspond to an incommensurate wave vector, yet no magnetic ordering has been observed in numerous experimental studies. All previous theoretical attempts, including the {\it state-of-the-art} dynamical mean-field theory approach, have failed to accurately describe the spin fluctuations, typically overestimating their strength and predicting the formation of an ordered state at finite temperatures.

In Ref.~\cite{Ruthenates}, we have elaborated a theoretical description of this multi-orbital system by self-consistently accounting for both local and non-local electronic correlations.
We demonstrate that a self-consistent treatment of spatial spin excitations suppresses their strength and reduces the electronic correlations in Sr$_2$RuO$_4$, ultimately eliminating the magnetic ordering predicted by {\it state-of-the-art} approaches.
Moreover, our results accurately reproduce the form of the spin susceptibility as deduced from experimental INS measurements, allowing us to detect an intriguing orbital dependence of the latter.

\begin{figure}[t!]
\centering
\includegraphics[width=0.7\linewidth]{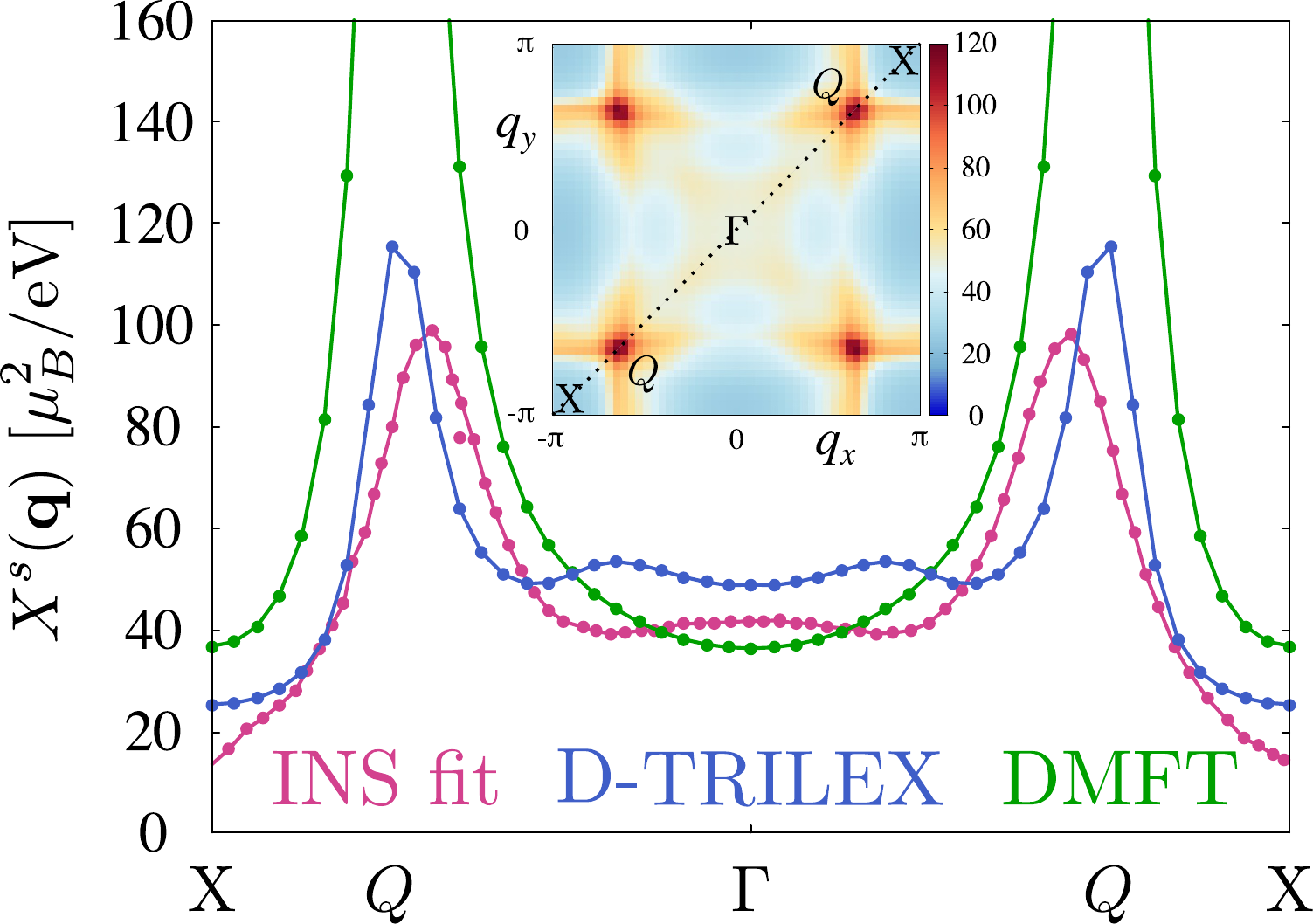}
\caption{The real part of the static spin susceptibility $X^{s}({\bf q})$ along the high symmetry path X-$\Gamma$-X (dashed line in the inset) of the first BZ.
The result is calculated using \mbox{D-TRILEX} (blue, ${T=145}$\,K) and is obtained by fitting the experimental INS data~\cite{PhysRevLett.122.047004} (magenta, ${T=150}$\,K). 
The result of the DMFT calculation is also shown for comparison (green, ${T=193}$\,K). 
The inset shows the \mbox{D-TRILEX} susceptibility in the ${(q_x,q_y,0)}$ plane. 
The momentum-space structure of the magnetic susceptibility exhibits peaks at the incommensurate wave vector ${Q=(3\pi/5,3\pi/5,0)}$, the dome-like background signal centered at $\Gamma$ and minima around the $M$ and $X$ points. The Figure is taken from Ref.~\cite{Ruthenates}. 
\label{fig:Ru}}
\end{figure}

To accurately describe magnetic fluctuations in Sr$_2$RuO$_4$, in Ref.~\cite{Ruthenates} we consider an effective three-band model corresponding to maximally localized $\{xz, yz, xy\}$ orbitals derived from DFT.
We account for the on-site electronic interaction that is parametrized in the Kanamori form. 
The value of the intra-orbital Coulomb repulsion, ${U=2.56}$\,eV, is chosen based on the constrained random-phase approximation (cRPA) analysis. 
A Hund's exchange coupling $J$, crucial for Sr$_2$RuO$_4$, is selected by evaluating results across different values of $J$. 
Specifically, we estimate the mass enhancement and spin susceptibility obtained for each $J$. 
We find that the correct mass enhancement is achieved for ${J\simeq0.35-0.40}$\,eV, and the accurate spin susceptibility for ${J\simeq0.30-0.35}$\,eV, leading to an optimal choice of ${J=0.35}$\,eV.

In order to obtain a reliable description of the feedback of electronic correlations on the spectral and magnetic properties of the compound we use the \mbox{D-TRILEX} approach. 
We perform calculations at a relatively low temperature ${T=145}$\,K, where experimental data are available for comparison. 
The inset in Fig.~\ref{fig:Ru} shows the real part of the static magnetic (${\omega=0}$) susceptibility ${X^{s}({\bf q})}$ obtained for the first BZ using \mbox{D-TRILEX}. 
The main part of Fig.~\ref{fig:Ru} displays a cut of the susceptibility along the X-$\Gamma$-X diagonal of the BZ (dashed line in the inset).
The results are calculated numerically using \mbox{D-TRILEX} (blue) and DMFT (green), and are compared to the INS result (magenta)~\cite{PhysRevLett.122.047004}.
At this temperature the magnetic fluctuations are already very strong. 
In fact, for the considered model DMFT predicts a SDW ordered state already at ${T\simeq145}$\,K.
For this reason, the DMFT result in Fig.~\ref{fig:Ru} is shown for a bit higher temperature ${T=193}$\,K.
The spin susceptibility of DMFT, calculated in the vicinity of the SDW transition, features large peaks at the incommensurate $Q$ vectors.
We find, that a self-consistent inclusion of the magnetic fluctuations beyond DMFT, using \mbox{D-TRILEX}, leads to a strong suppression of the SDW $Q$ peaks in a good agreement with the INS result, and no ordering is observed. 

The second important outcome of our results is related to the overall behavior of the spin susceptibility across the BZ. 
According to experimental measurements the magnetic signal can be decomposed into the sum of the SDW $Q$ peaks and a broad dome structure centered around the $\Gamma$ point, while DMFT calculations instead find a quasi-constant background signal besides the $Q$ peaks.
The \mbox{D-TRILEX} calculations reveal a significantly diminished spin susceptibility at the edges of the BZ, with a ``cross''-like structure in momentum space of higher intensity appearing at the center of the zone, visible in the inset of Fig.~\ref{fig:Ru}. 
The overall structure of the susceptibility agrees very well with the experimental results~\cite{PhysRevLett.122.047004}.

To conclude, we have studied the effect of magnetic fluctuations on the electronic correlations and the spin susceptibility of Sr$_2$RuO$_4$.
These excitations are found to be significant in this material, such that DMFT calculations predict an ordered SDW state that is not observed experimentally. 
We demonstrate that the self-consistent inclusion of spatial magnetic fluctuations suppresses their strength by reducing many-body correlations to the extent that no magnetic ordering is realized in Sr$_2$RuO$_4$. 
The overall behavior of the spin susceptibility in momentum space deduced from INS measurements is well reproduced by our calculations. 
We obtain finite peaks at the incommensurate SDW $Q$ vectors, a broad dome-shaped structure centered around the $\Gamma$ point and a diminished magnetic response at the edges of the BZ. 
We speculate that the predicted unusual dome structure may strongly affect the symmetry of the superconducting order parameter, as the spin fluctuations are considered as the most probable pairing mechanism for superconductivity in Sr$_2$RuO$_4$.

\subsubsection{Fingerprints of a charge ice state in the doped Mott insulator Nb$_3$Cl$_8$}
\label{sec:Nb3Cl8}

The interplay between strong electronic correlations and the inherent frustration of certain lattice geometries is a common mechanism for the formation of nontrivial states of matter. 
In Ref.~\cite{j6bj-gz7j}, we theoretically explored the collective electronic effects in the monolayer Nb$_3$Cl$_8$, a recently discovered triangular lattice Mott insulator. 
Our advanced many-body numerical simulations predict the emergence of a phase separation region upon doping this material. 
Notably, in close proximity to the phase separation, the static charge susceptibility undergoes a drastic change and reveals a distinctive bow-tie structure in momentum space. 
The appearance of such a fingerprint in the context of spin degrees of freedom would indicate the formation of a spin ice state. 
This finding allows us to associate the observed phase separation to a charge ice state, a state with a remarkable power law dependence of both the effective exchange interaction and correlations between electronic densities in real space.

To investigate the many-body effects in the monolayer Nb$_3$Cl$_8$, we use a single molecular orbital extended Hubbard model on an effective triangular lattice that was introduced in Ref.~\cite{grytsiuk2024nb3cl8} based on {\it ab-initio} calculations.
In Nb$_3$Cl$_8$, the Coulomb repulsion $U_{|j\text{-}j'|}$ is found to be extremely large compared to the non-interacting bandwidth and rather long-ranged~\cite{grytsiuk2024nb3cl8}: 
The local interaction is equal to ${U_{0}\simeq1.9}$\,eV and the interaction between the neighboring lattice sites $U_{1}$ is only approximately 2.5 times smaller than the local one.
For this reason, the undoped monolayer Nb$_3$Cl$_8$ resides in the Mott insulating state. 
To accurately describe of this state, as well as other manifestations of strong collective electronic behavior, we use the \mbox{D-TRILEX} approach.
This method is particularly useful for detecting various ordered phases that are determined by the divergence of the corresponding susceptibilities at the momentum that defines the wave vector of the ordering. 
The strength of the fluctuations driving the phase transition can be estimated by looking at the largest static (${\omega=0}$) dielectric function $\epsilon^{\varsigma}_{\omega}({\bf q})$ in the corresponding (charge or spin) channel.
The dielectric function is related to the susceptibility $X^{\varsigma}_{\omega}({\bf q})$ as: ${\epsilon^{\varsigma}_{\omega}({\bf q}) = \Pi^{\varsigma}_{\omega}({\bf q})/X^{\varsigma}_{\omega}({\bf q})}$ and shows how the polarization operator $\Pi^{\rm ch/sp}_{\omega}({\bf q})$ (irreducible with respect to the bare interaction ${U^{\rm ch/sp}({\bf q})}$ part of the susceptibility) is renormalized by the collective electronic fluctuations in the corresponding channel.
Thus, ${\epsilon=1}$ indicates the absence of the fluctuations, and ${\epsilon({\bf Q})=0}$ signals the formation of the ordered state with the ordering vector ${\bf q=Q}$.

We begin investigating many-body effects in the monolayer Nb$_3$Cl$_8$ with the half-filled case.
We perform calculations at two different temperatures, ${T=290\,\text{K}}$ and 145\,K.
At both temperatures, we confirm that the material lies very deep in the Mott insulating phase: The obtained electronic spectral function features two narrow and nearly dispersiveless Hubbard bands separated by a large gap of the order of $U_{0}$.
However, we do not observe any signature of notable magnetic or charge fluctuations. 
Indeed, the largest dielectric function in the spin channel ${\epsilon^{\rm sp}_{\omega=0}({\bf q})}$ is found at the wave vector ${{\bf q}=\text{K}=\{4\pi/3, 0\}}$, which corresponds to the 120$^{\circ}$ AFM type of spin fluctuations, in agreement with the findings of Ref.~\cite{grytsiuk2024nb3cl8}. 
However, by lowering the temperature from ${T=290}$\,K to 145\,K the spin dielectric function changes from ${\epsilon^{\rm sp}_{\omega=0}(\text{K})=0.97}$ to ${\epsilon^{\rm sp}_{\omega=0}(\text{K})=0.94}$, which indicates that the magnetic fluctuations are relatively weak.
Furthermore, a large extrapolated value of ${\epsilon^{\rm sp}_{\omega=0}(\text{K})=0.91}$ at ${T=0}$ indicates that the system does not exhibit any tendency toward the formation of the spin ordered state, at least at the considered temperatures.
Our calculations also do not reveal significant charge fluctuations.
The largest dielectric function in the charge channel corresponds to the zero momentum ${{\bf q}=\Gamma=\{0,0\}}$ and is very close to unity as well: ${\epsilon^{\rm ch}_{\omega=0}(\Gamma)=0.96}$ at ${T=290}$\,K and ${\epsilon^{\rm ch}_{\omega=0}(\Gamma)=0.97}$ at $T=145$\,K. 

\begin{figure}[t!]
\centering
\includegraphics[width=0.7\linewidth]{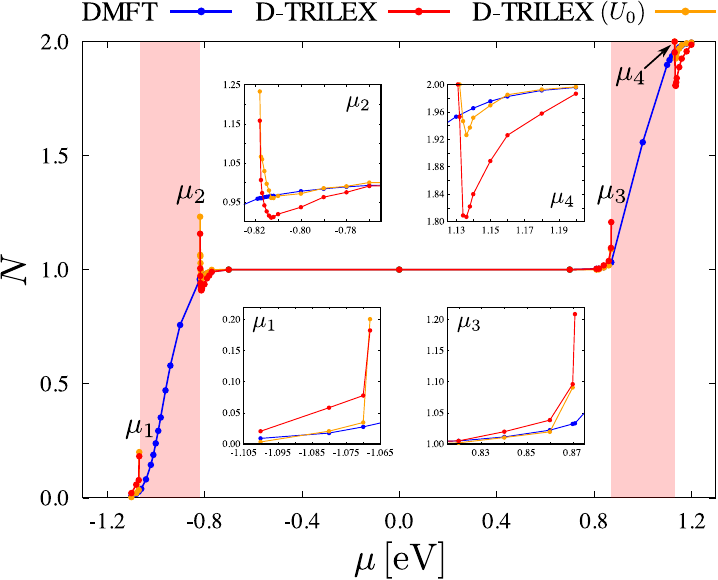}
\caption{Phase diagram. Dependence of the average electronic density $N$ on the value of the chemical potential $\mu$, where ${\mu=0}$ corresponds to the case of an undoped monolayer Nb$_3$Cl$_8$. Calculations are performed using DMFT with the local $U_0$ interaction (blue), \mbox{D-TRILEX} with the full long-range Coulomb potential $U_{|j\text{-}j'|}$ (red), and the ``\mbox{D-TRILEX} ($U_0$)'' method considering only the local interaction $U_0$ (orange). The chemical potentials $\mu_i$ correspond to the boundaries of the PS regions (shaded red areas). The result is obtained at ${T=290\,\text{K}}$. The insets show a more detailed behavior of the curves in the vicinity of PS regions. The Figure is taken from Ref.~\cite{j6bj-gz7j}.
\label{fig:Phase_diagram_NbCl}}
\end{figure}

Doping the system does not alter the strength of the magnetic fluctuations, which are already weak in the considered material at half filling. 
However, we observe a significant change in the charge fluctuations upon doping.
Figure~\ref{fig:Phase_diagram_NbCl} shows the evolution of the average electronic density $N$ as a function of the chemical potential calculated at ${T=290}$\,K.
Here, ${\mu=0}$ corresponds to the undoped case of a half-filled Mott insulator.
At ${|\mu|\lesssim0.8}$\,eV the chemical potential lies inside the gap and the system remains in the Mott insulating state with ${N=1}$.
At ${|\mu|\gtrsim0.8}$\,eV, taking into account only local correlations and the local Coulomb interaction $U_0$ within DMFT results in a gradual increase of the electronic density with increasing chemical potential (blue curve).
Importantly, no first-order phase transition is identified, which is consistent with other finite temperature single-site DMFT calculations of single-band models.
Additionally considering the non-local collective electronic fluctuations and the full long-range Coulomb potential $U_{|j\text{-}j'|}$ within the \mbox{D-TRILEX} approach drastically changes this picture (red curves). 
We find that in both electron (${N>1}$) and hole doped (${N<1}$) cases, the electronic compressibility ${\kappa =  \frac{1}{N^2}\frac{{\rm d} N}{{\rm d} \mu}}$ diverges almost immediately when the chemical potential reaches the Hubbard bands.
Importantly, this divergence occurs not only when approaching the Hubbard bands from half filling (${\mu_2=-0.819}$\,eV and ${\mu_3=0.871}$\,eV), but also from the unfilled (${\mu_1=-1.068}$\,eV) and fully filled (${\mu_4=1.131}$\,eV) sides.
The divergence of the compressibility indicates the appearance of the phase separation (PS) region, which is additionally confirmed by the divergence of the charge susceptibility at the zero momentum ${{\bf q}=\Gamma}$.
This divergence appears when the static dielectric function in the charge channel $\epsilon^{\rm ch}_{\omega=0}({\bf q})$ approaches zero.
Importantly, the divergence of the compressibility and the charge susceptibility occurs at the same values of the chemical potential, although these quantities are computed independently.
As Ref.~\cite{j6bj-gz7j} shows, extrapolating $\epsilon^{\rm ch}_{\omega=0}(\Gamma)$ to zero shows that the PS occurs at $19.5\pm1.5\%$ of doping.
At ${\mu_1<\mu<\mu_2}$ and ${\mu_3<\mu<\mu_4}$ the charge susceptibility remains divergent, defining the PS as regions of ``forbidden'' chemical potentials highlighted in red colour in Figure~\ref{fig:Phase_diagram_NbCl}.

\begin{figure}[t!]
\includegraphics[width=1\linewidth]{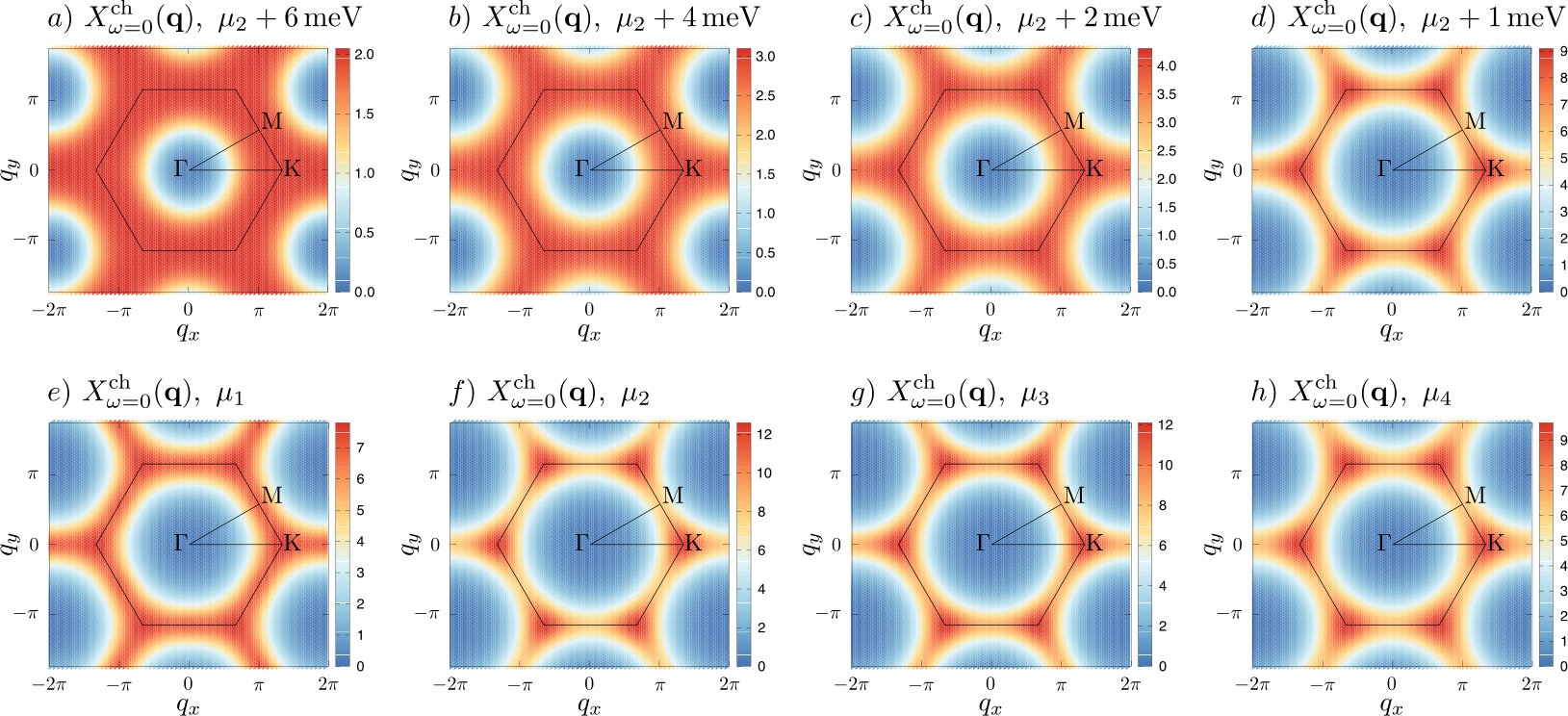}
\caption{Charge susceptibility. Top row: Evolution of the static charge susceptibility $X^{\rm ch}_{\omega=0}({\bf q})$ in the vicinity of the PS. The calculations are performed for ${\mu=\mu_2+6\,\text{meV}}$ (a), ${\mu_2+4\,\text{meV}}$ (b), ${\mu_2+2\,\text{meV}}$ (c) and ${\mu_2+1\,\text{meV}}$ (d). 
Bottom row: The static charge susceptibility obtained at different PS boundaries defined by the chemical potentials $\mu_1$ (e), $\mu_2$ (f), $\mu_3$ (g) and $\mu_4$ (h). 
All results are obtained at ${T=290\,\text{K}}$ and shown in the momentum space ${(q_x, q_y)}$. The first BZ is depicted by the black hexagon. The high-symmetry points $\Gamma$, K and M are labeled explicitly. The Figure is taken from Ref.~\cite{j6bj-gz7j}.
\label{fig:X_NbCl}}
\end{figure}

To understand the physical nature of the PS, let us examine the static charge susceptibility $X^{\rm ch}_{\omega=0}({\bf q})$ in the vicinity of $\mu_2$, as shown in the top row of Figure~\ref{fig:X_NbCl}.
We find that the divergence of the dielectric function is not visible in the susceptibility due to a very small value of the polarization operator at the $\Gamma$ point. 
Instead, the largest value of the charge susceptibility corresponds to momenta at the edge of the BZ depicted by the black hexagon.
Far from the PS, at ${\mu=\mu_2+6}$\,meV (a) and ${\mu=\mu_2+4}$\,meV (b), where the largest charge dielectric function is respectively equal to ${\epsilon^{\rm ch}_{\omega=0}(\Gamma)=0.406}$ and ${\epsilon^{\rm ch}_{\omega=0}(\Gamma)=0.436}$, the charge susceptibility is relatively small and its maximum is distributed over a rather large part of the BZ.
As the PS is approached more closely (c), at ${\mu=\mu_2+2}$\,meV (${\epsilon^{\rm ch}_{\omega=0}(\Gamma)=0.156}$) the maximum of the susceptibility begins to localize at the edge of the BZ, but the value of the susceptibility still remains rather small.
Finally, in close proximity to the PS, at ${\mu=\mu_2+1}$\,meV, ${\epsilon^{\rm ch}_{\omega=0}(\Gamma)=0.007}$ (d) and $\mu_2$, ${\epsilon^{\rm ch}_{\omega=0}(\Gamma)=0.002}$ (f) the susceptibility drastically increases and displays a distinctive ``bow-tie'' shape with the maximum at the K point and the pinch-point at the M point.

The appearance of the bow-tie structure~\cite{PhysRevLett.87.047205, doi:10.1126/science.1064761, PhysRevLett.93.167204, PhysRevB.71.014424, doi:10.1126/science.1177582} in the spin susceptibility is one of the most direct probes of the spin ice state~\cite{balents2010spin}.
Spin ice is a highly frustrated magnetic state without long-range magnetic order. 
It lacks a conventional order parameter, which makes identifying this elusive state significantly more challenging.
In the case of a spin ice, the bow-tie structure in momentum space is a consequence of the dipolar correlations between the magnetic moments in real space, which decay as a power law ${\sim1/R^{D}}$ with the distance $R$, where $D$ is the dimension of the system~\cite{PhysRevLett.93.167204, PhysRevB.71.014424, doi:10.1126/science.1177582}.
This power-law form of correlations is reminiscent of the behavior of hydrogen bonding interactions in water and indicates the absence of static magnetic order, suggesting that spin correlations persist over long distances. 
In contrast, in magnetic systems with long-range order, spin correlations decay exponentially with distance because the system reaches a stable magnetic configuration. 
If the power law is reproduced exactly, the susceptibility at the M point shows a singularity. 
Deviation from the power law rounds this singularity, resulting in a rapidly decaying value of the susceptibility in the ${\text{M}-\Gamma}$ direction, which corresponds to a finite correlation length of the spin ice state~\cite{balents2010spin}. 
By analogy with this state, we identify the observed bow-tie form of the charge susceptibility with a charge ice state.
Remarkably, the bottom row of Figure~\ref{fig:X_NbCl} demonstrates that this state is formed at all chemical potentials that define the boundaries of the PS regions and only in close proximity to the PS (${\epsilon^{\rm ch}_{\omega=0}(\Gamma)\lesssim0.01}$).
We find that the charge susceptibility does not exhibit a sharp singularity at the pinch-points (M). As shown in the top row of Figure~\ref{fig:X_NbCl}, approaching the PS state (at $\mu_2$) enhances the peak at the M point and reduces its width. Since the present calculations capture only the precursor of the charge ice state, one can expect the charge susceptibility near the M point to become significantly sharper deeper inside the PS region, corresponding to the fully developed charge ice state.

We have verified that the observed charge ice state is also present at lower temperatures.
At ${T = 145\,\text{K}}$, the PS state is found at approximately 8\% doping, and in its vicinity the charge susceptibility acquires a distinct bow-tie shape, qualitatively similar to that shown in the bottom row of Fig.~\ref{fig:X_NbCl}.
This result is consistent with the findings of Ref.~\cite{vandelli2024doping}, which reports that the phase boundary of the charge instability observed in the doped triangular lattice Mott insulator Si(111):Pb shifts to lower doping levels as the temperature decreases.

To conclude, we have investigated collective electronic instabilities in the monolayer Nb$_3$Cl$_8$. 
We have found that at half filling, the considered material lies deep in the Mott insulating phase and surprisingly does not reveal any tendency towards the formation of a charge or spin ordered state, contrary to some other Mott insulators with triangular lattice geometry~\cite{PhysRevLett.120.196402, vandelli2024doping}.
Upon doping, the system exhibits a region of PS, detected by a simultaneous divergence of the charge susceptibility (or dielectric function) at zero momentum $\Gamma$ and the electronic compressibility.
The critical doping required for the formation of the PS can be estimated through the extrapolation of the dielectric function to zero, resulting in an approximate value of 20\% doping at ${T=290\,\text{K}}$ and 8\% doping at ${T=145\,\text{K}}$.
This suggests that PS in monolayer Nb$_3$Cl$_8$ could potentially be realized experimentally, given the achievable doping levels in other two-dimensional systems, such as high-temperature superconducting cuprates.

We observe that away from the PS, the charge susceptibility is rather small and weakly momentum-dependent.  
However, in close proximity to the PS, the charge susceptibility dramatically increases and reveals a distinctive bow-tie pattern in momentum space, reminiscent of the magnetic susceptibility observed in the spin ice state.
Interestingly, Ref.~\cite{j6bj-gz7j} shows that the charge ice state is observed only when accounting for the effect of the full long-range Coulomb potential, while with only local interactions, the system reveals an ordinary PS state. 
This can be explained by the fact that geometrical frustration likely plays an important role in the development of the charge ice state, similarly to the spin ice state, as it prevents the formation of long-range order.
Incorporating the long-range Coulomb interaction enhances the frustration inherent in the triangular lattice, which, apparently, is not sufficient for the formation of the charge ice state if only the local interaction is considered. 
In this context, the single-band model for Nb$_3$Cl$_8$ provides a unique platform for exploring fascinating many-body effects driven by frustration.

\subsection{Application to superconductivity}

In Section~\ref{sec:SC_DB} we introduced the DF approach in the Nambu space representation in order to address the problem of superconductivity related to a $d$-wave pairing. 
A similar development can also be performed in the framework of the \mbox{D-TRILEX} approach.
To do so, we start with the DB action~\eqref{eq:SC_DB_action} with dual fermionic variables in the ${2\times2}$ Nambu subspace.
After that, we perform the usual transformation of the bosonic fields, described in Section~\ref{sec:PBDAction}, and get the following form of the partially bosonized dual action~\eqref{eq:DTRILEX_action}:
\begin{align}
{\tilde{\cal {S}}} =  -\mathrm{Tr} \sum_{1,2} \varphi^{*}_{1}  [ {\tilde{ \cal {G}}}  ]^{-1}_{12} \varphi^{\phantom{*}}_{2}  - \frac12\sum_{q,\varsigma}  b^{\varsigma}_{-q}
\left[\tilde{\cal W}^{\varsigma}_{q}\right]^{-1} b^{\varsigma}_{q}
+ \sum_{\substack{k,q,\varsigma,\\\sigma\sigma'}} \Lambda^{\varsigma}_{\nu\omega} f^{*}_{k\sigma}\sigma^{\varsigma}_{\sigma\sigma'} f^{\phantom{*}}_{k+q,\sigma'} b^{\varsigma}_{q}\,.
\label{eq:fbaction_SC_DT}
\end{align}

The \mbox{D-TRILEX} approach self-consistently incorporates the effects of spatial collective electronic fluctuations through ladder-type diagrams for the self-energy and polarization operator in the dual space.
In the presence of the superconducting field, the self-energy develops an anomalous contribution, while the normal part of the self-energy remains the same as in the paramagnetic case (see Section~\ref{sec:DT_diagrams}):
\begin{align}
\tilde{\Sigma}^{\uparrow\uparrow}_{k} 
= &-\sum_{q,\varsigma} 
\Lambda^{\varsigma}_{\nu\omega} \tilde{G}^{\uparrow\uparrow}_{k+q} \tilde{W}^{\varsigma}_{q} \Lambda^{\varsigma}_{\nu+\omega,-\omega}
+ 2\sum_{k'} 
\Lambda^{\rm c}_{\nu,0} \tilde{\cal W}^{\rm c}_{0} \Lambda^{\rm c}_{\nu',0} \, \tilde{G}^{\uparrow\uparrow}_{k'}\,,
\notag\\
\tilde{\Sigma}^{\uparrow\downarrow}_{k} 
= &-\sum_{q,\varsigma} 
\xi^{\varsigma}\Lambda^{\varsigma}_{\nu\omega} \tilde{G}^{\uparrow\downarrow}_{k+q} \tilde{W}^{\varsigma}_{q} \Lambda^{\varsigma}_{-\nu,-\omega}\,,
\label{eq:Sigma_DT}
\end{align}
where ${\xi^{m/d} = \pm1}$.
Despite the Green's function has an anomalous part, the polarization operator remains diagonal in the channel space, but receives an additional term originating from the anomalous Green's function:
\begin{align}
\tilde{\Pi}^{\varsigma}_{q}  
= 2\sum_{k} \left(\Lambda^{\varsigma}_{\nu+\omega,-\omega} \tilde{G}^{\uparrow\uparrow}_{k} \tilde{G}^{\uparrow\uparrow}_{k+q} \Lambda^{\varsigma}_{\nu\omega}
+ \xi^{\varsigma} \Lambda^{\varsigma}_{-\nu,-\omega} \tilde{G}^{\downarrow\uparrow}_{k} \tilde{G}^{\uparrow\downarrow}_{k+q} \Lambda^{\varsigma}_{\nu\omega}\right).
\label{eq:dual_pol_DT}
\end{align}
The dressed dual Green's function and the renormalized interaction are obtained via the corresponding Dyson equations:
\begin{align}
\big[\hat{\tilde{G}}_{k}\big]^{-1} &= \big[{\cal \hat{\tilde{G}}}_{k}\big]^{-1} - \hat{\tilde{\Sigma}}^{\phantom{*}}_{k},\\
\big[\tilde{W}^{\varsigma}_{q}\big]^{-1} &= \big[{\cal \tilde{W}}^{\varsigma}_{ q}\big]^{-1} - \tilde{\Pi}^{\varsigma}_{q}\,,
\end{align}
and the lattice Green's function is further obtained using the same Eq.~\eqref{eq:Nambu_GF} as in the DF approach. 
The normal ${\Sigma_{k} = \Sigma^{\uparrow\uparrow}_{k}}$ and anomalous ${S_{k} = \Sigma^{\uparrow\downarrow}_{k}}$ parts of the lattice self-energy can be obtained through the usual Dyson equation for the lattice Green's function.

In Section~\ref{sec:SC_DB}, the DF approach in the lattice DQMC form was applied to a hole-doped $t$-$t'$ Hubbard model, relevant for the high-temperature cuprate superconductors.
The obtained results there call for identifying the microscopic processes behind the formation of the two-gap structure in the superconducting spectral function and understanding the implications of these processes for the mechanism of the strong-coupling theory of $d$-wave superconductivity.
Our general answer to this question is that this mechanism is similar to Anderson RVB~\cite{Anderson2004} and kinetic $t_ \bot$ mechanism of multilayer pair-hole hopping~\cite{Anderson1993, Anderson1998}. 
The main idea is that, informally speaking, it is not energetically favorable for fermions in the normal phase to be ``bad'', that is, non-quasiparticle, so that they prefer to form, instead, a superconducting condensate. 
Anderson connected this energy balance with the interlayer hopping which is more difficult for non-Fermi-liquid state than for regular Fermi liquid. 
This seems to be not supported by further experimental development, and recent discovery of monolayer high-temperature cuprate superconductors~\cite{HTSC_1layer} probably closes the discussion. 
Contrary, we believe that this is in-plane NNN kinetic energy, which plays the role that Anderson attributed to the interlayer hopping. Despite that, we show that the pseudogap formation by itself competes with superconductivity, this is not the main effect of the NNN hopping. 
As was have demonstrated by exact diagonalization for a small, ${4 \times 4}$ periodic system, ${t'=-0.3}$ leads to a huge enhancement of the binding energy of two holes, and this is related to the peculiarities of energy spectrum of an isolated plaquette. We believe that this essentially strong-coupling physics provides the necessary missing component of superconducting glue. 
Indeed, in our point of view, the preformed superconducting pairs already exist in the small clusters of the order of ${4 \times 4}$ due to strong AFM correlations, and it is $t'$ that makes these hole pairs ``coherent'' even in the single Cu-O plane.
Another important effect comes from the large density of state at high-temperatures near the Fermi level for hole doping around ${\delta \simeq 15 \%}$ and ${t'=-0.3}$, which is related to a ``highly degenerate'' ground state of a small cluster~\cite{Danilov2022}. 
In this case, the formation of a pseudogap we can attribute to the physics of periodic Kondo problem or ``destructive interference phenomena''~\cite{Gunnarsson_2014}.
It would be very interesting to connect our microscopic approach to the phenomenological two-fluid model of cuprate superconductors~\cite{Ayres22}, which assumes the coexistence of two fermionic subsystems in the normal phase, a Fermi-liquid and a non-Fermi-liquid ones, where only the latter becomes superconducting at low temperatures.

To identify the precise mechanism by which ``bad'' electrons become superconducting, we perform \mbox{D-TRILEX} calculations and compare them to the DF lattice DQMC scheme.
Compared to the latter, the \mbox{D-TRILEX} method accounts for more diagrammatic contributions to the self-energy by including an infinite set of ladder-type diagrams, enabling an accurate treatment of collective charge and paramagnetic spin fluctuations (paramagnons) with arbitrary spatial range.
The trade-off, however, is that this approach cannot handle large lattice reference problems as efficiently as the DF lattice DQMC scheme.
We use a single-site DMFT impurity problem as the reference system for \mbox{D-TRILEX} calculations, which are performed for ${64\times64}$ ${\bf k}$-points in the Brillouin zone and 128 Matsubara frequencies.

\begin{figure}[t!]
\centering
\includegraphics[width=0.7\textwidth]{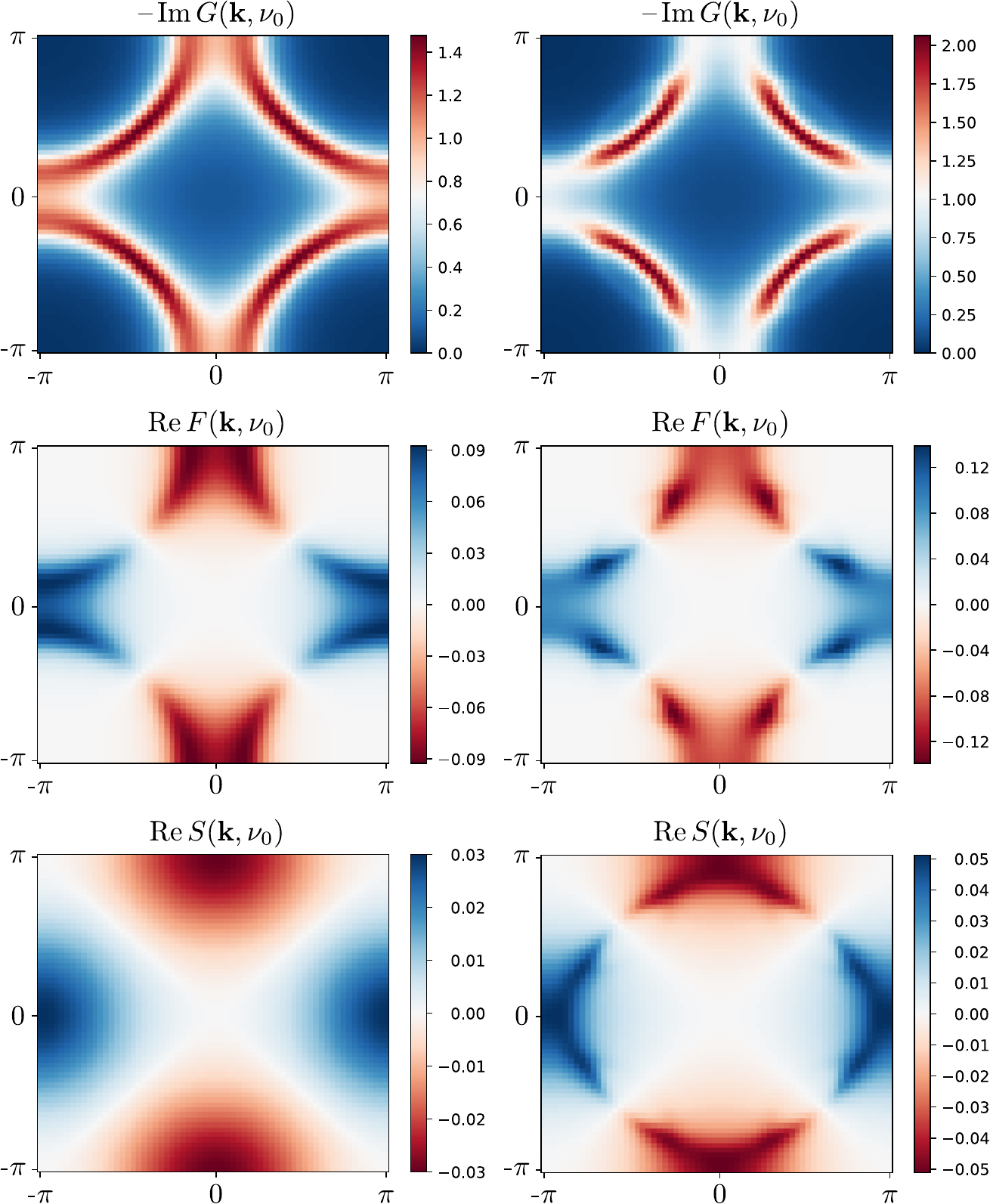}
\caption{Top panels: The imaginary part of the normal Green's function $G$. Middle panels: The real part of the anomalous Green's function $F$. Bottom panels: The real part of the anomalous self-energy $S$. The results are obtained using \mbox{D-TRILEX} at ${\beta=10}$ for ${\delta=16\%}$ of hall doping at the lowest Matsubara frequency ${\nu_0=\pi/\beta}$ as a function of momentum ${\bf k}$, comparing cases without (left column) and with (right column) the effective Higgs field ${h_{\rm AFM}=0.485}$. The Figure is taken from Ref.~\cite{Cuprates}.
\label{fig:PG}}
\end{figure}

In Figure~\ref{fig:PG}, we present the results obtained using \mbox{D-TRILEX} for the same model parameters as those in the DF lattice DQMC scheme, obtained at ${\beta=10}$ for ${\delta=16\%}$ hole doping and the ${h_{dw}=0.01}$ value of the superconducting $d$-wave field.
The left column shows the momentum-resolved normal $G({\bf k})$ (top panel) and anomalous $F({\bf k})$ (middle panel) contributions to the Nambu-Gor'kov lattice Green's function, as well as the anomalous lattice self-energy $S({\bf k})$ (bottom panel), all obtained for the lowest Matsubara frequency.
At this relatively large temperature, we find that the spatial magnetic fluctuations are still relatively weak and do not lead to the development of a pseudogap in the normal Green's function $G$.
The form of the anomalous Green's function $F$ in momentum space resembles the normal Green's function $G$, but with an additional $d$-wave-like symmetry, displaying zero response at the nodal point. 
The anomalous self-energy also exhibits a $d$-wave-like symmetry in momentum space, but in contrast to the Green's function $F$, it reveals a ${(\cos k_x - \cos k_y)}$ profile that closely resembles the form of the applied external $d$-wave superconducting field.

The results of the DF lattice DQMC scheme (Fig.~\ref{fig:Gk}), which clearly exhibit the pseudogap feature in $G$ and a substantial reduction in the intensity of $F$ at the antinodal point, are therefore not supported by the single-site \mbox{D-TRILEX} approximation.
The observed difference indicates that accounting for the effect of paramagnons alone is insufficient to reproduce the two-gap structure in the electronic spectra.
To identify the missing ingredient, we note the key difference between the two methods.
The DF lattice DQMC scheme utilizes an exactly solved lattice reference system, which efficiently captures the ``bad fermion'' behavior that emerges already at rather high temperatures.
This behavior manifests itself as the simultaneous appearance of the pseudogap at the antinodal point in the electronic spectrum and the formation of local magnetic moments (LMMs) in the system, composed of these ``bad'' electrons. 
These effects are entirely absent in the single-site impurity problem and cannot be fully accounted for by paramagnetic spin fluctuations within ladder-type diagrammatic approximations.
In the DF lattice DQMC scheme this effect exists already in the non-perturbative solution of the reference AFM Mott insulator system due to strong space-time correlations in the Monte-Carlo auxiliary spin-fields.

We note that at the considered temperature ${T=0.1}$ and doping level of 16\%, our \mbox{D-TRILEX} calculations do not reveal any divergence in the spin channel, indicating that the system does not develop LMMs through the formation of magnetic order. 
Nevertheless, in this parameter regime the spin fluctuations are already rather strong, which may suggest that the system could still form LMM in close proximity to the ordered state. 
However, applying the Landau-like criterion formulated in Ref.~\cite{PhysRevB.105.155151} indicates the absence of LMM.
A remaining possible source of LMM formation is the emergence of singlet states (local RVB physics), which cannot be captured by the single-site impurity problem and are very difficult to describe within a perturbative diagrammatic expansion.
This assumption would explain the difference between the cluster DF lattice DQMC and single-site \mbox{D-TRILEX} results.

The effect of the LMM formation can be incorporated into the \mbox{D-TRILEX} framework by introducing an effective Higgs condensate of paramagnons, corresponding to a static (${\omega=0}$) AFM (${{\bf q}=Q=\{\pi,\pi\}}$) spin-$z$ field $h_{\rm AFM}$ in \eqref{eq:fbaction_SC_DT}:
\begin{align}
\langle b^{z}_{Q,0}\rangle = \pm h_{\rm AFM}\,,
\end{align} 
which serves as the Higgs field~\cite{ScheurerE3665, PhysRevX.8.021048, PhysRevB.105.155151} that introduces the AFM ordered LMMs in the system. 
Averaging the dual Green's function, dressed by the Higgs field, over the positive and negative values of $h_{\rm AFM}$ effectively accounts for the singlet fluctuations and reproduces the paramagnetic case.
The averaging procedure results in an additional contribution to both the normal and anomalous parts of the self-energy:
\begin{align}
\hat{\tilde{\Sigma}}^{\rm AFM}_{k} = \Lambda^{z}_{\nu, 0}\hat{\tilde{G}}_{{\bf k}+Q,\nu}h^2_{\rm AFM}\Lambda^{z}_{\nu,0}\,.
\label{eq:Sigma_Higgs}
\end{align}
In this expression $\hat{\tilde{\Sigma}}^{\rm AFM}$ and $\hat{\tilde{G}}$ are ${2\times2}$ matrices in the Nambu space.
We note that a similar contribution, but only to the normal self-energy, has been considered in Refs.~\cite{ScheurerE3665, PhysRevX.8.021048}.
Recalling the \mbox{D-TRILEX} form for the self-energy~\eqref{eq:Sigma_DT}, the derived LMM contribution to the dual self-energy can be accounted for by adding the following contribution to the $z$ component of the renormalized interaction:
\begin{align}
\tilde{W}^{z}_{q} \to \tilde{W}^{z}_{q} - h^2_{\rm AFM} \delta_{{\bf q},Q} \delta_{\omega,0}\,.
\label{eq:SC_AFM_mode}
\end{align}
This additional term corresponds to an effective classical AFM mode and is similar to the one in Ref.~\cite{Irkhin1991}. 
Note that in our notations the renormalized interaction $\tilde{W}_{q}$ is negative.

The Higgs field plays a crucial role in the formation of a pseudogap, as discussed in Refs.~\cite{ScheurerE3665, PhysRevX.8.021048}.
Here, we further reveal its impact on the superconducting response.
In the right column of Figure~\ref{fig:PG}, we show the \mbox{D-TRILEX} results calculated in the presence of the Higgs condensate. 
The value of the field ${h_{\rm AFM}=0.485}$ is selected to ensure that the obtained results qualitatively align with those of the DF lattice DQMC scheme.
After accounting for the effect of the Higgs field, the momentum dependence of the Green's function closely matches the DF lattice DQMC results shown in the bottom panels of Figure~\ref{fig:Gk}. 
In particular, now the normal Green's function $G$ exhibits a pseudogap at the antinodal point, which further confirms that the formation of LMMs and the emergent ``bad fermion'' behavior are intrinsically connected processes. 
The appearance of the pseudogap leads to a reduction in spectral weight in the anomalous Green's function $F$, causing the extrema of $F$ to shift from the antinodal point towards the nodal point.
The inclusion of the Higgs field also leads to another significant modification.
The momentum dependence of the anomalous self-energy $S$ changes from the cosine form of the applied superconducting field, resulting from the scattering on the spatial magnetic fluctuations (bottom left panel), to a pattern resembling the Green's function with a momentum shift of ${{\bf k} \to {\bf k} + Q}$, where ${Q = \{\pi, \pi\}}$.
This behavior arises from the AFM correlations between the LMMs, which generate an effective classical AFM mode~\eqref{eq:SC_AFM_mode} that drives electronic scattering in the self-energy~\eqref{eq:Sigma_Higgs}.
We also observe that incorporating the effect of the Higgs field significantly enhances the superconducting response, leading to a substantial increase in the maximum values of the anomalous part of both the Green's function $F$ and self-energy $S$.

This result might look surprising since the formation of a pseudogap typically suppresses superconductivity by reducing the spectral weight at the Fermi surface. 
To investigate the role of the pseudogap in the superconducting response, we perform an additional \mbox{D-TRILEX} calculation in which the contribution of the Higgs field is included only in the normal part of the self-energy $\Sigma$. 
As shown in Ref.~\cite{Cuprates}, this results in a reduced superconducting response, consistent with expectations. 
Specifically, the maximum value of the anomalous self-energy $S$ decreases from ${\simeq0.03}$ (bottom left panel of Fig.~\ref{fig:PG}) to ${\simeq0.02}$, confirming that the formation of the pseudogap suppresses superconductivity. 
However, when the Higgs condensate is included in both the normal ($\Sigma$) and anomalous ($S$) self-energies (bottom right panel of Fig.~\ref{fig:PG}), the maximum value of the anomalous self-energy increases to ${\simeq 0.05}$.
This indicates that the AFM correlations of LMMs, which are responsible for the formation of the pseudogap, simultaneously enhance superconductivity.
By comparing the two cases with the pseudogap present, we conclude that the Higgs condensate contributes a bit more than 50\% to the superconducting response, in addition to the standard spin-fluctuation mechanism of electronic scattering on paramagnons. 
This estimation is consistent with the amount of mysterious contributions to the superconducting pairing found in C-DMFT calculations~\cite{Millis50}, establishing the LMMs formation and their AFM correlations as the missing strong-coupling mechanism of $d$-wave superconductivity. 
Note that bifacial effect of AFM correlations of LMMs on superconductivity can be considered
as an example of a more general tendency. The effect of bosonic degrees of freedom on
superconductivity is frequently twofold. On the one hand, they can produce additional glue
between electrons (that is, additional attractive interaction) which favors superconductivity.
On the other hand, they can lead to redistribution of spectral density decreasing the density
of states at the Fermi energy which may be harmful for superconductivity. In the weak-coupling
regime, the example of this twofold behaviour is provided by plasmonic superconductivity~\cite{Veld2023}.
We consider here the strong coupling situation, which means that direct analogies are not possible
but this example may be make the bifacial character of the effects under consideration less
surprising.

\subsection{Cluster extension of D-TRILEX}

In this Section, we extend the \mbox{D-TRILEX}, which until this point was introduced based on a single-site impurity problem, to a cluster reference system.
This framework allows us to consistently combine the exact treatment of short-range correlation effects within the cluster, with an efficient diagrammatic description of the long-range charge and spin collective fluctuations beyond the cluster.
We demonstrate the effectiveness of our approach by applying it to the one-dimensional nano-ring Hubbard model, where the low dimensionality enhances non-local correlations. Our results, presented in Ref.~\cite{fossati2025cluster}, show that the cluster extension of \mbox{D-TRILEX} accurately reproduces the electronic self-energy at momenta corresponding to the Fermi energy, in good agreement with the numerically exact quantum Monte Carlo solution of the problem, and outperforms significantly more computationally demanding approach based on the parquet approximation.
We show that the \mbox{D-TRILEX} diagrammatic extension drastically reduces the periodization ambiguity of cluster quantities when mapping back to the original lattice, compared to cluster dynamical mean-field theory (CDMFT).
Furthermore, we identify the CDMFT impurity problem as the main source of the translational-symmetry breaking.
To restore translational symmetry, it is necessary to perform a fully self-consistent solution of the problem by adjusting the cluster reference system. 

As the result, the cluster extension of \mbox{D-TRILEX} offers an efficient framework for a combined treatment of short- and long-range correlation effects.
Its modest computational cost makes it a promising tool for exploring symmetry-broken states associated with not only local (e.g., charge density wave or spin ordered states) but also non-local (e.g., $d$-wave superconductivity) order parameters.

\subsubsection{Diagonalization of the reference system}

The cluster reference problem has the same algebraic structure and can, in fact, be regarded as multi-orbital system~\eqref{eq:actionimp_app} in which the orbital indices $l$ correspond to cluster sites.  
However, the principal difficulty is that ordinary multi-orbital single-site impurity problems employ an orthogonal local basis, so both the on-site Hamiltonian and the hybridization function are diagonal.  
In a cluster formulation, the inter-site hopping necessarily introduces off-diagonal matrix elements in both objects. 
The presence of these terms drastically complicates the numerical solution of the impurity problem and frequently results in the fermionic sign problem~\cite{PhysRevB.41.9301, PhysRevB.85.201101, PhysRevB.92.195126}.
In order to avoid dealing with the sign problem, 
let us perform a basis transformation, $\mathcal{R}$, to diagonalize the Hermitian matrix of the hybridization function $\Delta^{ll'}_{\nu}$:
\begin{align}
\Delta_{\nu} =\mathcal{R}^{\dagger}_{\nu} \, D_{\nu} \,\mathcal{R}^{\phantom{\dagger}}_{\nu}\,,
\label{diag-hybr}
\end{align}
where $D_{\nu}$ is a diagonal matrix. 
This basis transformation~\eqref{diag-hybr} motivates the definition of new Grassmann variables:
\begin{align}
\overline{c}_{\nu\sigma{}l}^{\ast} = \sum_{l'} c_{\nu\sigma{}l'}^{\ast} \big[\mathcal{R}^{\dagger}_{\nu}\big]_{l'l}\,, ~~~~
\overline{c}_{\nu\sigma{}l} = \sum_{l'}\big[\mathcal{R}^{\phantom{\dagger}}_{\nu}\big]_{ll'} c_{\nu\sigma{}l'}\,.
\end{align}
The rotation matrix is unitary, since the hybridization function is Hermitian, ensuring that anti-commutation relations are preserved. 
The reference impurity problem~\eqref{eq:actionimp_app} becomes:
\begin{align}
{\cal S}_{\text{imp}}  = & \sum_{\nu,\sigma,l} 
\overline{c}_{\nu\sigma{}l}^{\ast} \left[i\nu + \mu - D^{ll}_{\nu}\right] \overline{c}_{\nu\sigma{}l}
+  \sum_{\{\nu\}, \omega} \sum_{\{l\},\{\sigma\}} {\cal U}^{\nu\nu'\omega}_{l_1 l_2 l_3 l_4}
\overline{c}_{\nu \sigma l_1}^{\ast} \overline{c}_{\nu+\omega, \sigma l_2} 
\overline{c}_{\nu'+\omega, \sigma' l_4}^{\ast} \overline{c}_{\nu', \sigma' l_3}\,,
\label{Seff-freq}
\end{align}
where the static electron-electron interaction transforms to a three-frequency-dependent object:
\begin{align}
{\cal U}^{\nu\nu'\omega}_{l_1 l_2 l_3 l_4} = \sum_{\{j\}} U^{\phantom{*}}_{j_1 j_2 j_3 j_4} \big[\mathcal{R}^{\phantom{\dagger}}_{\nu}\big]_{l_1j_1} \big[\mathcal{R}^{\dagger}_{\nu+\omega}\big]_{j_2l_2}
\big[\mathcal{R}^{\phantom{\dagger}}_{\nu'+\omega}\big]_{l_4j_4} \big[\mathcal{R}^{\dagger}_{\nu'}\big]_{j_3l_3}\,.
\end{align}
Handling such a complex interaction is beyond the capabilities of existing impurity solvers.
Thus, neither option provides a viable solution: avoiding the sign problem leads to an intractable interaction, while keeping the interaction static reintroduces the sign problem.

To address this problem, in Refs.~\cite{PhysRevB.92.195126, PhysRevB.85.201101} the author showed that for different types of clusters there exists a static (frequency-independent) transformation to an ``optimal'' single-particle basis for which the sign problem decreases significantly.
Yet, the sign problem cannot be completely  removed and still scales exponentially with the inverse of the temperature. 
In Ref.~\cite{fossati2025cluster}, we followed a similar idea and performed a basis transformation to diagonalize the local part of the single-particle Hamiltonian:
\begin{align}
\overline{\varepsilon}_{K} = \mathcal{R} \left[\sum_{K}\varepsilon_{K}\right] \mathcal{R}^{\dagger}\,.
\end{align}
This transformation minimizes the off-diagonal components of the hybridization function, as can be seen by applying the basis transformation directly to the hybridization function (see Ref.~\cite{fossati2025cluster} for a detailed derivation):
\begin{align}
\overline{\Delta}_{\nu}  = \mathcal{R} \, \Delta_{\nu} \, \mathcal{R}^{\dagger} 
= \mathcal{R} \left[ \sum_{K} \varepsilon_{K} \right] \mathcal{R}^{\dagger} + \mathcal{O} \left( \sum_{K} \mathcal{R} \, \frac{ \left( \varepsilon_{K} + \Sigma^{\rm imp}_{\nu} \right)^2}{i \nu + \mu} \, \mathcal{R}^{\dagger} \right),
\label{rotated-hybr}
\end{align}
where $\Sigma^{\rm imp}_{\nu}$ is the self-energy of the cluster impurity problem. 
However, the off-diagonal components of the hybridization function cannot be fully eliminated. 
In our approach, we take a further step by removing these off-diagonal terms through the flexibility in choosing the reference system for the dual diagrammatic expansion~\cite{BRENER2020168310}. 
In dual approaches, the impurity problem~\eqref{Seff-freq} is separated from the original lattice action~\eqref{eq:actionlatt} by adding and subtracting the hybridization function $\Delta^{ll'}_{\nu}$, which may be chosen arbitrarily in the band space $\{l, l'\}$. 
After performing the basis transformation $\mathcal{R}$ that diagonalizes the local part of the single-particle Hamiltonian $\varepsilon_{\bf K}$, we choose to work with a diagonal hybridization function, $\Delta^{ll}_{\nu}$. 
The off-diagonal components of the cluster self-energy are then generated via the diagrammatic expansion, which accounts for both momentum-independent and momentum-dependent contributions beyond the impurity self-energy $\Sigma^{\rm imp}_{\nu,ll}$. 
The proposed scheme is general and can be applied to arbitrary cluster problems. 
In what follows, we demonstrate its applicability using the one-dimensional nano-ring Hubbard model with a dimer reference system as a representative example.

\subsubsection{Basis transformation for the 1D Hubbard model}

We start with the one-dimensional Hubbard Hamiltonian:
\begin{equation} 
H = - t \sum_{j, \tau, \sigma} c_{j, \sigma}^{\dagger} c^{\phantom{\dagger}}_{j + \tau,\sigma} + U \sum_{j} n_{j\uparrow}n_{j\downarrow},\, 
\label{eq:Horiginal}
\end{equation}
where $j$ labels the atomic position, ${\tau=\pm1}$ denotes the nearest-neighbor site position difference, $t$ denotes the nearest-neighbor hopping integral, $c_{j, \sigma}^{(\dag)}$ is the annihilation (creation) operator for an electron at site $j$, with spin $\sigma$; $n_{j, \sigma} = c_{i, \sigma}^{\dagger}c^{\phantom{\dagger}}_{i, \sigma}$ is the density operator. 
For the remainder of this work, we set ${t = 1}$ and use it as the energy unit.
The chemical potential is set to ${\mu = U/2}$ so that the system is at half filling.

\begin{figure}[t!]
\centering
\includegraphics[width=0.5\linewidth]{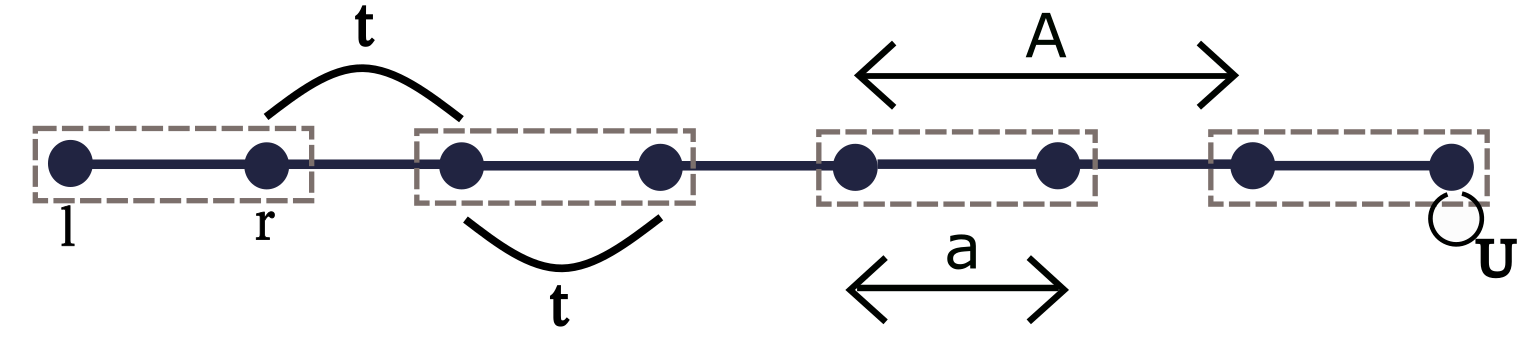}
\caption{Schematic representation of the one-dimensional Hubbard model. The lattice is tiled by identical two-site clusters, indicated by dashed boxes, consisting of the ``left'' (l) and ``right'' (r) sites. The nearest-neighbor hopping amplitude $t$ is considered the same within and between the clusters. The electrons interact via the on-site Coulomb repulsion \(U\). The distance between adjacent clusters is ${A = 2a}$, i.e., twice the lattice constant $a$. The Figure is taken from Ref.~\cite{fossati2025cluster}.
\label{fig:1D-Hubbard_drawing}}
\end{figure}

In the case of a two-site cluster formulation of the problem one can introduce the spinors:
\begin{align}
C_{I} = 
\begin{pmatrix}
l_{I}\\
r_{I}
\end{pmatrix}
~~\text{and} ~~
C^{\dag}_{I} = 
\left( l_{I}^{\dag} ~ r_{I}^{\dag} \right),
\label{original-cluster-variable}
\end{align}
where $l^{(\dagger)}_{I}$ and $r^{(\dagger)}_{I}$ are the annihilation (creation) operators for the electron on the left and right sites in the cluster $I$, respectively (see Fig.~\ref{fig:1D-Hubbard_drawing}). 
The local part of the single-particle Hamiltonian becomes:
\begin{align}
H_0^{\text{loc}} = -\sum_{I} C_{I}^{\dagger}
\begin{pmatrix}
0 & t\\
t & 0
\end{pmatrix} C_{I}\,.
\end{align}
It can be diagonalized by the following transformation to the bonding-antibonding basis: 
\begin{align}
\mathcal{R} =\frac{1}{\sqrt{2}}
\begin{pmatrix}
1 & 1\\
1 & -1
\end{pmatrix}.
\label{rotation-matrix}
\end{align} 
The full single-particle Hamiltonian has the following form:
\begin{align}
H_0 &=  -\sum_{I} C_{I}^{\dagger}
\Bigg\{ 
\begin{pmatrix}
0 & t\\
t & 0
\end{pmatrix}
C_{I} + 
\begin{pmatrix}
0 & t\\
0 & 0
\end{pmatrix}
C_{I-1} + 
\begin{pmatrix}
0 & 0\\
t & 0
\end{pmatrix} 
C_{I+1} \Bigg\} \notag \\
&= -\frac{t}{N_K} 
\sum_{K} C_{K}^{\dagger} 
\begin{pmatrix}
0 & 1 + e^{-iAK}\\
1 + e^{iAK} & 0
\end{pmatrix} 
C_{K}\,,
\label{H0-1D}
\end{align}
where $N_K$ is the number of $K$-points in the reduced BZ, ${A=2a}$ is the distance in real space between the neighboring clusters (vector of translation), and $a$ is the lattice constant.
Upon the basis transformation, the single-particle Hamiltonian becomes:
\begin{align}
\overline{H}_0  = \frac{1}{N_K} \sum_{K} \overline{C}^{\dagger}_{K} \,
\overline{\varepsilon}^{\phantom{\dagger}}_{K} \, \overline{C}^{\phantom{\dagger}}_{K}\,,
\label{H0-stag}
\end{align}
where:
\begin{align}
\overline{\varepsilon}_{K} = -t 
\begin{pmatrix}
1 + \cos(A K) & i \sin(A K)\\
-i \sin(A K) & - 1 - \cos(A K)
\end{pmatrix},
\end{align}
and we also introduced a new spinor:
\begin{align}
\overline{C} = \mathcal{R} \, C
= \frac{1}{\sqrt{2}}  
\left\{ 
\begin{pmatrix}
1\\
1
\end{pmatrix} 
l + 
\begin{pmatrix}
1\\
-1
\end{pmatrix}
r \right\}
= 
\begin{pmatrix}
a\\
b
\end{pmatrix}.
\label{new-fermionic-var}
\end{align}
The interaction part of the Hamiltonian:
\begin{align}
H_{U} = U \sum_{j} n_{j\uparrow} n_{j\downarrow}
\end{align}
transforms to the Kanamori-like form~\cite{10.1143/PTP.30.275} (${m,m'\in\{a,b\}}$):
\begin{align}
\overline{H}_U &= \frac{1}{2} \sum_{\substack{I,m\\ \sigma\sigma'}} {\cal U}\,n^{m}_{I, \sigma} n^{m}_{I, \sigma'}  
+ \frac{1}{2} \sum_{\substack{I,\sigma\sigma'\\m\neq m' }}\left({\cal U'} - {\cal J}\delta_{\sigma\sigma'}\right)n^{m}_{I, \sigma} n^{m'}_{I, \sigma'} \notag\\
&-{\cal J}\sum_{I}\left(a^{\dag}_{I\uparrow} a^{\phantom{\dag}}_{I\downarrow} b^{\dag}_{I\downarrow} b^{\phantom{\dag}}_{I\uparrow} + b^{\dag}_{I\uparrow} b^{\phantom{\dag}}_{I\downarrow} a^{\dag}_{I\downarrow} a^{\phantom{\dag}}_{I\uparrow}\right)
+{\cal J}\sum_{I}\left(a^{\dag}_{I\uparrow} a^{\dag}_{I\downarrow} b^{\phantom{\dag}}_{I\downarrow} b^{\phantom{\dag}}_{I\uparrow} + b^{\dag}_{I\uparrow} b^{\dag}_{I\downarrow} a^{\phantom{\dag}}_{I\downarrow} a^{\phantom{\dag}}_{I\uparrow} \right)
\label{interacting-part-H} 
\end{align}
with an effective intra- (${{\cal U}}$) and inter-band (${{\cal U'}}$) Coulomb interactions and Hund's exchange coupling (${\cal J}$) equal to ${{\cal U} = {\cal U'} = {\cal J} = U/2}$.

\subsubsection{Transformation to the original basis}

The lattice problem~\eqref{eq:Horiginal} in the bonding-antibonding basis:
\begin{align}
\overline{H} = \overline{H}_0 + \overline{H}_U
\label{eq:modelH}
\end{align}
is solved using the multi-band \mbox{D-TRILEX} approach as described in Section~\ref{sec:WF_DT}.
In order to obtain the self-energy and the Green's function in the original basis, we perform the following transformation:
\begin{align}
O_{K\nu} = \mathcal{R}^{\dagger} \, \overline{O}_{K\nu} \,\mathcal{R}
\end{align}
with $O_{K\nu}$ being the lattice self-energy $\Sigma_{K\nu}$ or the Green's function $G_{K\nu}$.
To obtain the lattice self-energy and Green's function corresponding to the single-site unit cell from the cluster quantities, we perform the following periodization step by imposing the translational invariance of the original lattice problem:
\begin{align}
\label{eq:L_def}
O^{\rm latt}_{k\nu} = \mathcal L_{k}[O^{ll'}_{k\nu}] = \frac{1}{N_{\rm c}}\sum_{ll'}e^{ik(r_{l}-r_{l'})}O^{ll'}_{k\nu}\,,
\end{align}
where $r_{l}$ is the position of the $l$-th atom in the unit cell and $k$ corresponds to the original (extended) BZ.
In the dimer case, this relation reduces to:
\begin{align}
O^{\rm latt}_{k\nu} = 
\tfrac12\left(O^{11}_{k\nu} + O^{22}_{k\nu}\right) + {\rm Re}\,O^{12}_{k\nu}\,\cos(ka) + {\rm Im}\,O^{12}_{k\nu}\,\sin(ka)\,.
\label{eq:Dimer_periodization}
\end{align}
The quantity $O^{ll'}_{k\nu}$ in the extended BZ can be obtained from the cluster quantity $O^{ll'}_{K\nu}$ in the reduced BZ using the periodicity in momentum space. 

\subsubsection{Application to a nano-ring Hubbard model} 

\begin{figure}[t!]
\centering
\includegraphics[width=0.65\linewidth]{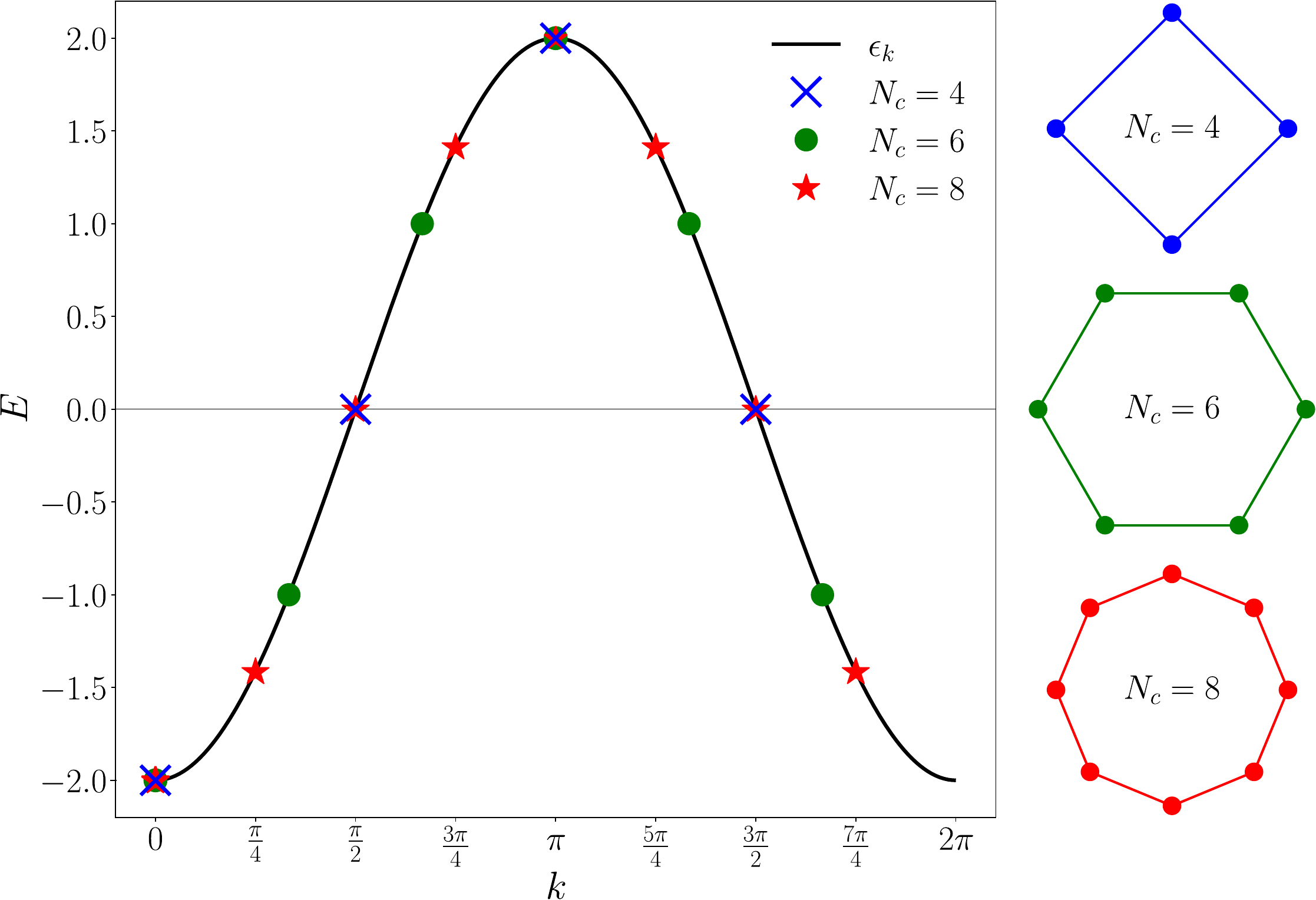}
\caption{The left panel shows the electronic spectral function $\varepsilon_k$ along the first Brillouin zone (lattice constant ${a = 1}$). The dashed curve represents the dispersion of an infinite one-dimensional chain, whereas the discrete symbols correspond to the finite number of lattice sites $N_{\rm c}$. 
The right column depicts the ring geometries: a four-site (blue, top), a six-site (green, middle), and an eight-site (red, bottom) rings.  
In the dispersion plot their spectra are indicated by blue crosses, green circles, and red stars, respectively. The Figure is taken from Ref.~\cite{fossati2025cluster}.
\label{fig:nanoring}}
\end{figure}

To demonstrate the performance of the \mbox{D-TRILEX} method, we apply it to the one-dimensional nano-ring Hubbard model. 
We focus on periodic chains with $N_{\text{c}} = 4$, 6, and 8 lattice sites, for which results can be directly compared to the exact Hirsch-Fye QMC solution and to the more involved D$\Gamma$A method in the ladder and parquet implementations, for which the data are available from Ref.~\cite{PhysRevB.91.115115}. 
For consistency with that work, all calculations are performed for ${t = 1}$, ${U = 2}$, and inverse temperature ${\beta = 10}$.

\paragraph{Insulating system, $\boldsymbol{N=6}$:}

We refer to the configuration with ${N_{\rm c}=6}$ sites as an insulating system, since its non-interacting spectral function exhibits an energy gap separating the occupied states at \mbox{$k = 0,~\pi/3,~5\pi/3$} from the unoccupied states at \mbox{$k = 2\pi/3,~\pi,~4\pi/3$}. This behavior is illustrated in Fig.~\ref{fig:nanoring}, where the ${N_{\rm c}=6}$ case is shown in green.

\begin{figure}[t!]
\centering
\includegraphics[width=0.49\linewidth]{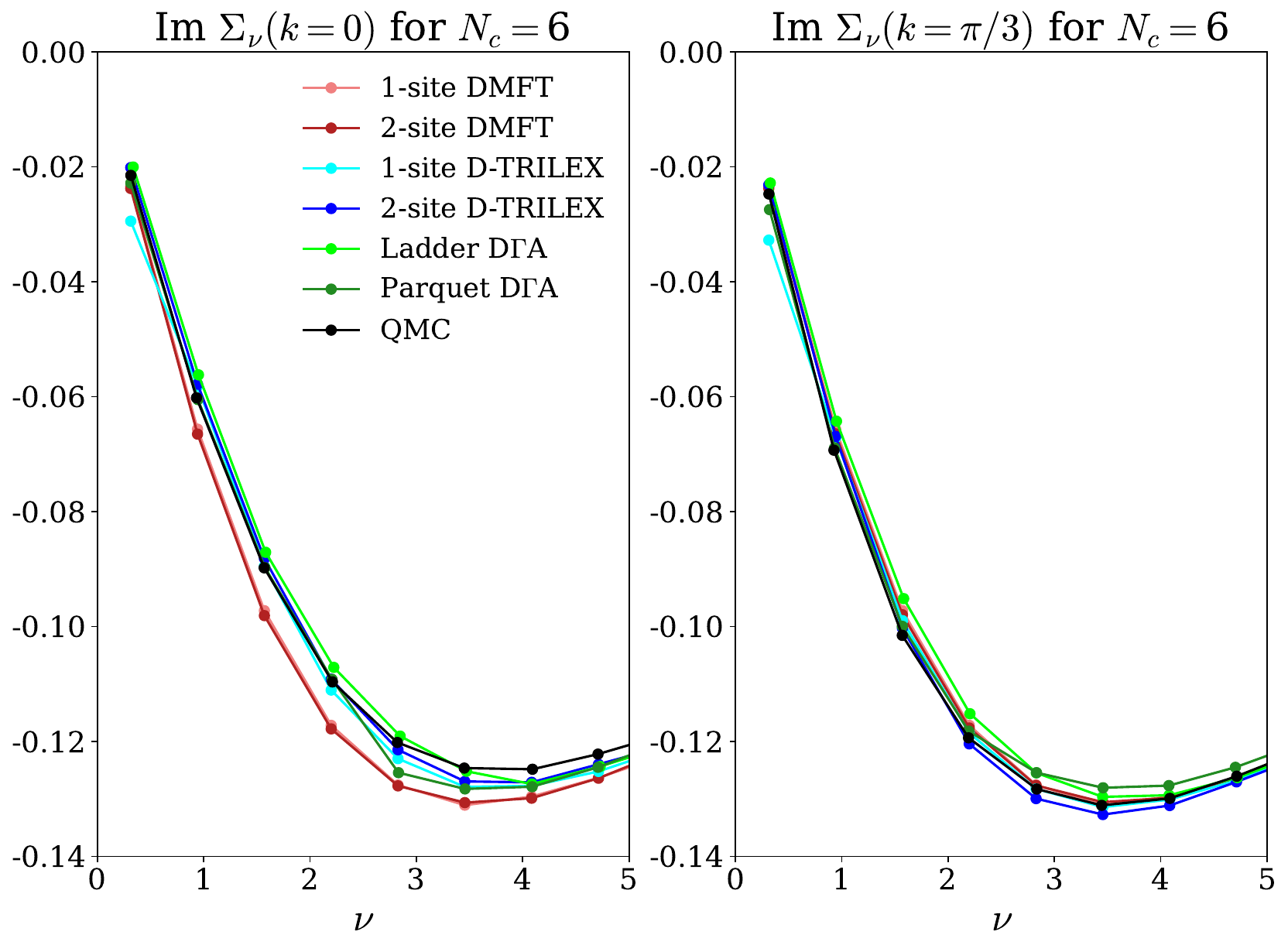}~~
\includegraphics[width=0.49\linewidth]{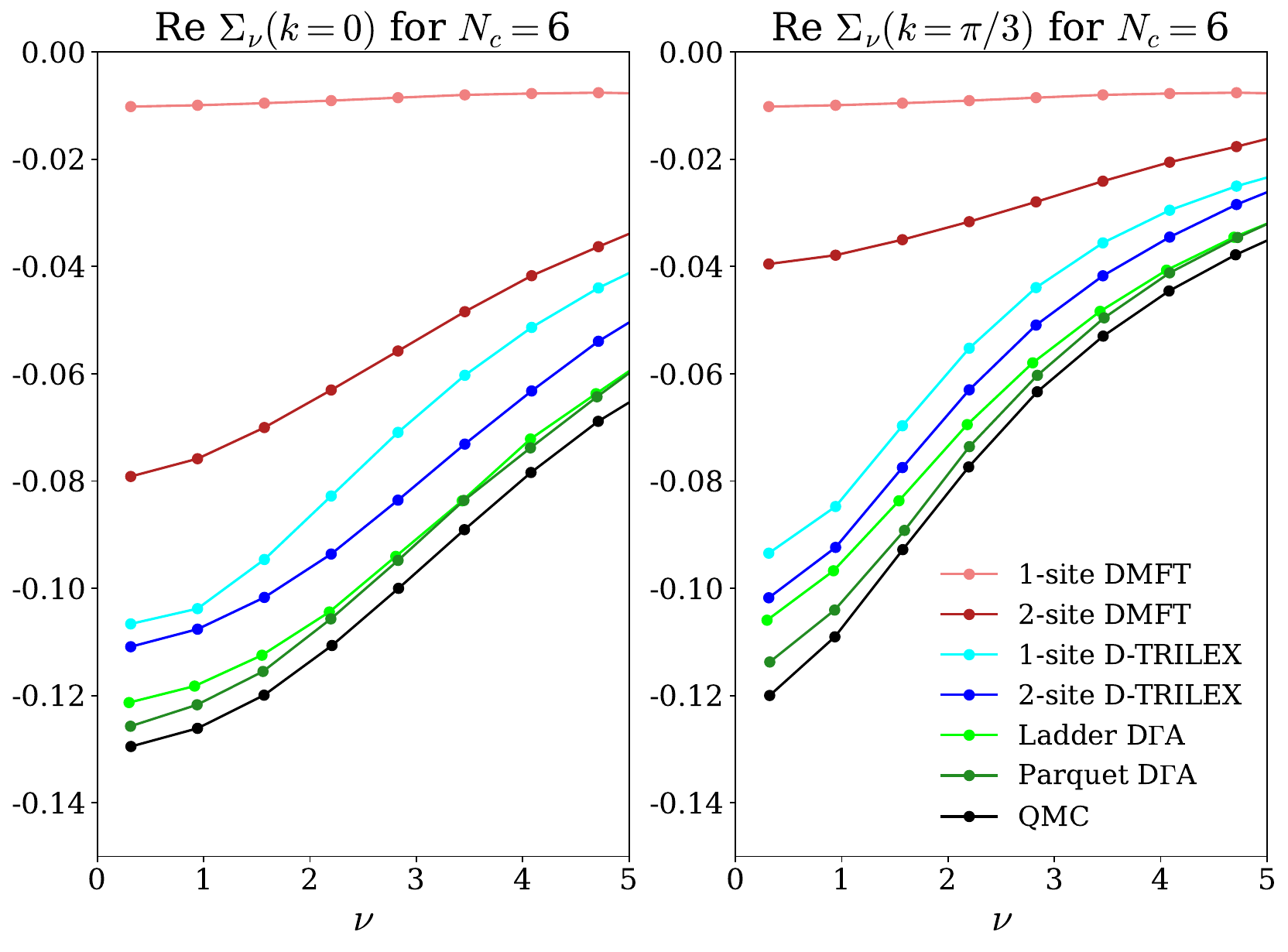}
\caption{The imaginary (two left panels) and real (two right panels) parts of the self-energy calculated as a function of Matsubara frequency $\nu$ at two momenta ${k=0}$ (left) and ${k=\pi/3}$ (right). The results are obtained for the case of ${N_{\text{c}}=6}$ at ${U = 2}$ and ${\beta = 10}$ using different methods indicated in the legend. The ladder D$\Gamma$A, parquet D$\Gamma$A, and exact QMC results are taken from Ref.~\cite{PhysRevB.91.115115}. The Figure is taken from Ref.~\cite{fossati2025cluster}.
\label{figs_N6}}
\end{figure}

In Fig.~\ref{figs_N6} we plot the imaginary part of the lattice self-energy calculated at ${k=0}$ (a) and ${k=\pi/3}$ (b) momenta as a function of Matsubara frequency $\nu$.
The results are obtained using the single-site DMFT (light red), two-site cluster DMFT (dark red), single-site \mbox{D-TRILEX} (light blue), two-site cluster \mbox{D-TRILEX} (dark blue) and are compared with the ladder D$\Gamma$A (light green), parquet D$\Gamma$A (dark green) and QMC (black) data of Ref.~\cite{PhysRevB.91.115115}.
The two-site cluster DMFT and \mbox{D-TRILEX} self-energies are periodized from the cluster space corresponding to the reduced BZ to the single-site form corresponding to the extended BZ using Eq.~\eqref{eq:L_def}.  
The small momentum-dependence of the imaginary part of the self-energy, that can be seen in Fig.~\ref{figs_N6}, allows for the DMFT results to be close to the exact QMC result. Both D$\Gamma$A and \mbox{D-TRILEX} methods improve upon DMFT and reproduce a slight change of the self-energy between the ${k=0}$ and ${k=\pi/3}$ points. 
We observe that the single-site \mbox{D-TRILEX} calculations slightly overestimate the self-energy at low frequencies, but the cluster extension of the method cures this problem and is in a very good agreement with the QMC result. 
Overall, we find that the imaginary part of the self-energy is reproduced with good accuracy by all methods in the insulating case of ${N_{\rm c}=6}$.

In contrast, the real part of the self-energy, shown in Fig.~\ref{figs_N6}, exhibits more differences among the considered methods. 
At the both ${k=0}$ and ${k=\pi/3}$ momenta the single-site \mbox{D-TRILEX} displays a nearly constant offset of comparable magnitude with respect to the exact QMC result, showing that the discrepancy is approximately momentum-independent.
The cluster \mbox{D-TRILEX} improves the real part of the self-energy at ${k=\pi/3}$ momentum close to the Fermi level, but the self-energy at the edge of the BZ (${k=0}$) become less accurate.
Interestingly, the real part of the cluster \mbox{D-TRILEX} self-energy at ${k=0}$ is very similar to that obtained from cluster DMFT.
We note that at ${k=0}$, the dominant contribution to $\text{Re}\,\Sigma$ originates from the off-diagonal cluster term $\Sigma_{12}$ (see Eq.~\eqref{eq:Dimer_periodization}).
The similarity between the cluster DMFT and \mbox{D-TRILEX} results suggests that the non-perturbative short-range correlation effects captured by the reference system contribute significantly to $\Sigma_{12}$. 
In the insulating case, non-local collective electronic fluctuations are expected to be weak, and therefore the additional perturbative diagrammatic corrections have only a minor impact on $\Sigma_{12}$.
We also note that, although the \mbox{D-TRILEX} results for the real part of the self-energy in the ${N_{\rm c}=6}$ case are less accurate than those of D$\Gamma$A, the self-energy itself remains very small compared to the electronic dispersion, ${|\text{Re}\,\Sigma|\ll|\varepsilon_{k}|}$.
Therefore, the observed discrepancy is not expected to have a significant impact.

\paragraph{Metallic systems, $\boldsymbol{N=4,\,8}$:}

In the cases of ${N_{\rm c}=4}$ and ${N_{\rm c}=8}$, which we refer to as metallic systems, the non-interacting spectral function exhibits a doubly degenerate state at the Fermi level, corresponding to the momenta ${k=\pi/2}$ and ${k=3\pi/2}$.
This behavior is illustrated by the blue and red colors in Fig.~\ref{fig:nanoring}, respectively.
This suggests that the self-energies at non-equivalent $k$-points may exhibit significant momentum-dependent variation, particularly when comparing values at the Fermi energy to those further away from it.
Single-site DMFT is not able to reproduce the momentum dependence of the self-energy, and more elaborate methods are required to capture these effects.

\begin{figure}[t!]
\centering
\includegraphics[width=0.49\linewidth]{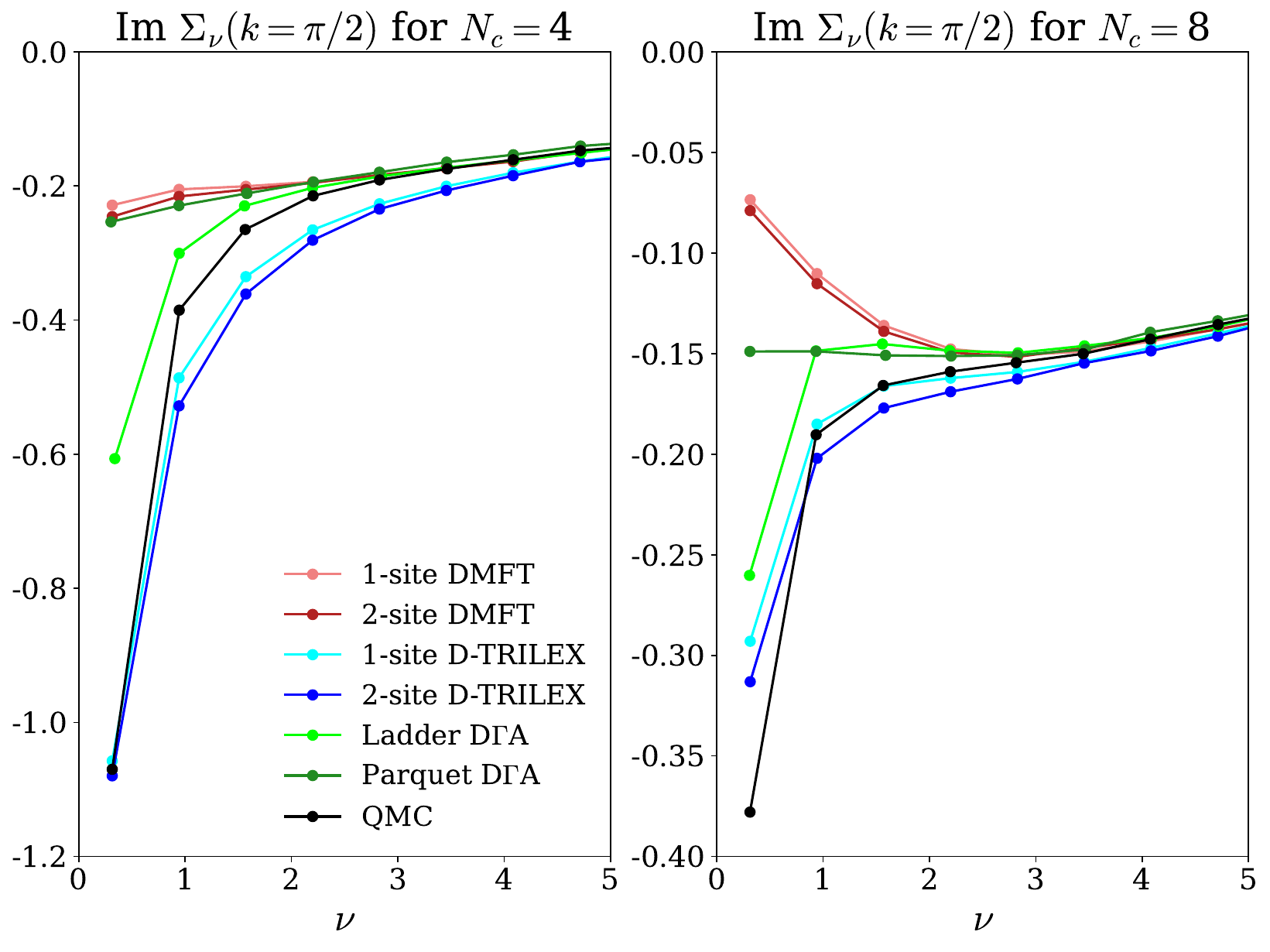}~~
\includegraphics[width=0.49\linewidth]{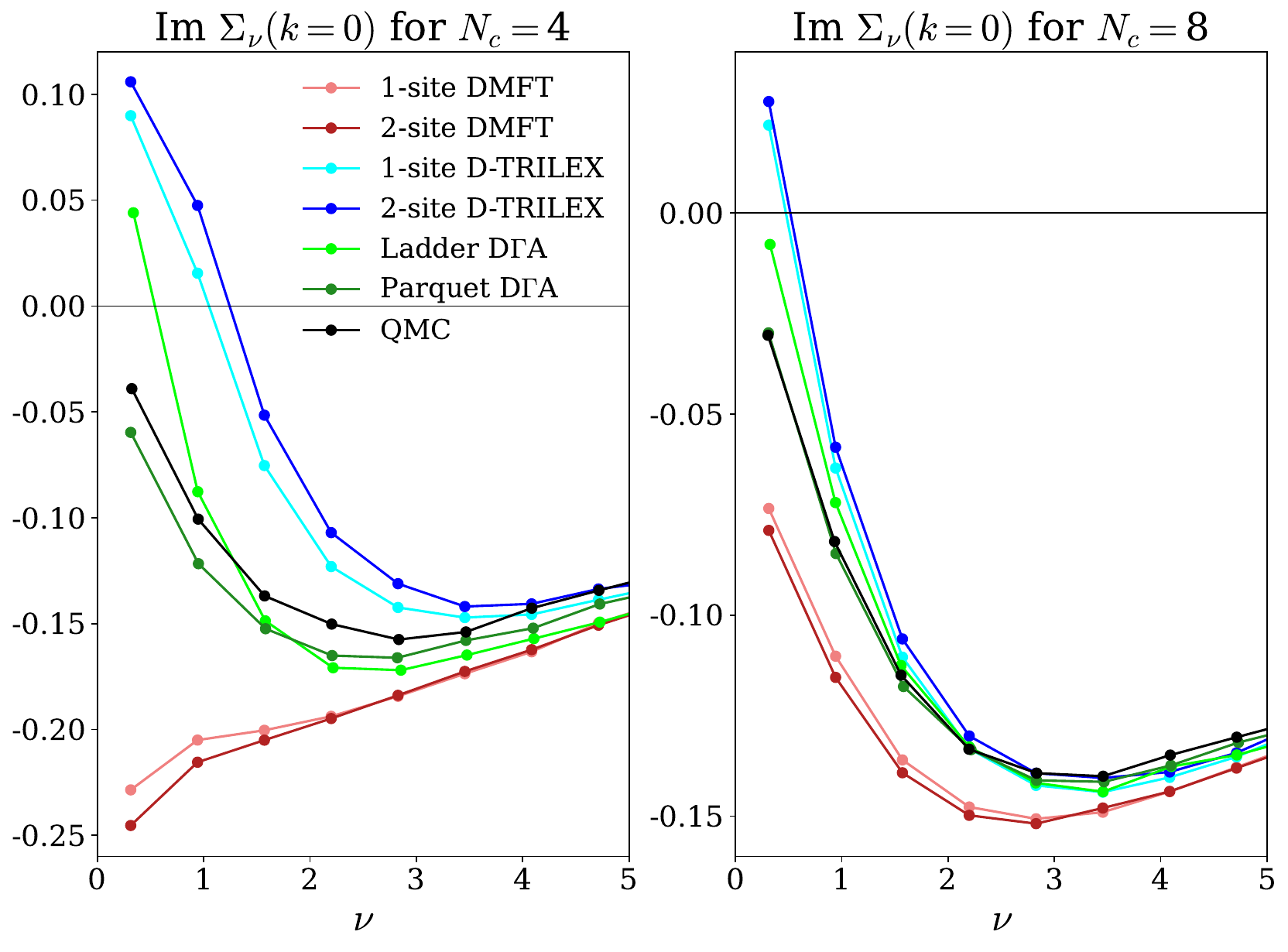}
\caption{The imaginary part of the self-energy as a function of Matsubara frequency $\nu$ at momenta ${k=\pi/2}$ and ${k=0}$ for the systems with $N_{\text{c}}=4$ and $N_{\text{c}}=8$ .  
All results are computed at $U = 2$ and $\beta = 10$ using the methods indicated in the legend.
The ladder D$\Gamma$A, parquet D$\Gamma$A, and exact QMC results are taken from Ref.~\cite{PhysRevB.91.115115}. The Figure is taken from Ref.~\cite{fossati2025cluster}.
\label{figs_Im}}
\end{figure}

In Fig.~\ref{figs_Im} we show the imaginary part of the self-energy calculated at ${k=\pi/2}$ and ${k=0}$ as a function of frequency for the case of ${N_{c}=4}$ and ${N_{c}=8}$. 
Note, that ${{\rm Re}\,\Sigma(k=\pi/2)=0}$ in our case.
The ${k=\pi/2}$ point corresponds to the Fermi energy, and the self-energy at ${k=\pi/2}$ is much larger than the one obtained at ${k=0}$.   
First, the exact ${{\rm Im}\,\Sigma(k=\pi/2)}$ provided by QMC for both ${N_{c}=4}$ and ${N_{c}=8}$ cases shows an insulating (divergent a ${\nu\to0}$) behavior.
Interestingly, the cluster DMFT result fails to reproduce this behavior and nearly coincides with the single-site DMFT result. This indicates that long-range collective electronic fluctuations play a crucial role in the metallic case, particularly at the Fermi energy.
Among all considered approaches, \mbox{D-TRILEX} provides the most accurate result for the self-energy at ${k=\pi/2}$. 
At ${N_{\rm c}=4}$, the single-site and cluster \mbox{D-TRILEX} results nearly coincide, with the cluster version being closer to the exact result at the lowest Matsubara frequency, while the single-site approximation is more accurate at higher frequencies. 
This trend persists for ${N_{\rm c}=8}$, although the difference between the single-site and cluster \mbox{D-TRILEX} results becomes more pronounced.
The ladder D$\Gamma$A captures the insulating behavior of the self-energy but is substantially less accurate than both versions of \mbox{D-TRILEX}. 
Surprisingly, we find that although the parquet D$\Gamma$A is a diagrammatic extension of the ladder version, it performs significantly worse: it fails to capture the insulating behavior of the self-energy and even remains at the level of DMFT for ${N_{\rm c}=4}$.
Ref.~\cite{PhysRevB.91.115115} attributes the better performance of ladder D$\Gamma$A at \mbox{$k=\pi/2$} to the so-called $\lambda$-correction, arguing that it effectively emulates an outer self-consistency by modifying the impurity problem. 
Our comparison challenges this interpretation, since the single-site \mbox{D-TRILEX} approach, which features a similar (ladder-like) diagrammatic structure, uses the same DMFT impurity problem as a reference for the diagrammatic expansion, and also does not perform an outer self-consistency loop, yet achieves better accuracy without employing the $\lambda$-correction.

\begin{figure}[b!]
\centering
\includegraphics[width=0.49\linewidth]{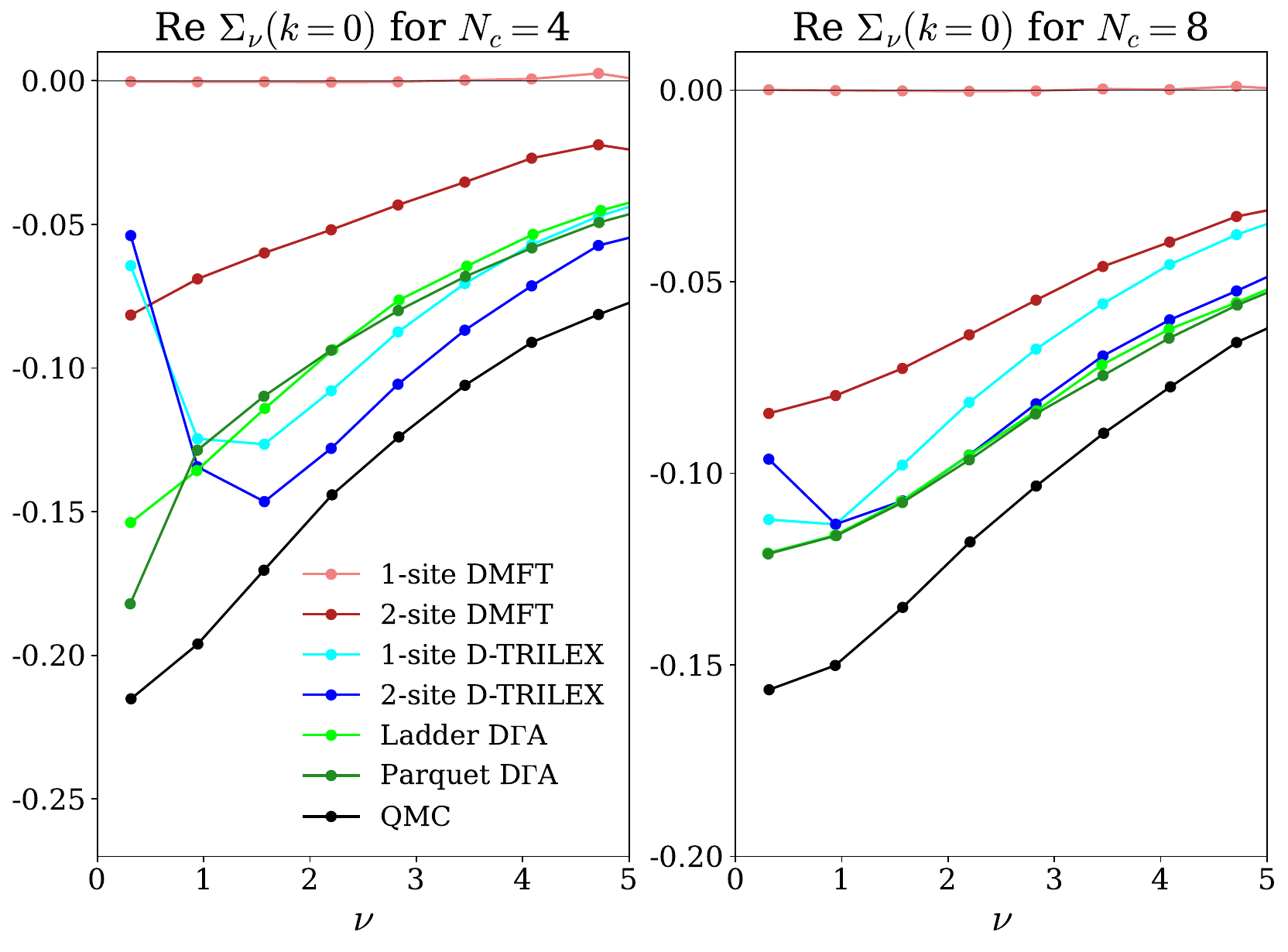} 
\caption{The real part of the self-energy as a function of Matsubara frequency $\nu$ at momentum ${k = 0}$ for lattices with ${N_{\text{c}} = 4}$ (left) and ${N_{\text{c}} = 8}$ (right).  
All results are computed at ${U = 2}$ and ${\beta = 10}$ using the methods indicated in the legend.
The ladder D$\Gamma$A, parquet D$\Gamma$A, and exact QMC results are taken from Ref.~\cite{PhysRevB.91.115115}. The Figure is taken from Ref.~\cite{fossati2025cluster}. \label{figs_Re}}
\end{figure}

\begin{figure}[t!]
\centering
\includegraphics[width=0.45\linewidth]{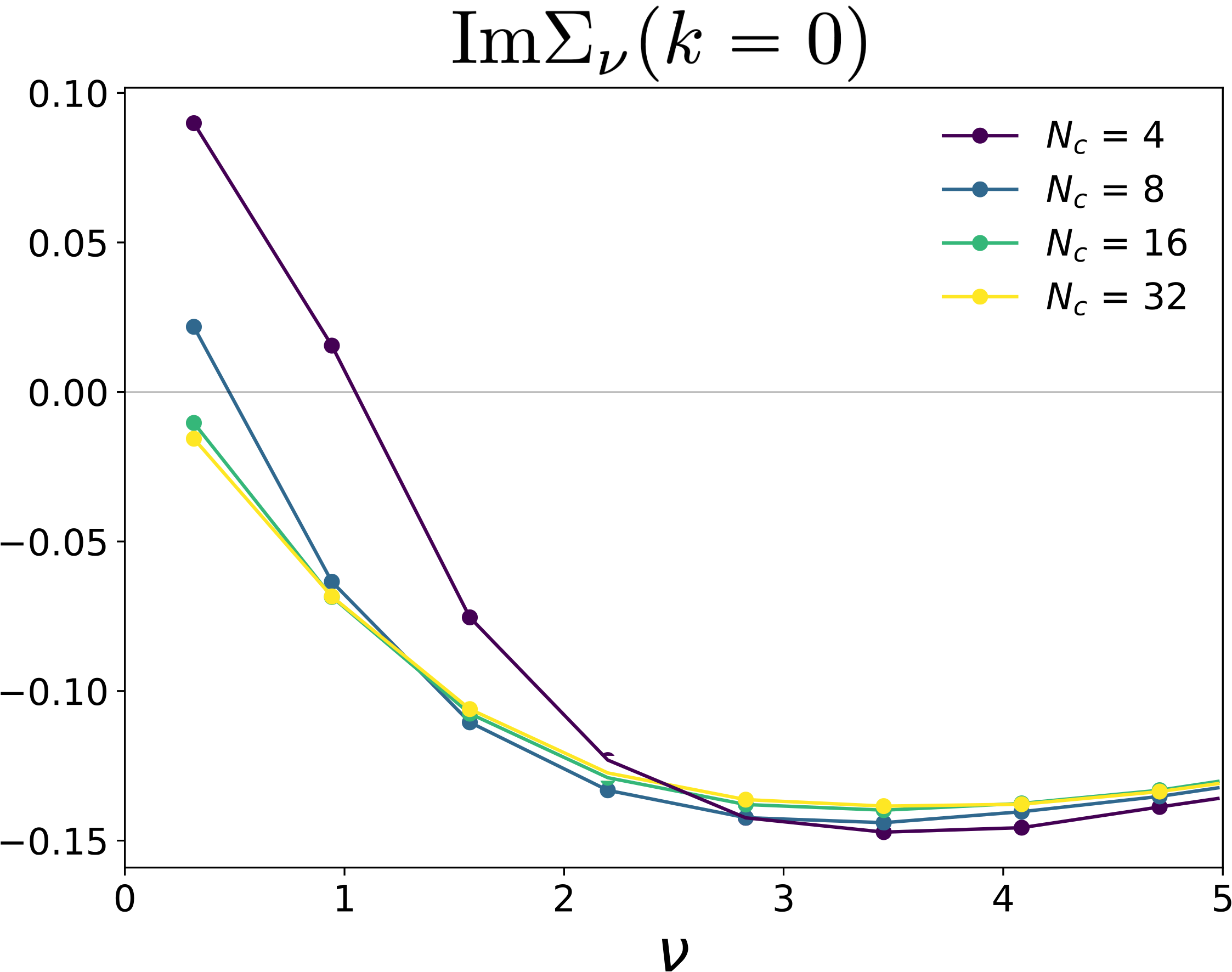}
\caption{Evolution of the imaginary part of the self-energy calculated at ${k=0}$, ${U=2}$, and ${\beta=10}$ for different number of lattice sites ${N_{\rm c}=4}$,~8,~16,~and~32 using the single-site \mbox{D-TRILEX} approach. The Figure is taken from Ref.~\cite{fossati2025cluster}. 
\label{fig-N-evol}}
\end{figure}

At the Brillouin zone center, ${k=0}$, the parquet D$\Gamma$A provides the most accurate description of both the imaginary (Fig.~\ref{figs_Im}) and real (Fig.~\ref{figs_Re}) parts of the self-energy, while both DMFT versions are the least accurate.
We note that ${{\rm Im}\,\Sigma(k=0)}$ for ${N_{\rm c}=4}$ predicted by the single-site and cluster \mbox{D-TRILEX} approaches shows a non-causal behavior for the two lowest Matsubara frequencies. 
The anomaly carries over to the real part of the self-energy, where the first two Matsubara points deviate strongly from the exact trend. 
This behavior improves with increasing the number of lattice sites to ${N_{\rm c}=8}$, but the imaginary part of the \mbox{D-TRILEX} self-energy still remains positive at the lowest Matsubara frequency. 
A similar behavior for ${{\rm Im}\,\Sigma(k=0)}$ is also found in Ref.~\cite{PhysRevB.91.115115} for the ladder D$\Gamma$A approach in the case of ${N_{\rm c}=4}$ and is attributed to the neglect of particle-particle diagrams inherent in the particle-hole ladder approximation.   
The current \mbox{D-TRILEX} implementation also omits particle-particle correlations.
However, the fact that the cluster extension of the single-site \mbox{D-TRILEX} does not lead to significant improvements suggests that the missing diagrammatic contributions are likely long-ranged and go beyond a perturbative ladder-like approximation.
Furthermore, Fig.~\ref{fig-N-evol} demonstrates that the non-causal behavior can be cured by increasing the number of lattice sites to ${N_{\rm c}=16}$, and the imaginary part of the single-site \mbox{D-TRILEX} self-energy calculated at ${k=0}$ for ${N_{\rm c}=32}$ nearly coincides with the one obtained for ${N_{\rm c}=16}$.
This result suggests that increasing the system size reduces the impact of correlations beyond the two-particle level that are neglected in the \mbox{D-TRILEX} and ladder D$\Gamma$A approaches.

In general, as in the insulating ${N_{\rm c}=6}$ case, the real part of the self-energy away from the Fermi energy remains relatively small compared to the electronic dispersion, ${|\text{Re}\,\Sigma(k=0)|\ll|\varepsilon_{k}|}$, so discrepancies with the exact QMC results are not expected to be significant for describing the electronic properties of the system. 
At the same time, we again observe a strong similarity between the cluster DMFT and \mbox{D-TRILEX} results, particularly in the ${N_{\rm c}=8}$ case. This behavior can be attributed to the fact that at ${k=0}$, the real part of the self-energy~\eqref{eq:Dimer_periodization}:
\begin{align}
{\rm Re}\,\Sigma_{k=0,\nu} = \tfrac12{\rm Re}\left[\Sigma^{11}_{k=0,\nu} + \Sigma^{22}_{k=0,\nu}\right] + {\rm Re}\,\Sigma^{12}_{k=0,\nu}
\end{align}
is dominated by the off-diagonal cluster component ${{\rm Re}\,\Sigma^{12}}$, which primarily arises from the non-perturbative short-range correlations captured by the reference system. The additional perturbative diagrammatic corrections to both ${{\rm Re}\,\Sigma^{ll}}$ and ${{\rm Re}\,\Sigma^{12}}$ are not strong enough to significantly alter the total ${\text{Re}\,\Sigma(k=0)}$ (see Fig.~\ref{figs_Re}).
The imaginary part of $\Sigma(k=0)$ originates only from the diagonal parts of the cluster self-energy:
\begin{align}
{\rm Im}\,\Sigma_{k=0,\nu} = \tfrac12{\rm Im}\left[\Sigma^{11}_{k=0,\nu} + \Sigma^{22}_{k=0,\nu}\right].
\end{align}
In this case, the diagrammatic contributions do not compete with the non-perturbative effects, resulting in better agreement with the exact QMC result (see Fig.~\ref{figs_Im}). However, at this $k$-point and for small cluster sizes, the correlation effects may exhibit a more intricate structure than what can be captured by simple two-particle ladder-like excitations, as discussed above.
Finally, at ${k=\pi/2}$ the imaginary part of the lattice self-energy reads:
\begin{align}
{\rm Im}\,\Sigma_{k=\pi/2,\nu} = \tfrac12{\rm Im}\left[\Sigma^{11}_{k=\pi/2,\nu} + \Sigma^{22}_{k=\pi/2,\nu}\right].
\end{align}
At this $k$-point, which corresponds to the Fermi level, particle-hole fluctuations are strong, and accounting for them via the ladder-type diagrams included in \mbox{D-TRILEX} substantially improves the self-energy, as demonstrated in Fig.~\ref{figs_Im}.

\subsubsection{Periodization and its impact on the lattice self-energy}

Let us compare two choices for the periodization of the cluster self-energy~\eqref{eq:L_def}: (i) applying the operation $\mathcal{L}_k$ to the self-energy, which is proportional to the inverse of the Green's function, i.e., to $G^{-1}$, and (ii) applying it to the Green's function $G$ and then computing the self-energy from the periodized Green's function by inverting the Dyson equation.
Since $\mathcal{L}_k$ is a linear operation while matrix inversion is not, the two procedures may yield different results within approximate methods:
\begin{align}
\text{(i)}~\Sigma_{k\nu}^{\text{latt}} 
& = \mathcal{L}_k[\mathbbm{1}(i\nu+\mu)] - \mathcal{L}_k \left[\varepsilon^{ll'}_{k}\right] - \mathcal{L}_k \left[G^{-1}_{k\nu}\right] \notag\\
& = i\nu+\mu - \varepsilon_{k} - \mathcal{L}_k \left[G^{-1}_{k\nu}\right],
\label{eq:Sigma_L_inverse} \\
\text{(ii)}~\Sigma_{k\nu}^{\text{latt}}
& = i\nu+\mu - \varepsilon_{k} - \left(\mathcal{L}_k\left[G_{k\nu}\right]\right)^{-1},
\label{eq:Sigma_L_green}
\end{align}
where $\varepsilon^{ll'}_{k}$ is the cluster version of the original dispersion $\varepsilon_{k}$.
In the following, we refer to Eq.~\eqref{eq:Sigma_L_inverse} and Eq.~\eqref{eq:Sigma_L_green} as the $\Sigma$- and $G$-periodization schemes, respectively.

\begin{figure}[b!]
\centering
\includegraphics[width=0.5\linewidth]{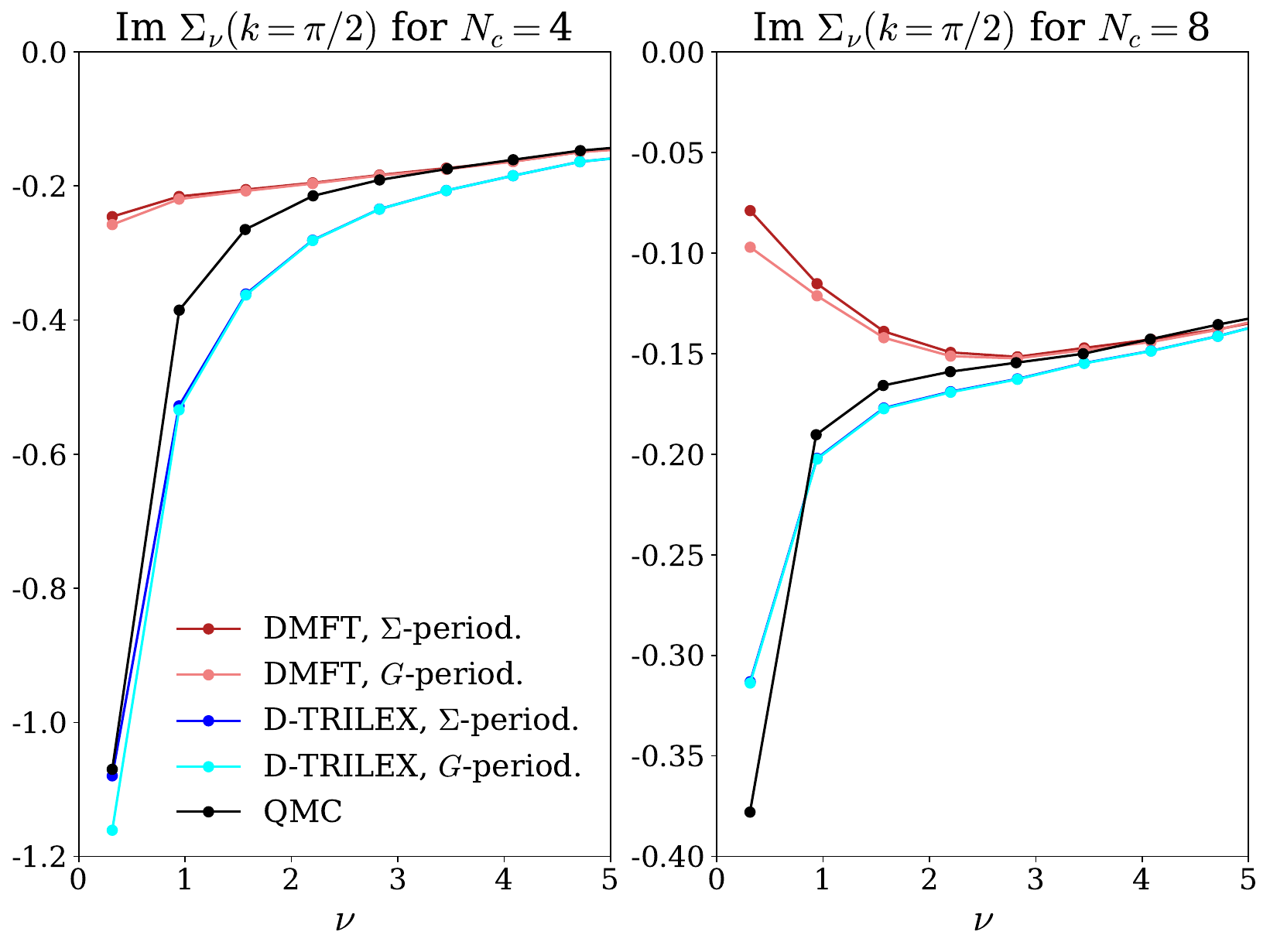}
\caption{The imaginary part of the lattice self-energy obtained at ${k=\pi/2}$ from the cluster DMFT (red colors) and cluster \mbox{D-TRILEX} (blue colors) methods using $\Sigma$- (light colors) and $G$- (dark colors) periodization schemes. The black curve corresponds to the exact QMC result taken from Ref.~\cite{PhysRevB.91.115115}. The Figure is taken from Ref.~\cite{fossati2025cluster}. 
\label{fig:periodization_comparison}}
\end{figure}

In the case of the two-cite cluster, the difference between the $\Sigma$- and $G$-periodization schemes can be obtained analytically.
In the particle-hole-symmetric case the two-site cluster Green function reduces to:
\begin{align}
G_{K\nu} =
\begin{pmatrix}
G^{11}_{K\nu} & G^{12}_{K\nu} \\[0.1cm]
G^{12}_{K\nu} & G^{11}_{K\nu} \\
\end{pmatrix},
\end{align}
where ${G^{11}_{K\nu}\in \mathbb C}$, and ${G^{12}_{K\nu}\in\mathbb R}$.
The difference between the two periodization schemes for the self-energy becomes:
\begin{align}
\label{eq:mismatch_formula}
\tilde{\Sigma}_{k\nu}^{\text{(ii)}} - \Sigma_{k\nu}^{\text{(i)}} 
&= \big( \mathcal L_k[G_{k\nu}] \big)^{-1} - \mathcal L_k\bigl[G^{-1}_{k\nu}\bigr] \\
&= \frac{ \left(G^{12}_{k\nu}\right)^{2} \sin^{2}(ka) }
  { \left[ \left(G^{12}_{k\nu}\right)^{2} - \left(G^{11}_{k\nu}\right)^{2} \right] \left[ G^{11}_{k\nu} + G^{12}_{k\nu}\cos(ka) \right] }\,.
\end{align}
Therefore, the two periodization schemes give identical results at ${k=0}$ and is the largest at ${ka=\pi/2}$, i.e. at the Fermi energy. 

Fig.~\ref{fig:periodization_comparison} compares the imaginary part of the self-energy obtained at ${k=\pi/2}$ (${a=1}$) using the $\Sigma$- and $G$-periodization schemes for ${N_{\text{c}} = 4}$ (left) and ${N_{\text{c}} = 8}$ (right) lattice sites.  
We find, that the two results calculated within the cluster DMFT scheme are already rather close to each other and become even closer in the cluster \mbox{D-TRILEX} approach. 
The discrepancy between the two periodization schemes in \mbox{D-TRILEX} are only noticeable at the lowest Matsubara frequency for ${N_{\rm c}=4}$, and for ${N_{\rm c}=8}$ the two periodization schemes already give identical results.

\subsubsection{Removing off-diagonal terms from the hybridization and partial restoration of the translation invariance}

Let us first justify the neglect of the off-diagonal contributions to the hybridization function. 
We perform CDMFT calculations for the two-site cluster impurity problem using the w2dynamics package~\cite{wallerberger2019}, which allows inclusion of the off-diagonal hybridization in the calculation of single-particle quantities. 
By carrying out CDMFT with the full hybridization function, we find that transforming the two-site cluster to the bonding-antibonding basis not only eliminates the fermionic sign problem but is also well justified numerically, as the off-diagonal hybridization components remain below the numerical noise across the entire explored parameter range.

\begin{figure}[b!]
\centering
\includegraphics[width=0.5\linewidth]{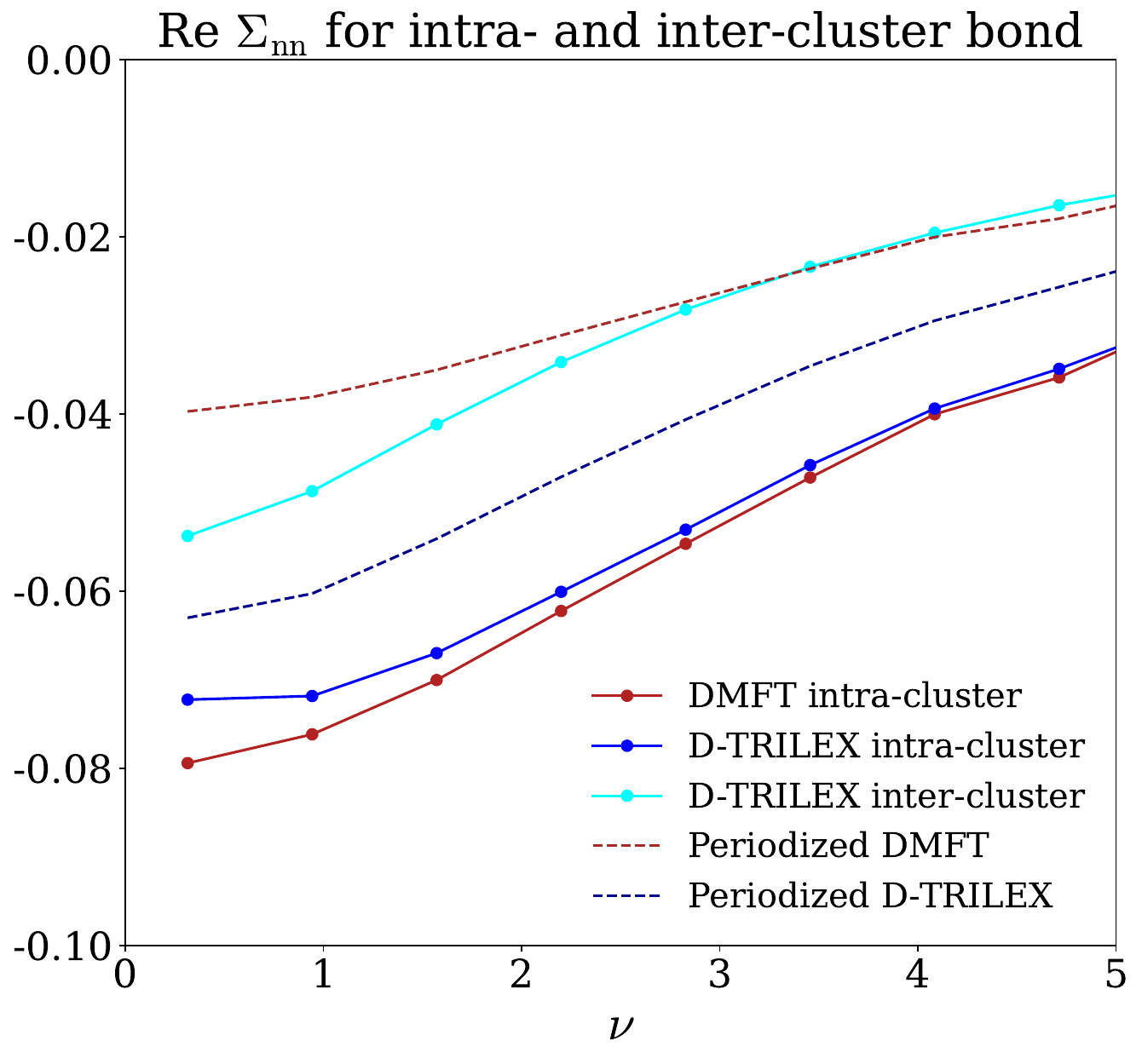}
\caption{The real part of the lattice self‐energy between the neighboring lattice sites calculated for the case of ${N_{\mathrm c}=8}$ as a function of Matsubara frequency $\nu$. The results are obtained for the intra-cluster DMFT (dark red), periodized DMFT (dashed red), inter-cluster \mbox{D-TRILEX} (light blue), intra-cluster \mbox{D-TRILEX} (dark blue), and periodized \mbox{D-TRILEX} (dashed blue) self-energies. The Figure is taken from Ref.~\cite{fossati2025cluster}. 
\label{fig:Sigma_comparison}}
\end{figure}

Let us now analysis if the diagrammatic contributions introduced beyond CDMFT within the cluster \mbox{D-TRILEX} scheme are able to, at least partially, restore the translational symmetry that got broken by introducing the cluster reference systems at the CDMFT step.
To this aim, in Fig.~\ref{fig:Sigma_comparison} we plot the inter- and intra-cluster self-energies corresponding to the nearest-neighbor lattice sites in real space for the case of ${N_{\rm c}=8}$ lattice sites. 
In a perfectly translationally invariant system these two self-energies should be identical, i.e. ${\Sigma^{\text{intra}}_{\nu} = \Sigma^{\text{inter}}_{\nu}}$, and also equal to a periodized self-energy between the neighboring sites.  
The difference between these quantities therefore provides a convenient metric for the degree to which translational symmetry is broken.

In CDMFT the inter-cluster self-energy is identically zero.
For this reason, the intra-cluster (solid red line) self-energy is far from the periodized one depicted by the dashed red line. 
In the cluster \mbox{D-TRILEX} approach, the inter-cluster contribution to the self-energy (light blue line) is generated diagrammatically, but it is still different from the intra-cluster (dark blue) and periodized (dashed blue) self-energies.
The discrepancy between these different self-energies substantially decreases in \mbox{D-TRILEX} compared to DMFT, which is already a good step in the direction of restoring the translational invariance. 
Remarkably, Fig.~\ref{fig:Sigma_comparison} clearly shows that the intra-cluster self-energies provided by the DMFT and \mbox{D-TRILEX} methods lie very close to each other.  
This close agreement can be explained by the fact that the correlations between the neighboring lattice sites in this small-scale one-dimensional system are strong and non-perturbative in nature, which is accounted for by the cluster impurity problem, as discussed above. 
Additional inclusion of perturbative particle-hole ladder-type contributions within \mbox{D-TRILEX} in not able to substantially change the intra-cluster self-energy (dark blue).
In this regard, the diagrammatic contribution to the inter-cluster self-energy (light blue) is such that the periodized self-energy (dashed blue) is in a good agreement with the exact result, as demonstrated above.   
Therefore, we argue that the CDMFT starting point for the diagrammatic expansion is the limiting factor that prevents the cluster \mbox{D-TRILEX} approach from restoring the translational invariance.
These findings imply that a fully self-consistent algorithm, with the outer self-consistency loop that updates the reference system to preserve the equivalence between the inter- and intra-cluster self-energies, would lead to a complete restoration of the translational invariance within the cluster \mbox{D-TRILEX} scheme.
However, performing such calculation will require recalculation of the two-particle quantities of the cluster impurity problem at every iteration of the outer-self-consistency loop, which is extremely expensive for the current implementation of the method.

\newpage

\section{Real-time dual $GW$ approach}
\label{sec:DGW}

The success of the multi-band implementation motivated us to formulate the time-dependent version of the \mbox{D-TRILEX} method. 
It is important to mention that the equilibrium \mbox{D-TRILEX} method is formulated in the Matsubara frequency space and accounts for the exact three-point (Hedin~\cite{GW1}) vertex corrections of the reference problem in the diagrams for the self-energy and the polarization operator. 
Calculating the three-point vertex out-of-equilibrium involves enormous computational resources and requires developing real-time impurity solvers to calculate this three-time-dependent object. 
In addition, computing the electronic self-energy and the polarization operator on an L-shaped Keldysh contour with these vertices involves four-time integration, which is extremely challenging. 
Unfortunately, with the available numerical tools, these complications prevent the efficient implementation of the time-dependent generalization of the \mbox{D-TRILEX} approach, making it prohibitively expensive computationally.

For this reason, in this Section we introduce a ``lighter'' version of the \mbox{D-TRILEX} method on the L-shaped Keldysh contour by approximating the three-point vertex function by its instantaneous (short-time limit) component.
With this approximation, the diagrammatic structure of the proposed method resembles that of the $GW$ theory~\cite{GW1, GW2, GW3}, so we call it the Dual-$GW$ (${D\text{-}GW}$) approach.
The developed framework provides a consistent diagrammatic extension of DMFT, which enables a simultaneous treatment of the spatial charge and spin fluctuations thus improving upon the $GW$+DMFT method~\cite{PhysRevLett.118.246402, PhysRevB.100.041111, PhysRevB.100.235117}, which is one of the most advanced real-time diagrammatic methods that have been implemented on the Keldysh contour.
Additionally, ${D\text{-}GW}$ is not restricted to a weak-coupling regime and is also able to account for the long-range Coulomb interaction, which is an important advantage over the time-dependent implementation of the TPSC+DMFT approach~\cite{PhysRevB.107.245137}.

\subsection{Formulation of the method}

\begin{figure}[t!]
\centering
\includegraphics[width=0.6\linewidth]{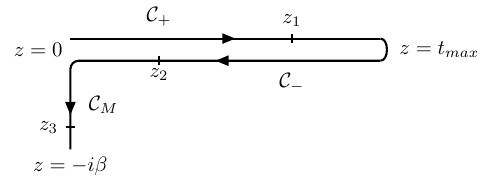}
\caption{Non-equilibrium time contour $\mathcal{C}$ in the complex time plane, consisting of the forward ($\mathcal{C}_+$), backward ($\mathcal{C}_-$) and imaginary ($\mathcal{C}_M$) time branches, $\mathcal{C} = \mathcal{C}_+ \cup \mathcal{C}_- \cup \mathcal{C}_M$, with general contour times $z_1$, $z_2$, $z_3 \in \mathcal{C}$. The Figure is taken from Ref.~\cite{vglv-2rmv}.
\label{fig:contour}}
\end{figure}

We start with the lattice action~\eqref{eq:actionlatt}.
It has the following form in the real-time ($z$) and lattice space ($i,j$) representation:
\begin{align}
{\cal S} = & -\int_{\cal C} dz \, \Bigg\{- \sum_{ij,\sigma} c^{*}_{i\sigma}(z) \left[\delta^{\phantom{*}}_{ij}(i\partial^{\phantom{*}}_{z}+\mu) - J^{\phantom{*}}_{ij}(z)\right] c_{j\sigma}^{\phantom{*}}(z) \notag\\
&+ \sum_{i} U n_{i\uparrow}(z) \, n_{i\downarrow}(z) + \frac{1}{2} \int_{\cal C} dz' \sum_{ij,\varsigma} \rho^{\varsigma}_{i}(z) V^{\varsigma}_{ij} \, \rho^{\varsigma}_{j}(z') \Bigg\}\,.
\label{eq:action_latt}
\end{align}
Note, that in this Section we use a bit different notations.
In particular, we use $J_{ij}$ is the hopping amplitude between the lattice sites $i$ and $j$. 
The time $z$ is defined on the Kadanoff-Baym contour $\mathcal{C}$ in the complex time plane (see Fig.~\ref{fig:contour}). 
The contour times ${z \in \mathcal{C}}$ runs from ${t=0}$ to the maximum simulation time $t_{max}$ along the real-time forward branch $\mathcal{C}_+$, back to ${t=0}$ along the backward branch $\mathcal{C}_-$, and then to $-i\beta$ along the imaginary-time branch $\mathcal{C}_M$ (for details see~\cite{Stefanucci:2013oq}). 
The partition function in the coherent-state path-integral formalism can be expressed as ${\mathcal{Z} = \int D[c^{*},c] \exp(i{\cal S})}$.

The electron coupling to the electrical field $\mathbf{E}_{p}(t)$ of light is introduced through the Peierls substitution. 
The induced vector potential ${\mathbf{A}(t) = - \int_0^t \mathbf{E}_{p}(\bar{t}) \, d\bar{t}}$, enters as a time-dependent momentum shift in the single particle dispersion ${\varepsilon\left(\mathbf{k},t\right) \equiv \varepsilon\left(\mathbf{k} - \frac{1}{\hbar} \mathbf{A}(t)\right)}$. We further set electric charge $e$, speed of light $c$ and lattice constant $a$ to unity ($e$ = $c$ = $a$ = 1). 

To derive the partially bosonized dual action~\eqref{eq:DTRILEX_action} in the real-time representation, we follow the steps described in Section~\ref{sec:PBDAction}.
In particular, we consider an effective DMFT-like impurity problem~\eqref{eq:actionimp_app} as the reference system by introducing a time-dependent fermionic hybridization function $\Delta(z,z')$.
After some path integral transformations, we arrive at
the partially bosonized dual action in the following form:
\begin{align}
{\cal S}_{fb}
=
&\int_{\cal C} dz \, dz' \sum_{ij,\sigma} f^{*}_{i\sigma}(z) \big[\tilde{\cal G}\big]^{-1}_{ij}(z,z')
f^{\phantom{*}}_{j\sigma}(z') 
+ \frac12 \int_{\cal C} dz \, dz' \sum_{ij, \varsigma} b^{\varsigma}_{i}(z) \big[\tilde{\cal W}^{\varsigma}\big]^{-1}_{ij}(z,z') b^{\varsigma}_{j}(z') \nonumber \\ 
& - \int_{\cal C} dz \, dz'  dz'' \sum_{i,\varsigma} \sum_{\sigma\sigma'} \Lambda^{\varsigma}_{zz'z''} f^{*}_{i\sigma}(z) \sigma^{\varsigma}_{\sigma\sigma'} f^{\phantom{*}}_{i\sigma'}(z') b^{\varsigma}_{i}(z'')\,.
\label{S_DB_fin}
\end{align}
The bare dual Green's function $\tilde{\cal G}$ and renormalized interaction $\tilde{\cal W}$ in the momentum-space representation explicitly read:  
\begin{align}
\tilde{\cal G}_{\bf k}(z,z') &= G_{\bf k}^{\rm DMFT}(z,z') - g(z,z')\,,
\label{eq:bare_G} \\
\tilde{\cal W}^{\varsigma}_{q}(z,z') &=  W^{\rm EDMFT}_{q,\varsigma}(z,z') -\delta_{\cal C}(z,z') U^{\varsigma}/2\,.
\label{eq:bare_w}
\end{align}
Here, $g(z,z')$ is the exact impurity Green's function and ${U^{d/m} = \pm U/2}$.
The DMFT Green's function $G_{\bf k}^{\rm DMFT}$ and the EDMFT-like renormalized interaction $W^{\rm EDMFT}_{q,\varsigma}(z,z')$ are given by the following Dyson equations:
\begin{align}
\big[G_{\bf k}^{\rm DMFT}\big]^{-1}(z,z') &= \big[{\cal G}_{\bf k}\big]^{-1}(z,z')-\Sigma^{\rm imp}(z,z'),
\label{eq:imp_G} \\
\big[W^{\rm EDMFT}_{\bf q,\varsigma}\big]^{-1}(z,z') &= \delta_{\cal C}(z,z')\big[U^{\varsigma}+V^{\varsigma}_{q}\big]^{-1} -\Pi^{\varsigma\,\rm imp}(z,z'),
\label{eq:imp_w}
\end{align}
where ${\cal G}_{\bf k}(z,z')$ is the bare lattice Green's function, and $\Sigma^{\rm imp}(z,z')$ and $\Pi^{\varsigma\,\rm imp}(z,z')$ are the self-energy and the polarization operator of the impurity problem. From the derived fermion-bosonic action~\eqref{S_DB_fin}, the self-energy and the polarization operator of the \mbox{D-TRILEX} theory in the momentum space representation can be found as follows:
\begin{align}
\tilde{\Sigma}_{{\bf k}}(z_1,z_2)
        &= i\int_{\cal C} \{dz'\} \sum_{{\bf q},\varsigma}  \Lambda^{\varsigma}(z_1,z',z'') \, \tilde{G}^{\phantom{*}}_{{\bf k+q}}(z'',z'''') \, \tilde{W}^{\varsigma}_{{\bf q}}(z''',z'''') \, \Lambda^{\varsigma}(z'''',z_2,z''') \notag\\
&~~~-2i\int_{\cal C} \{dz'\} \sum_{{\bf k'}} \Lambda^{\rm d}(z_1,z_2,z'') \, \tilde{\cal W}^{\rm d}_{{\bf q}=0}(z'',z''') \, \Lambda^{\rm d}(z''',z'''',z'') \, \tilde{G}^{\phantom{*}}_{{\bf k'}}(z'''',z''') \,,
\label{eq:Sigma_dual_actual}\\
\tilde{\Pi}^{\varsigma}_{{\bf q}}(z_1,z_2)
&= -2i \int_{\cal C} \{dz'\} \sum_{{\bf k}} \Lambda^{\varsigma}(z'',z',z_1) \, \tilde{G}^{\phantom{*}}_{{\bf k}}(z',z''') \, \tilde{G}^{\phantom{*}}_{{\bf k+q}}(z'''',z'') \, \Lambda^{\varsigma}(z''',z'''',z_2) \,,
\label{eq:Pi_dual_actual}
\end{align}
where the dressed dual Green's function and renormalized interactions on a Keldysh contour are defined as: 
\begin{align}
\tilde{G}_{\bf k}(z,z') &= -i\langle T_{\cal C} f^{\phantom{*}}_{\bf k,\sigma}(z) f^{*}_{\bf k,\sigma}(z')\rangle\,, \\
\tilde{W}^{\varsigma}_{\bf q}(z,z') &= -i\langle T_{\cal C} b^{\varsigma}_{\bf{q}}(z)b^{\varsigma}_{\bf{q}}(z')\rangle\,.
\end{align}
The diagrammatic structure of the dual self-energy and dual polarization of D-TRILEX can be found in Fig.~\ref{fig:DT_diagrams}.

We note that the vertex function $\Lambda^{\varsigma}$ can be defined in several ways, depending on the choice of the fermionic $B$ and bosonic $\alpha$ scaling parameters~\eqref{eq:Vertex_PH_app}.
However, as discussed in Section~\ref{sec:Scaling_parameters}, there exists a unique choice, ${B(z,z')=g(z,z')}$ and $\alpha(z,z')$ in the form of Eq.~\eqref{eq:alpha_app}, for which the asymptotic short-time (infinite frequency) limit of the three-point vertex is equal to unity. 
In the following, we adhere to this specific choice of the scaling parameters, which allows us to approximate the vertex by its instantaneous component ${\Lambda^{\varsigma}_{z_1z_2z_3} = \delta_{\cal{C}}(z_1,z_2) \delta_{\cal{C}}(z_2,z_3)}$.
After that, the \mbox{D-TRILEX} self-energy and polarization operator reduce to a simple $GW$-like form that we refer to as ${D\text{-}GW}$: 
\begin{align}
\tilde{\Sigma}_{{\bf k}}(z_1,z_2)
&= i \sum_{{\bf q},\varsigma} \tilde{G}^{\phantom{*}}_{{\bf k+q}}(z_1,z_2) \tilde{W}^{\varsigma}_{{\bf q}}(z_1,z_2) 
-2i \delta_{\cal C}(z_1,z_2)\sum_{\bf{k'}} \int_{\cal{C}} dz'\tilde{\cal W}^{d}_{{\bf q}=0}(z_1,z') \tilde{G}^{\phantom{*}}_{{\bf k'}}(z',z')\,,
\label{eq:Sigma_dual_para}\\
\tilde{\Pi}^{\varsigma}_{\bf q}(z_1,z_2)
&= -2i \sum_{\bf {k}} \tilde{G}_{\bf {k}}(z_1,z_2) \tilde{G}_{\bf {k+q}}(z_2,z_1)\,.
\label{eq:Pi_dual_para}
\end{align}
The dressed dual Green's function and renormalized interactions can be found by solving the following Dyson equations:
\begin{align}
\big[\tilde{G}_{\bf k}\big]^{-1}(z,z') = \big[\tilde{\cal G}_{\bf{k}}\big]^{-1}(z,z') -\tilde{\Sigma}_{\bf k}(z,z')\,,
\label{eq:dress_G} \\
\big[\tilde{W}^{\varsigma}_{\bf q}\big]^{-1}(z,z') = \big[\tilde{\cal W}^{\varsigma}_{\bf q}\big]^{-1}(z,z')-\tilde{\Pi}^{\varsigma}_{\bf q}(z,z')\,.
\label{eq:dress_w}
\end{align}

\begin{figure}[b!]
\vspace{1cm}
\includegraphics[width=1\linewidth]{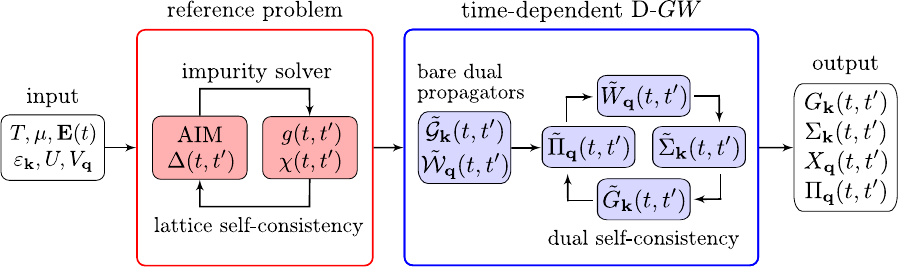}
\caption{The computational workflow of the non-equilibrium ${D\text{-}GW}$ method on Kelydsh contour $z=t \in C_{\pm,M}, z'=t' \in C_{\pm, M}$. 
The input consists of the model parameters: the initial temperature $T$, the chemical potential $\mu$, the applied electric field ${\bf{E}}_{p}(t)$, the electronic dispersion $\varepsilon_{\bf k}$, and the local $U$ and non-local $V_{\bf q}$ electronic interactions. The red box indicates the self-consistent solution of the time-dependent reference problem. This work considers the latter an Anderson impurity model (AIM) with a time-dependent hybridization function $\Delta(t,t')$ determined by the DMFT lattice self-consistency. The ${D\text{-}GW}$ formalism (blue box) takes the renormalized Green's function $g(t,t')$ and susceptibilities $\chi(t,t')$ of the reference problem are further used as inputs for constructing bare propagators for the dual fermionic $\tilde{\cal G}_{\bf k}(t,t')$ and bosonic $\tilde{\cal W}_{\bf q}(t,t')$ fields. The dressed time-dependent quantities in dual space: the polarization operator $\tilde{\Pi}_{\bf q}(t,t')$, the renormalized interaction $\tilde{W}_{\bf q}(t,t')$, the self-energy $\tilde{\Sigma}_{\bf k}(t,t')$, and the renormalized Green's function $\tilde{G}_{\bf k}(t,t')$ are determined through the ${D\text{-}GW}$ dual self-consistency (closed loop in the blue box). The physical quantities are obtained using the exact relations between the corresponding quantities in the dual and lattice spaces. The Figure is taken from Ref.~\cite{vglv-2rmv}.
\label{fig:cycle}}
\end{figure}

\subsubsection{Computational workflow}

The computational workflow of the ${D\text{-}GW}$ approach is illustrated in Fig.~\ref{fig:cycle}. 
The first step (red box) involves the self-consistent solution of the reference (DMFT impurity) problem, which, in the current implementation, is done using a lowest-order strong coupling perturbative method called the real-time non-crossing approximation (NCA). The formulation of real-time NCA on the Keldysh contour is detailed in Ref.~\cite{PhysRevB.82.115115, PhysRevB.96.165104}, while the real-time DMFT self-consistency implementation on a square lattice is described in Ref.~\cite{vglv-2rmv}.
The solution of the reference problem, subject to an external time-dependent perturbation ${{\bf{E}}_{p}(t)}$, yields the two-time dependent Green's function $g(t,t')$ and the susceptibilities $\chi^{\varsigma}(t,t')$. 
Together with the self-consistently obtained hybridization function $\Delta(t,t')$, these quantities are further used in the ${D\text{-}GW}$ diagrammatic expansion to construct the bare dual Green's function $\tilde{\cal G}_{\bf k}(t,t')$~\eqref{eq:bare_G} and renormalized interaction $\tilde{\cal W}^{\varsigma}_{\bf q}(t,t')$~\eqref{eq:bare_w}.
The dressed dual quantities are obtained via self-consistent solution of the corresponding Dyson equations~\eqref{eq:dress_G} and~\eqref{eq:dress_w} that involve the diagrammatic calculation of the dual self-energy $\tilde{\Sigma}_{\bf k}(t,t')$~\eqref{eq:Sigma_dual_para} and polarization operator $\tilde{\Pi}^{\varsigma}_{\bf q}(t,t')$~\eqref{eq:Pi_dual_para}. 
Ultimately, the dressed physical Green's function $G_{\bf k}(t,t')$ and susceptibilities $X^{\varsigma}_{\bf q}(t,t')$, as well as the self-energy $\Sigma_{\bf k}(t,t')$ and polarization operator $\Pi_{\bf q}(t,t')$ are obtained using the exact relations between the dual and lattice quantities.
The latter are explicitly defined in the next Section.

\subsubsection{Physical quantities}

\paragraph{Physical Green's functions:}

The first and most crucial quantity of interest is the lattice Green's function: 
\begin{align}
G_{\bf k}(z,z') = -i\langle T_{\cal C} c^{\phantom{\dagger}}_{\bf k,\sigma}(z) c^{\dagger}_{\bf k,\sigma}(z')\rangle\,,  
\label{eq:G_latt_def}
\end{align}
which is given by Eq.~\eqref{eq:Glatt_DT} in the integral form: 
\begin{gather}
\int_{\cal C} dz'' \, dz''' \left\{\delta_{\cal C}(z,z''') + \big[g(z,z'') + \tilde{T}_{\bf k}(z,z'')\big]\big[\Delta(z'',z''') - \overline{\varepsilon}_{\bf k}(z)\delta_{\cal C}(z'',z''')\big] \right\} G_{\bf k}(z''',z') = \notag\\
g(z,z') + \tilde{T}_{\bf k}(z,z')\,,
\label{eq:G_latt}
\end{gather}
where $\overline{\varepsilon}_{\bf k}(z) = \varepsilon_{\bf k}(z)-\varepsilon_{\rm loc}(z)$ and ${\tilde{T}_{\bf k}(z,z') = \int_{\cal C} dz'' \, dz''' \, g(z,z'') \tilde{\Sigma}_{\bf k}(z'',z''')g(z''',z')}$.\\
The lattice self-energy~\eqref{eq:Sigmalatt_DT}:
\begin{align}
\Sigma^{\rm latt}_{{\bf k}}(z,z') = \Sigma^{\rm imp}(z,z') + \overline{\Sigma}_{{\bf k}}(z,z')
\end{align}
consists of the impurity contribution $\Sigma^{\rm imp}(z,z')$ and the diagrammatic correction $\overline{\Sigma}_{{\bf k}}(z,z')$.
The latter can be obtained as follows:
\begin{align*}
\overline{\Sigma}_{{\bf k}}(z,z') = \tilde{\Sigma}_{{\bf k}}(z,z') - \int_{\cal C} dz'' \, dz''' \, \tilde{\Sigma}_{{\bf k}}(z,z'') g(z'',z''') \overline{\Sigma}_{{\bf k}}(z''',z')\,.
\end{align*}
Alternatively, the lattice Green's function, defined in Eq.~\eqref{eq:G_latt_def}, can be calculated from the diagrammatic correction to the impurity self-energy using the following Dyson equation:
\begin{align}
\big[G^{\phantom{1}}_{\bf{k}}\big]^{-1}(z,z') = \big[G^{\rm DMFT}_{\bf k}\big]^{-1}(z,z')- \overline{\Sigma}_{{\bf k}}(z,z')\,.
\end{align}
In the current implementation, the lattice Green's function is calculated using Eq.~\ref{eq:G_latt}, which is more efficient computationally.\\
The lattice susceptibility on a Keldysh contour is defined as:
\begin{align}
X^{\varsigma}_{\bf q}(z, z') =
-i\left\langle
T_{\cal C} \,
\rho^\varsigma_{\bf q}(z) 
\,
\rho^\varsigma_{\bf -q}(z') 
\right\rangle.
\end{align} 
To calculate this quantity, one has to obtain the lattice polarization operator~\eqref{eq:Pilatt_DT}:
\begin{align}
\Pi^{\varsigma}_{\bf q}(z,z') = \Pi^{\varsigma}_{\rm imp}(z,z') + \overline{\Pi}^{\varsigma}_{\bf q}(z,z'),
\end{align}
where the diagrammatic correction $\overline{\Pi}^{\varsigma}_{\bf q}(z,z')$ to the impurity polarization operator $\Pi^{\varsigma}_{\rm imp}(z,z')$ reads:
\begin{align}
\overline{\Pi}^{\varsigma}_{\bf q}(z,z') = \tilde{\Pi}^{\varsigma}_{\bf q}(z,z') - \int_{\cal C} dz'' \, \tilde{\Pi}^{\varsigma}_{\bf q}(z,z'') \frac{U^{\varsigma}}{2} \overline{\Pi}^{\varsigma}_{\bf q}(z'',z').
\end{align}
The lattice susceptibility can then be obtained via the Dyson equation:
\begin{align}
X^{\varsigma}_{\bf q}(z,z') = \Pi^{\varsigma}_{\bf q}(z,z') + \int_{\cal C} dz'' \, \Pi^{\varsigma}_{\bf q}(z,z'') \left(U^{\varsigma} + V^{\varsigma}_{\bf q}\right) X^{\varsigma}_{\bf q}(z'',z')\,.
\label{eq:chi_latt}
\end{align}

\paragraph{Equal-time quantities:}

The properties of the initial thermal equilibrium state is obtained by restricting the contour times $z, z'$ to the imaginary time contour $z = -i\tau, z' = -i\tau' \in \mathcal{C}_M$, $G_\mathbf{k}(\tau-\tau') = G_\mathbf{k}(z, z')$, while the non-equilibrium properties are obtained when restricting the contour times $z$ and $z'$ to the real-time branches,
giving the lesser and greater response function components:
\begin{equation}
G^{\lessgtr}_\mathbf{k}(t, t') =
G_{\mathbf{k}}(z, z')
\, , \quad
z = t \in \mathcal{C}_\pm
\, , \,\,
z' = t' \in \mathcal{C}_\mp
\,,
\end{equation}
which corresponds to the occupied and unoccupied states. 
The equal-time values of the propagators give physical observables, e.g., the time-dependent electron density ${\langle n_{\mathbf{k}\sigma}(t) \rangle = -iG^<_{\mathbf{k}}(t, t)}$ that determines the kinetic energy:
\begin{equation}
K(t) = \frac{2}{N_\mathbf{k}} \sum_{\mathbf{k}} \varepsilon_{\mathbf{k}}(t) \langle n_{\mathbf{k}\sigma} (t) \rangle
\, .
\label{eq:Ekin}
\end{equation}
The interaction energy within the Galitskii-Migdal formula is given by the convolution~\cite{Mahan}:
\begin{equation}
P(t) = 
-\frac{i}{N_\mathbf{k}} \sum_{\mathbf{k}} [\Sigma_{\mathbf{k}} \ast G_{\mathbf{k}}]^<(t, t)\,,
\label{eq:Eint}
\end{equation}
where $\Sigma_{\mathbf{k}}(z, z')$ is the single-particle self-energy and the contour product is given by $(A \ast B)(z, z') \equiv \int_\mathcal{C} d\bar{z} A(z, \bar{z})B(\bar{z}, z')$, see~\cite{RevModPhys.86.779}.
The combination gives the total energy:
\begin{equation}
E(t) = K(t) + P(t)
\, .
\label{eq:Etot}
\end{equation}
The interaction energy of the extended Hubbard model can be separated into local $E_{U}$ and non-local $E_{V}$ contributions, where the former is due to the Hubbard interaction $U$, and the latter comes from the non-local interaction $V$:
\begin{equation}
P(t) = E_{U}(t) + E_{V}(t)\,.
\label{eq:total_E}
\end{equation}
The local interaction energy is, in turn, given by ${E_U(t) = U D(t)}$, where the local double occupancy $D(t)$ can be determined using the local charge and spin susceptibilities $\chi^\varsigma$~\cite{PhysRevB.93.155162}:
\begin{align}
D(t) &\equiv 
\langle n_{i\uparrow}(t) n_{i\downarrow}(t) \rangle 
= 
\frac{1}{4}\left(
iX^{d,<}_{\rm loc}(t, t) 
- 
iX^{m,<}_{\rm loc}(t, t)
+
{\langle n^{d}(t) \rangle}^{2} 
\right).
\end{align}
The above double occupancy expression can be further used to decouple the local interaction energy between charge and spin channels ${E_{U} = E_{Ud} + E_{Um}}$, where: 
\begin{align}
E_{Ud} &= \frac{U}{4} \left( iX^{d,<}_{\rm loc}(t, t) + \langle n^{d}(t) \rangle^2 \right), \\ 
E_{Um} &= -\frac{U}{4} iX^{m,<}_{\rm loc}(t, t)\,.  
\end{align}
Finally, the non-local interaction energy can be obtained from the momentum-dependent charge susceptibility as follows:
\begin{equation}
E_{V}(t)= \frac12 \sum_{\bf q} V_{\bf q} iX^{d,<}_{\bf q}(t,t)\,.
\end{equation}
Since the interaction energy can be calculated using either the self-energy (via the Galitskii-Migdal formula) or the susceptibilities, the total energy of the photo-excited state can be evaluated in two distinct ways.
We denote the total energy as \(E_{\chi}\), when the interaction energy is calculated from the susceptibility, and as \(E_{Mig}\), when calculated from the self-energy. 
The non-equilibrium Green's function formalism allows us to access all components of the system's total energy. This capability enables us to study the non-equilibrium energy redistribution during and after photo-doping in the system.

\paragraph{Spectral functions:}

The non-equilibrium Green's function formalism provides access to real-time spectral information of electrons, as well as charge and spin degrees of freedom. The real-time generalization of the occupied and unoccupied electronic spectral functions is obtained using a partial Fourier transform:
\begin{equation}
A_\mathbf{k}^{\lessgtr}(t, \omega) = \frac{\pm\text{Im} \hspace{0.05cm}G^{\lessgtr}_{\mathbf{k}}(t,\omega)}{\pi} = 
\int_0^\infty \!\!\!\! d\bar{t} \, 
e^{i\omega \bar{t}} \, G^{\lessgtr}_\mathbf{k}(t + \bar{t}, t) 
\, ,
\end{equation}
which combines into the time-dependent electronic spectral function:
\begin{equation}
A_{\mathbf{k}}(t, \omega) = 
A^{>}_{\mathbf{k}}(t, \omega) + A^{<}_{\mathbf{k}}(t, \omega)
\, .
\end{equation}
At thermal equilibrium, the time dependence dropout and the fluctuation-dissipation theorem gives:
\begin{equation}
A_\mathbf{k}^<(\omega) = f_{e}(\omega) A_\mathbf{k}(\omega)
\, , ~\,
A_\mathbf{k}^>(\omega) = \left[ 1 - f_{e}(\omega) \right] A_\mathbf{k}(\omega)
\, ,
\end{equation}
where $f_{e}(\omega) = 1/(e^{\beta\omega} - 1)$ is the Fermi distribution function. 
This results in:
\begin{align}
F_{k}(\omega) \equiv \ln \left[ A^{<}_{k}(\omega) / A^{>}_{k}(\omega) \right] = -\omega \beta.
\end{align}
As used for electrons, a similar Fourier transform has been used to calculate the occupied and unoccupied spectral functions of the charge and spin-susceptibilities. 
They are defined as:
\begin{equation}
A^{d/m,\lessgtr}_\mathbf{q}(t, \omega) = \frac{-\text{Im} \hspace{0.05cm}X^{d/m,\lessgtr}_{\mathbf{q}}(t,\omega)}{\pi} = 
\int_0^\infty \!\!\!\! d\bar{t} \, 
e^{i\omega \bar{t}} \, X^{d/m,\lessgtr}_\mathbf{q}(t + \bar{t}, t) 
\, .
\end{equation}
One can calculate the retarded spectral functions of the charge and spin susceptibilities from the occupied and unoccupied functions, which are given by: 
\begin{equation}
A^{d/m}_{\mathbf{q}}(t, \omega) = 
A^{d/m,>}_{\mathbf{q}}(t, \omega) - A^{d/m,<}_{\mathbf{q}}(t, \omega)
\, .
\end{equation}
The fluctuation-dissipation theorem in thermal equilibrium relates the retarded spectral functions to their occupied and unoccupied susceptibilities through:
\begin{equation}
A^{d/m,<}_\mathbf{q}(\omega) = f_{b}(\omega) A^{d/m}_\mathbf{q}(\omega)
\, , ~\,
A^{d/m,>}_\mathbf{q}(\omega) = \left[ 1 - f_{b}(\omega) \right] A^{d/m}_\mathbf{q}(\omega)
\, ,
\end{equation}
where ${f_{b}(\omega) = 1/(e^{\beta\omega} + 1)}$ is the Bose distribution function. 
The occupied-to-unoccupied spectral susceptibilities ratio leads to a similar electron expression. To explore this out-of-equilibrium, we use the time-dependent generalization: 
\begin{align}
F_{k}(t, \omega) \equiv \ln \left[ \frac{A^{<}_{\mathbf{k}}(t, \omega)}{A^{>}_{\mathbf{k}}(t, \omega)} \right],
~~F^{d/m}_{q}(t,\omega) =\ln \left[ \frac{A^{d/m,<}_{\mathbf{q}}(t, \omega)}{A^{d/m,>}_{\mathbf{q}}(t, \omega)} \right],
\label{hkjwdlnsks}
\end{align}
which allows us to study the system's approach to thermalization. In the long-time limit, as ${t \rightarrow \infty}$, we find that $F_{k}(t, \omega)$ and $F^{d/m}_{q}(t,\omega)$ approach to  $-\beta_{\text{eff}}(t) \omega$, enabling us to determine the final effective inverse temperature $\beta_{\text{eff}}$. Additionally, it has been utilized to extract effective temperatures for quasi-particles, doublons, and charge and spin degrees of freedom during the relaxation of photo-excited states~\cite{PhysRevB.98.035113, PhysRevLett.117.096403, Stefanucci:2013oq}.

\subsection{Electron-magnon dynamics triggered by an ultrashort laser pulse}

To showcase the effectiveness of the ${D\text{-}GW}$ method, in Ref.~\cite{vglv-2rmv} we applied it to analyze the non-equilibrium dynamics of the single-band extended Hubbard model, focusing on the most challenging regime, i.e., near the Mott transition, on both its metallic and insulating sides.
Our simulations reveal that paramagnetic metals reach a thermal state within the simulation time, while Mott insulators do not, despite the same excitation protocol.
In Mott insulators, the pre-thermal state is characterized by photo-excited doublons and non-local charge and spin fluctuations, which reach a temperature indicative of a unique thermal state. 
This is different from that of quasi-particles. 
True thermalization occurs only when quasi-particles in the transient also achieve this unique thermal state temperature, which our simulations could not fully reach.
An intriguing finding is that even with AFM spin fluctuations, impact ionization- a nonlinear relaxation process for photo-excited charge carriers- remains highly favorable for extended systems. In this process, the time evolution of the occupied density of states illustrates the transfer of excess kinetic energy from the high-energy photo-excited doublons (holons) to the underlying antiferromagnetic spin fluctuations. This energy transfer can be directly measured using time-resolved photoemission spectroscopy.

\subsubsection{Numerical setup}

We consider a half-filled single-band extended Hubbard model on a square lattice with the nearest-neighbor hopping amplitude $J$, the on-site Coulomb potential $U$, and the nearest-neighbor Coulomb interaction $V$ between the charge densities. 
We use the NESSi implementation~\cite{SCHULER2020107484} for the solution of the Kadanoff-Baym equations on the Keldysh contour. Unless otherwise stated, we fix the initial equilibrium temperature of the systems at ${T=1/\beta=1/6}$ and pulse frequency to ${\omega_{p}=U}$. Calculations are performed for the periodic lattice of size ${20\times20}$. 
We select the Matsubara grid with ${\Delta \tau = 0.0066}$ (0.0060) and the real-time grid with ${\Delta t = 0.011}$ (0.010) for metals (Mott-insulators) to ensure that dynamical quantities converge with grid size.
The unit of time is set to $\hbar/J$. 

\subsubsection{Equilibrium calculations}

Before discussing the photo-excitation dynamics of the extended Hubbard model, let us first focus on the equilibrium results to determine the effect of non-local correlations, particularly spin fluctuations, on the electronic spectral function. 
These results will also provide us with further insights to identify the parameter space of interest for the photo-excitation dynamics.
Without the electric pulse ${\mathbf{E}_{p}(t)=0}$, the solution of the model on the $L$-shaped contour yields a time-translationally invariant electronic Green's function~\eqref{eq:G_latt} and susceptibilities~\eqref{eq:chi_latt}, which are the functions of time-difference ${(t-t')}$.
A straightforward Fourier transformation of these functions from real-time to frequency domain directly provides access to spectral information without requiring analytical continuation in finite-temperature Matsubara methods.

\begin{figure}[t!]
\centering
\includegraphics[width=0.7\linewidth, trim={10pt 10pt 10pt 10pt}, clip]{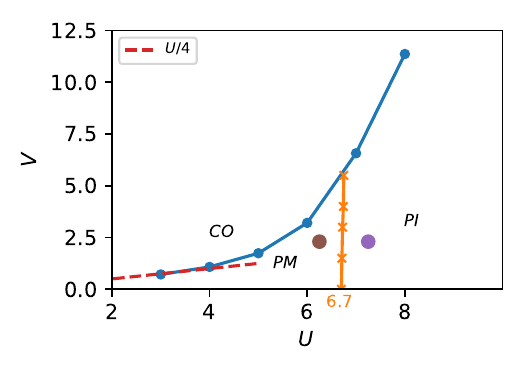}
\caption{Equilibrium phase diagram of the extended Hubbard model in the ${U\text{-}V}$ plane obtained using the real-time ${D\text{-}GW}$ method.
The blue line indicates the phase boundary between the charge-ordered (CO) and paramagnetic metal (PM) phases. 
The orange line depicts the phase boundary between the PM and the paramagnetic Mott insulator (PI). 
The red dashed line depicts the mean-field estimate ${V=U/4}$ for the CO phase boundary.
The brown (${U=6.25}$, ${V=2.3}$) and purple circles (${U=7.25}$, ${V=2.3}$) respectively indicate the parameters of a correlated metal and a Mott insulator at which calculations are carried out in this paper. The Figure is taken from Ref.~\cite{vglv-2rmv}.
\label{fig:fig1}}
\end{figure}

In Fig.~\ref{fig:fig1} we show the phase diagram of the extended Hubbard model at half-filling in the ${U\text{-}V}$ plane obtained using the developed real-time ${D\text{-}GW}$ method.  
At a critical temperature $T_N$, this model features the N\'eel transition to the antiferromagnetic state driven by strong spin fluctuations.
Above $T_N$, the long-range AFM order completely melts, but the short-range spin fluctuations persist to higher temperatures in the paramagnetic phase. Since the equilibrium temperature is chosen to be larger than AFM N\'eel temperature ($T>T_{N}$), we can not see such an AFM to paramagnetic transition in Fig.~\ref{fig:fig1}.
At a critical interaction $U_{c}$, the high-temperature paramagnetic phase exhibits the transition to the Mott insulating state, which is associated with the effect of local electronic correlations.
The formation of the insulating state driven by the magnetic fluctuations (Slater mechanism) and local electronic correlation (Mott scenario) can be distinguished by how the gap forms in the electronic spectral function, which is momentum-selective in the former case and momentum-independent in the latter case~\cite{PhysRevLett.132.236504}.

Following Ref.~\cite{PhysRevLett.132.236504}, we characterize the Mott transition through the simultaneous disappearance of the low-energy coherent peak in the electronic spectral function $A_{\bf k}(\omega)$
at the nodal (${\text{N}=(\pi/2,\pi/2)}$) and anti-nodal (${\text{AN}=(0,\pi)}$) points of the Fermi surface. 
At ${V=0}$, the real-time ${D\text{-}GW}$ study of the Hubbard model predicts Mott-transition at ${U_{c}=6.7}$. Usually, considering the non-local interaction $V$ results in an effective screening of the on-site potential $U$, which leads to an increase of the critical interaction $U_{c}$ for the Mott transition with increasing $V$ (see, e.g, Refs.~\cite{PhysRevB.66.085120, PhysRevB.87.125149, PhysRevB.90.195114}). 
However, Fig~\ref{fig:fig1} shows that the obtained phase boundary of the Mott transition is nearly independent of the value of $V$. We attribute this result to the lack of vertex corrections neglected in the simplified version of the ${D\text{-}GW}$ method. In addition to screening $U$, the nearest-neighbor interaction $V$ 
on a square lattice favors the formation of a charge-ordered (CO) phase.
The phase transition to the CO state can be detected by the divergence of the static charge susceptibility ${X^{d}_{{\bf q}}}$ at the ${{\bf q} = (\pi,\pi)}$ point. 
At small values of $U$ the phase boundary of the CO phase predicted by ${D\text{-}GW}$ matches with the mean-field result ${V = U/4}$ (dashed line).
At larger interactions, the phase boundary is shifted above this line, in agreement with previous studies.

\begin{figure}[t!]
\centering
\includegraphics[width=0.7\linewidth, trim={10pt 10pt 10pt 10pt}, clip]{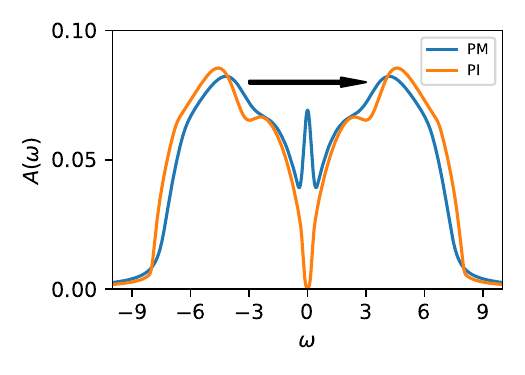}
\caption{Equilibrium electronic spectral functions of paramagnetic metals (PM) and paramagnetic Mott insulators (PI). The Figure is taken from Ref.~\cite{vglv-2rmv}. \label{fig:optics}}
\end{figure}

Let us focus on one of the most challenging regimes for theoretical analysis, namely the region  near the Mott and CO phase transitions. In this regime, local electronic correlations exhibit a non-perturbative character. 
Additionally, the system displays strong non-local charge and spin fluctuations on the metallic side.
To determine the effect of spatial collective electronic fluctuations, we calculate the local electronic spectral function ${A(\omega) = \frac{1}{N_k}\sum_{\bf k}A_{\bf k}(\omega)}$ for the two points depicted in Fig.~\ref{fig:fig1} by the circles.  
The brown circle represents the correlated metal (${U=6.25}$, ${V=2.3}$), and the purple one corresponds to the narrow-gap Mott insulator (${U=7.25}$,${V=2.3}$); see Fig.~\ref{fig:optics} for the corresponding equilibrium spectral functions. 
As expected, a three-peak structure with lower and upper Hubbard bands, along with a quasi-particle peak near the Fermi level, known as the Abrikosov-Suhl resonance~\cite{Abrikosov1965} in correlated metals, distinguishes them from Mott insulators, where a narrow gap near the Fermi level is identified. As we further focus on the effect of spin fluctuations, the interaction parameters are chosen to make short-range spin fluctuations more significant than the charge ones.

\begin{figure}[b!]
\centering
\includegraphics[width=1\linewidth]{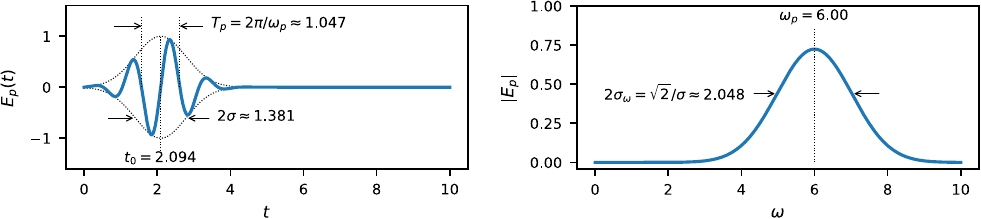} 
\caption{Time dependent electrical field $E_{p}(t)$ of the applied laser pump in time (left panel) and its frequency power spectra (right panel). The Figure is taken from Ref.~\cite{vglv-2rmv}.
\label{fig:pulse_new}}
\end{figure}

\subsubsection{Photo doping}
\label{sec:photo-doping}

\paragraph{Pump protocol:}

The electrical field of the applied laser pulse is directed along the lattice diagonal and has the form
\begin{equation}
    \mathbf{E}_{p}(t) = 
    \hat{n}_{xy} \cdot 
    E_0 \sin( \omega_p [t - t_0] ) \, e^{-\frac{(t - t_0)^2}{2\sigma^2}}
\end{equation}
with amplitude $E_0 = 1$, frequency $\omega_p = 6$, a Gaussian amplitude modulation centered at time $t_0 = 2.094$ with standard deviation $\sigma = 0.690$, and the unit vector $\hat{n}_{xy}$ in the positive x and y diagonal direction. This gives a period $T_p = 2\pi / \omega_p \approx 1.047$ and standard deviation $\sigma_\omega = 1/\sqrt{2 \sigma^2} \approx 1.024$ in frequency, see Fig.~\ref{fig:pulse_new}. Tuning the frequency ${\omega_p \approx U}$ to align with the local Hubbard interaction \(U\) centers the power spectra of the pump on the resonant excitation of electrons from the lower to the upper Hubbard band, as shown by the horizontal arrow in Fig.~\ref{fig:optics}. This alignment ensures that the energy quanta absorbed by the electrons, which are accelerated by the electric field, are sufficient to create local doubly occupied sites (doublons) and local empty sites (holons). In addition to the incoherent excitations, the electric pulse also influences the low-energy quasiparticle excitations commonly found in metals. These low-energy excitations can be observed in the single-particle spectral functions, as illustrated in Fig.~\ref{fig:optics}. 

We set the pulse amplitude ${E_0 = 1}$, which induces a density of 1.5\% for holons and doublons in the photo-excited state for the chosen metallic and Mott-insulating initial states through a non-linear process. The doublon density is defined by the change of its value from initial equilibrium value $\Delta D(t)= \frac{D(t=4)-D(t=0)}{D(t=0)}$. There are two primary excitation mechanisms responsible for producing doublon-holon excitations under an ac-field drive, both governed by the Keldysh parameter ${\gamma \equiv \frac{\omega_{p}}{E_0 \xi}}$, where $\xi$ characterizes the spatial correlation length of doublon-holon pair~\cite{Keldysh}. For the parameters of ${E_0}$ and ${\omega_{p}}$ selected in this study, the multi-photon absorption process (where ${\gamma \gg 1}$) is more dominant than quantum tunneling (where ${\gamma \ll 1}$). In the former case, the production rate of doublons demonstrates a power-law dependence on the electric field strength $E_0$. Additionally, the density of photoexcited doublons and holons in the multi-absorption process exhibits a momentum-dependent distribution, with doublons and holes primarily created at the edges of the Mott gap~\cite{PhysRevB.86.075148}.

\paragraph{Energy conservation in the photo-excited state:}

We analyze many-body theory in the context of an electric pulse by examining conservation laws. Among these, the most significant is the total energy, which is crucial for understanding the relaxation dynamics of photo-excited charge carriers in a closed quantum system. The relationship between the total energy $E(t)$ and an applied electric field $\mathbf{E}_{p}(t)$ is described by the equation ${\frac{dE(t)}{dt} = j(t) \cdot \mathbf{E}_{p}(t)}$, where $j(t)$ is the photo-induced electric current in the system. In the absence of an electric pulse, the total energy remains constant over time, as indicated by ${\frac{dE(t)}{dt} = 0}$. 

\begin{figure}[b!]
\centering
\includegraphics[width=1\linewidth]{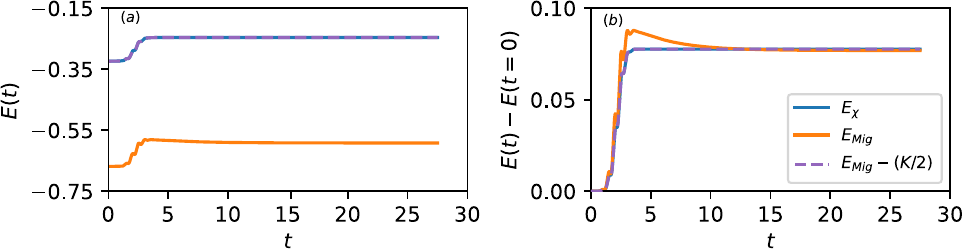}
\caption{(a) The total energy of the photo-doped metals within DMFT is plotted in the upper panel, and their change from the initial equilibrium state is plotted in the lower panel. The Figure is taken from Ref.~\cite{vglv-2rmv}.  
\label{fig:total_energies_dmft}}
\end{figure}

DMFT is known to be a conserving approximation, as described by Baym and Kadanoff, where the many-body self-energy is connected to the Green's function through the derivative of the Luttinger-Ward function~\cite{Baym61}.
We examined the conservation of the total energy in DMFT, as illustrated in Fig.~\ref{fig:total_energies_dmft}(a). 
As anticipated, the total energy calculated from the two-particle susceptibility, denoted as ${E_{\chi}}$, reached a constant value immediately after the pulse (for ${t < 4}$), confirming energy conservation. 
The total energy calculated from the electronic self-energy, labeled \(E_{Mig}\), also appears to be conserved after the pulse. However, there is a notable difference in the magnitude of the total energy derived from these two approaches. 
In Fig.~\ref{fig:total_energies_dmft}(b), we further plot the change in the total energy from the initial value  for clear identification of dynamics that occurs immediately after the pulse. 
We find that the quantity \(E_{Mig}\) violates energy conservation right after the pulse but maintains conservation for extended periods, while the ${E_{\chi}}$ becomes constant immediately when the pulse is switched of. 
DMFT is a conserving theory by construction, and both quantities should therefore be conserved and yield similar magnitudes.
However, our numerical calculations are based on the approximate (NCA) solver for the impurity problem, which seems to partially conserve the energy \(E_{Mig}\), leading to a different value compared to \(E_{\chi}\).
Empirically, we found that by subtracting half of the kinetic energy (\(K/2\)) from \(E_{Mig}\), we achieve perfect agreement between \(E_{\chi}\) and \(E_{Mig}\). 
From this result one can conclude that the violation of energy conservation stems from the NCA approximation, which apparently
provides a more accurate result for the energy when obtained through the susceptibilities rather than from single-particle quantities.
Using higher-order impurity solvers to restore the missing piece of self-energy in \(E_{Mig}\) could cure the problem, but this approach increases the numerical expense and complicates efforts to reach long-time scales. A similar discrepancy between \(E_{\chi}\) and \(E_{Mig}\) has also been observed for Mott-insulators within DMFT.

A conserving approximation for the ${D\text{-}GW}$ theory is possible, because in this approach the single (self-energy) and two-particle (polarization operator) quantities are also obtained from the functional ${\Phi[\tilde{G},\tilde{W}]}$ that corresponds to the partially bosonized dual action~\eqref{S_DB_fin}.
Since the transformation from the dual to the original space is exact, this should also provide a conserving description for the initial system~\cite{PhysRevB.79.045133}. 
However, there is an open question, which approach for calculating the energy should be used in this case.
Above, we have found that within the NCA framework the \(E_{\chi}\) approach to calculating the energy is more accurate.
However, on the contrary to DMFT, the self-energy in ${D\text{-}GW}$ has a non-local contribution.
This means that the consistent calculation of the two-particle susceptibilities should account for the vertex corrections corresponding to the collective fluctuations accounted for in the self-energy.
This procedure is prohibitively expensive numerically, and in the ${D\text{-}GW}$ approach, as well as in the other time-dependent diagrammatic methods, these corrections are neglected, and the polarization operator has a simple ``bubble'' form~\eqref{eq:Pi_dual_para}.
Therefore, the Galitskii-Migdal ($E_{Mig}$) way of calculating the energy through the self-energy is expected to be more accurate, but in this case we again encounter the issue that the NCA approximation misses some crucial contribution to the self-energy.
Within DMFT, this contribution is likely to be the half of the kinetic energy, as has been found above.
However, there is no grantee that the same contribution is missing in the ${D\text{-}GW}$ approach due to a non-trivial relation between the lattice and dual quantities in the theory.

\begin{figure}[t!]
\centering
\includegraphics[width=1\linewidth]{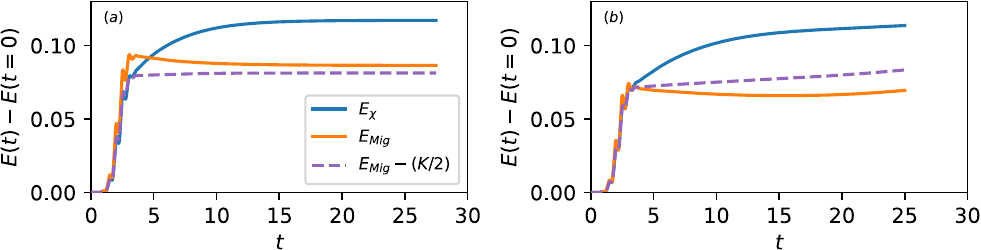}
\caption{The total energy change from the initial equilibrium state is plotted within ${D\text{-}GW}$ for metals and Mott-insulators in the upper and lower panels, respectively. The Figure is taken from Ref.~\cite{vglv-2rmv}. 
\label{fig:total_energies_DGW}}
\end{figure}

To investigate this, we examine total energy conservation, as illustrated in Fig.~\ref{fig:total_energies_DGW} for (a) metals and (b) Mott insulators. In contrast to DMFT, neither \(E_{\chi}\) nor \(E_{\text{Mig}}\) conserves energy immediately after the pulse. However, both approaches do conserve energy over longer time scales. 
By subtracting half of the kinetic energy from \(E_{\text{Mig}}\), we find that total energy conservation is maintained over an extended period in metals. 
In contrast, Mott insulators show a slight drift in long-term energy conservation (for ${t \geq 20}$) due to a drift in electron density, which can not be controlled with Matsubara and real-time grid sizes. Therefore, we cannot definitively determine whether the violation of energy conservation in ${D\text{-}GW}$ is inherited from the NCA approximation or arises from the diagrammatic approximation, including the instantaneous approximation for the three-point vertex $\Lambda$.

\paragraph{Relaxation dynamics of total energy components:}

Energy relaxation in photo-excited metals and narrow-gap Mott insulators typically occurs through intraband (within the Hubbard bands) and interband (between lower and upper Hubbard bands) relaxation processes involving charge carriers. A common example of these processes includes electron-electron scattering, scattering of electrons with charge and spin degrees of freedom, and impact ionization phenomena \cite{Yuta2023}. The latter refers to an interband relaxation process distinguished by increased doublon density, even after a pulse. A high-energy doublon can create additional low-energy doublon-holon pairs through particle-particle scattering. In contrast, intraband electron-electron scattering and scattering involving low-energy degrees of freedom do not change the doublon density and primarily lead to thermalization.

In DMFT, the potential energy is directly proportional to the number of doublons, which relax according to various processes based on the initial photo-excited state. Since DMFT does not consider non-local fluctuations, the potential energy gain in the photo-excited state is compensated by the loss of electronic kinetic energy. Hence, in the pure electronic models, they both relax simultaneously. However, this may not be the case in ${D\text{-}GW}$ theory, where non-local charge and spin degrees of freedom can affect their dynamics independently. 

\begin{figure}[t!]
\centering
\includegraphics[width=1\linewidth]{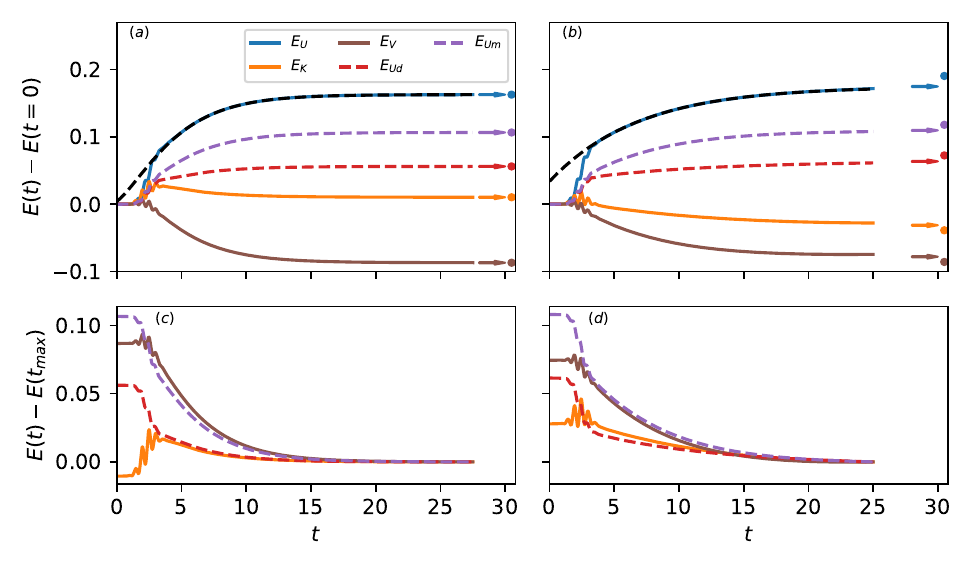}
\caption{The relaxation of total-energy components of the photo-doped metals(a) and Mott-insulators (b) is compared in ${D\text{-}GW}$ for a given excitation protocol. The arrows are the residual value of exponential fit at  $t=\infty$ for a given energy component. The circles are the unique thermal equilibrium state values obtained from the total energy check. The black dashed line is the double (single) exponential fit of $E_{U}$ in metals (Mott-insulator). The change in total-energy components from their maximum simulated time value is plotted in panels (c) and (d) for metals and Mott-insulators, respectively. The red (violet) dashed line in the lower panel represents -$E_{Ud}$ (-$E_{Um}$). The Figure is taken from Ref.~\cite{vglv-2rmv}.
\label{fig:energy_component_DGW}}
\end{figure}

To address this question, we plot the total energy components of metals and Mott insulators obtained from ${D\text{-}GW}$ in Fig.~\ref{fig:energy_component_DGW}\,(a) and (b), respectively. This analysis also illustrates how the system, subjected to the same excitation protocol but starting from different initial states, relaxes after the pulse. The initial observation is that all components of the total energy exhibit oscillations with a frequency of $\omega_p$ and are damped by the end of the pulse (for $t \sim 4$). For metals and Mott insulators, the local potential energy $E_U$, proportional to the number of doublons, increases during the pulse. This confirms that an electric pulse injects photo-excited charge carriers, known as doublons, into the system. In metals, we observe an exponential increase in $E_U$ after the pulse, followed by a saturation phase over a more extended period. This behavior is a smooth crossover of the doublon relaxation from inter-band to intra-band processes at $t > 15$. In contrast, we do not observe such a crossover in the Mott insulator, which likely happens at much longer timescales ($t > 25$). 

On the other hand, the electronic kinetic energy $E_K$ in the metals also increases during the pulse, but the increase is less significant compared to $E_U$. This delocalization behavior can be attributed to the disappearance of quasi-particles due to the rise in the effective temperature of the system, which can be identified in the single-particle spectral functions. After the pulse, the kinetic energy slightly decays and eventually reaches a constant value over a longer period. However, in Mott insulators, the kinetic energy during the pulse oscillates around its initial value before starting to decay. In the presence of the Mott gap in the initial state, the electrons in the Brillouin zone experience an oscillating electric field. As a result, the kinetic energy of Mott insulators neither localizes nor delocalizes during the pulse. 

Creating a single photo-excited doublon state increases the local potential energy by Hubbard interaction \(U\). On the other hand, the non-local interaction $V$ energetically favors the creation of doublon-holon pairs at neighboring lattice sites rather than random ones. Such an excitation disrupts the non-local density-density interaction among neighboring lattice sites (utmost four sites), which usually compensates for the increase in local potential energy.
As expected, the non-local potential energy \(E_V\), as shown in Fig.~\ref{fig:energy_component_DGW}, decreases during the pulse for both metals and Mott insulators to offset the local potential energy increase. 
Additionally, the close-packed doublon-holon pattern favored by $V$ strongly suppresses magnetic excitations in this region.
Consequently, we find that \(E_V\) mirrors the dynamics of local potential energy related to spin excitations \(E_{Um}\) as illustrated in Fig.~\ref{fig:energy_component_DGW}\,(c) and (d).
On the other hand, the creation of local charge excitations blocks the electron momentum, so \(E_{Ud}\) tracks the dynamics of electronic kinetic energy \(E_K\) for both metals and Mott insulators [see Fig.~\ref{fig:energy_component_DGW}(c) and (d)]. 
Since the local charge and spin potential energy are proportional to the Hubbard interaction \(U\), the number of local spin excitations created in the transient state is nearly double that of the local charge excitations. 

We extract the relaxation time scales of the total energy components using a function that characterizes relaxation dynamics over an extended time scale. In metals, we find that the relaxation dynamics can be well described by a double exponential function:
\begin{align}
f(t) = a(t=\infty) + b \exp(-t/\tau_{h}) + c \exp(-t/\tau_{l})
\end{align}
Fig.~\ref{fig:energy_component_DGW}\,(a) illustrates it for local potential energy. The relaxation dynamics involve two-time scales, denoted as \(\tau_{h}\) and \(\tau_{l}\). The microscopic origin of this two-time scale behavior stems from the interaction of a photo-excited high-energy doublon during the impact ionization process. As the high-energy doublon loses some of its kinetic energy, it creates an additional low-energy doublon-hole pair through particle-particle scattering, represented by the reaction: $\text{doublon}_{\text{high}} \rightarrow \text{doublon}_{\text{low}} + \text{doublon}_{\text{low}} + \text{holon}_{\text{low}}$. This particle-particle scattering mechanism accounts for the distinct relaxation time scales observed in the system. 

A similar pair-creation process can occur for holons in the photo-excited state. Due to particle-hole symmetry, the net effect of the impact ionization process is that each high-energy double-hole pair produces three low-energy double-hole pairs. Since the impact ionization process involves high-energy and low-energy doublons, one can expect a two-time scale behavior in the relaxation dynamics. These time scales are denoted by \(\tau_{h}\) and \(\tau_{l}\), corresponding to high-energy and low-energy doublons (holons), respectively. 

\begin{table}[t!]
\centering
\caption{Relaxation time scales of ${D\text{-}GW}$ energy components}~\\[-0.1cm]
\label{tab:DGW_taus}
\begin{tabular}{l  c  c  c  c  }
\hline\hline
& \multicolumn{2}{c} {Metal} & 
\multicolumn{2}{c} {Mott-insulator} \\
\hline
   &  $\tau_{h}$ & $\tau_{l}$ & $\tau_{h}$  & $\tau_{l}$ \\
\hline
$E_K$ & 1.20 & 3.58 & 9.40 & - \\
$E_{Ud}$ & 1.23 & 3.33 & 9.10 & - \\
$E_V$ & 2.00 & 3.14 & 6.00 & - \\
$E_{Um}$ & 2.04 &  3.13 & 6.00 & - \\
$E_{U}$ & 2.02 & 3.14 & 6.80 & - \\
\hline\hline
\end{tabular}
\end{table}

In Mott insulators, a single exponential function ${f(t) = a(t=\infty) + b \exp(-t/\tau_{h})}$ is sufficient to explain the relaxation dynamics, as shown in Fig.~\ref{fig:energy_component_DGW}(b). We believe that the lack of long-term data prevents us from determining the time scale \(\tau_{l}\) associated with low-energy doublons (holons). We summarize these time scales for metals and Mott insulators in Table~\ref{tab:DGW_taus}. The relaxation times of all energy components in metals and Mott insulators are divided into two groups: one comprising \(E_K\) and \(E_{Ud}\), while the other includes the remaining three components. It is worth noting that the relaxation time scales \(\tau_{h}\) for Mott insulators are an order of magnitude longer than those for metals.  

\begin{figure}[b!]
\includegraphics[width=1\linewidth]{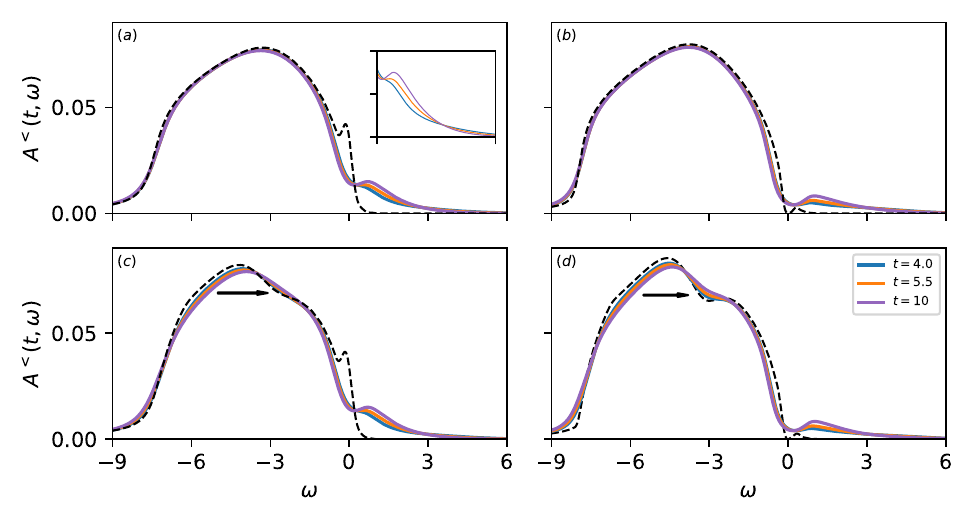}
\caption{The early (immediately after the pulse) time evolution of the occupied density of states for metals (left panels) and Mott-insulators (right panels) is compared between ${D\text{-}GW^{d}}$ (a,b) and ${D\text{-}GW}$ (c,d). The inset zooms in on the evolution of the occupied spectrum above the 
Fermi level. The dashed black lines indicate the occupied density of states of the initial equilibrium state. The horizontal arrows indicate the direction of transfer of spectral weight at the lower Hubbard band in the ${D\text{-}GW}$ density of states. The Figure is taken from Ref.~\cite{vglv-2rmv}.
\label{fig:spectrum_early}}
\end{figure}

\paragraph{Time-dependent spectral functions:}

\begin{figure}[t!]
\centering
\includegraphics[width=\linewidth]{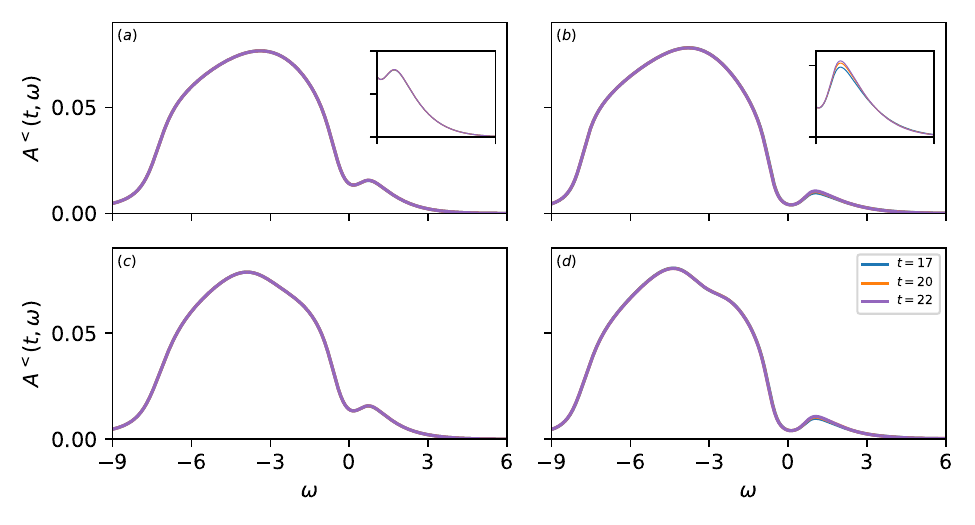}
\caption{The time evolution of the occupied density of states
at longer times ($t>15$) is compared between ${D\text{-}GW^{d}}$ (a,b) and ${D\text{-}GW}$ (c,d) for metals (left panels) and Mott insulators (right panels). The inset shows the evolution of the occupied spectrum above the Fermi level. The Figure is taken from Ref.~\cite{vglv-2rmv}.
\label{fig:spectrum_lat}}
\end{figure}

The integrated occupied density of states ${A^{<}(t,\omega)}$ above the Fermi level is closely related to the doublon number. Therefore, understanding their time evolution in the photo-excited state helps identify the intra- and interband relaxation regimes and their associated spectral features. In Fig.~\ref{fig:spectrum_early}, the time-resolved ${A^{<}(t,\omega)}$ for metals and Mott insulators is plotted immediately after the electric pulse (${t>4}$) for both ${D\text{-}GW}$ and a diagrammatic approximation ${D\text{-}GW^{d}}$. The latter includes only non-local charge fluctuations in the ${D\text{-}GW}$ electronic self-energy while disregarding spin fluctuations. This is achieved by independently setting the renormalized interactions \(W^{d/m}\) in the charge (d) and spin (m) channels to zero in Eq.~\eqref{eq:Sigma_dual_para}. This approach allows us to study the effects of spin fluctuations \mbox{D-$GW^{m}$} or charge fluctuations ${D\text{-}GW^{d}}$ on spectral functions separately. 

Immediately after the pulse, the photo-excited charge carriers in metals and Mott insulators relax primarily through an impact ionization phenomenon (see Fig.~\ref{fig:energy_component_DGW}). In the initial equilibrium state, the occupied density of states below the Fermi level, the lower Hubbard band, is filled. In contrast, the upper Hubbard band above the Fermi level is nearly unoccupied. A prominent quasi-particle peak at the Fermi level distinguishes a metallic state [Fig.~\ref{fig:spectrum_early}(a) and (c)] from a Mott insulating state [Fig.~\ref{fig:spectrum_early}(b) and (d)]. The applied electric pulse causes part of the spectrum to move from the lower to the upper Hubbard band. The low-energy quasi-particle peak completely melts in metals, whereas the gap fills up in Mott insulators. The time evolution of the photo-excited state after the pulse is similar for both metals and Mott insulators. 

A key characteristic of the impact ionization process in the occupied density of states is the transfer of spectral weight. This transfer occurs from the upper edge of the upper Hubbard band (approximately at ${\omega \sim 4}$) to the lower edge of the upper Hubbard band (around ${\omega \sim 1}$). This behavior is illustrated in the inset of Fig.~\ref{fig:spectrum_early}. Additionally, the ratio of spectral gain (integrated spectrum) at ${\omega_{gain} \sim 1}$ is nearly three times the spectral loss at ${\omega_{loss} \sim 4}$. This observation confirms that nearly three low-energy doublon-holon pairs are produced from a single high-energy doublon-holon pair~\cite{PhysRevB.90.235102}. Notably, in the ${D\text{-}GW}$ calculations, aside from the effect of impact ionization, spectral weight is transferred from the upper edge of the lower Hubbard band to its lower edge. The feature indicated by the arrow in the lower Hubbard band, as shown in Fig.~\ref{fig:spectrum_early}(c) and (d), which is absent in \mbox{D-$GW^{\rm d}$} states, can be attributed to a relaxation process involving magnetic fluctuations.

We further examine the time evolution of ${A^{<}(t,\omega)}$ over a longer duration, where the intraband relaxation of charge carriers dominates. In the case of metals, the ${A^{<}(t,\omega)}$ obtained from ${D\text{-}GW}$ and ${D\text{-}GW^{d}}$ approaches, as shown in Fig.~\ref{fig:spectrum_lat} (a) and (c), does not change further over time. The constant distribution of spectral weight at the upper Hubbard band (see the inset of Fig.~\ref{fig:spectrum_lat} (a)) indicates that the intraband relaxation does not alter the number of doublons. In Mott insulators, since the time scales for intraband relaxation have not yet been reached, a small fraction of spectral weight continues to transfer from high-energy to low-energy states above the Fermi level, as illustrated in the inset of Fig.~\ref{fig:spectrum_lat} (b). Interestingly, a relaxation process involving magnetic fluctuations at the lower Hubbard band in D-GW is absent over longer periods. They are only active immediately after the pulse (interband relaxation regime) [see Fig.~\ref{fig:spectrum_early}(c,d)]. 

In small cluster calculations, it has been observed that AFM fluctuations do not favor impact ionization~\cite{PhysRevB.102.245125}. This is because a high-energy doublon transfers its excess kinetic energy to the underlying spin background instead of generating an additional doublon-holon pair. However, the situation is less clear for extended systems. The time evolution of the occupied density of states in ${D\text{-}GW}$ indicates that impact ionization remains favorable even in magnetic fluctuations. We identified signs of energy transfer to the underlying antiferromagnetic spin fluctuations during impact ionization. This is evidenced by spectral weight transfer at the edges of the lower Hubbard band immediately after the pulse. This phenomenon can be measured using time-resolved photoemission spectroscopy. It has been demonstrated through equilibrium studies that the kinks at the lower edges of the Hubbard band are linked to electromagnetic coupling, as evidenced by various theoretical approaches \cite{PhysRevX.10.041023, PhysRevX.10.041023,wang2020emergence,PhysRevLett.108.076401,PhysRevB.92.075119, PhysRevLett.134.016502}. We have observed this feature in the equilibrium spectra presented in Fig.~\ref{fig:optics}(b).

\begin{figure}[b!]
\centering
\includegraphics[width=1\linewidth]{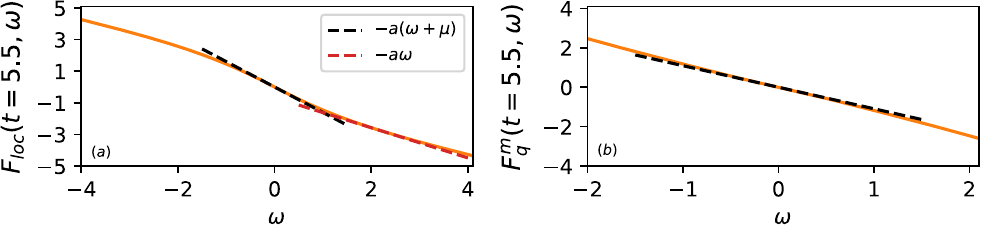}
\caption{The function related to the fluctuation-dissipation theorem is plotted for local electronic spectrum  (a) and ferromagnetic fluctuations at $\Gamma$ point in momentum space. The dashed lines are the linear fits of $F_{loc}(t,\omega)$ and $F^{m}_{q=\Gamma}(t,\omega)$. The Figure is taken from Ref.~\cite{vglv-2rmv}.
\label{fig:F_D}}
\end{figure}

\paragraph{Thermalization of photo-excited states:}

According to the eigenstate thermalization hypothesis, a closed quantum many-body system driven out of equilibrium will eventually reach thermalization over a longer period~\cite{PhysRevE.50.888, Deutsch_2018}. However, the evolution of a photo-excited state with non-local collective electronic fluctuations toward thermalization is not yet fully understood. We investigate thermalization by comparing the properties of a time-dependent state to a uniquely defined thermal equilibrium state. This thermal equilibrium state is determined by the condition that its energy matches the energy of a photo-excited state, given by the equation $E(t>4) = \frac{\text{Tr}[e^{-\mathcal{H}/T_{\text{eff}}} \mathcal{H}]}{\text{Tr}[e^{-\mathcal{H}/T_{eff}}]}$. This procedure results in a thermal equilibrium state with ${T_{\text{eff}} = 0.85}$ for metals and 0.95 for Mott insulators. The components of total energy obtained from these thermal equilibrium states are plotted in Fig.~\ref{fig:energy_component_DGW} for both metals and Mott insulators. The perfect agreement of the residual energy components $a(t=\infty)$ with the thermal equilibrium state confirms the thermalization of the photo-excited metals, as shown in Fig.~\ref{fig:energy_component_DGW}(a). In contrast, the discrepancy between the thermal values and the photo-excited state highlights the pre-thermal nature of the transient state for Mott insulators, as observed in DMFT~\cite{PhysRevB.90.235102}.

\begin{figure}[t!]
\centering
\includegraphics[width=1\linewidth]{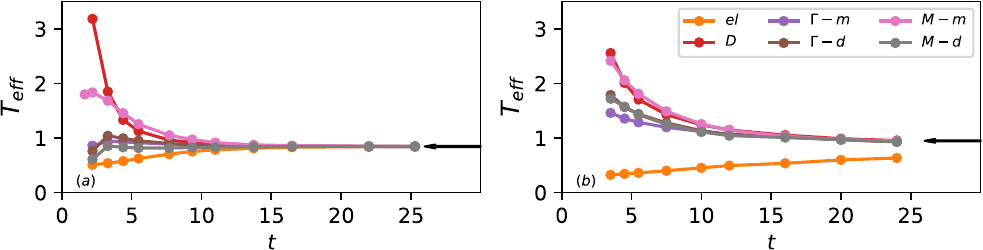}
\caption{The time-dependent effective temperature of electrons ($el$), doublons ($D$), charge ($d$) and spin fluctuations ($m$) are obtained from the slope of a linear fit to $F_{loc}(t,\omega)$, $F^{d/m}_{q}(t,\omega)$ for metals (a) and Mott-insulators (b). The charge and spin fluctuations extracted at two different $q=(\Gamma, M)$ points in momentum space. The black arrow is the effective temperature of the unique thermal state obtained from the total energy check. The Figure is taken from Ref.~\cite{vglv-2rmv}.
\label{fig:T_eff}}
\end{figure}

The total energy procedure determines whether the photo-excited electronic state reaches thermal equilibrium over time. To understand how and when a system with non-local fluctuations attains a thermal state, we analyzed the dynamics of the effective temperature for electrons and charge and spin fluctuations near the Fermi level (with $\omega \sim 0$). In Fig.~\ref{fig:F_D}, we present the distribution function for the local electronic spectral function $F_{loc}(t,\omega)$ and spin susceptibility at the ferromagnetic wave vector $F^{m}_{q=\Gamma}(t,\omega)$ defined in Eq.~\eqref{hkjwdlnsks}.
The electronic distribution function $F_{loc}(t,\omega)$ in Fig.~\ref{fig:F_D}(a) shows linear behavior around the Fermi level and transitions to a shifted linear behavior at high frequencies (between $\omega \sim 1.5$ and $4$). This latter behavior corresponds to the frequency range of spectral weight transfer in $A^{<}(t,\omega)$ at the upper Hubbard band, as shown in Fig.~\ref{fig:spectrum_early}. As mentioned earlier, the total spectral weight in this frequency range indicates the doublon number. Therefore, fitting a (shifted) linear function at the Fermi level allows us to determine the effective temperature of the (doublon) quasi-particles. Due to symmetric conditions, holons have the same temperature as doublons. This can be similarly derived from a linear fit of $F_{loc}(t,\omega)$ below the Fermi level. In Fig.~\ref{fig:F_D}(b), a linear fit of ${F^{m}_{q=\Gamma}(t,\omega)}$ 
near $\omega=0$ yields the effective temperature of FM fluctuations at the specified time scale. The same procedure has also been applied to charge fluctuations at the \(\Gamma\) and $M$ points, where the latter represents the AFM order in a given channel. The corresponding effective temperatures for metals and Mott insulators are shown in Fig.~\ref{fig:T_eff}(a) and (b), respectively.

On the other hand, the electric pulse injects a finite amount of energy into the photo-excited system, distributed among the quasi-particles, doublons, charge carriers, and spin fluctuations. This process raises their temperature from the initial equilibrium temperature of ${T = 1/\beta = 0.16}$. The increase in electronic kinetic energy within metals (see Fig.~\ref{fig:energy_component_DGW}(a)) during the electric pulse is attributed to a rise in the quasi-particle temperature, as illustrated in Fig.~\ref{fig:T_eff}(a). This increase in temperature, in turn, leads to a suppression of the quasi-particle lifetime. This effect is also reflected in the time evolution of the occupied density of states near the Fermi level (see Fig.~\ref{fig:spectrum_early}(a) and (c)), where the spectral weight is significantly diminished during the pulse.  

In addition to the dynamics of quasi-particles, an electric pulse triggers the creation of photo-excited doublons and AFM fluctuations, which rapidly increases their temperature. In metals (see Fig.~\ref{fig:T_eff}(a)), other collective fluctuations, such as charge and FM, do not respond similarly. This discrepancy is primarily due to itinerant electrons, which significantly contribute to AFM fluctuations in comparison to FM fluctuations. In Mott insulators, electronic correlations are primarily local, independent of momentum. As a result, an electric pulse activates all collective dynamics, with doublons and AFM fluctuations remaining the most prominent, as illustrated in Fig.~\ref{fig:T_eff}(b). 

Over time, the doublons and AFM fluctuations in metals decrease temperature by effectively transferring their energy to the quasi-particle excitations. This process further raises the temperature of the damped quasi-particles. This energy transfer is nearly complete by the time scale of approximately ${t \sim 17}$, and all fluctuations, including charge and FM, equilibrate at the same temperature. It has been illustrated in Fig.~\ref{fig:T_eff}(a). The perfect alignment of this temperature with the unique thermal state (indicated by the black arrow in Fig.~\ref{fig:T_eff}(a)) around ${t \sim 17}$ confirms that this is the typical time scale for thermalization in metals.

In Mott insulators (see Fig. \ref{fig:T_eff}(b)), all collective fluctuations transfer their energy to quasi-particle excitations as the temperature decreases, eventually reaching a specific temperature around ${t \sim 20}$. This temperature corresponds to a distinct thermal state (indicated by the black arrow), but it does not align with the temperature of the quasi-particles. In the Mott state, there are initially no quasi-particles present. However, when an electric pulse is applied, it shifts the spectral weight from the lower to the upper Hubbard band, effectively filling the Mott gap (see Fig. \ref{fig:spectrum_early} and \ref{fig:spectrum_lat}).

The finite spectral weight at the Fermi level in the transient state indicates a behavior similar to that of a bad-metal, characterized by highly damped quasi-particles. An ineffective energy transfer between collective fluctuations and quasi-particle excitations results in a temperature mismatch, which slows down the thermalization process in Mott insulators. Thermalization occurs when the temperatures of the collective fluctuations and quasi-particles become equal. The inability to reach a single temperature state suggests the system does not entirely lose its initial correlations within the simulated time frame. As a result, much longer durations are needed to observe thermalization in Mott insulators.

\subsection{Non-local correlation effects in conductivity of the Hubbard model}

Conductivity is one of the most direct probes of electronic systems, yet its theoretical description remains challenging in the presence of strong non-local correlations. 
DMFT is one of the most widely used method for addressing transport properties of correlated systems.
However, DMFT neglects spatial correlations, which frequently leads to notable discrepancies between DMFT predictions with experimental transport data for real materials, such as the resistivity of various ruthenate compounds~\cite{PhysRevLett.116.256401, PhysRevMaterials.7.093801}.
The non-local correlations can significantly influence transport properties by modifying the electronic spectral function and giving rise to complex multi-electron scattering processes, known as vertex corrections, which can strongly impact the conductivity. 
In contrast, non-local vertices, which are of key
importance, cannot be captured by the local DMFT framework and are challenging to include into the theory.

In Ref.~\cite{Conductivity}, we systematically investigated the impact of non-local correlations on the conductivity of the single-band Hubbard model using the ${D\text{-}GW}$ approach.
This method is ideally sited for this purpose, as it enables a consistent real-time description of local electronic correlations and long-range collective charge and spin fluctuations across weak- and strong-coupling regimes, offering direct access to single- and two-particle observables in real frequencies. By focusing on the region near the Mott transition, we find that the impact of non-local correlations on the conductivity differs between the correlated metallic and Mott insulating phases. 
We demonstrate that incorporating non-local correlations in both the electronic spectral function and vertex corrections is crucial for accurately describing the DC and optical conductivity in the metallic phase. 
The crossover between the metallic and Mott insulating phases is marked by a vanishing contribution of vertex corrections to the DC conductivity, although the DC conductivity itself remains finite at high temperatures. 
At the same time, non-local vertex corrections remain essential for describing the optical conductivity in the Mott insulating regime, despite the fact that this regime is dominated by local electronic correlations.

\begin{figure}[t!]
\centering
\includegraphics[width=0.7\columnwidth]{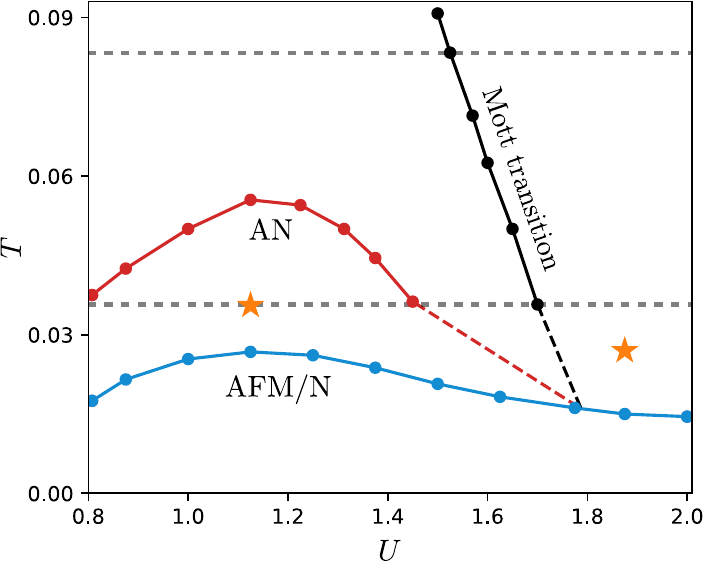}
\caption{Phase diagram of the half-filled single-band Hubbard model calculated using ${D\text{-}GW}$ in the $T$-$U$ plane. The blue curve represents the N\'eel transition and corresponds to the divergence of the spin susceptibility at the ${{\bf q} = (\pi, \pi)}$ point. The red curve indicates the gap opening at the AN point.
The Mott transition (black curve) is characterized by a simultaneous gap opening at the N and AN points 
and is 
extrapolated to low temperatures by the dashed black line. The red dashed line is the extension of AN curve according to findings in Ref.~\cite{PhysRevLett.132.236504}. 
The orange stars indicate the two points at which the calculations are performed in Fig.~\ref{fig:fig2} and Fig.~\ref{fig:fig3}\,(a,\,b).
The dashed gray lines depict the temperatures at which the the optical conductivity scans are performed in Fig.~\ref{fig:fig3}\,(c). The Figure is taken from Ref.~\cite{Conductivity}.  
\label{fig:phase}}
\end{figure}

In Ref.~\cite{Conductivity}, we considered the half-filled Hubbard model on a square lattice with the nearest-neighbor hopping ${t=0.25}$.
The electron Fermi surface (FS) of the half-filled model has a perfect nesting.
At weak coupling, this gives rise to the strong antiferromagnetic fluctuations of itinerant electrons.
Upon lowering the temperature, these fluctuations lead to a momentum-selective opening of a gap in the electronic spectrum. The gap appears first at the antinodal (${\text{AN} = (\pi,0)}$) point of the FS, then gradually extends across the FS, eventually reaching the nodal (${\text{N} = (\frac{\pi}{2},\frac{\pi}{2})}$) point, as the system undergoes the N\'eel transition to the ordered AFM state.
At large interaction strengths, the system undergoes a transition into a Mott insulating state, which is driven by the local electronic correlations.
The weak and strong-coupling limits are connected by a rather broad crossover regime with coexisting itinerant fluctuations and local correlations~\cite{PhysRevLett.132.236504}. 
The phase diagram of the Hubbard model obtained from ${D\text{-}GW}$ is shown in Fig.~\ref{fig:phase} and is similar to the one of Ref.~\cite{PhysRevLett.132.236504}. 
The AFM  transition is represented by a blue curve.
Above this curve, in the paramagnetic phase the red curve depicts the opening of the gap at the AN point.
The crossover to the Mott insulating state is shown by the black line and is characterized by a simultaneous opening of the gap at all ${\bf k}$-points of the FS~\cite{PhysRevLett.132.236504}.

The identified correlation effects are expected to strongly influence transport properties of the system.
The study of electronic transport is usually conducted within linear response theory, where conductivity $\sigma_{\alpha\beta}(t,t')$ is defined as the linear coefficient relating the induced current $\langle j_{\alpha}(t)\rangle $ to an applied electromagnetic probe field $A_{\beta}(t')$, where ${\alpha, \beta \in \{x, y, z\}}$ label the direction in real space. 
The current operator is defined by the relation $J(r)$=$-c\delta H/\delta A(r)$. Here, $c$ is the velocity of light, $H$ is the Hubbard Hamiltonian, and $A(r)$ is the vector potential. In the Peierls approximation, the electric current in the long wave-length limit is given by ${\langle j(t) \rangle = \langle \frac{1}{V} \int d^d r J(r) e^{iqr}\rangle_{q\rightarrow 0}}$ and $V$ is the number of ${\bf k}$-points in the Brillouin zone. In the momentum representation 
\begin{equation}
\langle j_{\alpha}(t) \rangle= \frac{-ie}{V}\sum_{{\bf k},\sigma} v^{\alpha}_{\bf k}(t) G^{<}_{{\bf k}\sigma}(t,t)\,,
\end{equation}
where the current vertex ${v^{\alpha}_{\bf k}(t) = \frac{1}{\hbar} \partial_{k_{\alpha}} {\varepsilon}_{{\bf k}-\frac{e}{\hbar c} {\bf A}(t)}}$. 
In case of a square lattice with nearest-neighbor hopping $t$, the electronic band dispersion within Peierl's approximation has the following form: ${\varepsilon_{k-\frac{e}{\hbar c} A(t)} = -2J*[cos(k_{x}-\frac{e}{\hbar c} A(t))+cos(k_{y}-\frac{e}{\hbar c} A(t))]}$.
Since we are interested in the linear current response to the weak probe field, we define the current-current correlation function called susceptibility as:
\begin{align}
{\chi_{\alpha \beta}(t,t') = \frac{\partial j_{\alpha}(t)}{\partial A_{\beta}(t')}\Big|_{A=0}}\,,
\label{eq:chi}
\end{align}
where $G^{<}_{\bf k}$ is the lesser component of the electronic Green's function. In the chosen gauge ${E(t)=-\partial_{t} A(t)}$, the susceptibility $\chi_{\alpha \beta}(t,t')$ is related to the optical conductivity $\sigma_{\alpha \beta}(t,t')$ by:   
\begin{equation}
\sigma_{\alpha \beta}(t,t') = -c \int^{\infty}_{t'} \chi_{\alpha \beta} (t,\bar{t}) d \bar{t}\,.
\end{equation}
In equilibrium, the conductivity \(\sigma_{\alpha \beta}(t,t')\) depends solely on the time difference and thus the frequency-dependent conductivity $\sigma_{\alpha \beta}(\omega)$ can be obtained through a straightforward Fourier transform:
\begin{align}
\text{Re} \left[\sigma_{\alpha \beta}(\omega)\right] = \int^{t_{max}}_0 \sigma_{\alpha \beta}(t - t') e^{i\omega (t - t')} \, d(t - t')\,, 
\end{align}
where $t_{max}$ is maximum simulated time. The susceptibility can be calculated by taking the derivative in Eq.~\eqref{eq:chi}, where the vector potential appears in both the velocity and Green's function. 
It gives a diagrammatic and paramagnetic contribution to the susceptibility:
\begin{equation}
 \chi_{\alpha \beta} (t,t') = \chi_{\alpha \beta}^{\rm dia} (t,t') + \chi_{\alpha \beta}^{\rm pm}(t,t')
\end{equation}
with 
\begin{equation}
\chi_{\alpha \beta}^{\rm dia} (t,t') = \frac{-ie}{V} \sum_{{\bf k},\sigma} \frac{\partial v^{\alpha}_{{\bf k}}(t)}{\partial A_{\beta}(t')} G^{<}_{{\bf k}\sigma}(t,t)~~~\text{and}~~~
\chi_{\alpha \beta}^{\rm pm} (t,t') = \frac{-ie}{V} \sum_{{\bf k},\sigma} v^{\alpha}_{{\bf k}}(t) \frac{\partial G^{<}_{{\bf k}\sigma}(t,t)}{\partial A_{\beta}(t')}\,.
\end{equation}
The diamagnetic contribution can be further simplified and reads:
\begin{equation}
\chi_{\alpha \beta}^{\rm dia} (t,t') = \frac{-i\chi_0}{\hbar V} \sum_{{\bf k},\sigma} \partial_{k_\alpha}  \partial_{k_\beta} \varepsilon_{{\bf k}} (t) G^{<}_{{\bf k}\sigma}(t,t) \delta(t-t')\,,
\end{equation}
where the prefactor ${\chi_0 = \frac{e^2}{\hbar c}}$. 
The paramagnetic contribution is found from the variation of the lattice Dyson equation: 
\begin{align}
\frac{\partial G_{{\bf k}\sigma}(t_1,t_2)}{\partial A_{\beta}(t)} 
=  - \iint \{dt'\} \,
G_{{\bf k}\sigma}(t_1,t'_1) \frac{\partial G^{-1}_{{\bf k}\sigma}(t'_1,t'_2)}{\partial A_{\beta}(t)}G_{{\bf k}\sigma}(t'_2,t_2)\,,
\end{align}
\begin{align}
\frac{\partial G^{-1}_{{\bf k}\sigma}(t_1,t_2)}{\partial A_{\beta}(t)} 
= \frac{\partial \left[G^{-1}_{0,{\bf k}\sigma}(t_1,t_2) - \Sigma_{{\bf k}\sigma}(t_1,t_2) \right]}{\partial A_{\beta}(t)}
= - \delta(t_1,t_2) \delta(t_1,t) \frac{e}{\hbar c} v^{\beta}_{\bf k}(t) - \frac{\partial \Sigma_{{\bf k}\sigma}(t_1,t_2)}{\partial A_{\beta}(t)}\,.
\end{align}
Considering the self-energy in the form: 
\begin{align}
\Sigma_{\bf k\sigma}(t_1,t_2) = i\iint dt_3\,dt_4 \sum_{\bf k',\sigma'} \gamma^{\sigma\sigma'}_{{\bf kk'}}(t_1,t_2,t_3,t_4)G_{{\bf k'}\sigma'}(t_3,t_4)
\label{eq:Sigma_SM}
\end{align}
and assuming that the vertex function does not change upon applying a small field, gives in the linear response approximation:
\begin{align}
\frac{\partial \Sigma_{{\bf k}\sigma}(t_1,t_2)}{\partial A_{\beta}(t)} = i\iint dt_3\,dt_4 \sum_{\bf k',\sigma'} \gamma^{\sigma\sigma'}_{{\bf kk'}}(t_1,t_2,t_3,t_4) \frac{\partial G_{{\bf k'}\sigma'}(t_3,t_4)}{\partial A_{\beta}(t)}\,.
\end{align}
Note that considering the self-energy in the exact Schwinger-Dyson form immediately identifies $\gamma$ as the exact two-particle irreducible vertex in the particle-hole channel.

\begin{figure}[t!]
\centering
\includegraphics[width=0.7\linewidth]{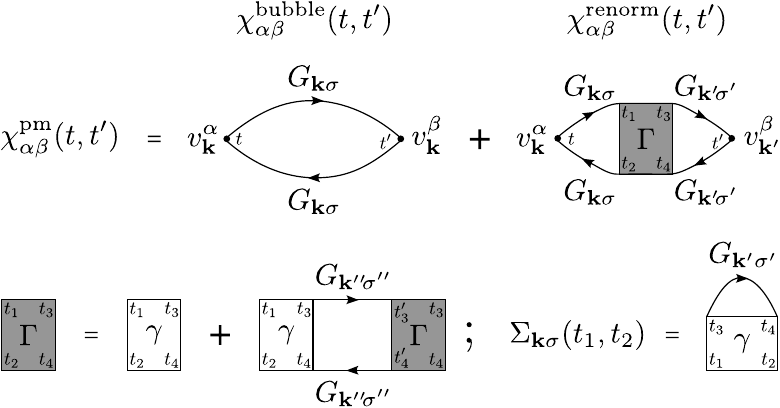}
\caption{Diagrammatic representation for the paramagnetic susceptibility ${\chi^{\rm pm}_{\alpha\beta}}$, vertex function ${\Gamma^{\sigma\sigma'}_{\bf k,k'}}$, and self-energy $\Sigma_{{\bf k}\sigma}$. The Figure is taken from Ref.~\cite{Conductivity}. 
\label{fig:diagrams_conductivity}}
\end{figure}

Collecting all terms gives the final result for the paramagnetic susceptibility function:
$\chi^{\rm pm}_{\alpha \beta}(t,t') = \chi^{\rm bubble}_{\alpha \beta}(t,t') + \chi^{\rm renorm}_{\alpha \beta}(t,t')$,
where the first, so called, ``bubble'' term:
\begin{align}
\chi^{\rm bubble}_{\alpha \beta}(t,t') = \frac{-i\chi_0}{V}\sum_{{\bf k},\sigma} v^{\alpha}_{\bf k}(t) G_{{\bf k}\sigma}(t,t')G_{{\bf k}\sigma}(t',t) v^{\beta}_{\bf k}(t')
\label{eq:bubble_SM}
\end{align}
is expressed only through the Green's functions and velocities, and the second, renormalized, term:
\begin{align}
&\chi^{\rm renorm}_{\alpha \beta}(t,t') = \iint \{dt_{i}\}\sum_{{\bf k,k'},\sigma,\sigma'} v^{\alpha}_{\bf k}(t) X^{0}_{{\bf k}\sigma}(t,t,t_1,t_2) \Gamma^{\sigma\sigma'}_{{\bf k,k'}}(t_1,t_2,t_3,t_4) X^{0}_{{\bf k'}\sigma'}(t_3,t_4,t',t') v^{\beta}_{\bf k'}(t')
\label{eq:renorm_SM}
\end{align}
additionally contains all possible multi-electron scattering processes through the correction $\Gamma$.
Importantly, the vertex correction $\Gamma$ is directly related to the scattering processes considered in the self-energy~\eqref{eq:Sigma_SM} via $\gamma$:
\begin{align}
\Gamma^{\sigma\sigma'}_{{\bf k,k'}}(t_1,t_2,t_3,t_4) &= \gamma^{\sigma\sigma'}_{{\bf k,k'}}(t_1,t_2,t_3,t_4) \notag\\
&+ \iint \{dt'\} \sum_{\bf k'',\sigma''} \gamma^{\sigma\sigma''}_{{\bf k,k''}}(t_1,t_2,t'_1,t'_2) X^{0}_{{\bf k''}\sigma''}(t'_1,t'_2,t'_3,t'_4)\Gamma^{\sigma''\sigma'}_{{\bf k'',k'}}(t'_3,t'_4,t_3,t_4)\,,
\end{align}
where 
\begin{align}
X^{0}_{{\bf k}\sigma}(t_1,t_2,t_3,t_4) = i G_{{\bf k}\sigma}(t_1,t_3)G_{{\bf k}\sigma}(t_4,t_2)\,.
\end{align}
In particular, in ${D\text{-}GW}$ the $\gamma$ vertex in the self-energy contains the charge and spin fluctuations via the renormalized charge and spin interaction $\tilde{W}^{\varsigma}$, which results in the vertex correction $\Gamma$ that is related to a multi-electron scattering on these charge and spin fluctuations.
Importantly, if the $\Gamma$ vertex is momentum-independent, the renormalised contribution to susceptibility~\eqref{eq:renorm_SM} disappears, because the integrand becomes odd in momenta ${\bf k}$ and ${\bf k'}$.
The diagrammatic representations of the susceptibility, the vertex function, and the self-energy are shown in Fig.~\ref{fig:diagrams_conductivity}.

In our calculations, the ``bubble'' contribution to optical conductivity involves the susceptibility based on both diamagnetic and paramagnetic terms. 
Traditional ``full'' calculations require a renormalized susceptibility, necessitating a solution to the Bethe-Salpeter equation; however, this approach is computationally intensive. To address this challenge, we adopt a different strategy. We measure the linear current response of the system by applying a weak probe field and then compute the optical conductivity using the equation:
${\sigma(\omega) = J_{\rm pr}(\omega)/E_{\rm pr}(\omega)}$.
In this formulation, we apply a time-dependent probe field of the form 
${E_{\rm pr}(t) = E_0 \frac{(t - t_d)}{t^2_c} e^{-(t - t_d)^2 / (2t^2_c)}}$, where $t_c$ represents the pulse duration and $t_d$ is the pulse delay. The probe amplitude $E_0$ is kept sufficiently weak so that $\sigma(\omega)$ can be measured in the linear response limit, ensuring it does not depend on the amplitude of the probe. The probe current $J_{\rm pr}(\omega)$ measured under this weak pulse captures all vertex corrections associated with multi-particle scattering processes, which is a significant advantage of real-time methods~\cite{PhysRevLett.103.047403, PhysRevB.78.205119, PhysRevB.93.195144, vglv-2rmv, PhysRevB.100.235117, Jaksa}. Details of the numerical calculations can be found in Ref.~\cite{Conductivity}.

\begin{figure}[t!]
\centering
\includegraphics[width=0.7\columnwidth]{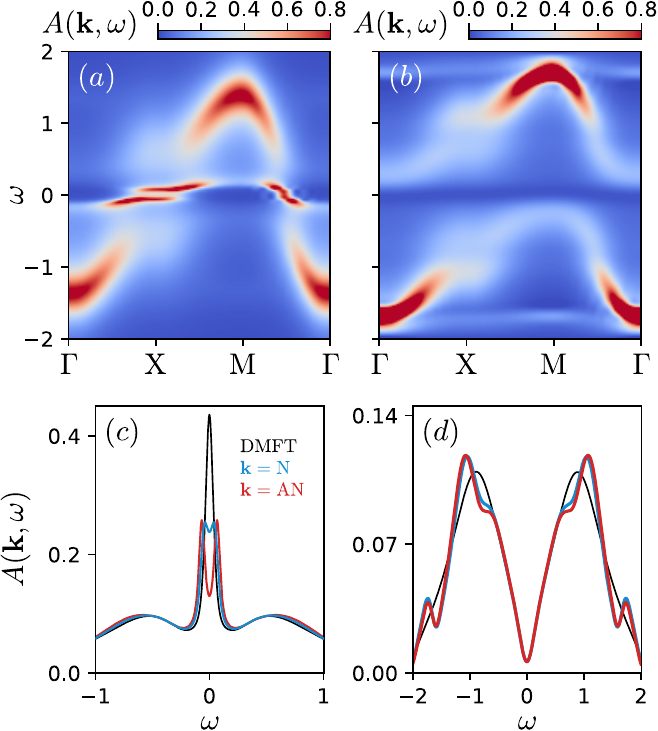}
\caption{The momentum-resolved spectral function ${A({\bf k},\omega)}$ calculated close to the N\'eel transition (orange stars in Fig.~\ref{fig:phase}) using ${D\text{-}GW}$ along the high-symmetry path of the Brillouin zone (${\Gamma=(0,0)}$, ${\text{X}=\text{AN}=(\pi,0)}$, ${\text{M}=(\pi,\pi)}$) in the metallic (${U=1.125}$, ${T = 0.036}$, panel (a)) and Mott insulating (${U=1.875}$, ${T = 0.027}$, panel (b)) phases. Panels (c) and (d) show the corresponding spectral functions at the $N$ (blue) and $AN$ (red) points in comparison with DMFT (black). The Figure is taken from Ref.~\cite{Conductivity}.
\label{fig:fig2}}
\end{figure}

Single-particle excitations primarily determine the optical conductivity, so accounting for an accurate electronic spectral function is essential for calculating the conductivity. 
Strong magnetic fluctuations can extend to rather high temperatures above the AFM transition and can drastically modify the electronic spectral function.
Thus, in the metallic phase, they can result in the opening of a gap at some part of the Fermi surface, as illustrated in Fig.~\ref{fig:fig2}\,(a). 
The comparison of spectra at the N and AN points with DMFT in Fig.~\ref{fig:fig2}\,(c) shows that in ${D\text{-}GW}$ this gap opening is momentum-dependent.
In turn, DMFT does not account for spatial fluctuations, resulting in a clear quasi-particle peak at the Fermi level. 
In the Mott insulating case, strong magnetic fluctuations affect the electronic spectrum mainly at high energies, resulting in the splitting of Hubbard bands at the N and AN points, as illustrated in Fig.~\ref{fig:fig2}\,(b).
A three-peak structure of the Hubbard band obtained on both sides of the Mott-gap using ${D\text{-}GW}$ signifies the role of spatial spin fluctuations in comparison with DMFT in Fig.~\ref{fig:fig2}\,(d).We note that a similar structure of Hubbard bands is found in variational Monte-Carlo calculations~\cite{PhysRevX.10.041023, singh2022unconventional} for the same model, and also in \mbox{D-TRILEX} calculations for the two-orbital model in the presence of strong magnetic fluctuations~\cite{PhysRevLett.129.096404}.

\begin{figure}[t!]
\centering
\includegraphics[width=0.7\columnwidth]{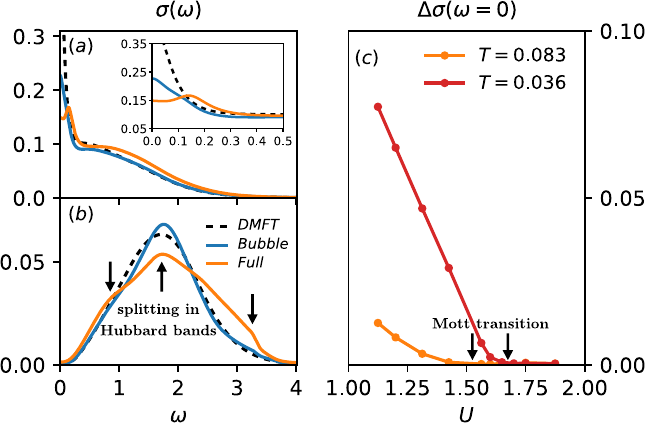}
\caption{The optical conductivity ${\sigma(\omega)}$ calculated close to the N\'eel transition (orange stars in Fig.~\ref{fig:phase}) in the metallic (${U = 1.125}$, ${T = 0.036}$, panel (a)) and Mott insulating (${U = 1.875}$,  ${T = 0.027}$, panel (b)) phases. 
The results are obtained using DMFT (dashed black), ``bubble'' ${D\text{-}GW}$ (blue), and ``full'' ${D\text{-}GW}$ (orange). 
The inset in (a) shows the low-frequency behavior of the conductivity. The arrows in panel (b) indicate the peaks and kinks originating from the optical transitions between the splittings in Hubbard bands due to strong magnetic fluctuations seen
in Fig.~\ref{fig:fig2}\,(d).
Panel (c) shows the difference in the DC conductivity $\Delta{\sigma(\omega=0)}$ between the bubble and full results. The scans are performed for ${T=0.083}$ and ${T=0.036}$ depicted in Fig.~\ref{fig:phase} by the dashed gray lines. 
The arrows in panel (c) indicate critical interactions ${U = 1.525}$ (${T=0.083}$) and ${U = 1.7}$ (${T=0.036}$) for the Mott transition shown in black dots in Fig.~\ref{fig:phase}. The Figure is taken from Ref.~\cite{Conductivity}.
\label{fig:fig3}}
\end{figure}

One can naively expect that if a method accurately captures the single-particle spectral features, the bubble approximation for the current-current correlation function would be sufficient to describe the optical conductivity qualitatively.
To demonstrate that this assumption does not hold, we calculate the optical conductivity in both the metallic and Mott insulating regimes, and compare the results obtained from DMFT, ${D\text{-}GW}$ within the ``bubble'' approximation, and the ``full'' ${D\text{-}GW}$ calculation that includes vertex corrections.
This comparison allows us to disentangle the different impacts of correlations on the conductivity, as DMFT accounts only for local correlations, the bubble ${D\text{-}GW}$ approximation additionally includes the effect of non-local correlations, but only on the electronic spectral function, and the full ${D\text{-}GW}$ scheme further incorporates non-local correlations via the vertex corrections.
The results obtained for the metallic case are shown
in Fig~\ref{fig:fig3}\,(a). 
In DMFT, we observe a finite spectral weight at low energy, corresponding to the Drude peak, followed by incoherent excitations at higher energies. 
In the bubble ${D\text{-}GW}$ approximation, we find that the Drude weight, corresponding to the DC conductivity, is significantly suppressed, while the optical (high-energy) part of the conductivity remains essentially unchanged compared to DMFT. 
In the full ${D\text{-}GW}$ calculation, the DC conductivity is further suppressed relative to the bubble approximation, and an additional peak appears at ${\omega \simeq 0.15}$.
This peak reflects the formation of a gap in the electronic spectral function (Fig.~\ref{fig:fig2}\,(c)) driven by strong magnetic fluctuations. 
Remarkably, this peak is not present in the bubble approximation
, although the later accounts for the correct electronic spectral function through the Green's functions.
The high-energy part of the conductivity is also noticeably modified upon considering vertex corrections in the metallic case. 

The transition from metal to Mott insulator is driven by local electronic correlations. 
Therefore, one may expect that vertex corrections to the conductivity are negligible, as the locality of the self-energy leads to local vertex corrections that, by symmetry, drop out of the current-current correlation function.
In this case, the bubble approximation 
should provide an accurate description of the optical conductivity in the Mott insulating phase. 
However, it is evident from the splitting of Hubbard bands (Fig.~\ref{fig:fig2}\,b,\,d), 
that non-local spin fluctuations can strongly affect the electronic spectral function even in the Mott phase. 
In order to illustrate how spin fluctuations influence conductivity, in Fig.~\ref{fig:fig3}\,(b) we compare results obtained using DMFT (black dashed curve), bubble ${D\text{-}GW}$ approximation (blue curve), and full ${D\text{-}GW}$ scheme (orange curve).
Apart from low energy, the optical conductivity from the bubble approximation differ significantly from the full calculation.
The full ${D\text{-}GW}$ results show peaks and kinks at three marked energies (indicated by black arrows), resembling those of the single-particle spectral function illustrated in Fig.~\ref{fig:fig2}\,(d).
DMFT, instead, shows only one peak corresponding to resonant excitations between Hubbard bands.
Surprisingly, the bubble approximation, although based on the correct form of the electronic spectral function, does not show a clear three-peak form of the conductivity observed in the full ${D\text{-}GW}$ calculation. 
It only reveals a ``resonant'' transition between the Hubbard bands, and a slight change in the curvature at frequencies, where the full calculation finds two additional peaks. 
We note that experimental measurements on iridate compounds reveal a similarly peaked structure in the optical conductivity~\cite{PhysRevLett.101.076402, PhysRevB.80.195110, seo2017infrared, 10.1063/1.4870049}. 
In Ref.~\cite{Cassol} these peaks were related to the splitting of Hubbard bands due to antiferromagnetic fluctuations that are strong in these materials~\cite{Lenz_2019}. 
However, the bubble approximation for conductivity, that accounts for the band splitting, captured only part of this feature by producing a less intense precursor of the experimentally measured peaks~\cite{Cassol}.
Therefore, accurately reproducing the optical conductivity in the Mott phase necessary requires considering non-local vertex corrections related to electronic scattering on spatial magnetic fluctuations. 

At the same time, we observe that the low-frequency part of the conductivity in the Mott phase does not vary significantly across different approximations. 
The DC conductivity, ${\sigma(\omega = 0)}$, is primarily governed by the electronic spectral weight at the Fermi energy. 
Accordingly, the opening of the Mott gap should be directly reflected in the DC conductivity. 
Since this gap opening is momentum-independent (i.e., local), one can expect the role of vertex corrections in the DC conductivity to vanish upon entering the Mott insulating phase.
In Fig.~\ref{fig:fig3}\,(c), we show the difference in DC conductivity, ${\Delta\sigma(\omega = 0)}$, between the bubble and full ${D\text{-}GW}$ results for two different temperatures, plotted as a function of interaction strength $U$. 
We find that ${\Delta\sigma(\omega = 0)}$ decreases linearly with increasing $U$ in the metallic phase and vanishes at the critical interaction strength indicated by the black arrows, where the Mott transition occurs.
We note that at high temperatures, the first-order Mott transition evolves into a smooth crossover, during which the quasiparticle peak at the Fermi energy transforms into a minimum, while the electronic density at the Fermi level remains finite due to thermal fluctuations. 
Consequently, the DC conductivity in the finite-temperature Mott regime is not zero, as illustrated in Ref.~\cite{Conductivity}.
We have confirmed that the critical interaction values for the Mott transition, as identified via the DC conductivity, are consistent with those obtained from the electronic spectral function.
These result may have significant implications for understanding strange metal behavior near the Mott transition, as our findings demonstrate that the DC conductivity can be accurately described by the simple bubble approximation, provided the correct electronic spectral function is used.

\newpage
\section{Effective exchange models for charge and spin densities}
\label{sec:Exchange}

In some cases, when a direct treatment of collective electronic fluctuations is challenging, the initial interacting fermionic problem can be replaced by a simplified bosonic one. 
This approach is commonly used, for example, to describe spin degrees of freedom (d.o.f.) via the Heisenberg model or charge ordering in alloys through an effective Ising model. 
Deriving such effective models is conceptually and technically non-trivial.
The existence of an effective ``bosonic'' description for spins relies on the so-called adiabatic approximation, which assumes that spin d.o.f. are much slower and have lower energy than electronic excitations. 
This separation of timescales allows magnetic excitations to be disentangled from the single-electron fluctuations. 
In contrast, describing charge d.o.f. is more complicated, as the corresponding adiabatic approximation does not exist, making the existence of a distinct charge dynamics and the corresponding classical Hamiltonian uncertain. 

The most common way to introduce an effective spin model for an interacting electronic problem is based on a Schrieffer-Wolff transformation~\cite{CHAO1977163, Chao_1977, PhysRevB.37.9753, spalek2007tj}, which, strictly speaking, is justified only at integer filling in the limiting case of a very large interaction between electrons. Already $t$-$J$ or $s\text{-}d$ exchange models~\cite{vonsovsky1974magnetism} that are frequently used to describe the physics of a doped Mott insulator cannot be easily mapped onto a pure spin Hamiltonian. 
Moreover, spin degrees of freedom in the transformed problem are described in terms of composite fermionic variables and not in terms of physical bosonic fields as would be desirable for pure spin models.
This results in a need to introduce artificial constraints in order to conserve the length of the total spin.
In addition, one also has to assume that the average value of these composite fermionic variables that define the local magnetization is nonzero.
The latter is hard to justify in a paramagnetic regime, where, generally speaking, it should also be possible to introduce a Heisenberg-like spin model.

Even though already deriving an effective spin problem for interacting electrons is not an easy task, one must do more than that and find a way to introduce a correct equation of motion for spin degrees of freedom.
For localized spins, the classical equation for the spin precession can be obtained by evaluating path integrals over spin coherent states in the saddle-point approximation~\cite{Inomata_book, Auerbach_book, Schapere_book}. 
In this approach, the kinetic term that describes the rotational dynamics of spins originates from the topological Berry phase, for which the conservation of the length of the total spin on each site is a necessary condition. 
For this reason, generalizing the formalism of spin-coherent states to itinerant electronic problems is mathematically a highly non-trivial task.
Nevertheless, finding a way to derive the equation of motion for the local magnetic moment in the framework of electronic problems is crucially important for a correct description of the full spin dynamics of the system. 
Indeed, studying classical spin Hamiltonians allows one to describe only a uniform precession of the local magnetic moment.
Taking into account dissipation effects, e.g. Gilbert damping, requires to couple classical spins to itinerant electrons~\cite{Sayad_2015, PhysRevLett.117.127201}.
In addition, considering classical spins disregards quantum fluctuations of the modulus of the local magnetic moment~\cite{doi:10.1146/annurev-conmatphys-031214-014350} that have been observed in recent experiments~\cite{PhysRevLett.100.205701, merchant2014quantum, jain2017higgs, PhysRevLett.119.067201, hong2017higgs, PhysRevLett.122.127201}.
In analogy with high-energy physics, these fast fluctuations are usually described in terms of a massive Higgs mode~\cite{PhysRevLett.13.321, HIGGS1964132, PhysRevLett.13.508, PhysRevLett.13.585}, while slow spin rotations are associated with Goldstone modes that originate from the broken rotational invariance in spin space.

The problem of describing the physics of the local magnetic moment in the framework of interacting electronic models was intensively studied in late 1970th -- early 1980th~\cite{PhysRevB.16.4032, PhysRevB.16.4048, PhysRevB.16.4058, PhysRevB.19.2626, PhysRevB.20.4584, 19791504, doi:10.1143/JPSJ.49.178, doi:10.1143/JPSJ.49.963, Hasegawa_1983, Edwards_1982, EDWARDS1983213}. 
In these works the local moments were formally introduced into the Hubbard model by using the Hubbard-Stratonovich transformation and making use of a static approximation for the introduced decoupling fields. 
Note that the static approximation in the Hubbard model is closed conceptually to the disordered local moment approach~\cite{oguchi1983magnetism, Pindor_1983, Gyorffy_1985, Staunton_1986, PhysRevLett.69.371, PhysRevB.67.235105} within the density functional theory.
As a result, initial translationally invariant system of interacting electrons is replaced by a single-particle problem of electrons moving in a random magnetic field acting on spins. Fluctuations in the direction of these fields are taken into account thus allowing to go beyond a mean-field approach. For the case of the Hubbard or $s$-$d$ exchange models at Bethe lattices, one can build the effective classical spin Hamiltonian taking into account both Anderson superexchange and Zener double exchange of essentially non-Heisenbergian character~\cite{auslender1982effective, AUSLENDER1982387}.  
This approach allowed one to go far beyond Stoner picture of itinerant-electron magnetism and clarified several important questions such as the origin of Curie-Weiss law for magnetic susceptibility above Curie temperature but it did not result in a complete quantitative theory of magnetism of itinerant electrons. In particular, it does not work at low temperatures where magnon-like dynamical excitations play a crucial role. An attempt to add these effects and to come to an unified picture in a phenomenological way was made by Moriya and collaborators which is summarized in the book~\cite{moriya_book}. 
Several important questions remained yet unsolved, 
e.g., the role of dynamical fluctuations that are known to be responsible for the Kondo effect~\cite{PhysRev.158.570} was not clarified. 

There were also many attempts to address the problem of the spin dynamics of interacting electrons. 
To get the Berry phase, one usually follows a standard route that consists in introducing rotation angles for a quantization axis of electrons~\cite{PhysRevLett.65.2462, PhysRevB.43.3790, doi:10.1142/S0217979200002430, DUPUIS2001617}.
These angles are considered as path integral variables to fulfill rotational invariance in the spin space.
In this case, the Berry phase term appears as an effective gauge field that, however, is coupled to fermionic variables instead of a spin bosonic field.
Considering purely electronic problems makes it difficult to disentangle spin and electronic degrees of freedom.
For this reason, until very recently it was not possible to connect the Berry phase to a proper bosonic variable that describes the modulus of the local magnetic moment.
For the same reason, it was also not possible to introduce a proper Higgs field to describe fluctuations of the modulus of the magnetization.
Indeed, in electronic problems this field is usually introduced by decoupling the interaction term~\cite{sachdev2008quantum, PhysRevX.8.011012, ScheurerE3665, PhysRevX.8.021048, PhysRevX.10.041057}.
First, such decoupling field does not have a clear physical meaning and its dynamics does not necessary correspond to the dynamics of the local magnetic moment.
Also, in actual calculations this effective Higgs field is usually treated in a mean-field approximation assuming that it has a non-zero average value, which is non-trivial to justify in a paramagnetic phase. 
One should also keep in mind that, although the decoupling of the interaction term is a mathematically exact procedure, it can be performed in many different ways, which leads to a famous Fierz ambiguity problem~\cite{PhysRevD.68.025020, PhysRevB.70.125111, Jaeckel} if the decoupling field is further treated in a mean-field approximation.

Despite all these challenges, the dual transformations make it possible to develop a general theory that provides a complete description of charge and spin dynamics in strongly correlated electronic systems, including exchange interactions and equations of motion.
In this Section, we demonstrate how an effective action written in terms of physical bosonic variables can be rigorously derived starting from a pure electronic problem.
We illustrate that the introduced effective bosonic problem allows one to obtain all kinds of exchange interactions between charge and spin densities, and in the case of well-developed collective fluctuations reduces to an effective Heisenberg (for spin d.o.f.)~\cite{PhysRevLett.121.037204} and Ising (for charge d.o.f.)~\cite{PhysRevB.99.115124} models. 
Remarkably, the resulting effective exchange interactions between the charge and magnetic densities have a simple form that can be efficiently used in realistic calculations for multiband systems. 
Importantly, we show that this derivation can be performed without assuming that the average magnetization is non-zero and without imposing any constraints such as artificial magnetic fields.
Therefore, the formalism remains valid beyond the strongly localized regime, enabling the description of a wide range of magnetic systems with well-defined local magnetic moments. 
Further, we show that the corresponding equation of motion for the spin bosonic action correctly describes the dissipative rotational dynamics of the local magnetic moment via the Berry phase and Gilbert damping term, and also takes into account the Higgs fluctuations of the modulus of the magnetic moment.
We also introduce a physical criterion for the formation of the local magnetic moment in the system and show that this approach is applicable even in the paramagnetic regime.
In the case of charge d.o.f., we demonstrate that the resulting effective Ising model allows one to describe phase transition to the charge ordered state.

\subsection{Back transformation from the dual space to physical bosonic variables}
\label{sec:Heisenberg_Ising}

To introduce a consistent theory of spin and charge dynamics, we will mainly follow the route presented in Refs.~\cite{PhysRevLett.121.037204, PhysRevB.105.155151, RevModPhys.95.035004}.
We start with a general action for a multi-orbital extended Hubbard model~\eqref{eq:actionlatt}, as a particular example of the strongly-correlated electronic problem that possesses charge and spin dynamics:
\begin{align}
{\cal S} = &-\int_{0}^{\beta} d\tau \sum_{jj',\sigma\sigma',ll'} \hspace{-0.1cm} c^{*}_{j\tau\sigma{}l} \left[\delta_{jj'}\delta_{\sigma\sigma'}\delta_{ll'}(-\partial_{\tau}+\mu)-\varepsilon^{\sigma\sigma'}_{jj'll'}\right]
c_{j'\tau\sigma'l'}^{\phantom{*}} \notag\\
&+\frac12\int_{0}^{\beta} d\tau \, \Bigg\{\sum_{j,\sigma\sigma',\{l\}} \hspace{-0.1cm} U^{\phantom{*}}_{l_1 l_2 l_3 l_4} c^{*}_{j\tau\sigma{}l_1} c^{\phantom{*}}_{j\tau\sigma{}l_2} c^{*}_{j\tau\sigma'{}l_4} c^{\phantom{*}}_{j\tau\sigma'{}l_3} +\sum_{jj',\varsigma,\{l\}} V^{jj'\varsigma}_{l_1l_2l_3l_4} \rho^{\varsigma}_{j\tau{}l_1l_2}\rho^{\varsigma}_{j'\tau{}l_4l_3} \Bigg\},
\label{eq:s9_action_latt}
\end{align}
For convenience, we work in the real (lattice sites) space $j^{(\prime)}$ and imaginary time $\tau$ representation.
We also exclude the interaction in the particle-particle channel (${V^{\vartheta}=0}$) and use ${\varsigma=c}$ and ${\varsigma=s}$ to label the charge and spin channels, respectively.
The remaining notations are consistent with the ones introduced in Section~\ref{sec:Lattice_action}.

We note that the exchange interactions between spin and charge densities in the bosonic problem that we aim to derive are non-local, while the dynamics of the magnetic moment is usually described by local Berry and Higgs terms. 
For this reason, it is useful to explicitly separate local and non-local correlations in the system. 
In addition, the effective Heisenberg and Ising models are expressed in terms of bosonic variables, whereas the charge and spin d.o.f. in the original model~\eqref{sec:Lattice_action} correspond to composite fermionic variables $\rho^{\varsigma}$. In this regard, the formalism based on the dual action introduced in Section~\ref{sec:Action_Dual} provides an ideal starting point for deriving an effective exchange model. 
In this formalism, the local correlation effects are contained in the reference impurity problem~\eqref{eq:actionimp_app}, which is integrated out, leading to an effective dual action~\eqref{eq:DB_action_app} that describes only the remaining (non-local) correlations. 
Furthermore, the dual action already contains the true bosonic variables $\varphi^{\varsigma}$, introduced as decoupling fields for the original composite fermionic variables $\rho^{\varsigma}$.
The only drawback of the dual action in the context of deriving effective exchange models is that the dual bosonic variables $\varphi^{\varsigma}$, introduced through the Hubbard-Stratonovich transformation, do not have a clear physical meaning. 
Therefore, one must find a way to transform these dual variables back into the physical ones associated with the charge and spin densities.

To derive the dual action~\eqref{eq:DB_action_app}, we follow a route similar to that described in Section~\ref{sec:Action_Dual}.
In particular, we introduce the reference system~\eqref{eq:actionimp_app} in the form of the DMFT impurity problem.
In general, the impurity problem can be considered either in a polarized~\cite{PhysRevLett.121.037204} or in a non-polarized~\cite{PhysRevB.105.155151} form, which corresponds to an ordered or paramagnetic solution for the problem, respectively.
At present, we stick to a non-polarized local reference system, which allows one to describe a regime of the system where the average local magnetization is identically zero $\langle n^{s}_{ll'}\rangle_{\rm imp}=0$.
In this case, the hybridization function $\Delta^{ll'}_{\tau\tau'}$ is spin independent, and can be determined from the self-consistent condition ${\frac12\sum_{\sigma}G^{\tau\tau'll'}_{jj\sigma\sigma}=g^{ll'}_{\tau\tau'}}$~\cite{PhysRevB.105.155151} that equates the spin diagonal, local part of the interacting lattice Green function $G^{\tau\tau'll'}_{jj\sigma\sigma}$ and the interacting Green function of the local reference problem $g^{ll'}_{\tau\tau'}$.
As the next step, we rewrite the remaining part of the lattice action~\eqref{eq:actionrem_app} in terms of new fermionic ${c^{(*)} \to f^{(*)}}$ and bosonic ${\rho^{\varsigma}\to\varphi^{\varsigma}}$ variables by means of the Hubbard-Stratonovich transformations~\eqref{eq:HSf} and~\eqref{eq:HSb}. After these transformations, the lattice action ${\cal S}'[c^{(*)},f^{(*)},\varphi^{\varsigma}]$~\eqref{eq:Sprime_action} depends on two fermionic and one bosonic variables.
Original Grassmann variables $c^{(*)}$ are contained only in the local part of the lattice action, which includes the impurity problem~\eqref{eq:actionimp_app}, and thus can be integrated out.

Before integrating out the reference problem, it is worth recalling that isolating local correlation effects is essential for correctly describing the dynamics of spin degrees of freedom.
In general, spin dynamics might have a non-trivial form, since it involves a combination of a slow spin precession and fast Higgs fluctuations of the modulus of the local magnetic moment.
For this reason, it is more convenient to treat these two contributions separately.
In electronic systems, the Berry phase term that describes the uniform spin precession is commonly obtained by transforming original electronic variables to a rotating frame~\cite{PhysRevLett.65.2462, PhysRevB.43.3790, doi:10.1142/S0217979200002430, DUPUIS2001617}.
This can be achieved by introducing a unitary matrix in the spin space :
\begin{align}
R_{j\tau} = 
\begin{pmatrix}
\cos(\theta_{j\tau}/2) & - e^{-i\phi_{j\tau}}\sin(\theta_{j\tau}/2) \\
e^{i\phi_{j\tau}}\sin(\theta_{j\tau}/2) & \cos(\theta_{j\tau}/2)
\end{pmatrix}
\label{eq:Rmatrix}
\end{align}
and making the corresponding change of variables ${c_{j\tau{}l} \to R_{j\tau} c_{j\tau{}l}}$ in the action~\eqref{eq:Sprime_action}, where ${c_{j\tau{}l} = (c_{j\tau{}l\uparrow}, c_{j\tau{}l\downarrow})^{T}}$. 
Rotation angles $\Omega_{R} = \{\theta_{j\tau}, \phi_{j\tau}\}$ are considered as site $j$ and time $\tau$ dependent variables.
Introducing an additional functional integration over them allows one to preserve the rotational invariance in the spin space.
As a consequence, the modified lattice action~\eqref{eq:Sprime_action} takes the following form: ${{\cal S}'[c^{(*)}, f^{(*)}, \varphi^{\varsigma},\Omega_{R}]}$.

The Berry phase arises from the local impurity problem~\eqref{eq:actionimp_app} that upon rotation becomes~\cite{PhysRevB.105.155151}:
\begin{align}
{\cal S}_{\rm imp} &\to {\cal S}_{\rm imp} + 
\int_{0}^{\beta} d\tau \,\Tr_{\sigma} c^{*}_{i\tau} R^{\dagger}_{i\tau} \dot{R}^{\phantom{*}}_{i\tau} c^{\phantom{*}}_{i\tau}
= {\cal S}_{\rm imp} +
\int_{0}^{\beta} d\tau \sum_{\varsigma} {\cal A}^{\varsigma}_{i\tau} \, \rho^{\varsigma}_{i\tau}\,,
\label{eq:s9_Berry_term}
\end{align}
where $\dot{R}^{\phantom{*}}_{i\tau} = \partial_{\tau} R^{\phantom{*}}_{i\tau}$ and ${\cal A}^{\varsigma}_{i\tau}$ is an effective gauge field introduced as ${R^{\dagger}_{i\tau} \dot{R}^{\phantom{*}}_{i\tau} = \sum_{\varsigma} {\cal A}^{\varsigma}_{i\tau}\sigma^{\varsigma}}$.
The explicit form of the rotation matrix~\eqref{eq:Rmatrix} implies that ${\cal A}^{c}_{i\tau}=0$.
Note, that we have omitted the orbital indices for simplicity.
They will be explicitly restored at a later stage. The explicit derivation in the multi-orbital form can be found in Ref.~\cite{PhysRevB.105.155151}.
The ${z}$ component of an effective gauge field ${\cal A}^{s}_{j\tau}$ has the desired form of the Berry phase term ${{\cal A}^{z}_{j\tau} = \frac{i}{2} \dot{\phi}_{j\tau} (1-\cos\theta_{j\tau})}$.
To exclude other components of the gauge field from consideration, one usually assumes that the rotation angles $\Omega_{R}$ correspond to the spin-quantization axis of electrons. 
In this case, the composite fermionic variable in the spin channel $\rho^{s}$ is replaced by its $z$ component $\rho^{z}$ which is coupled to the ``correct'' component of the gauge field ${\cal A}^{z}_{j\tau}$.
Proceeding in this direction leads to several problems.
Associating rotation angles with the spin-quantization axis is non-trivial to formulate in a strict mathematical sense.
In Refs.~\cite{doi:10.1142/S0217979200002430, DUPUIS2001617} it was done introducing a slave boson approximation.
However, there is no guarantee that the {\it average magnetization} on a given lattice site will also point in the $z$ direction.
Indeed, the spin-quantization axes on different sites may point in different directions, which may induce an effective mean magnetic field that will change the direction of the magnetization on a given site.
In particular, this does not allow one to replace the composite fermionic variable $\rho^{z}$ by its average value in the Berry phase term~\eqref{eq:s9_Berry_term}.
Moreover, in the paramagnetic phase this replacement does not make sense, because the average magnetization in this case is identically zero.
Finally, in Eq.~\eqref{eq:s9_Berry_term} the effective gauge field ${\cal A}^{s}_{j\tau}$ is coupled to a composite fermionic variable $\rho^{s}$ instead of a proper vector bosonic field that describes fluctuations of the local magnetic moment.
This representation of spin degrees of freedom does not conserve the length of the total spin, which is a necessary condition for a correct description of a spin precession.

We emphasize that the rotation angles also cannot be associated with the direction of the dual bosonic field for spin degrees of freedom $\varphi^{s}$.
This field enters the lattice action~\eqref{eq:Sprime_action} as an effective quantum magnetic field that polarises the electrons~\cite{PhysRevLett.121.037204, PhysRevB.105.155151} and is frequently associated with the Higgs field~\cite{sachdev2008quantum, PhysRevX.8.011012, ScheurerE3665, PhysRevX.8.021048, PhysRevX.10.041057}.
However, this effective bosonic field is introduced as the result of a Hubbard-Stratonovich transformation~\eqref{eq:HSb} and does not have a clear physical meaning.  
Moreover, even if it would be possible to associate $\varphi^{s}$ with the physical Higgs field, its dynamics would not necessarily correspond to the dynamics of the local magnetic moment.
All these observations suggest that the idea to describe the spin precession in terms of rotation angles is very appealing, but one has to find a way to relate these angles to the direction of the local magnetic moment and not to the spin-quantization axis or to the effective Higgs field.
 
The composite charge and spin variables also transform under the spin rotation and become:
\begin{align}
\rho^{\varsigma}_{i\tau} \to \sum_{\varsigma'} {\cal U}^{\varsigma\varsigma'}_{i\tau}\rho^{\varsigma'}_{i\tau}\,,
\end{align}
where ${\cal U}^{\varsigma\varsigma'}_{i\tau}$ satisfies:
\begin{align}
R^{\dagger}_{i\tau} \sigma^{\varsigma} R^{\phantom{*}}_{i\tau} 
= \sum_{\varsigma'} {\cal U}^{\varsigma\varsigma'}_{i\tau}\sigma^{\varsigma'}\,.
\label{eq:B_coupling}
\end{align}
It can be shown that ${\cal U}^{ss'}$ is a unitary matrix, i.e. ${[{\cal U}^{-1}]^{ss'}=[{\cal U}^{\rm T}]^{ss'}}$, and that ${{\cal U}^{cs}_{i\tau}=0}$ and ${{\cal U}^{cc}_{i\tau}=1}$.
The last equality originates from the fact that the charge density $n^{c}_{i\tau}$ is invariant under rotation in the spin space.
After transforming the original electronic variables $c^{(*)}$ to a rotating frame they can finally be integrated out, which results in the dual boson action $\tilde{S}_{\rm DB}[f^{(*)},\varphi^{\varsigma},\Omega_{R}]$~\eqref{eq:DB_action}.
We recall, that in this action the bare propagators for the fermionic $f^{(*)}$ and bosonic $\varphi^{\varsigma}$ variables are purely non-local and explicitly depend on rotation angles $\Omega_{R}$~\cite{PhysRevB.105.155151}.   
All local correlations are absorbed in the interaction part of the fermion-boson action $\tilde{\cal F}[f^{(*)},\varphi^{\varsigma},\Omega_{R}]$~\eqref{eq:Wfull}, which, as in the DB approach, is truncated at the two-particle level and contains the exact four-point (fermion-fermion) $\Gamma$ and three-point (fermion-boson) $\Lambda^{\varsigma}$ vertex functions of the reference impurity problem~\eqref{eq:actionimp_app}.

Integrating out the reference system not only disentangles local and non-local correlations, but also allows one to get rid of composite fermionic variables $\rho^{\varsigma}$ that are no longer present in the dual boson action ${\tilde{\cal S}_{\rm DB}[f^{(*)},\varphi^{\varsigma},\Omega_{R}]}$. 
Now, charge and spin degrees of freedom are described by a proper bosonic field $\varphi^{\varsigma}$ that has a well-defined propagator and a functional integration over them. 
Moreover, in this action the gauge field ${\cal A}^{s}_{j\tau}$ is coupled (up to a certain multiplier) to the spin component of this bosonic field $\varphi^{s}$~\cite{PhysRevB.105.155151}. 
However, as discussed above, the bosonic variable $\varphi^{\varsigma}$ does not have a clear physical meaning.
The way of introducing a physical bosonic variable was proposed in Ref.~\cite{PhysRevLett.121.037204} and was inspired by works~\cite{doi:10.1142/S0217979200002430, DUPUIS2001617} where a similar transformation was performed for fermionic fields.
The idea consists in introducing a source field $j^{\varsigma}$ for the {\it original composite fermionic variable} $\rho^{\varsigma}$ that describes fluctuations of charge and spin densities (see Section~\ref{sec:Action_Dual} for details). 
Then, after obtaining the dual boson action~\eqref{eq:DB_action} one performs one more Hubbard-Stratonovich transformation ${\varphi^{\varsigma}\to\bar\rho^{\varsigma}}$ that makes $j^{\varsigma}$ the source field for the resulting {\it physical bosonic field} $\bar\rho^{\varsigma}$.
Further, unphysical bosonic fields $\varphi^{\varsigma}$ are integrated out, which leads to the fermion-boson action:
\begin{align} 
{\cal S}
= 
&-\int_{0}^{\beta} \{d\tau_{i}\} \sum_{ij,\sigma\sigma'} f^{*}_{i\tau_1\sigma}g^{-1}_{\tau_1\tau_2}
\left( [\tilde{\varepsilon}^{-1}]^{\sigma\sigma'}_{ij,\tau_2\tau_3} - \delta_{ij}R^{\phantom{\dagger}}_{i\tau_2}g^{\phantom{*}}_{\tau_2\tau_3}R^{\dagger}_{i\tau_3}\right)
g^{-1}_{\tau_3\tau_4}f^{\phantom{*}}_{j\tau_4\sigma'} \notag\\
&+ \frac12\iint_{0}^{\beta} d\tau\,d\tau' \sum_{i,\{\varsigma\}} \bar{\rho}^{\,\varsigma_1}_{i\tau} {\cal U}^{\varsigma_1\varsigma_2}_{i\tau} \left[\chi^{\varsigma_2}\right]^{-1}_{\tau\tau'} [{\cal U}_{i\tau'}^{-1}]^{\varsigma_2\varsigma_3} \bar{\rho}^{\,\varsigma_3}_{i\tau'}
+ \frac12 \int_{0}^{\beta} d\tau \sum_{ij,\varsigma} \bar{\rho}^{\,\varsigma}_{i\tau} V^{\varsigma}_{ij} \, \bar{\rho}^{\,\varsigma}_{j\tau} \notag\\
&+ \int_{0}^{\beta} \{d\tau_{i}\} \Tr_{\sigma} \sum_{i,\varsigma\varsigma'} f^{*}_{i\tau_1}
R^{\phantom{\dagger}}_{i\tau_1}\sigma^{\varsigma}R^{\dagger}_{i\tau_2}
f^{\phantom{*}}_{i\tau_2} \Lambda^{\varsigma}_{\tau_1\tau_2\tau_3} [{\cal U}_{i\tau_3}^{-1}]^{\varsigma\varsigma'} \bar{\rho}^{\,\varsigma'}_{i\tau_3} \notag\\
&+ \int_{0}^{\beta} d\tau \sum_{i} \Bigg\{ \sum_{ss'}{\cal A}^{s}_{i\tau}[{\cal U}_{i\tau}^{-1}]^{ss'} \bar{\rho}^{\,s'}_{i\tau}
+ \sum_{\varsigma} j^{\,\varsigma}_{i\tau}\,\bar{\rho}^{\,\varsigma}_{i\tau} \Bigg\}.
\label{eq:S3}
\end{align}
Note, that we used the following choice for the fermionic scaling parameter: ${B_{\tau\tau'}=g_{\tau\tau'}}$. The bosonic scaling parameter vanishes within the last Hubbard-Stratonovich transformation.

The source field $j^{\,\varsigma}_{i\tau}$ enters the new lattice problem~\eqref{eq:S3} only multiplied by the new bosonic field $\bar{\rho}^{\,\varsigma}_{i\tau}$. 
This means that all correlation functions written in terms of original composite variables $\rho^{\varsigma}_{i\tau}$, e.g. the susceptibility~\eqref{eq:Xsource}, identically coincide with the ones where the $\rho^{\varsigma}_{i\tau}$ variables are replaced by the corresponding bosonic fields $\bar{\rho}^{\,\varsigma}_{i\tau}$.
Therefore, the introduced fields $\bar{\rho}^{\,\varsigma}_{i\tau}$ have the same physical meaning as the composite variables $\rho^{\varsigma}_{i\tau}$ that describe fluctuations of charge and spin densities.
To emphasise this point hereinafter we omit the bar over the $\bar{\rho}^{\,\varsigma}_{i\tau}$.

Importantly, the derived fermion-boson action~\eqref{eq:S3} has a simpler form compared to the dual boson action~\eqref{eq:DB_action}. 
Indeed, the interaction part of the fermion-boson action~\eqref{eq:S3} contains only the three point vertex function $\Lambda^{\varsigma}$.
The four-point vertex $\Gamma$ that is present in the dual boson theory is approximately cancelled by the counterterm that is generated during the last Hubbard-Stratonovich transformation~\cite{PhysRevLett.121.037204, PhysRevB.100.205115}. 
Note, that the bosonic Hubbard-Stratonovich transformation ${\varphi^{\varsigma}\to\bar\rho^{\varsigma}}$ is different from the one introduced in the \mbox{D-TRILEX} approach~\eqref{eq:HSb_DT}.
Therefore, the counterterm has a bit different, and in fact more approximate, form than the partially-bosonized approximation for the four-point vertex~\eqref{eq:Gamma_approx_app}.
On the other hand, the Hubbard-Stratonovich transformation used here was aimed at obtaining physical bosonic fields $\bar\rho^{\varsigma}$, which is not possible to combine with a more accurate approximtion for the four-point vertex.

We note that at this point all parameters of the fermion-boson action~\eqref{eq:S3}, including the coupling of the gauge field ${\cal A}^{s}_{j\tau}$ to the bosonic field $\bar\rho^{s}$, explicitly depend on the rotation angles $\Omega_{R}$.
From the very beginning, these angles are introduced to account for the spin precession explicitly.
For this reason, $\Omega_{R}$ should be related to the direction of the local magnetic moment, which in the fermion-boson action is defined by a bosonic vector field $\bar\rho^{s}$.  
It is convenient to rewrite the latter in spherical coordinates as ${\rho^{s}_{j\tau{}ll'} = M^{\phantom{*}}_{j\tau{}ll'} e^{s}_{j\tau}}$, where $M_{j\tau{}ll'}$ is a scalar field 
that describes Higgs fluctuations of the modulus of the orbitally-resolved local magnetic moment. 
In this expression we assume that the multi-orbital system that exhibits a well-developed magnetic moment is characterised by a strong Hund's exchange coupling that orders spins of electrons at each orbital in the same direction.
Therefore, the direction of the local magnetic moment in the system is
defined by the orbital-independent unit vector $\vec{e}_{j\tau}$, e.g. described by a set of polar angles ${\Omega_M=\{\theta'_{j\tau},\varphi'_{j\tau}\}}$ associated with this vector.
It has been shown in Ref.~\cite{PhysRevB.105.155151} that taking the path integral over rotation angles $\Omega_{R}$ in the saddle-point approximation allows one to equate these two sets of angles ${\Omega_R=\Omega_M}$ that from now on define the direction of the local magnetic moment.
From this point on, we consider the integral over $\vec{e}_{i\tau}$ taken and the two sets of rotation angles are set equal to one another $\Omega_{R}=\Omega_{M}$.
This results in the following relation for the unit vector field:
\begin{align}
R^{\dagger}_{i\tau} \, \vec{\sigma} \cdot \vec{e}^{\phantom{*}}_{i\tau} \, R^{\phantom{*}}_{i\tau} 
= \sigma^{z}\,.
\end{align}
Using the relation~\eqref{eq:B_coupling}, one also finds: 
\begin{align}
\sum_{s'} [{\cal U}_{i\tau}^{-1}]^{ss'} e^{s'}_{i\tau} =  \delta^{\phantom{*}}_{s,z} \,,
\end{align}
which immediately yields the Berry phase term:
\begin{align}
\sum_{ss'}{\cal A}^{s}_{i\tau}[{\cal U}_{i\tau}^{-1}]^{ss'} \rho^{s'}_{i\tau} =
\sum_{ss'}{\cal A}^{s}_{i\tau}[{\cal U}_{i\tau}^{-1}]^{ss'} e^{s'}_{i\tau} M^{\phantom{*}}_{i\tau} = 
{\cal A}^{z}_{i\tau}M^{\phantom{*}}_{i\tau}\,,
\end{align}
where ${{\cal A}^{z}_{i\tau} = \frac{i}{2} \dot{\varphi}_{i\tau} (1-\cos\theta_{i\tau})}$.
Also, one straightforwardly gets:
\begin{align}
\sum_{\{s\}}\rho^{\,s_1}_{i\tau} {\cal U}^{s_1s_2}_{i\tau} \left[\chi^{s_2}\right]^{-1}_{\tau\tau'} [{\cal U}_{i\tau'}^{-1}]^{s_2s_3} \rho^{\,s_3}_{i\tau'} = M^{\phantom{*}}_{i\tau} \left[\chi^{z}\right]^{-1}_{\tau\tau'} M^{\phantom{*}}_{i\tau'}\,.
\end{align}
The remaining rotation matrices can be excluded from the action~\eqref{eq:S3} by assuming the adiabatic approximation that characteristic times for electronic degrees of freedom are much faster than for spin ones. In this framework $g^{\phantom{\dagger}}_{\tau_2\tau_3}$ changes much faster than the rotation matrices $R^{\phantom{\dagger}}_{i\tau_2}$. 
This approximation results in:
\begin{gather}
\iint^{\beta}_{0} d\tau_2 \, d\tau_3 \, g^{-1}_{\tau_1\tau_2} R^{\phantom{\dagger}}_{i\tau_2} g^{\phantom{\dagger}}_{\tau_2\tau_3} R^{\dagger}_{i\tau_3} g^{-1}_{\tau_3\tau_4} \simeq 
\int^{\beta}_{0} d\tau_3 \, \delta^{\phantom{*}}_{\tau_1\tau_3} R^{\phantom{\dagger}}_{i\tau_3}R^{\dagger}_{i\tau_3}g^{-1}_{\tau_3\tau_4} 
= g^{-1}_{\tau_1\tau_4}\,,
\end{gather}
because the local reference system~\eqref{eq:actionimp_app} is considered non-polarized, and its exact Greens function $g_{\tau\tau'}$ is diagonal in the spin space. 
A similar trick can be performed for the three-point vertex function, which leads to (see Ref.~\cite{PhysRevB.105.155151}):
\begin{gather}
\Tr_{\sigma}\sum_{ss'} f^{*}_{i\tau_1}
R^{\phantom{\dagger}}_{i\tau_1}\sigma^{s}R^{\dagger}_{i\tau_2}
f^{\phantom{*}}_{i\tau_2} \Lambda^{\hspace{-0.05cm}s}_{\tau_1\tau_2\tau_3} [{\cal U}_{i\tau_3}^{-1}]^{ss'} \rho^{s'}_{i\tau_3} 
\simeq 
\sum_{s,\sigma\sigma'} f^{*}_{i\tau_1\sigma}
\sigma^{s}_{\sigma\sigma'}
f^{\phantom{*}}_{i\tau_2\sigma'} \Lambda^{\hspace{-0.05cm}s}_{\tau_1\tau_2\tau_3}\,
\rho^{\,s}_{i\tau_3}\,.
\end{gather}
As a result, the fermion-boson action~\eqref{eq:S3} takes the form of an effective $t$-$J$ or $s\text{-}d$ exchange model~\cite{vonsovsky1974magnetism} that describes local charge and spin moments $\rho^{\varsigma}$ coupled to itinerant electrons $f^{(*)}$ via the local fermion-boson vertex function $\Lambda^{\varsigma}$. 
\begin{align} 
{\cal S}_{s\text{-}d}
= 
&-\iint_{0}^{\beta} d\tau \, d\tau' \sum_{ij,\sigma\sigma'}  
f^{*}_{i\tau\sigma} \left[\tilde{\cal G}^{-1} \right]^{\sigma\sigma'}_{ij,\tau\tau'}
f^{\phantom{*}}_{j\tau'\sigma'} \notag\\
&+ \iiint_{0}^{\beta} d\tau_1 \, d\tau_2 \, d\tau_3 \sum_{i,\varsigma,\sigma\sigma'} f^{*}_{i\tau_1\sigma}\sigma^{\varsigma}_{\sigma\sigma'}
f^{\phantom{*}}_{i\tau_2\sigma'} \,\Lambda^{\varsigma}_{\tau_1\tau_2\tau_3}\, \rho^{\varsigma}_{i\tau_3} \notag\\
&+ \frac12 \int_{0}^{\beta} d\tau\,\sum_{ij,\varsigma} \rho^{\varsigma}_{i\tau} V^{\varsigma}_{ij} \, \rho^{\varsigma}_{j\tau}
+ \int_{0}^{\beta} d\tau \sum_{i} {\cal A}^{z}_{i\tau} M^{\phantom{*}}_{i\tau} \notag\\
&- \frac12 \iint_{0}^{\beta} d\tau\,d\tau' \sum_{i} \Bigg\{ \rho^{c}_{i\tau} \left[\chi^{c}\right]^{-1}_{\tau\tau'} \rho^{c}_{i\tau'} 
+ M^{\phantom{*}}_{i\tau} \left[\chi^{z}\right]^{-1}_{\tau\tau'} M^{\phantom{*}}_{i\tau'} \Bigg\} .
\label{eq:S4}
\end{align}
Moreover, in this action the gauge field ${\cal A}^{s}_{j\tau}$ is coupled to the spin component of the physical bosonic field $\bar\rho^{s}$ as desired for a correct description of the rotational dynamics of the local magnetic moment~\cite{PhysRevB.105.155151}. 
The source fields $j^{\,\varsigma}$ have been introduced only to identify correct bosonic variables and were excluded from the action~\eqref{eq:S4}.

With the choice of the scaling parameter ${B_{\tau\tau'}=g_{\tau\tau'}}$ the bare Green's function of the action~\eqref{eq:S4} is given by the difference between the DMFT (here we assume ${G^{\rm DMFT}=G}$) and the impurity Green's functions (see Section~\ref{sec:Scaling_parameters}):
\begin{align}
\tilde{\cal G}^{\sigma\sigma'}_{ij,\tau\tau'} = G^{\sigma\sigma'}_{ij,\tau\tau'} - \delta^{\phantom{*}}_{ij}\delta^{\phantom{*}}_{\sigma\sigma'} g^{\phantom{*}}_{\tau\tau'}\,.
\label{eq:s9_G_dual}
\end{align}
Therefore, this quantity is dressed only in the local impurity self-energy, which in our case is obtained from the self-consistent DMFT calculation.
In the absence of the SOC the local part of the bare dual Green's function is identically zero due to DMFT self-consistency condition.
Therefore, the performed set of transformations separates spatial electronic fluctuations described by fields $f^{(*)}$ from local correlation effects that are accounted for by the reference system~\eqref{eq:actionimp_app}.
Another advantage of the introduced fermion-boson action~\eqref{eq:S4} is that it allows one to obtain not only standard correlation functions, such as the electronic Green's function and the (charge, spin, etc.) susceptibility, but also various exchange interactions between charge and spin densities that cannot be calculated directly from the initial electronic problem~\eqref{eq:actionlatt}.

The bosonic problem that describes the behavior of charge and spin densities can be obtained from the fermion-boson action~\eqref{eq:S4} by integrating out fermionic fields $f^{(*)}$.
The fermion-boson action is Gaussian in terms of these fields, so this integration can be performed exactly.
The final result for the bosonic action in the multi-orbital form is the following~\cite{PhysRevB.105.155151}:
\begin{align} 
{\cal S}_{\rm boson}
= 
&- \mathrm{Tr} \ln \left[ \big[\tilde{\cal G}^{-1}\big]^{\tau\tau'll'}_{jj'\sigma\sigma'}  - \delta^{\phantom{*}}_{jj'} \int^{\beta}_{0} d\tau_1 \sum_{\varsigma,l_1l'_1} \, \sigma^{\varsigma}_{\sigma\sigma'} \Lambda^{\varsigma\,\tau\tau'\tau_1}_{ll'l_1l'_1} \, \bar\rho^{\varsigma}_{j\tau_1l'_1l_1} \right] \notag\\
&+\frac12\int_{0}^{\beta} d\tau \sum_{jj',\varsigma,\{l\}} \bar\rho^{\varsigma}_{j\tau{}ll'} \, V^{jj'\varsigma}_{ll'l_1l'_1} \, \bar\rho^{\varsigma}_{j'\tau{}l'_1l_1}
+ \int_{0}^{\beta} d\tau \sum_{j} {\cal A}^{z}_{j\tau} {\cal M}^{\phantom{*}}_{j\tau}
\notag\\
&- \frac12 \iint_{0}^{\beta} d\tau\,d\tau' \sum_{j,\{l\}}
\Bigg\{
\bar\rho^{c}_{j\tau{}ll'} \left[\chi^{c\,-1}\right]^{\tau\tau'}_{ll'l_1l'_1} \bar\rho^{c}_{j\tau'l'_1l_1}
+ M^{\phantom{*}}_{j\tau{}ll'} \left[\chi^{z\,-1}\right]^{\tau\tau'}_{ll'l_1l'_1} M^{\phantom{*}}_{j\tau'l'_1l_1}
\Bigg\}.
\label{eq:s9_Boson_action}
\end{align}
Importantly, in this action the modulus of the total magnetic moment ${{\cal M}_{j\tau} = \sum_{l}M^{\phantom{*}}_{j\tau{}ll}}$ is coupled only to the $z$ component of the effective gauge field ${\cal A}^{z}_{j\tau}$ that gives exactly the desired Berry phase term.
Other components of the gauge field disappear upon associating rotation angles with the direction of the local magnetic moment.

\subsection{Exchange interactions in many-body theory and relation to other approaches}

Before introducing the explicit expression for the exchange interaction it is worth noting that an unambiguous definition for this quantity does not exist. The exchange interactions are internal parameters of the model and thus depend on the particular form of the considered Hamiltonian.
In its turn, the latter crucially depends on the downfolding scheme used to map the interacting electronic problem onto an effective bosonic (i.e., spin) model.
For instance, it has been shown that considering small local variations from the ordered magnetic state leads to the bilinear exchange interaction that depends on the magnetic configuration, and the resulting spin Hamiltonian also contains higher-order non-linear exchange interactions that are not negligible \emph{a priori}~\cite{auslender1982effective, AUSLENDER1982387}.
On the other hand, one can try to map the interacting electronic problem onto a global Heisenberg-like spin model with only bilinear exchange interaction. In this case, the value of the bilinear exchange might be different compared to the one of the non-linear spin model.

The mapping on the general and bilinear forms of the spin Hamiltonian are both useful. 
The form that contains non-linear exchange interactions better reproduces the spectrum of spin waves~\cite{PhysRevB.64.174402}.  
On the other hand, the Heisenberg Hamiltonian is a standard model for atomistic spin simulations and gives reasonable thermodynamic properties of the system~\cite{eriksson2017atomistic}. In order to establish connection between different definitions for the exchange interaction, we start with the bosonic action~\eqref{eq:s9_Boson_action} derived above.
In this action local and non-local correlation effects are completely disentangled by construction of the theory.
The first line in Eq.~\eqref{eq:s9_Boson_action} describes non-local exchange interactions between charge $\bar\rho^{c}$ and spin $\bar\rho^{s}$ densities.
The first term in this expression is responsible for all possible kinetic exchange processes (including higher-order ones) mediated by electrons.
This can be illustrated by directly expanding the logarithm function to all orders in $\bar\rho^{\varsigma}$ variables.
Since this expansion is performed in terms of the bosonic variables that correspond to charge and magnetic densities, the resulting bilinear and non-linear exchange interactions are well defined.
This expansion is essentially different from the one performed in terms of rotation angles in DFT-based formalisms. Indeed, the latter is based on the magnetic force theorem, which cannot be used for derivation of higher-order expansion terms in the rotation angle. 
The situation is similar to that in the problem of calculations of elastic moduli of solids in DFT: whereas the first-order variations with respect to deformation are very simple and can be calculated according to the local force theorem, the second-order variations contain a lot of additional terms related to the differentiation of the double-counting contributions~\cite{zein1984}. 
At the same time, the effective bosonic action discussed here is based on formally exact transformations. 

The bilinear exchange interaction $J^{\varsigma\varsigma'}_{jj'}$ is given by the second order of the expansion:
\begin{gather}
J^{\varsigma\varsigma'\tau\tau'}_{jj'll'l''l'''} = 
\int_{0}^{\beta} \{d\tau_i\} \sum_{\{\sigma_i\},\{l_i\}}
\Lambda^{*\,\varsigma\,\tau\tau_1\tau_2}_{ll'l_1l_2} \, \tilde{\cal G}^{\tau_1\tau_3l_1l_3}_{jj'\sigma_1\sigma_3} \, \tilde{\cal G}^{\tau_4\tau_2l_4l_2}_{j'j\sigma_4\sigma_2} \, \Lambda^{\varsigma'\tau_3\tau_4\tau'}_{l_3l_4l''l'''}\,,
\label{eq:s9_J}
\end{gather}
where a ``transposed'' three-point vertex function
${\Lambda^{*\,\varsigma\,\tau_1\tau_2\tau_3}_{l_1l_2l_3l_4} = \Lambda^{\varsigma\,\tau_3\tau_2\tau_1}_{l_4l_3l_2l_1}}$ is introduced to simplify notations.
The diagonal part of the bilinear exchange interaction is given by the Heisenberg exchange $J^{ss}_{jj'}$ for spin~\cite{PhysRevLett.121.037204, PhysRevB.105.155151} and the Ising interaction $J^{cc}_{jj'}$ for charge~\cite{PhysRevB.99.115124} densities.
The latter will be discussed in details in Section~\ref{Section10}.
The non-diagonal ${J^{s\neq{}s'}_{jj'}}$ components give rise to the Dzyaloshinskii-Moriya and the symmetric anisotropic interactions (see, e.g., Ref.~\cite{PhysRevB.52.10239}) that may appear in the system due to spin-orbit coupling.
These kinetic exchange interactions compete with the bare non-local electron-electron interaction $V^{\varsigma}_{jj'}$ that plays a role of a direct exchange between charge and spin densities.
This makes the total, non-local bilinear exchange interaction to have the form:
\begin{align}
{\cal I}^{\varsigma\varsigma'}_{jj'} = J^{\varsigma\varsigma'}_{jj'} + \delta^{\phantom{*}}_{\varsigma\varsigma'}\, V^{\varsigma}_{jj'}\,.
\label{eq:s9_exch}
\end{align}
Importantly, the non-local interaction $V^{\varsigma}_{jj'}$ enters the bosonic problem in the same way as it was introduced in the initial lattice action~\eqref{eq:s9_action_latt}.
We also note that the direct spin-spin interaction $V^{s}_{jj'}$ usually has the opposite sign to the kinetic interaction $J^{ss}_{jj'}$. 
More involved interactions~\cite{auslender1982effective, AUSLENDER1982387}, e.g. the ring~\cite{PhysRevB.47.11329, PhysRevB.59.1468, PhysRevLett.83.5122}, the chiral three-spin~\cite{PhysRevLett.93.056402, bauer2014chiral, PhysRevB.95.014422, grytsiuk2020topological, zhang2020imprinting, PhysRevB.103.L060404} 
and the four-spin~\cite{PhysRevB.76.054427,heinze2011spontaneous, paul2020role} exchange interactions can be obtained by expanding the first term in Eq.~\eqref{eq:s9_Boson_action} to higher orders in the $\rho^{\varsigma}$ variable. For calculations of bilinear exchange interactions~\eqref{eq:s9_J} in a realistic material context see Ref.~\cite{vandelli2024doping}.

At this step we can establish a relation between the bilinear exchange interactions derived using a magnetic force theorem and a quantum many-body path-integral technique. In this case it is convenient to work in the Matsubara fermionic $\nu$ and bosonic $\omega$ frequency representation. 
To simplify expressions we further omit orbital indices that can be restored trivially.
First, we note that the three-point vertex function $\Lambda^{\varsigma}$ for the zeroth bosonic frequency can be obtained from single-particle quantities: 
\begin{align}
\Lambda^{s}_{\nu,\omega=0} = \triangle^{s}_{\nu} + \chi^{s\,-1}_{\omega=0}
\label{eq:s9_vertex_sigma}
\end{align}
by varying the self-energy of the local reference problem~\eqref{eq:actionimp_app} with respect to the magnetization~\cite{PhysRevB.105.155151}
\begin{align}
\triangle^{s}_{\nu} = \partial{}\Sigma^{\rm imp}_{\nu}/\partial{}M_{\omega=0}\,.
\label{eq:s9_triangle}
\end{align}
In the ordered phase, where the spin rotational invariance is broken, this variation can be approximated as: 
\begin{align}
\triangle^{s}_{\nu} = \frac{\Sigma^{\rm imp}_{\nu\uparrow\uparrow} - \Sigma^{\rm imp}_{\nu\downarrow\downarrow}}{2\langle M \rangle} \,.
\label{eq:s9_triangle2}
\end{align}
This relation is justified by local Ward identities and the fact that in the regime of a well-developed magnetic moment the renormalized fermion-fermion interaction (four-point vertex function) does not depend on fermionic frequencies~\cite{PhysRevLett.121.037204}.
Therefore, in Eq.~\eqref{eq:s9_vertex_sigma} the $\triangle^{s}_{\nu}$ term describes the spin splitting of the self-energy due to polarization of the system. In turn, $\chi^{s\,-1}_{\omega=0}$ can be seen as a kinetic self-splitting effect, because ${\chi^{s}_{\omega}}$ is the exact spin susceptibility of the reference system.  
In magnetic materials with a relatively large value of the magnetic moment the kinetic contribution can be neglected.
Indeed, in this case the spin splitting of the self-energy is determined by the Hund's exchange coupling.
The latter is much larger than the inverse of the spin susceptibility, for which the estimation $\chi^{s\,}_{\omega=0}\sim T^{-1}$ holds due to Curie-Weiss law~\cite{moriya_book}.
Then, the static exchange interaction ${J^{ss'}_{jj'}(\omega=0) = \int d\tau'\,J^{ss'}_{jj'}(\tau-\tau')}$ (see Ref.~\cite{PhysRevB.105.155151} for discussions) reduces to the form:
\begin{align}
J^{ss'}_{jj',\omega=0} = 
\sum_{\nu,\{\sigma\}} \triangle^{s}_{j\nu} \, \tilde{\cal G}^{\sigma_1\sigma_3}_{jj'\nu} \,
\triangle^{s'}_{j'\nu} \,
\tilde{\cal G}^{\sigma_4\sigma_2}_{j'j\nu}\,,
\label{eq:s9_J_w0}
\end{align}
that under the approximation~\eqref{eq:s9_triangle2} coincides with the expression that for the ordered phase was derived in Refs.~\cite{liechtenstein1984exchange, liechtenstein1985curie, liechtenstein1987local, katsnelson2000first, cardias2020dzyaloshinskii} using the magnetic force theorem.
The magnetic force theorem can also be applied in a paramagnetic phase.
In the Hubbard I approximation this was done in Ref.~\cite{PhysRevB.94.115117}, and the result coincides with Eq.~\eqref{eq:s9_J_w0}, where the relation~\eqref{eq:s9_triangle} is calculated numerically exactly.
It should be emphasized that in Eq.~\eqref{eq:s9_J}, and consequently in Eq.~\eqref{eq:s9_J_w0}, the vertex function~\eqref{eq:s9_vertex_sigma} and thus the self-energy~\eqref{eq:s9_triangle} are given by the {\it local} reference system~\eqref{eq:actionimp_app}. Moreover, the Green function~\eqref{eq:s9_G_dual} that enters the expression for the exchange interaction is also dressed only in the local self-energy. The spin splitting $\triangle^{s}$ obtained from the non-local self-energy was introduced in Ref.~\cite{SECCHI2016112}. 
However, the corresponding exchange interaction is formulated in terms of bare (non-interacting) Green functions and can be derived considering only the density-density approximation for the interaction between electrons. For these reasons, the limit of applicability of this approach and the relation to other methods remain unclear.  

In addition, if one fully neglects the dependence on the fermionic frequency in Eq.~\eqref{eq:s9_vertex_sigma}, the vertex function can be approximated by the inverse of the local bare polarization $\Lambda^{s} \simeq \chi^{0\,-1}_{\omega=0}$, where ${\chi^{0}_{\omega} = \sum_{\nu}g_{\nu}g_{\nu+\omega}}$.
Then, the exchange interaction~\eqref{eq:s9_J} reduces to the form of an effective bare non-local susceptibility, as was derived in Ref.~\cite{Antropov_2003}:
\begin{align}
J^{ss'}_{jj',\omega=0} = \chi^{0\,-1}_{\omega=0}\,\tilde{X}^{0}_{jj',\omega=0}\,\chi^{0\,-1}_{\omega=0}\,,
\end{align}
where ${\tilde{X}^{0}_{jj'\omega} = \sum_{\nu}\tilde{\cal G}_{jj'\nu} \,
\tilde{\cal G}_{j'j\nu+\omega}}$.

One can also establish a relation between the results of the introduced many-body theory result and the bilinear exchange interaction that can be deduced from the lattice susceptibility $X^{\varsigma\varsigma'}_{jj'}$ using the following expression:
\begin{align}
\bar{J}^{\varsigma\varsigma'}_{j\neq{}j'} = \delta_{jj'}\delta_{\varsigma\varsigma'}\left[\chi^{\varsigma} \right]^{-1} - \left[X^{-1}\right]^{\varsigma\varsigma'}_{jj'}\,.
\label{eq:s9_susceptibility}
\end{align}
This expression was used in the works in Refs.~\cite{Antropov_2003, PhysRevB.91.195123, PhysRevB.96.075108, PhysRevB.99.165134} to estimate the magnetic exchange interaction based on the DMFT approximation for the spin susceptibility~\cite{RevModPhys.68.13}.
One can find that this form for the bilinear exchange interaction~\eqref{eq:s9_susceptibility} can also be obtained from the derived above many-body theory if the non-linear action~\eqref{eq:s9_Boson_action} is approximated by the Gaussian form
\begin{align}
\bar{\cal S} = - \frac12 \iint_{0}^{\beta} d\tau\,d\tau' \sum_{jj',\varsigma\varsigma'} \bar\rho^{\varsigma}_{j\tau} \left[X^{-1}\right]^{\varsigma\varsigma',\tau\tau'}_{jj'} \bar\rho^{\varsigma'}_{j'\tau'}\,.
\label{eq:s9_Sapprox}
\end{align}
Since the bosonic variables $\bar{\rho}^{\varsigma}$ correspond to the charge and magnetic densities, the quantity $X^{\varsigma\varsigma',\tau\tau'}_{jj'}$ is nothing more than the lattice susceptibility~\cite{PhysRevLett.121.037204, PhysRevB.99.115124, PhysRevB.105.155151}.
More accurately this approximation can be done using Peierls-Feynman-Bogoliubov variational principle~\cite{PhysRev.54.918, Bogolyubov:1958zv, feynman1972}.
Comparing the two actions~\eqref{eq:s9_Boson_action} and~\eqref{eq:s9_Sapprox} shows that in this case the bilinear exchange interaction should indeed be given by the relation~\eqref{eq:s9_susceptibility}. 

Effectively, this procedure corresponds to the mapping of the spin problem~\eqref{eq:s9_Boson_action} that contains all possible exchange interactions onto an effective Heisenberg problem that accounts only for the bilinear exchange.
It should be emphasised that for this reason it would be incorrect to relate two expressions for the bilinear exchange  introduced in Eqs.~\eqref{eq:s9_J} and~\eqref{eq:s9_susceptibility}. Indeed, equating these two quantities corresponds to truncating the expansion of the logarithm in the bosonic action~\eqref{eq:s9_Boson_action} at the second order in terms of $\bar{\rho}$ variables.
In other words, it means neglecting the effect of the higher-order exchange interactions on the lattice susceptibility and, consequently, on the bilinear exchange interaction $\bar{J}$.
Taking this effect into account will obviously modify the expression~\eqref{eq:s9_J} for the bilinear exchange interaction.
In particular, it will result in dressing the Green's functions $\tilde{G}$ by the non-local self-energy and in the renormalization of one of the two vertex functions, $\Lambda$, by collective non-local fluctuations in Hedin's fashion~\cite{GW1}.

These observations confirm the statement that we made at the beginning of this Section, namely that the expression for the exchange interaction strongly depends on the form of the considered spin model.
If one is limited to the simplest approximation with only bilinear form of the exchange interaction, then the latter should be calculated via the Eq.~\eqref{eq:s9_susceptibility} provided that consistent calculation for the lattice susceptibility is possible. 
For instance, using the DMFT form of the susceptibility might already be questionable, because it accounts for the renormalization of the vertex function (in the ladder approximation) but disregards the non-local self-energy. 
At the same time, if a more accurate model that contains the bilinear and the non-linear exchange interactions is considered, these interactions should be computed in the form given by the action~\eqref{eq:s9_Boson_action}. 
In this case, the bilinear interaction is given by Eq.~\eqref{eq:s9_J} or its approximation~\eqref{eq:s9_J_w0}.
Calculating it via the lattice susceptibility~\eqref{eq:s9_susceptibility} would be incorrect, because it would lead to a double-counting problem for the higher-order interactions, since some contribution of them is already taken into account in the lattice susceptibility. 
The difference between the two forms for the bilinear exchange interaction can also serve as a measure of the importance of the non-linear exchange processes in the system.

\subsection{Equation of motion for the local magnetic moment}

The second line in the bosonic action~\eqref{eq:s9_Boson_action} contains only local contributions that describe dynamics of charge and spin degrees of freedom. 
The first term in this line accounts for the Higgs fluctuations of the modulus of the charge $\rho^{c}$ and spin $M$ moments around their average value.
This can be seen by formally expanding the time-dependence of the moments in powers of ${\tau-\tau'}$.
For the local magnetic moment this gives:
\begin{align}
{\cal S}_{\rm Higgs} &= - \frac12 \iint_{0}^{\beta} d\tau\,d\tau' \sum_{j}
M^{\phantom{*}}_{j\tau} \left[\chi^{z\,-1}\right]_{\tau\tau'} M^{\phantom{*}}_{j\tau'} \notag\\
&\simeq - \frac12\int_{0}^{\beta} d\tau \sum_{j} \left\{\chi^{z\,-1}_{\omega=0} \, M^2_{j\tau} + \frac{\partial^{2}\chi^{z\,-1}_{\omega}}{2\,\partial\omega^2}\Big|_{\omega=0} \, \dot{M}^{2}_{j\tau} \right\}.
\label{eq:s9_Higgs}
\end{align}
The first order difference in time vanishes, because the exact local susceptibility $\chi^{\varsigma}_{\omega}$ is the even function of the frequency $\omega$.
The Lagrangian equation for this action immediately gives the standard equation of motion for a simple harmonic oscillator ${\ddot{M}_{j\tau} + \lambda^2 M_{j\tau} = 0}$, where ${\lambda^2 = - 2\chi^{z\,-1}_{\omega=0}/\left.\left(\partial^{2}_{\omega}\chi^{z\,-1}_{\omega}\right)\right|_{\omega=0}}$.
Note that in our definition the susceptibility $\chi^{\varsigma}_{\omega}$ is negative.
However, this expansion has to be performed with ultimate care.
Indeed, Higgs fluctuations of the modulus of the local magnetic moment are fast, and the spin susceptibility is strongly non-local in time~\cite{PhysRevB.105.155151}. 
For this reason, there is no uniform justification that the Higgs fluctuations can be accurately described using an equal-time term (second line of Eq.~\eqref{eq:s9_Higgs}) instead of the full non-stationary in time local part of the lattice action (first line of Eq.~\eqref{eq:s9_Higgs}).

The last term in the bosonic action~\eqref{eq:s9_Boson_action} that contains the effective gauge field ${\cal A}^{z}_{j\tau}$ accounts for the rotational spin dynamics.
It has been shown in Ref.~\cite{PhysRevB.105.155151} that after averaging over fast Higgs fluctuations the equation of motion for the bosonic action reduces to the standard Landau-Lifshitz-Gilbert form. 
To illustrate this, we replace the scalar field $M_{j\tau}$ by its constant non-zero average value $\langle M_{j\tau}\rangle=2S$ and introduce ${\vec{S}_{j\tau} = S\,\vec{e}_{j\tau}}$.
The spin part of the action becomes:
\begin{align}
{\cal S}_{\rm spin} = \int^{\beta}_0 d\tau\sum_{j} \left( i\dot{\varphi}_{j\tau}(1-\cos\theta_{j\tau}) \,S - \vec{S}_{j\tau}\cdot\vec{h}_{j\tau} \right),
\label{eq:s9_action_spin}
\end{align}
where we explicitly rewrote the gauge field in terms of rotation angles.
Components of the effective magnetic field $\vec{h}_{j\tau}$ can be expressed via the bilinear exchange interaction and the effective magnetic field that appears due to spin-orbit coupling~\cite{PhysRevB.105.155151}:
\begin{align}
h^{s}_{j\tau} = -~4\int_0^{\beta} d\tau'\sum_{j',s'}\mathcal{I}^{ss'}_{jj'}(\tau-\tau') \, S^{s'}_{j'\tau'} + h^{{\rm soc}\,s}_{j\tau}\,.
\label{eq:s9_h_eff}
\end{align}
In the general case, the equation of motion for the non-stationary spin action~\eqref{eq:s9_action_spin} is a complex set of integro-differential equations. 
However, one can make use of the fact that the interaction between spins is determined by the super-exchange processes mediated by electrons~\eqref{eq:s9_J} and thus decays fast on the time scales of inverse band width.
Instead, the time-dependence of the angle variables $\varphi_{j\tau}$ and $\theta_{j\tau}$ is slow, because the spin precession is slow in time~\cite{Sayad_2015, PhysRevLett.117.127201, PhysRevLett.125.086402}.
Contrary to the case of Higgs fluctuations, this allows one to expand the time-dependence of the spin variable $S^{s'}_{j'\tau'}$ in Eq.~\eqref{eq:s9_h_eff} up to the first order in powers of ${\tau-\tau'}$, which allows to write:
\begin{align}
h^{s}_{j}(t) = &-4\sum_{j',s'} I^{\mathrm{R}\,ss'}_{jj'}(\Omega=0) \, S^{s'}_{j'}(t) + h^{{\rm soc}\,s}_{j}(t)
-4\sum_{j',s'}\frac{\partial}{\partial\Omega}\left. {\rm Im} \, I^{\mathrm{R}\,ss'}_{jj'}(\Omega)\right|_{\Omega=0}\dot{S}^{s'}_{j'}(t)\,.
\label{eq:s9_h_stationarry}
\end{align}
With this expression for the effective magnetic field the spin problem~\eqref{eq:s9_action_spin} becomes stationary in time, and the corresponding equation of motion for this action takes the standard Landau-Lifshitz-Gilbert form:
\begin{align}
\dot{\vec{S}}_{j}(t) = -\, \vec{h}_{j}(t)\times\vec{S}_{j}(t)\,.
\label{eq:s9_eq_motion}
\end{align}
This expression can be derived by making analytical continuation that transforms the imaginary-time exchange interaction ${{\cal I}^{ss'}_{jj'}(\tau-\tau')}$ to a retarded function ${I^{{\rm R}\,ss'}_{jj'}(t-t')}$ in real time $t$. 
In turn, $I^{{\rm R}\,ss'}_{jj'}(\Omega)$ is a Fourier transform of the retarded exchange interaction to real frequency $\Omega$.
This transformation allows one to obtain the Gilbert damping, which is described by the last term in the effective magnetic field~\eqref{eq:s9_h_stationarry}.
A similar expression for the Gilbert damping was derived in Refs.~\cite{Sayad_2015, PhysRevLett.117.127201} for the case of a classical spin coupled to the system of conduction electrons.
Note that the Gilbert damping cannot be obtained in the imaginary-time representation, because the exchange ${{\cal I}^{ss'}_{jj'}(\tau-\tau')}$ is an even function of time. 
Physically, this means that dissipation effects cannot be visible in the equilibrium formalism. 

There are several restrictions for the derived Landau-Lifshitz-Gilbert equation of motion that have to be discussed.  
Eq.~\eqref{eq:s9_eq_motion} describes the spin precession that is assumed to be slow in time compared to electronic processes in the system.
The corresponding effective magnetic field~\eqref{eq:s9_h_stationarry} thus takes into account only the low-frequency part of the exchange interaction.
In general, the exchange term~\eqref{eq:s9_J} has a non-trivial frequency dependence and even diverges at high frequencies, because it is given by a non-local part of the inverse of the lattice susceptibility~\eqref{eq:s9_susceptibility}. 
Non-adiabatic effects that correspond to high-frequency behavior of the exchange interaction are not taken into account by the Eq.~\eqref{eq:s9_eq_motion}. The latter can only be described using the derived bosonic action~\eqref{eq:s9_Boson_action} that has no restriction on the regime of frequencies, but is non-stationary in time.
Another important point is that the Higgs and the Berry phase terms, in the form they enter the bosonic action~\eqref{eq:s9_Boson_action}, can be obtained only after associating the rotation angles with the direction of the local magnetic moment.
As discussed above, this can be done taking the path integral over rotation angles in the saddle-point approximation.
However, this approximation can be justified only for the case of a large  magnetic moment~\cite{PhysRevB.105.155151}.
In practice, it means that the classical Landau-Lifshitz-Gilbert equation of motion is applicable only in the multi-orbital case, where the large value of the local magnetic moment is provided by a strong Hund's coupling.
If the magnetic moment is small, spin dynamics in the system is governed by quantum fluctuations.
In this case, the local magnetic moment can still be well-defined, but its behavior can no longer be described in terms of classical equations of motion.

\subsection{Local magnetic moment formation}
\label{sec:LMM_formation}

The Landau-Lifshitz-Gilbert equation of motion~\eqref{eq:s9_eq_motion} makes physical sense only for a non-zero value of the average magnetic moment $\langle M \rangle$. 
In the ordered phase this is ensured by a non-zero average value of the magnetization.
Defining $\langle M \rangle$ in a paramagnetic regime is much more problematic, because in this case the average magnetization is identically zero.
For this reason, the value of $\langle M \rangle$ is commonly estimated from the static (equal-time) spin susceptibility as:
\begin{align}
3\chi^{z}_{\tau\tau} = \langle M^2 \rangle \simeq \langle M \rangle \big(\langle M \rangle +2 \, \big)\,.
\label{eq:s9_XlocS}
\end{align}
However, this approximation gives quite large and almost temperature-independent value for the magnetic moment even in the high-temperature regime where the moment is not yet formed~\cite{PhysRevB.105.155151}.
Taking into account dynamical screening effects changes the value of the average moment, but it still remains substantially larger compared to the one measured experimentally~\cite{PhysRevLett.104.197002, PhysRevB.86.064411, PhysRevLett.125.086402}.
This result can be explained by the fact that the local spin susceptibility simultaneously accounts for correlations of the local magnetic moment and for spin fluctuations of itinerant electrons.
These two contributions to the susceptibility cannot be easily disentangled.

\begin{figure}[t!]
\centering
\includegraphics[width=0.7\linewidth]{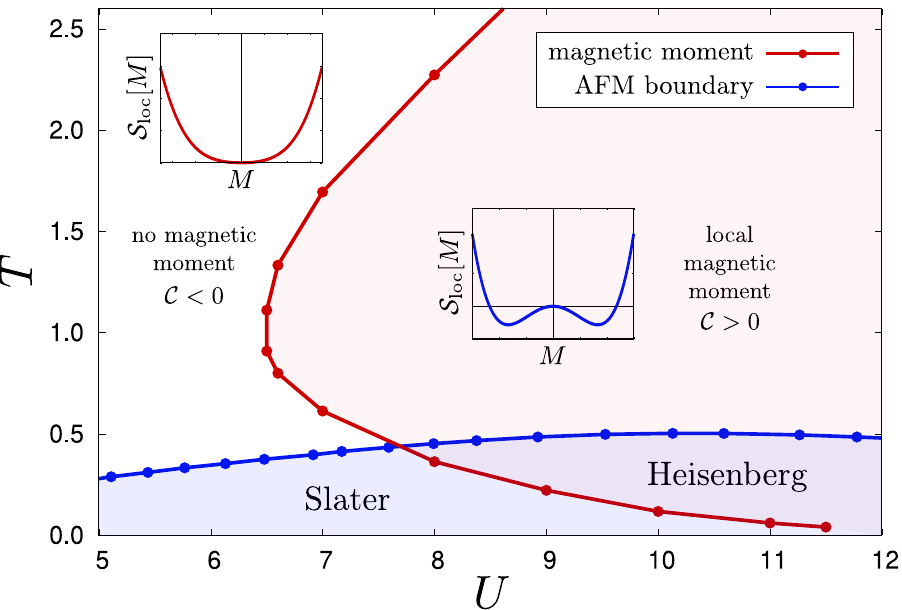}
\caption{\label{fig:Section9_phase} Phase diagram for the 3D Hubbard model as a function of temperature $T$ and local Coulomb interaction $U$. Red (light grey) line corresponds to the criterion~\eqref{eq:s9_condition_C} for the formation of the local magnetic moment. Blue (dark gray) line depicts the N\'eel phase boundary obtained in Ref.~\cite{PhysRevB.92.144409}. The insets show the local free energy~\eqref{eq:s9_action_local} as a function of the magnetic moment in two regimes, when it does not exist to the left of the red (light grey) line and where it is already formed shown by the red (light grey) shaded area. The Figure is taken from Ref.~\cite{RevModPhys.95.035004}.}
\end{figure}

In Ref.~\cite{PhysRevB.105.155151} the average value of the magnetic moment was proposed to obtain from the free energy of the local problem that describes the behavior of the magnetic moment.
The action of this local problem:
\begin{align} 
{\cal S}_{\rm loc}
= 
&-\Tr\ln \left[[g^{-1}]_{\tau\tau'} \delta^{\phantom{*}}_{\sigma\sigma'} + \int^{\beta}_{0} d\tau_1 \sum_{\varsigma} \sigma^{\varsigma}_{\sigma\sigma'} \Lambda^{\varsigma}_{\tau\tau'\tau_1} \rho^{\varsigma}_{\tau_1} \right]
- \frac12 \iint_{0}^{\beta} d\tau\,d\tau' \sum_{\varsigma}\rho^{\varsigma}_{\tau} \left[\chi^{\varsigma\,-1}\right]_{\tau\tau'} \rho^{\varsigma}_{\tau'} 
\label{eq:s9_action_local}
\end{align}
can be derived by excluding the contribution of itinerant electrons from the local reference system~\eqref{eq:actionimp_app}.
The resulting problem reminds of the bosonic action~\eqref{eq:s9_Boson_action}, where the non-local Green function $\tilde{\cal G}$ is replaced by the full local Green function $g$.
In the introduced local problem~\eqref{eq:s9_action_local} the magnetic moment appears as a result of a spontaneous symmetry breaking.
According to Landau phenomenology~\cite{LL_V} the latter corresponds to the change of the free energy from a paraboloid-like form with a minimum at ${\langle M \rangle=0}$ to a Mexican-hat potential characterized by a continuous set of minima at ${\langle M \rangle\neq0}$ (see insets in Fig.~\ref{fig:Section9_phase}).
Remarkably, the resulting value for the average local magnetic moment appears to be substantially smaller than the one deduced from the local spin susceptibility~\eqref{eq:s9_XlocS}. 

The change of the form of the free energy can be captured by the sign change of its second variation with respect to the the local magnetic moment:
\begin{align}
-\frac{\partial^2{\cal S}_{\rm loc}[\rho^{s}]}{\partial\rho^{s}_{\tau}\,\partial\rho^{s}_{\tau'}} = \left[\chi^{s\,-1}\right]_{\tau\tau'} - J^{\rm loc}_{\tau\tau'}\,.
\label{eq:s9_local_criterion}
\end{align}
The right-hand side of this equation can be seen as a self-exchange between the local magnetic moments, because it is given by the inverse of the local susceptibility with subtracted contribution of itinerant electrons.
The latter is described by a local analog of the kinetic exchange interaction~\eqref{eq:s9_J}:
\begin{align}
J^{\rm loc}_{\tau\tau'} = 
\int^{\beta}_{0} \{d\tau_i\} \sum_{\sigma} \Lambda^{*\,s}_{\tau\tau_1\tau_2} g^{\sigma}_{\tau_1\tau_3} g^{\sigma}_{\tau_4\tau_2} \Lambda^{s}_{\tau_3\tau_4\tau'}\,.
\label{eq:s9_J_loc}
\end{align}
It is important to emphasize that the local magnetic moment exists only at relatively long times compared to single-electron processes.
In the static limit the moment is screened by Kondo effect or by intersite exchange-induced spin flips.
For this reason, formation of the local magnetic moment in the system corresponds to the symmetry breaking at intermediate time scales.
Consequently, as has been shown in Ref.~\cite{PhysRevB.105.155151}, the second variation of the local free energy~\eqref{eq:s9_local_criterion} changes sign at any times except ${\tau=\tau'}$.
Therefore, the formation of the local moment is not a real physical transition and should be considered as a crossover effect.
The static contribution to the local problem~\eqref{eq:s9_action_local} is contained in the inverse of the local susceptibility ${\chi^{s\,-1}_{\tau\tau'} = (\Pi^{s\,{\rm imp}}_{\tau\tau'})^{-1} - \delta_{\tau\tau'}U^{s}}$.
It is given by the bare local interaction in the spin channel ${U^{s}=-U/2}$.
In this expression $\Pi^{s\,{\rm imp}}_{\tau\tau'}$ is the exact polarization operator of the reference system~\eqref{eq:actionimp_app}. 
The criterion for the local magnetic moment formation can thus be obtain by explicitly excluding this static contribution from Eqs.~\eqref{eq:s9_action_local} and~\eqref{eq:s9_local_criterion}.
The corresponding condition written in the frequency space is that:
\begin{align}
{\cal C} = \big(\Pi^{s\,{\rm imp}}_{\omega=0}\big)^{-1} - J^{\rm loc}_{\omega=0} = 0\,.
\label{eq:s9_condition_C}
\end{align}
This expression illustrates that when the effective self-exchange becomes diamagnetic ${({\cal C}>0)}$ the system acquires a magnetic moment.
The derived criterion~\eqref{eq:s9_condition_C} can be approximately related to the first variation of the local electronic self-energy with respect to the magnetization. This fact suggests that the formation of the local magnetic moment is energetically favorable when this variation is negative, which minimizes the energy of electrons.

Applying the derived criterion~\eqref{eq:s9_condition_C} to interacting electronic systems shows that the local magnetic moment develops at temperatures well above the phase transition to the ordered state~\cite{PhysRevB.105.155151}.
At the same time, the moment can be formed only above a relatively large critical value of the local Coulomb interaction $U$, which for the case of a half-filled single-orbital cubic lattice exceeds half of the bandwidth.
The corresponding result is shown in Fig.~\ref{fig:Section9_phase}, where the blue (dark grey) line corresponds to the N\'eel phase boundary, and the red (light grey) line is obtained from the condition~\eqref{eq:s9_condition_C}.
At low temperatures the red (light grey) line determines the point at which the local magnetic moment disappears.
In the regime of large interactions this is related to Kondo screening~\cite{Hewson_book, PhysRevLett.126.056403}.
At small $U$, the local magnetic moment is destroyed by local spin fluctuations, which corresponds to the regime of valence fluctuations of the Anderson model~\cite{Hewson_book}.
The low-temperature branch of the red (light grey) line splits the ordered phase into two parts, which allows one to distinguish between Slater~\cite{PhysRev.82.538, PhysRevB.94.125144} and Heisenberg regimes of spin fluctuations.

To summarise, the path-integral formalism allows us to derive the bosonic problem~\eqref{eq:s9_Boson_action} that describes spin dynamics of itinerant electronic systems.
The non-local part of this problem gives a general form for all kinds of magnetic exchange interactions.
Upon certain approximations, the derived expression for the bilinear exchange~\eqref{eq:s9_J} reduces to the result that was originally introduced in a completely different framework of DFT.
Apart from deriving the magnetic interactions, the path-integral formalism makes it possible to introduce the equation of motion for spin degrees of freedom.
It was shown that for a relatively large value of the magnetic moment its slow rotational dynamics is described by a standard Landau-Lifshitz-Gilbert equation, and fast Higgs fluctuations can be taken into account by the local non-stationary in time contribution to the bosonic problem.  
Deriving the criterion for the formation of the local magnetic moment completes the path-integral formulation of the theory of magnetism and magnetic interactions.

\subsubsection{Application to the adatom system Si(111):Pb}

\begin{figure}[b!]
    \centering
    \includegraphics[width=0.7\linewidth]{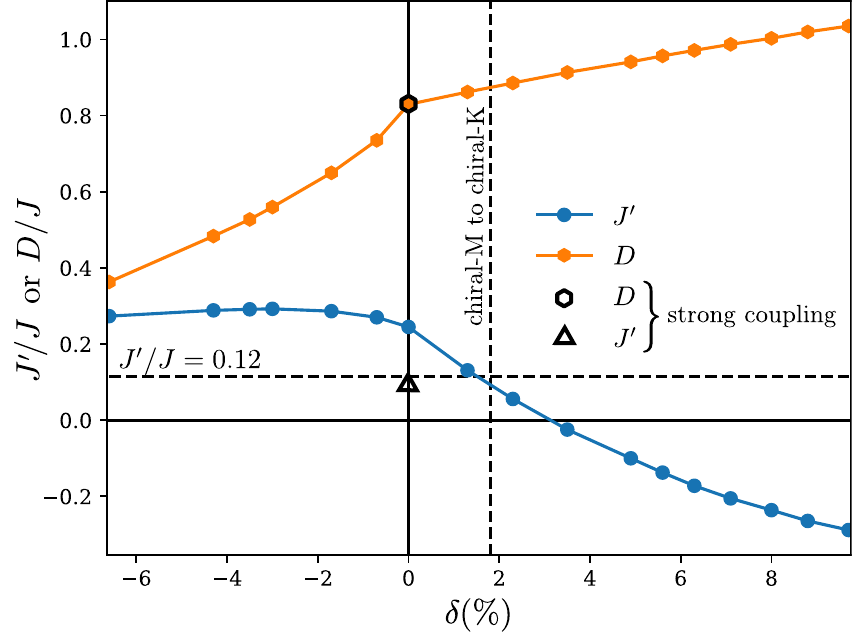}
     \caption{Magnetic exchange interactions as a function of doping. The orange line depicts the value of the nearest-neighbor Dzyaloshinskii-Moriya interaction $D/J$. The blue line corresponds to the next-nearest-neighbor exchange interaction $J'/J$. Both quantities are normalized by the value of the nearest-neighbor exchange $J$. The results are obtained at ${T=50 \; {\rm K}}$. The black hexagon and the black dot represent the values of $D/J$ and $J'/J$ obtained at half-filling in Ref.~\cite{PhysRevB.94.224418} using the strong-coupling approximation. The vertical dashed black line at ${\delta=1.8 \%}$ indicates the transition from the \mbox{chiral-M} to the \mbox{chiral-K} phases according to our calculations. The horizontal dashed line at ${J'/J=0.12}$ represents the prediction for the M to K transition in the $J$-$J'$ Heisenberg model obtained from Monte Carlo calculations in Ref.~\cite{Ramazanov2011}. The Figure is taken from Ref.~\cite{vandelli2024doping}.}
    \label{fig:exchange_int}
\end{figure}

In Section~\ref{sec:SiPb}, we discussed the phase diagram of the adatom system Si(111):Pb.
We found that the system hosts various charge- and spin-ordered phases, revealed by calculating the corresponding susceptibilities using \mbox{D-TRILEX}.
The formation of spin-ordered states in this system can alternatively be investigated by evaluating the exchange interaction, as discussed in Section~\ref{sec:Exchange}.
To this aim, we consider the following effective Heisenberg-like classical spin Hamiltonian with bilinear magnetic exchange interactions:
\begin{align}
H = J\sum_{\langle ij \rangle}{\bf S}_i \cdot {\bf S}_j + J'\sum_{\langle \langle ij \rangle \rangle}{\bf S}_i \cdot {\bf S}_j + D\sum_{\langle ij\rangle} {\bf z} \cdot \left({\bf S}_i \times {\bf S}_j\right)\,.
\label{eq:Hspin}
\end{align}
In this expression, $J$ and $J'$ are the nearest-neighbor ${\langle ij \rangle}$ and the next-nearest-neighbor ${\langle\langle ij \rangle\rangle}$ exchange interactions, respectively.
$D$ is the nearest-neighbor Dzyaloshinskii-Moriya interaction (DMI), which appears due to the spin-orbit coupling. 
We have also calculated the symmetric anisotropy, but we omit it for simplicity as it hardly affects the following considerations. 
The value of its only non-zero component is ${\Gamma_{yy} \approx 0.5D}$ in the whole range of $\delta$ considered here.    

Fig.~\ref{fig:exchange_int} shows the evolution of $J'$ and $D$, normalized by the value of $J$, as a function of doping.
Remarkably, we find that the magnitude of $D$ in Si(111):Pb is of the order of the nearest-neighbor exchange interaction $J$, which is very unusual for magnetic systems.
Moreover, $D$ and $J$ even become equal in the electron-doped case.
At half-filling the value of $D/J$ coincides with the one obtained in Ref.~\cite{PhysRevB.94.224418} using the strong-coupling approximation.
This fact confirms that the half-filled Si(111):Pb material lies in the strong-coupling regime.
Further, we observe that the ratio $D/J$ has an approximately linear dependence on doping with different slopes in the hole- and electron-doped regimes. 
In the hole-doped case, $D/J$ substantially decreases upon increasing the doping.
Instead, in the electron-doped regime, $D/J$ slowly increases with increasing $\delta$. 
This behavior explains the formation of the chiral SDW orderings in the regime of doping levels ${\delta\gtrsim-7\%}$, where DMI is rather strong (${D/J\gtrsim0.4}$) to be able to shift the Bragg peaks in the spin structural factor from a commensurate to an incommensurate position.

While DMI is responsible for the formation of chiral spin structures, the change in the ratio $J'/J$ with doping explains the transformation of the magnetic ordering from the M- to K-type, as observed in our calculations. 
The magnitude of $J'$ is rather small compared to $J$ and $D$, but it is not negligible.
In addition, we find that the actual value of the more distant, next-nearest-neighbor exchange interaction $J'$ is substantially larger than the one predicted by a strong-coupling estimate~\cite{PhysRevB.94.224418}.
An important feature is that the ratio $J'/J$ is nearly constant in the hole-doped regime,
while in the electron-doped case it substantially decreases and even changes sign.
We attribute this variation of $J'/J$ to the shift of the Bragg peaks in the spin structural factor from M to K, which is consistent with Monte Carlo calculations for the $J$-$J'$ Heisenberg model on a triangular lattice performed in Ref.~\cite{Ramazanov2011}.
It has been shown there, that the transition from a row-wise (${{\bf Q}=\text{M}}$) to a N\'eel (${{\bf Q}=\text{K}}$) magnetic order occurs for ${J'/J \simeq 0.12}$.
As shown in Fig.~\ref{fig:exchange_int}, this result coincides with our estimate for the transition point between the \mbox{chiral-M} to \mbox{chiral-K} SDW orderings.
In this figure, the horizontal dashed black line depicts the ${J'/J = 0.12}$ value, and the vertical dashed black line marks the mean-point between the closest doping levels that correspond to \mbox{chiral-M} and \mbox{chiral-K} SDW orderings.

\subsection{Effective exchange interaction for charge degrees of freedom}
\label{Section10}

Another interesting feature of correlated materials that can be potentially described by a corresponding bosonic model is charge ordering.
In electronic systems this phenomenon attracts a considerable attention since the discovery of the Verwey transition in magnetite Fe$_3$O$_4$~\cite{VERWEY1941979, doi:10.1063/1.1746466, Mott}.
Further, effects similar to the Verwey transition have been observed in many other materials, such as the rare-earth compound Yb$_4$As$_3$~\cite{PhysRevB.71.075115, 0295-5075-31-5-6-013, doi:10.1080/0001873021000057114}, transition metal MX$_2$~\cite{PhysRevB.89.235115, ritschel2015orbital, ugeda2016characterization} and rare-earth R$_3$X$_4$~\cite{FURUNO1988117, IRKHIN199047, doi:10.1080/01418638008221893} chalcogenides (${\rm M = V, Nb, Ta}$; ${\rm R = Eu, Sm}$; ${\rm X = S, Se}$), Magn\'eli phase Ti$_4$O$_7$~\cite{doi:10.1080/01418638008221887, doi:10.1080/01418638008221888, EYERT2004151, 0953-8984-18-48-022}, vanadium bronzes Na$_x$V$_2$O$_5$ and Li$_x$V$_2$O$_5$~\cite{doi:10.1080/0001873021000057114, doi:10.1080/01418638008221890}. 
In these materials the charge ordering is driven by the strong non-local Coulomb interaction and/or the electron-phonon mechanism.
Both these interactions effectively reduce the strength of the local Coulomb repulsion~\cite{PhysRevLett.111.036601, PhysRevB.94.165141, PhysRevB.52.4806, PhysRevLett.94.026401, PhysRevLett.99.146404} and may even result in an effective attraction between electrons.
Describing these effects in the framework of {\it ab initio} electronic models requires to use very advanced many-body approaches, such as the quantum Monte-Carlo technique~\cite{PhysRevB.90.085146, PhysRevB.89.205128, Buividovich:20174n}, the $GW$ method combined with the extended dynamical mean-field theory~\cite{PhysRevB.87.125149, PhysRevB.95.245130}, the dynamical cluster approximation~\cite{PhysRevB.95.115149, PhysRevB.97.115117, PhysRevB.99.245146}, or the dual theories~\cite{PhysRevB.90.235135, PhysRevB.94.205110, van2018competing, PhysRevB.102.195109, stepanov2021coexisting}.
These theoretical calculations require significant numerical efforts, which additionally motivates reformulating the original electronic problem in terms of effective bosonic variables.

Similarly to magnetism, the charge ordering is characterised by the local order parameter, the onsite electronic density. This ordering appears as the result of a spontaneous symmetry breaking of a discrete lattice symmetry contrary to the case of a magnetic ordering, which is associated with breaking of a continuous $SU(2)$ symmetry.
For this reason, effective models formulated in terms of scalar bosonic variables are more suitable for addressing this problem.
In particular, Ising-like models are frequently used for describing the ordering in alloys~\cite{PhysRevB.70.125115, PhysRevB.72.104437, PhysRevB.79.054202, PhysRevLett.105.167208, PhysRevB.83.104203}.
In this framework, one deals with a configuration energy written in terms of effective interactions $V^{(n)}_{\alpha}$ for clusters of order $n$ and type $\alpha$.
For the case of a binary alloy $A_{c}B_{1-c}$ with the concentration $c$ the configuration energy can be written as:
\begin{align}
H_{\rm conf} &= \sum_{p}V^{(2)}_{p}\sum_{i,j\in{}p}\sigma_{i}\sigma_{j} 
+ \sum_{t}V^{(3)}_{t}\sum_{i,j,k\in{}t}\sigma_{i}\sigma_{j}\sigma_{k}
+ \sum_{q}V^{(4)}_{q}\sum_{i,j,k,l\in{}q}\sigma_{i}\sigma_{j}\sigma_{k}\sigma_{l} + \ldots\,,
\end{align}
where scalar variables $\sigma_{i}$ take the value $-1$ or $+1$ depending whether $A$ or $B$ atom occupies the site $i$.
Parameters for this microscopic model can be derived from {\it ab initio} energy calculations within the framework of DFT~\cite{PhysRevB.27.5169, 0305-4608-13-11-017, ducastelle1991order, 0034-4885-71-4-046501}.
To this aim, one can apply, e.g., a generalized perturbation theory~\cite{doi:10.1080/00318087508228689, Gautier_1975, Giner_1976, ducastelle1976generalized, Treglia_1978, Ducastelle_1980, PhysRevB.36.4630, doi:10.1080/13642819708205703}.
In this approach effective cluster interactions $V^{(n)}_{\alpha}$ can either be obtained by calculating the corresponding $n$-point correlation functions (see, e.g., Refs.~\cite{PhysRevB.66.024202, PhysRevB.83.104203}) or from the single-electron energy using the force theorem~\cite{mackintosh1980electrons}.  
In the latter case, the variation of the concentration of atoms of a given kind is considered as a perturbation. This seems to be very different from consideration of small spin rotations, that has been used successfully in the case of magnetism. Nevertheless, the resulting pair interaction between sites $j$ and $j'$ is given by the expression:
\begin{align}
V^{(2)}_{jj'} = -\frac{2}{\pi}  {\Im}\int^{E_{F}}_{-\infty} dE  \Delta{}t_j \, \tilde{G}_{jj'}(E) \, \Delta{}t_{j'} \, \tilde{G}_{j'j}(E)\,,
\label{eq:s10_exchange_alloys}
\end{align}
which very closely resembles the magnetic exchange interaction~\eqref{eq:s9_J_w0}, which can be derived using the magnetic force theorem. 
Here, ${\Delta{}t_{j} = (t^{A}_{j} - t^{B}_{j})/2}$ is the difference between single-site scattering matrices for $A$ and $B$ type of atoms, and $\tilde{G}_{jj'}(E)$ is the partial interatomic Green's function of the reference system provided by a random alloy. 

As in the case of magnetism, using the force theorem does not allow one to rigorously determine limits of applicability of the theory.
In this regard, deriving effective Ising-like models in the many-body framework should be beneficial.
In the context of interacting electronic problems this has been achieved in Refs.~\cite{PhysRevB.99.115124, PhysRevB.105.155151}.
The corresponding derivation was discussed in Section~\ref{sec:Heisenberg_Ising} leading to an effective bosonic problem~\eqref{eq:s9_Boson_action}.
It is important to note that introducing the bosonic model for charge degrees of freedom does not require imposing the adiabatic approximation that separates time- and energy-scales of single- and two-particle fluctuations in the magnetic case~\cite{PhysRevB.99.115124}.

All possible interactions between the electronic densities at different lattice sites can be obtained by expanding the logarithm in Eq.~\eqref{eq:s9_Boson_action} in terms of the bosonic field $\rho^{c}$ that describes fluctuations of the charge densities $n$ around their average values.
The explicit form for the pair interaction is given by Eqs.~\eqref{eq:s9_J} and~\eqref{eq:s9_exch}. 
The tree-point vertex function $\Lambda^{c}$ that enters the kinetic exchange~\eqref{eq:s9_J} represents a remormalized local coupling between electronic and charge degrees of freedom.
Thus, this vertex can be seen as a single-site scattering matrix, which makes the many-body expression for the exchange interaction~\eqref{eq:s9_J} very similar to the pair cluster interaction derived in the context of alloys~\eqref{eq:s10_exchange_alloys}.

Mapping the quantum bosonic problem for electronic densities~\eqref{eq:s9_Boson_action} onto a classical Ising-like model can be justified only in the regime of well-developed charge fluctuations.
In a broken symmetry (charge ordered) phase, the electronic density at a given lattice site strongly differs from the average density of the system.
This allows one to replace the bosonic variable $\rho^{c}_{j}$ at each site $j$ by its average value $\langle \rho^{c}_{j} \rangle$, which reduces the quantum bosonic action~\eqref{eq:s9_Boson_action} to a classical Ising-like Hamiltonian.
In the normal phase the average density on each lattice site is uniform, which makes it difficult to introduce the corresponding classical problem and complicates determining the regime of applicability of this approach.

\begin{figure}[t!]
\centering
\includegraphics[width=0.7\linewidth]{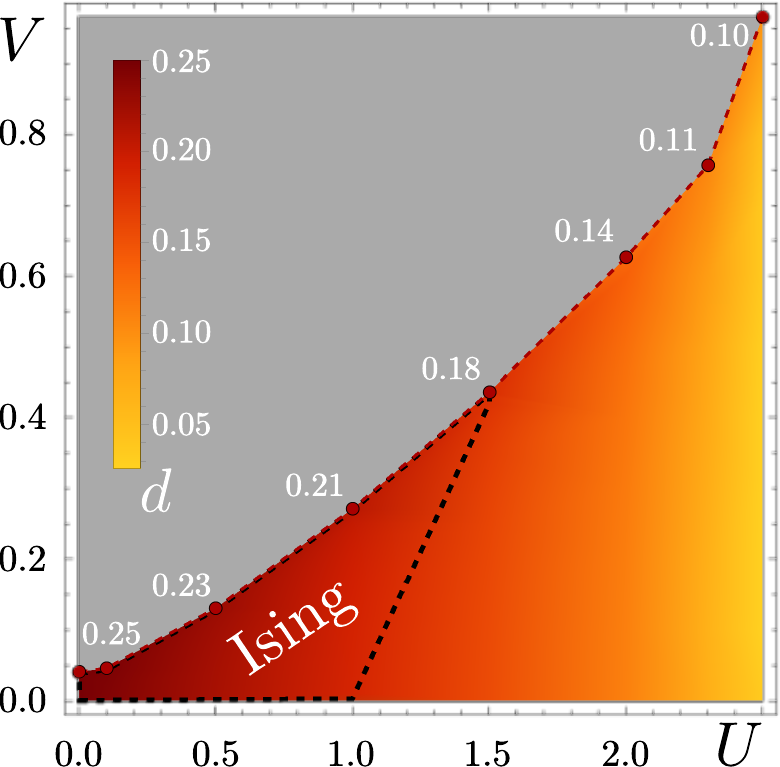}
\caption{Double occupancy of the extended Hubbard model shown on the $U$-$V$ phase diagram. Calculations are performed in the normal phase, where the value of the double occupancy $d$ is depicted by color. The (light) gray color depicts the charge ordered phase. The black dashed line surrounds the area of the large double occupancy $d\gtrsim70\%\,d_{\rm max}$, where charge excitations can be described by an effective Ising model.
Values of Coulomb interactions $U$ and $V$ are given in units of half of the bandwidth $4t=1$, $t$ is the nearest-neighbor hopping amplitude. The inverse temperature was set to ${T^{-1}=50}$. The Figure is taken from Ref.~\cite{PhysRevB.99.115124}. 
\label{fig:Section10_phase}}
\end{figure}

In Ref.~\cite{PhysRevB.99.115124} the double occupancy $d=\langle n_{\uparrow} n_{\downarrow} \rangle$ of the lattice site was proposed as a measure of the strength of the charge fluctuations in the normal phase. 
The double occupancy for a particular case of the extended Hubbard model on a square lattice is shown in Fig.~\ref{fig:Section10_phase}.
The result is obtained at half filling, where the maximum value of the double occupancy is ${d_{\rm max}=0.25}$.
In this model the charge ordered phase (light grey area) is driven by the nearest-neighbor Coulomb interaction, $V$.
If the latter defeats the onsite Coulomb repulsion, $U$, the electronic density forms a checkerboard pattern on the lattice made of alternating doubly occupied and empty sites. 
For a given value of $U$ the maximum value of the double occupancy appears at the boundary between the normal and ordered phases, depicted by a dashed red line.
This fact confirms that the strongest charge fluctuations in the normal phase emerge in the region close to the phase transition to the ordered state.
However, the value of the double occupancy is not uniformly distributed along the phase boundary and decreases with the increase of the local Coulomb interaction. 
It has been shown in Ref.~\cite{PhysRevB.99.115124} that strong charge fluctuations drastically suppress the frequency dependence of the effective local electron-electron interaction (two-particle irreducible four-point vertex function).
The value of the double occupancy at which the effective local interaction is nearly frequency independent and coincides with the actual Coulomb interaction, $U$, was estimated as $d\gtrsim70\%\,d_{\rm max}$.
This condition defines the Ising regime of the system, depicted by the black dashed line in Fig.~\ref{fig:Section10_phase}, where charge fluctuations are indeed well-developed. Interestingly, this regime is not limited to small values of the local interaction, $U$, that for some values of $V$ exceed half of the bandwidth.

In the Ising regime of the normal phase the quantum action~\eqref{eq:s9_Boson_action}, expanded to the second order in the charge bosonic field $\rho^{c}$, can be mapped onto an effective classical Ising Hamiltonian:
\begin{align}
{\cal S}_{\rm charge} = \frac12 \int d\tau \sum_{jj'} \rho^{c}_{j\tau} \, {\cal I}^{c}_{jj'} \rho^{c}_{j'\tau} ~\rightarrow~ H_{\rm Ising} = \frac{1}{2}\sum_{jj'} {\cal J}_{jj'}\sigma_{j}\sigma_{j'}
\end{align}
with ${\cal I}^{c}_{jj'}$ defined in Eq.~\eqref{eq:s9_exch} and the effective Ising exchange interaction ${{\cal J}_{jj'} = {\cal I}^{c}_{jj'}/d}$.
This can be achieved by replacing the charge bosonic variable $\rho^{c}$ by an effective charge density variation, which is given by the square root of the double occupancy ${d = \langle n_{\uparrow} n_{\downarrow} \rangle}$: ${\rho^{c}\to\sqrt{d}}$.
Note that determining the effective charge density can be performed more accurately by finding the minimum of the local free energy in the same way as is done for estimating the value of the local magnetic moment (see discussion in Section~\ref{sec:LMM_formation}).
However, using the two particle correlation function (the double occupancy) to define the average density in the case of charge degrees of freedom is also well justified, contrary to the case of magnetism, where the magnetic phase corresponds to the ordering of single-particle quantities (local magnetizations). Since charge ordering is realised through the formation of double occupations, one needs to characterize this state from two-particle observables.

In Ref.~\cite{PhysRevB.99.115124}, we showed that the critical temperature of the introduced effective Ising model, ${T_{c} = 2{\cal J}/\ln(\sqrt{2}+1)}$, agrees well with the transition temperature between the normal and charge-ordered phases of the extended Hubbard model on a square lattice obtained using much more elaborate dynamical cluster approximation method~\cite{PhysRevB.97.115117}, even though our calculations were performed in the unbroken-symmetry phase.

\subsubsection{Application to Nb$_3$Cl$_8$}

\begin{figure}[b!]
\centering
\includegraphics[width=0.7\linewidth]{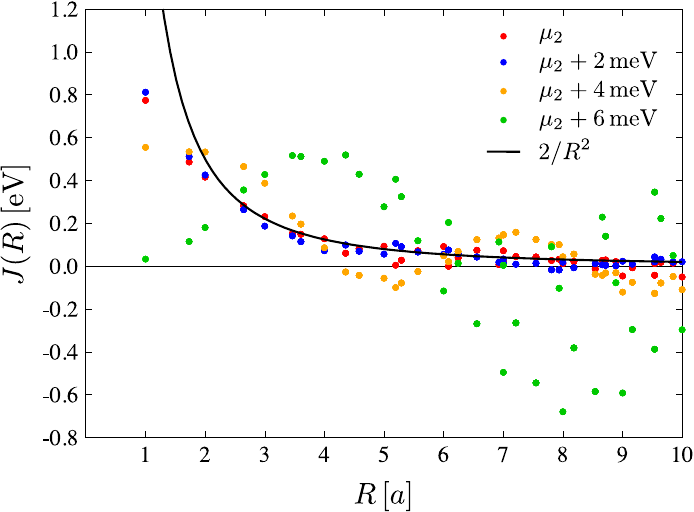}
\caption{Effective charge exchange interaction. The effective exchange interaction between the charge densities ${J(R)}$ calculated as a function of the real space distance $R$ in units of the lattice constant $a$. The results are obtained at ${T=290\,\text{K}}$ for ${\mu=\mu_2+6\,\text{meV}}$ (green), ${\mu_2+4\,\text{meV}}$ (orange), ${\mu_2+2\,\text{meV}}$ (blue) and ${\mu_2}$ (red). The fit function $2/R^2$ is depicted by the solid black line. The Figure is taken from Ref.~\cite{j6bj-gz7j}.  
\label{fig:J}}
\end{figure}

In Section~\ref{sec:Nb3Cl8}, we discussed the formation of a charge-ice state revealed by a distinct bow-tie form of the charge susceptibility calculated using the \mbox{D-TRILEX} approach. 
Remarkably, the fingerprint of this state can also be found in the behavior of the effective exchange interaction between the charge densities.
To this effect, we map the correlated electronic system in the vicinity of the charge instability onto an effective Ising model, as discussed above:
\begin{align}
H_{\rm eff} = \frac12\sum_{jj'}J_{jj'} \, \rho_{j} \, \rho_{j'}\,,
\label{eq:Ising}
\end{align}
where ${\rho_{j}=n_{j}-N}$ is the difference of the electronic density on the site $j$ from the average occupation of the system $N$.
The effective charge-exchange interaction ${J_{jj'} = {\cal I}^{c}_{jj'}}$, calculated from Eqs.~\eqref{eq:s9_J} and~\eqref{eq:s9_exch} near the phase separation (PS) point corresponding to the chemical potential $\mu_2$, is shown in Fig.~\ref{fig:J} as a function of the real-space distance $R$. 
We find that further from the PS, at ${\mu=\mu_2+6}$\,meV (green) the charge exchange interaction is highly frustrated.
It has both, positive and negative values and the amplitude of $J(R)$ does not decay even at the distance of 10 lattice constants $a$. 
Reducing the chemical potential to ${\mu=\mu_2+4}$\,meV (orange) suppresses the frustration and the long-range tail of the charge-exchange interaction.
Remarkably, only in a close proximity to the PS (blue and red), the long-range tail of the charge-exchange interaction can be accurately fitted by the power law dependence of the distance ${J(R)\simeq2/R^2}$, as depicted by the black line.
This result is consistent with the behaviour of the charge susceptibility that reveals the bow-tie structure only at the PS boundary (Figure~\ref{fig:X_NbCl}).
We note that considering this power law form for the dipolar coupling between the magnetic moments was found important for modeling the spin-ice behaviour in Ising pyrochlore magnets~\cite{PhysRevLett.83.1854, PhysRevLett.84.3430, doi:10.1126/science.1064761, PhysRevLett.101.037204, castelnovo2008magnetic,  jaubert2009signature, Bramwell_2020}. 
On the other hand, in some cases the power-law correlations between the magnetic moments in real space can also be realized by considering only the nearest-neighbour exchange interaction~\cite{doi:10.1126/science.1064761}. 
Therefore, it is remarkable that the formation of the charge-ice state is associated with the power law form of both the correlations between the charge densities, reflected in the bow-tie form of the charge susceptibility, and the effective charge-exchange interaction.

\newpage
\section{Conclusions and Outlook}
\label{sec:Conclusions}

We have shown that dual theories provide a powerful and versatile framework for addressing a wide range of correlated electronic problems. 
Originally developed to obtain accurate solutions of the single-band (extended) Hubbard model within the dual fermion (boson) approximation, the approach has since been extended to equilibrium multi-band systems and time-dependent single-band problems through the \mbox{D-TRILEX} and \emph{D-GW} frameworks, respectively. 
Moreover, the dual formalism enables an efficient description of spin and charge dynamics in terms of effective Heisenberg- and Ising-like models, offering a simplified yet accurate treatment of charge and spin subsystems in regimes where the original interacting electronic problem is prohibitively difficult to solve.

The dual techniques rely on introducing a suitable interacting reference system. 
Integrating out this system renormalizes the original lattice quantities by the corresponding correlations of the reference problem, thereby reducing the effective strength of correlations in the resulting theory. 
This procedure enables a consistent separation of local electronic correlations from spatial collective fluctuations, allowing both to be treated appropriately. 
Local correlations, which often give rise to important non-perturbative many-body effects, are fully captured by the exactly solvable reference system. 
In contrast, the dominant non-local collective fluctuations can often be represented by the relevant bosonic modes (charge, spin, orbital, etc.), which are efficiently handled within diagrammatic approximations. 
This framework is implemented rigorously by transforming the original electronic degrees of freedom into effective dual fermionic and bosonic variables. 
We have shown that this transformation is flexible and can be adapted to the specific problem at hand. 
Within the DF/DB approach, the dual variables do not carry direct physical meaning but are introduced in a form that facilitates numerical implementation. 
In \mbox{D-TRILEX}, the specific form of the dual bosonic fields is determined by the transformation that implements the partially bosonized approximation for the four-point vertex function.
Finally, in effective exchange models, the bosonic fields are introduced so as to represent physical charge and spin degrees of freedom directly.

The developed approach represents one of the most advanced frameworks for describing non-local correlation effects in quantum many-body systems. 
At the same time, there remains room for further improvement, and several directions for such developments can be identified:

The DF/DB approach is the most accurate among the dual theories, but it also has the most complex diagrammatic structure, which limits further developments. 
In particular, several attempts have been made to extend the method to a parquet set of diagrams~\cite{PhysRevB.101.075109, PhysRevB.101.165101, PhysRevB.102.195131, PhysRevB.102.235133}, which captures the complex scattering of bosonic modes among themselves, but the resulting schemes are computationally prohibitive.
Considering a small cluster of atoms as a reference system combined with ladder-type diagrammatics is also computationally demanding~\cite{hafermann2008, PhysRevB.84.155106, PhysRevB.97.125114, BRENER2020168310}, since it requires solving the Bethe-Salpeter equation in the multi-band context.
In this regard, a promising direction is to use interacting lattices with tuned parameters -- for instance, the half-filled particle-hole symmetric lattice problem employed in the DF lattice DQMC scheme~\cite{DFQMC} -- as the reference system, which can be solved exactly for relatively large lattice sizes due to the absence of a sign problem.
In this case, the dual perturbation expansion becomes perturbative in terms of the original variables, making it sufficient to restrict the diagrammatic series to the lowest-order terms. 
An important next step would be to incorporate non-local Coulomb interactions into the existing DF lattice DQMC method, which is currently limited to local interactions, by adopting the strategy of the DB approach.

The development of \mbox{D-TRILEX} has made it possible to investigate the effects of spatial collective electronic fluctuations in realistic multi-band systems.
Currently, \mbox{D-TRILEX} can self-consistently describe the interplay between charge, spin, and orbital degrees of freedom in the normal (non-symmetry-broken) phase. 
Incorporating non-local vertex corrections consistent with those in the momentum-dependent \mbox{D-TRILEX} self-energy, which captures electron scattering on charge and spin fluctuations, would enable the study of unconventional superconductivity, nematicity, and transport properties in correlated materials.
Further developments of the cluster extension of the \mbox{D-TRILEX} method are expected to be important for addressing symmetry-broken states, such as charge- and spin-density-wave phases, and for investigating the influence of non-perturbative short-range correlation effects, such as singlet formation, on the system's properties.
Another promising direction is addressing the effects related to the spin-orbit coupling and the electron-phonon coupling, that have already been implemented in the \mbox{D-TRILEX} framework~\cite{vandelli2024doping, stepanov2021coexisting, stepanov2023charge} but were not explored in detail yet. 

The ability of the \emph{D-GW} method to treat different collective electronic fluctuations (in particular, magnetic excitations) on an equal footing makes it one of the most advanced theoretical methods for time-dependent correlated problems to date. 
This approach is especially valuable for transport calculations~\cite{Conductivity}, as it directly computes the corresponding response functions, thereby implicitly including non-local vertex corrections that are otherwise difficult to account for directly.
A promising direction is to extend the method to multi-band systems, in order to study how the interplay between charge, spin, and orbital degrees of freedom -- particularly relevant in the context of Hund's metals~\cite{Hunds_metals1, Hunds_metals2} -- is affected by time-dependent perturbations.
The main bottleneck of this approach lies in existing real-time impurity solvers (see, e.g., NESSi package~\cite{SCHULER2020107484}), which currently lag behind equilibrium continuous-time quantum Monte Carlo techniques~\cite{PhysRevB.72.035122, PhysRevLett.97.076405} due to the extreme complexity of the real-time problem.
Furthermore, a more accurate description of spatial collective electronic fluctuations out of equilibrium would require incorporating the exact fermion-boson vertex functions of the reference problem, which are currently treated using an instantaneous approximation within the \emph{D-GW} approach.
We believe that both challenges can be addressed using a novel tensor train technique~\cite{doi:10.1137/090752286, DOLGOV2020106869, PhysRevX.12.041018, 10.21468/SciPostPhys.18.3.104}, which has already been implemented in the context of strong-coupling impurity solvers~\cite{eckstein2024, PhysRevB.111.125120} and in the compression of vertex functions~\cite{PhysRevX.13.021015, PhysRevResearch.7.023087, frankenbach2025compressinglocalvertexfunctions}.
The \emph{D-GW} approach has great potential for analyzing ultrafast excitations and studying emergent states of matter, such as light-induced magnetism and superconductivity.
Furthermore, the method can be extended to compute more complex response functions beyond conductivity, which is essential for interpreting Raman and resonant inelastic X-ray scattering (RIXS) data~\cite{PhysRevB.106.165106, PhysRevB.111.045147, Werner_2021, PhysRevB.103.115136}.

The methodology for deriving effective exchange interactions within the dual framework is well established. 
The next step is to apply it to realistic materials and compare the results against conventional DFT and DMFT calculations. 
While the exchange interactions obtained within the dual approach reduce to those of DFT or DMFT in certain limits, such as the atomic limit or the highly spin-polarized case, they provide a more general and accurate description in correlated systems, capturing the renormalization of Green's functions and three-point vertex functions by non-perturbative local correlations.

\newpage
\section{References}
\bibliographystyle{apsrev4-1}
\bibliography{Ref.bib}

\end{document}